# "O Desenvolvimento do Heliômetro do Observatório Nacional e aplicação ao estudo do Sistema Sol-Terra"

Eugênio Reis Neto

TESE SUBMETIDA AO CORPO DOCENTE DO PROGRAMA DE PÓS-GRADUAÇÃO EM GEOFÍSICA DO OBSERVATÓRIO NACIONAL COMO PARTE DOS REQUISITOS NECESSÁRIOS PARA OBTENÇÃO DO GRAU DE DOUTOR EM GEOFÍSICA.

Orientador: Dr. Alexandre Humberto Andrei

À minha família e amigos, que sempre me apoiaram e

à memória de meu pai

.

# Agradecimentos

Ao professor e amigo Dr. Alexandre Humberto Andrei, pela orientação desta tese, pelos ensinamentos, pela amizade, pelos momentos divertidos e, principalmente, por ter sempre acreditado em minha capacidade de conduzir este projeto.

Ao professor e amigo Dr. Victor de Amorim D'Ávila, pela amizade, companheirismo, e por toda a sua capacidade e conhecimento técnico imprescindíveis para este projeto instrumental.

Um obrigado muito especial para a querida amiga Dra. Jucira Lousada Penna, por seu carinho, amizade e apoio ao longo de todos estes anos de convívio no Observatório Nacional.

À caríssima Dra. Adriana Válio Roque da Silva, por sua valiosa orientação no começo do meu doutorado e por sua compreensão e amizade.

Ao colega e amigo Sergio Calderari Boscardin por toda a colaboração e por estar sempre pronto a me ajudar.

Aos talentosíssimos Luiz Carlos Oliveira, Alissandro Coletti, Victor Matias, Ricardo Dunna, José Carlos Diniz, Sergio Lomonaco e a todos os integrantes do NGC-51, por sua inestimável ajuda desde o início do projeto, além da amizade e a Marcomede Rangel Nunes por ter me apresentado ao grupo.

Agradeço também à Coordenação de Pós-Graduação do Observatório Nacional e seus integrantes por toda a ajuda e apoio, em especial ao Dr. Andres Reinaldo Rodriguez Papa.

Gostaria de agradecer à Coordenação de Aperfeiçoamento de Pessoal de Nível Superior (CAPES) pela concessão da bolsa de doutorado, ao Museu de Astronomia e Ciências Afins (MAST), ao Espaço Ciência Viva, à Fundação de Apoio ao Desenvolvimento da Computação Científica (FACC), à Financiadora de Estudos e Projetos (Finep), por ter dado o suporte financeiro para o desenvolvimento e construção do Heliômetro do Observatório Nacional e ao Conselho Nacional de Desenvolvimento Científico e Tecnológico (CNPq), por ter me concedido bolsa DTI para trabalhar no início deste projeto.

Um agradecimento especial ao Observatório Nacional pelo apoio institucional, bem como pela ajuda de todos os seus pesquisadores, técnicos e colegas da pós-graduação.

Por fim, um agradecimento aos meus colegas de pós-graduação do Centro de Rádio Astronomia e Astrofísica Mackenzie (CRAAM).

Nada te perturbe,
nada te amedronte.
Tudo passa,
a paciência tudo alcança.
A quem tem Deus nada falta.
Só Deus basta!

*Santa Teresa D´Ávila*

A vida sem humor não tem graça

*Eug*

# Resumo

O presente trabalho traz o desenvolvimento e a construção do Heliômetro do Observatório Nacional. Com este instrumento se visa monitorar as variações do diâmetro solar, de modo comensurável com a precisão das observações embarcadas na próxima geração de satélites solares, e com duração com o ciclo solar de 11 anos.

Uma revisão do método heliométrico é feita e são construídos e testados protótipos de 4 diferentes implementações. O instrumento definitivo tem uma objetiva em diedro, formado pela hemi-secção de um espelho parabólico. Toda uma metodologia própria foi criada, desde a confecção do espelho até a mecânica do telescópio. Os materiais que compõem o instrumento têm estabilidade térmica e mecânica em dex-7. A quantidade de peças ópticas é minimizada e a qualidade de suas superfícies é superior a $\lambda/12$. Foram desenvolvidos softwares originais de coleta e análise automática das imagens. O projeto foi desenvolvido através de acréscimos experimentais sobre um heliômetro de testes. Com sua última versão completamente desenvolvida, foi realizada uma campanha observacional de 9 dias, derivando mais de 70.000 imagens heliométricas do Sol. Estes resultados indicam precisão de 0,5 segundos de arco, sem viés instrumental, e limitada pela modelização atmosférica. Portanto, como objetivado, uma acurácia de 0,005 segundos de arco pode ser atingida.

Para o projeto, a antiga cúpula da luneta foto-equatorial, no campus do Observatório Nacional e cedida pelo Museu de Astronomia e Ciências Afins, foi inteiramente reformada e adaptada para a utilização do novo instrumento solar. A reforma é atestada pelos órgãos de preservação do patrimônio.

Um estudo da inter-relação entre o diâmetro solar e as medidas geomagnéticas foi feito. Uma correspondência entre os picos das séries ligadas à atividade solar, quais sejam, a do semi-diâmetro solar, dos *flares*, e da contagem de manchas e os picos negativos da série da intensidade do campo geomagnético é encontrada. No entanto esta correspondência é complexa, requerendo modos diferenciados de resposta e fase dependendo da etapa do ciclo de atividade solar. A interpretação direta da evidência observacional, indica que o semi-diâmetro solar parece apresentar variações significativas precedendo a correspondentes variações do campo geomagnético. O efeito é mais evidente na fase descendente do ciclo solar.

# Abstract

This work presents the development and construction of the Heliometer of the Observatório Nacional/MCT. This instrument is designed to monitor changes on the solar diameter, to the accuracy of the observations of the next-generation of solar satellites, and on the duration of the 11 years solar cycle.

A review is made of the heliometric method and the building and testing of 4 prototypes is described. The instrument has a split-mirror objective in dihedral formed by the hemi-section of a parabolic mirror. Original methodology was developed, from the making of the mirror to the mechanics of the telescope. The materials that form the instrument have thermal and mechanical stability to dex-7. The number of optical parts is minimized and their quality is greater than $\lambda/12$. Original software for the automated collection and analysis of the images was developed. All along the project was developed by actual experimental increments on a trial heliometer. With its latest version fully developed, we conducted an observational campaign of 9 days, deriving more than 70,000 heliometric images of the Sun These results indicate a precision of 0.5 arcseconds, with no instrumental bias, and limited only by the provisional atmospheric modeling. Therefore, as planned in the beginning, an accuracy of 0.005 arcseconds can be achieved.

The old dome of the photo-equatorial refractor, in the Observatório Nacional campus, owned by Museu de Astronomia e Ciências Afins, was courteously granted to this project, and completely renovated and adapted for the use of the new solar instrument. The reform conforms the demands of the organs of preservation heritage.

A study was made of the correlation between the solar diameter and the geomagnetic field intensity. A correspondence between the peaks of the series related to solar activity, namely, the semi-diameter variation, the flares index, and the sunspots counts, and the number of negative peaks on the intensity of the geomagnetic field is found. It must be cautioned though that it is a complex correspondence, requiring different modes of response and phase depending on the stage of the cycle of solar activity. The straightforward interpretation of observational evidence indicates that the semi-diameter of the Sun seems to vary significantly prior to the corresponding variations of the geomagnetic field. The effect is most evident in the downward phase of the solar cycle.

# Sumário

# Capítulo 1 - Introdução

Esta tese é fundamentalmente de natureza instrumental. Seu progresso, dia a dia, foi centrado quase exclusivamente na tarefa do desenvolvimento e construção do Heliômetro do Observatório Nacional.

Este telescópio não é um instrumento comum e não existem heliômetros à venda no mercado. Para se obter um, é preciso que ele seja construído a partir de um projeto exclusivo.

É sobre o desenvolvimento de um telescópio deste tipo que versa esta defesa. Entretanto é oportuno, como introdução, apresentar um panorama da interação Sol-Terra, no contexto do clima espacial, pois aí reside a motivação científica para o Heliômetro. Com ele serão feitas medidas angulares do diâmetro solar semelhantemente àquelas que serão feitas pelos satélites PICARD e SDO (*Solar Dynamics Observatory*), embora em outra sistemática. Por outro lado, são medidas conceitualmente diferentes daquelas realizadas pelo Astrolábio Solar, as quais são medidas de trânsito. Como se apresenta no escopo desta introdução, as medidas do diâmetro solar, em forma de longas séries temporais, são peça importante para os modelos do Sol. Por conseqüência, permitem agregar de forma significativa ao entendimento do sistema Sol-Terra e, em particular, para as pesquisas realizadas no Observatório Nacional, à modelagem das variações do campo geomagnético, sobretudo associadas a tempestades geomagnéticas.

Há tanto aspectos puramente científicos quanto motivações práticas para produzir uma melhor compreensão das variações do sistema Sol-Terra e identificar suas causas. Além da própria Terra, o Sistema Solar, e mesmo fora dele, constituem o ambiente cósmico em que vive a humanidade. Por isso, a busca pela pesquisa e aplicação de ferramentas tecnológicas para garantir um melhor entendimento deste ambiente é perene.

## *1.1 - Modelo do Sol*

As estruturas do Sol são esquematizadas na figura 1.1, na qual estão representadas as camadas do Sol, a temperatura e a densidade de partículas, a partir do centro.

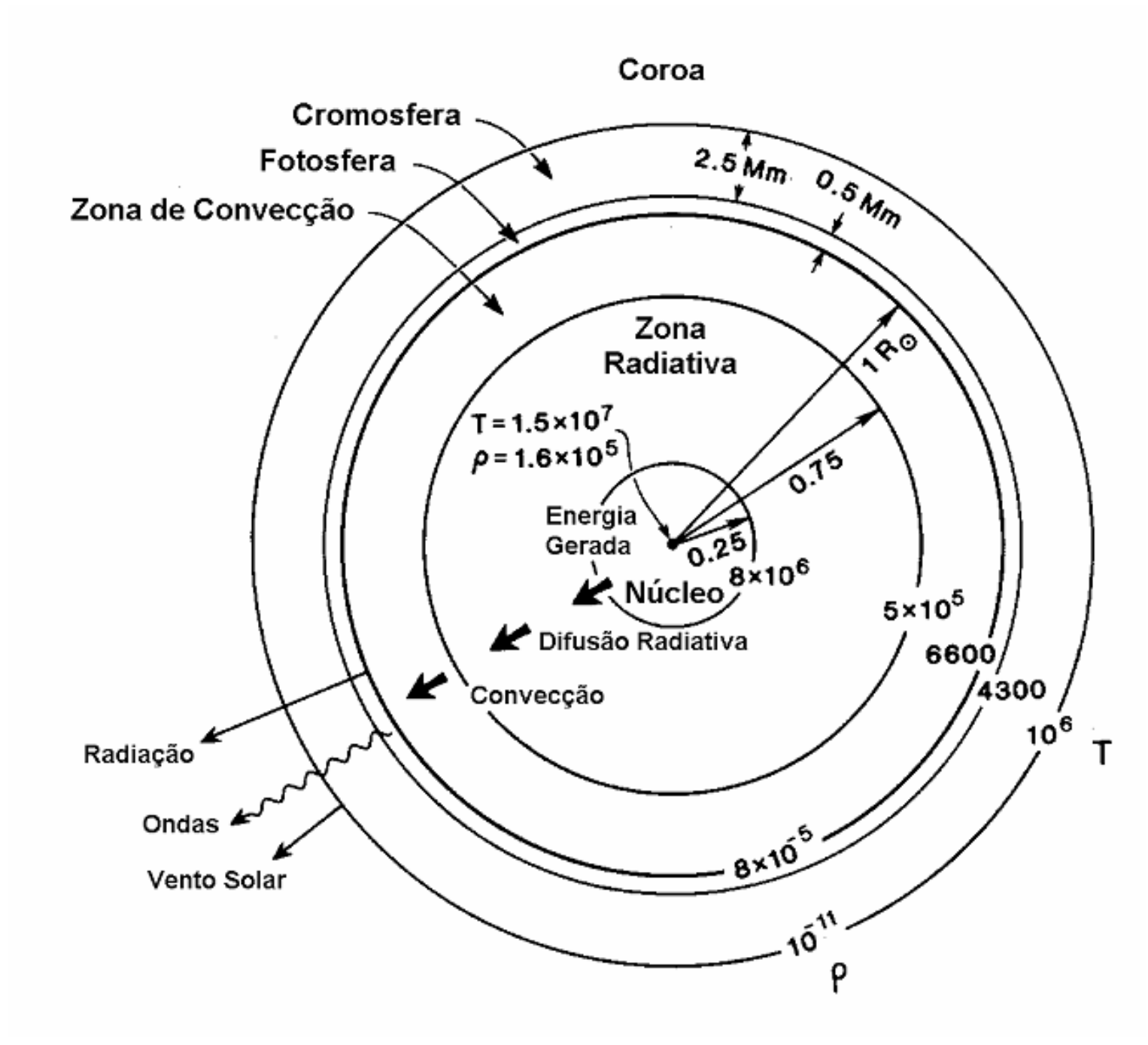

Fig. 1.1: Esquema das camadas do Sol, com suas respectivas temperaturas e densidades (Kivelson e Russell, 1995 *apud* Dal Lago, 1999, p. 32).

O interior é formado pelas seguintes regiões: Núcleo, se estendendo até 0,25 raio solar, onde a energia é gerada através de reação de fusão nuclear, apresentando uma temperatura da ordem de $1,5\times10^7$ K e densidade $\rho\sim1,6\times10^5$ $m^{-3}$; Zona Radiativa, se estendendo no intervalo de 0,25 a 0,75 raio solar, onde a energia gerada no núcleo é dissipada através do processo de difusão radiativa, tendo uma temperatura no intervalo de $8\times10^6$ K a $1,5\times10^6$ K; Zona de Convecção, estendendo-se no intervalo de 0,75 a 1 raio solar, com temperatura no intervalo de $5\times10^5$ a 6600 K e onde a energia é dissipada sob a forma de convecção. A Atmosfera Solar, em função de suas características e processo físicos, é convencionalmente constituída de três grandes regiões: Fotosfera, com espessura da ordem de $0,5\times10^6$ m, ou 0,5 Mm, e temperatura da ordem de 6600 K; Cromosfera, com espessura em torno de 2,5 Mm, temperatura no intervalo de 4300 a $10^6$ K e densidade 10-11$m^{-3}$; Coroa, acima da cromosfera e sem limite exterior, apresenta uma temperatura da ordem de $10^6$ K. A coroa é a região da atmosfera solar de maior importância para os fenômenos físicos do meio interplanetário, pois sua elevada temperatura faz com que os prótons, elétrons e alguns íons, como o $^4He^{++}$, que constituem a Coroa, tenham velocidades térmicas muito altas, superiores à velocidade de escape do Sol e assim se evadam para o espaço interplanetário, na forma do Vento Solar (Dal Lago, 1999**)**.

## *1.2 - Vento Solar*

Estudos detalhados do vento solar só foram possíveis quando satélites artificiais passaram a medir diretamente as propriedades do espaço interplanetário, o que inclui por exemplo medidas da intensidade do campo magnético interplanetário, da densidade e velocidade do vento solar. Entretanto, muito antes disso já havia indícios de um fluxo de partículas emitidas pelo Sol, em particular por meio de medidas de raios cósmicos e pela observação da cauda de cometas que penetravam no Sistema Solar. Outras evidências relacionadas com o vento solar incluem as auroras (Maciel, 2005).

O vento solar se divide em duas componentes: o vento rápido, mais uniforme, com velocidades que atingem os 800 km/s e proveniente dos buracos coronais polares (regiões de baixa densidade, se estendendo a partir dos dois pólos do Sol) e o vento lento, com cerca de 300 km/s, mais irregular, proveniente de estruturas mais próximas do equador solar (Lima, 1999). Estas emissões contínuas confinam e distorcem as linhas de força do campo geomagnético, comprimindo-as no lado iluminado pelo Sol, e formando uma espécie de cauda na direção oposta ao Sol. Elas são moduladas fortemente pela atividade solar, de modo que ejeções energéticas transientes podem lançar grandes quantidades de material constituinte da coroa a altas velocidades, da ordem de 2000 km/s (Plunkett e Wu, 2000). Tais eventos estão fortemente relacionados a fenômenos geomagnéticos, tais como tempestades geomagnéticas, e à variabilidade do clima espacial.

A atividade solar, regulada pelo ciclo de 11 anos, se traduz no crescimento e decrescimento do número de manchas solares.

## *1.3 - Ciclo Solar*

O ciclo solar advém da reorganização da energia do campo magnético solar, à medida que este gradualmente passa de um estado de polarização, decai e, finalmente, passa por uma inversão, quando os pólos magnéticos se intercambiam, ao fim de 11 anos. À medida que as linhas do campo magnético se tornam cada vez mais entrelaçadas durante o máximo solar, os campos magnéticos do Sol se tornam mais inomogêneos. Isso leva a aumentos localizados da atividade na superfície do Sol, representada por mais e mais manchas solares (regiões de campos magnéticos intensos e complexos), levando, ultimamente, a mais freqüente ocorrência de *flares* (explosões solares) e *CMEs* (ejeções de massa coronal). Eventos impulsivos, como ejeções de massa coronal, explosões solares, ou fluxos de alta velocidade no vento solar, ocorrem em escalas de tempo de minutos a horas ou mesmo dias.

Um *flare* é em geral visível como um flash brilhante de raios-X numa região solar ativa. O *flare* dura alguns minutos, mas libera uma quantidade imensa de energia (até $10^{25}$ J liberados entre 100-1000 segundos**)**. Durante os *flares* o Sol pode ser 1000 vezes mais brilhante em raios-X do que em sua intensidade normal. É exatamente com base no fluxo de raios-X que a intensidade dos *flares* é categorizada. *Flares* intensos podem aumentar grandemente os níveis da ionização da Ionosfera da Terra, causando problemas para as comunicações, bem como para mecanismos de navegação em baixa freqüência.

A *CME* ocorre quando uma proeminência acima da superfície do Sol entra em erupção e envia milhões de toneladas de material para o espaço. Esta nuvem de partículas carregadas é geralmente confinada dentro de um campo magnético (ou bolha magnética), expandindo-se e viajando através do Sistema Solar a velocidades de cerca de 200 km/s e até 2000 km/s. Quando dirigida para a Terra uma *CME* normalmente chega de 2-3 dias após a erupção, mas em casos excepcionais, pode chegar em menos de 24 horas. A chegada da CME no ambiente terrestre e a tempestade geomagnética deflagrada podem causar efeitos desastrosos, com falhas de satélite, e *blackouts* de cidades ou regiões inteiras. As *CMEs* podem gerar eventos de partículas energéticas (SEP), em que prótons altamente energéticos podem chegar à Terra em frações de velocidades quase luminais. A energia das partículas atinge até 1 GeV, mas decai em apenas poucas horas. A maioria dos modelos de *CME* supõe que a energia é armazenada no campo magnético coronal durante um longo período de tempo e de repente lançado por alguma instabilidade ou perda de equilíbrio do campo. *Flares* também obtém energia a partir do campo magnético e têm energia suficiente para acionar a *CME*, no entanto apenas 60% dos *CMEs* revelam associação com *flares* (Andrews, 2003), de modo que não há consenso sobre o mecanismo de liberação.

## 1.4 - Tempestades magnéticas

A tempestade geomagnética é uma perturbação temporária da magnetosfera da Terra causada pelo distúrbio no clima espacial, modificando as correntes elétricas da ionosfera.

A queda da intensidade do campo magnético durante a fase principal de uma tempestade geomagnética é normalmente precedida por um breve aumento, causado por uma intensificação da magnetopausa que ocorre com o aumento da pressão do vento solar. Este fenômeno é conhecido como "início súbito" (*Sudden Storm Commencement*) e marca a fase inicial de uma tempestade.

Tempestades geomagnéticas são classificadas como recorrentes e não-recorrentes. Tempestades recorrentes ocorrem a cada 27 dias (período ligado a rotação do Sol) e mais frequentemente na fase de declínio do ciclo solar. Tempestades não-recorrentes, por outro lado, ocorrem mais frequentemente próximo ao máximo solar.

A atividade geomagnética pode ser dividida em duas categorias principais, tempestades e subtempestades. Tempestades são iniciadas quando a transferência de energia a partir do vento solar para a magnetosfera leva à intensificação de correntes elétricas ao longo das linhas do campo.

Define-se a ocorrência de uma tempestade magnética através de índices geomagnéticos. Os mais freqüentemente utilizados para determinar a intensidade e a duração das tempestades geomagnéticas são os índices, *Planetarische Kennziffer* ou índice planetário (Kp), *Auroral Electrojet* (AE) e o *Disturbance Storm-Time* (Dst).

O índice Kp é obtido a cada período de 3 horas (3-h) e é derivado de 13 observatórios em latitudes subaurorais, entre 44° e 60º de latitude, principalmente no Hemisfério Norte. Ele é obtido como o valor médio dos níveis de perturbação obtidos nos 13 observatórios (Prestes, 2002).

O índice AE é derivado das variações geomagnéticas da componente horizontal, observada em observatórios selecionados ao longo da zona auroral no Hemisfério Norte. Ele é obtido através da superposição em mesma escala de amplitude e tempo dos vários magnetogramas obtidos nos diferentes observatórios. As variações são medidas de uma linha de base determinada para cada observatório. O nível de campo magnético de tempo calmo deve ser determinado para cada um deles e subtraído dos valores medidos, deixando assim somente o valor da perturbação (Prestes, 2002).

O índice Dst monitora a variação da corrente de anel, que é uma corrente que circula a Terra próxima ao equador magnético fluindo na direção oeste, no cinturão de radiação de Van Allen na Magnetosfera (Daglis *et al.*, 1999). Como o campo produzido pela corrente de anel equatorial é quase paralelo ao eixo do dipolo, seu efeito é predominante na componente H e mais forte em baixas latitudes (Prestes, 2002). Este índice é um número proporcional ao grau de perturbação do campo geomagnético, obtido por uma cadeia de magnetômetros localizados na região equatorial, ao longo do globo terrestre dado, entre outras unidades, em nanotesla (nT). Antes do início da tempestade, o índice Dst apresenta um pico de intensidade que é conhecido por fase inicial ou início súbito (*sudden commencement*). Após esse pico desenvolve-se a fase principal (*main phase*) da tempestade que é caracterizada pela brusca queda no valor da intensidade do índice. Após alcançar o mínimo, o índice começa a subir de valor, cuja fase é conhecida por fase de recuperação ou *recovery phase*, até atingir aproximadamente o valor quiescente, que ocorre quando não há tempestade (Yamashita, 1999**)**. Estas fases estão mostradas na figura 1.2.

É a partir do índice Dst que é definido a ocorrência das tempestades magnéticas. A classificação de uma tempestade magnética segundo sua intensidade utilizando o índice Dst é mostrada na tabela 1.1.

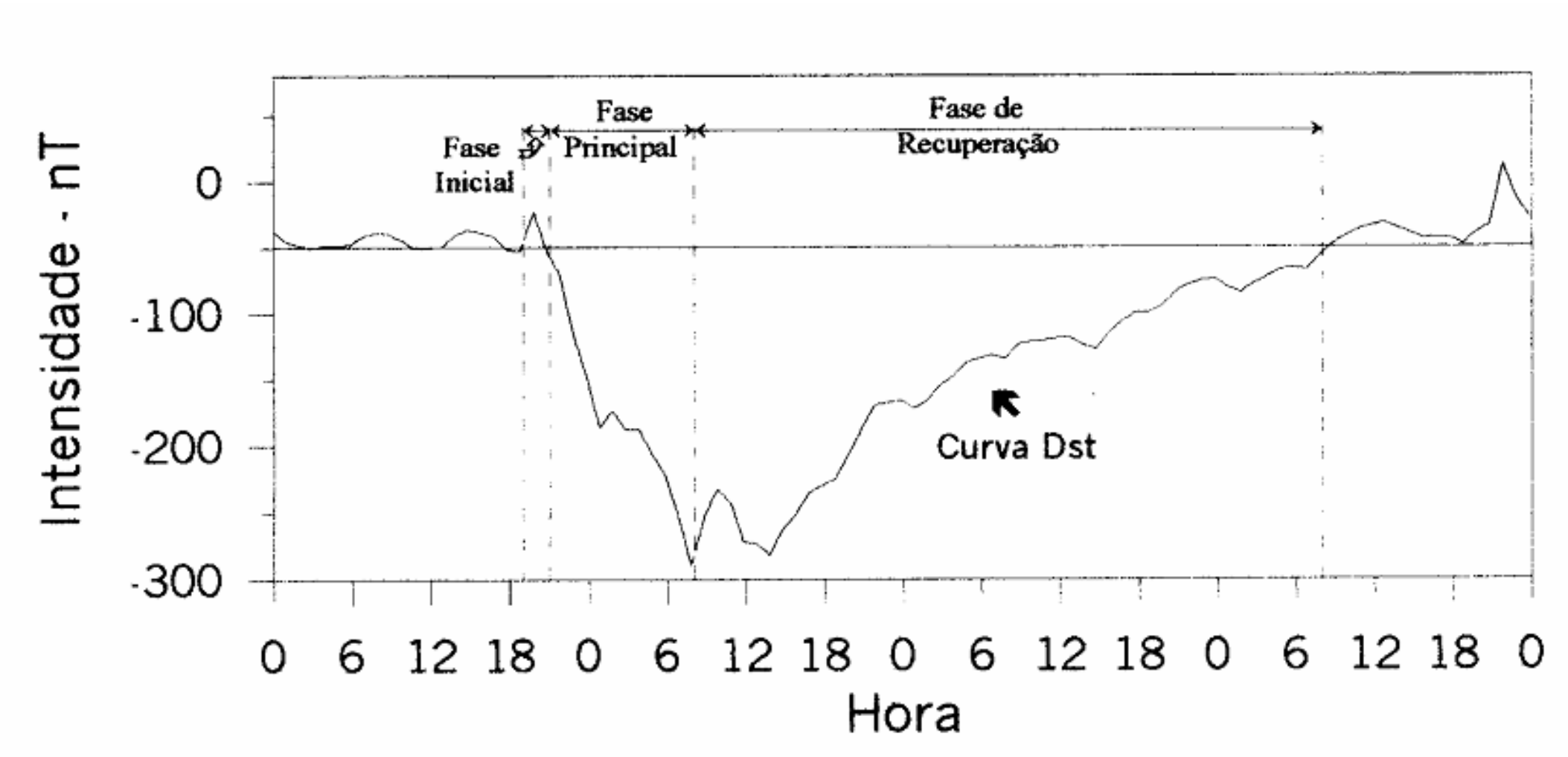

Fig. 1.2: Curva Dst para os dias 5-8 de setembro de 1982 com uma intensa tempestade magnética e suas fases características (Yamashita, 1999).

Tabela 1.1:Classificação de uma tempestade magnética (Fedrizzi, 2003).

| **Condição da Tempestade Geomagnética** | **Valor do índice Dst (nT)** |
|---|---|
| Fraca | -30 nT a -50 Nt |
| Moderada | -50 nT a -100 nT |
| Intensa | -100 nT a -250 nT |
| Muito Intensa | < -250 nT |

## *1.5 - Variações do diâmetro fotosférico*

Uma vez estabelecida a relação entre o diâmetro da fotosfera e a atividade solar, e fazendo a suposição de que a variação do raio solar não é completamente caótica, como demonstrado pelos ciclos solares, conhecendo-se o diâmetro se pode deduzir a luminosidade solar. As diversas medidas do diâmetro solar, observado a partir do solo e em órbita, mostram a importância desta quantidade, porém evidenciam o papel da Atmosfera. Observado a partir do solo, o diâmetro se distribuí por vários décimos de segundo de arco. À parte a dispersão das medidas, a observação concomitante do diâmetro por satélites tornará possível separar as variações de origem solar de pressupostas variações de origem atmosférica. Antes de extrair estas informações sobre efeitos atmosféricos, ao contrário de meramente pressupô-los sem embasamento físico concreto, é necessário compreender todos os processos de alteração, de espalhamento, de absorção e de turbulência. Com o Heliômetro, cuja determinação se baseia numa medida angular, como nos satélites, medições simultâneas do solo e em órbita estão previstas, o que tornará possível estudar o efeito intrínseco de turbulência, levadas em conta as diferenças de abertura efetiva. Será possível estimar a parte da degradação da frente de onda induzida pela turbulência atmosférica, pelo conhecimento da localização das camadas turbulentas responsável pela degradação das imagens caracterizada pelo parâmetro de Fried, a dimensão externa da janela de coerência espacial, o domínio do isoplanatismo, o tempo característico da evolução da frente de onda e o perfil de turbulência.

O esforço investigativo multi-disciplinar nas últimas duas década produziu melhorias significativas na compreensão dos processos físicos dentro de cada um dos domínios Sol-Terra, e permitiu avançar efetivamente e melhor compreender o domínio como um todo. O lançamento de satélites, como YOHKOH, SOHO e UARS, resultou em melhorias sem precedentes na compreensão dos processos físicos no Sol e no vento solar, os quais geram a variável energia solar impactando a Terra. Estas novas observações e os modelos desenvolvidos a partir delas, têm facilitado o progresso na previsão de potenciais perturbações geoespaciais, no impacto sobre os campos magnéticos interplanetário e a Magnetosfera, de choques produzidos por *CMEs*, na variação da composição e atividade química na Mesosfera, sobre as variações de ozônio atribuíveis à radiação solar UV e prótons energéticos e, enfim, das variações da irradiância solar total, possivelmente associadas às alterações climáticas. O papel do domínio ultravioleta é considerado na Estratosfera, e propagado para a Troposfera nos modelos climáticos. Sua variabilidade deve ser medida no que corresponde à fotoquímica do ozônio, frente às condições de temperatura e da dinâmica da Estratosfera. À par dos avanços observacionais, o desenvolvimento de modelos numéricos, auto-consistentes, tridimensionais, e tempo-dependentes, do sistema Termosfera-Ionosfera permitem investigar em detalhes os efeitos do acoplamento dos níveis mais altos e mais baixos da Atmosfera e de suas camadas neutra e ionizada, e confrontar estes modelos com aqueles que descrevem o sistema Sol-Terra.

## *1.6 - Questões em aberto*

Há muitas questões em aberto sobre o sistema Terra-Sol, e sua contribuição para as alterações climáticas globais e no espaço, entendidas de forma mais ampla que os dramáticos, porém pontuais, extremos reportados na mídia. Algumas destas alterações estão relacionadas com a variação temporal de tempestades solares, outras com escalas de tempo mais longas da atividade solar. Que mecanismos conduzem à quase-periodicidade de 11 anos no ciclo de atividade solar? Como o fluxo magnético é sintetizado na região ativa e dispersado por toda a superfície solar? Como a reconexão magnética em pequenas escalas reorganiza a topologia de campo em grande escala e sistemas atuais e quão é significante no aquecimento da coroa e aceleração do vento solar? Onde surgem as variações observadas na irradiância solar e no UV extremo, e como elas se relacionam com os ciclos de atividade magnética? Que configurações de campo magnético levam às *CMEs*, erupções filamentares, e *flares*, para produzir partículas energéticas e radiação? Como se podem ligar os processos que produzem ejeções de massa coronal, com sua transferência através da Heliosfera, sua interação com a Magnetosfera, e finalmente com a produção de tempestades geomagnéticas que afetam a Atmosfera? Como se podem identificar evidências de variações de longo prazo da luminosidade solar relacionada à atividade solar e aos impactos resultantes sobre a Terra, em comparação com outros mecanismos de mudança climática? Pode a estrutura e dinâmica do vento solar perto da Terra ser determinada a partir da configuração do campo magnético e estrutura atmosférica perto da superfície solar? Quando a atividade solar ocorre, é possível fazer previsões precisas e confiáveis sobre sua propagação e a perturbação no clima espacial? Em que medida a Magnetosfera e Ionosfera-Termosfera são moduladas pela atividade solar em diversas escalas de tempo, incluindo o ciclo solar, e como as variações impulsionadas por tão diferentes processos interagem com aqueles de natureza essencialmente terrestre ou mesmo antrópica? Como conciliar as respostas aparentemente não-lineares da média e baixa Atmosfera com a atividade solar, identificar os mecanismos físicos, em comparação com influências antropogênicas, e estimar as futuras alterações de ozônio?

## *1.7 - Motivação Científica*

Estas interrogações deixam claro a relevância científica de contribuir para o acompanhamento sistemático do diâmetro solar, numa escala de tempo comensurável com o ciclo de atividade, a qual, portanto, não pode ser atingida pela presente tecnologia de satélites científicos de monitoramento. Ao mesmo tempo, fica claro que, em relação às medidas atualmente efetuadas com o Astrolábio Solar, se deva atingir um maior patamar de precisão (permitindo maior detalhamento na modelagem de eventuais termos de natureza atmosférica), maior quantidade de medidas (permitindo extrair precisão através de um processo de média em condições semelhantes do ponto de vista observacional e do fenômeno estudado), maior agilidade no recobrimento de heliolatitudes (permitindo endereçar a questão do achatamento e da forma solar), de capacidade para observação mais ágil de regiões ativas (o que é parcialmente realizado pela observação em heliolatitudes), e finalmente utilizando um método diferente daquele do Astrolábio Solar (para permitir comparação dos vieses instrumentais e passar a dispor de duas séries efetivamente independentes).

O Heliômetro alcança todos estes objetivos. Sua descrição detalhada é o objeto do restante desta tese que segue a seguinte estrutura:

- O capítulo 2 traz o histórico do projeto, de um astrolábio aperfeiçoado até a idéia da construção de um heliômetro e de sua fundamentação;
- O capítulo 3 traz a construção de protótipos, com diferentes metodologias, para a escolha da versão final do instrumento.
- No capítulo 4 é mostrado o desenvolvimento da metodologia de confecção.
- O capítulo 5 traz a construção do heliômetro.
- No capítulo 6 estão o desenvolvimento do programa de aquisição e análise das imagens, assim como as primeiras medidas feitas com o heliômetro protótipo.
- No capítulo 7 é apresentado um estudo piloto (objeto de publicação já submetida) da inter-relação entre o diâmetro solar, medido pelo Astrolábio CCD e as medidas geomagnéticas, consubstanciando a aplicação científica que será trazida pelas medidas do Heliômetro do Observatório Nacional.
- As conclusões são apresentadas no capítulo 8.

# Capítulo 2 - Um novo instrumental

## *2.1 - Primeiras idéias*

Desde 1977 o Observatório Nacional desenvolveu um programa de medidas envolvendo o diâmetro do Sol com o Astrolábio Danjon. Aquelas medidas visavam o posicionamento astrométrico da órbita aparente do Sol no referencial terrestre. Ou seja, a definição da eclíptica verdadeira. Como o centro do Sol não é observável pelo Astrolábio Danjon, limbos opostos eram observados, de modo que a média das medidas fornecia o posicionamento buscado do centro solar em relação ao horizonte local. O interesse daquelas medidas quanto à Astronomia Fundamental, em si, foi ultrapassado por outros métodos. No entanto a análise daquelas observações já indicava resíduos sistemáticos, simétricos em relação aos bordos inferior e superior. Ou seja, reconciliáveis com variações de longo prazo do diâmetro solar aparente. A precisão das observações diurnas com Astrolábio resultava em acurácia da ordem de 0,05" para os parâmetros orbitais. Mesmo que o tempo típico da constância do diâmetro solar seja, e é, muito menor que aquele relativo aos parâmetros orbitais, ainda assim a precisão instrumental intrínseca era competitiva para o monitoramento de variações do diâmetro.

Diante disto, em 1997 o instrumento passou por aprimoramentos e foi dotado de uma câmara CCD para a aquisição digital dos dados, para medidas da variação do diâmetro solar. Para uma descrição detalhada, ver Penna *et al.* (1996). O principal aprimoramento instrumental foi a adoção de um prisma objetivo de ângulo variável, capaz de observar o Sol entre 26° e 56° de distância zenital. Com isto, diversas observações diárias do Sol tornaram-se possíveis, varrendo uma gama de heliolatitudes de vários graus. O registro digitalizado da passagem, além de contribuir para o aumento da precisão intrínseca e a remoção do viés do observador, permitiu estender uma observação isolada por vários minutos, filtrando muito efetivamente as componentes de turbulência e homogeneizando as observações dos dois limbos (que por princípio já são feitas à mesma distância zenital). Aqui, é útil enfatizar que o princípio da medida com o Astrolábio Solar é de trânsito – passagem de pontos sucessivos por um almicantarado fixo, isto é, conceitualmente semelhante à observação durante um eclipse total, porém a maior precisão pela melhor definição do almicantarado em comparação com a cartografia do bordo lunar.

Em 1999 entra em funcionamento na França o DORAYSOL (Definition and Observation of the Solar Radius), um astrolábio aperfeiçoado (fig. 2.1). Nele, a parte óptica transformava-se num pequeno telescópio.

### 2.1.1 - DORAYSOL

O DORAYSOL foi construído com o objetivo de continuar a experiência de F. Laclare (Chollet, 1991), gerando uma série de medidas do raio solar, analisando suas variações aparentes em paralelo com medidas que serão feitas do espaço pelo satélite Picard (Damé *et al.*, 1999, Damé *et al.*, 2001).

Este instrumento, que está localizado na estação Calern do Observatoire de la Côte d´Azur, França, pode ser considerado uma nova geração de astrolábio (Delmas *et al.*, 2006), mas mantendo as mesmas características metrológicas de seu predecessor. Ele é um astrolábio refletor, que conta com um prisma variável (que permite observações em diferentes distâncias zenitais, aumentando o número de observações por sessão). Filtros permitem fazer observações em diferentes comprimentos de onda (mas não foram implementados até esta data). Uma lúnula vertical colocada na frente de um obturador rotativo permite que sejam gravadas alternadamente as imagens, direta e refletida, evitando a superposição real delas no CCD, que poderia causar problemas de saturação e *blooming*.

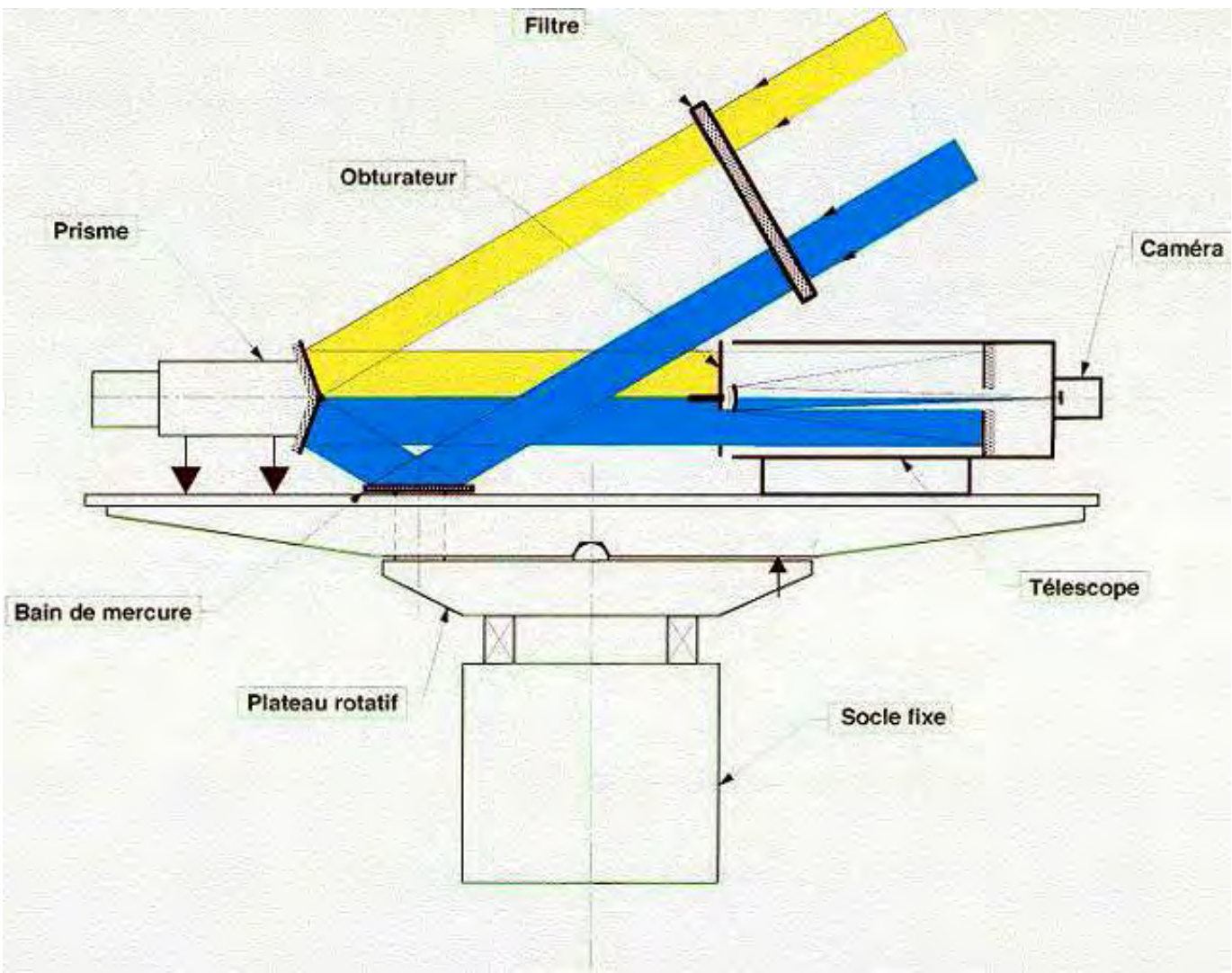

Fig. 2.1: Esquema do DORAYSOL (Delmas *et al.*, 2006).

Entre 1999, quando entrou em funcionamento, até 2003, este instrumento realizou medidas do diâmetro solar em paralelo com o ASTROSOL (Astrolábio Solar de Calern), localizado no mesmo sítio, a 20 m de distância. A figura 2.2 mostra a comparação entre as séries do ASTROSOL e DORAYSOL.

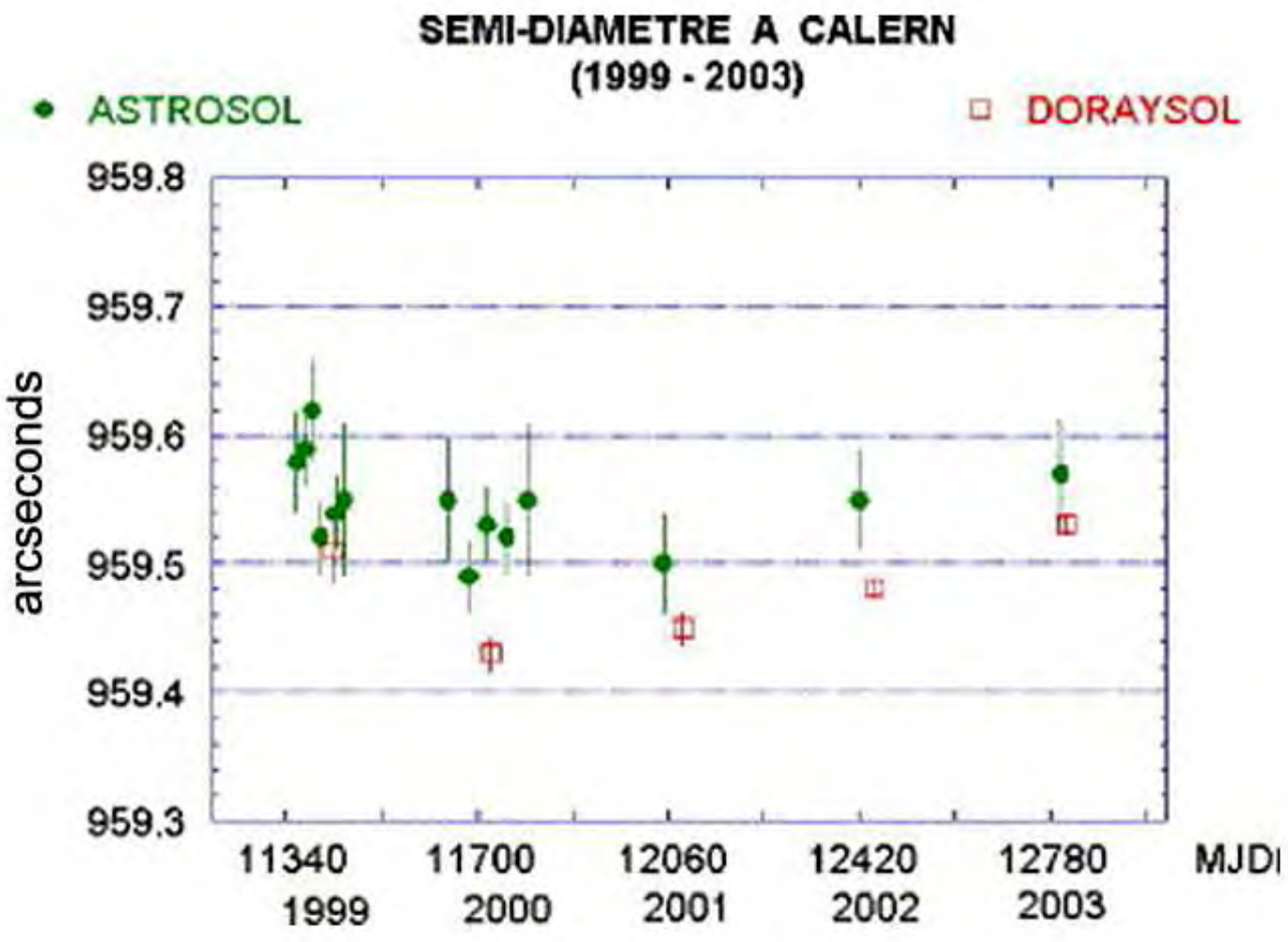

Fig. 2.2: Comparação entre as séries realizadas simultaneamente pelos instrumentos ASTROSOL e DORAYSOL (Delmas *et al.*, 2006).

O remover da superposição de imagens solares, já seriamente afetadas individualmente pela observação diurna, é um ponto importante que foi considerado no projeto do Heliômetro.

## 2.1.2 - Astrolábio vertical

Um acordo de cooperação científica entre a Universidade Estadual de Feira de Santana (Uefs) e o Observatório Nacional em 2004, permitiu a vinda de outro astrolábio Danjon para o Observatório Nacional. Utilizando-o com o mesmo propósito da equipe francesa, a equipe do Rio viu a oportunidade de fazer alterações no projeto óptico original deste astrolábio de forma a aperfeiçoá-lo, obtendo um novo instrumento cujas medidas solares ganhassem em precisão.

O primeiro projeto (fig. 2.3) considerado para uma nova montagem foi concebido por Bourget (2004).

Em destaque, algumas notáveis modificações:

- A substituição do prisma de ângulo variável original por um conjunto com outra geometria de articulação[1];
- As lúnulas de entrada teriam seu desenho invertido, acompanhando a nova configuração do prisma (fig. 2.4). Esta nova geometria teria vantagem sobre a geometria original da lúnulas, pois o alongamento da figura de difração ocorreria na direção horizontal, sentido oposto ao da medida do diâmetro;
- A substituição da objetiva de 1 m de distância focal e do redutor focal (que aumenta esta distância para 3 m) por um espelho primário. Este espelho teria a distância focal equivalente a este conjunto e seria localizado numa abertura dentro do pilar astronômico, figura 2.5.

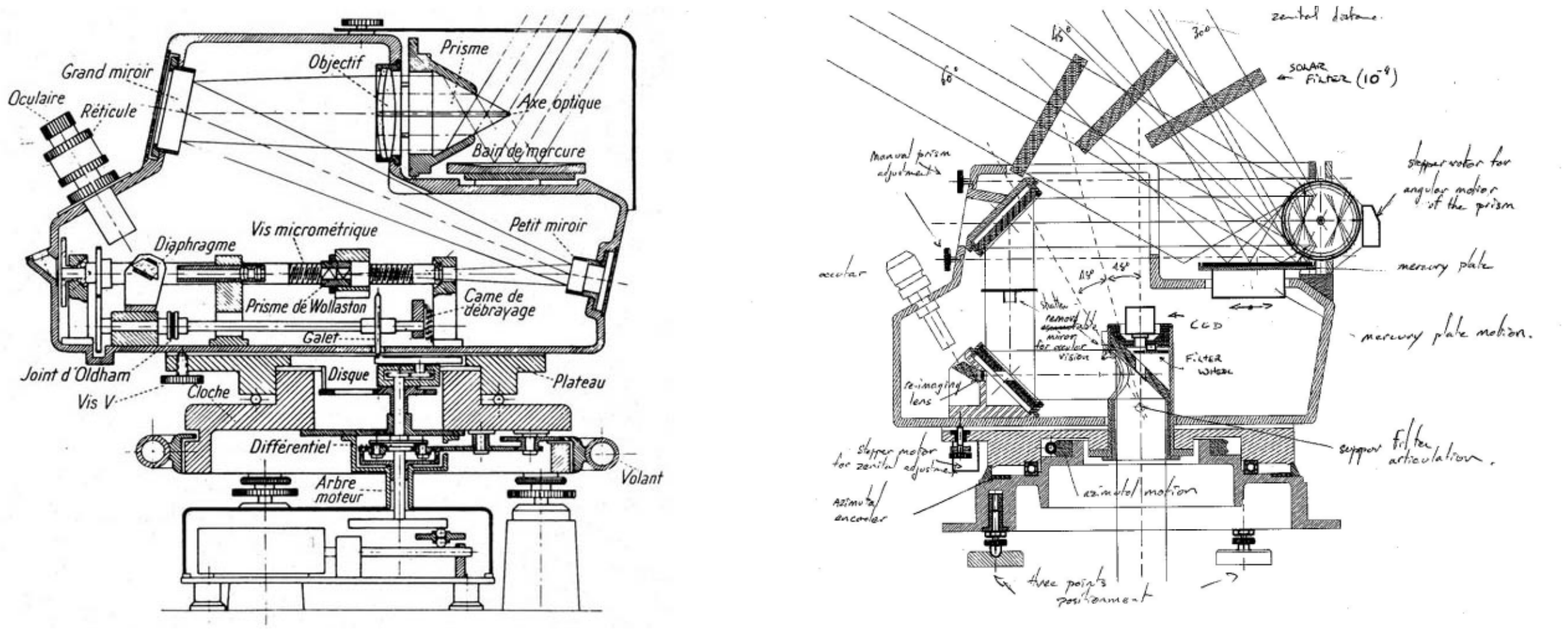

Fig. 2.3: Comparação entre o esquema do Astrolábio Danjon e o esboço do projeto do novo astrolábio vertical (Bourget, 2004).

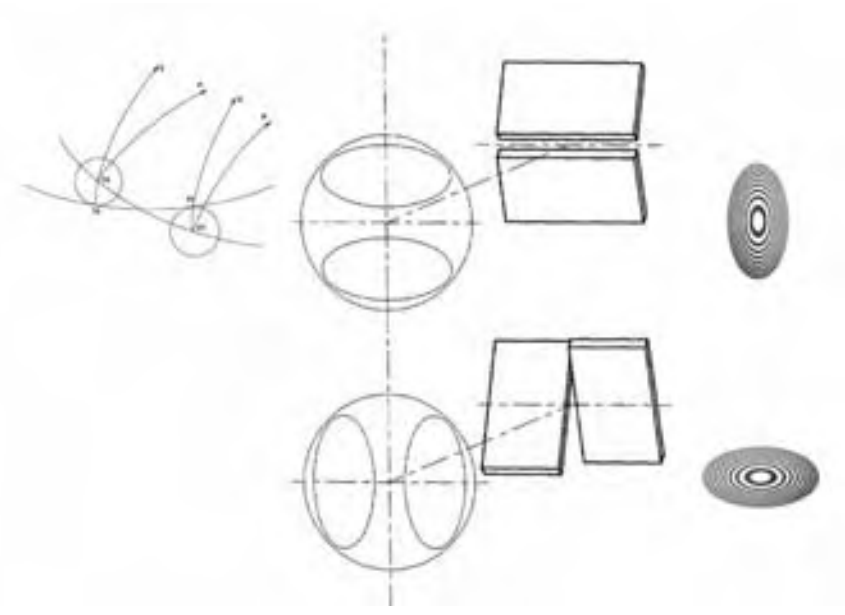

Fig. 2.4: À Esquerda, vemos o esquema da medida vertical do diâmetro solar feito com o astrolábio. Ao centro, a comparação entre o conjunto lúnula/prisma variável original e o proposto. À direita, o efeito das respectivas lúnulas de entrada na figura de difração (Bourget, 2004).

---

- [1] O protótipo construído pelo Dr. Pierre Bourget no Observatório Nacional serviu de base para o prisma variável utilizado no DORAYSOL (Delmas, *et al.*, 2006).

A imagem da figura 2.5 traz o esquema de instalação da nova montagem do astrolábio solar na antiga cúpula da luneta foto-equatorial do Museu de Astronomia e Ciências Afins (MAST).

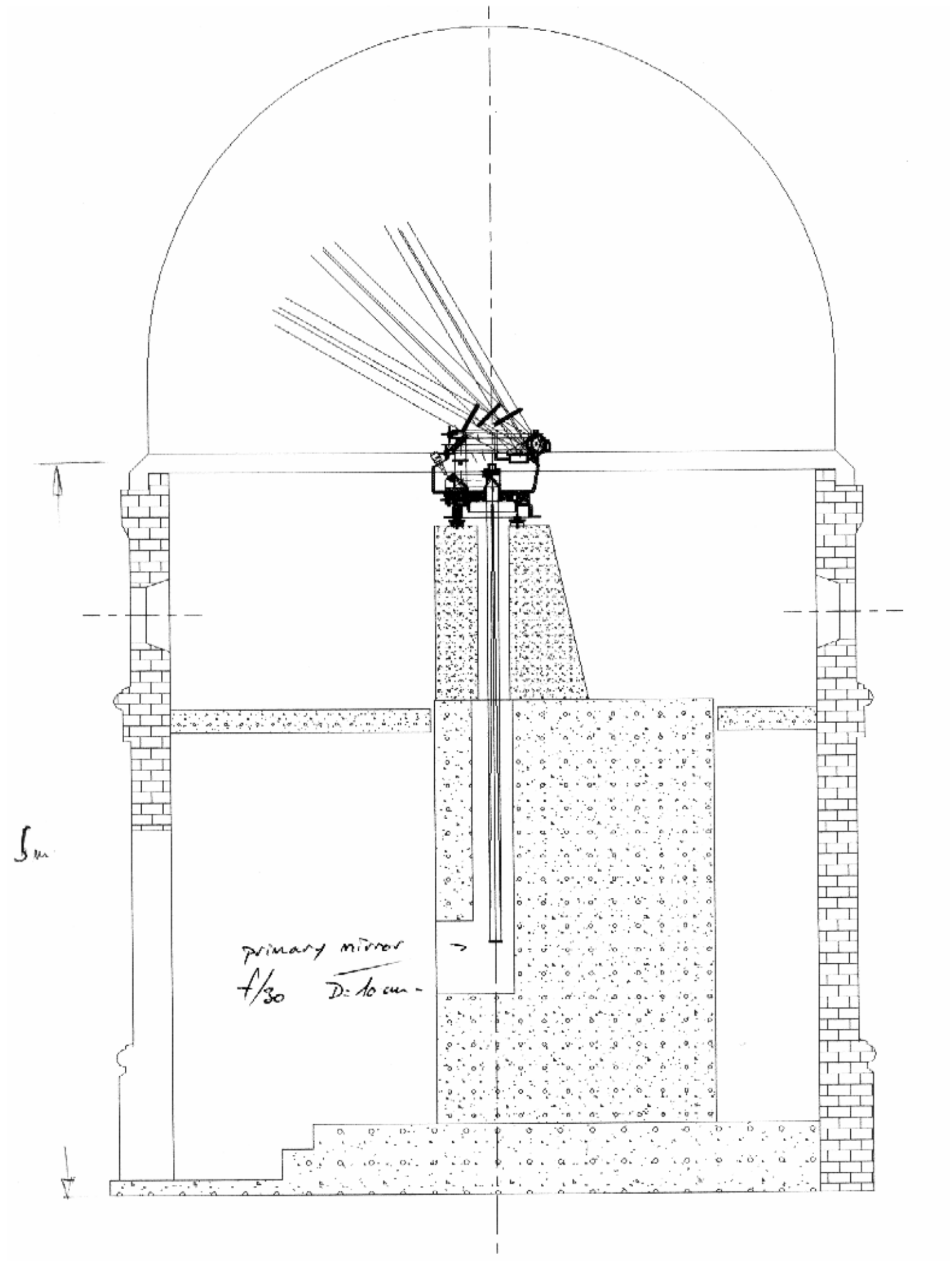

Fig. 2.5: Primeiro projeto para utilização do Astrolábio de Feira de Santana e da cúpula (Bourget, 2004).

Sua localização, à 5 m de altura do solo, também colaboraria para se obter imagens mais estáveis, pois a esta altura o ar é menos turbulento. O caminho óptico vertical através de camadas de ar estratificadas dentro do pilar reduziria também a turbulência da imagem.

Este astrolábio solar refletor trabalharia em paralelo, e próximo, ao astrolábio solar refrator do Observatório Nacional, obtendo séries independentes. Dentro do âmbito deste projeto a cúpula recebeu sua primeira reforma, figura 2.6.

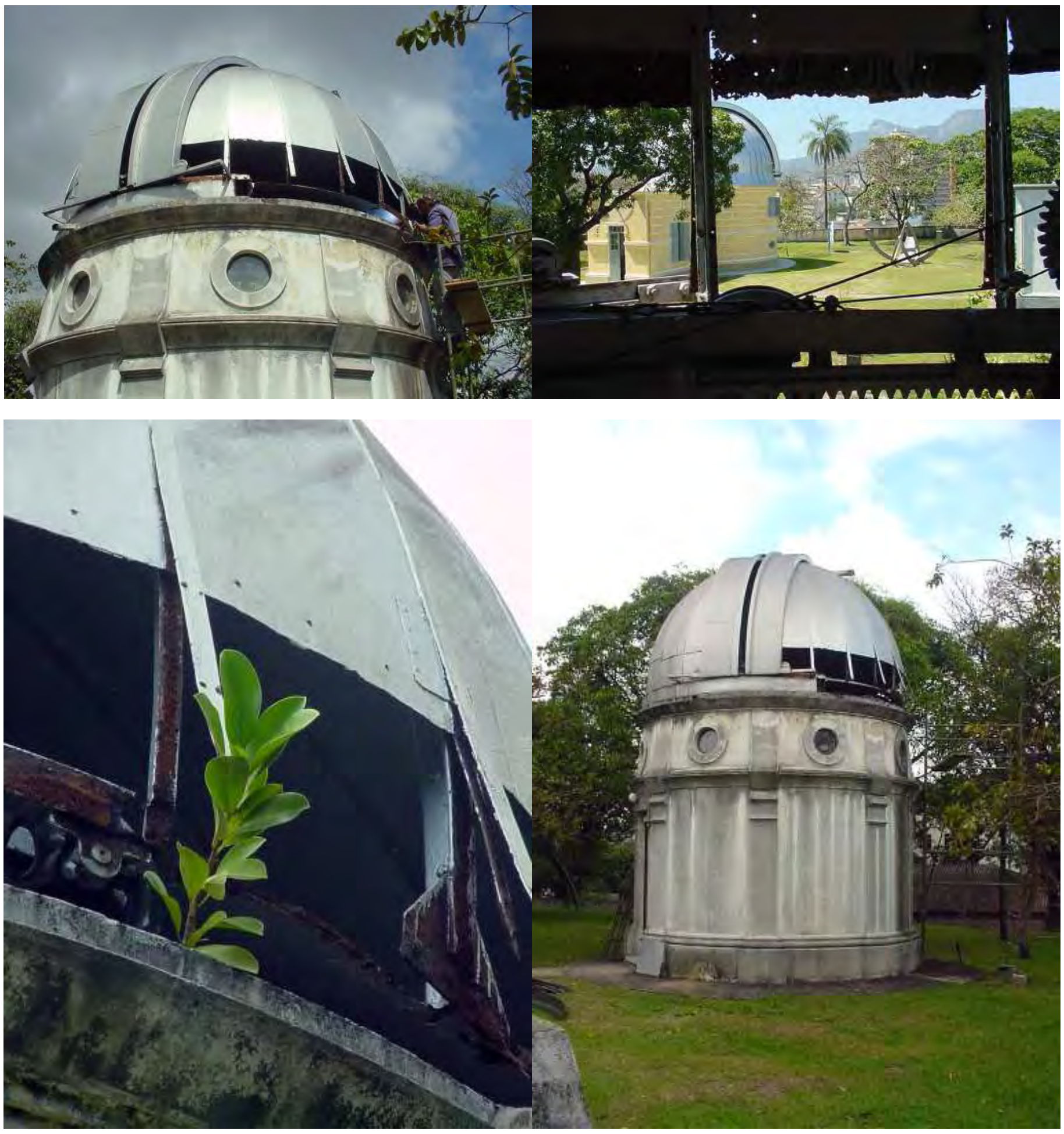

Fig. 2.6: Registro da primeira reforma da cúpula do novo instrumento.

Este projeto teve o mérito, não desprezível, de mobilizar a equipe para o desenvolvimento instrumental. E deixou como herança concreta a cúpula da luneta foto-equatorial, no campus do Observatório Nacional e cedida pelo Museu de Astronomia e Ciências Afins (MAST), instituição co-irmã, reformada e adaptada para a utilização de um novo instrumento solar. Reforma esta autorizada pelos órgãos de preservação do patrimônio.

### 2.1.3 - O Astrolábio heliométrico

Ao final da Introdução, apresentamos diversos objetivos para um novo instrumento solar. Ao mesmo tempo em que o Astrolábio possui uma comprovada precisão e robustez metrológica, ele não preenche diversos itens pretendidos para o novo instrumento. Inicialmente, tirando partido do instrumento cedido pela Universidade Estadual de Feira de Santana, tentou-se combinar a qualidade do Astrolábio com a flexibilidade do método heliométrico.

A técnica das alturas iguais, apesar de seus atributos, apresenta algumas dificuldades e limitações (como o banho de mercúrio que é utilizado para definir o almicantarado de observação, que é de manuseio difícil e representa um risco ao operador e ao meio ambiente). É sabido que a imagem refletida no banho de mercúrio possui sempre uma qualidade inferior à da imagem direta, o que pode significar uma materialização imperfeita do almicantarado e prejudicar diretamente os resultados.

O método das alturas iguais faz com que o diâmetro solar só possa ser medido ao longo do vertical do observador, o que estreita severamente a faixa de heliolatitudes observada. Vale também reafirmar que a medida do diâmetro solar através da referência ao almicantarado é uma medida de trânsito, enquanto que se deseja implementar uma série conceitualmente independente.

A solução proposta consistiu na utilização do segundo Astrolábio Danjon para desenvolver um outro instrumento. Uma mudança radical do método de medidas, pois ao invés de se fazer medidas dinâmicas do diâmetro vertical aparente do Sol, a proposta era fazer medidas angulares do diâmetro, não necessariamente apenas os verticais.

O astrolábio seria transformado num heliômetro clássico (mostrado na figura 2.10), com sua objetiva cortada ao meio e, estas metades, deslocadas linearmente.

Sem o prisma de ângulo variável, nem o banho de mercúrio, a imagem do Sol deveria ser levada até a objetiva heliométrica do instrumento. A solução foi incorporar um celostato ao projeto. Deste modo, o instrumento poderia permanecer fixo em sua base. Na figura 2.7 vemos um esboço do projeto do Astrolábio Heliométrico.

Logo se tornou evidente que à adaptar um instrumento, mesmo de alta qualidade metrológica, melhor seria desenvolver um instrumento próprio, o Heliômetro do Observatório Nacional.

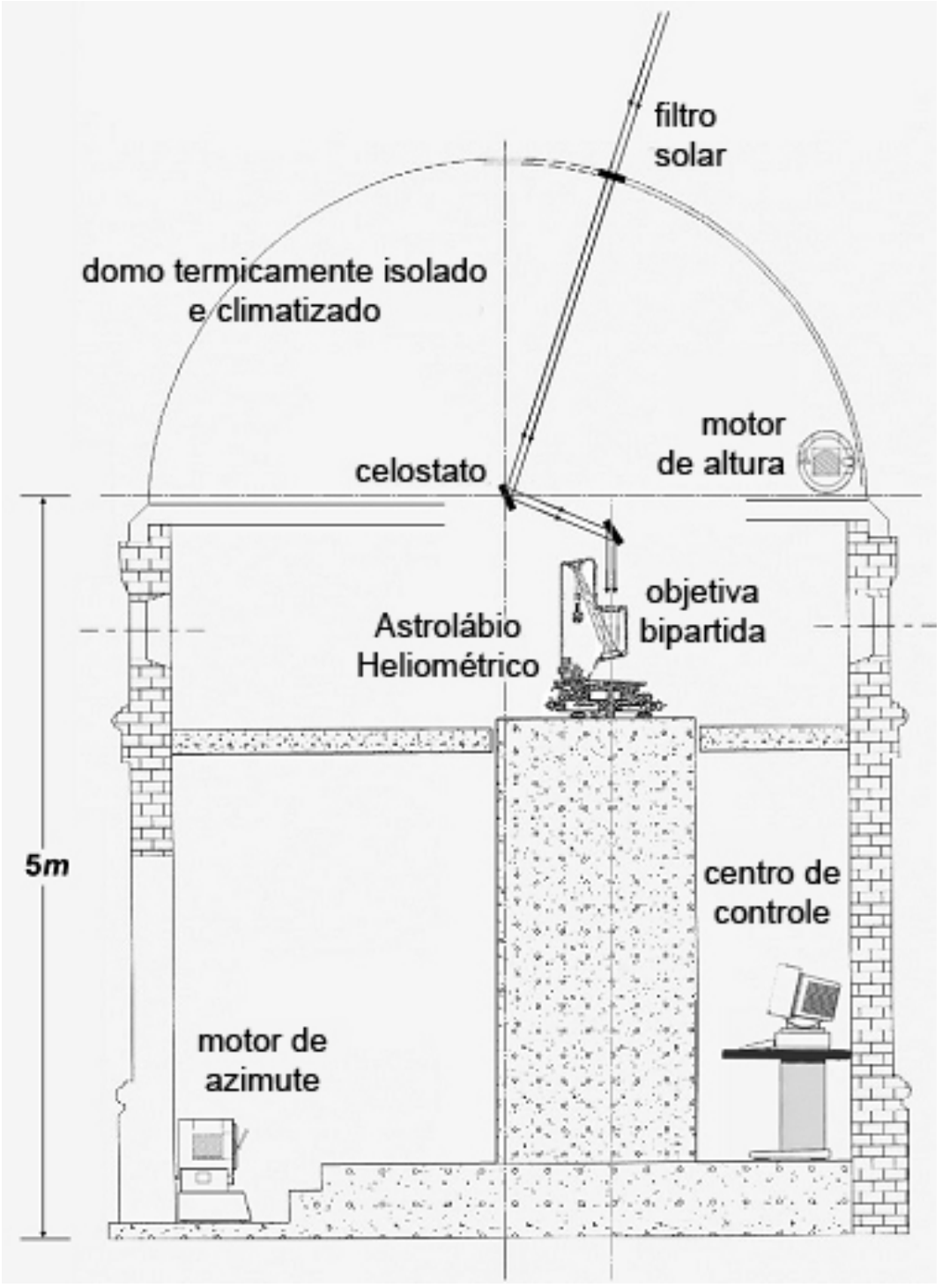

Fig. 2.7: Esboço do projeto do astrolábio heliométrico do ON em sua cúpula (adaptação de Bourget, 2004).

O instrumento seria colocado na vertical, sobre o pilar. O domo da cúpula seria motorizado e o interior da cúpula, climatizada. O filtro solar ficaria numa janela móvel, presa à trapeira da cúpula, e também seria motorizado. Com isso, previa-se a automatização das medidas. A sala de controle do domo e do instrumento ficaria no 1º andar do prédio.

Toda esta concepção foi posteriormente mantida para o projeto do Heliômetro do Observatório Nacional.

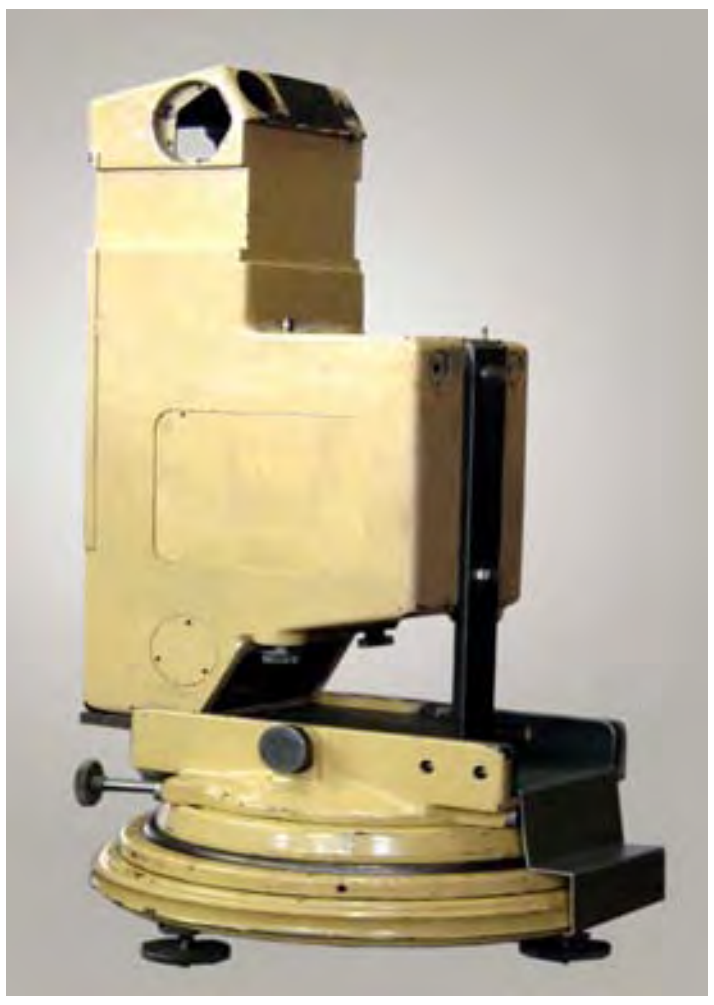

Fig. 2.8: Corpo do Astrolábio Danjon sendo adaptado em um heliômetro.

Um modelo de celostato foi construído em madeira (fig. 2.9), como parte do estudo do projeto.

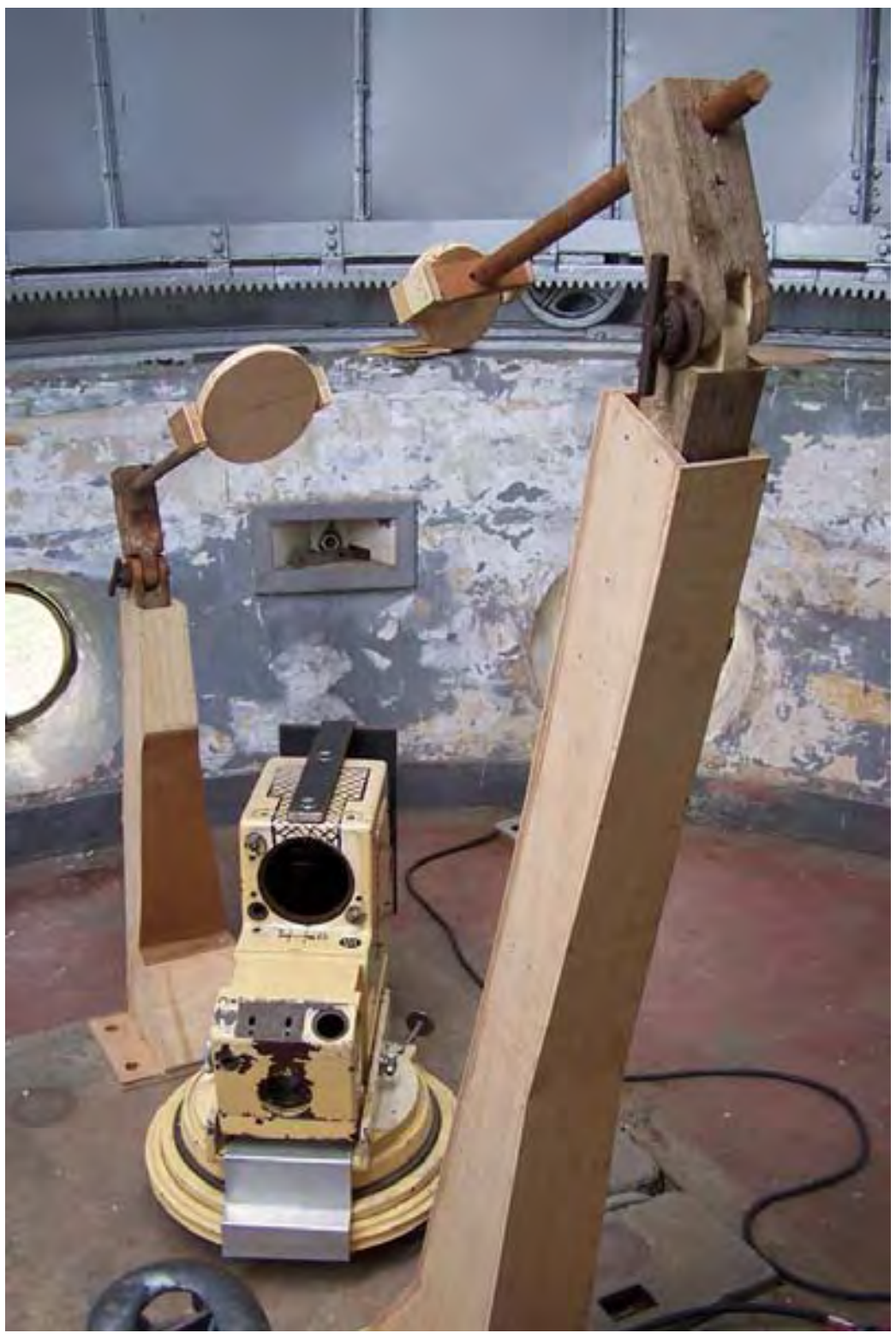

Fig. 2.9: Corpo do Astrolábio e o modelo, em madeira, do celóstato, dentro da cúpula.

## 2.2 - A fundamentação do heliômetro

O princípio de observação de campos diferentes projetados sobre o mesmo plano focal é de capacidade metrológica expressa na metodologia das missões astrométricas Hipparcos e GAIA (Bastian, U. e Schilbach, E., 1996).

No método heliométrico uma objetiva é bi-seccionada e, após um pequeno deslocamento linear, cada um das hemi-objetivas resultantes aponta para dois campos próximos. A razão por trás da construção de uma objetiva dividida ao meio é a de ser capaz de se poder mover uma metade em relação à outra, em uma direção perpendicular ao eixo óptico, por distâncias bem precisas. As duas imagens formadas, então, se sobrepõem no plano focal da ocular. Ao mover-se uma parte da objetiva com respeito à outra, as imagens irão mover-se de modo análogo (Manly, 1991).

Este tipo de instrumento foi primeiramente proposto em 1675 pelo astrônomo dinamarquês Ole Rømer e construído em 1748 pelo astrônomo francês Pierre Bouguer, que montou um telescópio solar com duas objetivas, que foram colocadas lado a lado e que, conseqüentemente, formavam duas imagens dos objetos observados. Através de uma escala colocada no foco em comum, media-se a distância de contato entre os limbos solares. Mas tarde, em 1754, John Dollond (1706-1761) sugeriu o corte da objetiva, como forma de duplicar a imagem (fig. 2.10).

Por volta de 1800, Fraunhofer, conhecido por ter descoberto as linhas de absorção no espectro da luz solar, que permitiu conhecer a composição química do sol e de estrelas, construiu uma série de heliômetros. Em 1840, a partir das observações heliométricas, Bessel conseguiu, pela primeira vez, determinar a paralaxe de uma estrela (61 Cygni).

O método heliométrico de medidas angulares foi posteriormente largamente ultrapassado pelo uso de um micrômetro bifiliar ou de um micrômetro de dupla imagem, adaptado à ocular dos instrumentos, fazendo com que os heliômetros deixassem de ser usados, apesar de que estes últimos, com suas objetivas cortadas fossem capazes de medir separações angulares muito amplas (Manly, 1991).

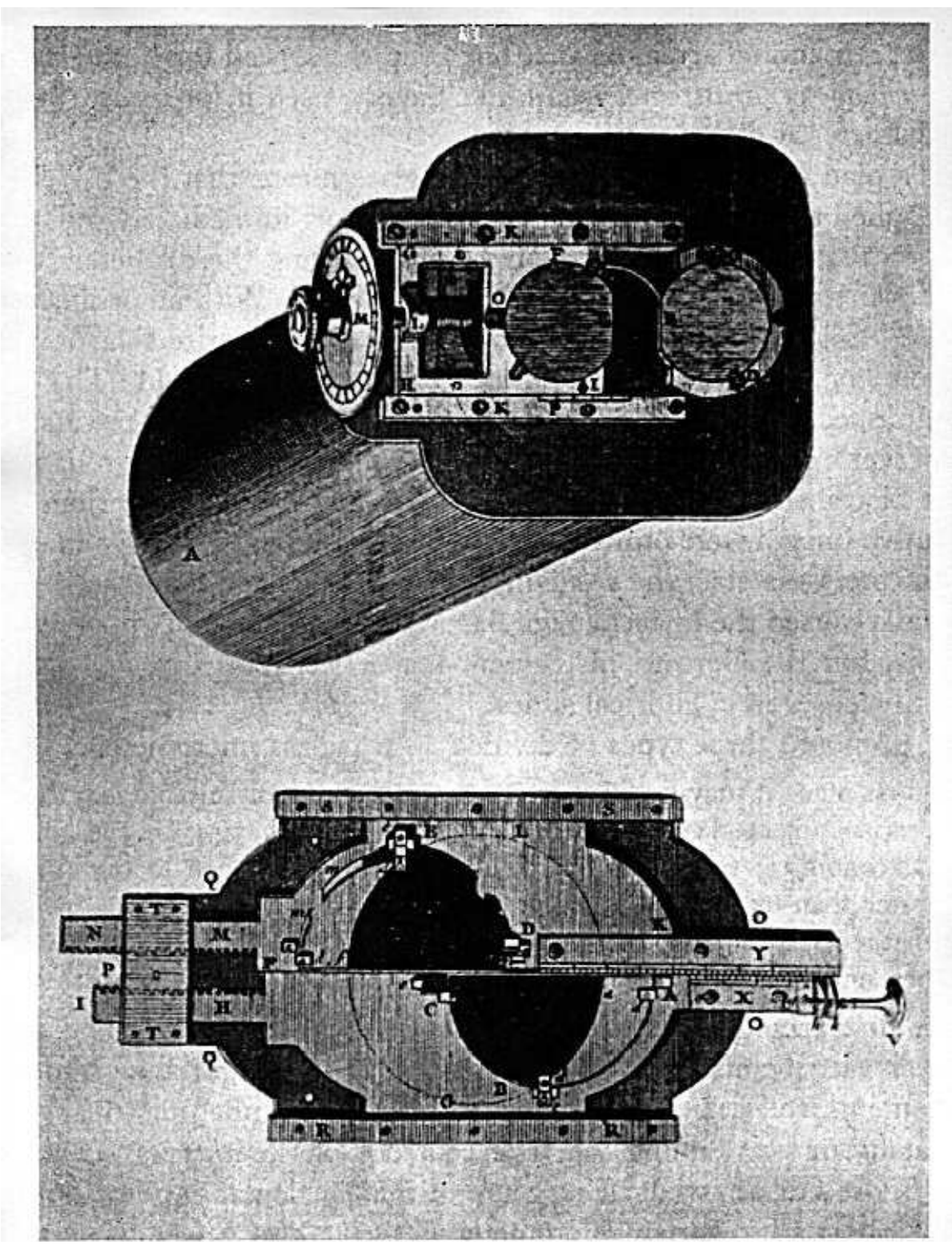

Fig. 2.10: No alto, o desenho do heliômetro proposto por Bouguer. Embaixo, o de Dollond (Brayebrook Observatory, 2009).

### 2.2.1 - O método

A dificuldade básica em se medir um pequeno ângulo por meios ópticos vem do fato de que a medida é diretamente dependente da posição do plano de observação em relação ao plano focal do instrumento, independendo ser este refrator ou refletor (d'Ávila, *et al.*, 2009).

Heliômetro Refrator

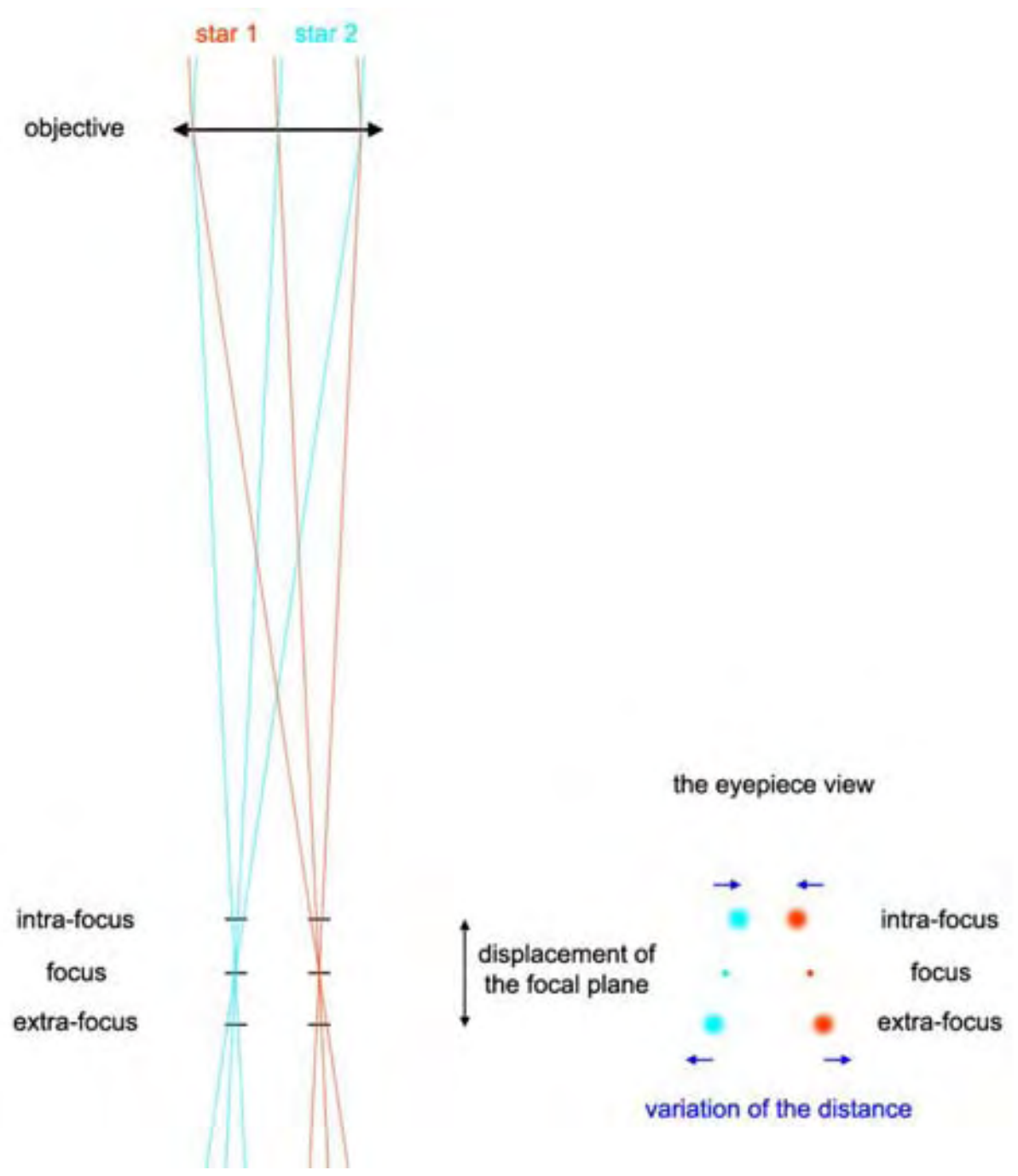

Fig. 2.11: Esquema da dependência da distância angular entre duas estrelas com o plano de observação. As estrelas, além de saírem do foco, deslocam-se uma em relação à outra (d'Ávila, *et al.*, 2009).

A solução apresentada pelo heliômetro refrator consiste em bisseccionar a objetiva e separá-las perpendicularmente ao seu eixo óptico, de forma a duplicar o campo. O uso de uma pupila de entrada apropriada, como visto na figura 2.12 (d'Ávila *et al.*, 2009), possibilita a eliminação da dependência da medida com a estabilidade do plano focal, figura 2.13.

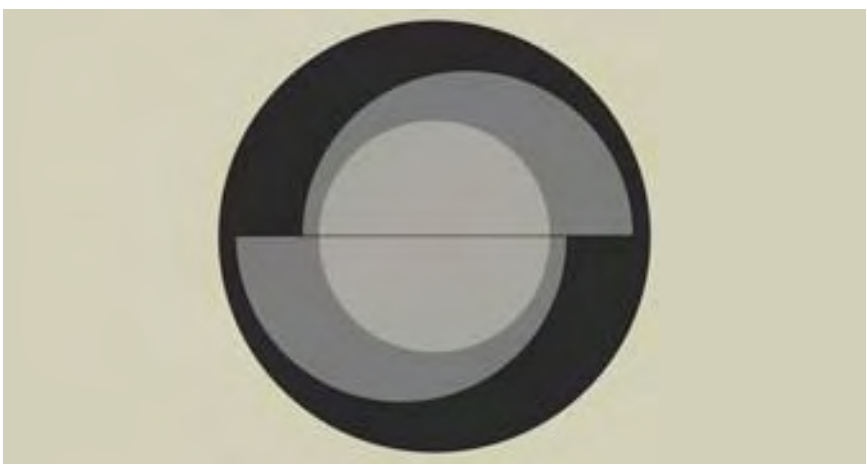

Fig. 2.12: Esquema da separação linear da objetiva seccionada e da pupila de entrada.

Na figura 2.13, as imagens das duas estrelas foram trazidas para superposição, no centro do campo, pelo deslocamento conveniente das hemi-objetivas.

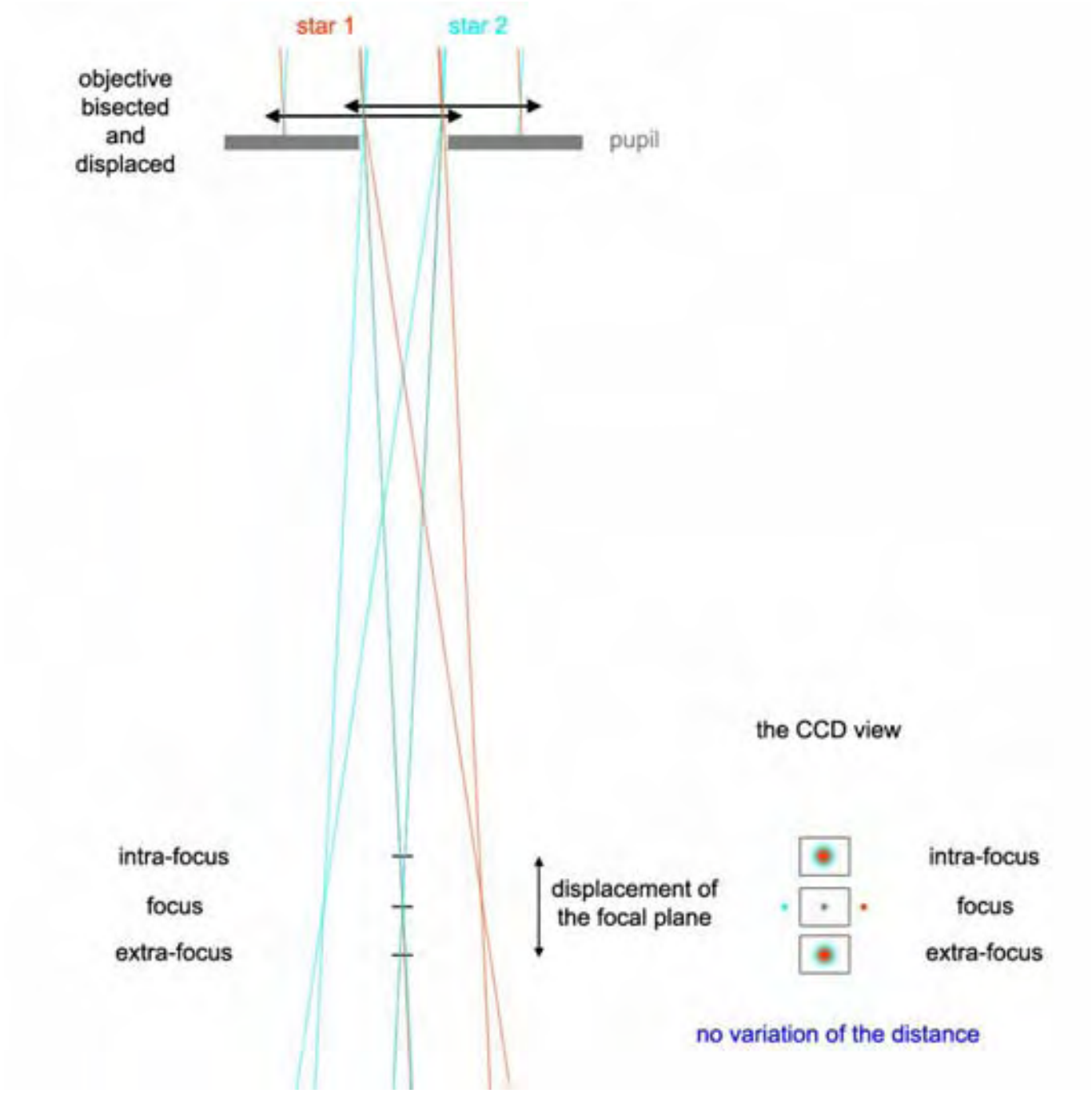

Fig. 2.13: Esquema do uso de uma pupila de entrada para tirar a dependência da distância angular entre duas estrelas com o plano de observação. As estrelas saem do foco juntas e não mais se deslocam uma em relação à outra (d'Ávila, *et al.*, 2009).

Esta solução clássica tem suas desvantagens: o fato de lentes serem suscetíveis às deformações térmicas e mecânicas e a dependência da distância focal com o comprimento de onda da luz.

## Heliômetro refletor

Um heliômetro refletor possui o espelho principal bisseccionado (fig. 2.14), ao invés da objetiva.

Cada metade do espelho pode ser deslocada linearmente ou angularmente.

O deslocamento angular, ou inclinação, é a configuração que possuiu a mesma característica da montagem anterior, ou seja, a independência da separação angular com a distância ao plano focal, mas sem a necessidade de uma pupila.

Além disso, o uso de espelhos:

- elimina o problema da aberração cromática;

- garante sua estabilidade mecânica, se forem fabricados com material cerâmico de baixíssimo coeficiente de dilatação térmica;

Mais uma vantagem estratégica: existem fabricantes de espelhos de excelência no Brasil. O mesmo não se pode falar das objetivas.

A desvantagem do uso de espelhos reside no fato de que a separação angular entre as imagens é, agora, muito mais sensível a qualquer variação do diedro entre os hemi-espelhos. Por isso, neste projeto, a estabilidade mecânica é importantíssima.

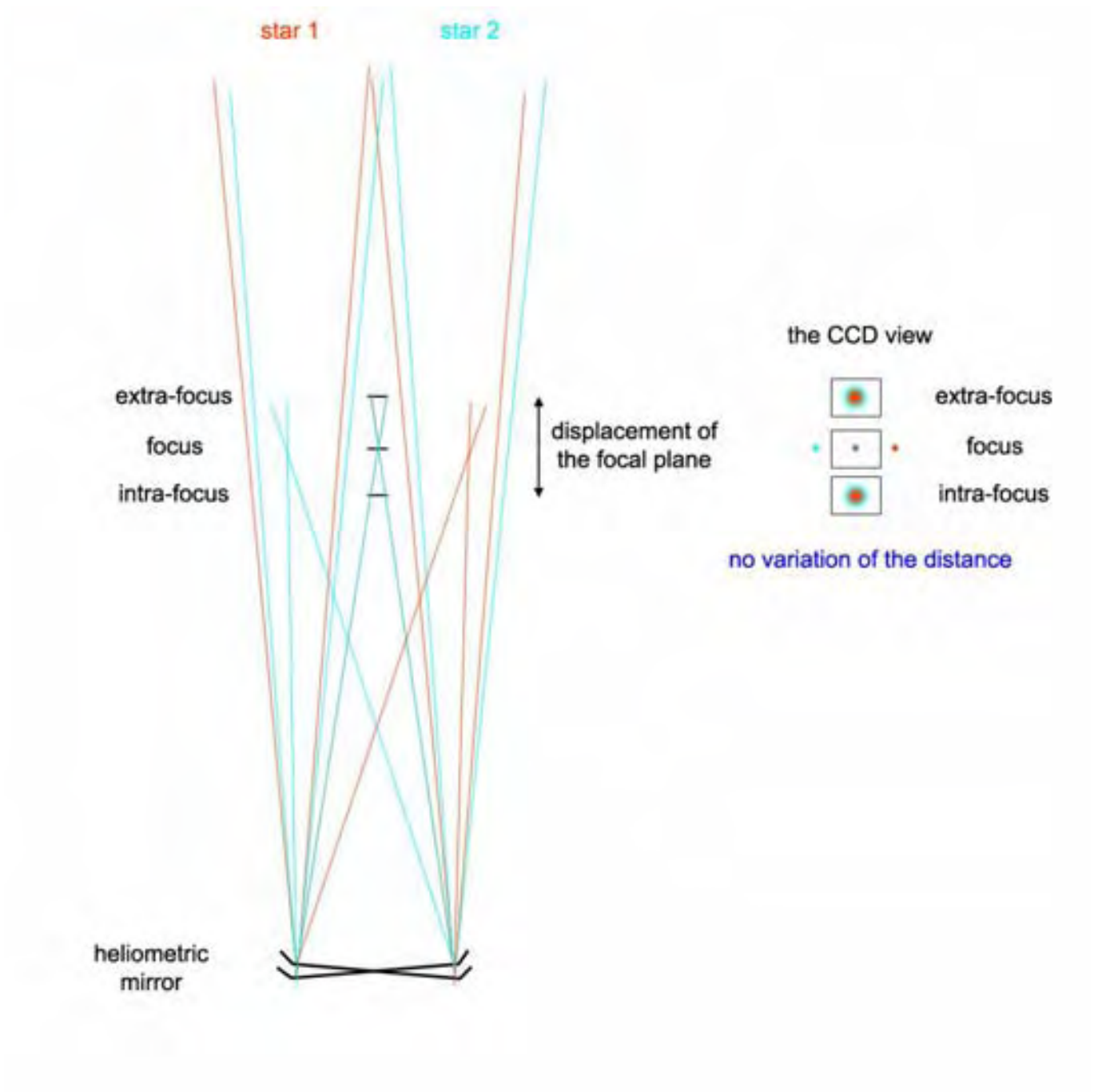

Fig. 2.14: Esquema do espelho heliométrico As imagens deslocadas para o centro do campo mantêm sua superposição independentemente do deslocamento do plano de observação. As estrelas saem do foco juntas, mas não se deslocam uma em relação à outra (d'Ávila, *et al.*, 2009).

## 2.3 – Heliômetros históricos e astrométricos

A seguir são descritos alguns heliômetros refratores que foram montados e usados para medições de paralaxes estelares e observações de trânsitos planetários e de modernos satélites astrométricos que utilizam o mesmo princípio heliométrico.

A figura 2.15, superior, apresenta a objetiva bi-partida do Heliômetro construído por Peter Dollond (1730-1820). Nesta foto podemos ver o parafuso micrométrico, para mover as partes longitudinalmente e a pupila. Este instrumento foi um de vários encomendados pela Royal Society de Londres, para ser levado em expedições para observação do trânsito de Vênus, em 1769, para a determinação acurada da distância do Sol (Science Museum/Science & Society, 2009).

Na parte inferior da figura 2.15, vemos também a objetiva bi-partida do maior instrumento deste tipo já construído, o Heliômetro do Observatório Kuffner, em Viena. Este heliômetro foi construído em 1893 e começou a operar em 1896, medindo paralaxes estelares. Entre 1917 e 1947, sua objetiva e mecânica foram danificadas, e sua valiosa objetiva heliométrica substituída por uma lente comum. Somente em 1999, uma reforma recuperou suas históricas peças originais (Kuffner Observatory, 2009).

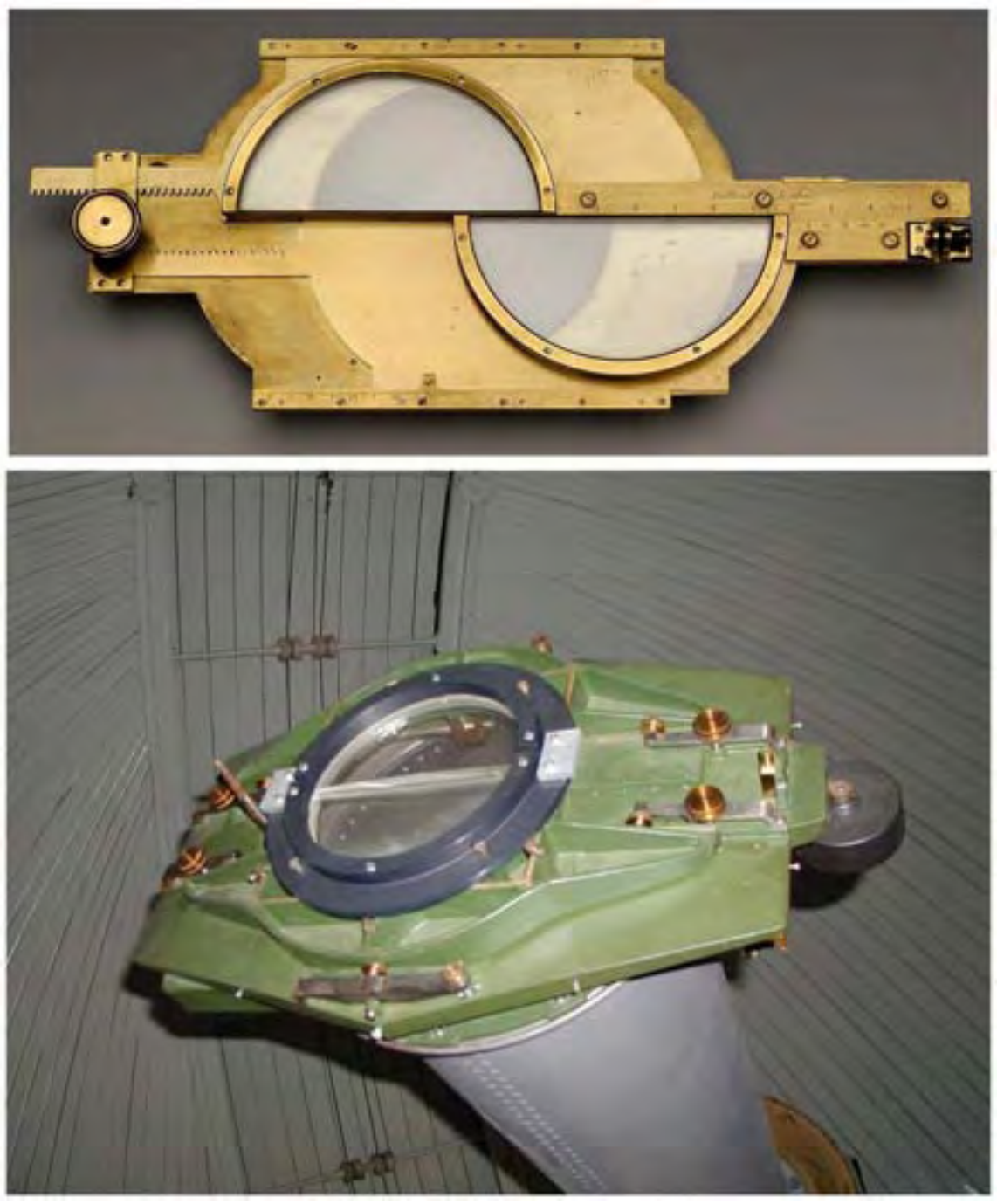

Fig. 2.15: Heliômetro de Dollond e o Heliômetro do Observatório de Kuffner, Áustria, 1893 (Science & Society Picture Library, 2009 e Kuffner Observatory, 2009).

E por fim, a figura 2.16 mostra o Heliômetro Yale, em New Haven, nos Estados Unidos. Este heliômetro foi o primeiro do tipo na América e, na época era o maior do mundo, com um diâmetro de 6 polegadas. O primeiro uso científico feito com ele foi a observação do trânsito de Vênus, em 1882 (Hoffleit, 2009)

.

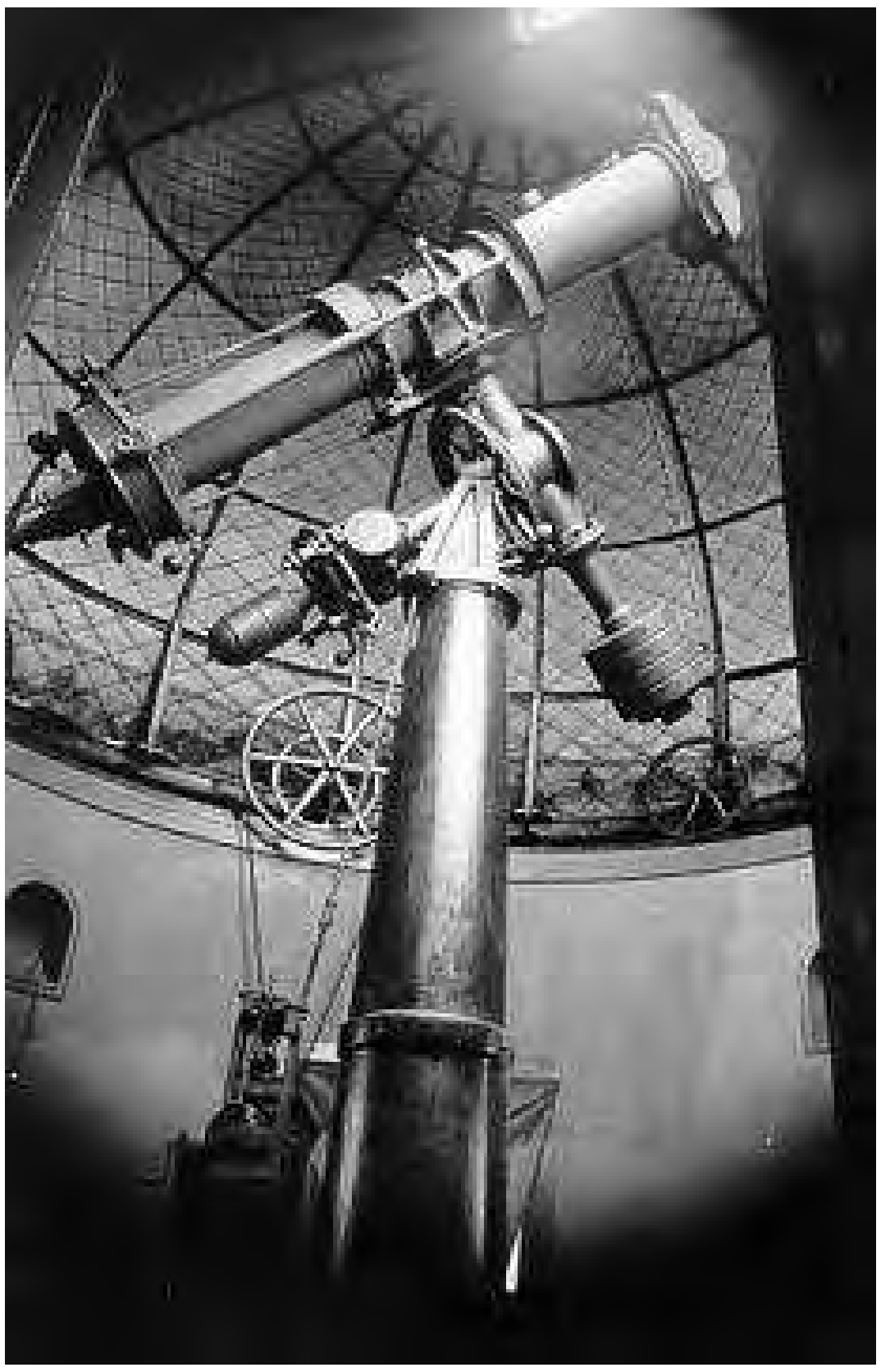

Fig. 2.16: Heliômetro de Yale, de 1880, semelhante ao usado por Bessel. Com este instrumento, de 1883 até 1910, foram obtidas as melhores medidas de paralaxes (mais de 200) antes do advento da astrometria fotográfica (Yale University, 2009).

O princípio heliométrico, ou seja, a divisão de um elemento óptico para combinar a luz de duas direções diferentes, foi o passo crucial no desenho do satélite Hipparcos (**Hi**gh **P**recision **Par**alax **Co**llecting **S**atellite). Este satélite foi concebido com o objetivo de medir, com grande precisão, posições, movimentos próprios e paralaxes de milhares de objetos celestes. Sua missão teve início em agosto de 1989 e término em março de 1993 (European Space Agency, 2009).

O projeto do telescópio para o Hipparcos é mostrado esquematicamente na figura 2.17. A luz de duas pupilas de entrada é combinada em um único caminho óptico por um espelho cortado ao meio e colado com suas metades em diedro (Leeuwen, 2007).

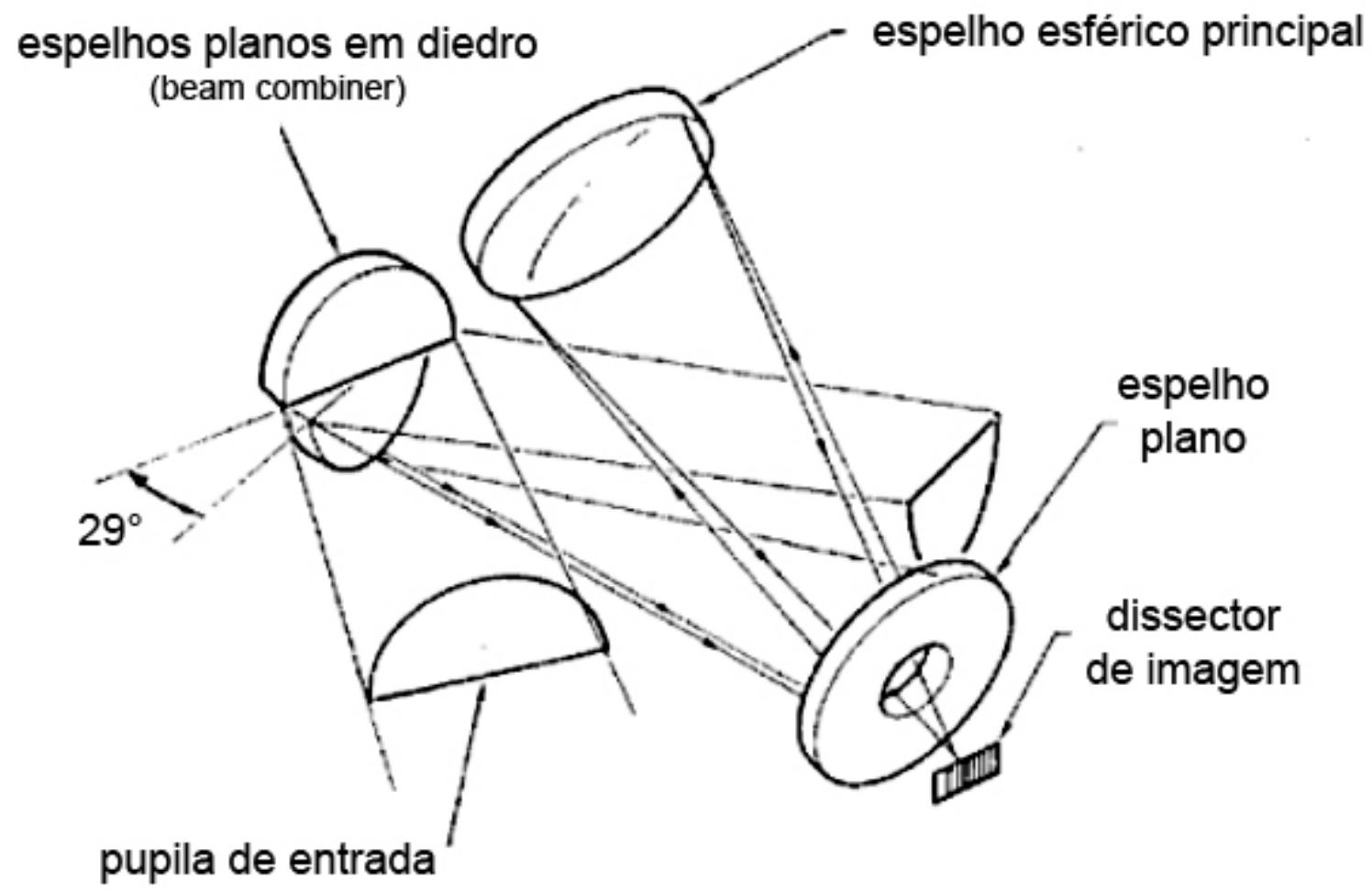

Fig. 2.17: Esquema do espelho heliométrico do satélite Hipparcos (adaptação de Leeuwen, 2007).

Como Hipparcos e ampliando seus objetivos, para incluir velocidades radiais, espectroscopia e fotometria, o satélite Gaia (com previsão de lançamento em 2012) também vai observar o céu de duas direções simultaneamente: ângulo de 106,5° (Hog *et al.*, 1997). Para isso, contará com um sistema de espelhos tipo-Hipparcos. Na figura 2.18 vemos um esquema de sua configuração óptica: um *beam combiner* (S1) juntamente com um telescópio gregoriano aplanático (S2 e S3). O restante do caminho óptico até o plano focal F não é mostrado nesta figura.

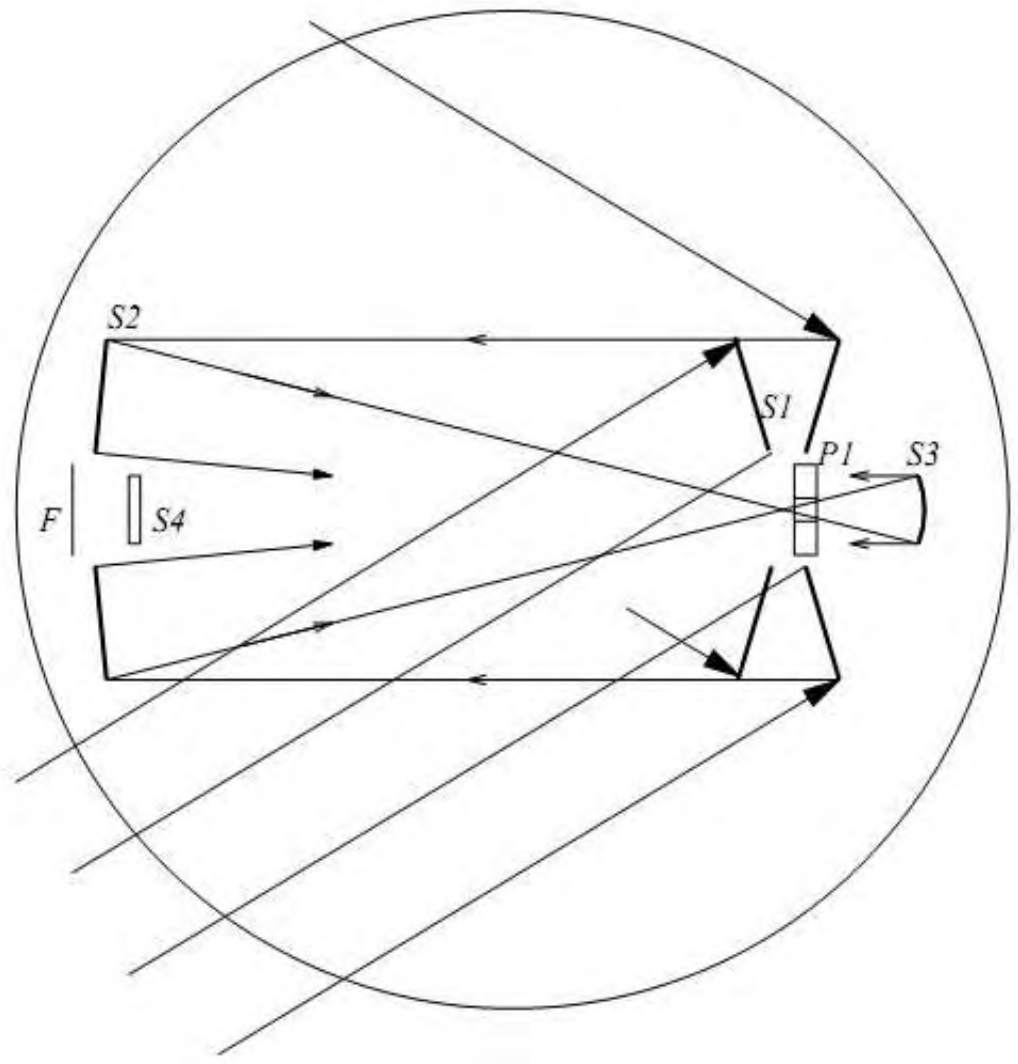

Fig. 2.18: Esquema do caminho óptico heliométrico do satélite Gaia (Hog *et al.*, 1997).

E para citar apenas mais um satélite astrométrico que usará o mesmo princípio heliométrico, temos o Nano-JASMINE. Seu telescópio possui dois espelhos planos posicionados em ângulos diferentes (99,5º), na frente do espelho primário (JASMINE team, 2009).

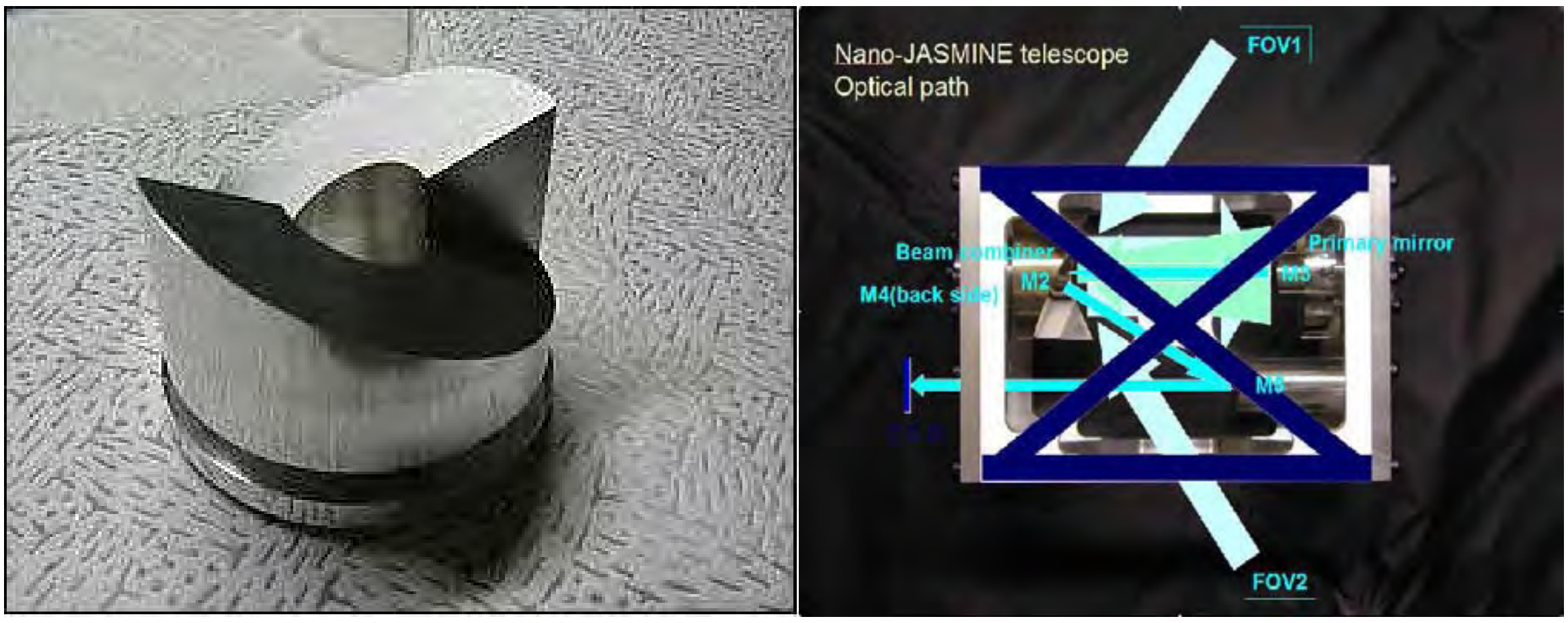

Fig. 2.19: À esquerda, os espelhos planos em diedro do satélite Nano-JASMINE. À direita, o caminho óptico do telescópio (JASMINE team, 2009).

## *2.4 - Heliômetro do Pic-du-Midi*

Ao contrário dos heliômetros clássicos, o Heliômetro do Pic-du-Midi não duplica imagens. Sua objetiva acromática (de 500 mm de diâmetro e 6500 mm de distância focal) projeta o disco do Sol (de ~60 mm de diâmetro) no plano focal e as imagens de limbos opostos são transportadas, através de prismas, para junto do eixo óptico, onde um sistema rotativo faz com que, alternadamente, as imagens caiam num CCD linear.

Detalhes sobre o Heliômetro do Pic-du-Midi e seus resultados podem ser encontrados na literatura (Rösch e Yerle, 1983 e Rösch *et al.*, 1996).

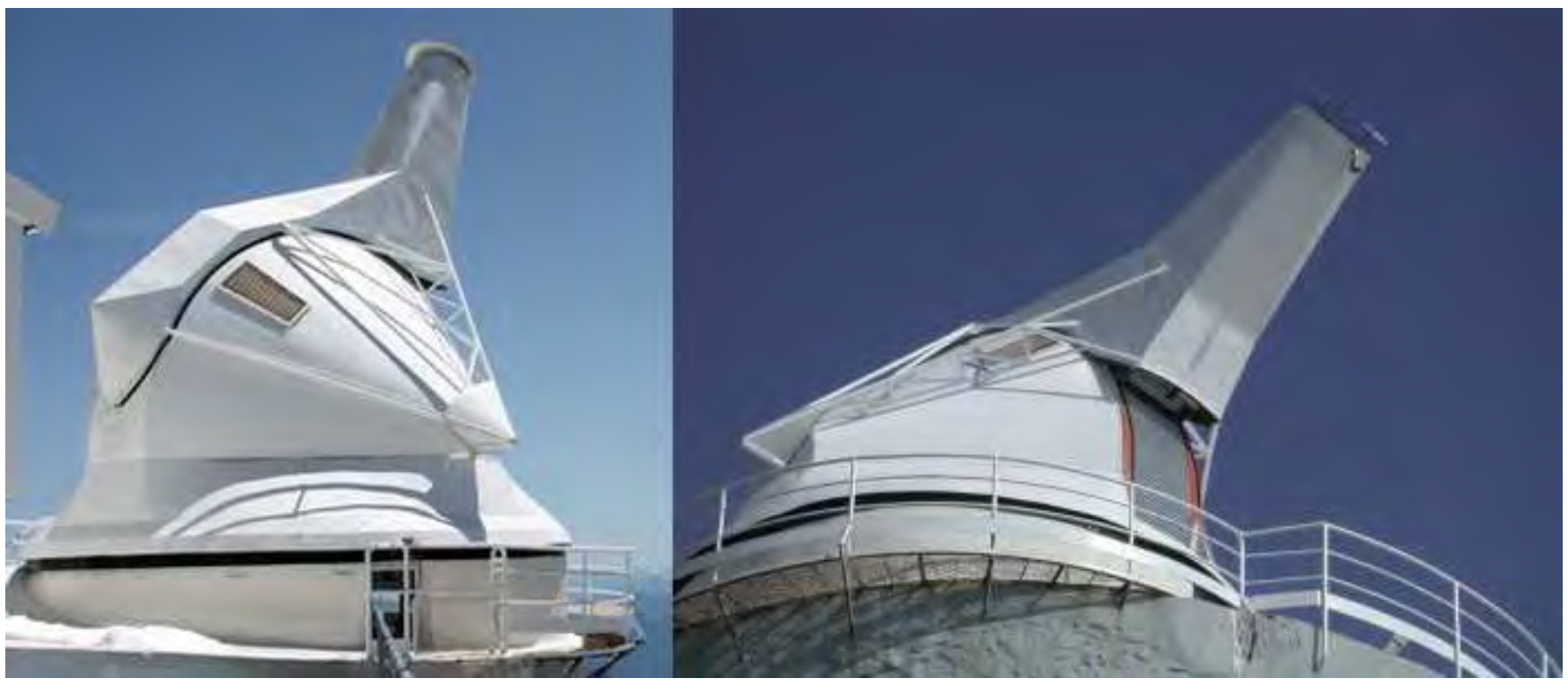

Fig. 2.20: Luneta Jean-Rösch, onde o heliômetro está instalado (Equipe Astrométrie et Métrologie Solaires, 2009).

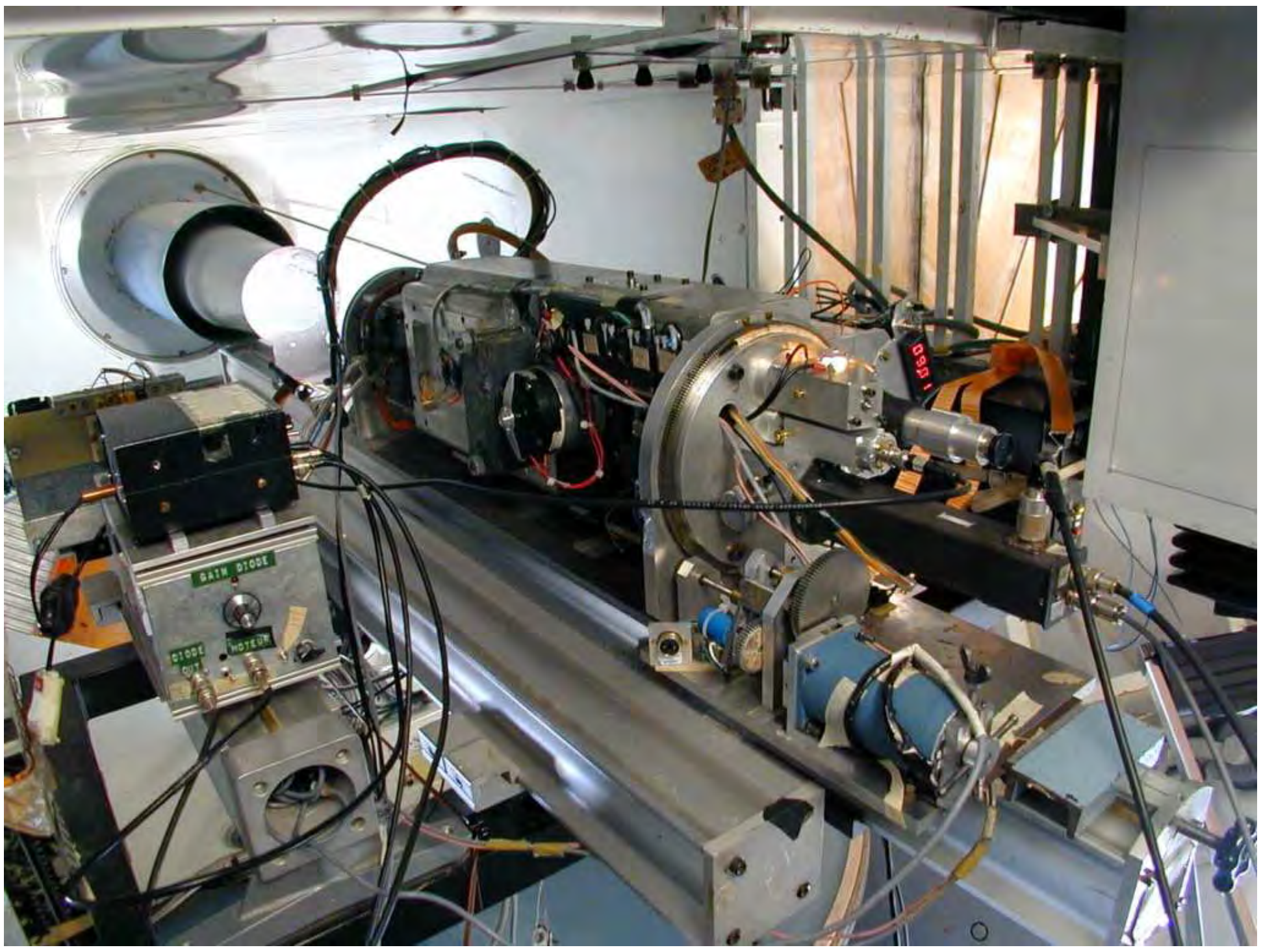

Fig. 2.21: O Heliômetro do Pic-du-Midi (Equipe Astrométrie et Métrologie Solaires, 2009).

O sistema pode girar sobre seu eixo óptico para medir o diâmetro solar em diversas heliolatitudes.

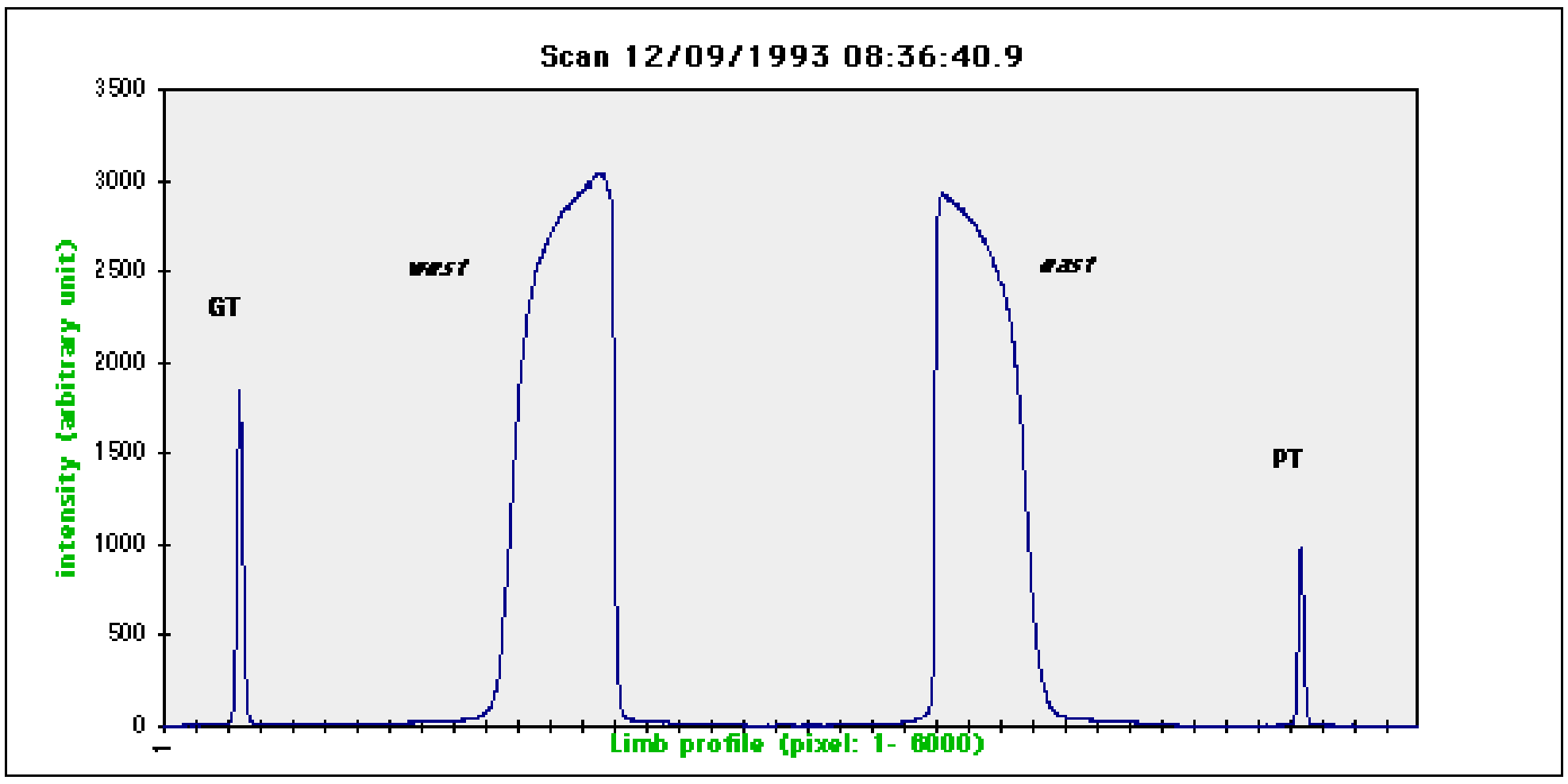

Fig. 2.22: Perfil típico do limbo solar obtido pelo Heliômetro do Pic-du-Midi (Rösch *et al.*, 1996).

O diâmetro do Sol medido é considerado como a distância entre os pontos de inflexão das curvas de intensidade (Rösch e Yerle, 1983 e Rösch *et al.*, 1996).

## *2.5 - Instrumentos de monitoramento do diâmetro solar*

O Heliômetro do Pic-du-Midi, o DORAYSOL, o Monitor de Imagens Solares de Calern, o telescópio Solar de Lucerna e mais o Astrolábio de Calern e o Astrolábio do Observatório Nacional, que lidera a Rede de Monitoramento Solar, são todos instrumentos baseados em solo, para a observação do diâmetro solar.

Entre as medidas espaciais estão as obtidas através das imagens do Satélite SOHO (Emilio *et al.*, 2001) e dos dados do satélite RHESSI (Fivian, *et al.*, 2004).

Espera-se para 2010 o lançamento dos micro-satélites Picard e SDO. Com estes satélites farão medidas absolutas:

- da irradiância solar total;
- do espectro;
- do diâmetro e a forma do Sol;
- de Heliossismologia.

Estas séries de medidas espaciais, quando comparadas às séries feitas pelos instrumentos de solo, servirão para o aprimoramento dos modelos atmosféricos, que serão usados para verificar a influência da atmosfera nas medidas do Sol.

# Capítulo 3 - Construção e testes dos protótipos

Existem diversas maneiras de, opticamente, fazer-se a duplicação do campo de um telescópio. Neste capítulo, serão mostrados os instrumentos protótipos construídos e testados a fim de que se tomasse a decisão técnico-científica de qual tipo de heliômetro seria construído, e como ele seria construído.

## *3.1 - As quatro técnicas heliométricas estudadas*

1. Uso de prismas montados após a ocular do instrumento (mesmo princípio duplicador usado em microscópios binoculares);
2. Disposição de dois espelhos planos, na frente da objetiva do instrumento, contidos em planos que formam um diedro;
3. Corte de um espelho telescópico ao longo de seu diâmetro;
4. Corte da objetiva ao longo de seu diâmetro (método clássico).

A idéia do uso do corpo do Astrolábio Danjon foi abandonada e para cada uma das técnicas supracitadas, um protótipo foi construído.

O número do protótipo corresponde à técnica utilizada para a duplicação das imagens.

### 3.1.1 - Protótipo nº1

O princípio binocular de microscópio (dois prismas mais um prisma divisor de 50%) foi utilizado para duplicar as imagens.

Os prismas foram montados num suporte, de forma a girarem em torno do seu eixo óptico, a fim de trazer as imagens próximas uma da outra.

O sistema foi montado num telescópio Sky-Watcher de 1000 mm de distância focal focal, figura 3.1.

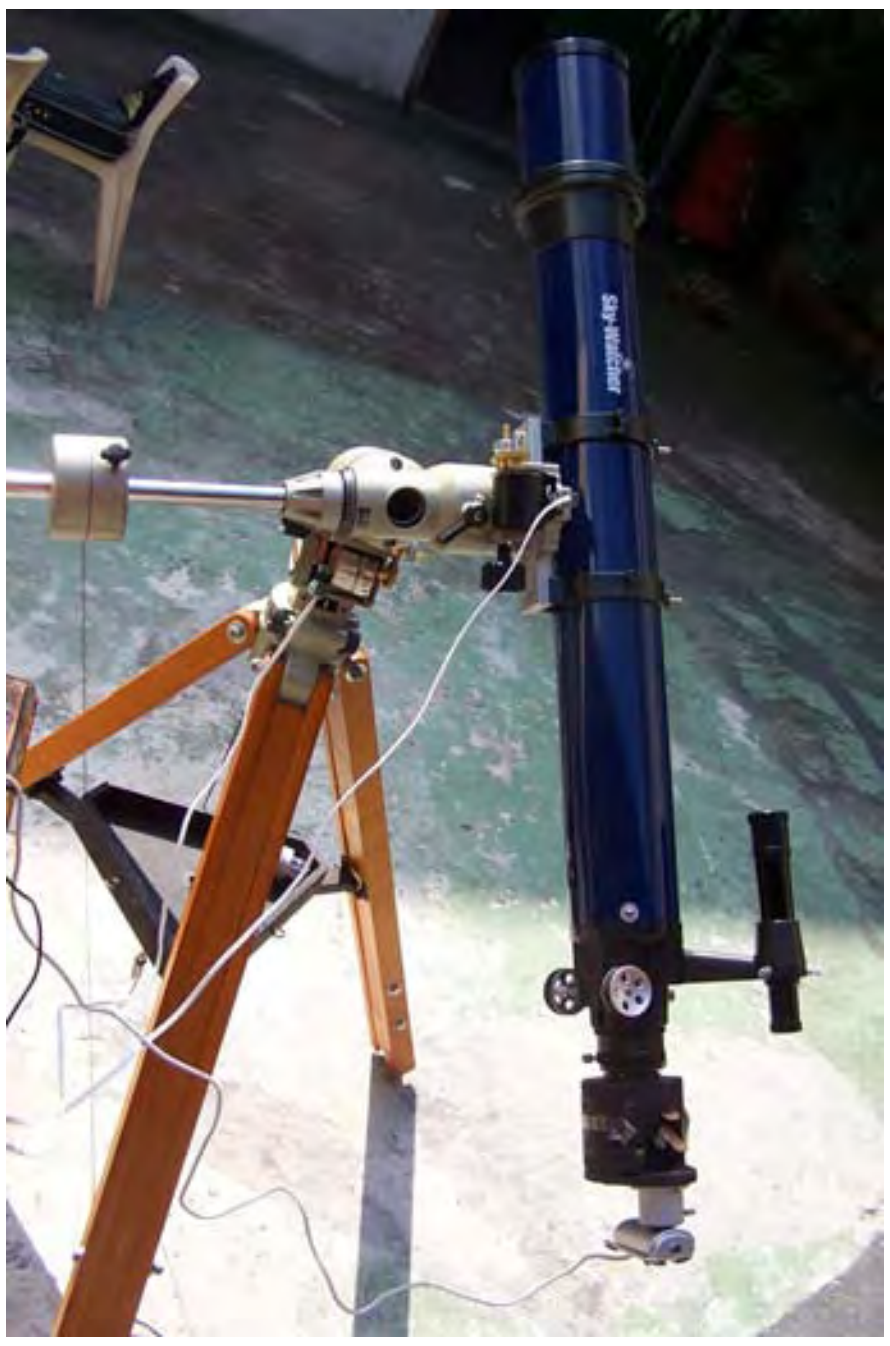

Fig. 3.1: Esquema da montagem dos prismas e do caminho óptico.

Nem todas as partes ópticas destes prismas puderam ser montadas, por isso os caminhos ópticos das imagens têm comprimentos diferentes após a duplicação, figura 3.2.

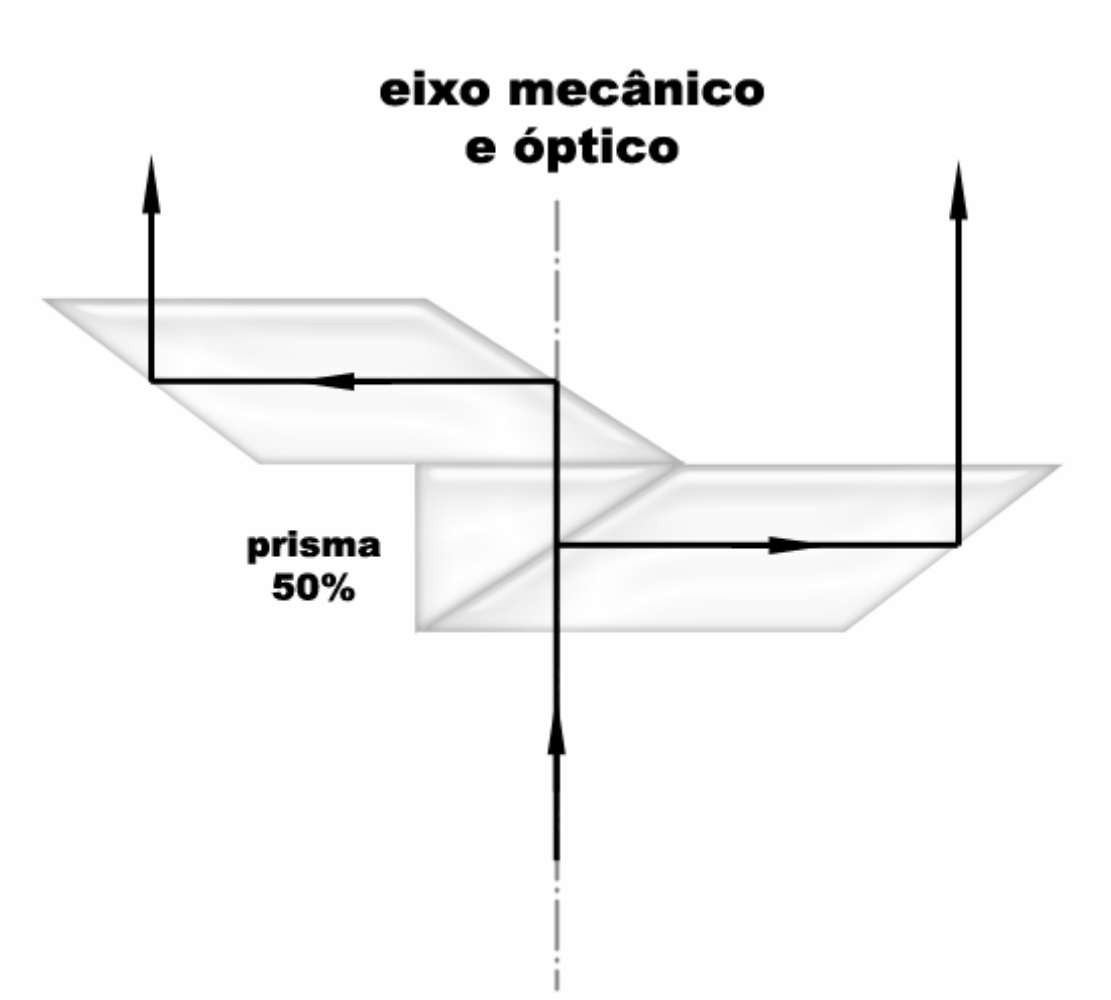

Fig. 3.2: Prismas e webcam encaixados em um telescópio.

Com isso, as imagens dos discos solares ficaram, visivelmente, com diâmetros diferentes, figura 3.3.

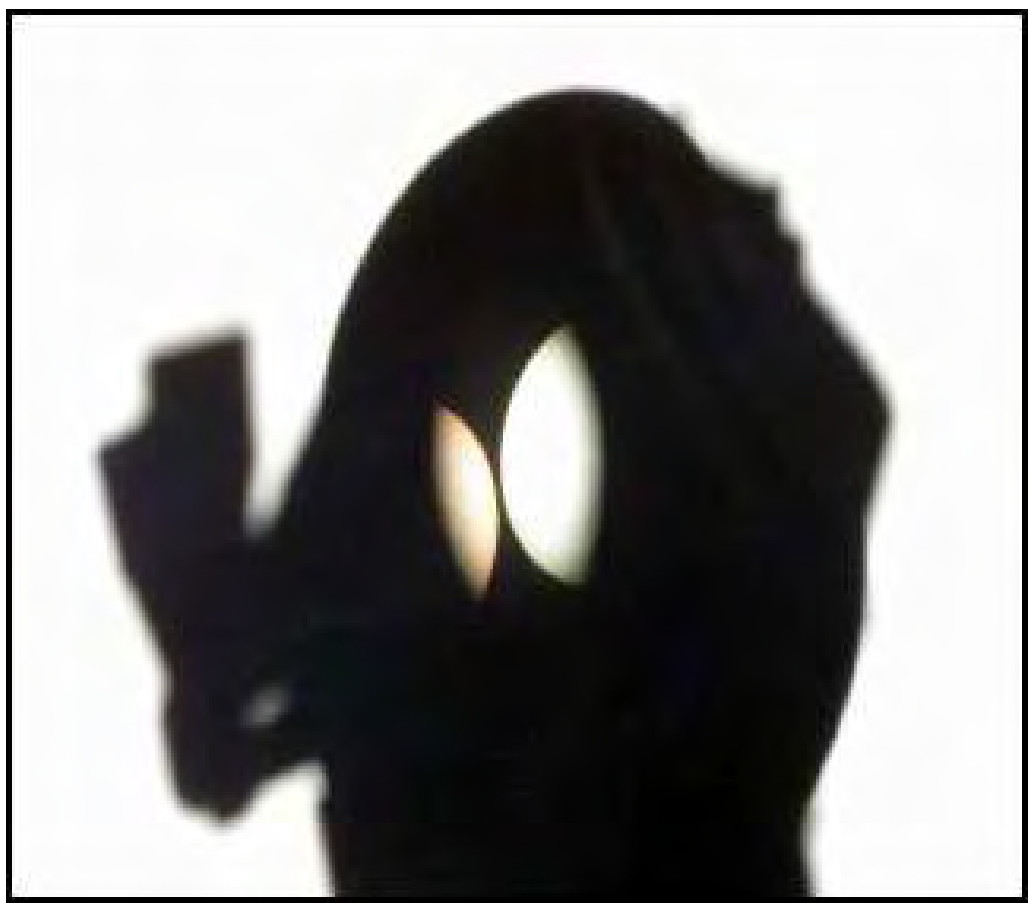

Fig. 3.3: Imagem da luneta de 500 mm, com os prismas. Os discos solares duplicados estão sendo projetados num anteparo.

Os discos solares também se apresentaram com cores diferentes devido à deterioração da aluminização do prisma divisor de 50%.

Pelos problemas apresentados nesta montagem, nenhuma sessão de aquisição de imagens foi feita com este primeiro protótipo.

Este método de duplicação tem a vantagem de poder ser adaptado a qualquer telescópio.

### 3.1.2 - Protótipo nº2

Dois espelhos planos, montados em diedro, fornecem a um telescópio, imagens já duplicadas do disco do Sol.

Estes espelhos foram montados numa base de madeira, fixados com uma fita adesiva dupla-face emborrachada, figura 3.4.

Parafusos localizados atrás da base faziam o ajuste do ângulo do diedro. O plano da base foi ajustado num ângulo de 45º e preso a uma estrutura, que poderia ser adaptada na frente de um telescópio.

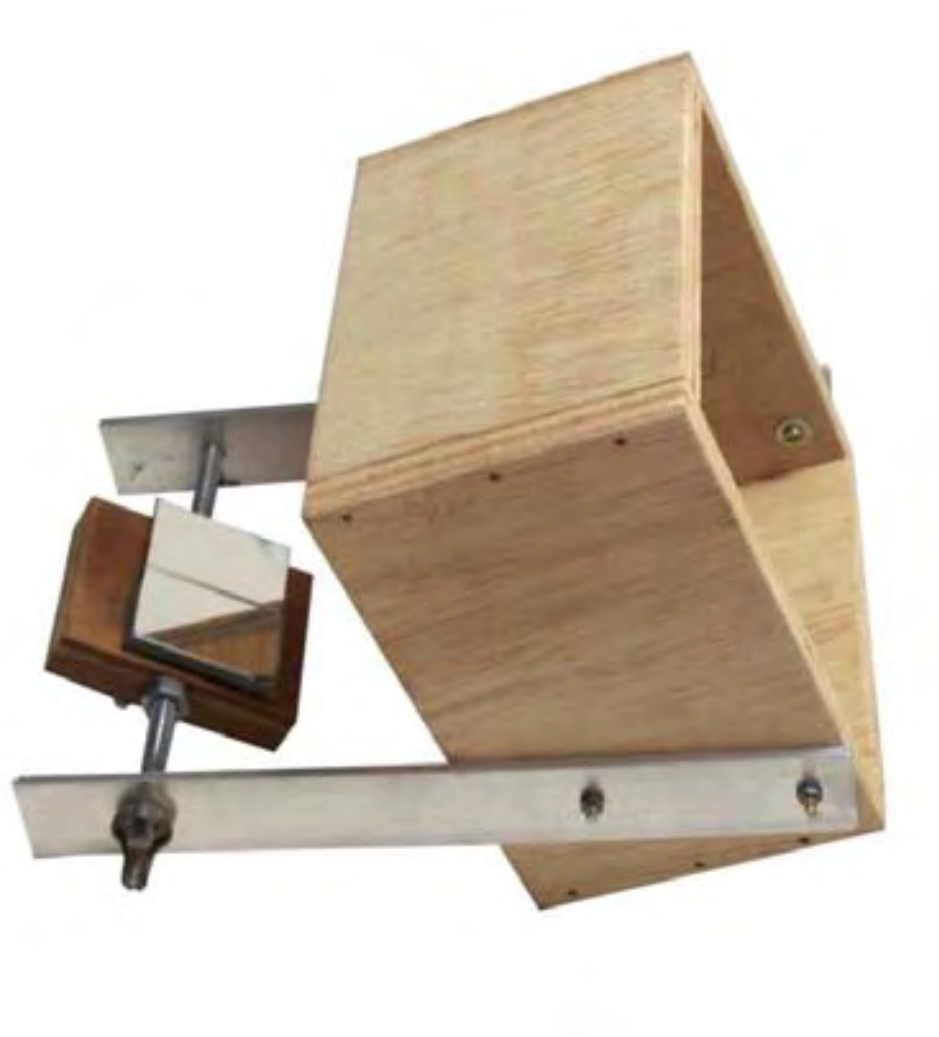

Fig. 3.4: Detalhes da montagem dos espelhos duplicadores.

O sistema duplicador foi montado na frente da objetiva de um telescópio refrator (luneta Sky-Watcher) de 1000 mm de distância focal, figura 3.5.

O filtro solar utilizado foi um *Baader Planetarium*. Este filtro é constituído por uma película (polímero) extremamente fina, cuja precisão da janela óptica, segundo o fabricante, é de cerca de $\lambda/10$.

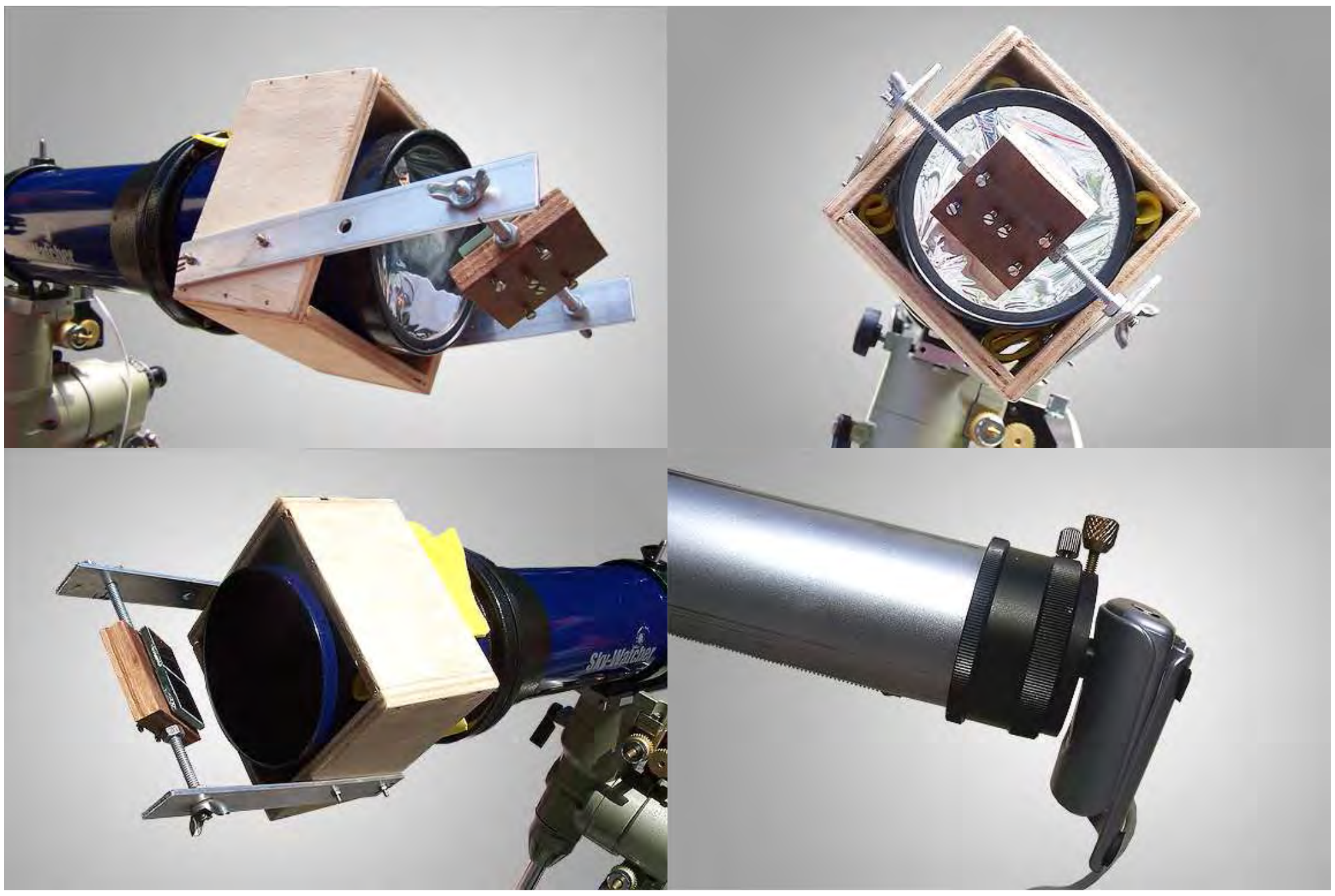

Fig. 3.5: Detalhe da montagem dos espelhos duplicadores, do filtro solar e da WebCam.

Com este protótipo foram obtidas as primeiras imagens heliométricas deste projeto, no dia 10 de julho de 2006, através de uma câmera *Philips PCVC840K ToUcam*. Foram obtidas 32 imagens, em seqüência e com acompanhamento, entre 12h17min e 12h26 min, no formato *BMP*, com 352×288 *pixels*, figura 3.6.

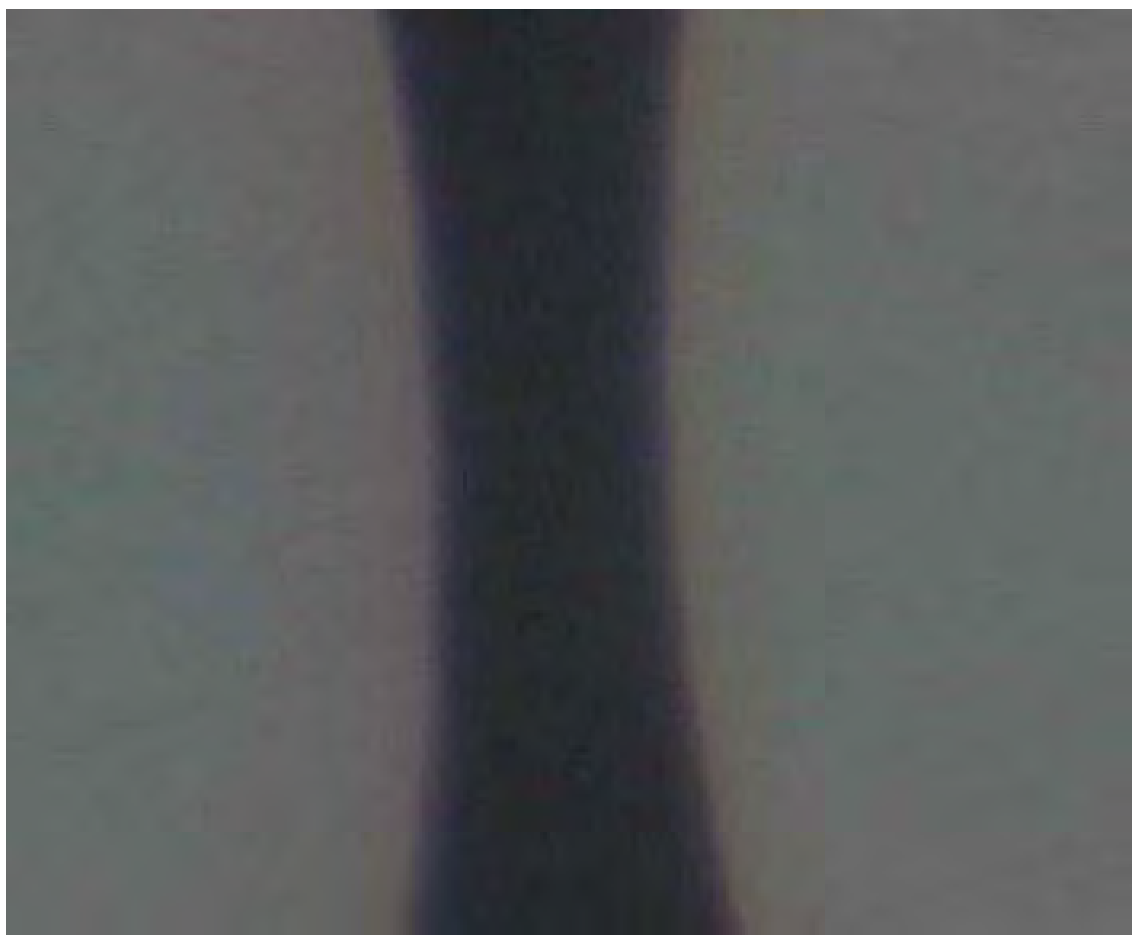

Fig. 3.6: Uma das 32 imagens dos discos solares duplicados, obtidas com o protótipo nº2.

As imagens dos discos duplicados são de baixa resolução, mas isso não foi impedimento para o desenvolvimento das rotinas de análise (comentadas mais adiante no Capítulo 6).

Mais uma vez este método de duplicação se mostrou viável, com a mesma vantagem do método anterior de poder ser adaptável em qualquer telescópio.

### 3.1.3 - Protótipo nº3

#### 3.1.3.1 - Versão A

Neste método foi testada a duplicação da imagem do Sol através do afastamento linear, perpendicular ao corte, das metades de um espelho, mantendo os eixos ópticos paralelos entre si.

Para um espelho de 1 m de distância focal, o afastamento linear de 9 mm já é o suficiente para um deslocamento maior do que 0,5º, separando as imagens dos discos solares, figura 3.7.

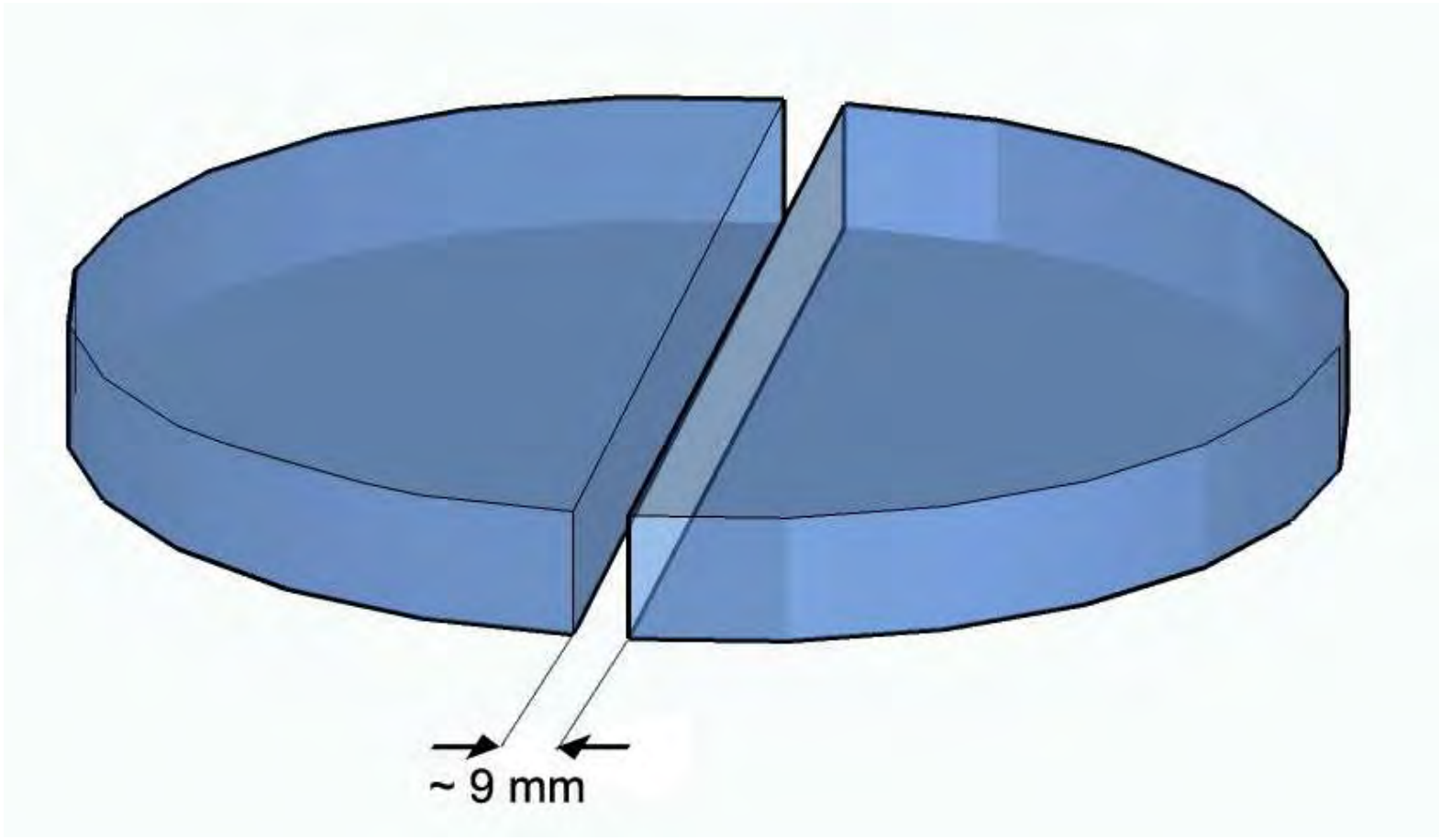

Fig. 3.7: Esquema da configuração do espelho duplicador.

Ao contrário dos dois primeiros protótipos, que poderiam ser adaptados em vários instrumentos de observação astronômica, para este protótipo, um instrumento próprio teve de ser construído e uma montagem dobsoniana foi escolhida, uma vez que não há necessidade de acompanhamento equatorial, figura 3.8. Ela também é mais adequada pela praticidade da construção e, conseqüente, baixo custo. A câmara CCD foi posicionada diretamente no plano focal do hemi-espelhos, simplificando ainda mais sua montagem.

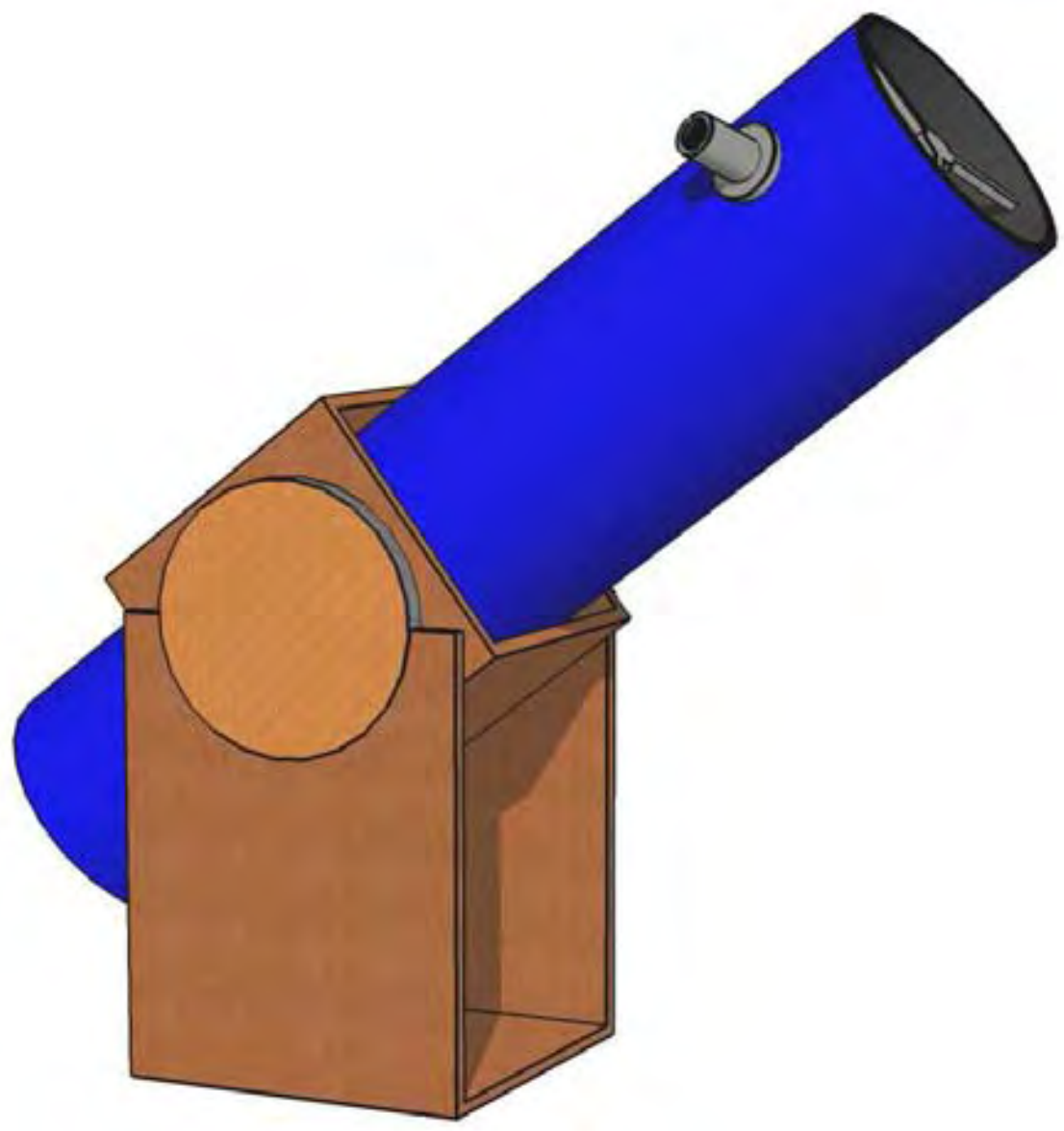

Fig. 3.8: Esquema de uma montagem dobsoniana.

### 3.1.3.2 - Preparação do espelho

Primeiro um disco de vidro plano e comum foi cortado ao meio e remontado, e em seguida fixado numa forma para desbaste, lapidação e polimento (fig. 3.9). A distância focal realizada foi de 960 mm, próxima portanto daquela planejada para o Heliômetro definitivo, de 1000 mm, à exemplo do Astrolábio Danjon, adequada após o pupilamento, ao tamanho da janela de coerência atmosférica.

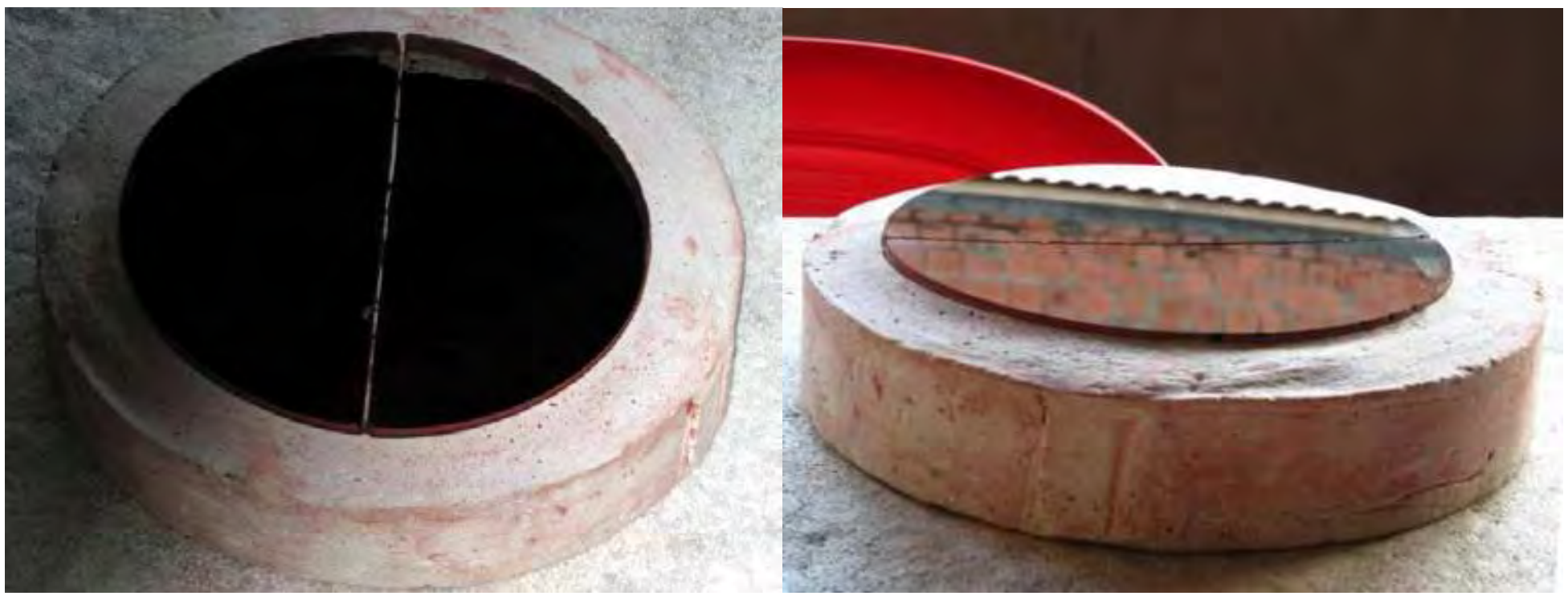

Fig. 3.9: Espelho em sua forma, já polido.

Este processo, no entanto, resultou no aparecimento de pequenas depressões, evidenciadas pelos Testes de Foucault e de Ronchi, nas bordas do espelho junto ao corte (fig. 3.10), demonstrando que esta metodologia de fabricar-se um espelho com partes já cortadas não é indicada para a fabricação do espelho definitivo.

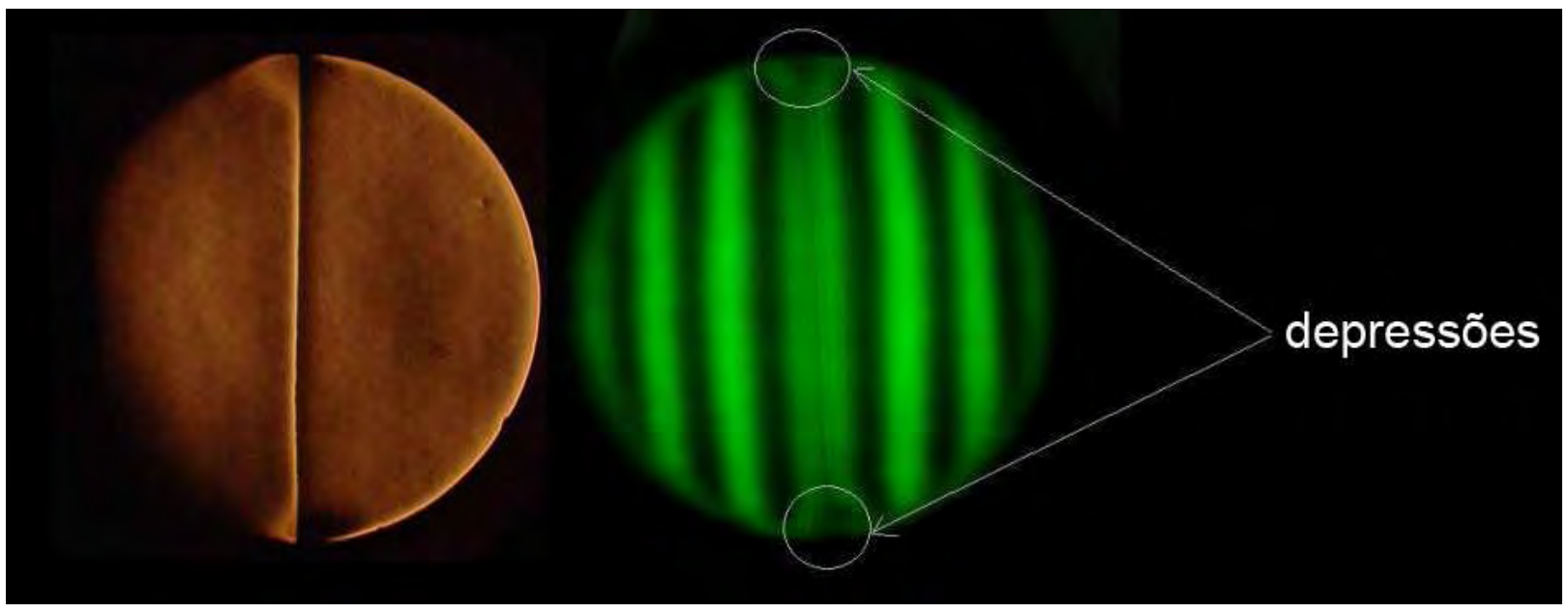

Fig. 3.10: Teste de Foucault e de Ronchi para avaliação da superfície.

A planicidade do fundo do espelho evita o desnível entre as bordas e a região do corte, resultando em dois eixos ópticos, se cruzando em convergência quando o espelho é montado sobre sua base plana, figura 3.11. O processo de desbaste foi repetido até atingir esta condição.

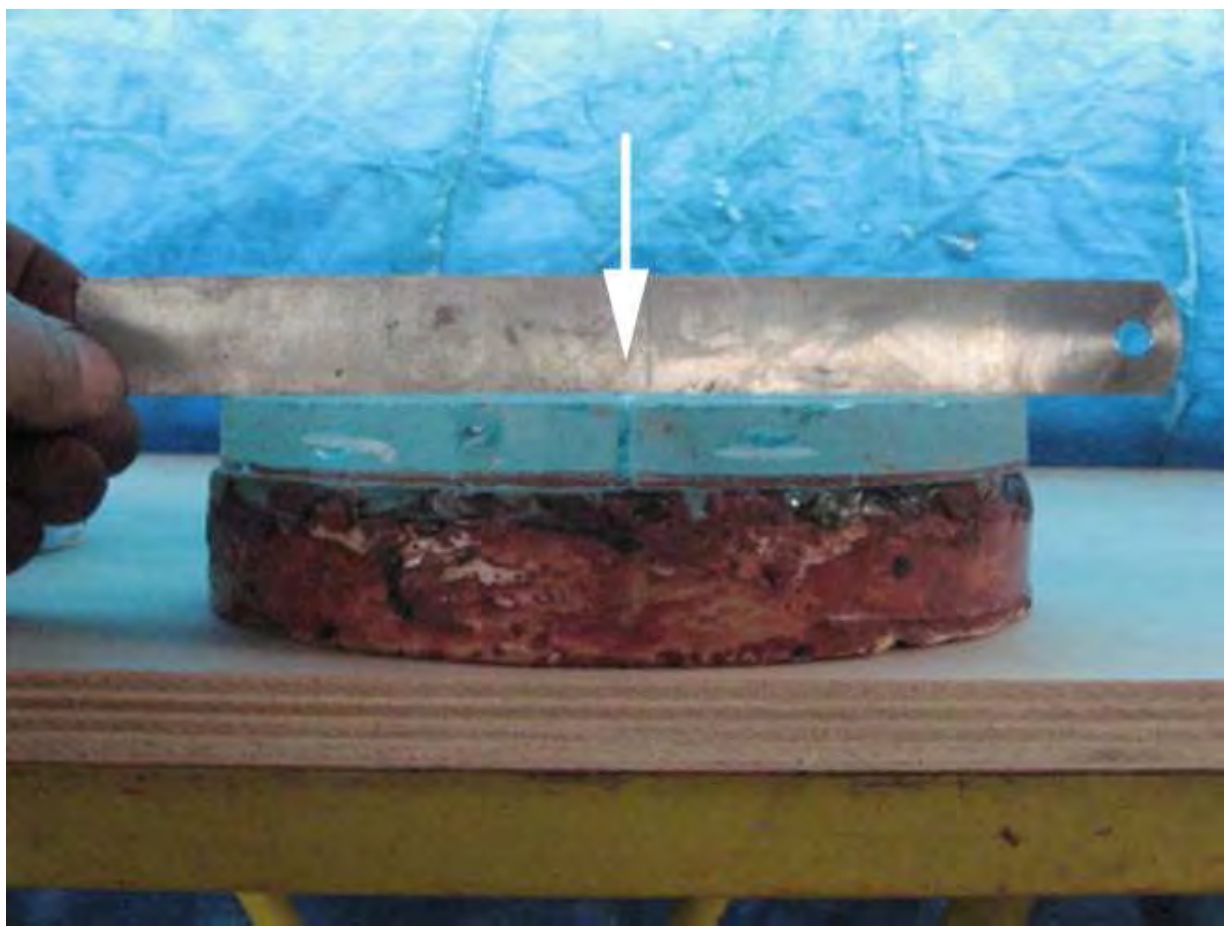

Fig. 3.11: Espelho sobre a ferramenta. A seta indica o desnível evidenciado pela régua.

Para tanto, as metades do espelho foram pousadas sobre sua própria ferramenta de polimento de forma a recompor o espelho. O conjunto foi colocado dentro de outra forma e fixado com gesso branco comum (fig. 3.12).

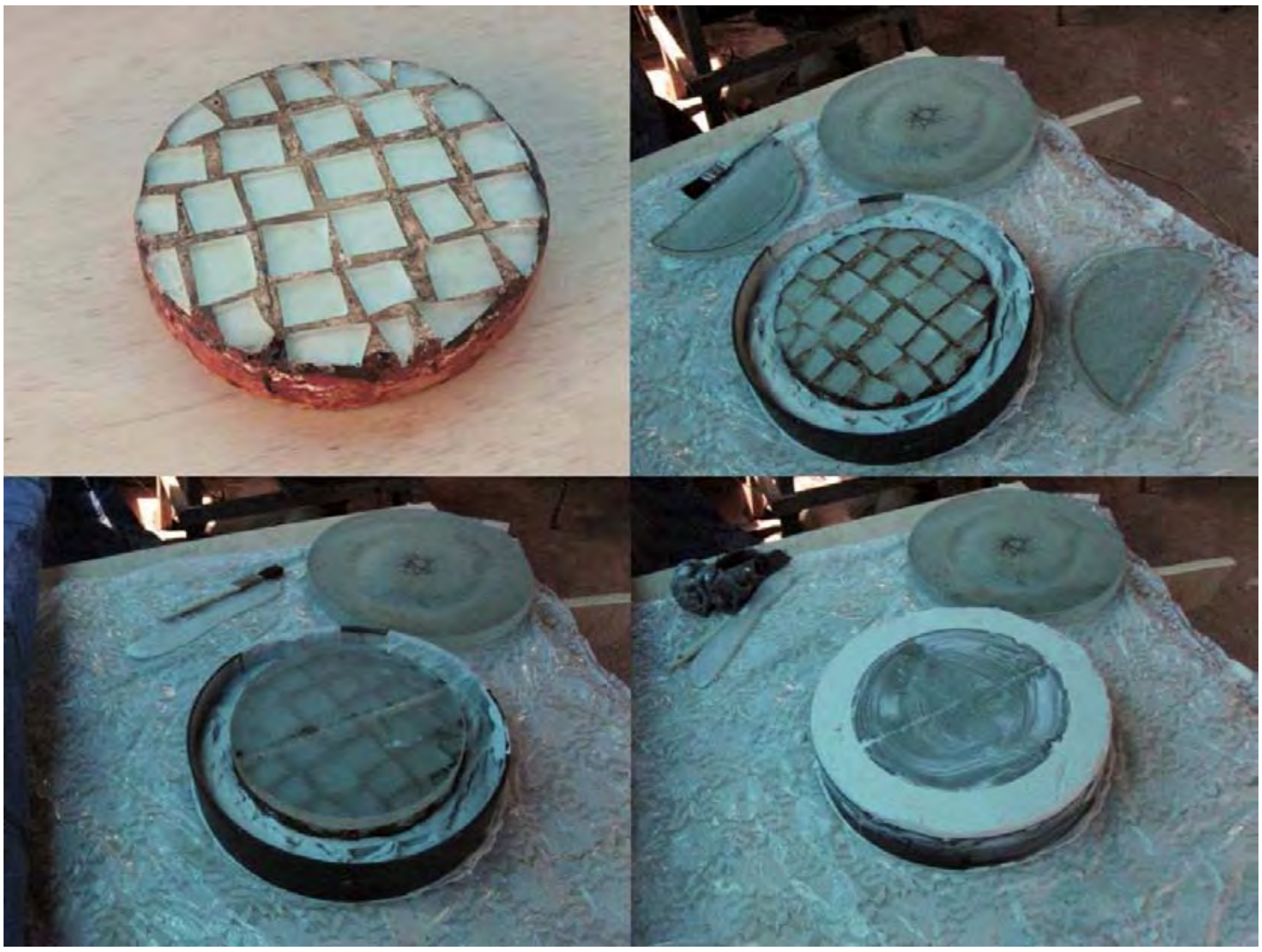

Fig. 3.12: Seqüência de fixação do espelho sobre sua ferramenta de polimento.

Após a secagem do gesso o fundo do espelho foi esmerilhado contra sua base, até que o esferômetro utilizado para medir a planicidade não mais acusasse curvaturas, figura 3.13.

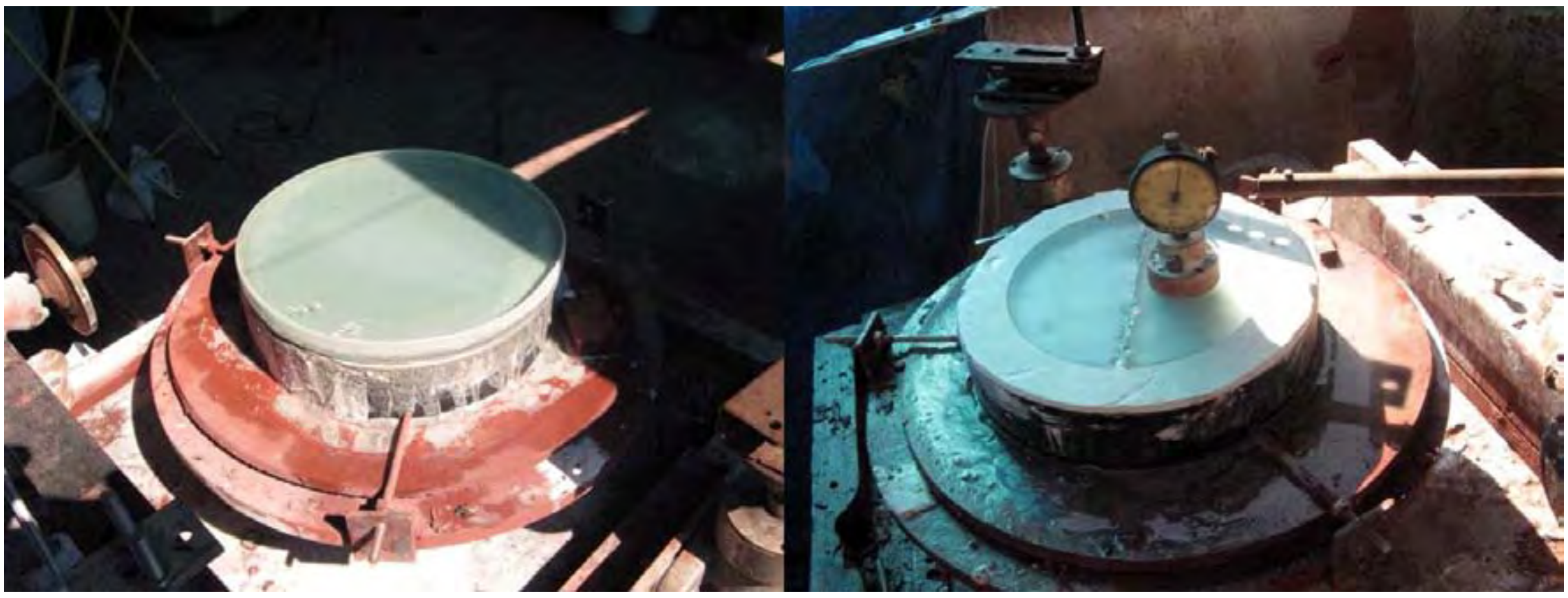

Fig. 3.13: À esquerda se vê a base de desbaste sobre o fundo do espelho e à direita, a planicidade do fundo do espelho testada com um esferômetro.

Na etapa seguinte, passou-se à construção do suporte do espelho, denominada célula. Este suporte possui parafusos com molas que permitem a colimação óptica do espelho dentro do tubo do instrumento, figura 3.14.

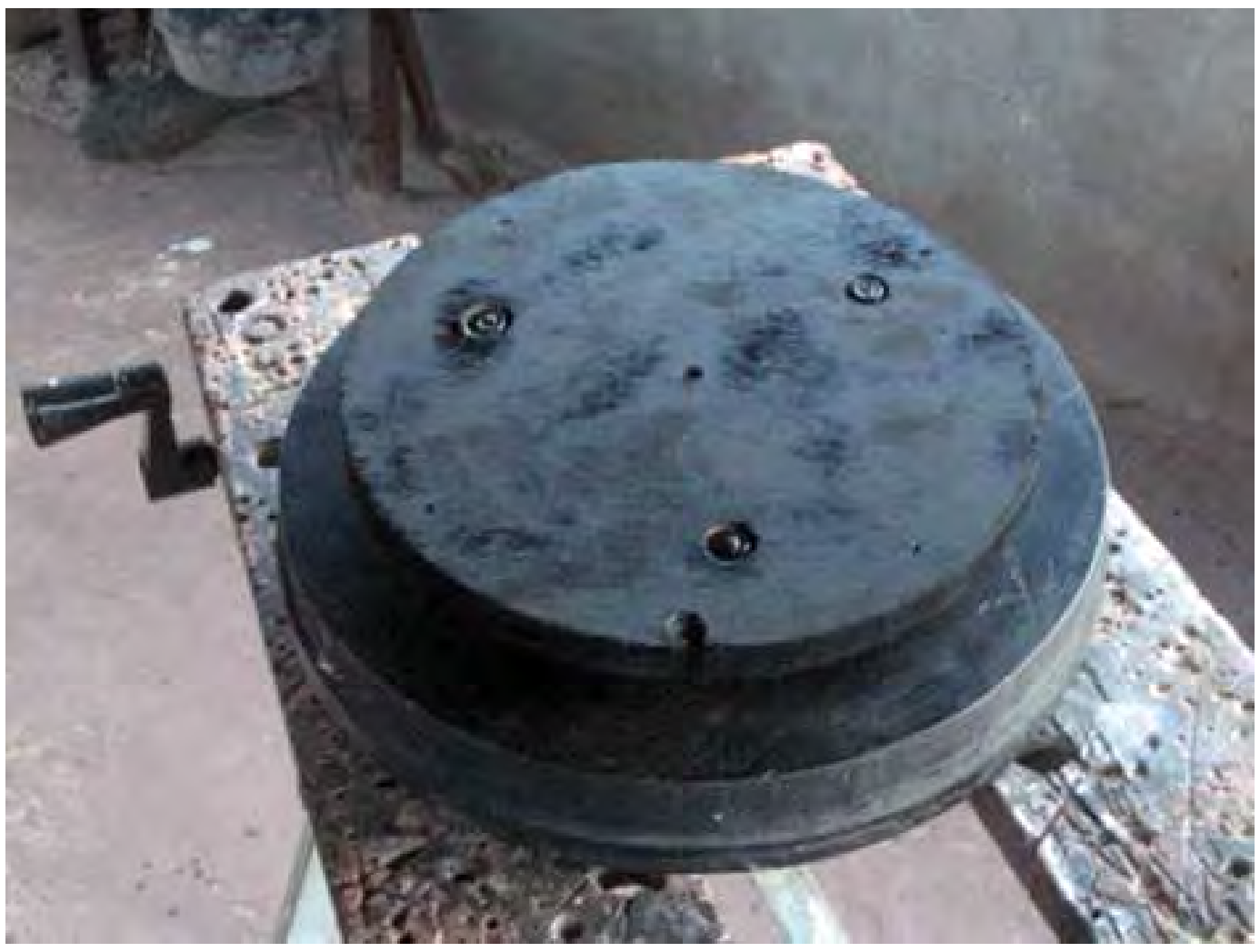

Fig. 3.14: Suporte do espelho construído de madeira.

Em seguida a base do espelho foi fixada ao suporte e os hemi-espelhos, ainda soltos, posicionados sobre ela, figura 3.15.

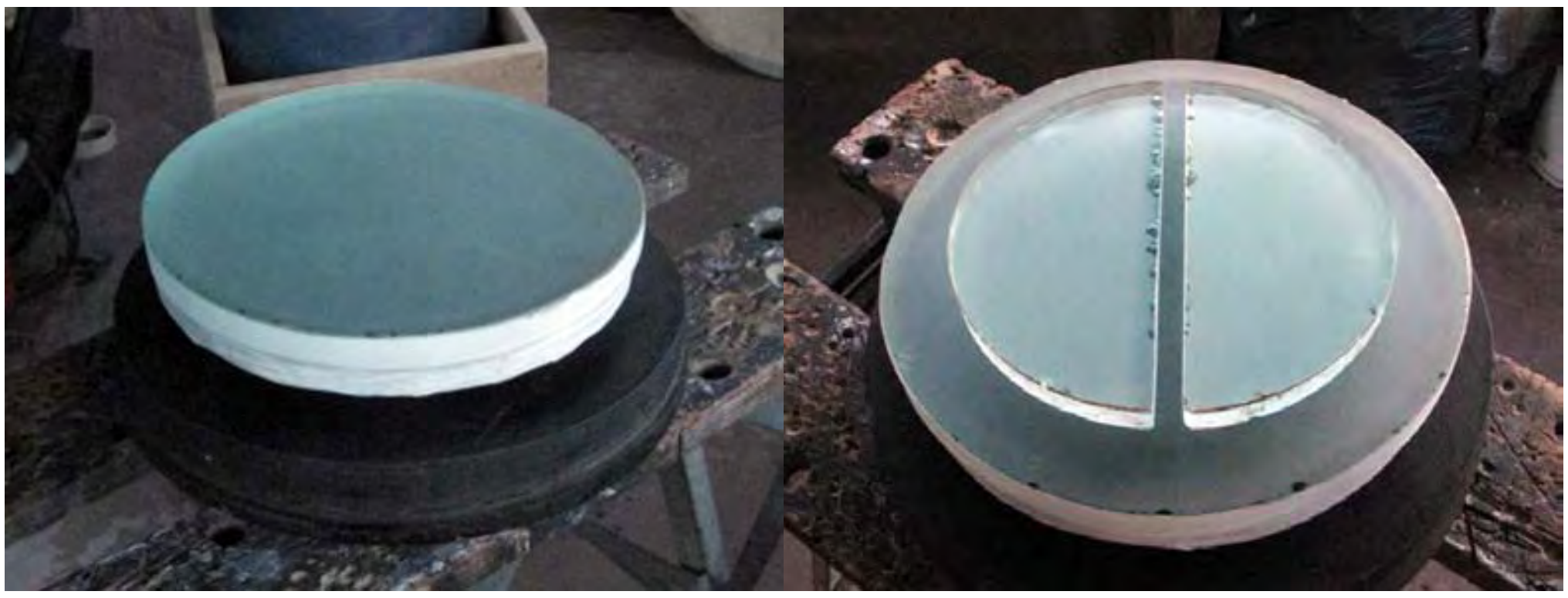

Fig. 3.15: À esquerda, fita foi usada para fixação da base ao suporte. À direita, as metades do espelho pré-posicionadas sobre a base.

Com peças auxiliares de suporte, em madeira, o espelho foi ajustado à base na configuração desejada.

Em um teste de projeção do disco solar sobre um anteparo verificou-se que um afastamento lateral linear de apenas 5 mm, entre as metades do espelho, era o suficiente para duplicar e separar os discos solares (fig. 3.16). Isto significava que apesar do fundo dos hemi-espelhos estarem planos, existia entre eles um deslocamento angular de ~0,1º, que compensava o afastamento linear menor do que o calculado.

Depois de ajustado em sua configuração final, o espelho foi fixado à base. Uma máscara cobriu cerca de 1 cm em torno das bordas das metades para minimizar os efeitos das depressões na lapidação.

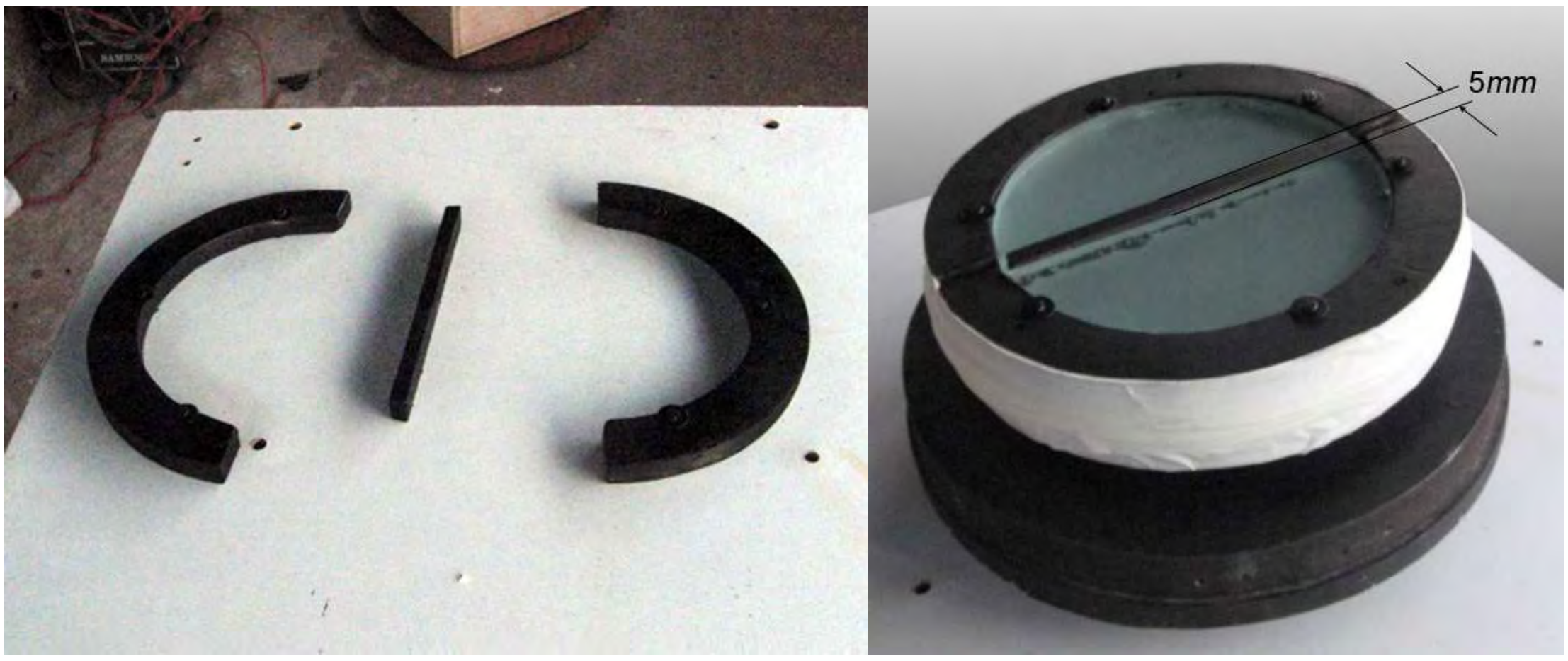

Fig. 3.16: Detalhe das peças de suporte dos hemi-espelhos e da montagem completa da célula.

O instrumento foi apoiado horizontalmente e a célula do espelho foi fixada ao tubo. O alinhamento, com precisão de 1 mm, foi efetuado visualmente com o auxílio de uma fonte *laser*, figuras 3.17 e 3.18.

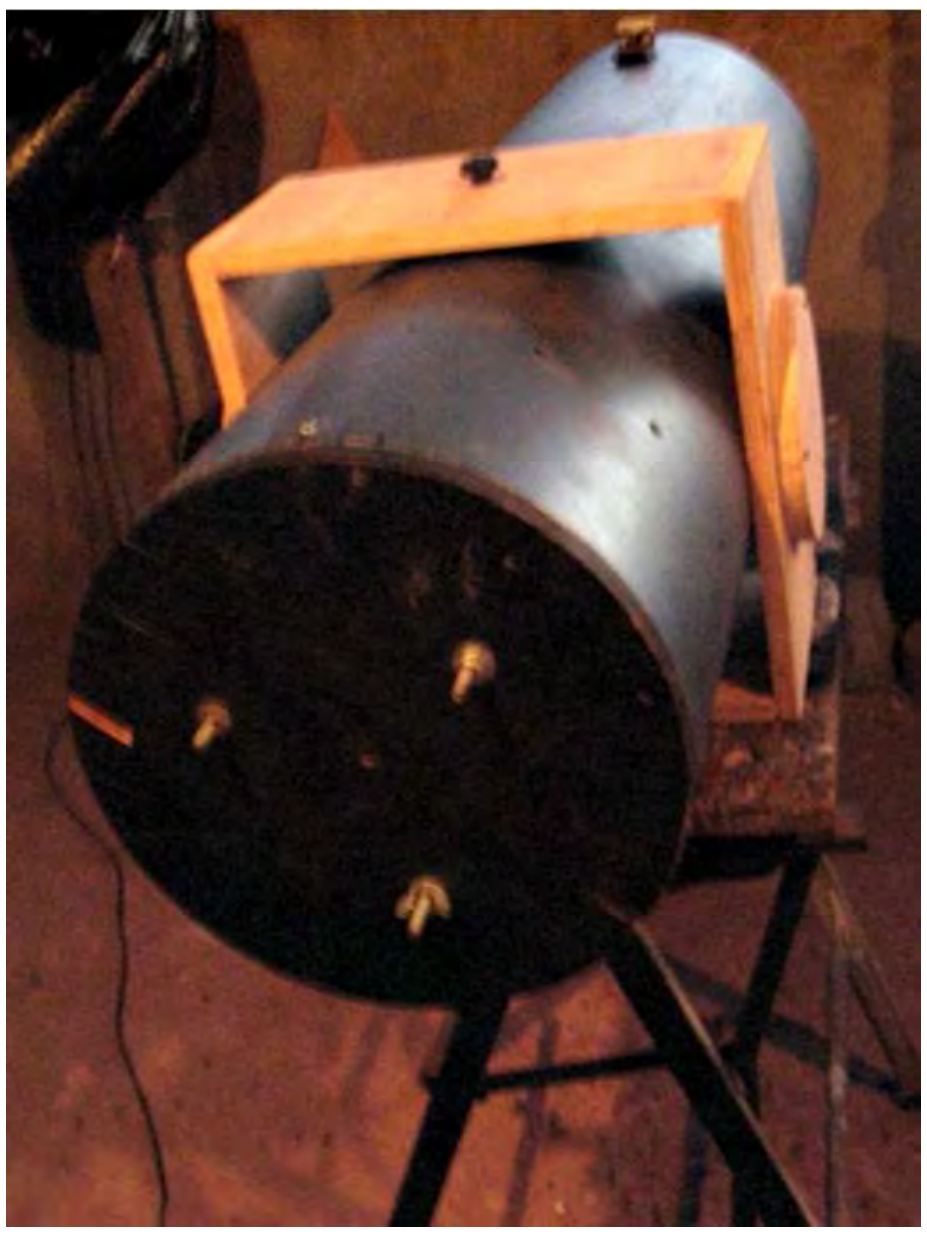

Fig. 3.17: Detalhe da parte de trás do instrumento, ainda na fase de sua montagem.

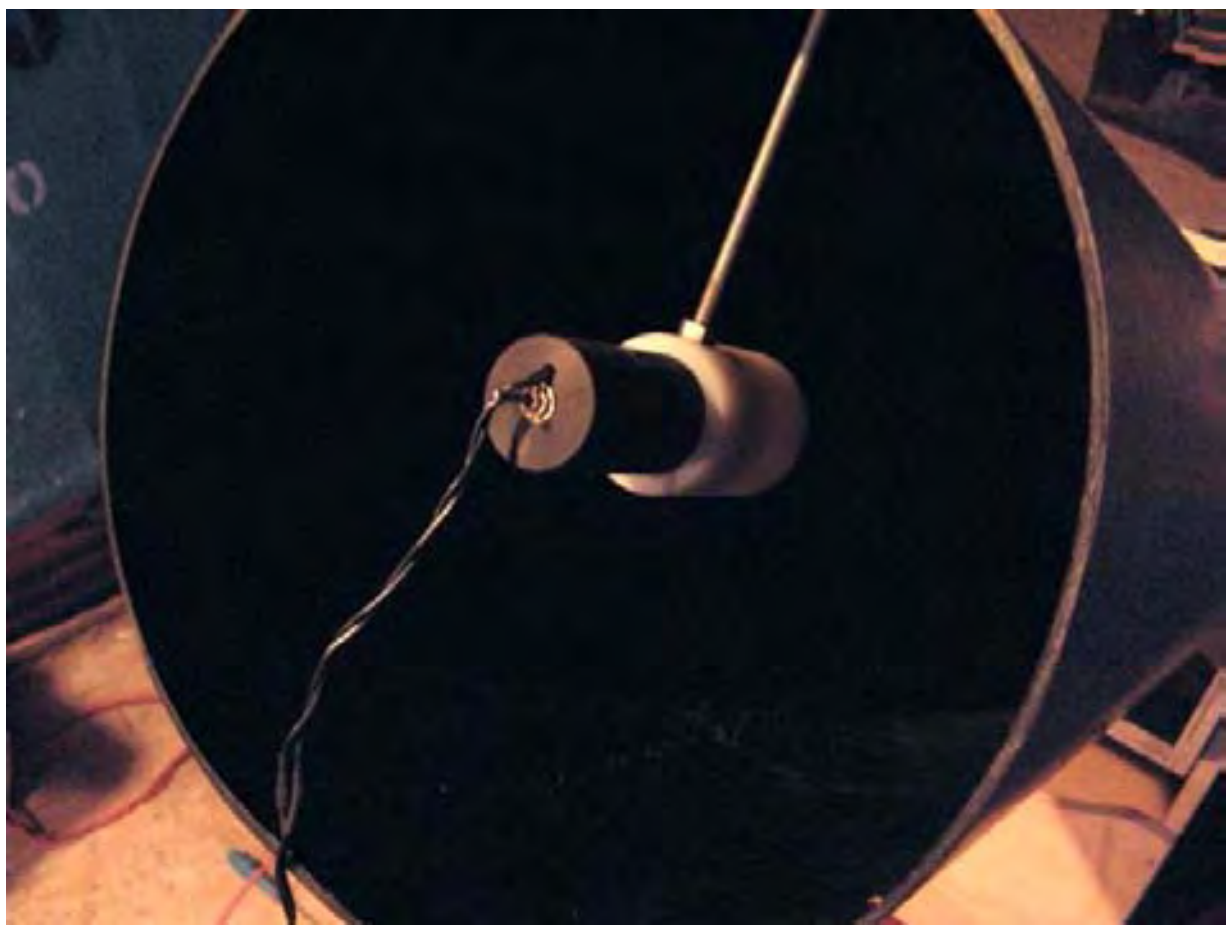

Fig. 3.18: Detalhe da entrada do tubo do instrumento. A fonte laser para o alinhamento pode ser vista encaixada no suporte montado para câmera CCD.

Localizado próximo à altura do plano focal do espelho heliométrico, foi montado o focalizador (fig. 3.19). Seu curso de 10 mm é suficiente para colocar a câmara na posição exata. O telescópio praticamente montado pode ser visto na figura 3.20.

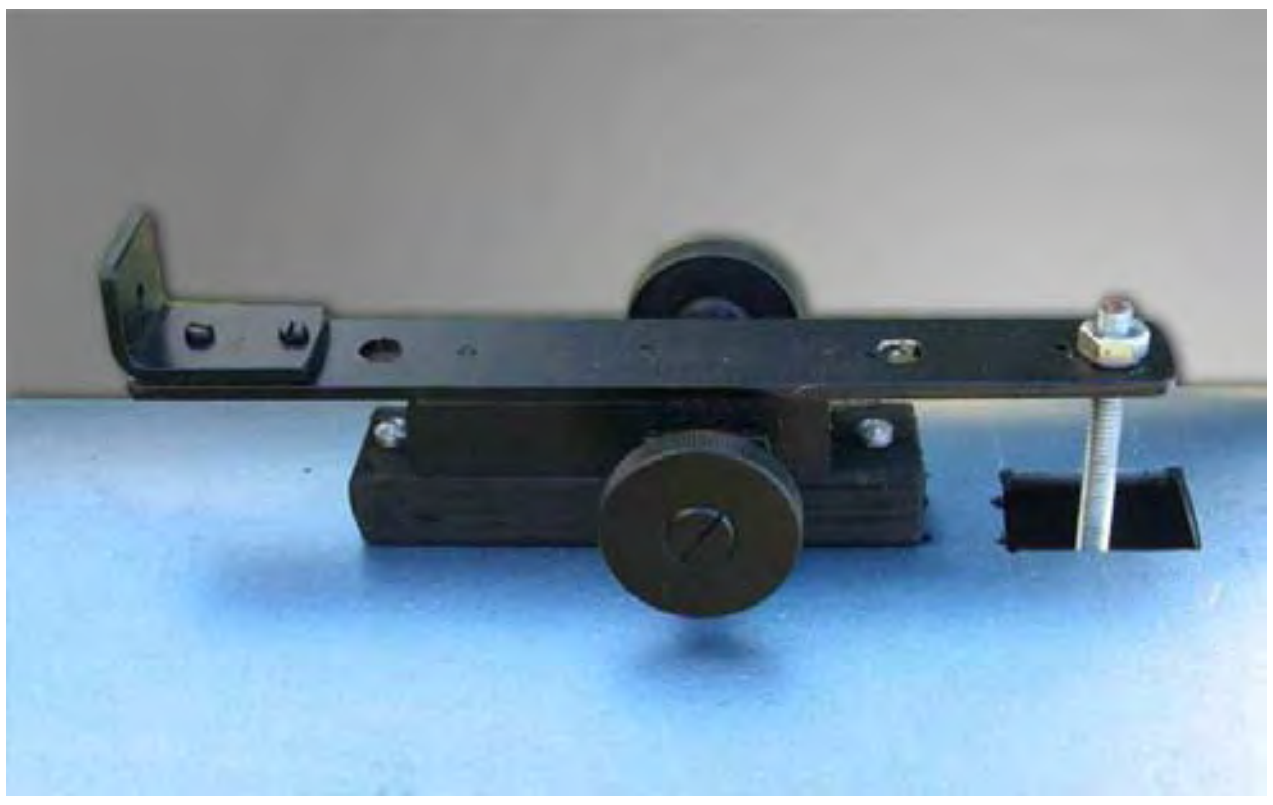

Fig. 3.19: Detalhe do focalizador do instrumento.

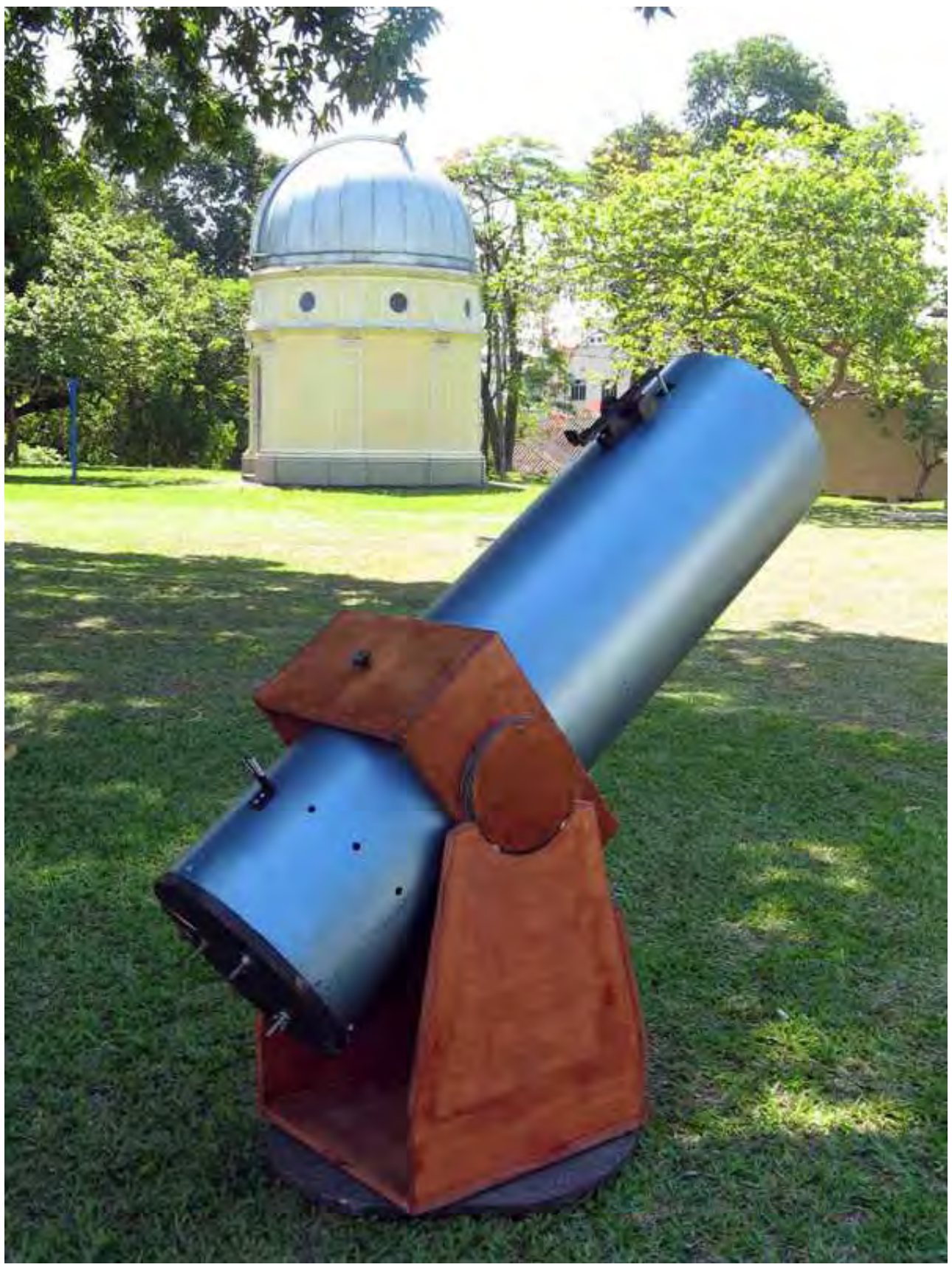

Fig. 3.20: Protótipo nº 3A, no campus do Observatório Nacional, preparado para testes.

O filtro solar utilizado foi um *Baader Planetarium*. Uma *webcam Philips ToUCam Pro 740K* fornecia as imagens, que eram gravadas diretamente em disco, figura 3.21.

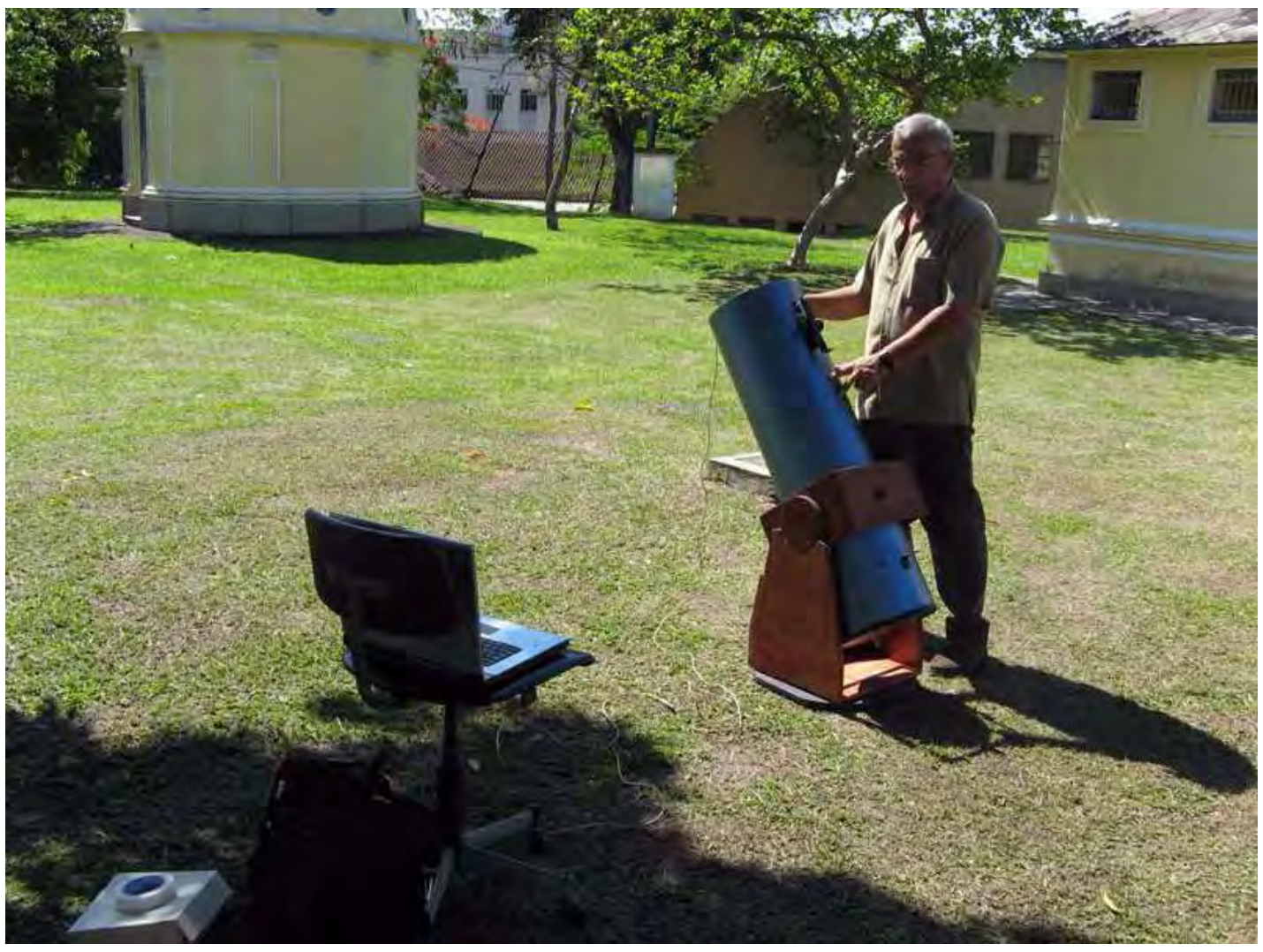

Fig. 3.21: Sessão teste de aquisição de imagens heliométricas do Sol, em 24 de novembro de 2006. O notebook próximo ao telescópio armazenava as imagens.

Para se implementar uma das principais características do heliômetro final, que é a de poder observar o disco do Sol em diferentes heliolatitudes numa mesma sessão observacional, este protótipo foi modificado para permitir que seu tubo girasse sobre seu eixo óptico. Solidariamente ao tubo, também giram o espelho e a câmara CCD. O resultado deste giro é a rotação dos discos solares em sentidos opostos.

Uma escala de referência foi colocada externamente ao tubo, graduada de -90º até 90º, com marcações a cada 10º. A precisão angular aqui não era importante no momento. O zero da escala foi orientado para coincidir com a linha de corte do espelho, figuras 3.22 a 3.25.

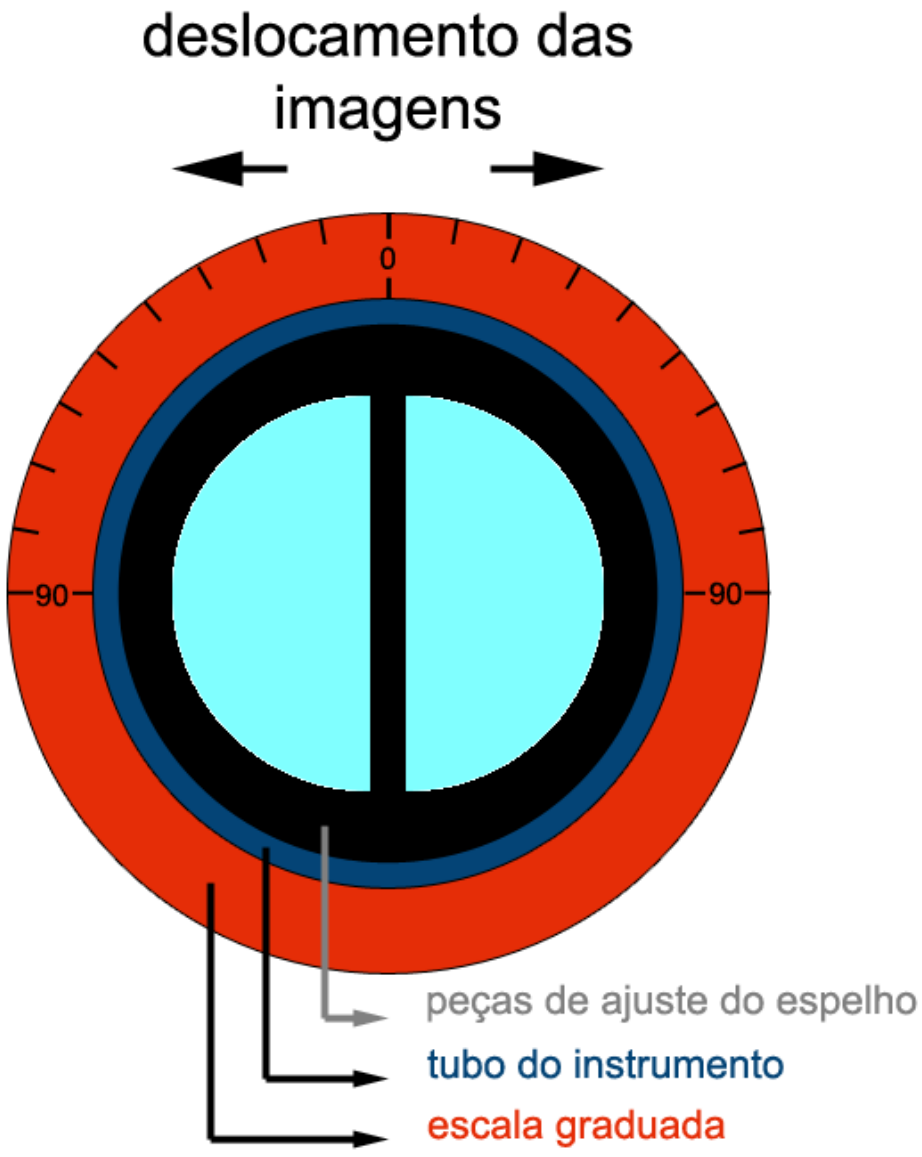

Fig. 3.22: Vista de topo do esquema de montagem da escala graduada em relação ao espelho do protótipo.

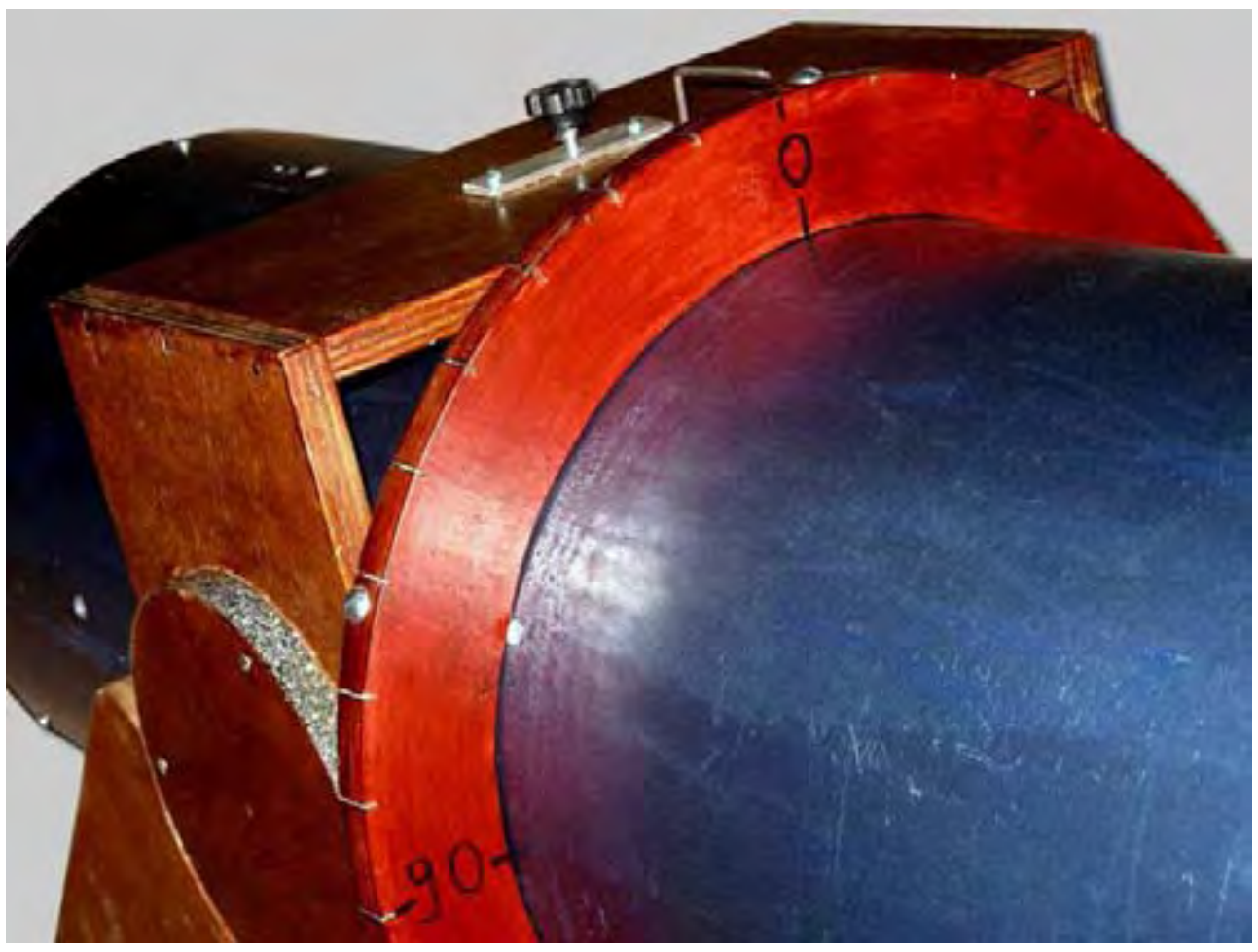

Fig. 3.23: Detalhe da escala graduada fixada ao tubo do instrumento.

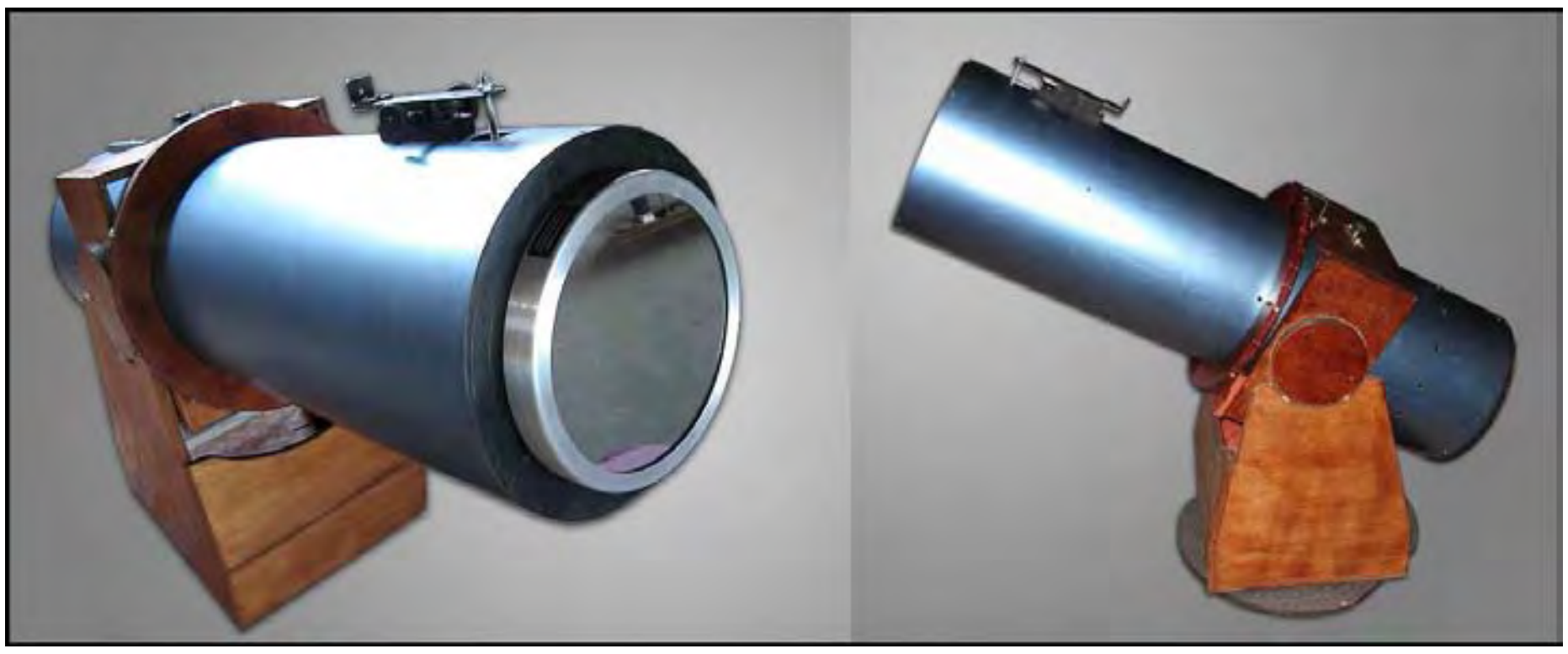

Fig. 3.24: Vistas da montagem do protótipo. À esquerda, um filtro solar *Thousand Oaks* pode ser visto fixado à boca do tubo do instrumento.

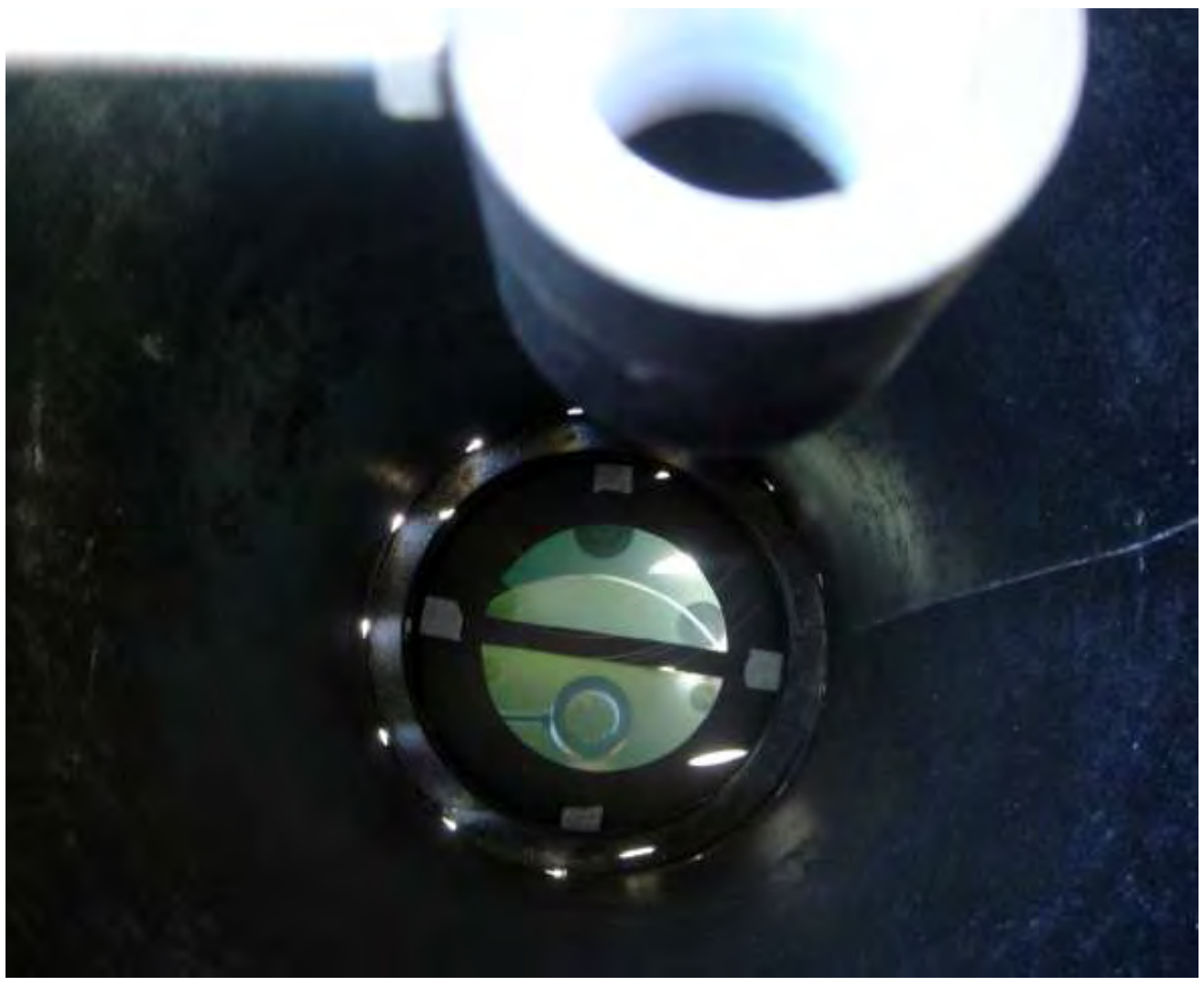

**Fig. 3.25: Vista da parte interna do tubo do instrumento onde se vê o espelho partido, com sua máscara, refletindo o suporte da câmera.**

Os resultados obtidos com este protótipo serão comentados mais adiante, quando serão comparados aos resultados alcançados por este mesmo espelho, mas montado em outra configuração para a duplicação das imagens.

### 3.1.3.3 - Versão B

Nesta versão testou-se uma nova configuração para o espelho heliométrico, já mostrada na figura 3.7. Assim, ao invés de duplicar a imagem do disco do Sol, deslocando linearmente os hemi-espelhos, eles foram deslocados angularmente, com o eixo de rotação perpendicular ao corte.

O objetivo é testar, na prática, a independência da separação angular, no sentido do desdobramento das imagens, com a distância ao plano focal. Isto deve se refletir numa maior estabilidade das imagens e, conseqüentemente, numa melhora nos resultados das medidas.

Esta configuração foi designada como 'configuração em X', figura 3.26.

Os experimentos prévios permitiram certificar um ajuste visual do ângulo, através da projeção dos discos solares num anteparo. Utilizando calços os hemi-espelhos foram posicionados na configuração desejada, ou seja, os eixos ópticos foram deslocados paralelamente ao plano do corte de pouco mais de 0,5°.

Finalmente, o conjunto foi fixado à base, com pontos de silicone e cola de composição plástica.

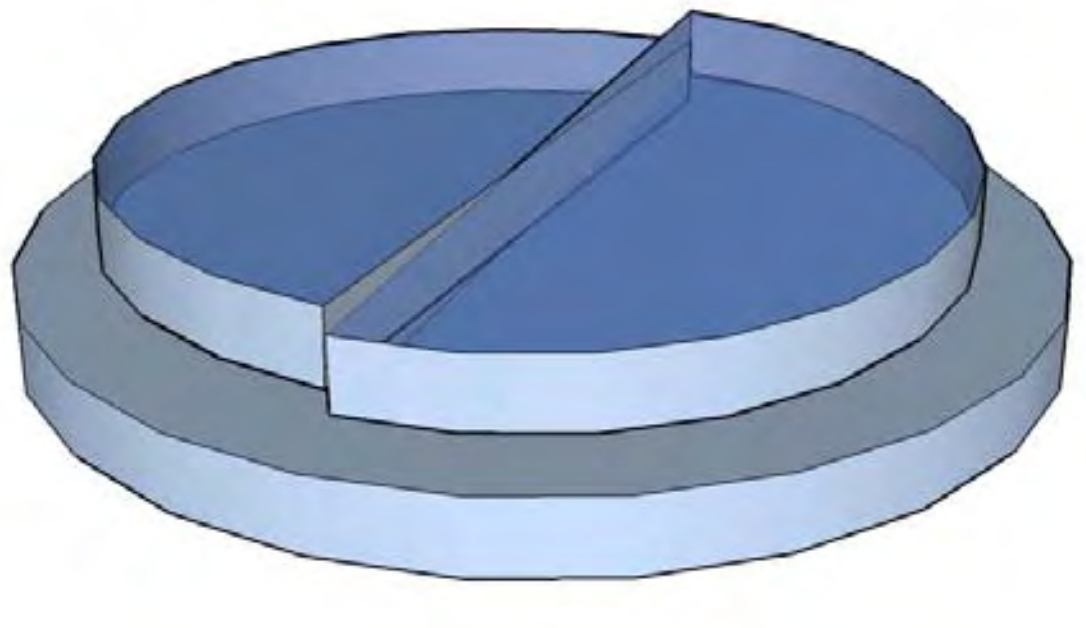

Fig. 3.26: Esquema dos espelhos, sobre a base, na configuração em X. A angulação foi exagerada para melhor visualização.

### 3.1.3.4 - resultados comparativos entre a versão A e B

O método heliométrico utilizado tanto na versão A, quanto na versão B do heliômetro refletor, funcionou como o esperado, gerando imagens duplas do disco solar.

No dia 16 de dezembro de 2006, uma sessão observacional foi realizada, ainda com a versão A do instrumento. A presença da mancha solar 930 (*AR 10930*) e de sua localização disponível de fontes especializadas (pouco menos de 10º ao sul do equador solar) permitiu se fazer a conversão entre a escala de ângulo do instrumento e a heliolatitude aproximada da medida. A tabela 3.1 traz os resultados obtidos com a experiência examinado a distância mínima entre os bordos que é a grandeza fundamental para a medida do diâmetro, como será apresentado extensamente no capítulo seguinte.

Tabela 3.1: Valores das distâncias mínimas em função da heliolatitude.

| Heliolatitude Aproximada (º) | Nº de imagens | distância mínima (*pixel*) | σ (*pixel*) |
|---|---|---|---|
| 0 | 51 | 160,22 ± 0,23 | 1,68 |
| 10 | 69 | 160,23 ± 0,17 | 1,40 |
| 20 | 47 | 161,28 ± 0,22 | 1,51 |
| 30 | 70 | 160,76 ± 0,17 | 1,46 |
| 40 | 69 | 161,41 ± 0,21 | 1,78 |
| 50 | 80 | 161,20 ± 0,24 | 2,17 |
| 60 | 87 | 161,78 ± 0,25 | 2,33 |
| 70 | 63 | 162,15 ± 0,36 | 2,88 |
| 80 | 68 | 161,18 ± 0,21 | 1,70 |
| 90 | 21 | 159,94 ± 0,34 | 1,56 |

O gráfico da figura 3.27 mostra os resíduos da distância mínima entre os bordos, em função da heliolatitude observada.

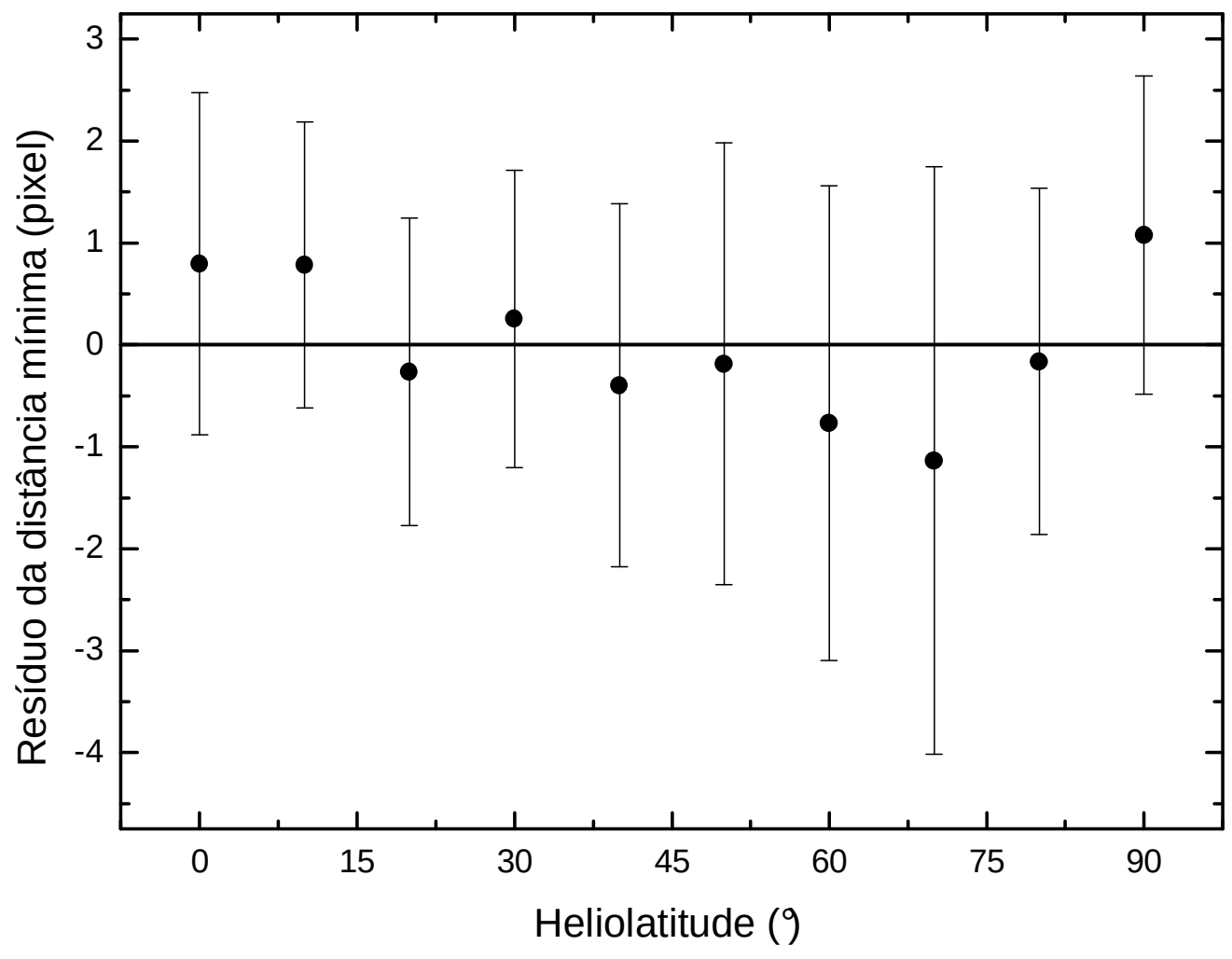

Fig. 3.27: Resíduos da distância mínima medida entre os bordos dos discos solares para diferentes heliolatitudes. A barra de erro corresponde à dispersão das medidas.

Desta experiência pode-se afirmar que o mecanismo de medidas em heliolatitudes implementado funciona como idealizado: ao ser girado sobre seu eixo, os discos circulam em sentidos opostos, mudando a heliolatitude observada. No entanto, não era esperado que esta versão experimental fornecesse a precisão necessária para fazer este tipo de medida.

Na tabela 3.2, podemos comparar os resultados de duas sessões, para cada uma das versões, em condições observacionais semelhantes. A mudança no valor da distância mínima medida não é relevante, pois a configuração dos hemi-espelhos mudou, mas a dispersão das medidas na versão B (angular) caiu à metade.

Isto confirma uma estabilidade maior para medidas feitas com espelhos em configuração em X, indicando este modelo como o melhor candidato para a montagem final do instrumento.

Tabela 3.2: Comparação entre os resultados das duas versões do heliômetro refletor. A escala de placa é de ~1"/*pixel*.

| Protótipo versão | nº de imagens | distância mínima (*pixel*) | σ (*pixel*) |
|---|---|---|---|
| A | 108 | 156,98 ± 0,35 | 1,53 |
| B | 114 | 93,75 ± 0,07 | 0,76 |

### 3.1.4 - Protótipo nº4

O objetivo da construção deste protótipo foi usar o método clássico e histórico para a duplicação das imagens. Para tanto, um dubleto acromático, usado como objetiva, foi cortado ao longo do seu diâmetro e remontado de forma descentrada, com um pequeno deslocamento. O ajuste deste deslocamento foi visual, menor de 2 mm, em conformidade com a distância focal daquela objetiva (200 *mm*), para um deslocamento angular de cerca de 0,5°.

Um anel foi especialmente fabricado para adaptar esta objetiva ao corpo do instrumento, figura 3.28.

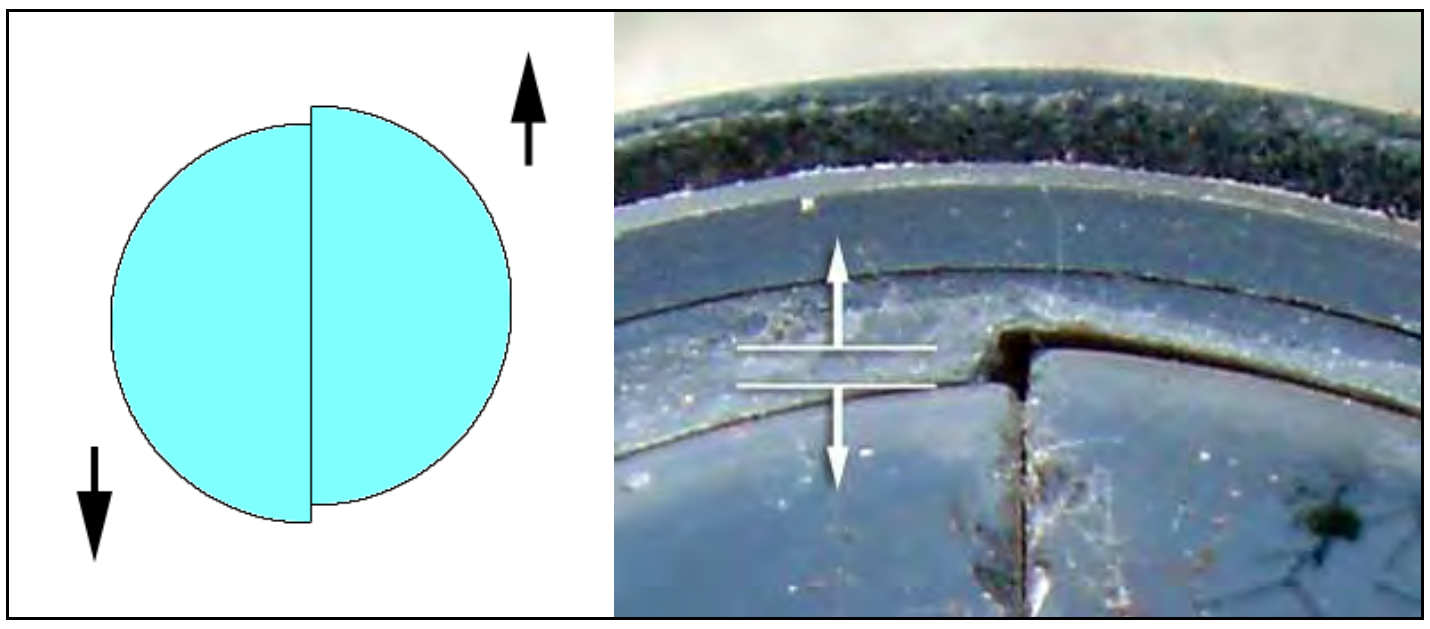

Fig. 3.28: Esquema do deslocamento das metades da lente e, em detalhe, o deslocamento real dado às partes.

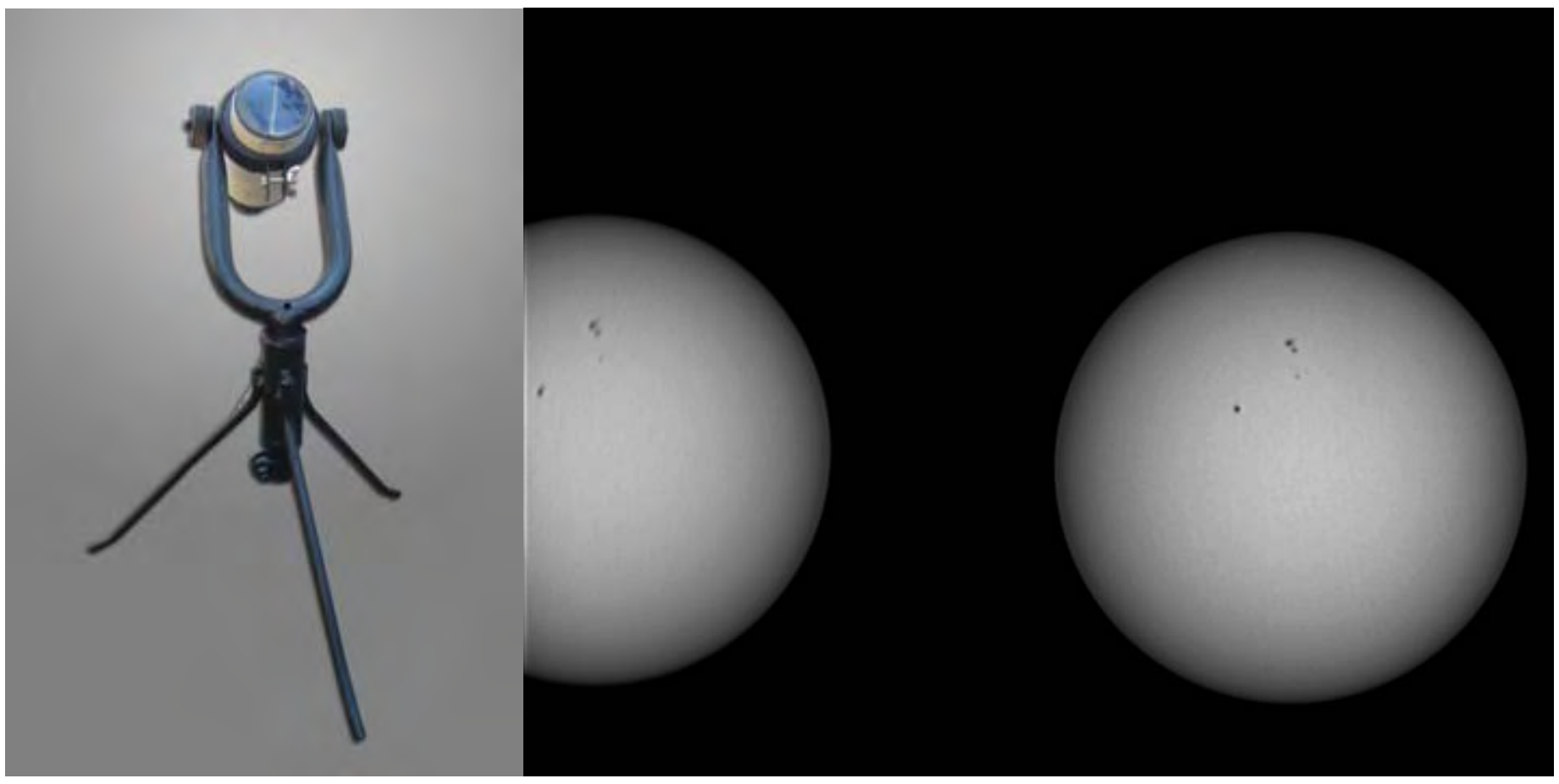

Fig. 3.29: À esquerda, o protótipo nº4 montado no seu tripé. À direita, uma das imagens duplas do disco solar obtida com este experimento.

As imagens deste heliômetro refrator têm boa qualidade óptica (fig. 3.29), mas devido à pequena distância focal desta objetiva (5× menor do que a planejada para o Heliômetro definitivo, de 1000 mm) a série de imagens feitas por este instrumento não foi utilizada para gerar medidas heliométricas.

## 3.2 - Conclusão dos testes dos protótipos - a escolha do método

Todos os protótipos testados funcionaram, gerando imagens duplas do Sol astronomicamente exploráveis. Ou seja, como esperado, todos têm qualidade e podem ser usados para fornecer resultados científicos. No cômputo dos resultados, porém, o protótipo 3B (um instrumento refletor, com espelho principal cortado e configuração dos hemi-espelhos em X) foi o protótipo escolhido para ser construído, para o Heliômetro definitivo, pelas seguintes razões:

- Um heliômetro refrator tem desvantagens: lentes são suscetíveis às deformações térmicas e mecânicas e existe a dependência da distância focal com o comprimento de onda da luz. Além disso, estas lentes objetivas são peças compostas (dubletos ou tripletos) o que faz com que seja muito difícil, depois de cortadas, estabilizar mecanicamente diversas peças ao mesmo tempo;

- Por ser refletor, não existe o problema de aberrações cromáticas;

- A configuração em X forneceu imagens mais estáveis.

- O instrumento pode ser construído com um mínimo de superfícies ópticas envolvidas (espelho principal + espelho secundário + câmara CCD, ou simplesmente espelho principal + câmara CCD);

- Existem disponíveis substratos para a fabricação de espelhos com baixíssimo coeficiente de dilatação térmica;

- E, uma grande vantagem estratégica, existem no Brasil fabricantes de espelhos com experiência em projetos internacionais que podem fabricar superfícies de grande qualidade óptica e com conhecimento técnico para ajudar a desenvolver a metodologia para: a fabricação, o corte, a fixação e a montagem de um espelho heliométrico cientificamente aprovado.

# Capítulo 4 – Desenvolvimento da metodologia de confecção

Neste capitulo são apresentadas as três etapas fundamentais no desenvolvimento da metodologia, cujo emprego é utilizado na construção do Heliômetro final descrito no capítulo 5.

A extensão deste capítulo é requerida para detalhar os diversos testes efetuados para criar e aperfeiçoar tais metodologias. Note-se que, o Heliômetro cuja construção é objeto desta tese constitui-se num instrumental inovador, requerendo assim soluções metodológicas e científicas igualmente inovadoras.

O capitulo é divido em três sub-itens principais, quais sejam: o Desenvolvimento do espelho do Heliômetro; a fixação mecânica dos hemi-espelhos; e a montagem final do Heliômetro. Toda a metodologia descrita a seguir se constitui em processo experimental original.

## *4.1 - Desenvolvimento do espelho do heliômetro*

Na construção do Heliômetro definitivo, um passo inicial era desenvolver e testar diferentes métodos de confeccionar e fixar, em laboratório, o espelho heliométrico na configuração angular desejada.

### 4.1.1 - Processos de corte do hemi-espelhos

#### 4.1.1.1 - Corte à serra

A experiência com o espelho heliométrico que foi preparado já cortado (fig. 3.10), evidenciou que está técnica não era a indicada, pois surgiam defeitos em sua superfície que não poderiam ser sanados ou homogeneizados. A opção mais indicada, portanto, foi cortar um espelho já preparado. No entanto, era importante verificar se o corte em si não deformaria sua superfície óptica, assim como a liberação de possíveis tensões internas também não o faria.

Após testes com diferentes equipamentos de corte, verificou-se que o corte é mais adequadamente feito utilizando-se uma serra diamantada.

Com o primeiro espelho testado, porém, durante o corte manual, uma das metades lascou, inviabilizando este espelho para uso instrumental. Entretanto este acidente acabou sendo muito útil para estudos dos testes ópticos de Ronchi e Foucault[2], pois serviu para como gabarito da sensibilidade destes testes. A figura 4.1 mostra o detalhe deste corte.

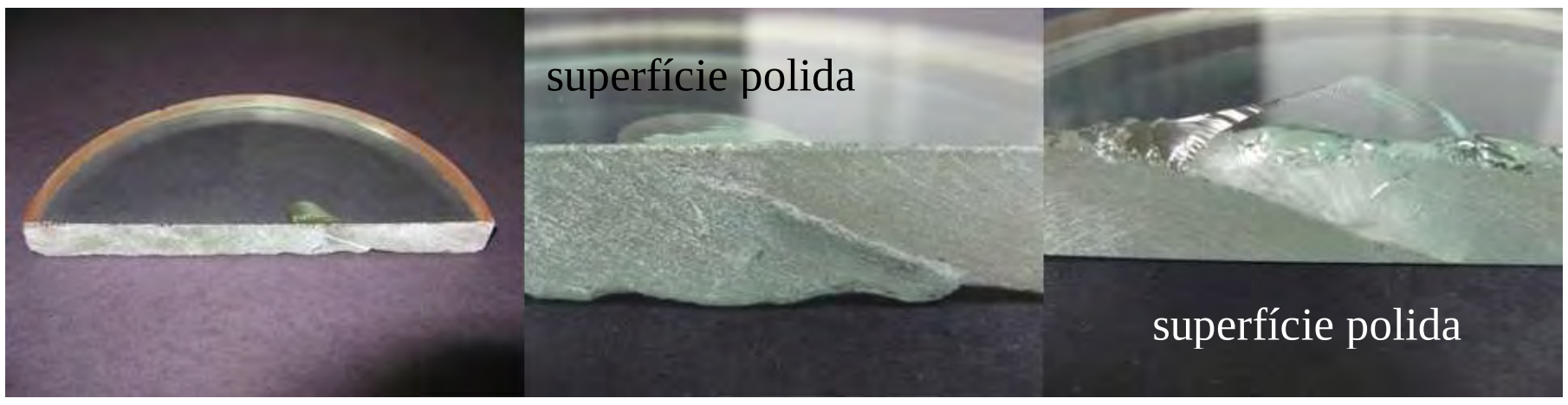

Fig. 4.1: Composição mostrando a metade do espelho cortado e o detalhe da trinca, que não chegou à superfície polida.

O teste de Ronchi realizado nos hemi-espelhos mostrou que, na metade que não sofreu danos, a curvatura do espelho não se alterou pelo fato dele ter sido cortado ao meio com uma serra. Isto pode ser visto pelas sombras retas ao longo do corte do espelho (fig. 4.2, à esquerda). A deformação na curvatura da superfície do espelho trincado é evidente.

Este mesmo teste foi repetido ao longo de vários dias e as imagens de Ronchi dos hemi-espelhos não se alteraram, mostrando que não houve deformações posteriores por relaxamento de tensões internas do vidro.

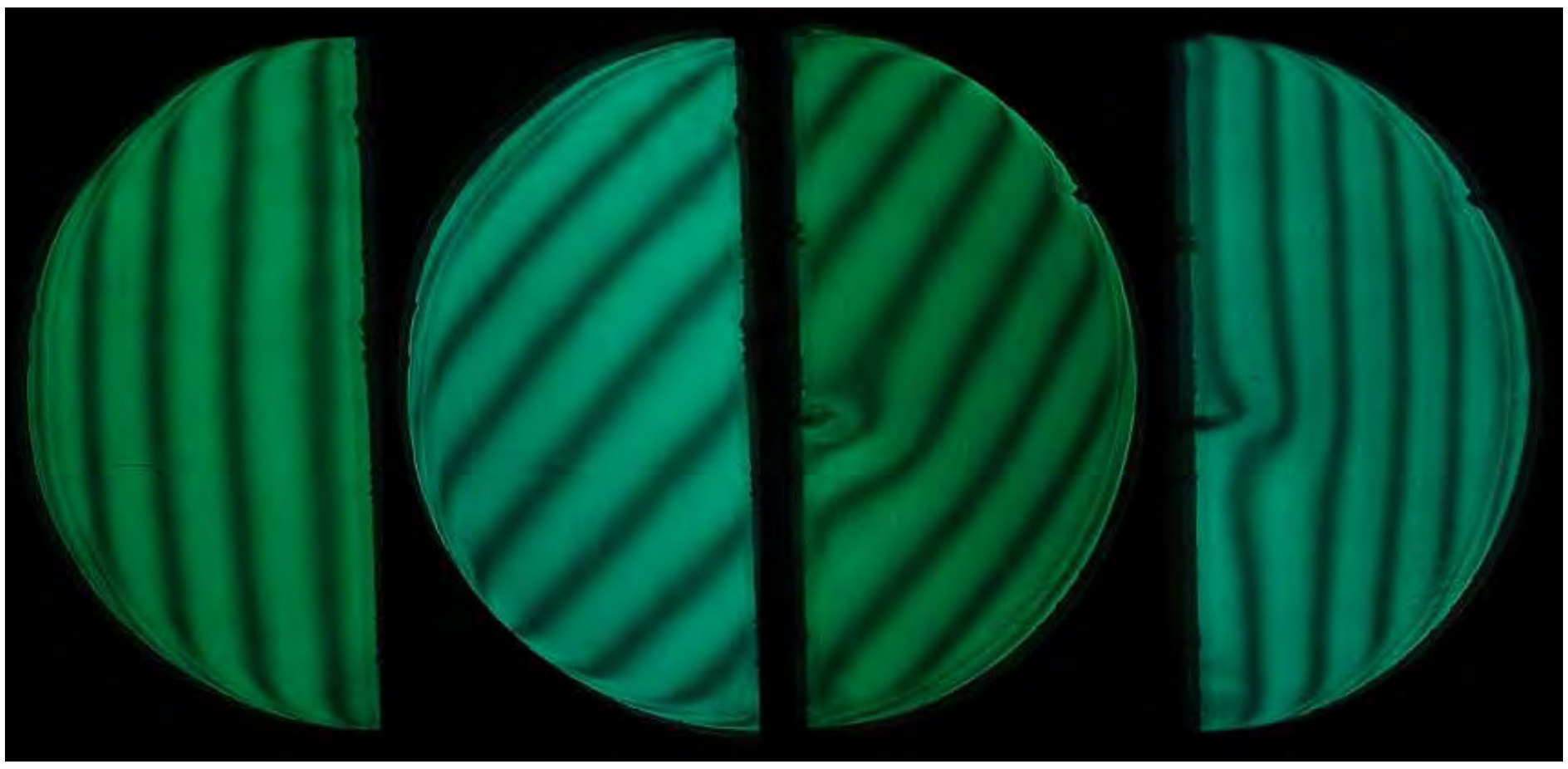

Fig. 4.2: Imagem do Teste de Ronchi nas metades do espelho, realizado dois dias após o corte. A deformação na superfície da metade trincada é visível nas imagens à direita.

[2] Estes testes, de grande sensibilidade qualitativa, permitem avaliar a qualidade de uma superfície óptica (Texereau, 1961).

### 4.1.1.2 - Corte à diamante

Características físicas do espelho usado neste teste:

- Diâmetro = 180 mm
- Distância focal = 1003 mm
- Relação focal f/D = 5,57
- Superfície parabólica

Antes porém do corte ser feito, imagens dos testes ópticos de Ronchi e de Foucault foram registradas para posterior avaliação do método de corte e exame de sua superfície óptica antes e depois do processo, figuras 4.3, 4.4 e 4.5.

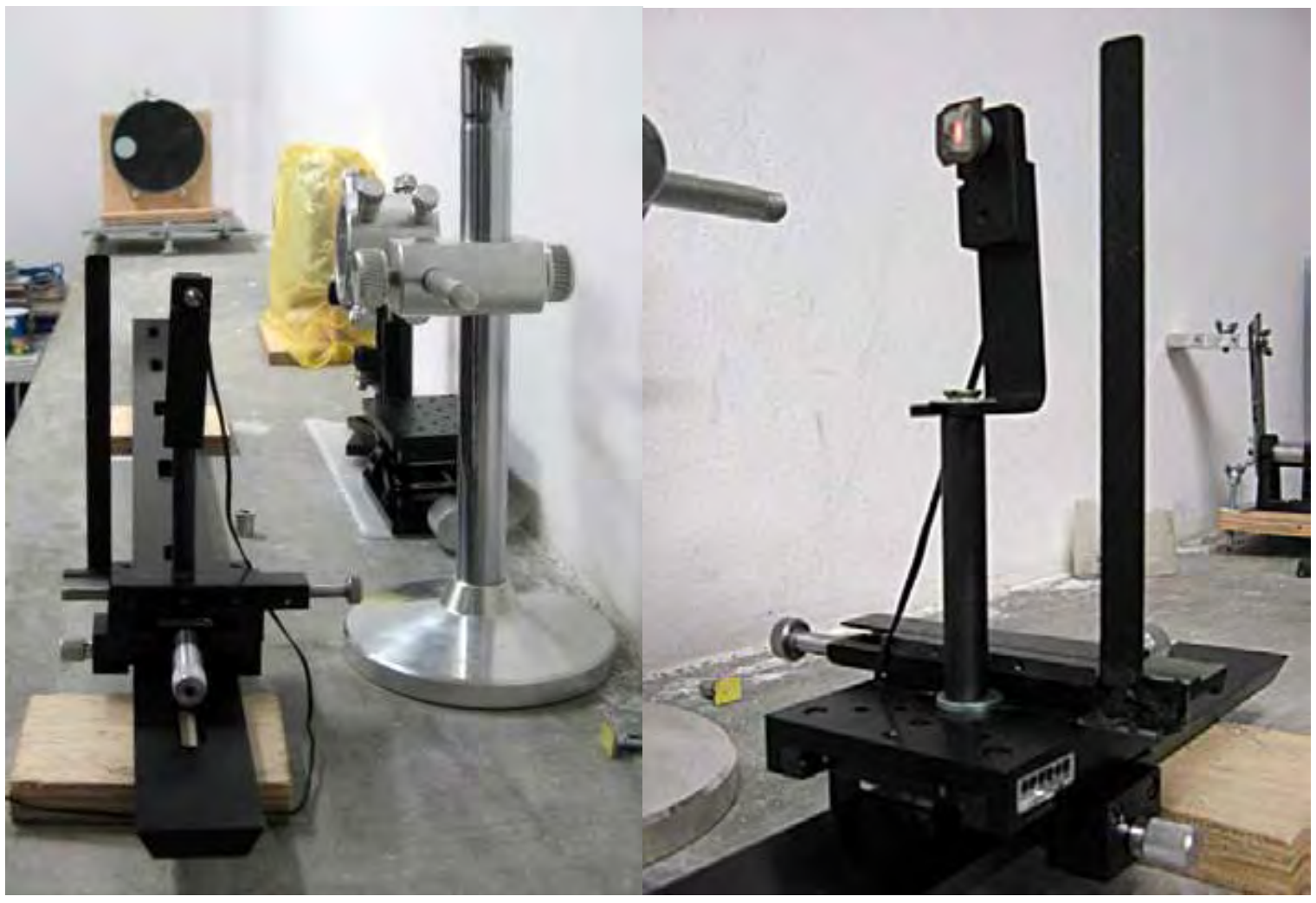

Fig. 4.3: À esquerda, espelho posicionado ao fundo da bancada para o teste. À direita, detalhe da fenda (fonte de luz) e da faca, que serve também como suporte para a tela de Ronchi.

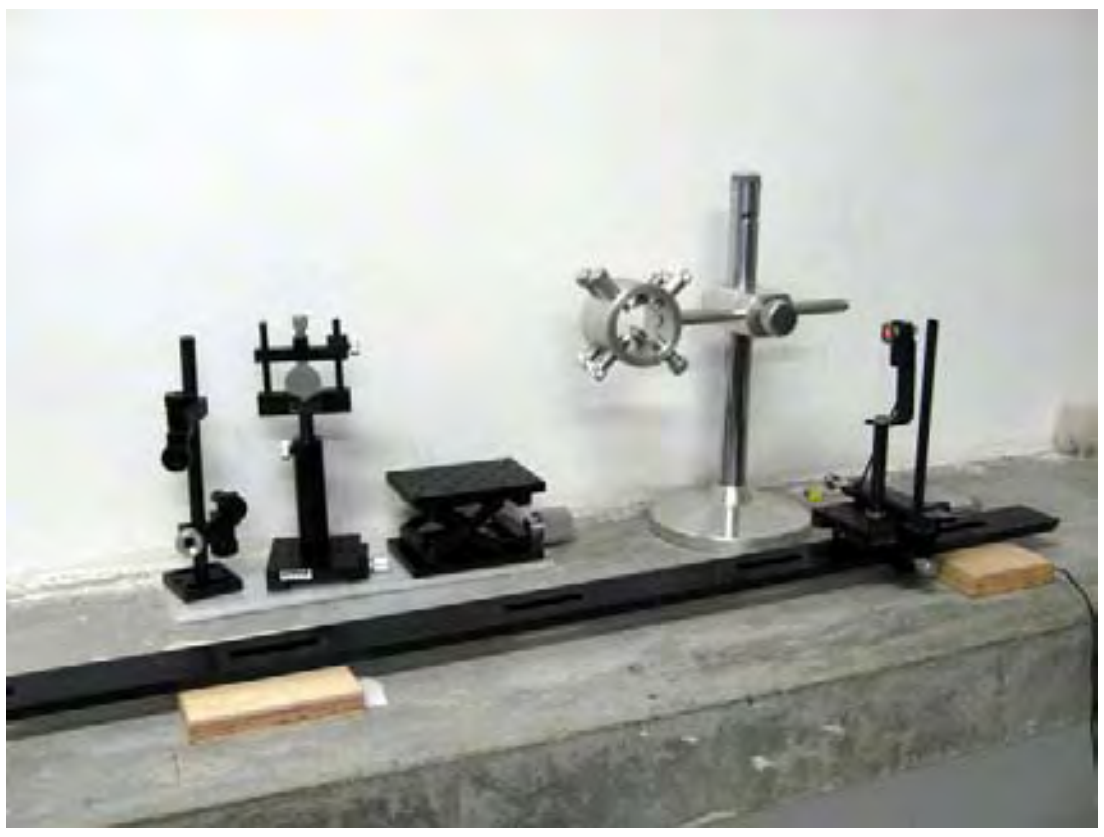

Fig. 4.4: Vista lateral da bancada óptica.

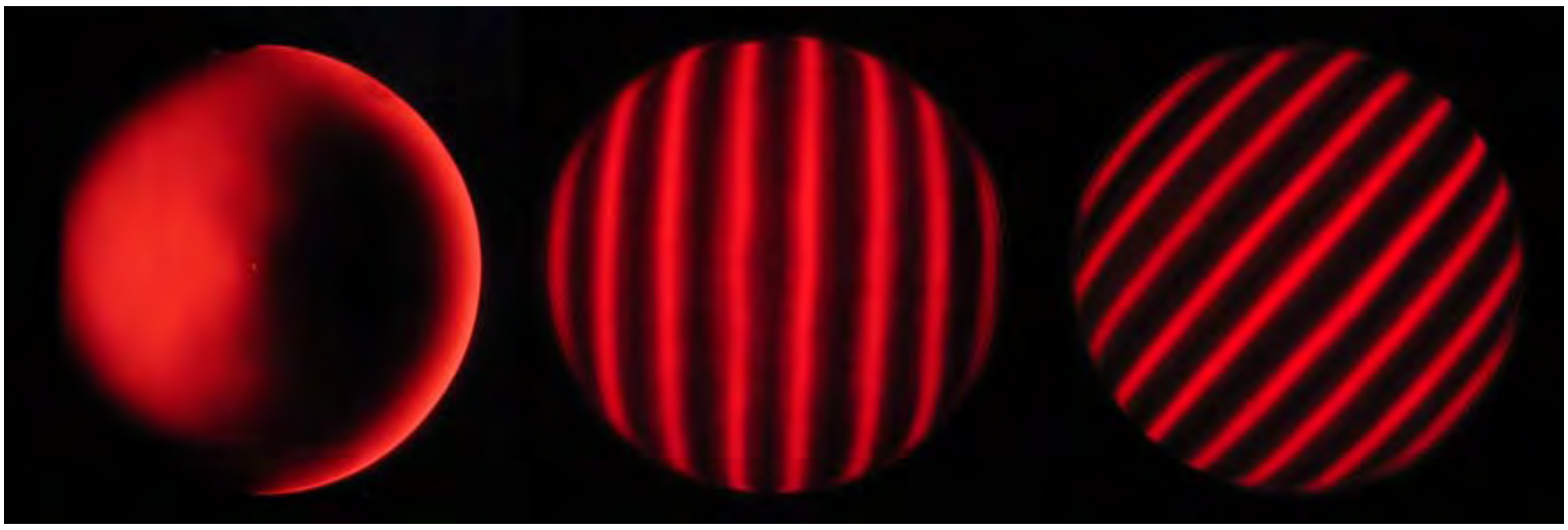

Fig. 4.5: Imagens dos testes de Foucault e Ronchi do espelho antes de ser seccionado, mostrando a excelente qualidade de sua superfície.

O corte deste espelho foi feito com ponta de diamante. O sulco foi aberto na superfície não parabolizada do espelho. As imagens das figuras 4.6 e 4.7 trazem a seqüência do corte.

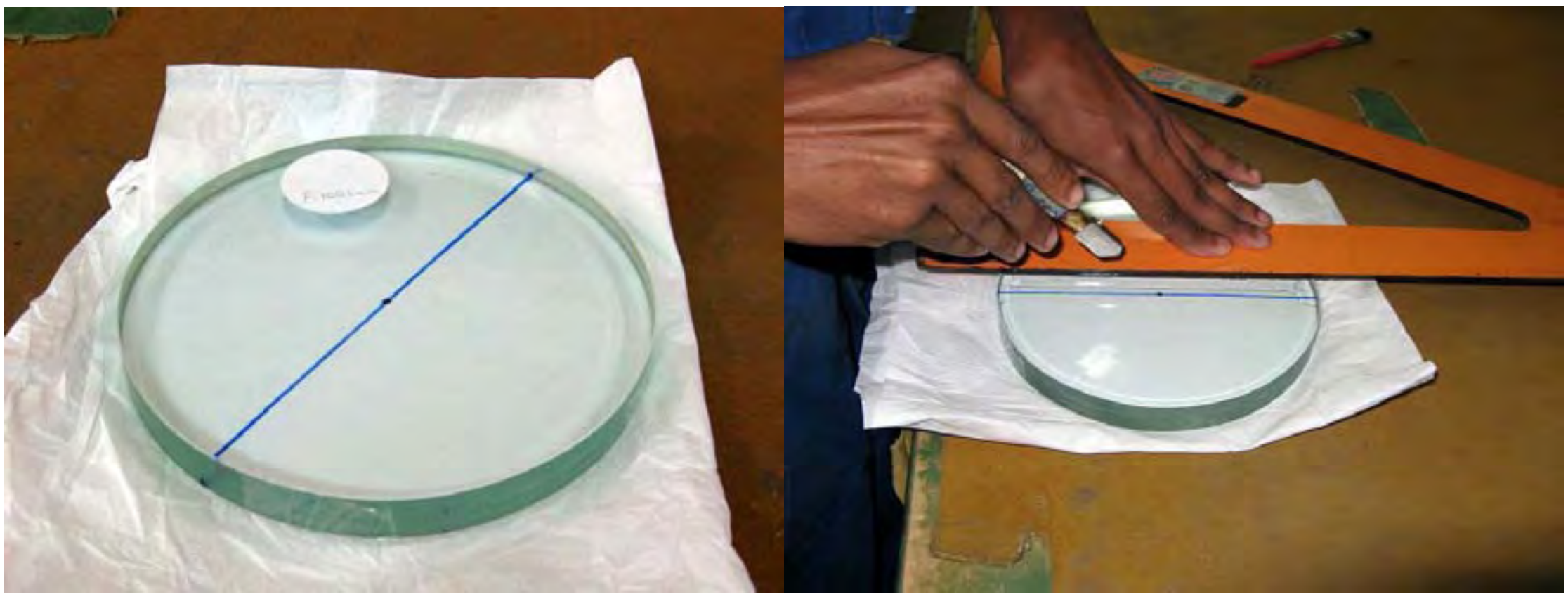

Fig. 4.6: Risco para o corte do espelho.

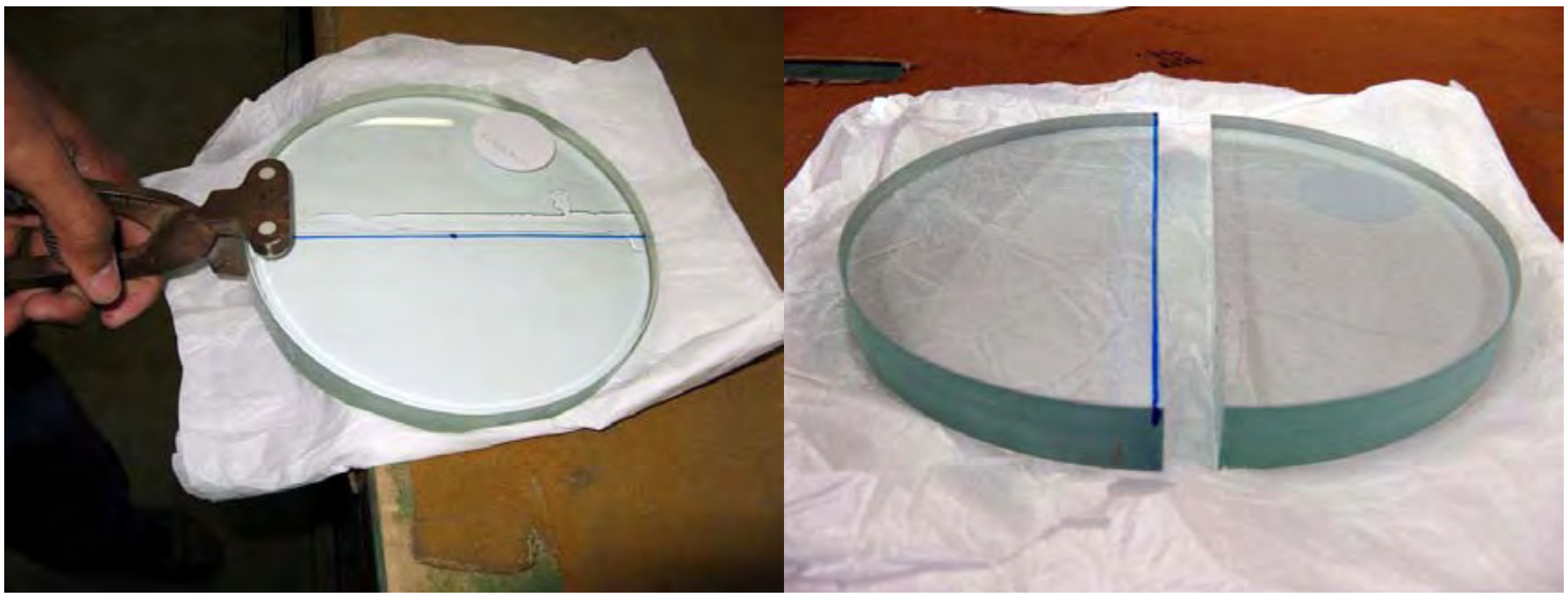

Fig. 4.7: Corte do espelho.

O corte é aproximadamente diametral, com um deslocamento de cerca de 2 mm do centro do disco. Posteriormente as superfícies laterais foram desbastadas para igualar as metades, deixando entre elas um espaçamento de cerca de 3 mm.

Os testes ópticos pós-corte revelaram que houve uma pequena deformação do material próximo à linha do corte, resultante da pressão exercida quando da separação, como mostram a figura 4.8.

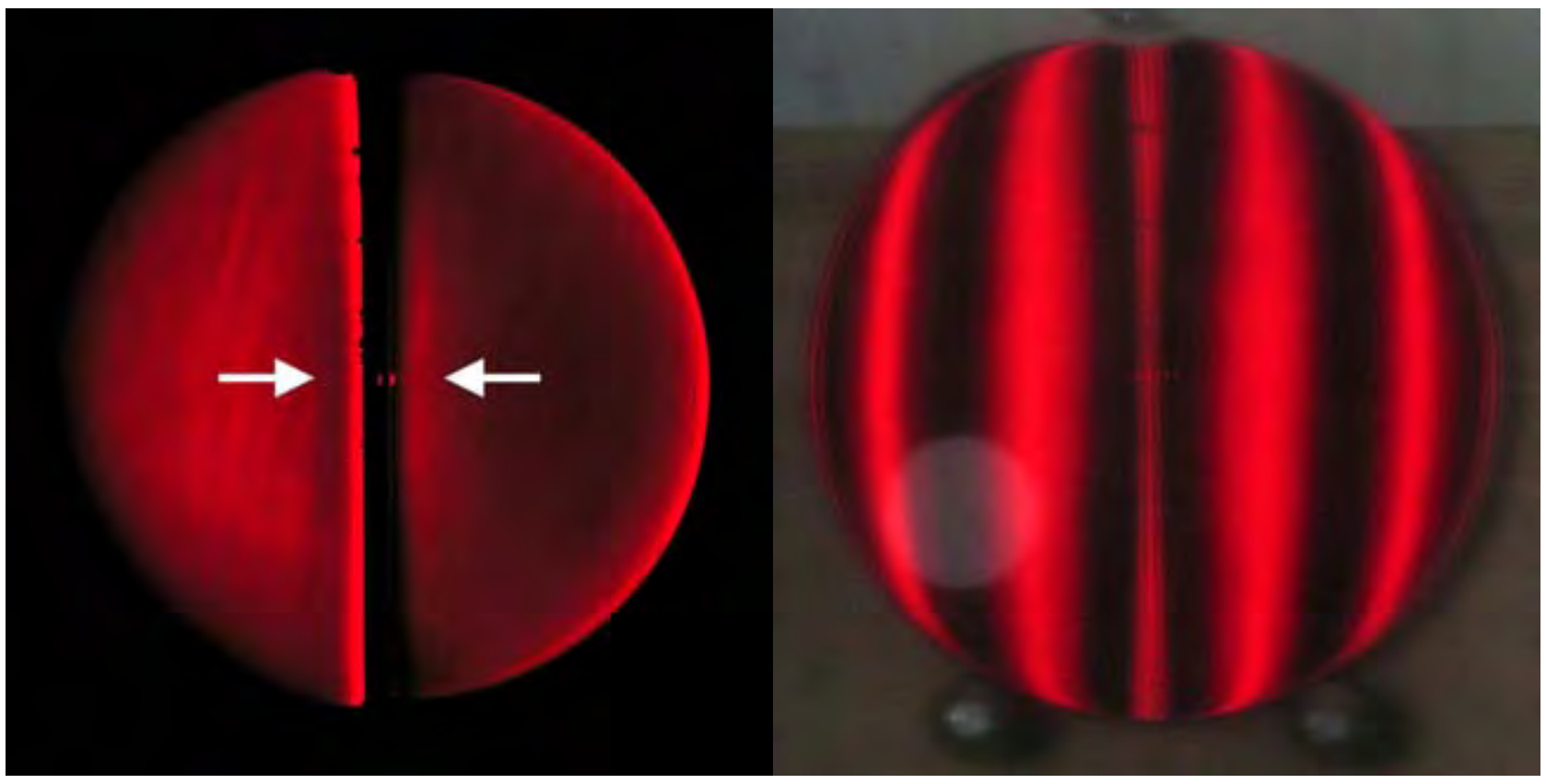

Fig. 4.8: Imagens dos testes de Foucault e Ronchi do espelho depois de seccionado.

No teste de Foucault (fig. 4.8, à esquerda), a pequena sombra e o pequeno brilho nas superfícies (indicados pelas setas) mostram que a região do espelho, ao longo do corte, possui agora uma ligeira elevação (que não pode ser avaliada quantitativamente neste teste). Na imagem da direita, a distorção das sombras no teste de Ronchi, em direção ao centro do disco, indicam a mesma coisa. Como o corte e as bordas dos discos são mascarados, estas elevações não causam distorções nas imagens.

Em conclusão, o corte à serra, garantido o cuidado que o processo requer, atende à exigências do projeto, pois não deforma a superfície do espelho. A comparação com os testes pré-corte fez que este seja o método para o corte do espelho heliométrico definitivo.

### 4.1.2 - Montagem do espelho na configuração em X

O objetivo aqui é desenvolver a metodologia de se obter hemi-espelhos prismáticos, ou seja, peças que ao serem assentadas sobre um plano óptico, já se apresentem com sua configuração heliométrica em X.

A etapa do estudo consiste em:

1. montar os hemi-espelhos numa configuração de forma que a imagem de uma fonte teste fosse desdobrada em duas, distanciadas verticalmente, no plano focal, de pouco mais de 20 mm;
2. fixar o conjunto nesta configuração;
3. Com o conjunto fixado, desbastar o fundo do espelho, juntamente com sua base, para deixar ambas as superfícies planas, preservando-se o diedro.

A montagem experimental teve os seguintes elementos, figura 4.9:

- base em madeira para o suporte vertical do espelho;
- tira de vidro, com o mesmo comprimento do diâmetro e ~3mm de espessura, colocada para preencher o espaço entre as metades;
- uma metade apoiada em três pontos de altura fixa e a outra em um ponto de altura fixa (o central) e dois pontos de altura variável, de forma que apenas esta parte do espelho poderia ser movida para se fazer o ajuste do diedro;
- uma tira de borracha colocada em volta do conjunto para segurá-lo à base, mantendo-o pressionado contra os pontos de apoio;

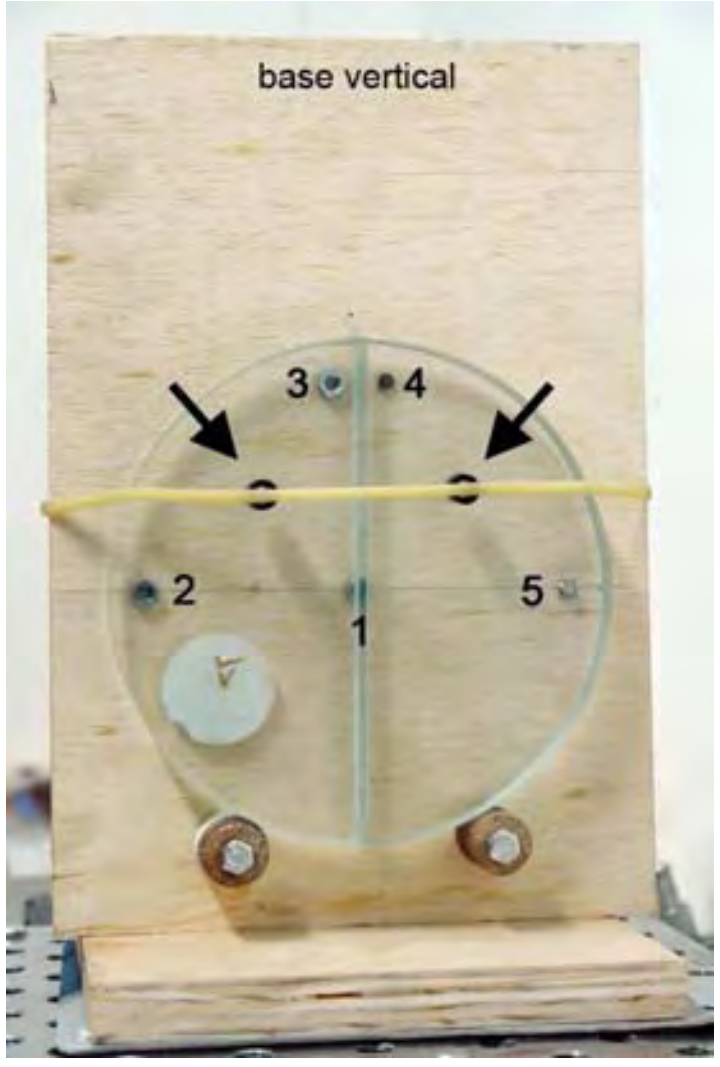

Fig. 4.9: Espelho fixado na sua base vertical.

- conforme mostrado na figura 4.9, os pontos 1, 2 e 3 são os pontos de apoio fixos e os pontos 4 e 5 os variáveis. As setas indicam o local onde foram colocados pequenos calços de borracha para pressionar as metades contra os pontos de apoio;
- o conjunto alinhado de forma que o corte do espelho fique na vertical, segundo materializado por fio de prumo;
- uma régua de aço fixada à faca do Foucault e também alinhada verticalmente, figura 4.10;
- a fonte teste e a régua ficam posicionadas no plano focal do espelho, de forma que as imagens da fonte, de cada hemi-espelho, se formam sobre ela, figura 4.10;
- os parafusos são ajustados até as imagens se distanciarem verticalmente de 20 mm, correspondente ao do diedro igual ao ângulo aparente do Sol (0,5°), figura 4.11.

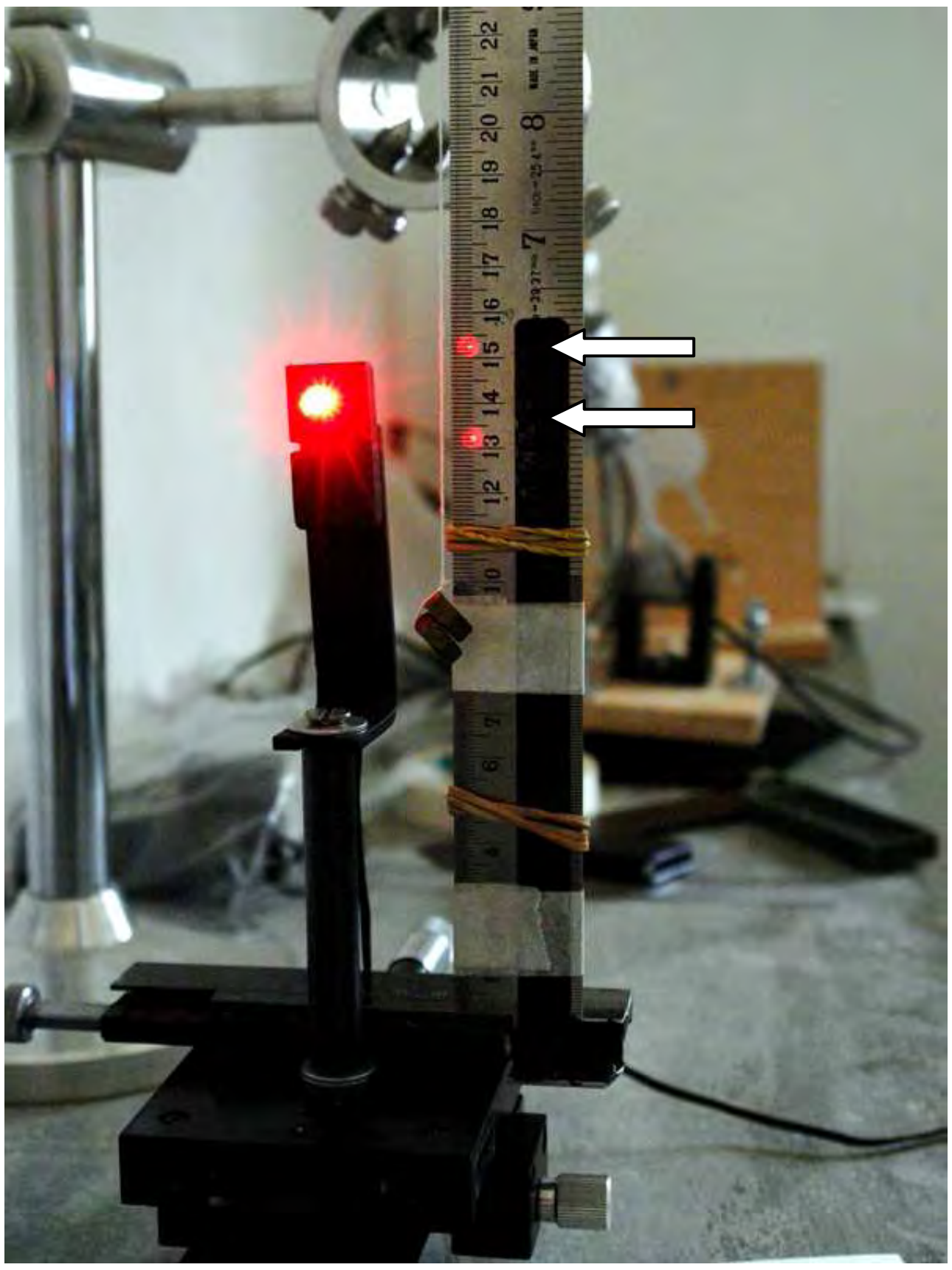

Fig. 4.10: Aspectos da montagem do laboratório. Um led vermelho foi utilizado como fonte extensa, posicionada em relação ao espelho de forma que suas imagens fossem projetadas no plano focal, sobre a régua graduada, indicado pelas setas.

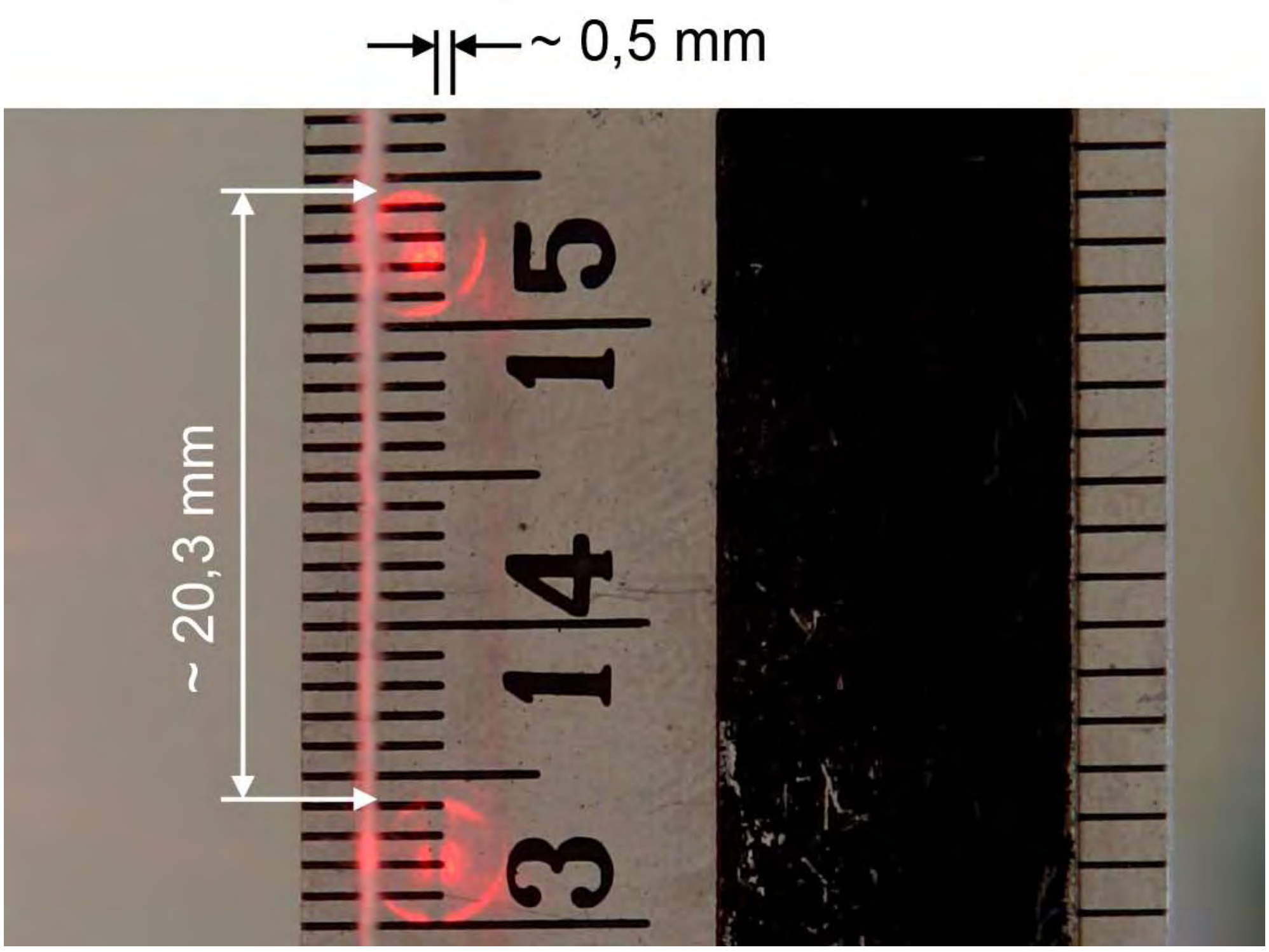

Fig. 4.11: Detalhe das imagens sobre a régua. O fio de prumo pode ser visto como uma linha na frente da escala da régua e pela sua sombra, o deslocamento horizontal pôde ser avaliado.

Obs.: Este deslocamento horizontal (chamado de configuração em V, figura 4.12) das imagens é indesejável, mas é impossível eliminá-lo completamente nesta base. Uma base aperfeiçoada fundamentada nesta metodologia será apresentada no Cap.6. com este problema solucionado.

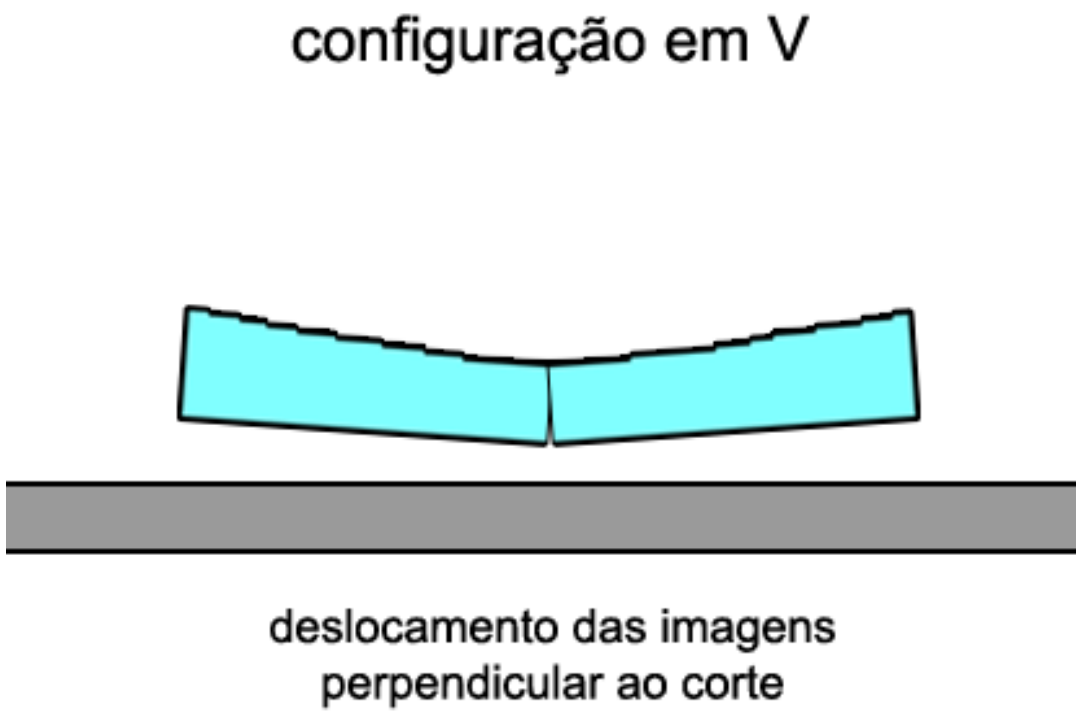

Fig. 4.12: Esquema da configuração em V. A imagem do espelho da esquerda encontra-se à direita da linha paralela ao corte. O oposto ocorre com o espelho da direita.

### 4.1.3 - Desbaste do fundo dos hemi-espelhos

#### 4.1.3.1 - Fixação do diedro com gesso

Uma vez encontrada a melhor configuração, as metades e a tira de preenchimento foram pré-fixadas utilizando-se pasta fixadora, para garantir que ao serem retiradas da bancada, juntamente com a base, não haveria deslocamento relativo entre elas, figura 4.13.

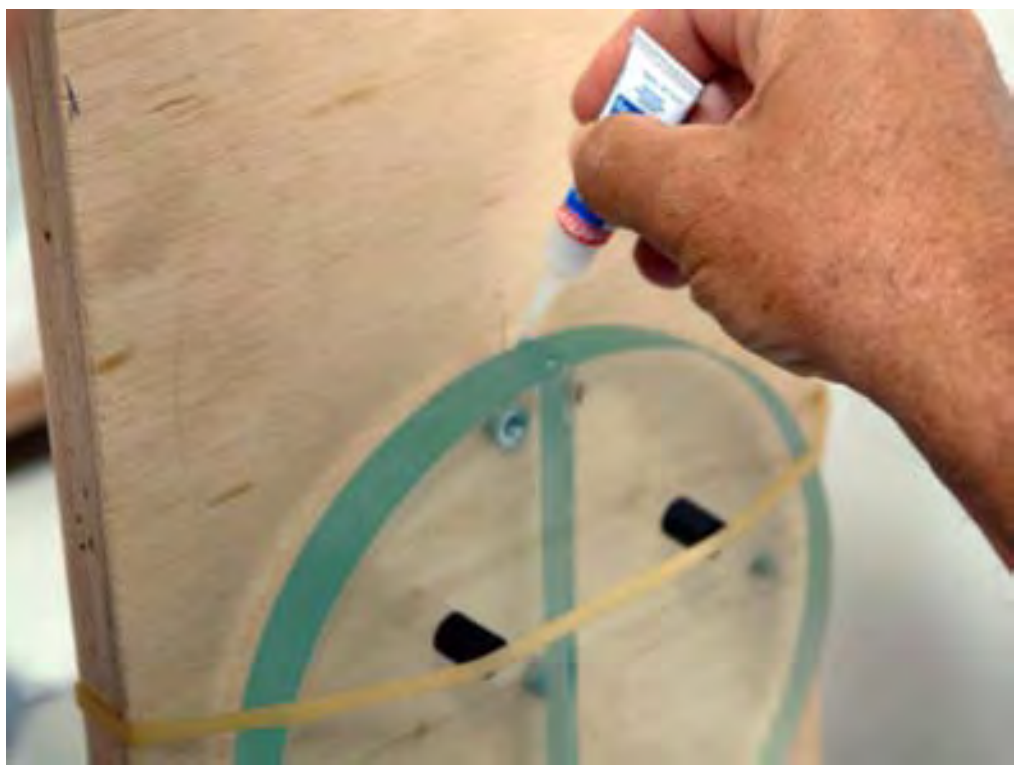

Fig. 4.13: Aplicação de cola instantânea entre a tira de preenchimento e as metades.

A figura 4.14 ilustra o resultado obtido para a configuração através da projeção dos discos solares num anteparo.

Antes do conjunto ser retirado da base para o desbaste, a região do corte é protegida contra o pó abrasivo com uma fita, figura 4.15.

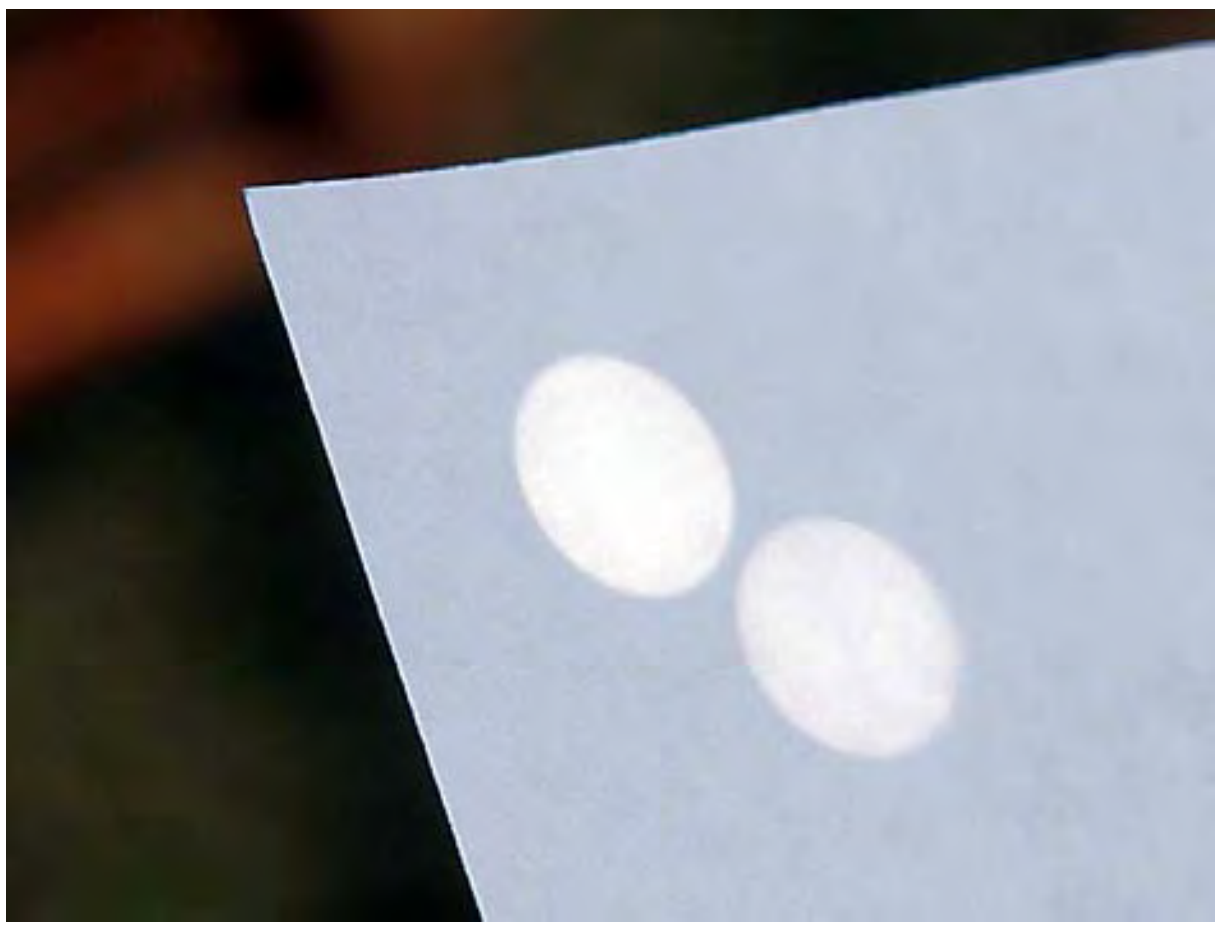

Fig. 4.14: Discos solares duplicados projetados numa folha de papel. O afastamento entre os discos era aceitável.

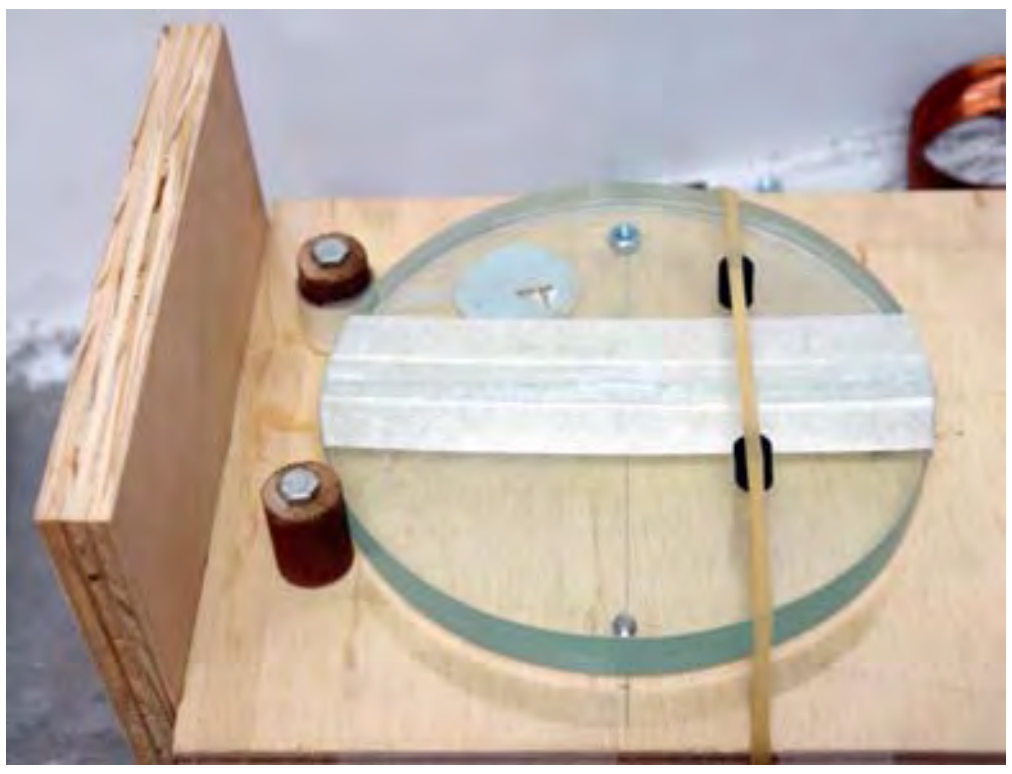

Fig. 4.15: Detalhe da proteção do corte com uma fita.

Ainda sobre a base, todo o conjunto é imobilizado com a aplicação de gesso, cobrindo-o completamente, figura 4.16.

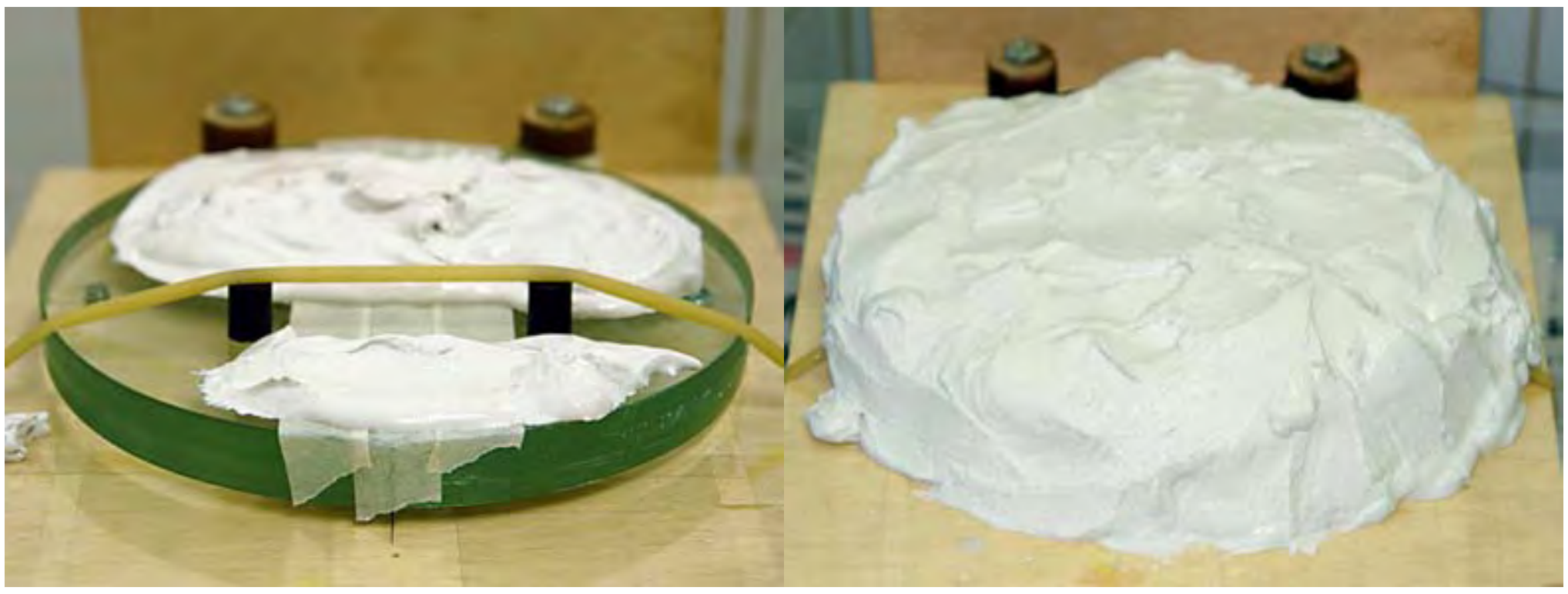

Fig. 4.16: Engessamento do espelho.

Finalmente, conforme detalhado na figura 4.17, o bloco é retirado da base de madeira para que a parte de trás do conjunto seja trabalhada, a fim de se obter uma superfície plana.

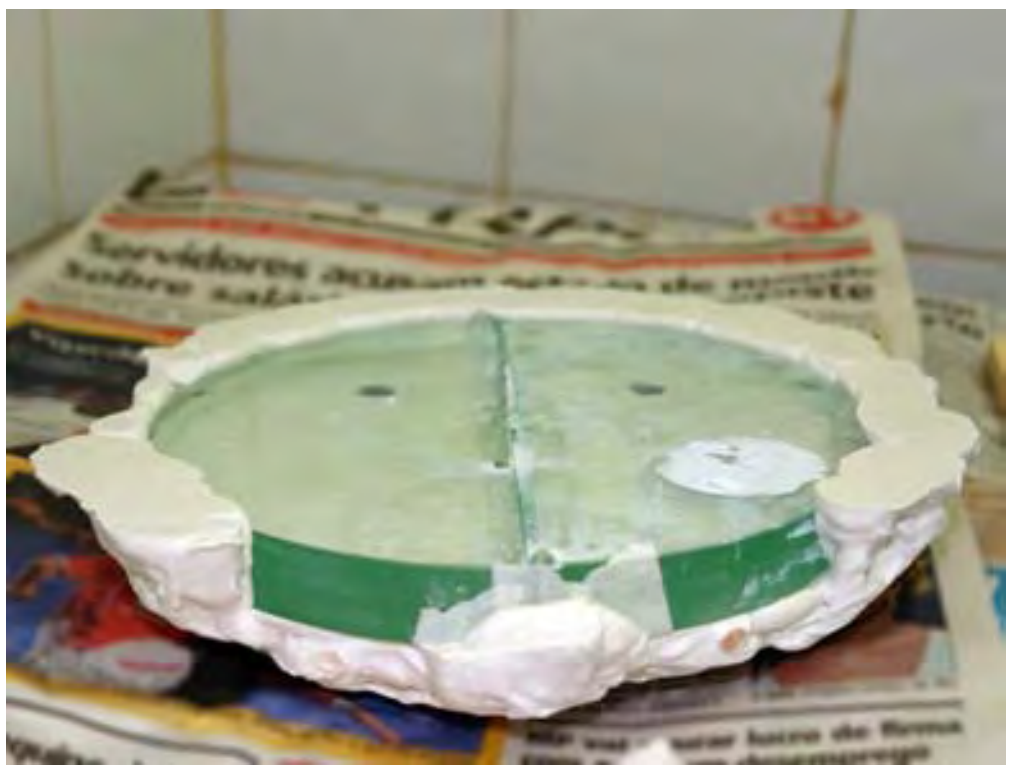

Fig. 4.17: Bloco a ser desbastado. O fundo do espelho está virado para cima.

A operação de desbaste consiste em atritar parte de trás do espelho (tecnicamente, o "bloco") sobre o outro bloco fixo numa bancada (tecnicamente, a "ferramenta").

No que se segue, apresentamos uma descrição esquemática do processo. Note-se que se trata de produzir a peça óptica fundamental do Heliômetro.

- sobre a ferramenta, acrescenta-se abrasivos em pó (carbureto de silício ou óxido de alumínio) com um pouco de água, como lubrificante, figura 4.18;
- faz-se movimentos aleatórios de vaivém, repondo-se o abrasivo quando este se desgasta e girando-se a bancada para evitar direções privilegiadas de desbaste, figura 4.19.

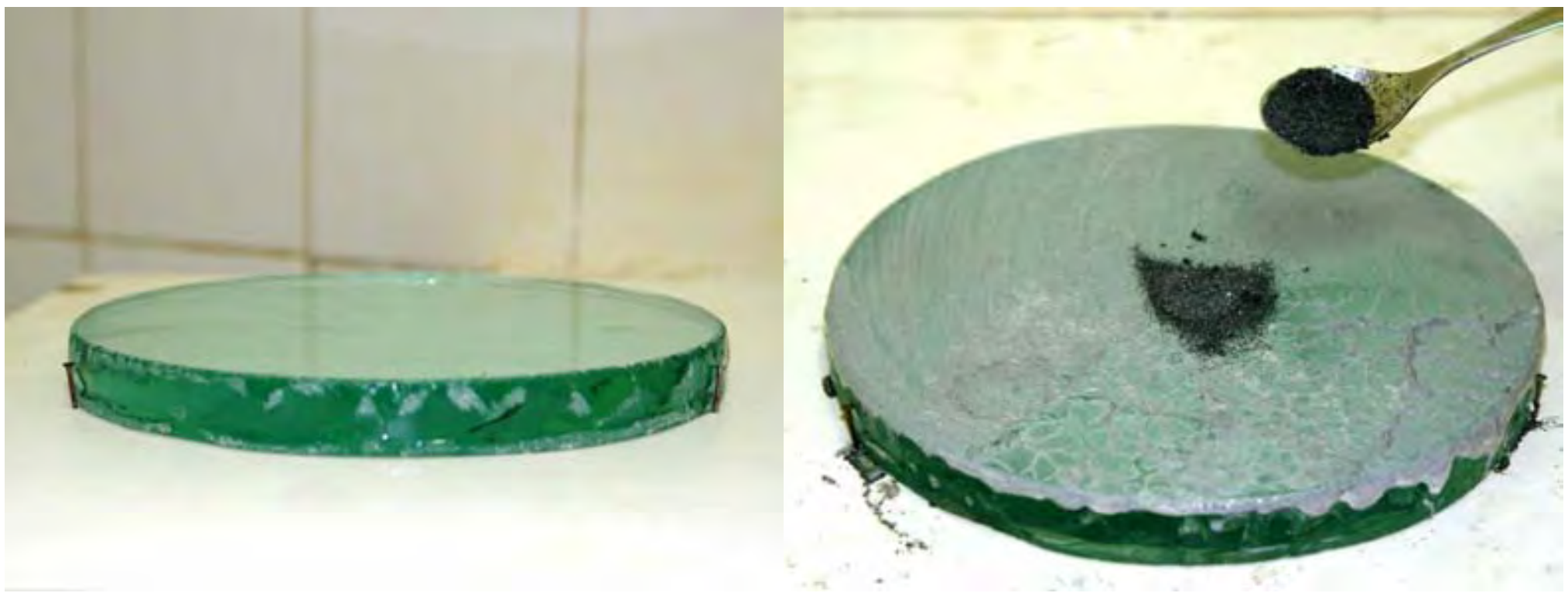

Fig. 4.18: Detalhe da ferramenta fixada e do abrasivo em pó depositado em sua superfície.

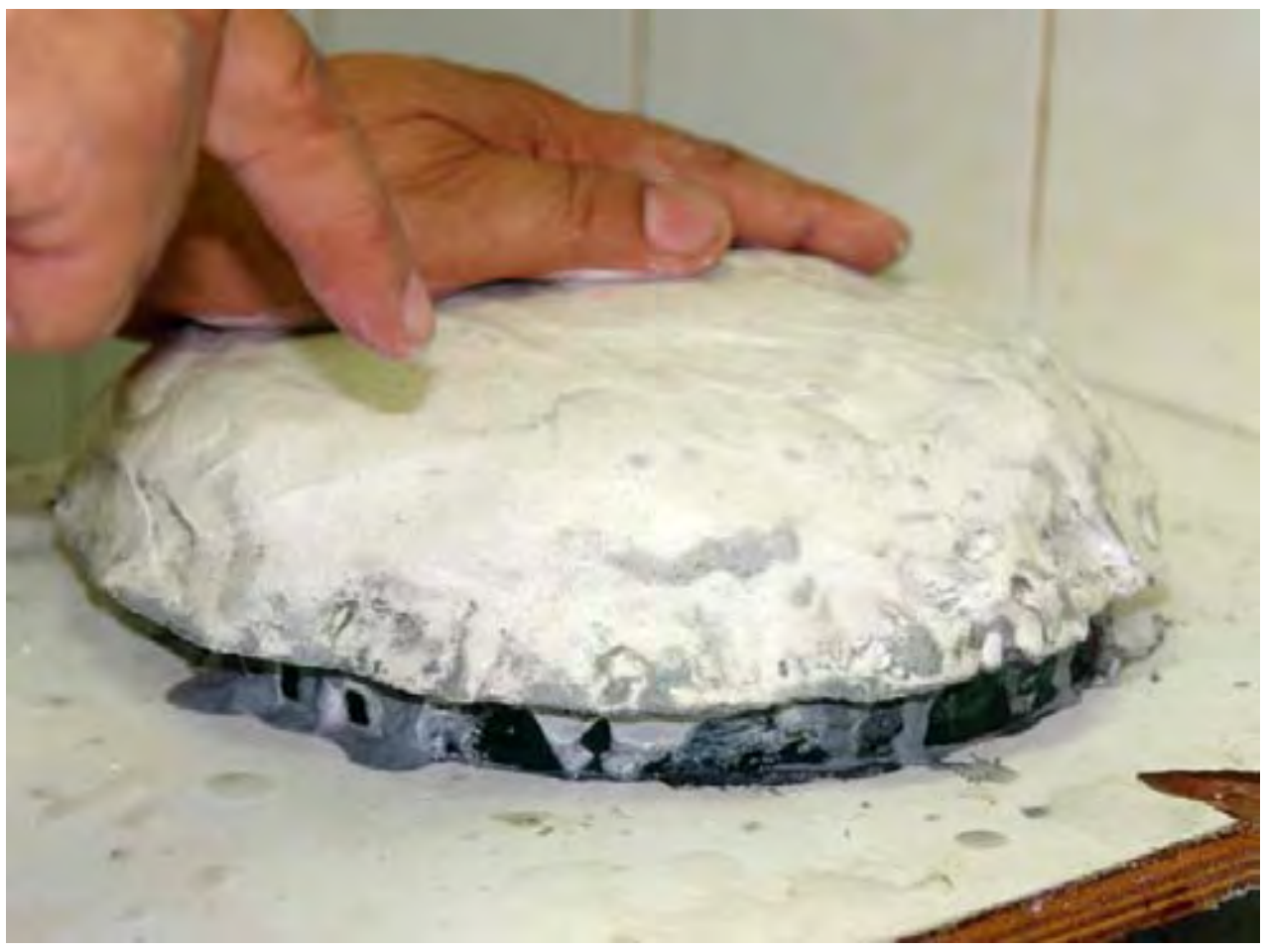

Fig. 4.19: Detalhe do desbaste do bloco.

Neste processo, existe uma tendência natural de que as regiões do meio do bloco e na borda da ferramenta sejam as mais desgastadas, formando superfícies côncavas e convexas, figura 4.20.

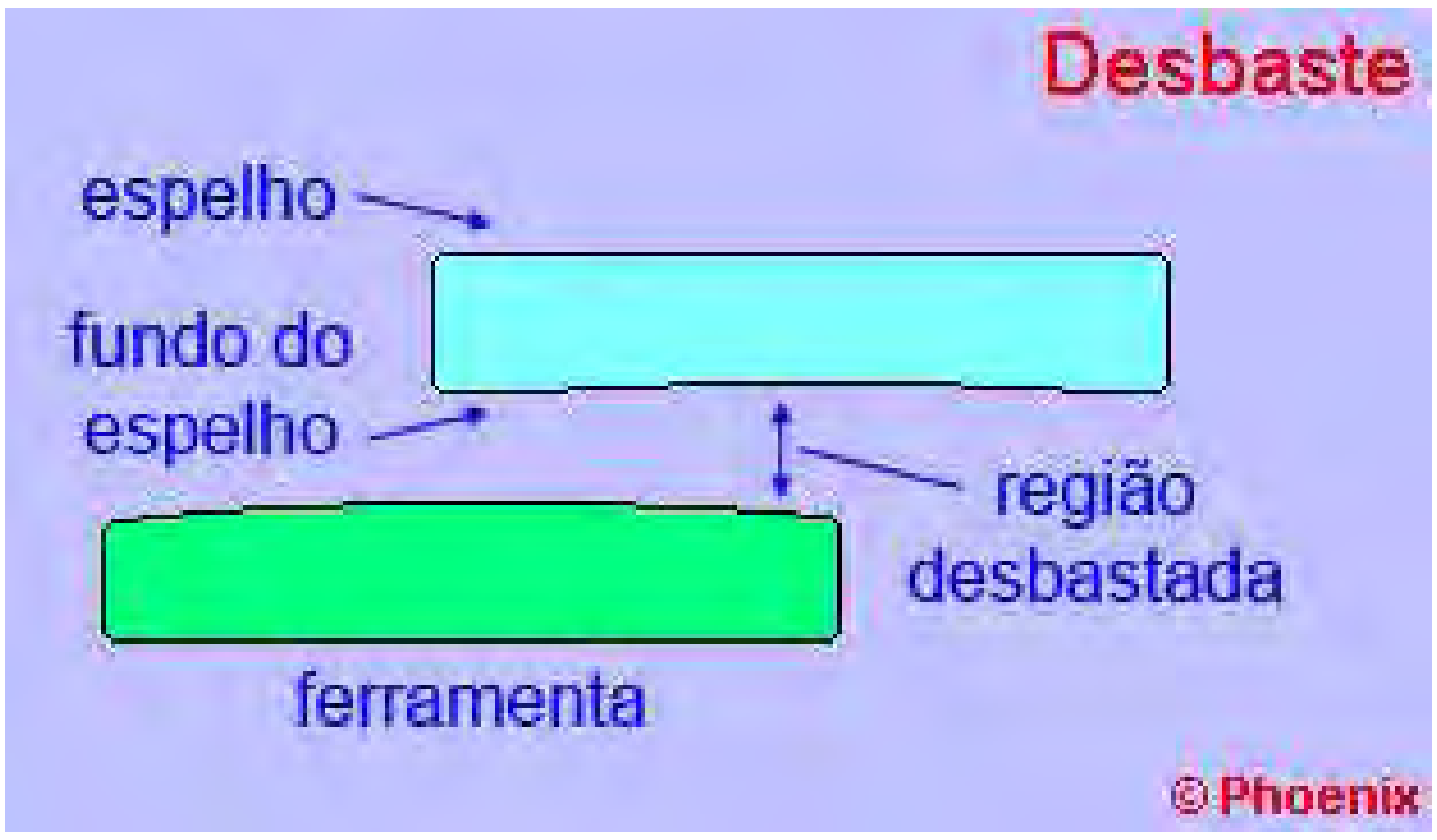

Fig. 4.20: Esquema do desbaste.

De modo a anular este efeito, é necessário que bloco e ferramenta tenham suas posições invertidas. Para isso, mais gesso é aplicado ao bloco de modo que este sirva de base para a ferramenta, figura 4.21.

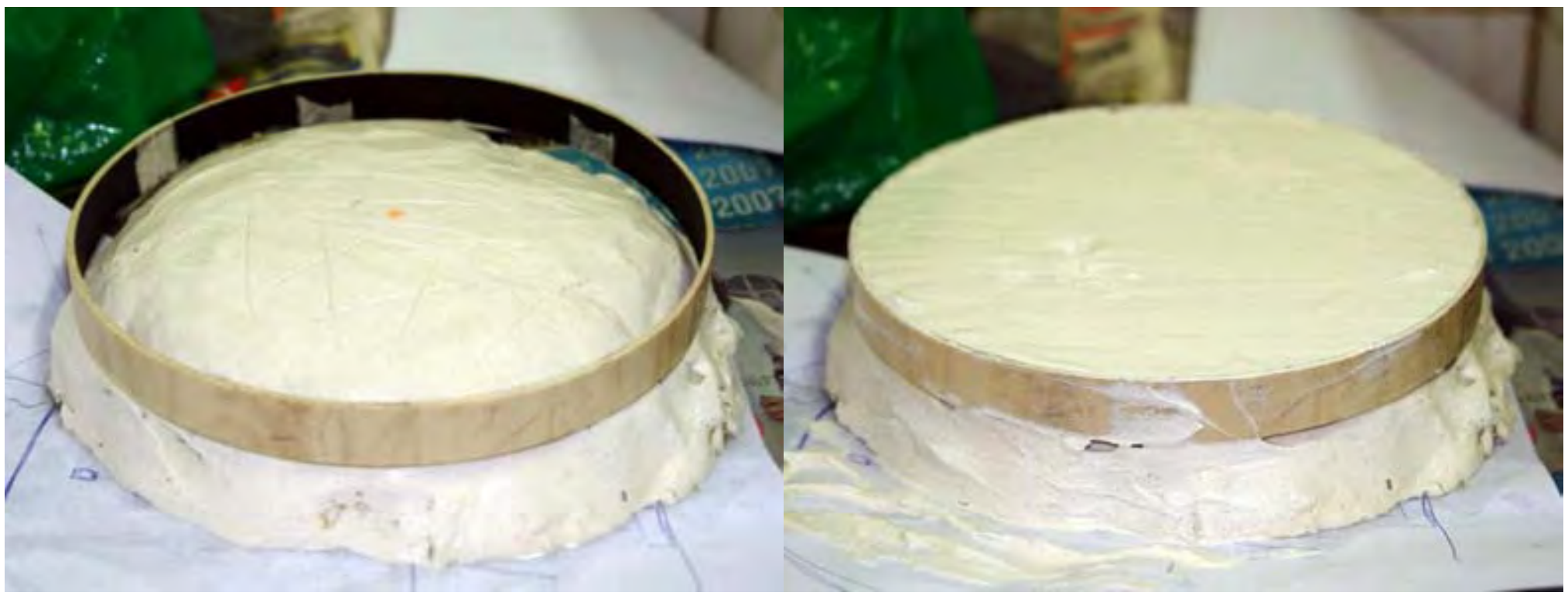

Fig. 4.21: Ajuste do bloco para servir de base para a ferramenta.

O bloco, então, é fixado a uma bancada (em madeira) e é repetido o procedimento de desbaste, figura 4.22.

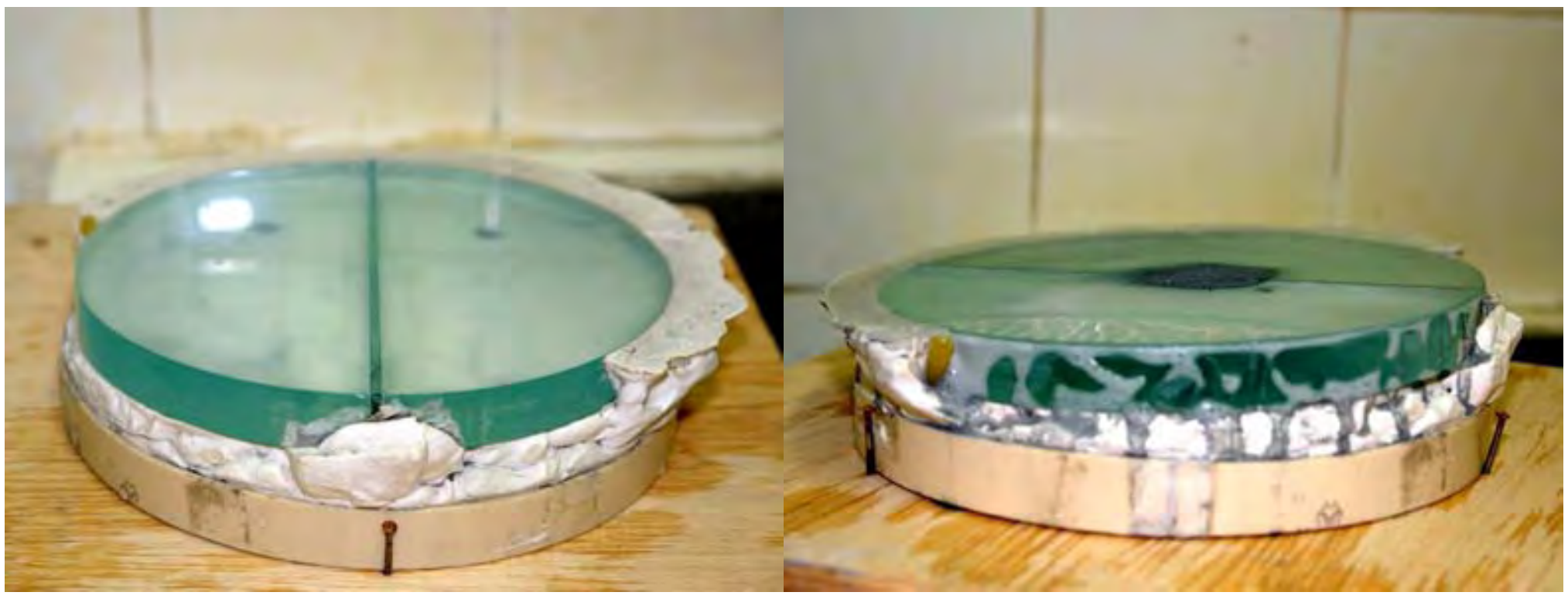

Fig. 4.22: Bloco fixado à bancada e com o depósito do abrasivo.

Para se obter a precisão desejada para as superfícies planas dos hemi-espelhos do Heliômetro, foram necessárias 4 sessões alternadas de desbaste (invertendo-se as posições entre bloco e ferramenta) e em cada sessão, 5 aplicações de abrasivo #80 (carbureto de silício com grãos de ~165µm de diâmetro).

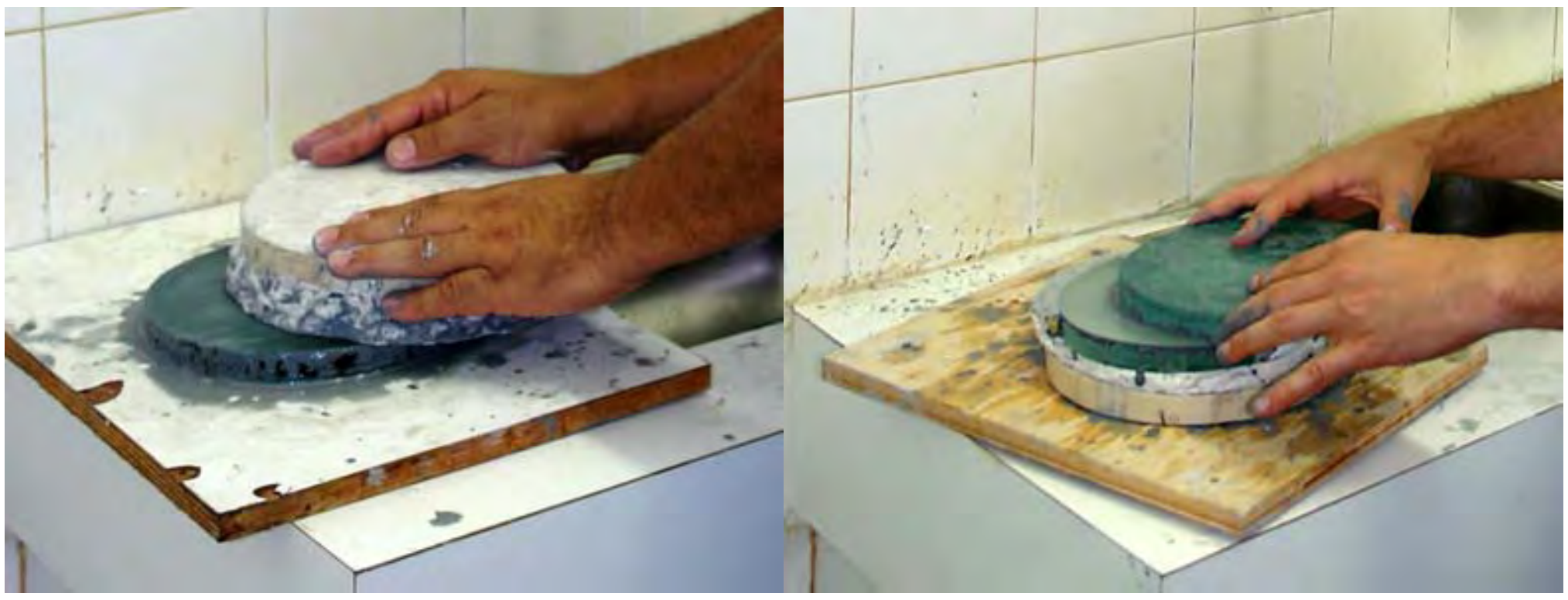

Fig. 4.23: Sessões alternadas de desbaste (bloco sobre a ferramenta e a ferramenta sobre o bloco).

Depois, bloco e ferramenta são lavados para eliminar os resíduos de abrasivo #80 e estas mesmas operações de desbaste são processados com abrasivos cada vez mais finos: #120 (carbureto de silício com grãos de ~125µm de diâmetro) e #320 (carbureto de silício com grãos de ~35µm de diâmetro), figura 4.23.

No final do processo, quando as metades foram acamadas à sua base, o deslocamento das imagens em X permaneceu inalterado, mas havia resultado um pequeno deslocamento horizontal oposto (chamado de configuração em A) ao deslocamento anterior ao desbaste, figura 4.24.

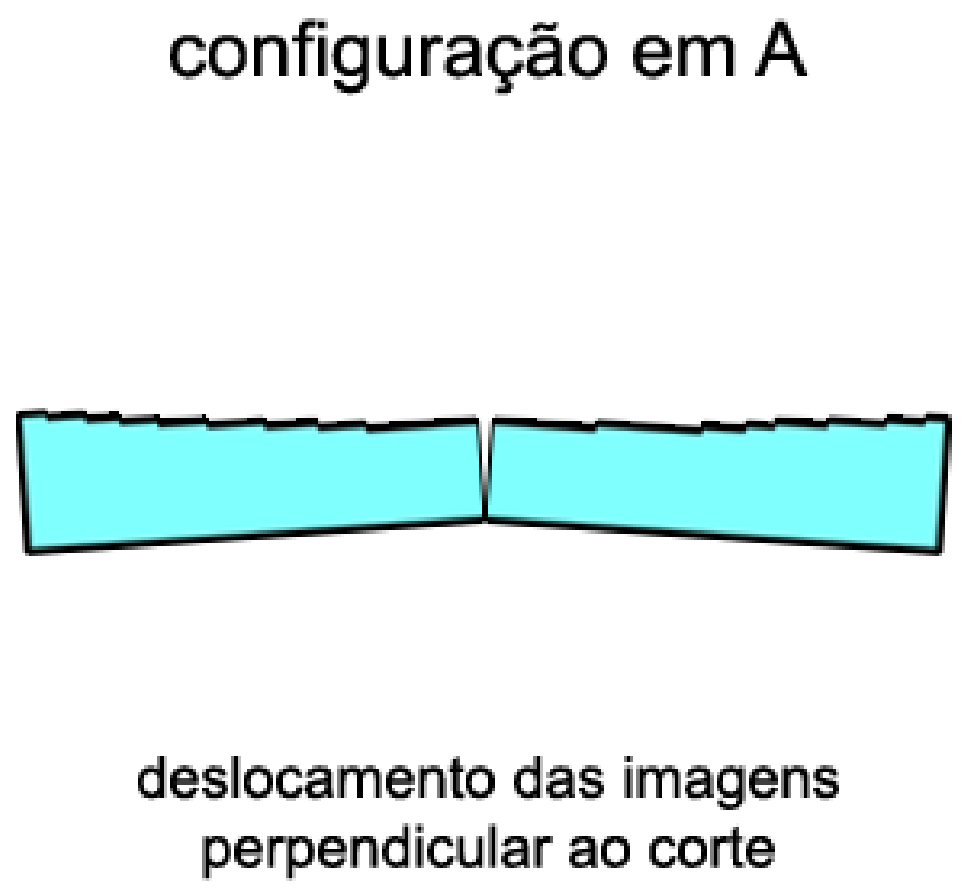

Fig. 4.24: Esquema da configuração em A. A imagem do espelho da esquerda encontra-se à esquerda da linha paralela ao corte. O oposto ocorre com o espelho da direita.

Para verificar se este defeito de deslocamento estava sendo introduzido pelo método de fixação (gesso), foi montada uma nova base, agora na horizontal, onde o espelho pudesse se assentar pelo seu próprio peso, sem a necessidade de ser seguro verticalmente.

Esta mesa-base foi construída em alumínio, com parafusos de ajuste de rosca fina com molas travadoras, pegadores laterais para fixação do espelho e pés com alturas variáveis, para o ajuste de nível, figura 4.25.

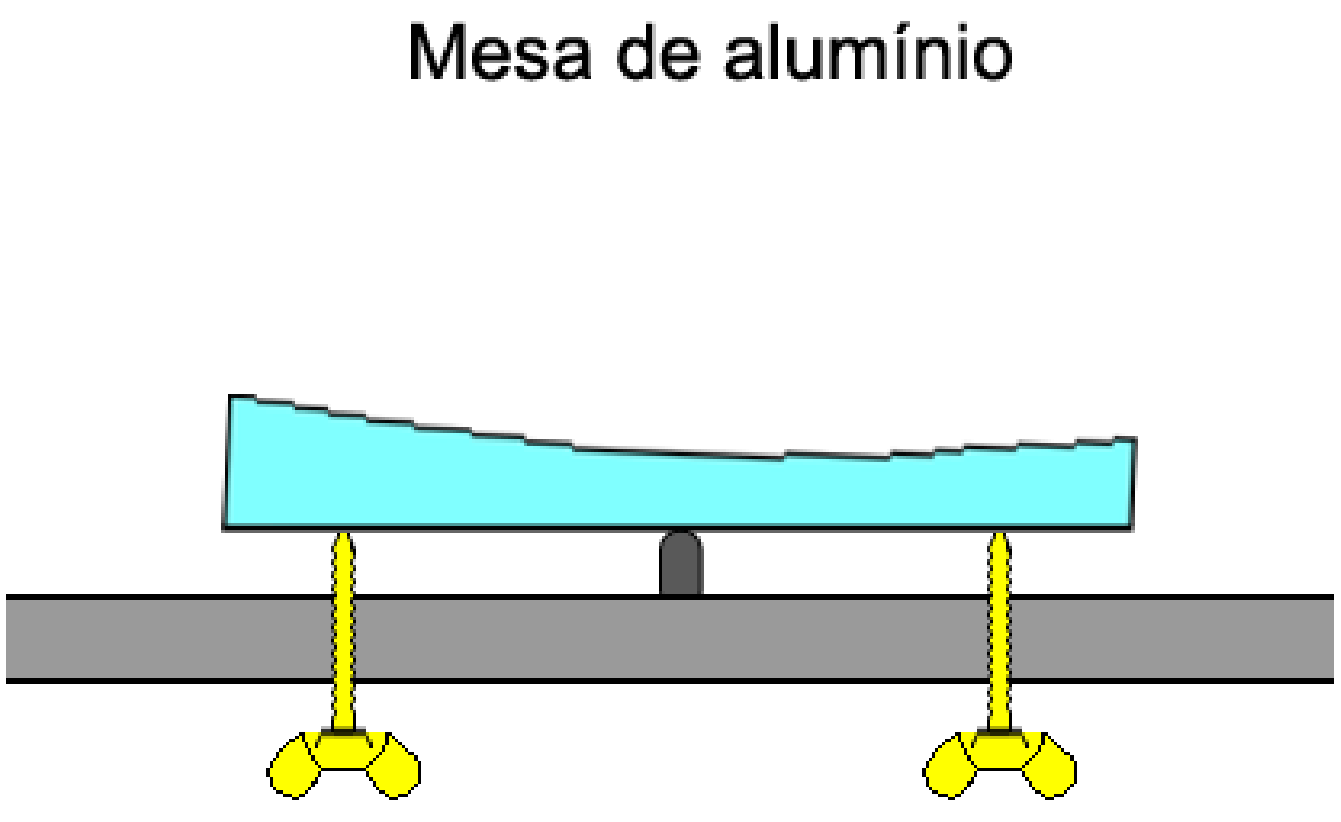

Fig. 4.25: Esquema simplificado da mesa de alumínio pra ajuste da configuração em X. As metades já foram desbastadas anteriormente.

Uma bancada foi construída em metalon, em forma de tronco de pirâmide, para que o espelho a ser trabalhado pudesse ficar numa base horizontal, refletindo as imagens de uma fonte, verticalmente acima dele, sobre uma régua metálica no plano focal. A figura 4.26 mostra uma foto da estrutura da bancada vertical, ainda na fase de construção e acabamentos, enquanto a mesa-base de alumínio estava sendo confeccionada na oficina do Observatório Nacional.

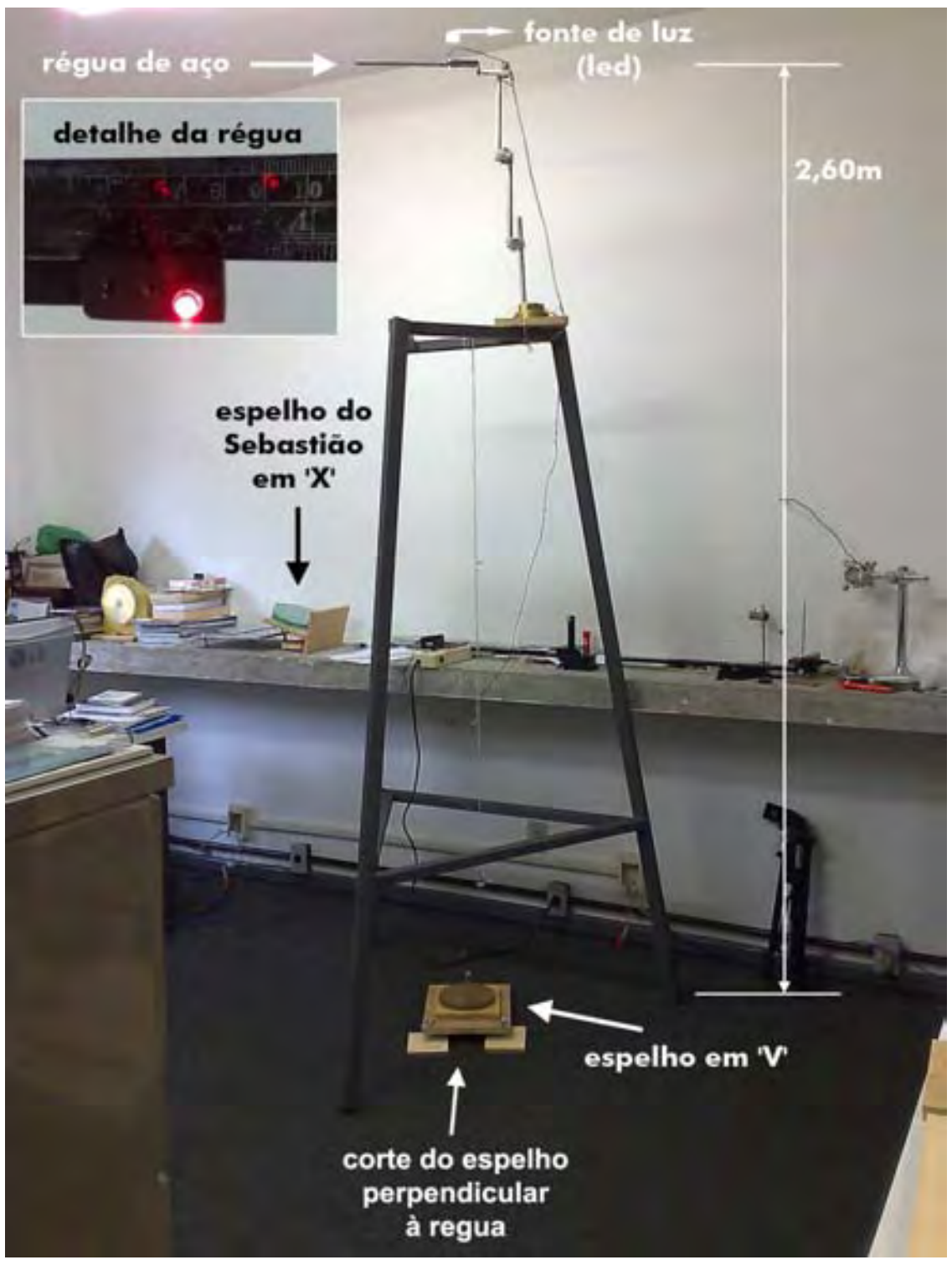

Fig. 4.26: Teste da bancada vertical com um espelho de ~1300mm de distância focal. A fonte e suas imagens sobre a régua aparecem no detalhe.

Para permitir a leitura precisa da régua, uma câmara foi instalada próxima, de modo que sua imagem pudesse ser vista num monitor. A bancada ganhou também uma base em madeira para o apoio da mesa-base de alumínio. Pegadores laterais ajustáveis contêm os movimentos laterais dos hemi-espelhos, figura 4.27.

A mesa-base de alumínio é centralizada com a fonte, usando-se um fio de prumo, de forma que a fonte ficasse diretamente acima do apoio fixo central do espelho. Uma linha de referência é traçada sobre a mesa-base e uma segunda régua de metal é fixada à mesa-base, paralelamente a esta linha. e com um espelho plano colocado sobre a base de madeira, de modo a refletir a régua junto à fonte, a mesa é alinhada até que a imagem refletida da régua de cima fique paralela à régua de baixo (fig. 4.28). O espelho heliométrico é, então, colocado sobre os parafusos de ajuste da mesa-base, com seu corte alinhado à esta linha de referência.

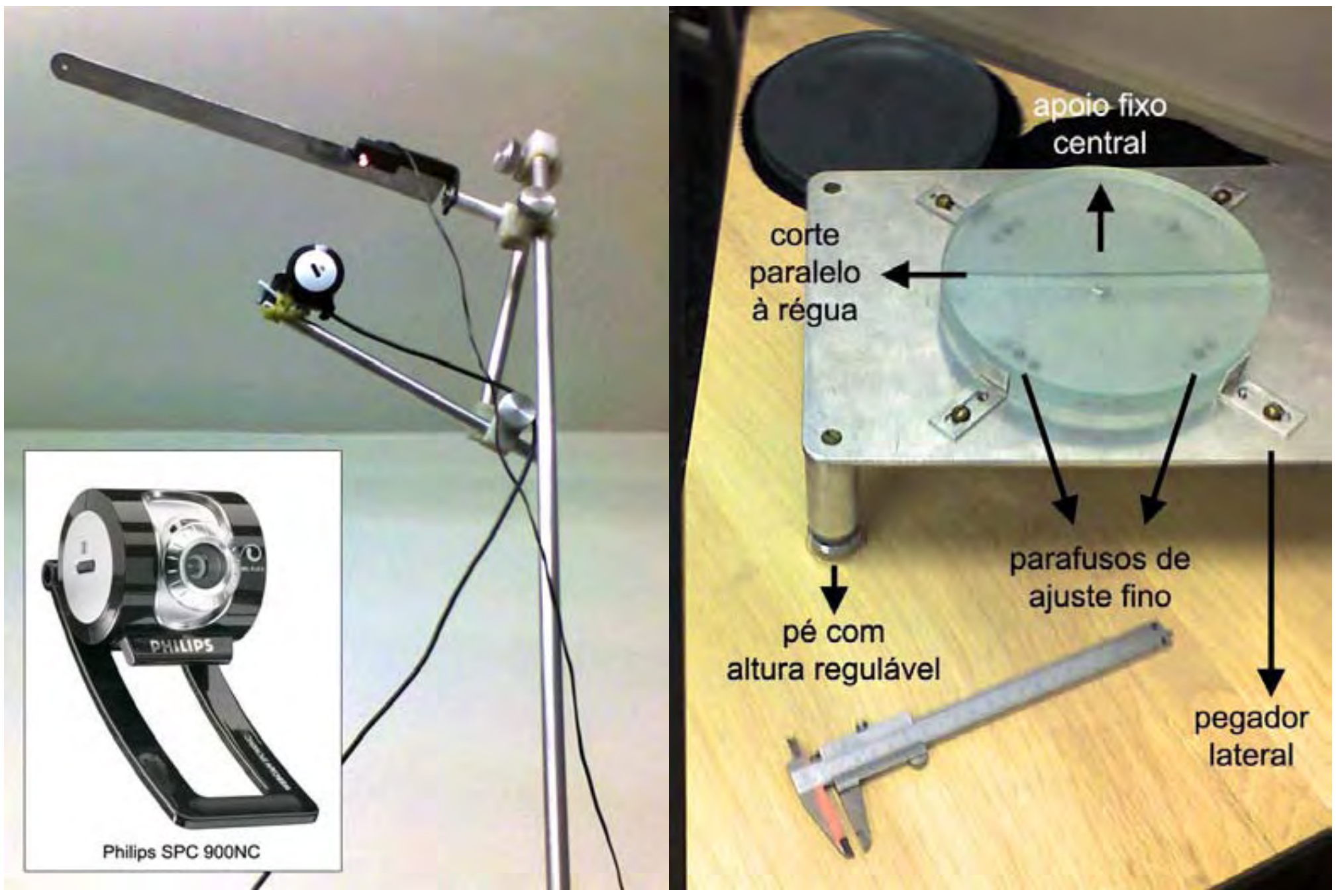

Fig. 4.27: À esquerda, detalhe da webcam montada junto à régua e da mesa de alumínio no alto da bancada. À direita, o espelho sobre a mesa-base. Sob o espelho, podem ser vistos o ponto de apoio fixo e os parafusos de ajuste.

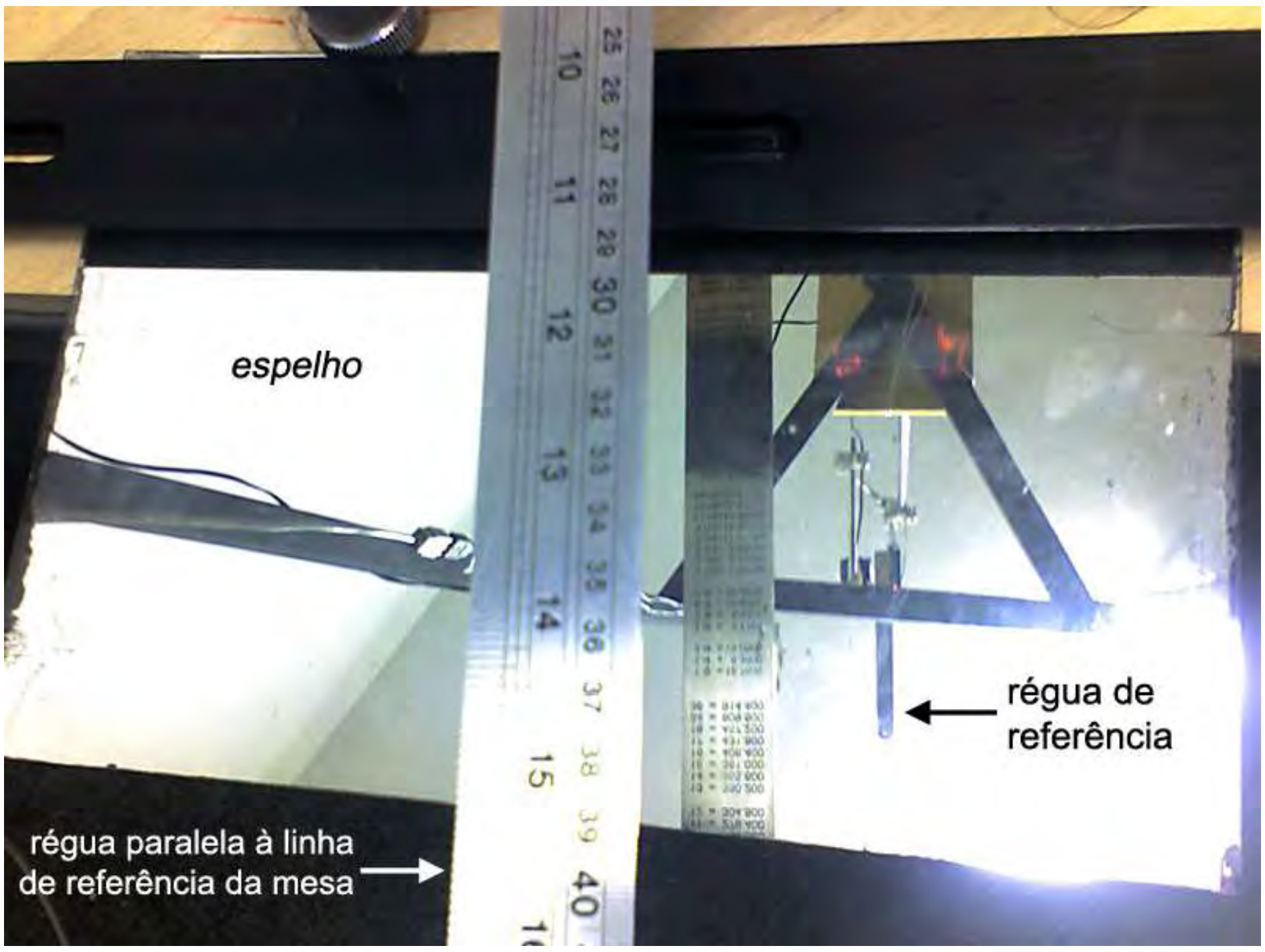

Fig. 4.28: Esquema do alinhamento da mesa. O conjunto era considerado alinhado quando a imagem da régua junto à webcam era visualmente paralela à régua de referência da mesa.

### 4.1.3.2 - Testes de fixação da configuração

A imagem da figura 4.29 mostra a evolução da configuração após um intervalo de 18 horas. Percebeu-se que houve um pequeno deslocamento conjunto para fora da régua, proveniente de acomodações no aparato instrumental, mas como a distância relativa entre as imagens permaneceu a mesma, a configuração do espelho não se alterou, podendo o teste ser prosseguido.

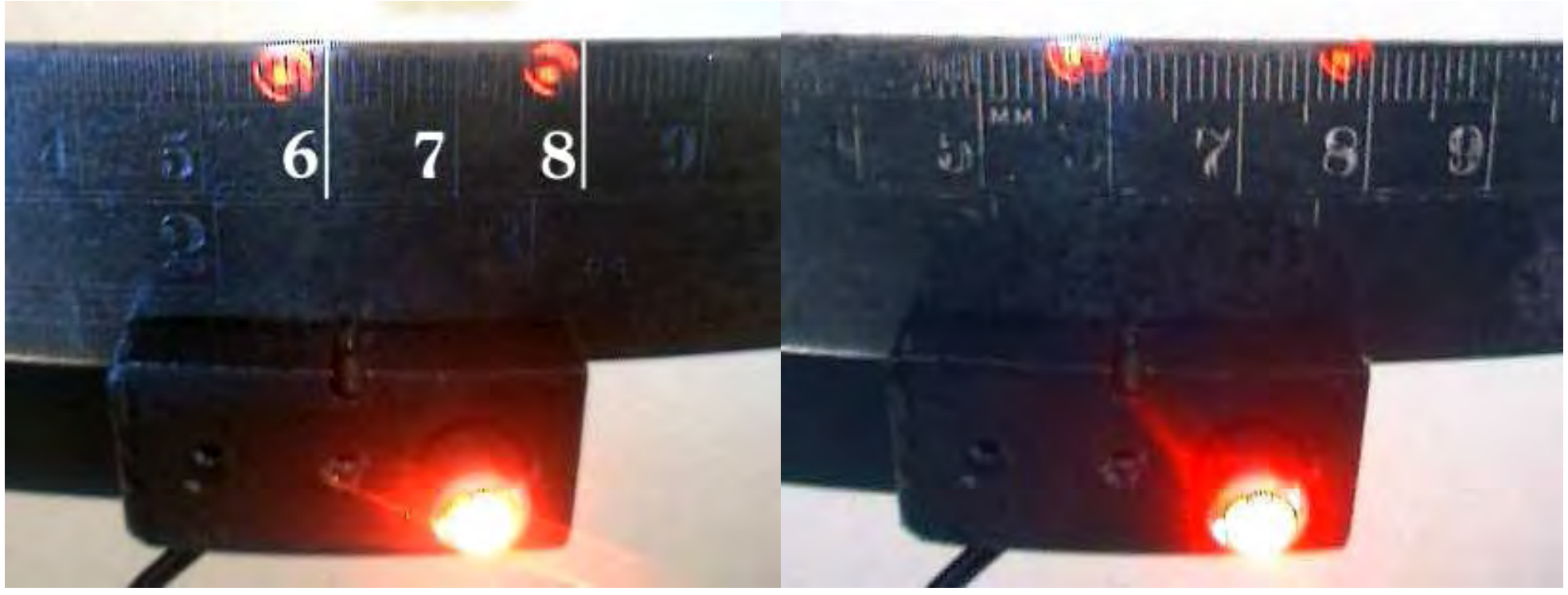

Fig. 4.29: Imagens da régua, mostrando que houve um pequeno deslocamento do conjunto mesa/bancada.

O gesso branco comum foi substituído por gesso pedra especial, tipo IV, por causa de suas características físicas: elevada resistência à compressão, flexão e baixíssima expansão.

Um anel foi colocado no centro do disco, para se preservar esta superfície refletora, de forma que se pudesse monitorar a evolução da configuração do espelho durante a aplicação e cura do gesso. Um anel de PVC serviu para conter o gesso e preservar a mesa de alumínio, figura 4.30.

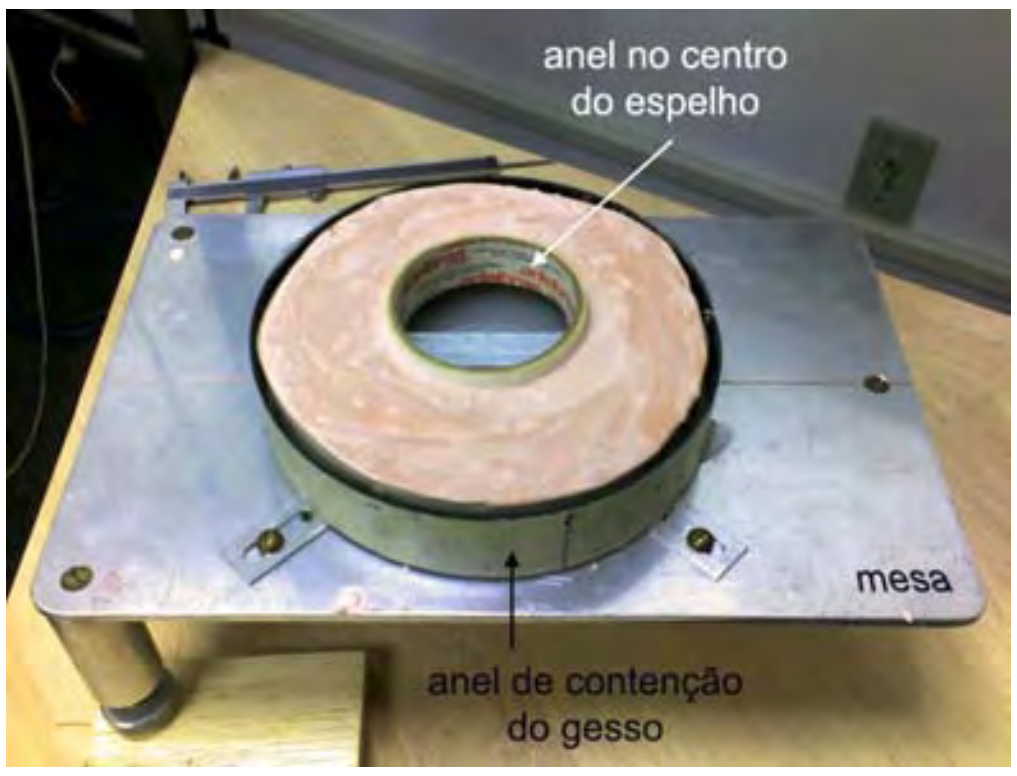

Fig. 4.30: Gesso pedra melhorada aplicado sobre o espelho.

Durante a cura do gesso, percebeu-se que a configuração em X do espelho modificava-se lentamente em direção à configuração em V, sinal de que o gesso estava encolhendo durante o processo e levantando as superfícies.

Este teste foi repetido mais duas vezes, tomando-se o cuidado de, antes de um novo teste, se retirar completamente o gesso anterior, alinhar-se o espelho e ajustar-se o diedro na configuração somente em X, com o uso dos parafusos da mesa-base. Em cada teste, diferentes modos de aplicação do gesso sobre as superfície foram experimentadas e em todas elas, após a cura, o diedro apresentou a configuração residual em V.

A imagem da figura 4.31 mostra a diferença de configuração do espelho heliométrico depois de 24 h de cura. Vê-se que apesar da distância vertical entre as imagens ter se mantido, uma separação lateral de 2,5 mm é verificada, indicando uma pequena configuração em V.

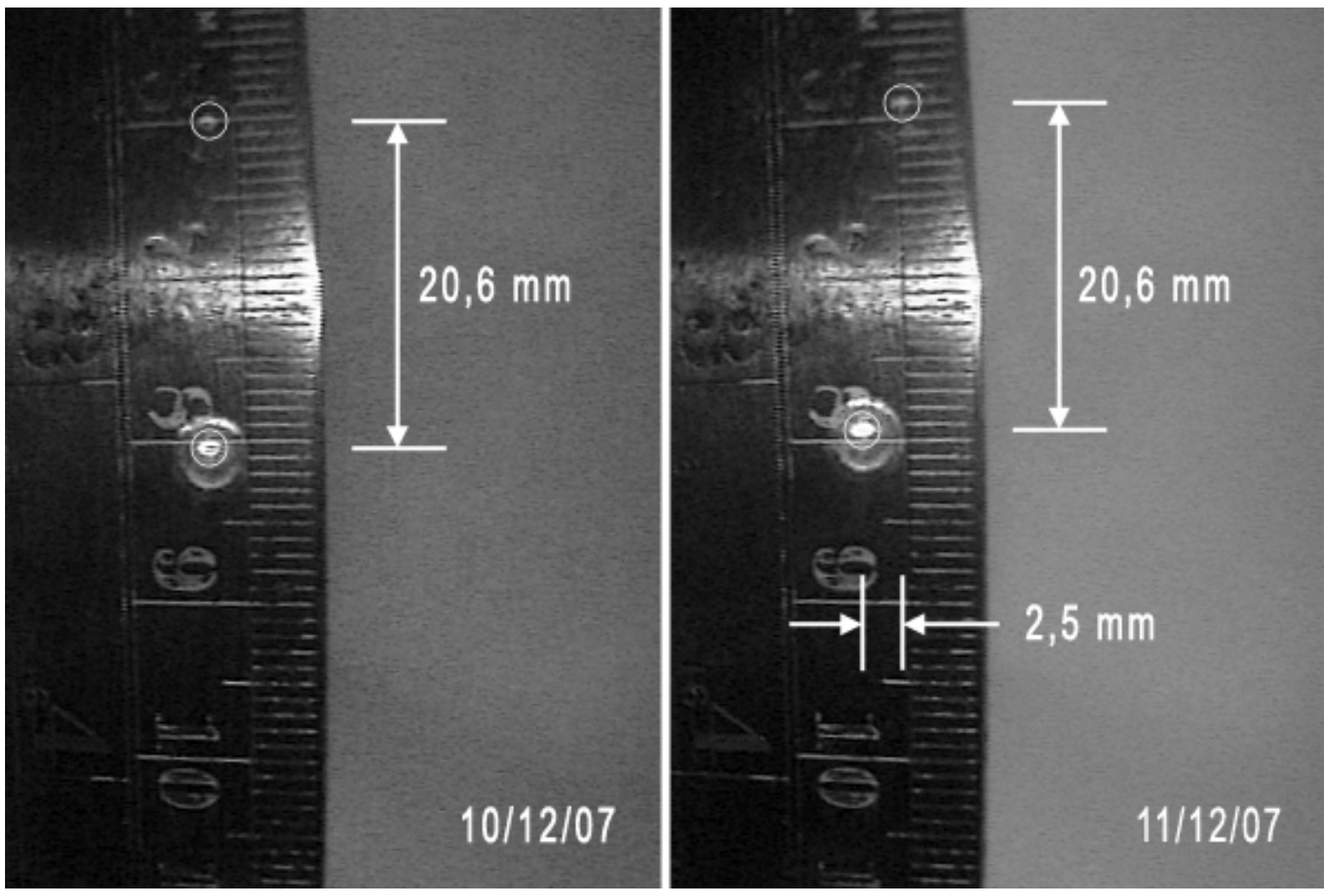

Fig. 4.31: Antes e depois da secagem da película do gesso pedra especial. Os pequenos círculos brancos indicam o centro das imagens.

Em conclusão, esta técnica de fixação para a fabricação de hemi-espelhos prismáticos com gesso não atende às especificações do projeto, pois confirmou-se que o gesso, quando cura, modifica seu volume, deslocando as peças de suas posições originais.

### 4.1.3.3 - Mudança na configuração da mesa

Em resposta ao problema apresentado no item anterior, uma nova metodologia foi desenvolvida para a fixação do diedro. Esta consiste em se ajustar a configuração com os hemi-espelhos fixados à própria mesa-base de alumínio de forma a se poder desbastar o fundo das peças sem que elas precisem ser removidas.

A mesa-base foi invertida, e os hemi-espelhos posicionados sob ela, de modo que os parafusos de ajuste fino, agora trabalham contra a superfície parabolizada, e não mais sobre o fundo das peças.

Aberturas foram feitas na mesa para permitir a passagem da luz para o monitoramento da configuração, figura 4.32.

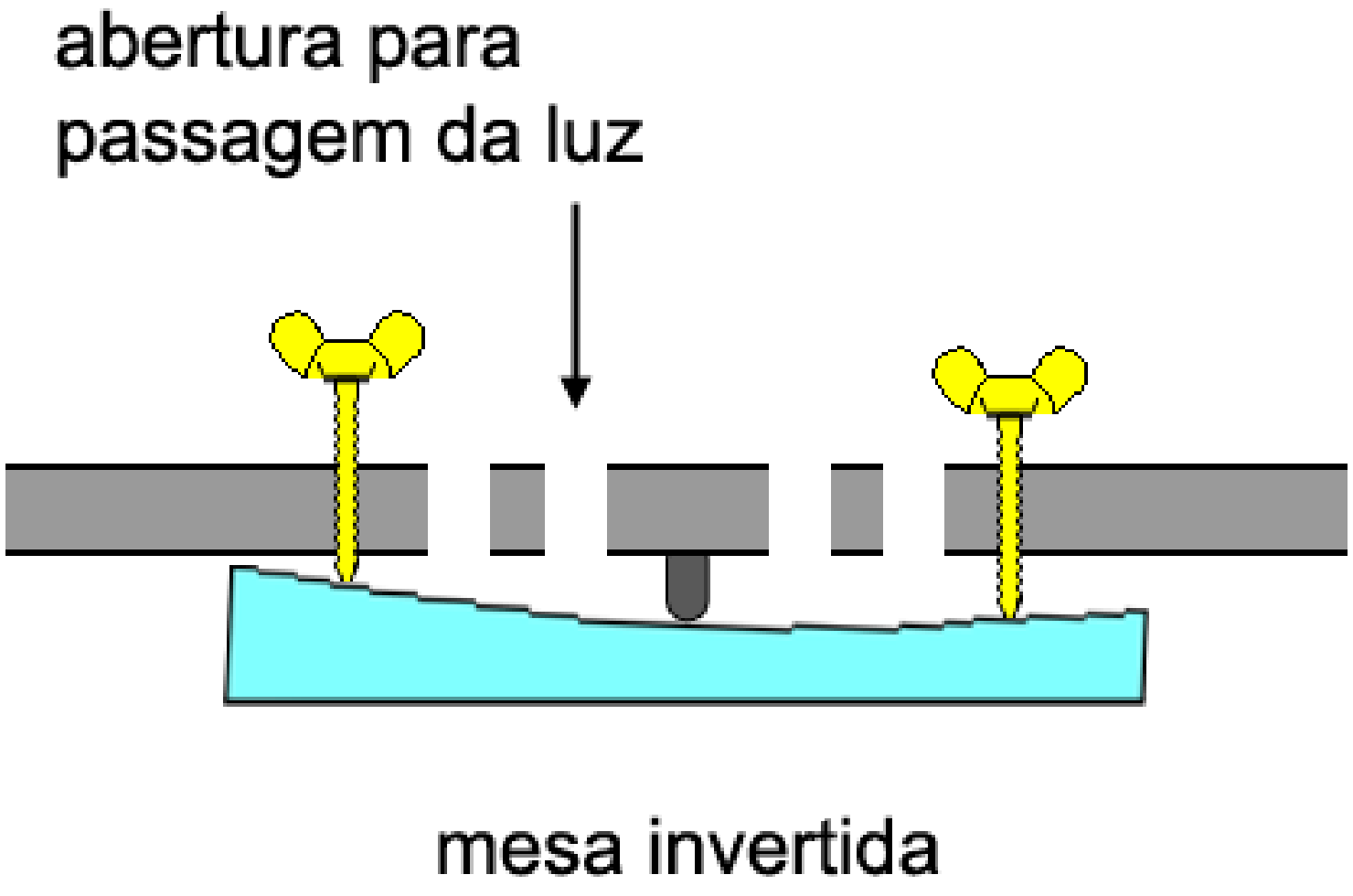

Fig. 4.32: Esquema simplificado da mesa invertida. O espelho era forçado contra os parafusos através de tiras de borracha (que não aparecem no desenho acima).

Pequenos calços de vidro foram colados com resina epóxi na lateral do espelho (dois em cada metade), de forma que eles ficassem entre o fundo e a superfície do espelho, e deixados secar por mais de 48 horas. Alças foram passadas nestes calços e por buracos na mesa. Tiras de borracha foram passadas pelas alças e esticadas a fim de que o espelho ficasse seguro e sempre pressionado contra os parafusos de ajuste. Os suportes laterais do espelho foram apertados, mas ainda permitindo que o espelho pudesse se movimentar.

O ajuste do diedro é feito do mesmo modo anterior: as imagens da fonte são monitoradas pela câmara e os parafusos são ajustados até que a configuração em X é alcançada. Feito isto, os suportes laterais são fixados, e o conjunto é retirado, preso à mesa-base para que o fundo das peças possa ser trabalhado, figura 4.33.

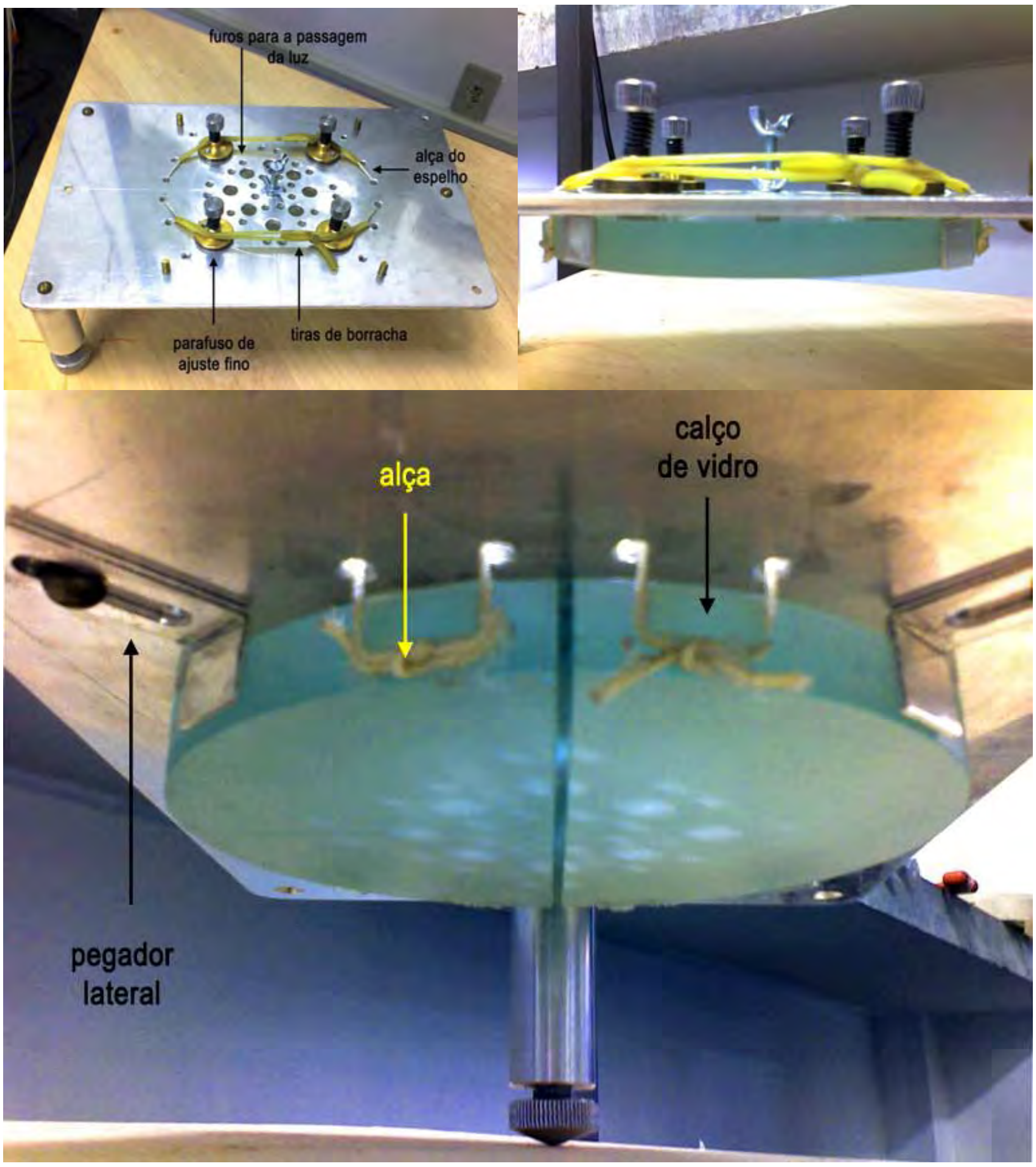

Fig. 4.33: Diversas vistas da mesa invertida com o espelho já fixo a ela.

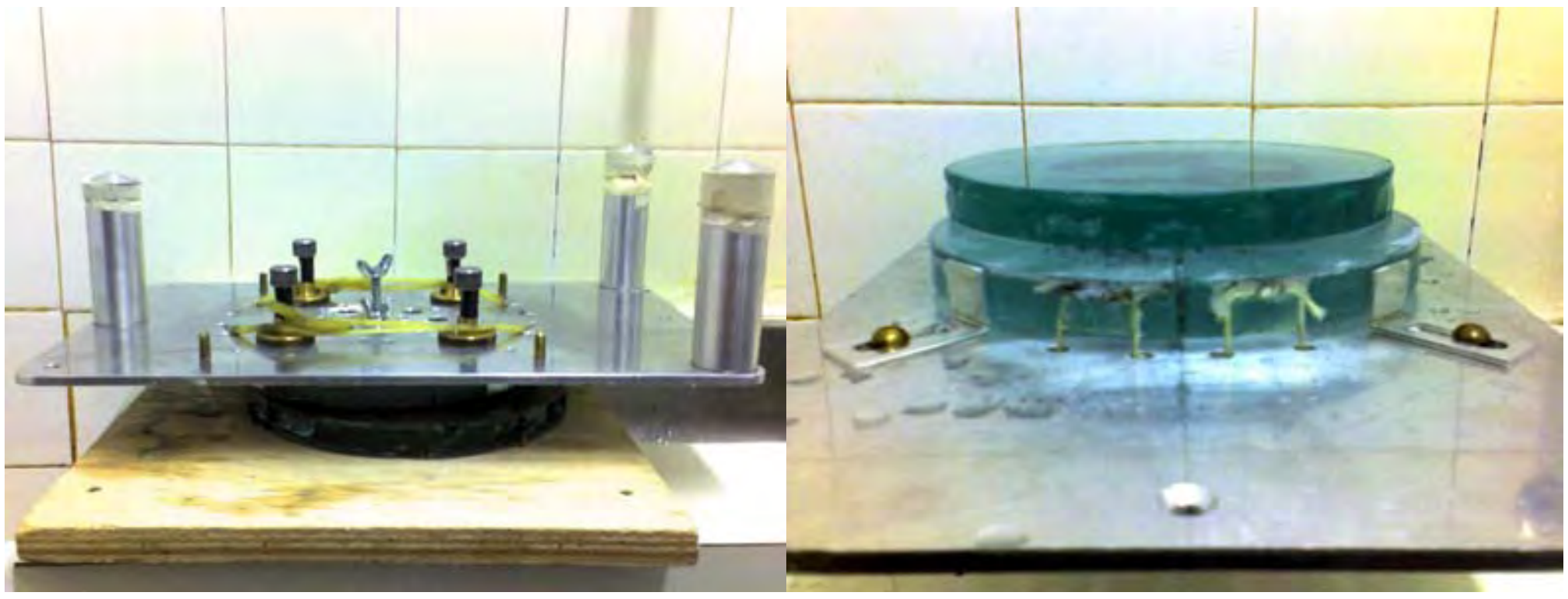

Fig. 4.34: Desbaste do espelho, fixado à mesa. Mesa sobre a base, na primeira foto e base sobre a mesa, na segunda.

Após 8 sessões de desbastes alternados (fig. 4.34), as superfícies estavam planas, com as metades acamando-se perfeitamente sobre a base, sem resíduo aparente de configurações transversais, quer em V ou em A.

Em conclusão, se os hemi-espelhos estiverem fixados, em diedro, em uma mesa-base especialmente preparada para isso e as peças forem desbastadas sem serem removidas desta mesa, o resultado atende à especificação do heliômetro.

### 4.1.3.4 - Teste óptico

Para se verificar as condições finais da superfície óptica dos hemi-espelhos, a imagem do teste de Ronchi foi registrada (fig. 4.35). Ela mostra que depois de todos os desbastes, a superfície do espelho não sofreu deformações, permanecendo com suas características inalteradas após o espelho ser cortado (vide fig. 4.8).

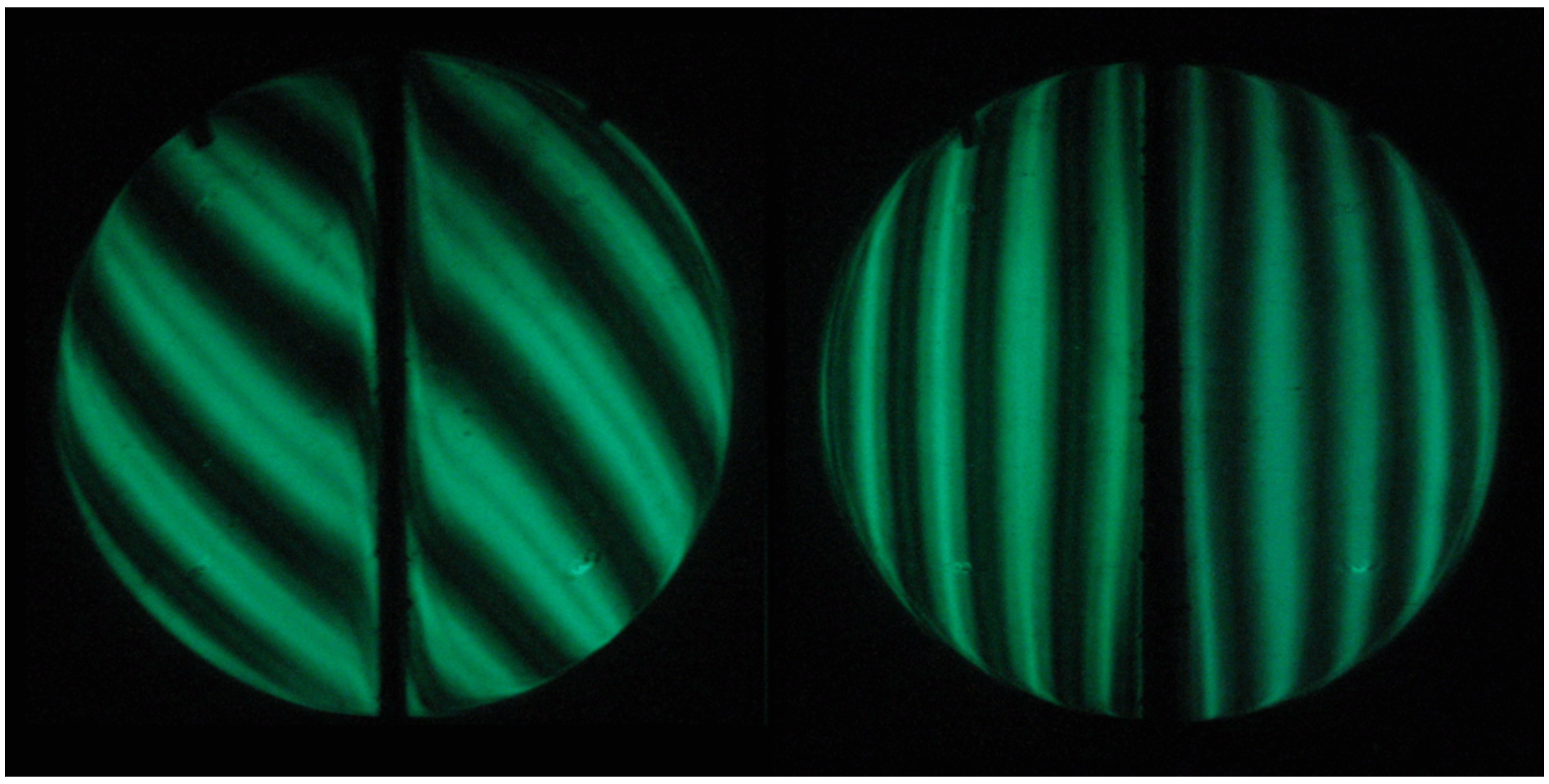

Fig. 4.35: Teste de Ronchi nos hemi-espelho após o desbaste de ajuste da configuração em X.

## *4.2 - A fixação mecânica dos hemi-espelhos*

### 4.2.1 - Fixação dos hemi-espelhos sobre sua base plana

Cada hemi-espelho é apoiado à base sobre 3 calços e molas metálicas transportam a força vertical sobre estes calços, pressionado-o contra a base. Para servir de apoio para as molas dos hemi-espelhos, um sulco é aberto na lateral da base. A figura a 4.36 traz a concepção do projeto.

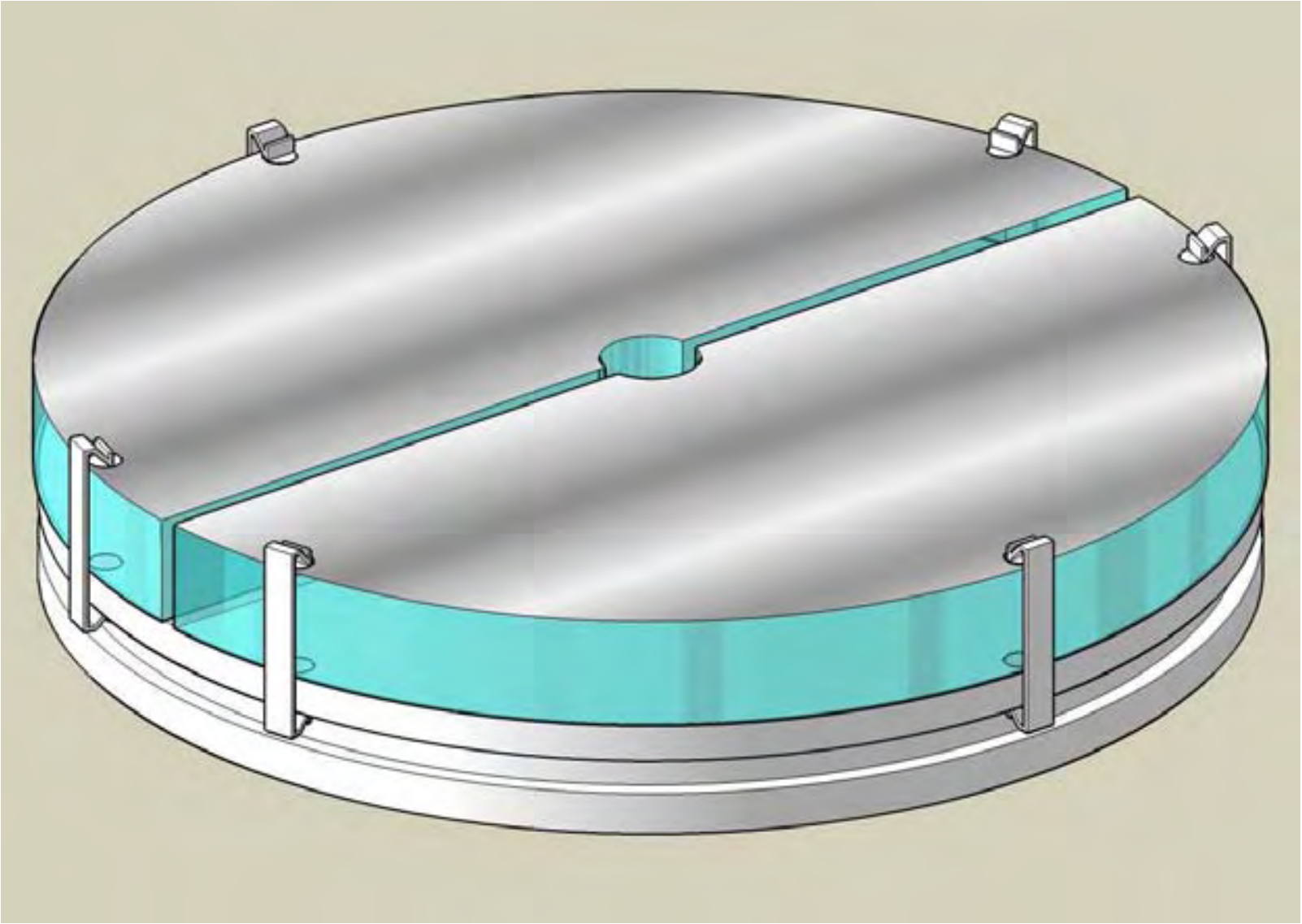

Fig. 4.36: Projeto de fixação vertical dos hemi-espelhos sobre sua base.

Para garantir que os hemi-espelhos não se movam horizontalmente, são utilizados 3 calços laterais. Dois passivos, apoios somente, e um ativo, com efeito mola, figura 4.37.

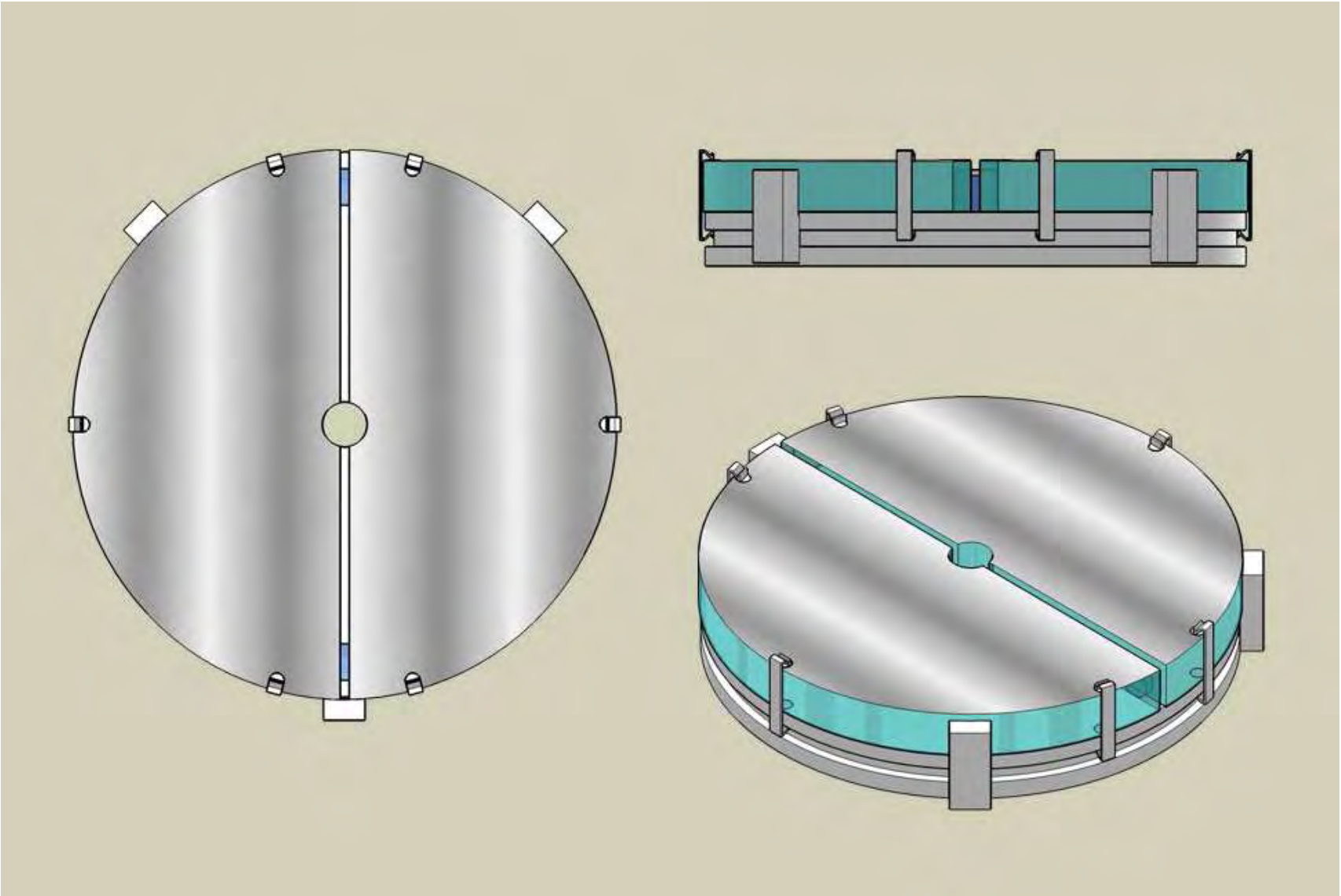

Fig. 4.37: Vistas do projeto de apoios verticais e horizontais do espelho heliométrico protótipo. Pequenos espaçadores serão usados para recuperar o perímetro circular do espelho.

Os calços laterais servem para imobilizar lateralmente a base. Esta, por sua vez, fica apoiada sobre 3 calços, figura 4.38.

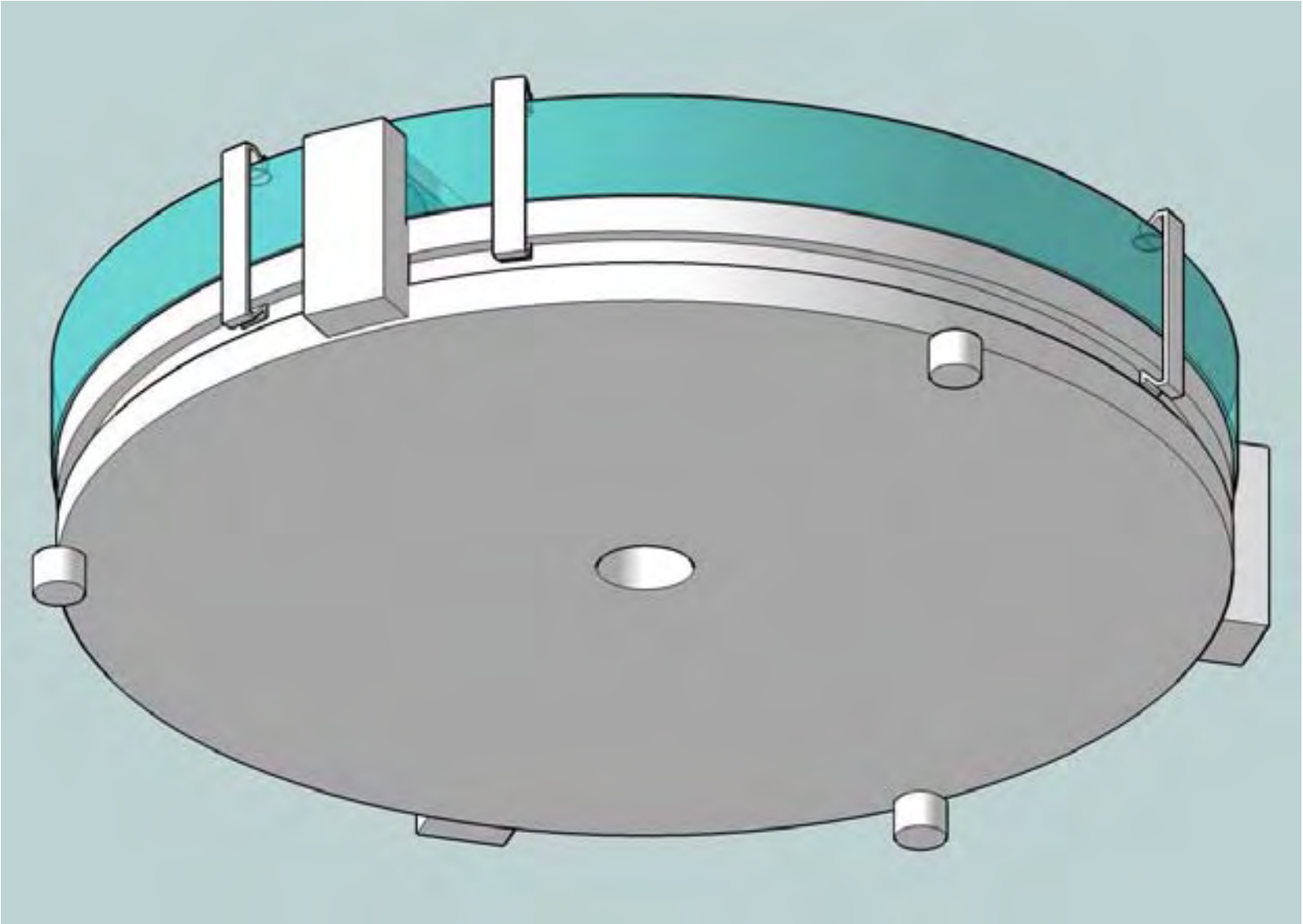

Fig. 4.38: Projeto dos calços inferiores da base do espelho heliométrico.

Para forçar verticalmente a base sobre os calços de apoio, 3 hastes compressoras, com efeito mola, ficam encaixadas no sulco lateral da base. Estas hastes não tocam nos hemi-espelhos, de forma que a força compressora age apenas sobre a base, figura 4.39.

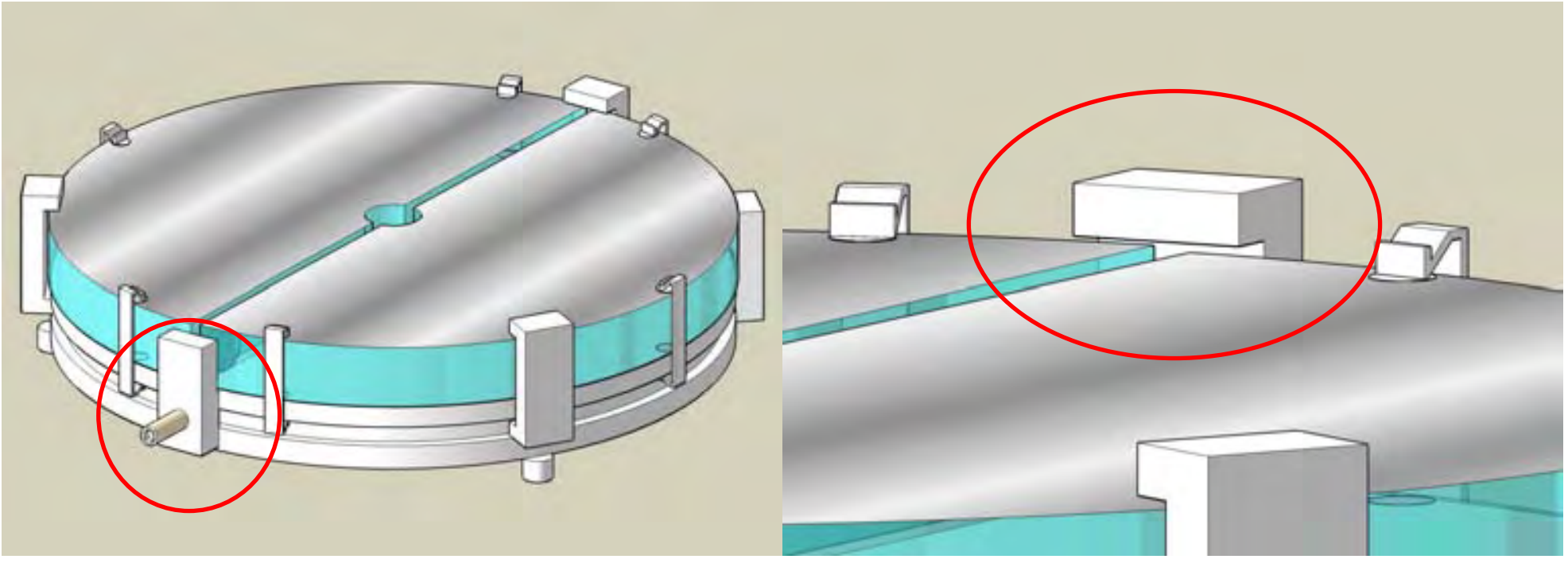

Fig. 4.39: Na imagem à esquerda, o projeto das hastes compressoras da base, com destaque para um pino de compressão lateral, que deverá ficar preso à célula. À direita, detalhe mostrando que as hastes não tocam os hemi-espelhos.

### 4.2.2 - Projeto da célula suporte para o espelho protótipo

Todo conjunto descrito no item 4.2.1 fica dentro de uma célula, feita em alumínio torneado. Parafusos nas laterais fazem a imobilização do conjunto dentro da célula, e uma abertura maior, em frente a um dos batentes laterais serve de apoio para um pino móvel que faz uma compressão ativa, empurrado por uma placa metálica flexível, figura 4.40.

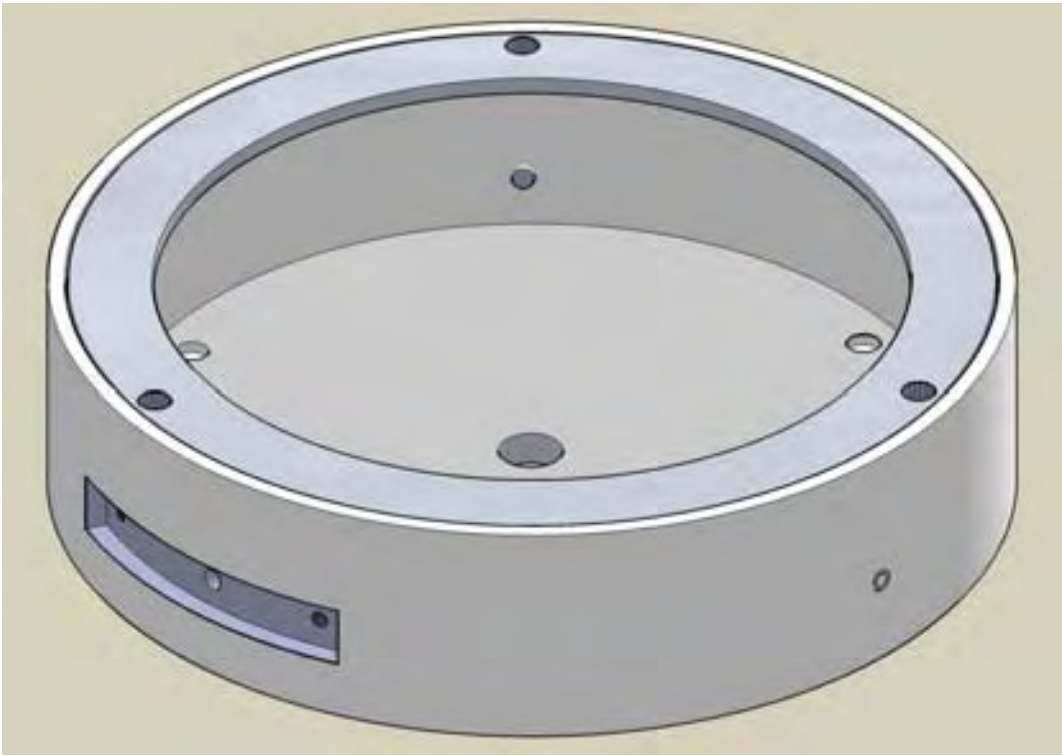

Fig. 4.40: Projeto da célula de alumínio apara o espelho heliométrico protótipo. A abertura lateral maior servirá para acomodar a placa metálica que fará uma compressão horizontal ao conjunto base/hemi-espelhos.

Um aro superior faz o papel da tampa da célula com o principal objetivo, quando fixado à célula, de pressionar verticalmente as hastes compressoras, figura 4.41.

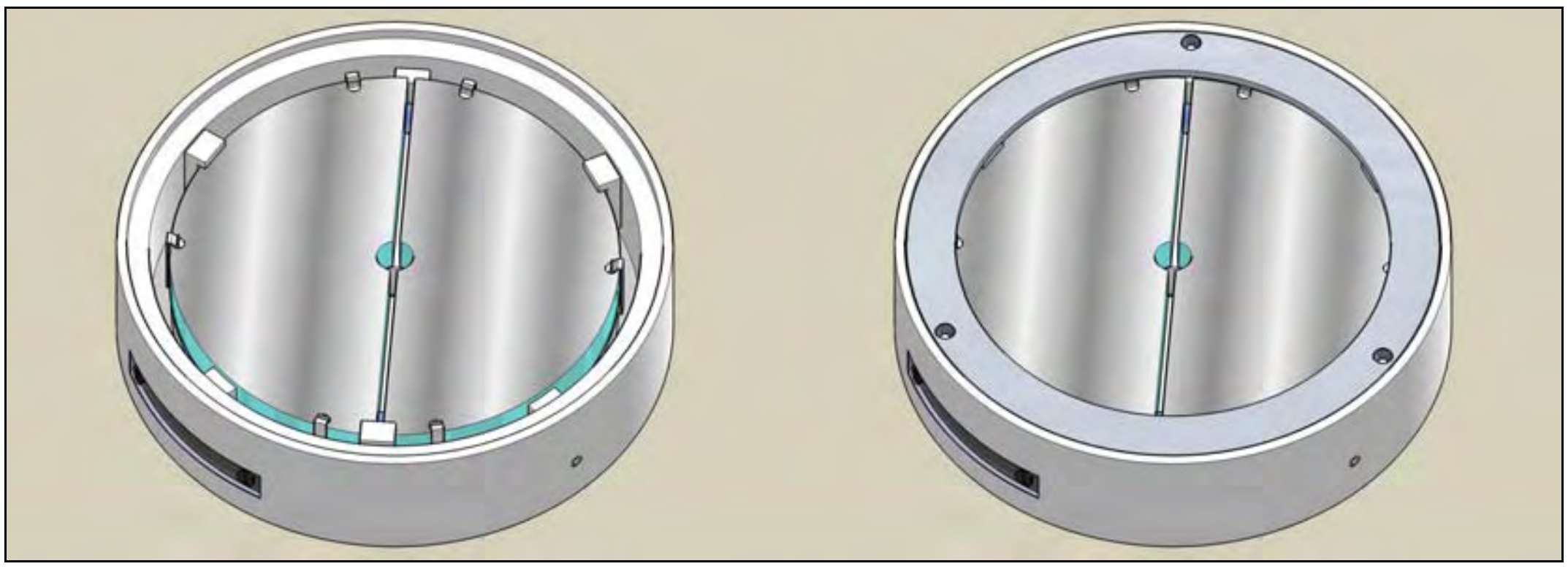

Fig. 4.41: Projeto completo da célula do espelho heliométrico

Esta célula fica presa ao tubo do instrumento através de uma base de colimação, também fabricada em alumínio.

A colimação da célula é feita através de um conjunto de parafusos "push-pull", figura 4.42.

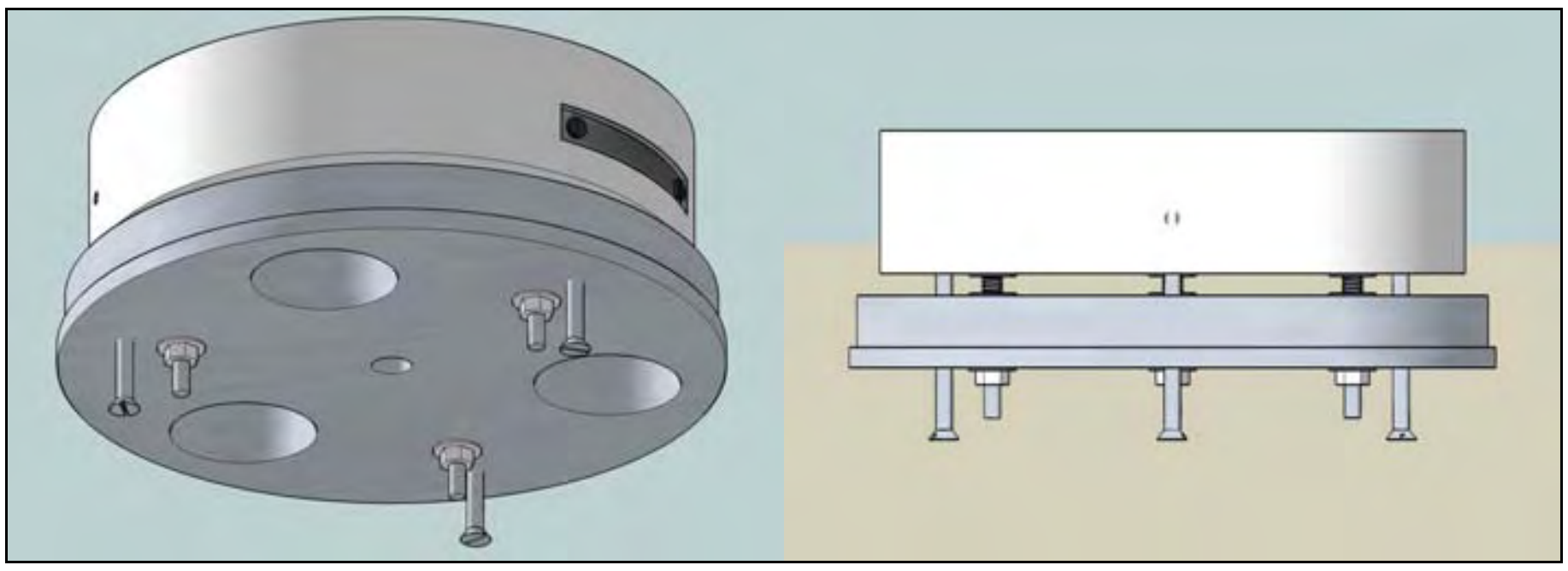

Fig. 4.42: Vistas do projeto da base de colimação. Os três pares de parafusos "push-pull" servirão para colimar o espelho dentro do tubo do telescópio. Os furos na base servem para aliviar o peso.

### 4.2.3 - Seqüência de montagem do espelho heliométrico em sua célula

1. Colocação de lamínulas de mica sobre a base para servirem como os 3 pontos de apoio para cada um dos hemi-espelhos, figura 4.43.

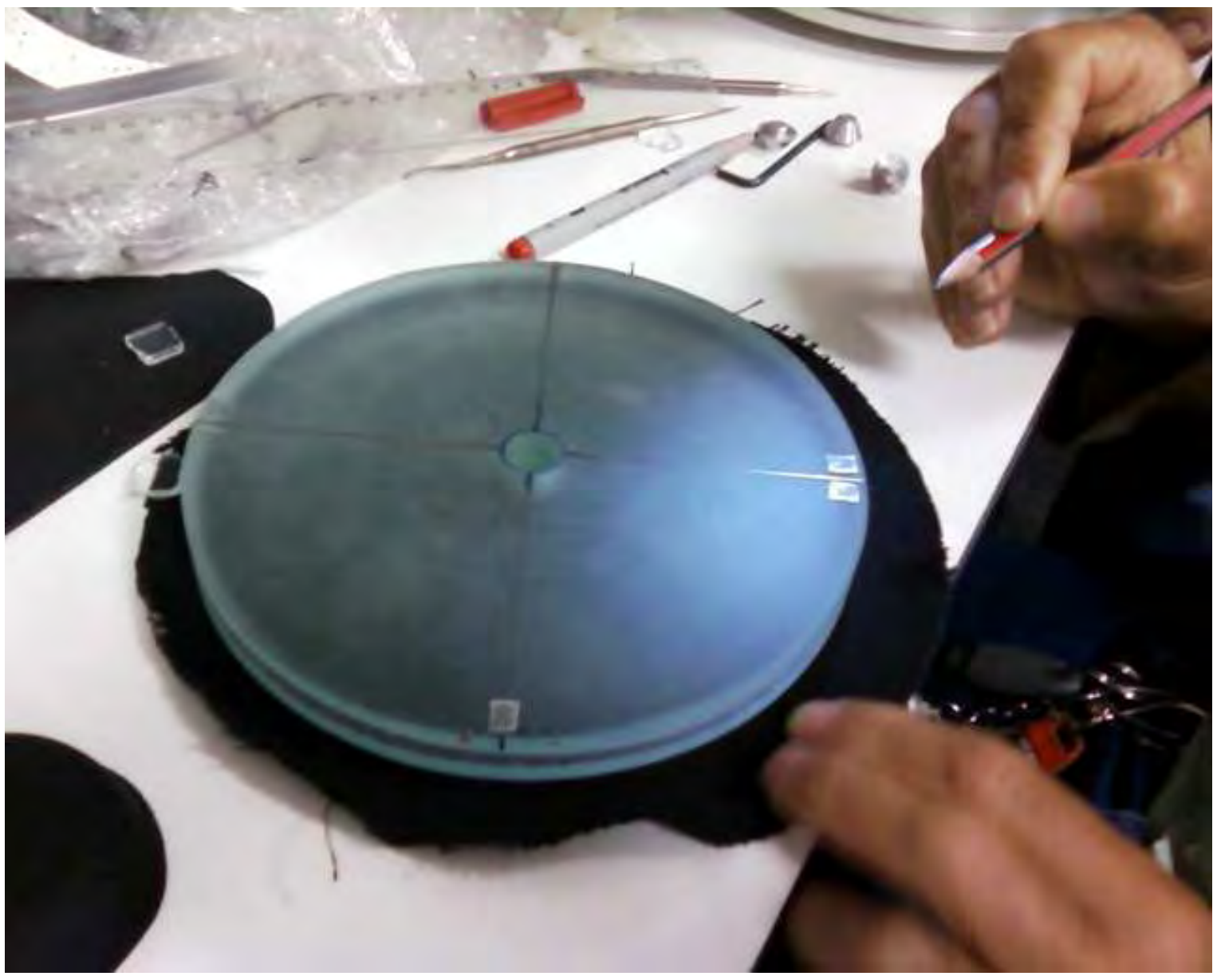

Fig. 4.43: Calços dos hemi-espelhos sendo colocados sobre a base.

2. Os hemi-espelho são posicionados sobre os calços.
3. As molas metálicas são encaixadas de forma a exercerem força verticalmente sobre os calços. Lamínulas de mica são interpostas entre as molas e a superfície para proteger a metalização, figura 4.44.

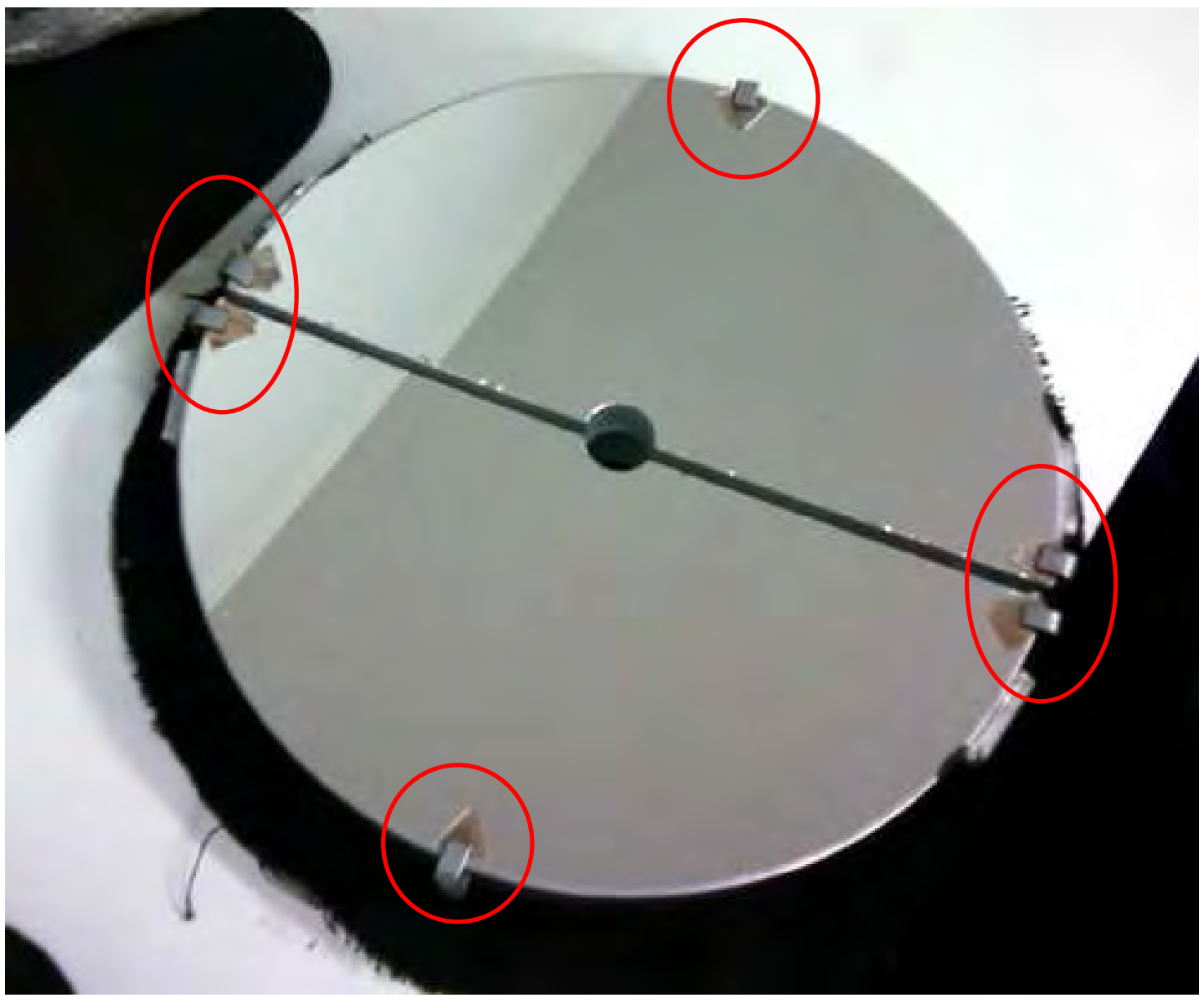

Fig. 4.44: Hemi-espelhos fixados à base. Destaque para os pontos de contato entre as molas e a superfície.

4. Os calços laterais e as hastes compressoras são posicionados em seus lugares, figura 4.45.

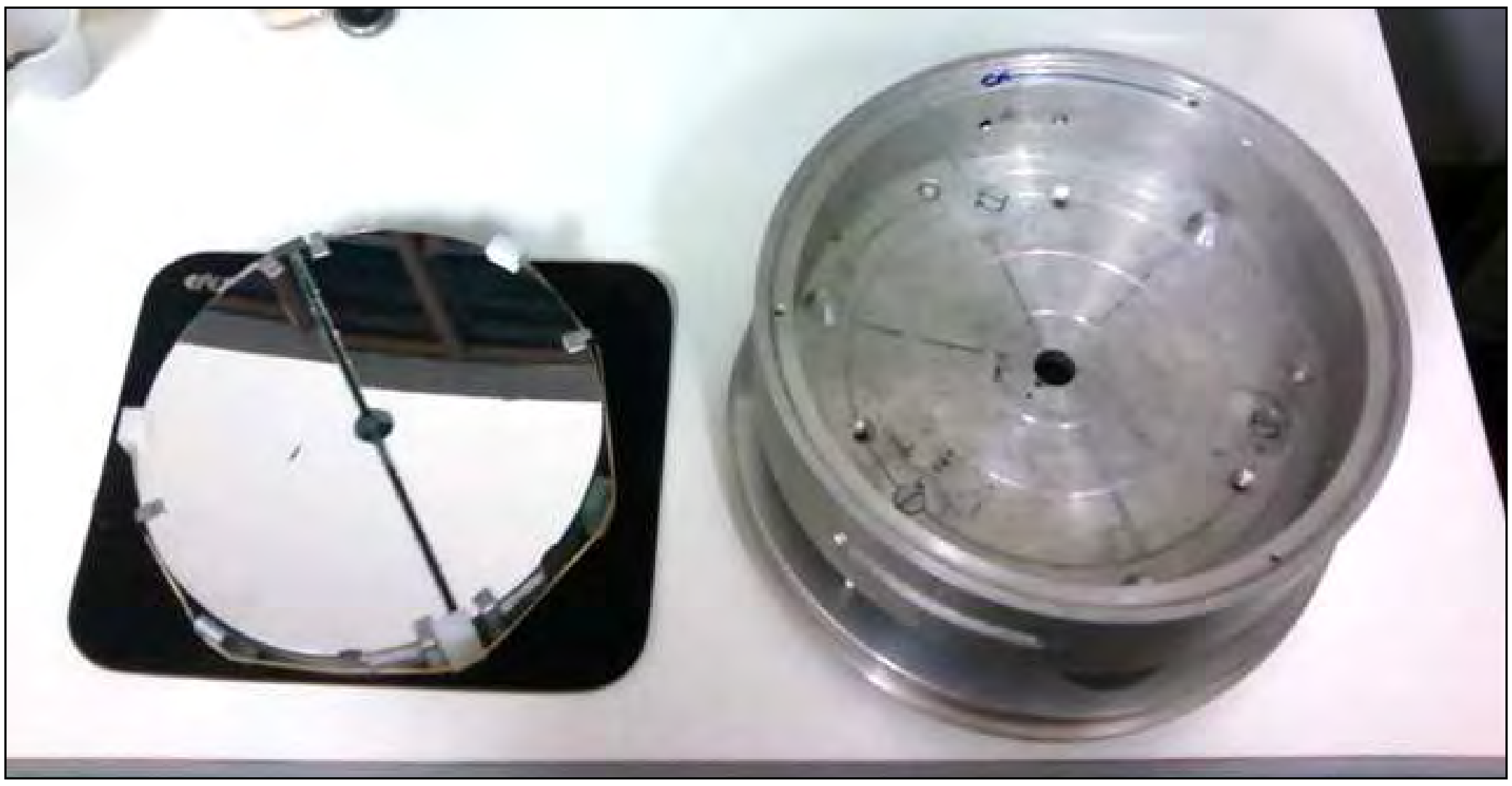

Fig. 4.45: Espelho heliométrico pronto para sua célula.

5. Com uma pinça, especialmente manufaturada, o conjunto todo é seguro através do furo central. O espelho é colocado e posicionado dentro da célula, figura 4.46 e 4.47.

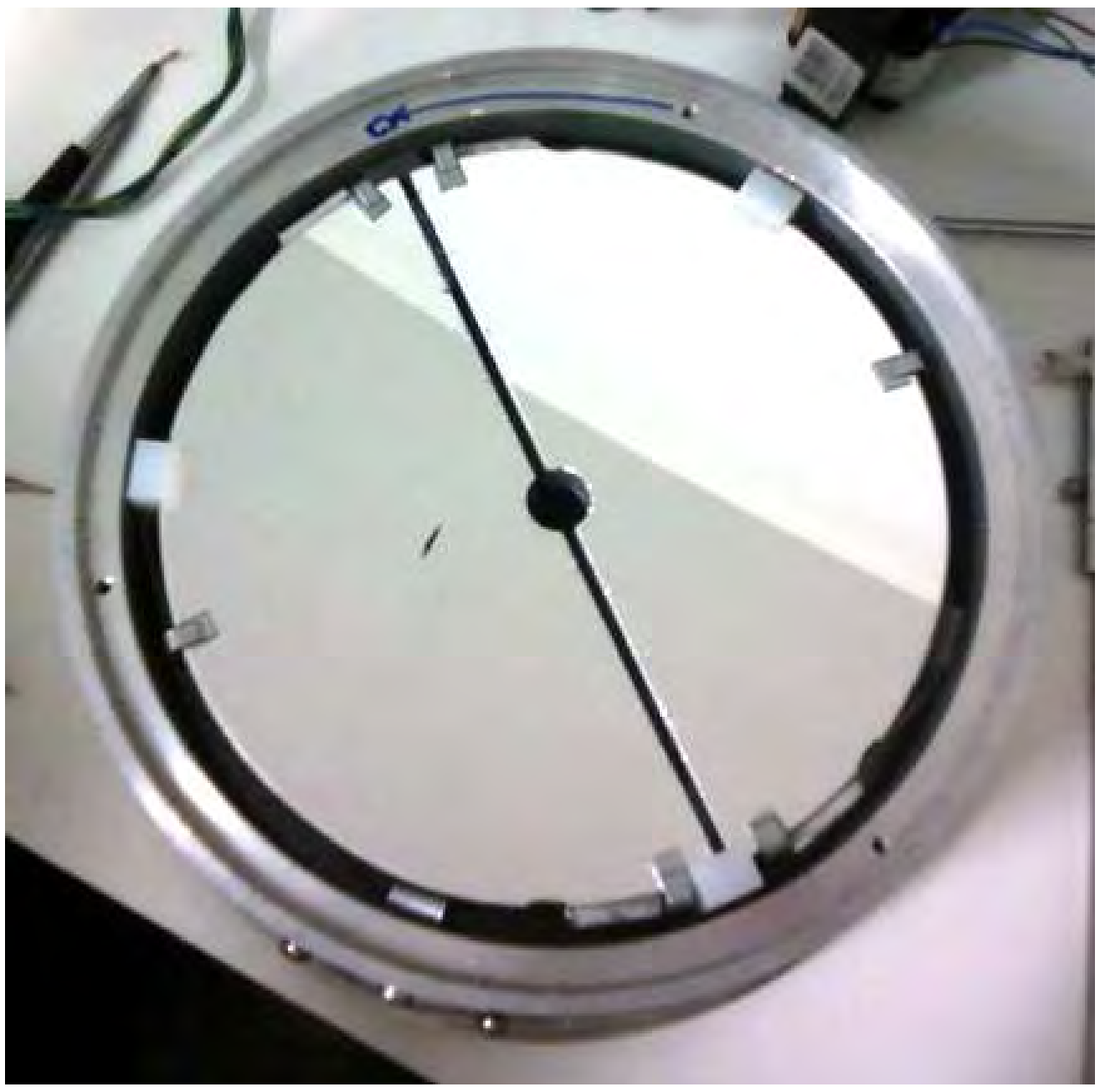

Fig. 4.46: Espelho heliométrico dentro de sua célula.

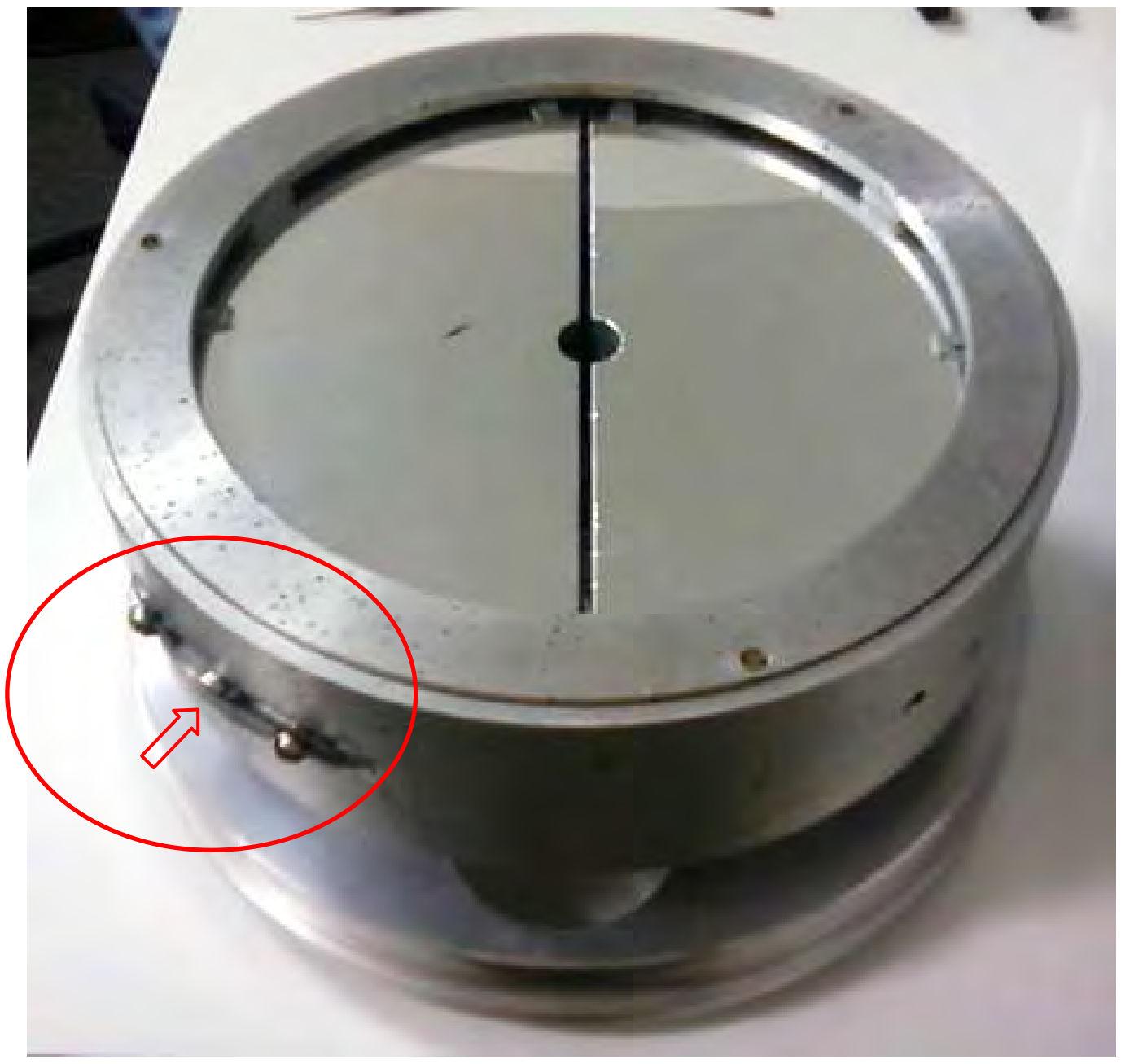

Fig. 4.47: Espelho Heliométrico completamente montado dentro de sua célula, já presa à base de colimação. Destaque para a placa metálica flexível forçando o pino móvel horizontalmente.

## 4.3 - Desenvolvimento do teste de auto-colimação

O objetivo destes ensaios é desenvolver um método para se monitorar as futuras medidas heliométricas. Criar um padrão de laboratório que pudesse ser rotineiramente medido, utilizando-se o espelho heliométrico e o mesmo programa de análise das suas imagens.

A idéia é que qualquer variação, seja por qual motivo for, na medida do padrão, poderá estar relacionada a uma variação correspondente nas medidas. Estas variações poderão ser modeladas e usadas para a correção de uma série de medidas solares.

Para se observar este padrão através do espelho heliométrico, as condições de observações reais deveriam ser recriadas, ou seja, a imagem deste padrão deveria estar infinitamente distante do espelho.

A solução para isso foi colocar este padrão perto do espelho heliométrico e, ao mesmo tempo, no foco de um espelho parabólico, localizado acima do espelho heliométrico, de forma que os raios refletidos por este fossem paralelos ao eixo óptico. O espelho parabólico utilizado tinha 1250 mm de distância focal e foi colocado no alto da bancada vertical. Na base da bancada foi posicionado o espelho heliométrico e no seu centro, um pequeno disco iluminado. Até então, os espelhos estavam soltos, apoiados apenas em sua base, figura 4.48.

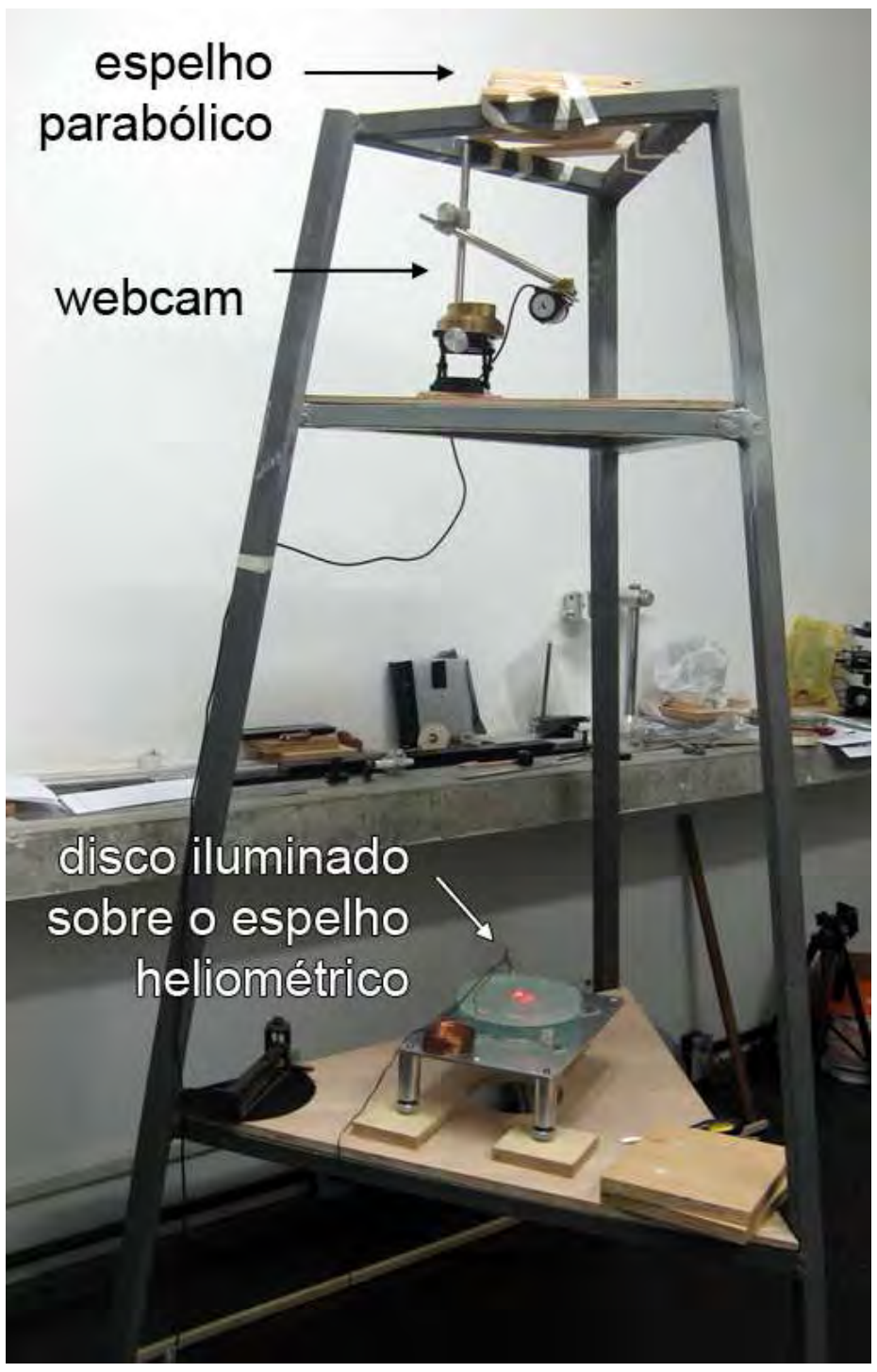

Fig. 4.48: Primeira montagem experimental do teste de colimação do Heliômetro.

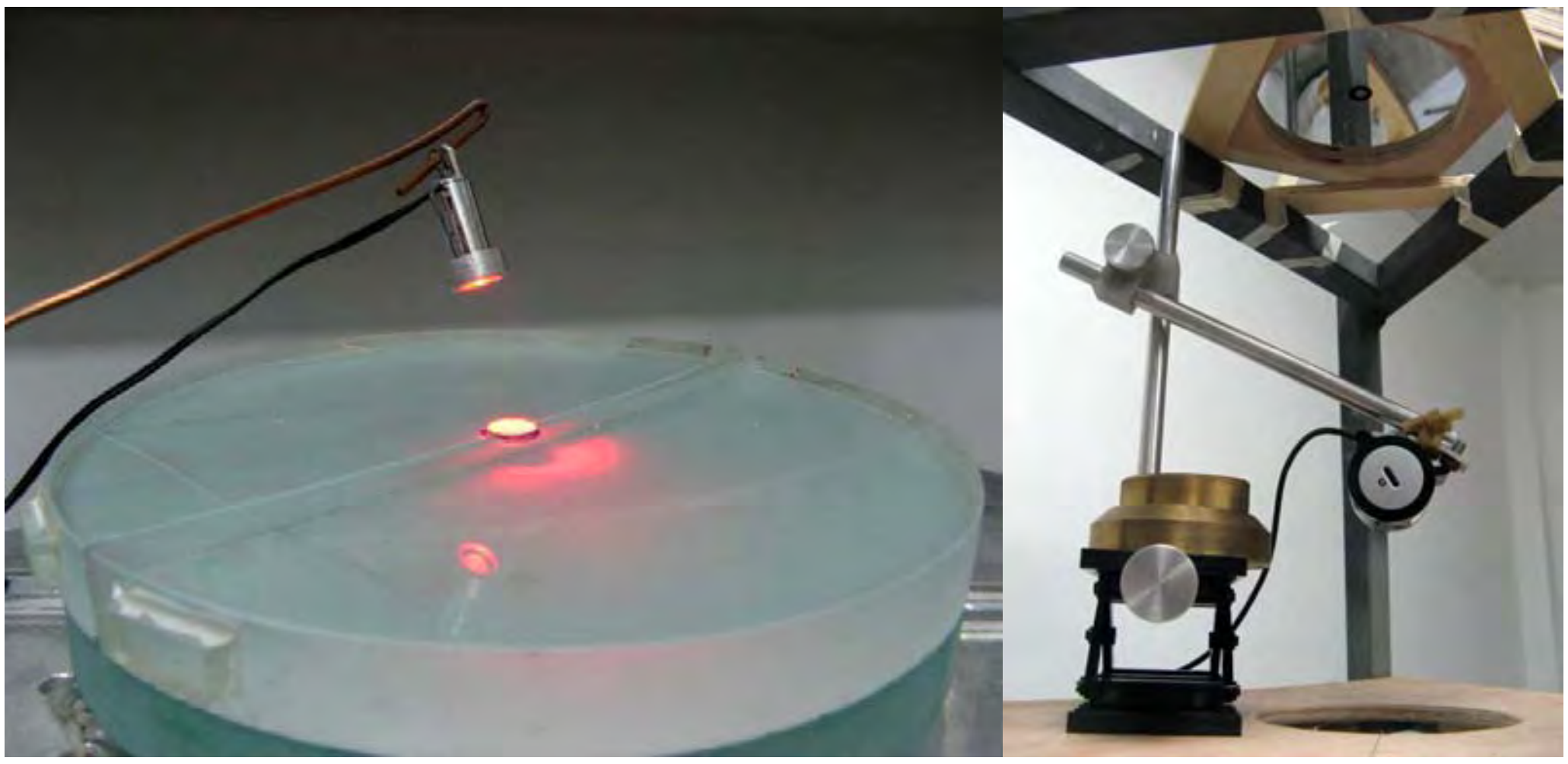

Fig. 4.49: Detalhes da primeira montagem.

A câmara CCD foi posicionada no plano focal dos hemi-espelhos (fig. 4.49). As imagens obtidas pela câmara CCD não tinham brilho suficiente, por isso o disco foi substiuído pelo próprio led de iluminação, figura 4.50.

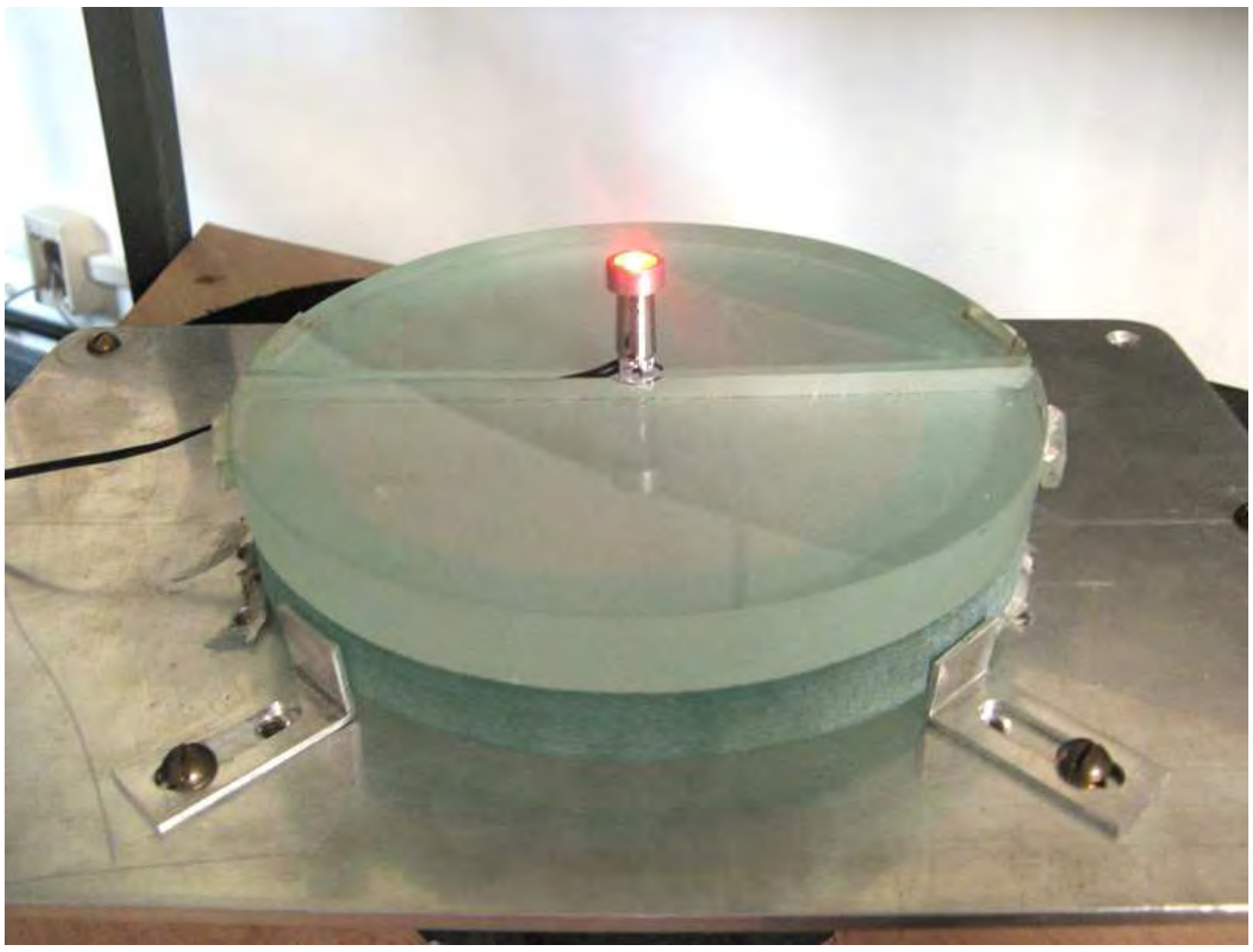

Fig. 4.50: Segunda montagem do teste.

Tanto o espelho heliométrico, quanto sua base, são furados no centro para que o suporte para o padrão de teste seja encaixado e iluminado por trás, figura 4.51.

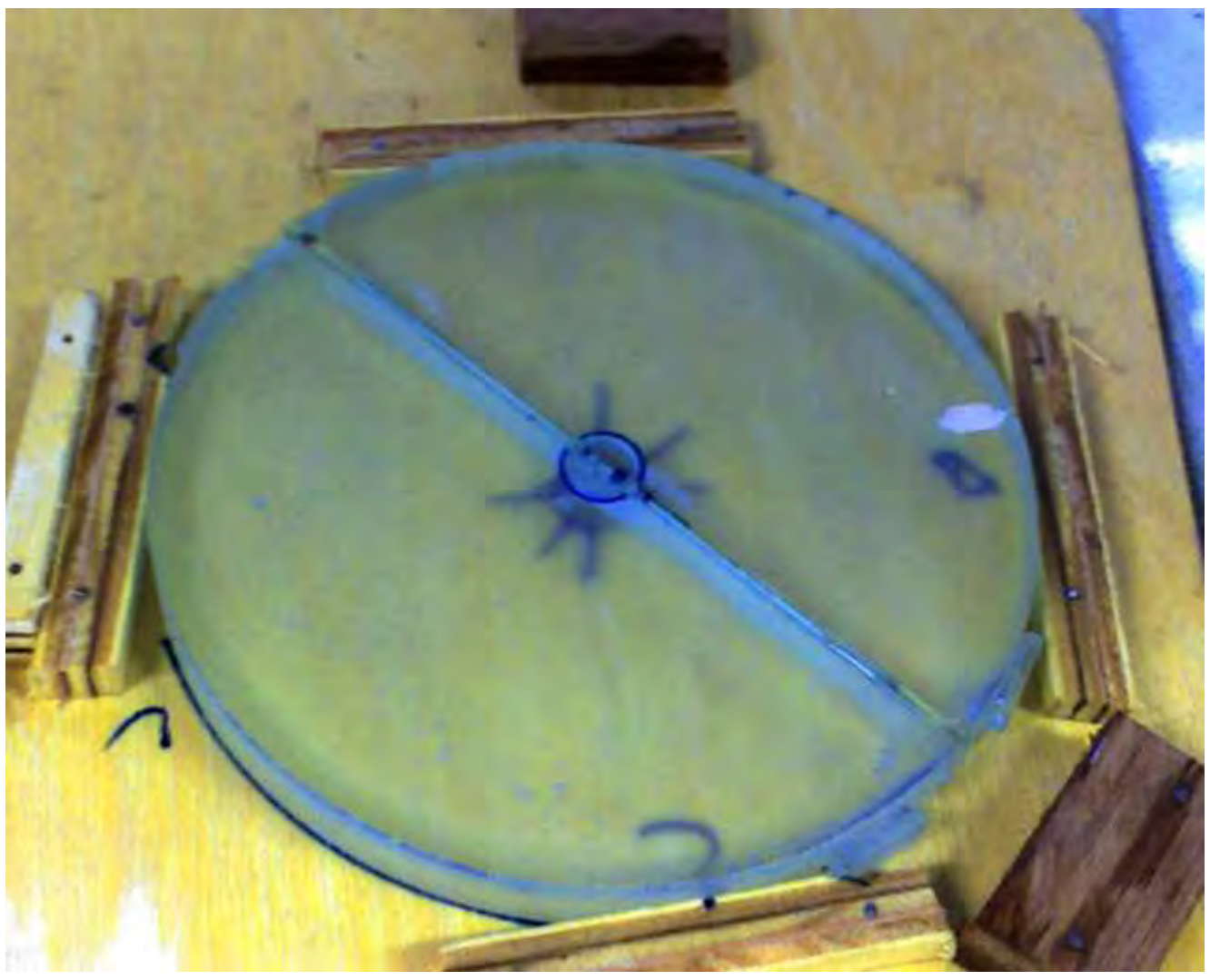

Fig. 4.51: Espelho Heliométrico, fixado a uma base de madeira, pronto para ser furado.

Para melhorar ainda mais a qualidade da imagem do teste, nesta etapa os hemi-espelhos já estão aluminizados. O diedro já está mecanicamente fixo e acomodado em sua célula de contenção.

### 4.3.1 - Montagem do procedimento de auto-colimação

O disco fosco e o led foram substituídos por um pequeno cilindro de vidro despolido, mimetizando o disco solar. Este cilindro é fixado a um suporte oco para que possa ser iluminado por trás, de forma que a superfície do cilindro brilhe. Este suporte, por sua vez, fica apoiado sobre uma peça metálica que cruza o espelho, paralelamente ao corte, figura 4.52.

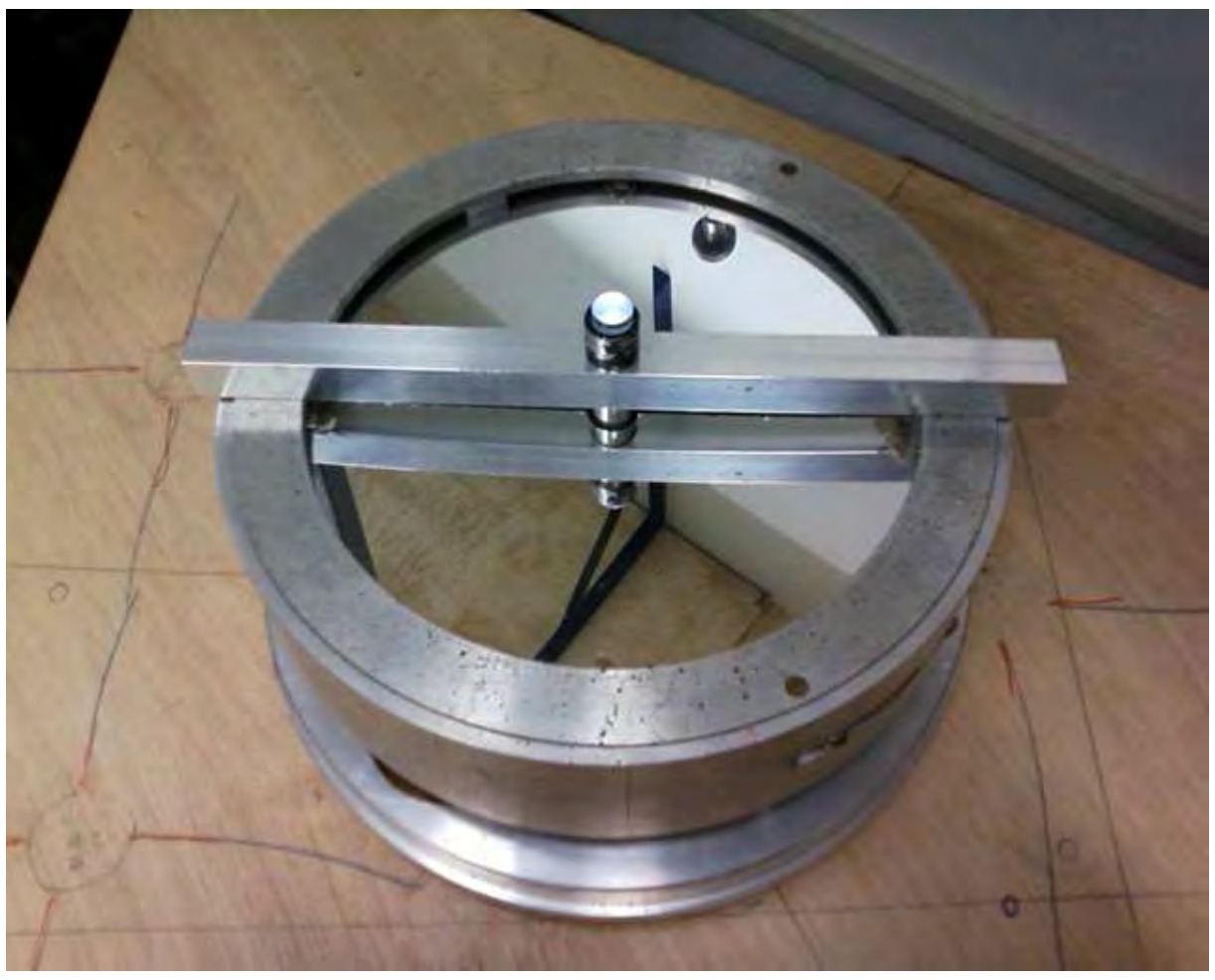

Fig. 4.52: Espelho heliométrico sobre a bancada de teste. A superfície do cilindro de vidro brilha, iluminada posteriormente por um led.

Os espelhos são alinhados de forma que a imagem duplicada do disco de vidro se forme no plano focal do heliômetro, figura 4.53.

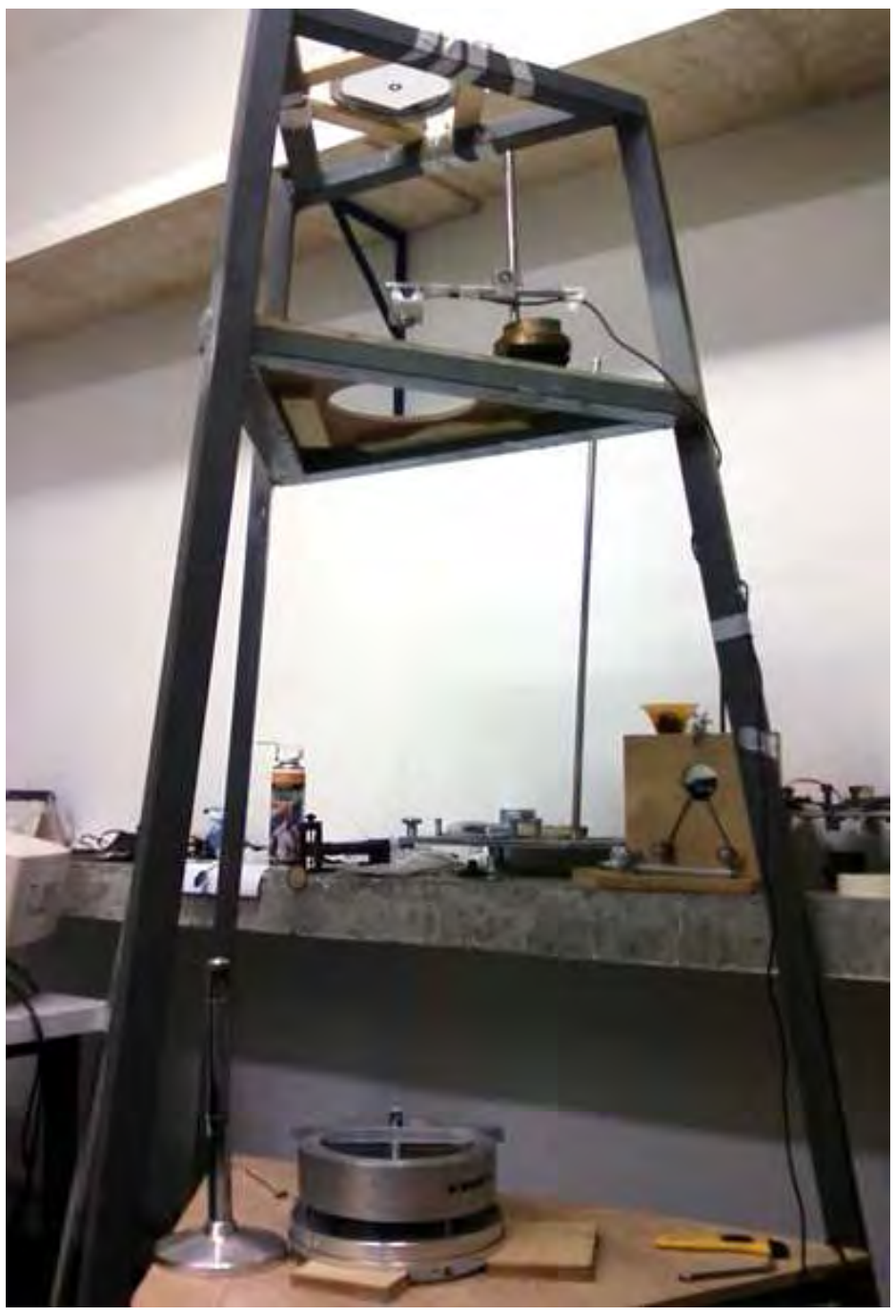

Fig. 4.53: Montagem experimental para o desenvolvimento do teste de auto-colimação.

A materialização do eixo do diedro é obtida através de uma linha de nylon esticada, passando pelo centro do corte do espelho, figura 4.54.

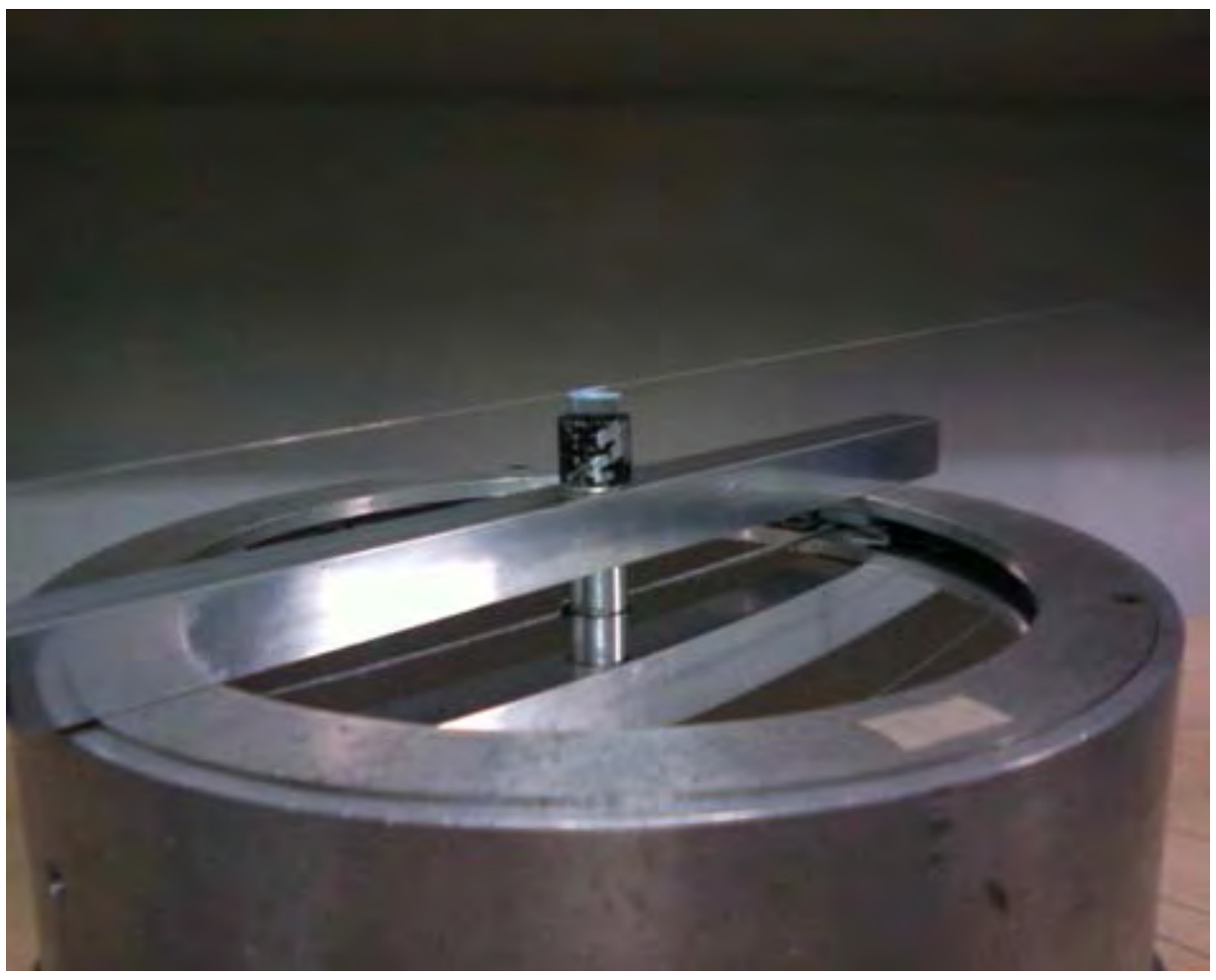

Fig. 4.54: Detalhe da linha de nylon passando sobre o cilindro de vidro

Esta linha tem dupla função: serve para o alinhamento da matriz do CCD com a direção do deslocamento das imagens e serve para se monitorar qualquer deslocamento angular transversal (configuração em A ou em V).

Um suporte em madeira foi construído para que a webcam possa ser movida circularmente, mantendo-se no centro óptico do sistema, figura 4.55.

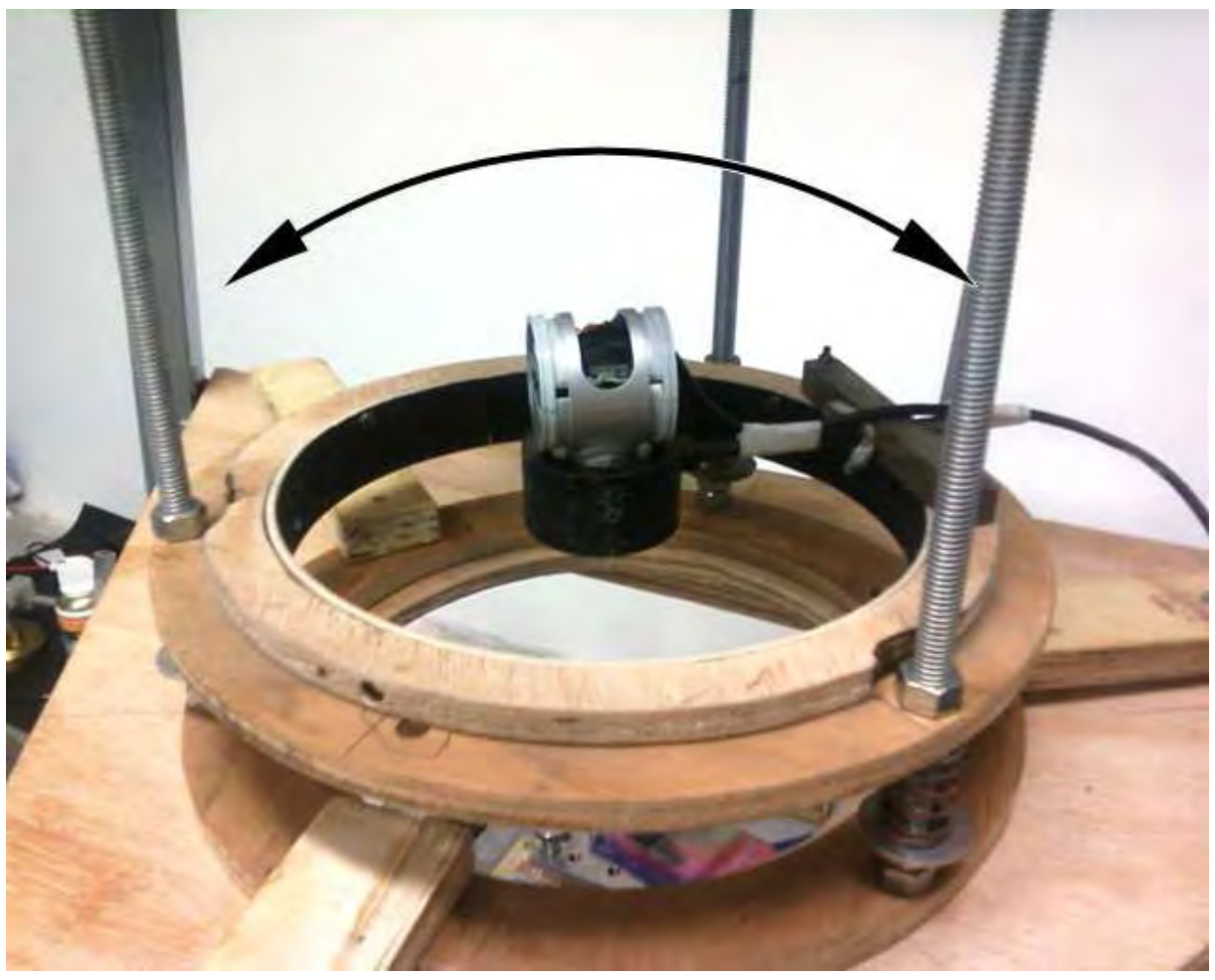

Fig. 4.55: Webcam no foco do espelho heliométrico, presa ao suporte As seta indica o movimento que a câmara podia fazer em torno do centro óptico do espelho. Parafusos de colimação também foram acrescentados ao suporte para ajuste fino do foco.

A figura 4.56 retrata a primeira imagem de auto-colimação obtida com resolução e qualidades suficientes para se avaliar eficácia do procedimento. Nela podemos ver o disco de vidro duplicado, reproduzindo a geometria da medida do disco solar, e a linha de nylon passando sobre o disco.

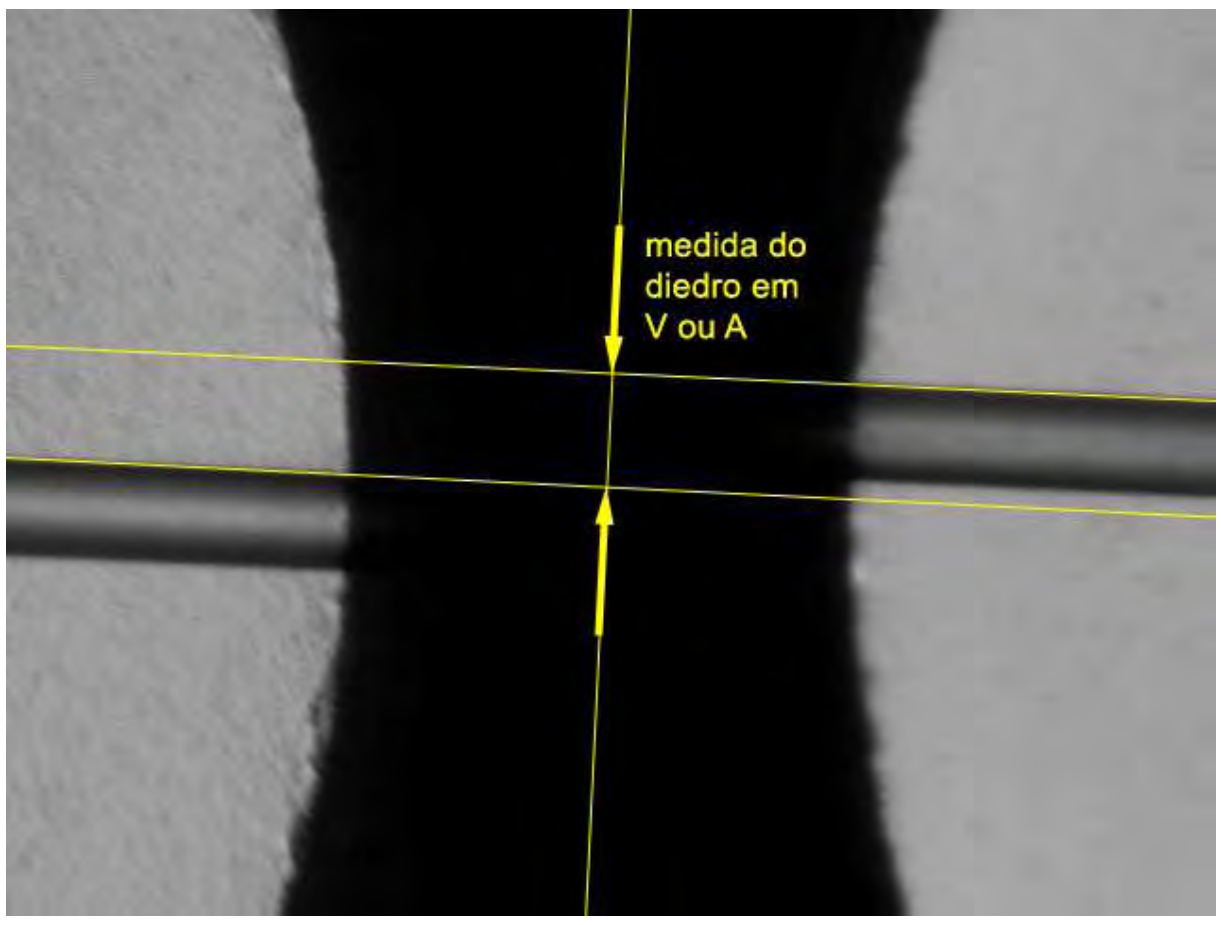

Fig. 4.56: Imagem do teste de auto-colimação. A linha de nylon pode ser vista nas bordas dos discos.

### 4.3.2 - Conclusões

A imagem da linha de referência sobre a superfície do cilindro é conveniente para alinhar a matriz do CCD com a direção do deslocamento das imagens heliométricas. O desalinhamento da imagem do fio é a medida do deslocamento transversal ao corte (diedro em A ou em V), figura 4.57.

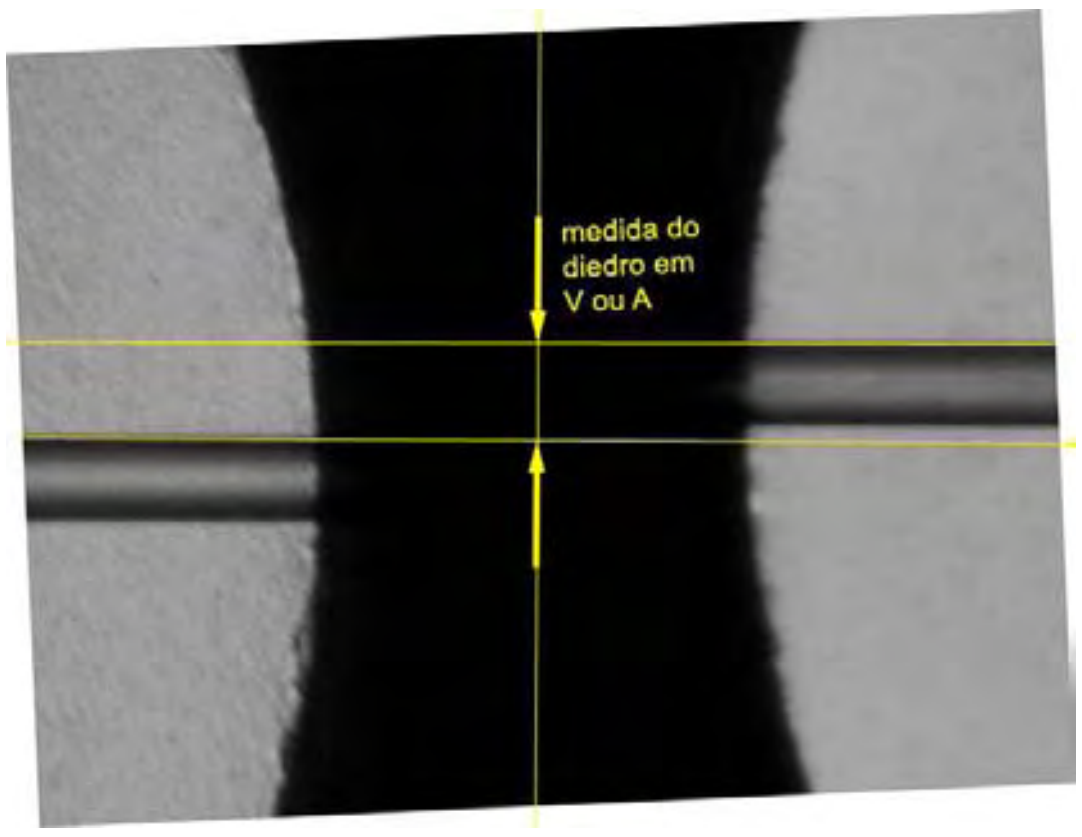

Fig. 4.57: Imagem do teste simulando o alinhamento do CCD com o deslocamento horizontal das imagens.

Nota-se que a imagem do fio permanece desalinhada, pois o espelho do heliômetro de testes ao ser montado dentro de sua célula resultou com um diedro que não é puramente em X. Se não houvesse o diedro em A (ou em V) as imagens estariam perfeitamente alinhadas.

Nesta imagem, a separação entre os centros dos discos, ou seja o diedro, vale cerca de 2173" (o mínimo seria de 1980"), e o deslocamento transversal (diedro em A ou V) vale cerca de 74", que equivale ~3,5% do diedro em X (a tolerância para o heliômetro final foi definida em 1%).

No heliômetro final a linha de nylon é substituída por um fio da ordem de décimos de milímetro.

No heliômetro final o cilindro do padrão é confeccionado com material termicamente estável (de preferência o mesmo que servirá de base para o espelho final) para evitar mudanças em suas dimensões com a variação da temperatura.

Note que, como o espelho de auto-colimação é posicionado na entrada do telescópio, sua distância focal é fator determinante para o comprimento final do tubo do telescópio.

A idéia de se colocar dois espelhos face a face originou outro teste, o de comparação. No teste de comparação, o espelho parabólico é substituído por um espelho heliométrico.

Deste modo a estabilidade dos dois heliômetros pode ser monitorada. Os detalhes e resultados destes ensaios estão no Apêndice I.

## *4.4 - Montagem do telescópio heliométrico*

### 4.4.1 - Projeto da montagem equatorial

Para servir de suporte temporário do heliômetro de testes foi utilizado um tubo de PVC, de 250 mm de diâmetro e 1200 mm de comprimento.

O tubo do telescópio fica dentro de um suporte que permite que o instrumento gire sobre seu eixo, e assim, variar o ângulo de desdobramento do campo em relação à vertical.

Este suporte é sustentada por dois braços presos à uma montagem equatorial germânica convencional com estrutura super-reforçada e adequada para suportar instrumentos com cerca de quinze quilos. Esta montagem tem corpo em aço galvanizado e eixos em aço maciço. Ela usa rolamentos nos dois eixos (polar e de declinação) ao invés de buchas de bronze, oferecendo mais suavidade nos movimentos, mesmo sob grandes cargas. Apresenta ainda eixo polar alongado para acoplamento do sistema motorizado de acompanhamento, figura 4.58 até figura 4.61.

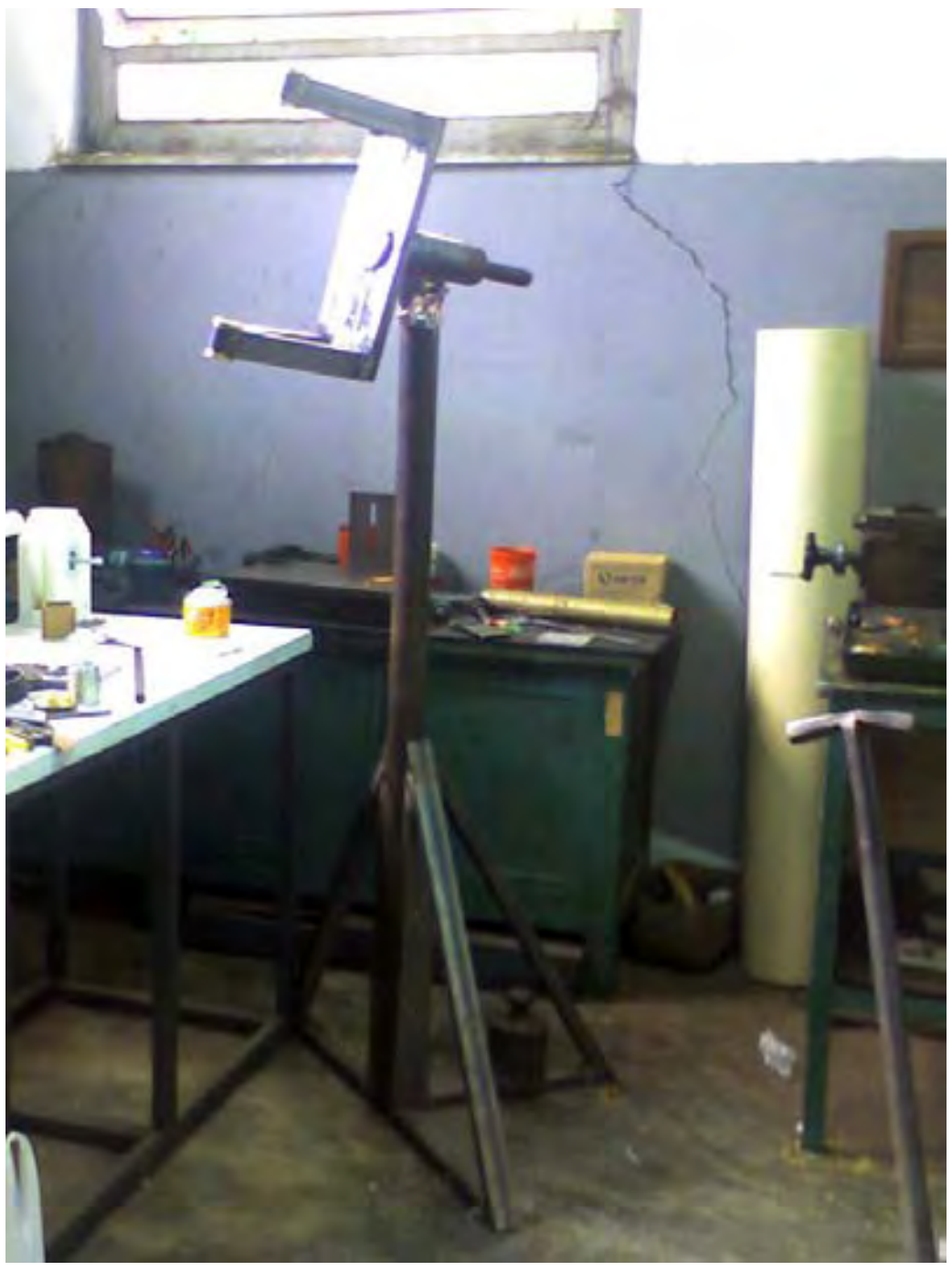

Fig. 4.58: Montagem equatorial de garfo em fase de construção na oficina.

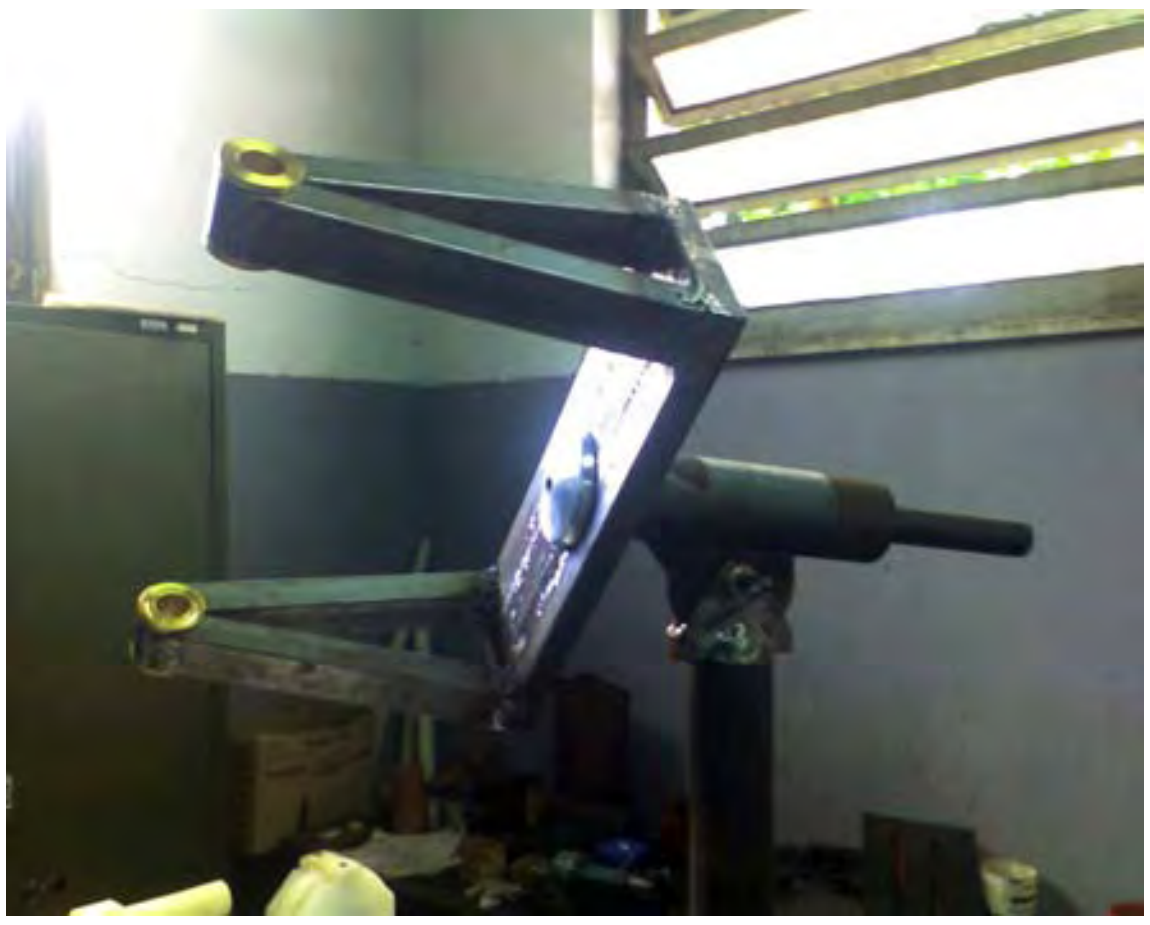

Fig. 4.59: Detalhe do garfo.

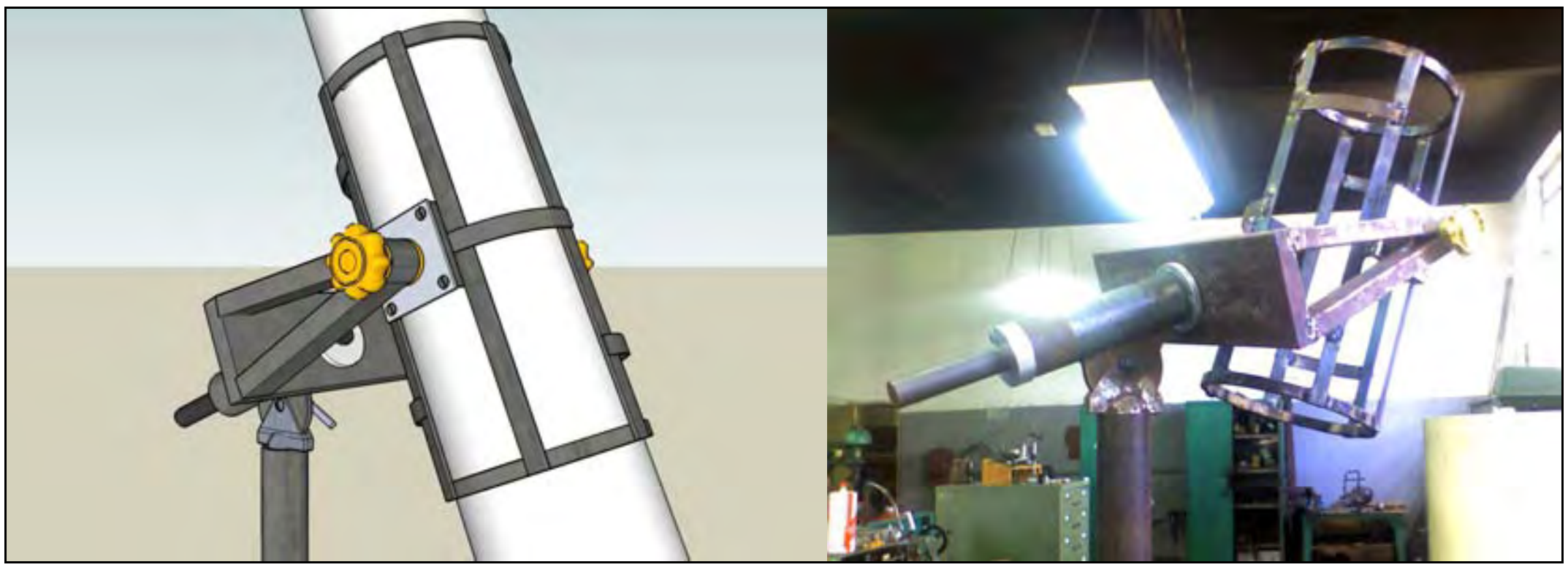

Fig. 4.60: Comparação entre o projeto do suporte e sua montagem na oficina.

Esta configuração de montagem foi escolhida, pois permite que o Heliômetro observe o Sol em todas as alturas, inclusive no zênite.

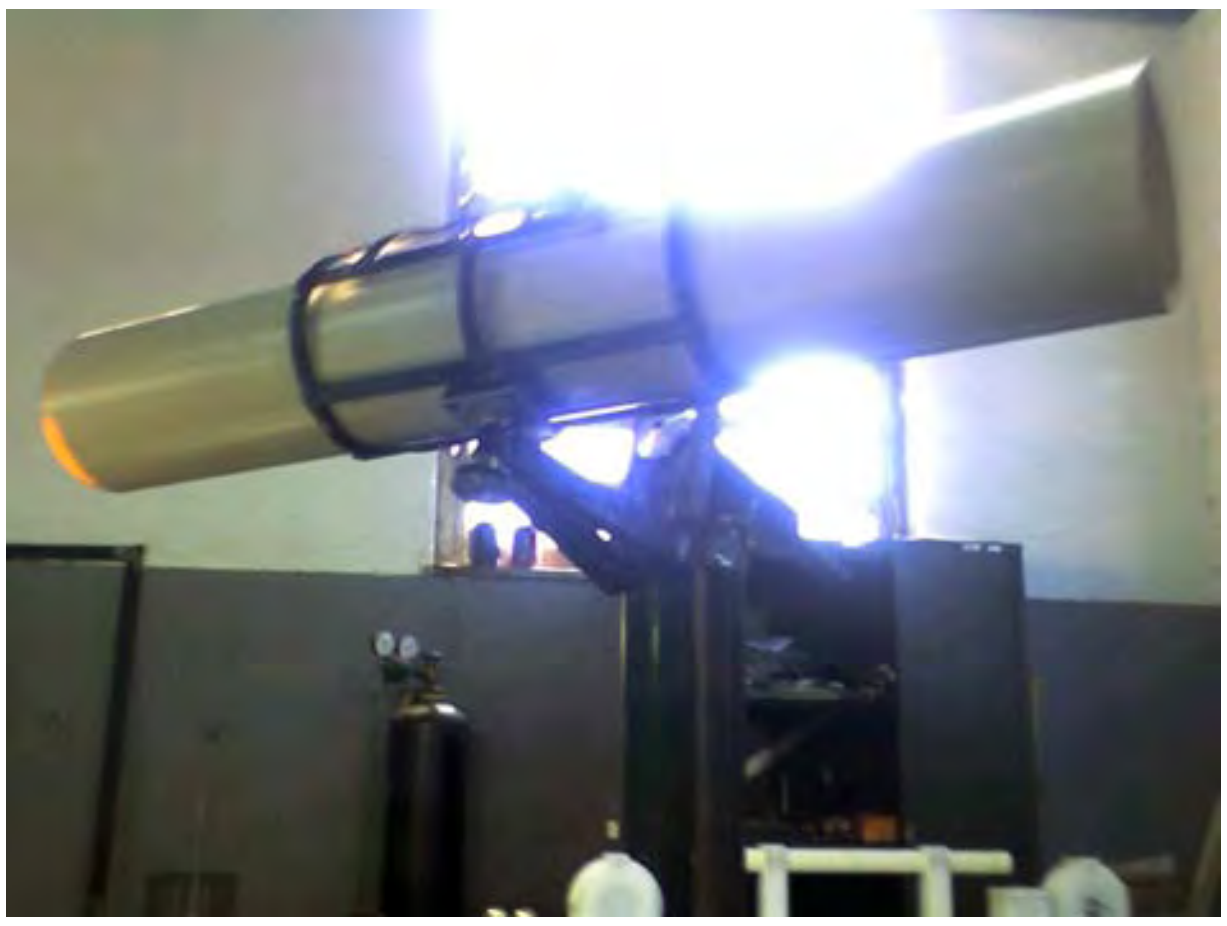

Fig. 4.61: Tubo do instrumento dentro do suporte.

## 4.4.2 - Primeira luz

Na entrada do tubo é usado um filtro solar da *Thousand Oaks Optical*, figura 4.62.

Este filtro tem grande durabilidade e é confeccionado por deposição sobre vidro float. Sua moldura de alumínio facilita a fixação ao tubo.

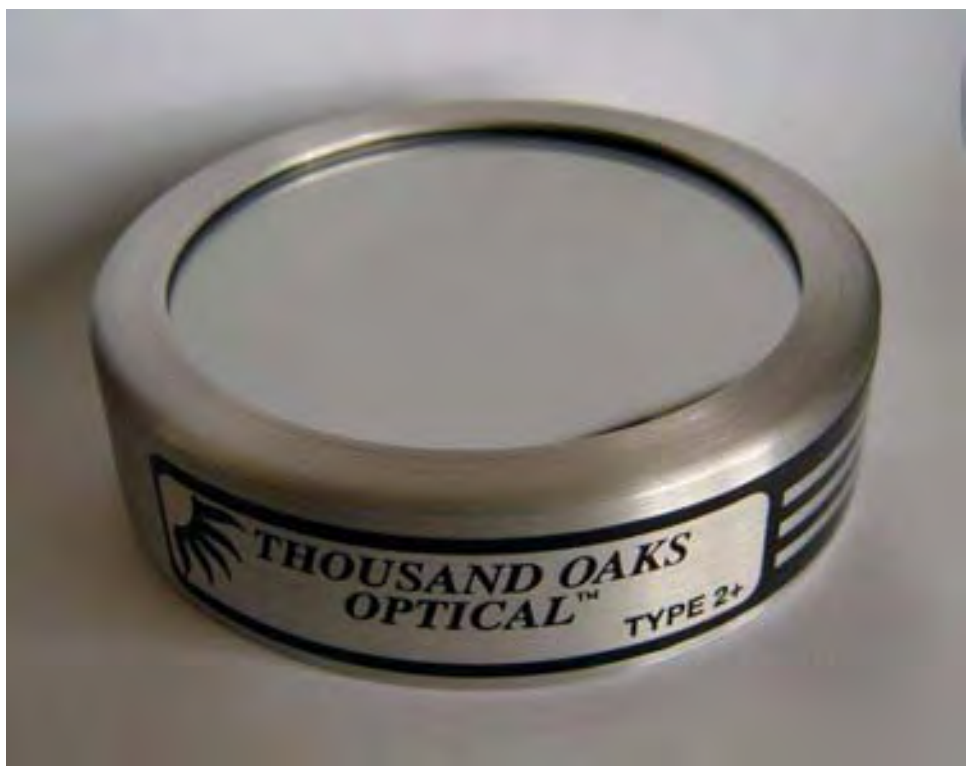

Fig. 4.62: Filtro solar.

No começo de dezembro de 2008 o Heliômetro de testes foi colocado em sua montagem equatorial, já na cúpula, para os primeiros testes, aproveitando a passagem zenital do Sol neste período, figuras 4.63, 4.64 e 4.65.

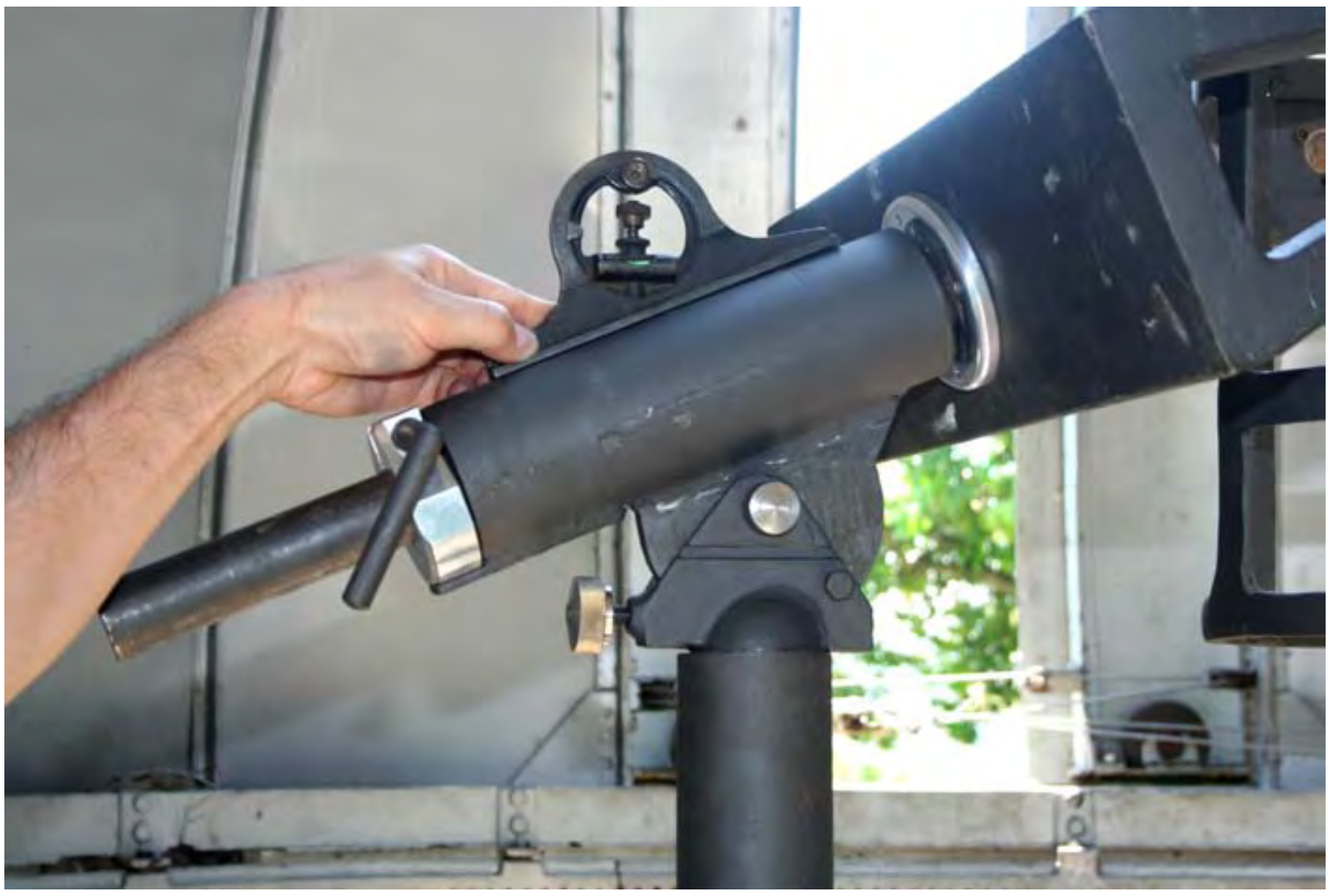

Fig. 4.63: Ajuste do eixo polar da montagem equatorial.

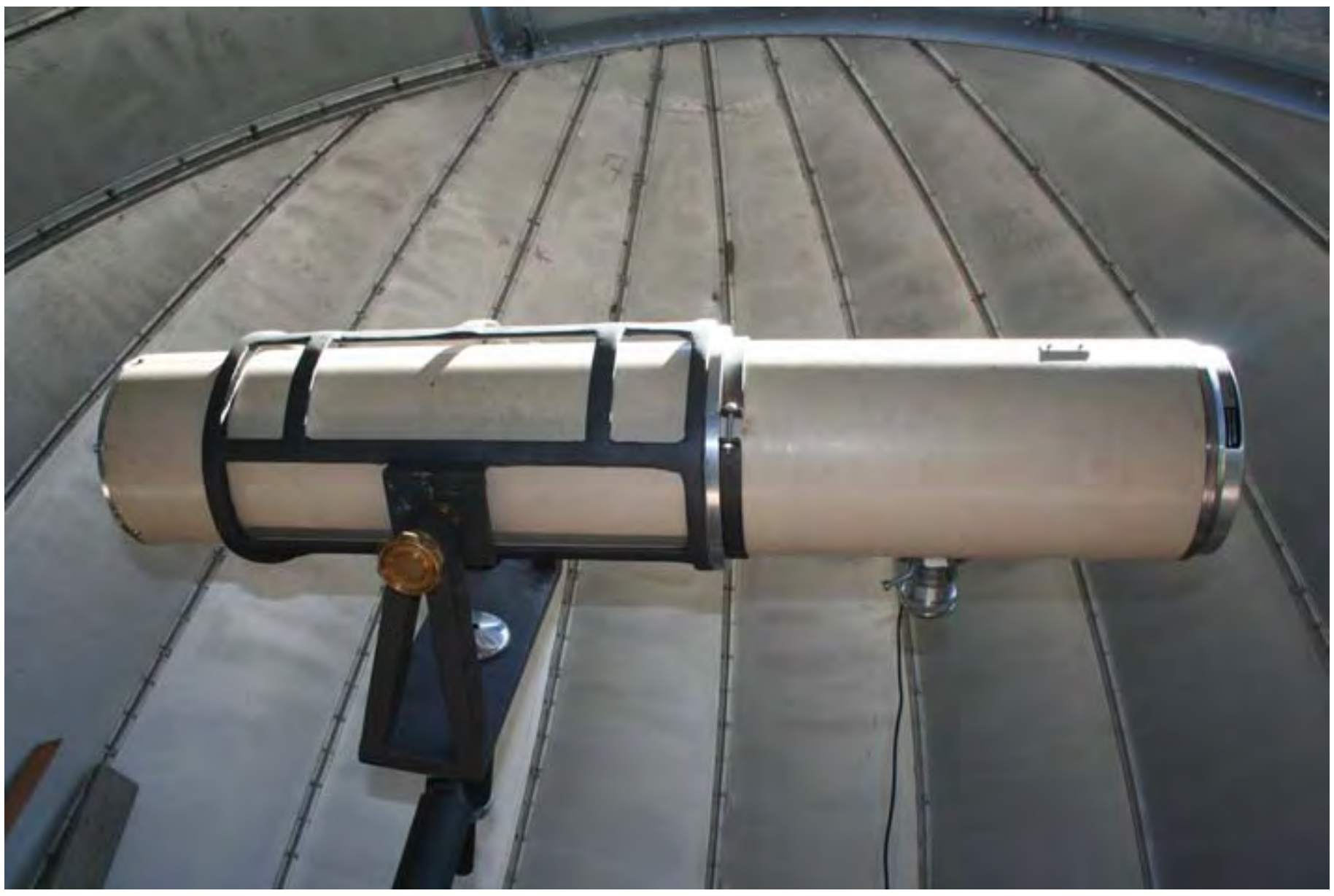

Fig. 4.64: Heliômetro de testes posicionado dentro do suporte.

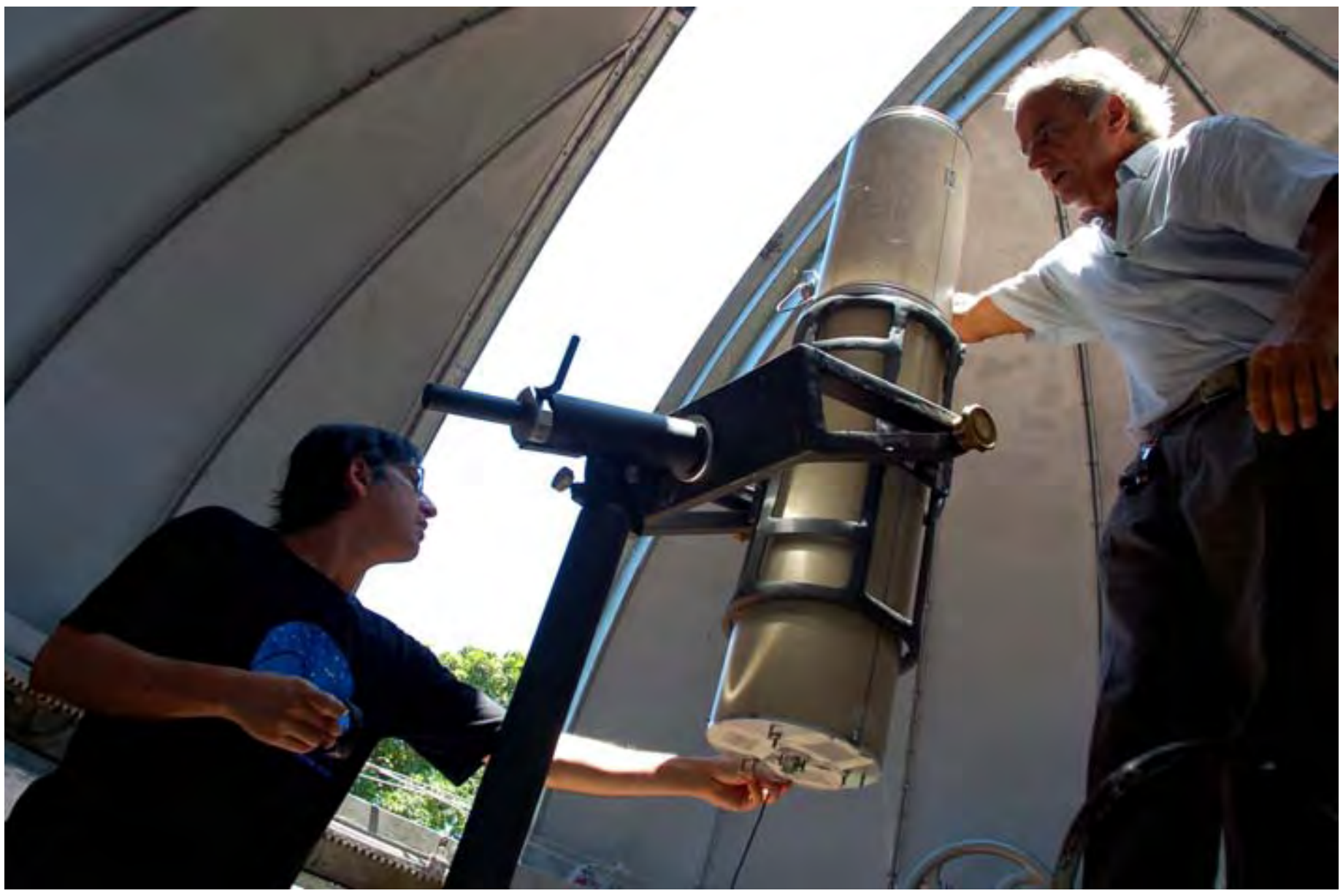

Fig. 4.65: Primeiro teste observacional do Heliômetro protótipo.

As sessões do dia 8, 9 e 10 de dezembro de 2008 evidenciaram as modificações que deveriam ser feitas para aperfeiçoar o instrumento e melhorar as observações (fig. 4.66 até fig. 4.73):

- colocar um contra-peso para o conjunto de focalização;
- adaptar um sistema de ajuste fino para os movimentos de declinação e ascensão reta da montagem.

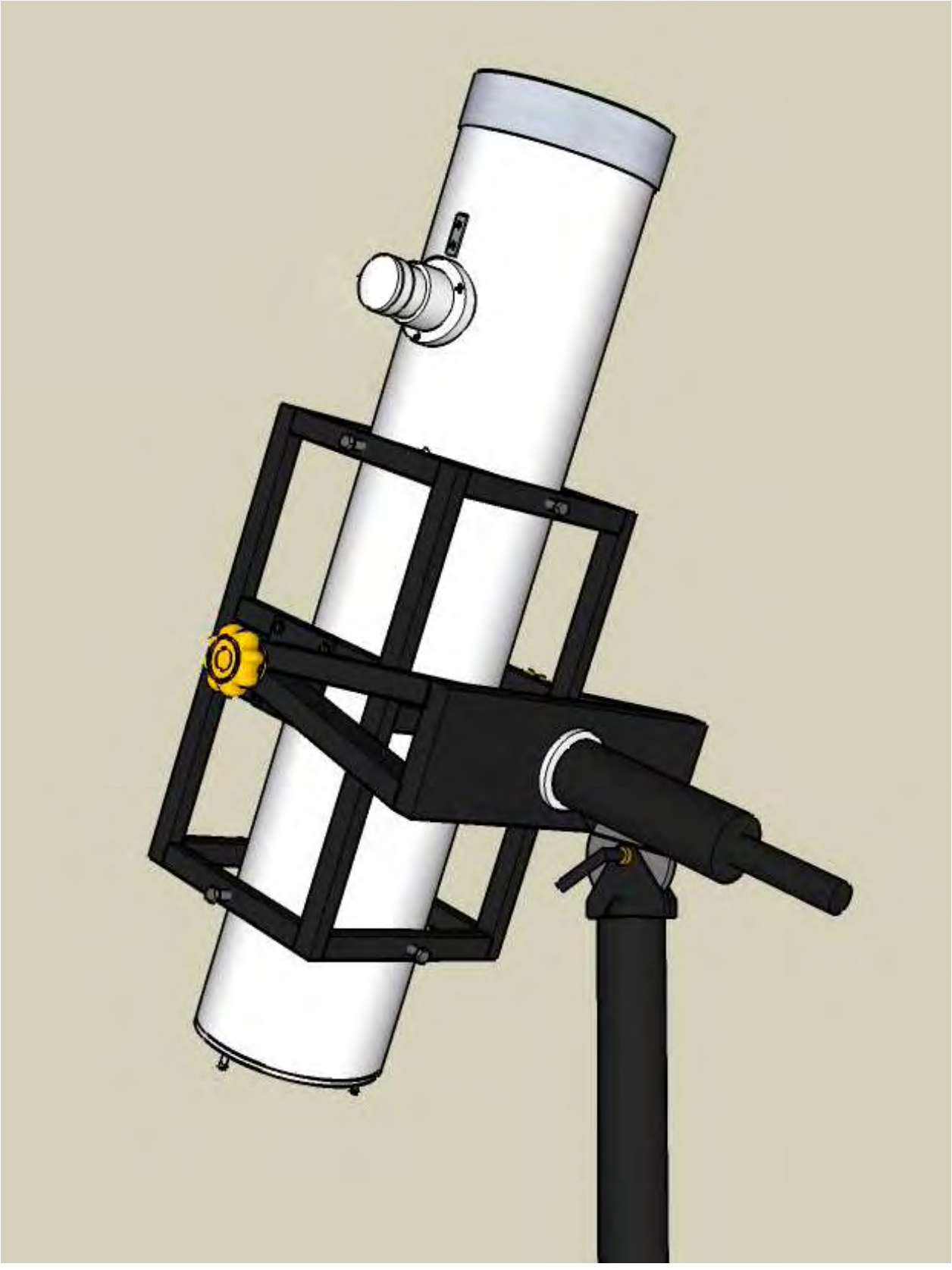

Fig. 4.66: Projeto do novo suporte do tubo.

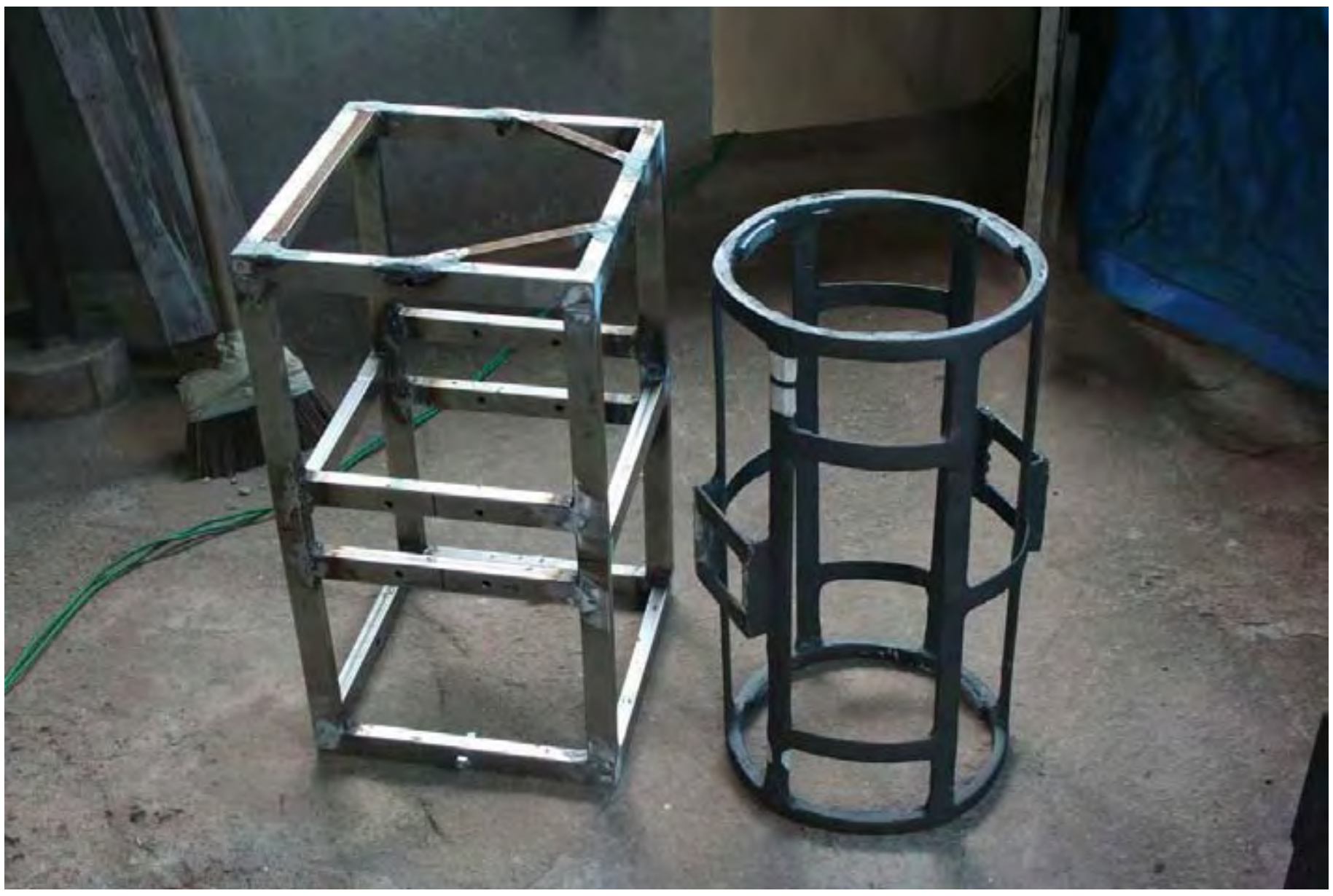

Fig. 4.67: Novo suporte do tubo ao lado do antigo.

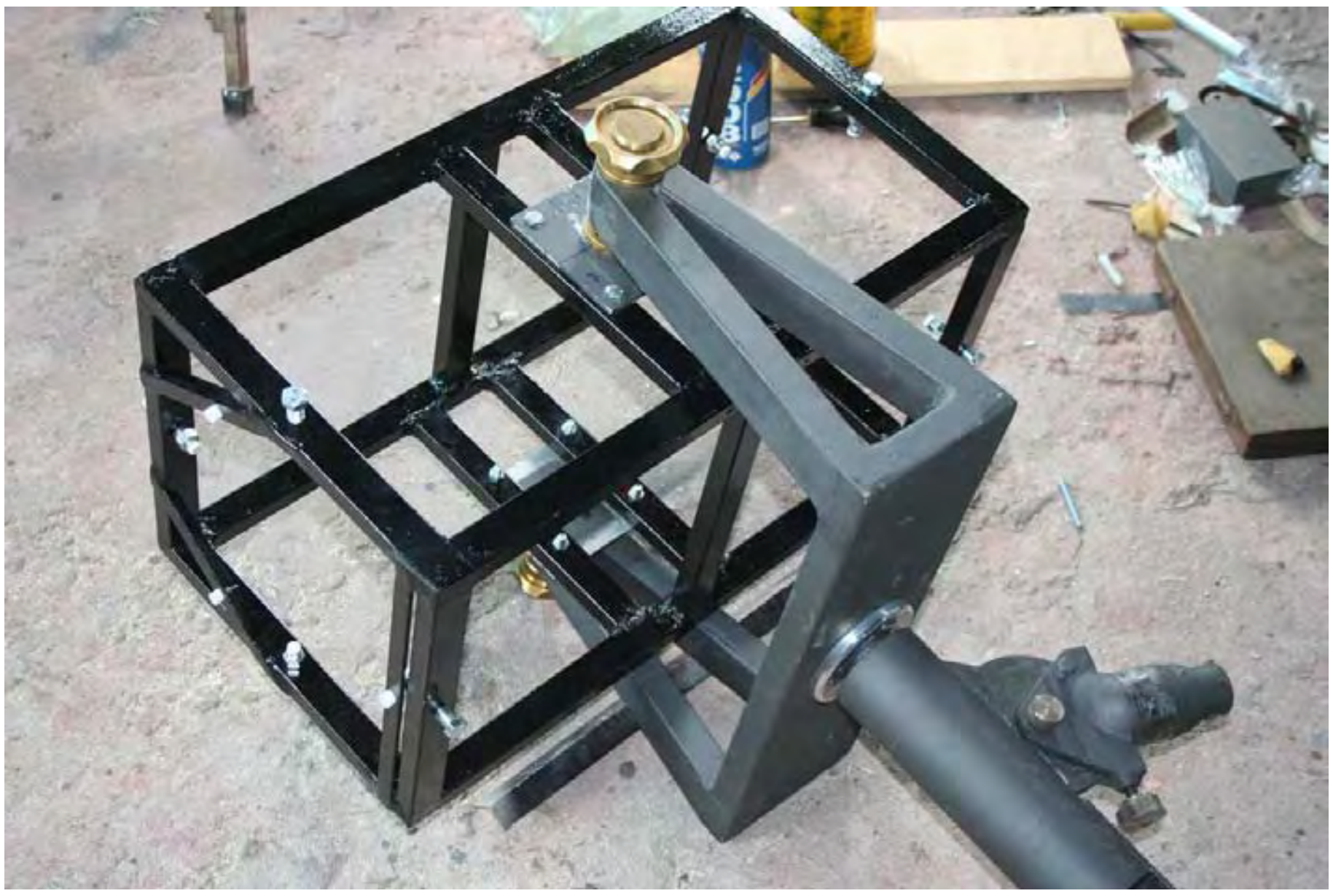

Fig. 4.68: Novo suporte do tubo encaixado no garfo da montagem.

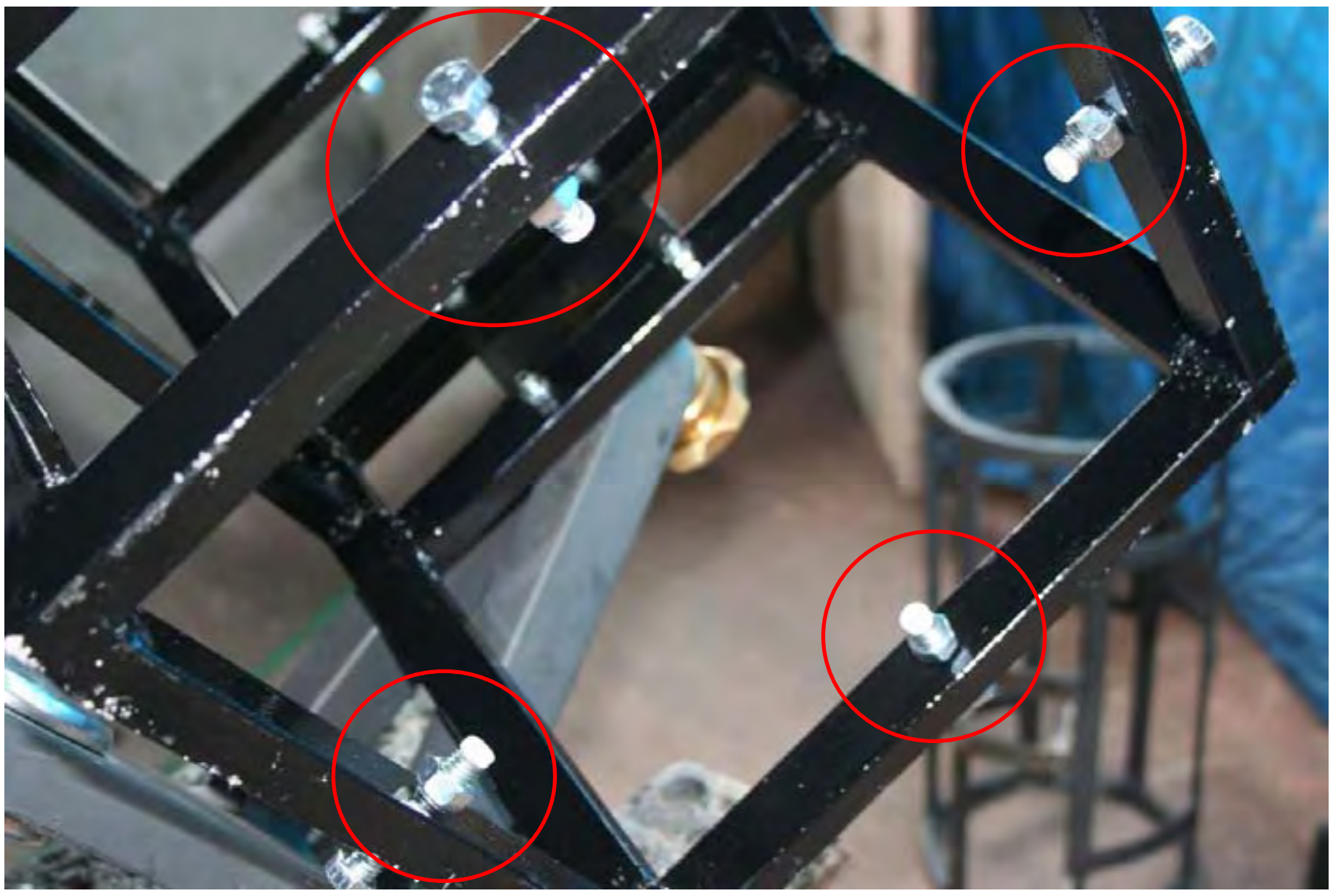

Fig. 4.69: Detalhe dos parafusos de colimação. As pontas de nylon facilitarão o giro do tubo em seu interior.

O Heliômetro foi então recolocado na cúpula, com novo suporte, com o contra-peso do suporte da câmara e com uma coroa dentada, adaptada ao eixo polar, para movimentos mais suaves e precisos em ascensão reta, figura 4.70 e 4.72.

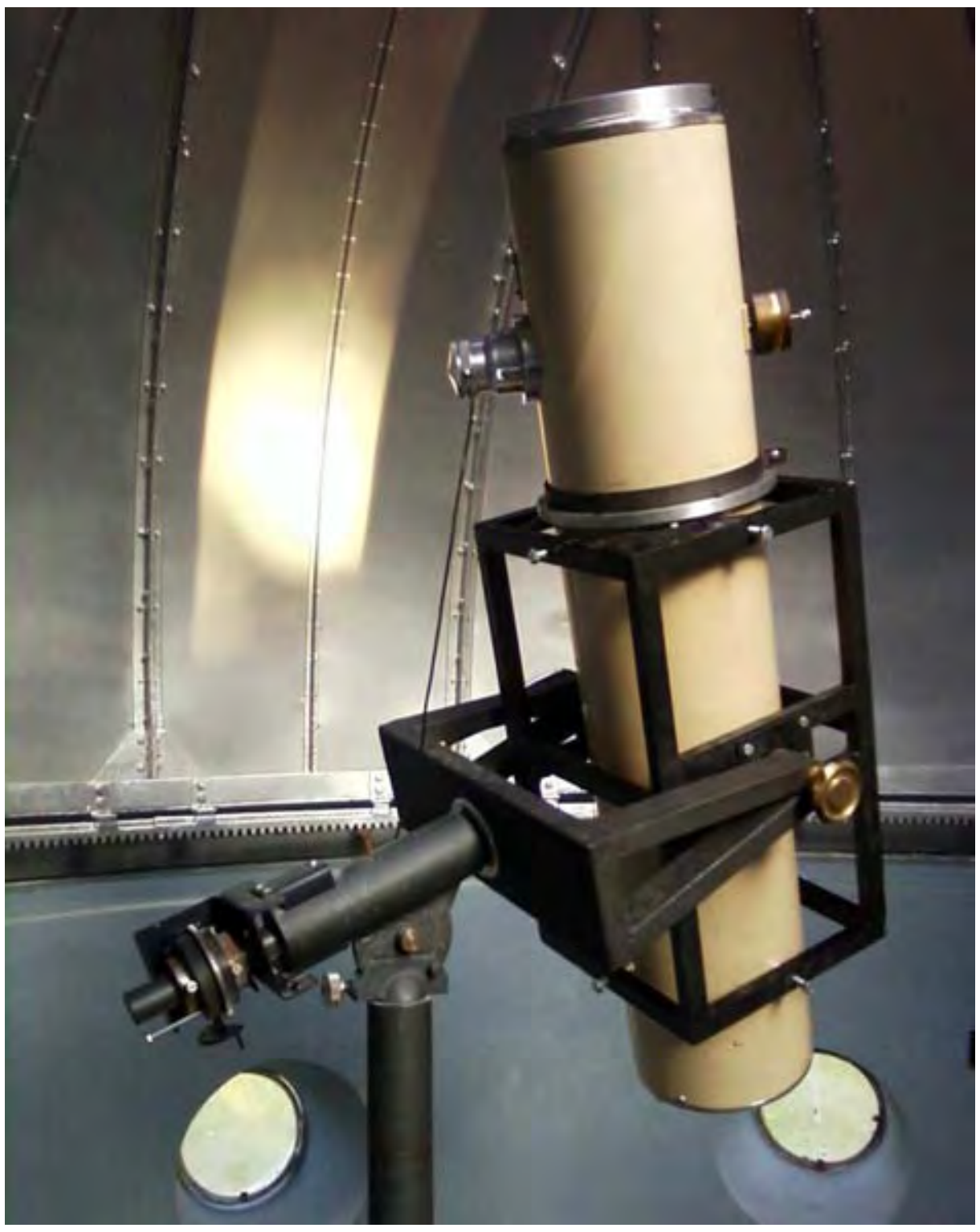

Fig. 4.70: Montagem equatorial em sua nova configuração.

No eixo de declinação também foi adaptada uma peça com a mesma finalidade, permitir um ajuste fino, figura 4.71.

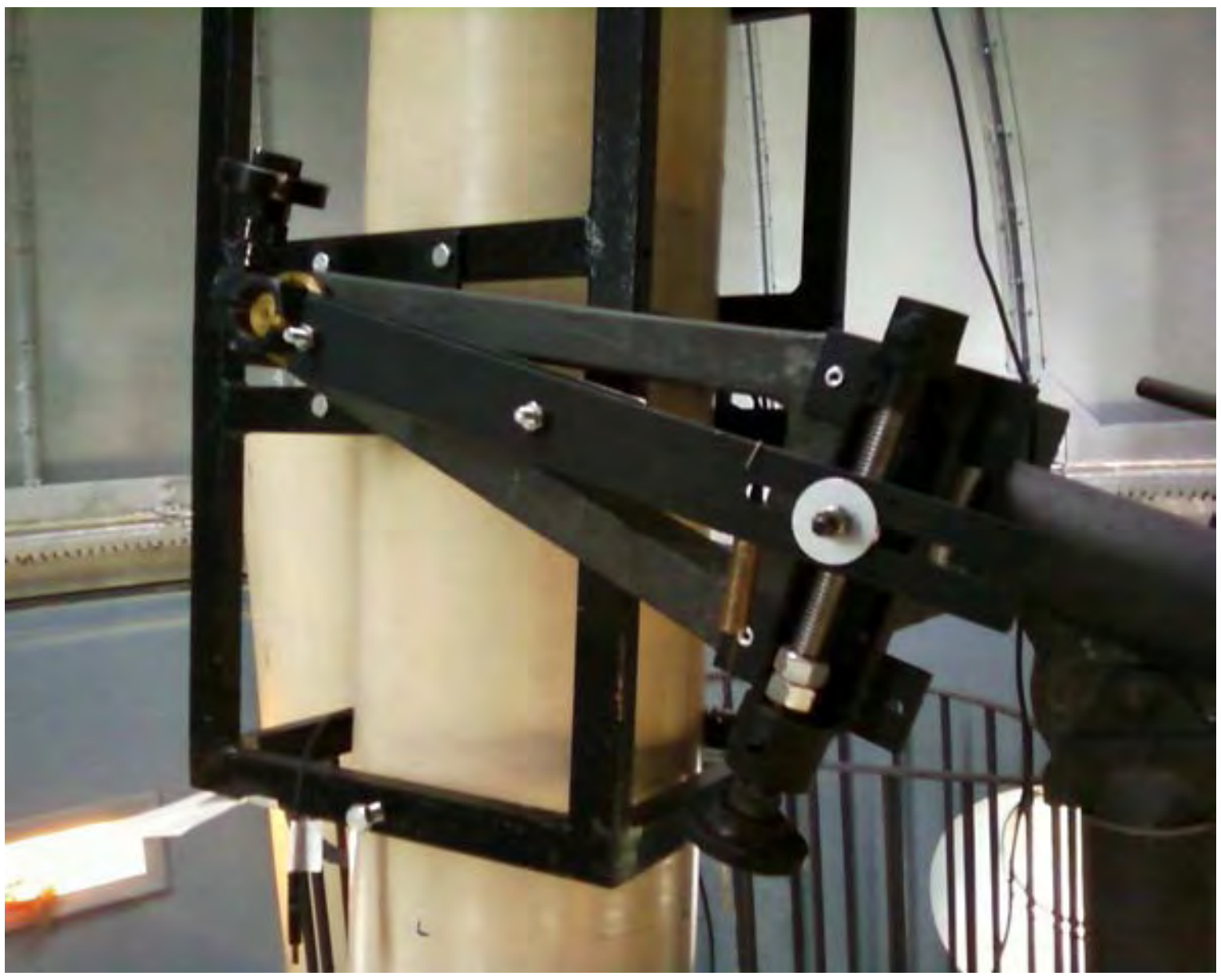

Fig. 4.71: Peça para ajuste fino dos movimentos em declinação da montagem equatorial.

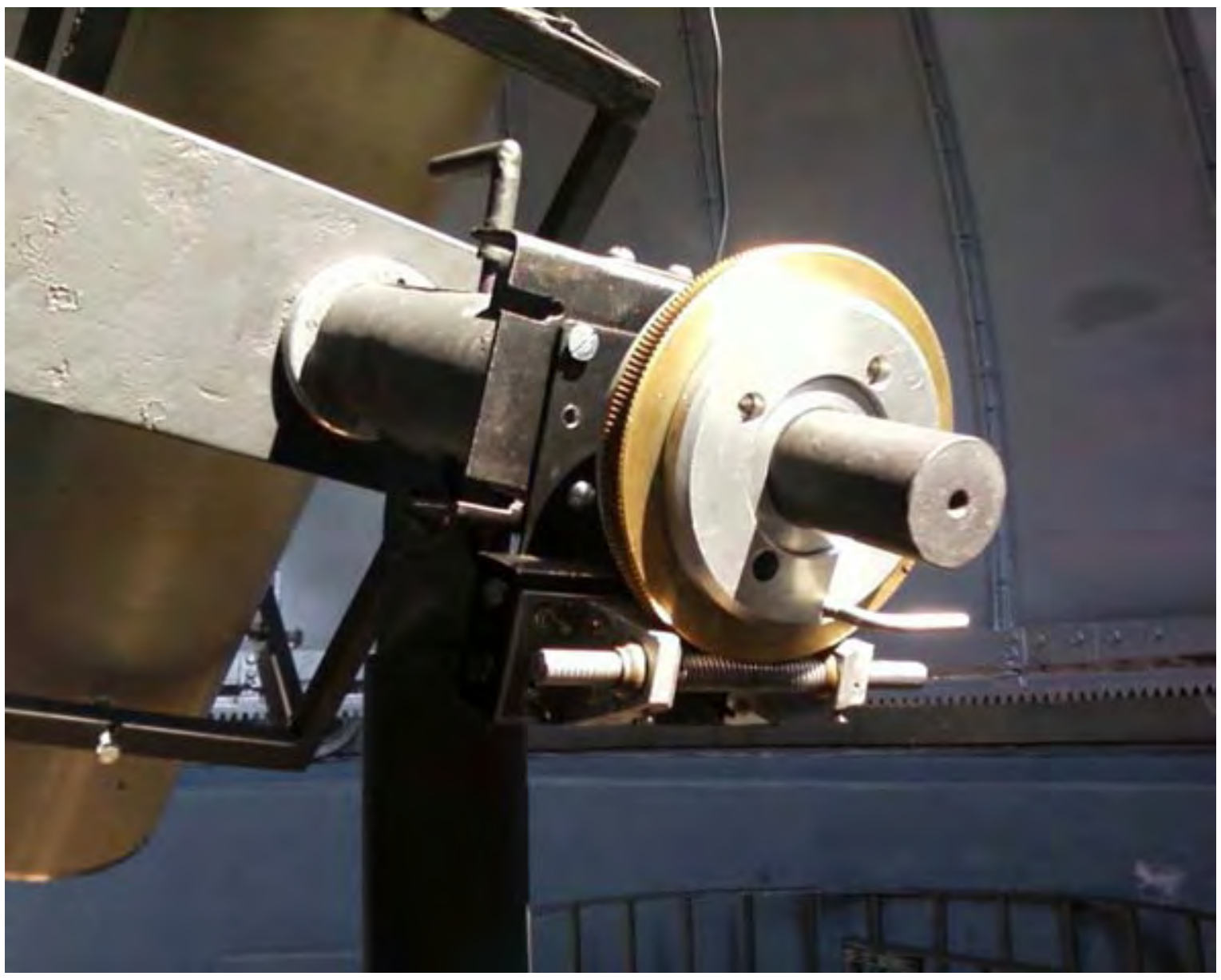

Fig. 4.72: Peça para ajuste fino do movimento em ascensão reta da montagem.

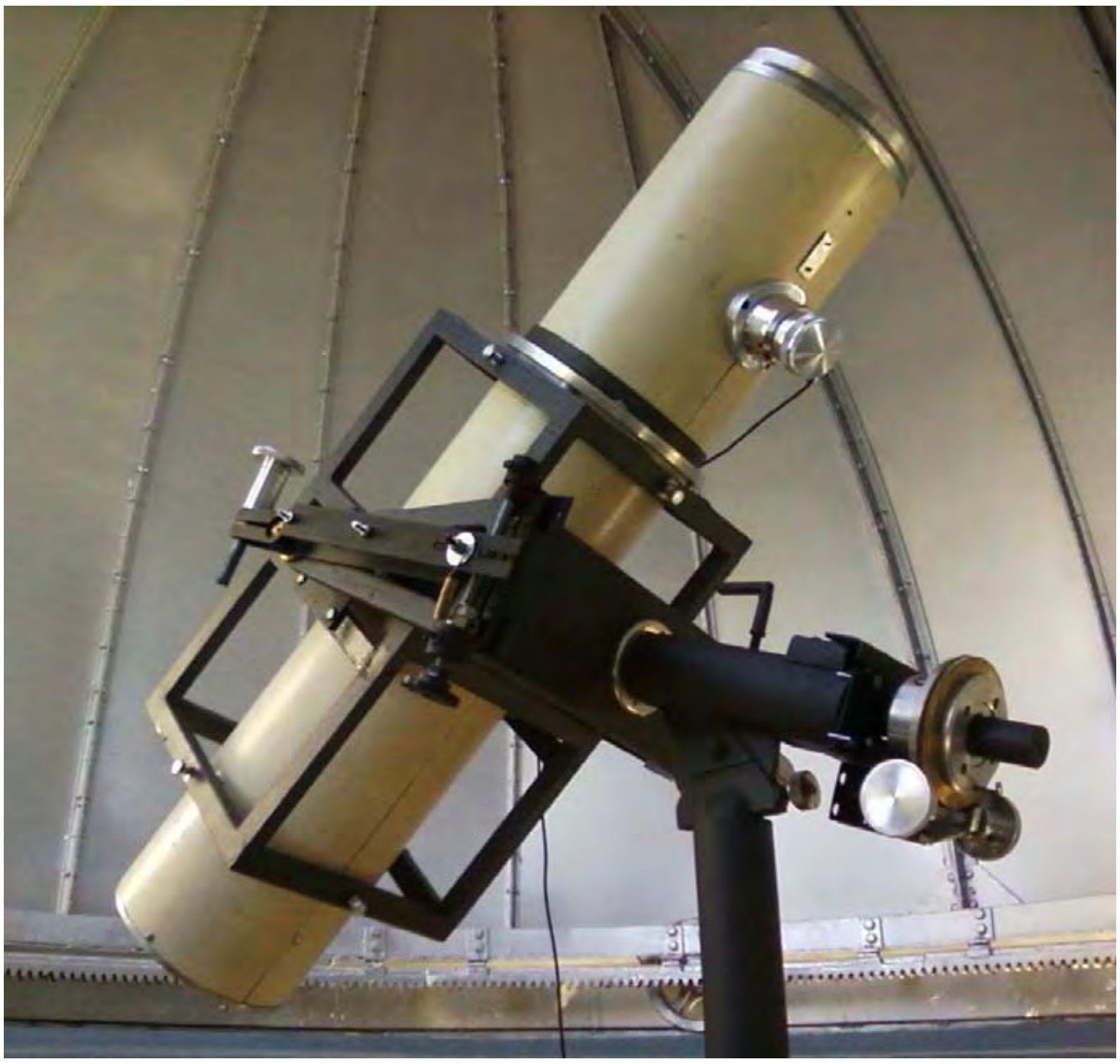

Fig. 4.73: Configuração final da montagem equatorial do Heliômetro protótipo.

Com esta configuração (fig. 4.73), em maio de 2009 foi realizada a 1ª campanha observacional deste Heliômetro.

Os resultados desta campanha, assim como sua análise estão no capítulo 6.

# Capítulo 5 - Construção do Heliômetro

## 5.1 - Partes ópticas

### 5.1.1 - Escolha do material e das dimensões do espelho

A menos que o coeficiente de dilatação térmica seja próximo de zero, para toda a gama de temperaturas a que o espelho será submetido, ele irá se deformar quando as condições térmicas mudarem. Em primeira ordem, o raio de curvatura do espelho é afetado pela alteração do volume, conseqüentemente sua distância focal se modificará. Para minimizar ao máximo este efeito, o material usado na fabricação do espelho precisa ser termicamente estável.

Comercialmente, podem ser encontrados materiais cerâmicos com coeficientes de expansão térmica perto de zero, como: o Zerodur (da alemã Schott), o Astrosital (fabricado na Rússia) e o ClearCeram-Z e ClearCeram-Z-HS (de japonesa Ohara).

Todos estes materiais possuem características semelhantes, de modo que pela facilidade de aquisição para o projeto, este último foi escolhido.

O material é homogêneo e o valor médio do coeficiente de expansão térmica do CCZ-HS, certificado de fábrica, entre 0ºC e 50ºC, é de $0,0 \pm 0,2 \times 10^{-7}/°C$ (fig. 4.51, do fabricante), o que significa que, num espelho construído com este material, a variação esperada da distância focal com a temperatura é desta mesma ordem de grandeza, ou seja, desprezível.

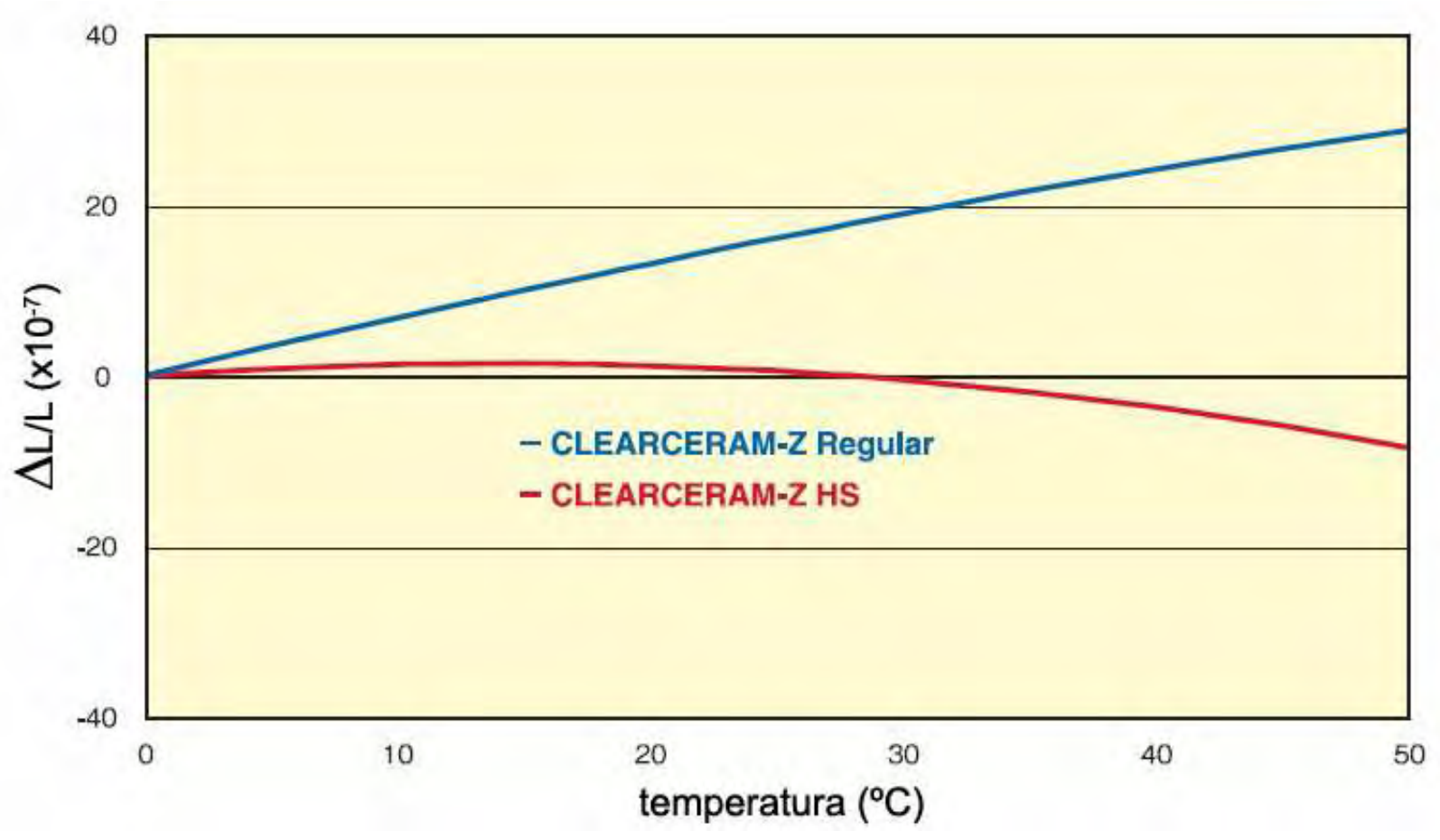

Fig. 5.1: Comparação entre a variação do coeficiente de expansão térmica com a temperatura do CCZ-Regular e CCZ-HS, entre 0ºC e 50ºC.

No projeto de um espelho objetivo, outra preocupação relevante é quanto às suas dimensões. Existe uma proporção mínima entre o diâmetro de um espelho e sua espessura, para que ele não se flexione quando submetido ao seu próprio peso.

A relação de Couder (1932) permite calcular a espessura mínima que um disco precisa ter para que não apresente flexões ópticas relevantes quando estiver horizontalmente apoiado:

$\frac{R^4}{h^2} \leq 1000$, onde $R$ e $h$ são o raio e a espessura do disco expressos em cm.

Empiricamente a proporção $2R/h$, para que esta deflexão seja desprezada, é de 6/1.

Outro fator levado em consideração na escolha das dimensões de um espelho é a turbulência atmosférica. Existe um limite para o diâmetro de um espelho, acima do qual não há mais aumento na resolução espacial da imagem. Este limite pode ser estimado pelo "Fator de Fried" ($r_0$), que mede a qualidade do "seeing" da atmosfera.

Ele pode ser entendido como o diâmetro da pupila circular para que uma imagem no limite de difração e uma imagem limitada pelo "seeing" tenham a mesma resolução angular.

O limite de difração de um telescópio vale ~$1{,}22\frac{\lambda}{D}$, onde $D$ é o diâmetro do espelho.

Considerando um "seeing" típico de 1" de arco, temos:

$$1{,}22\frac{\lambda}{D} = 1{,}22\frac{\lambda}{r_0} = 4{,}85 \times 10^{-6}\,rad$$

Para $\lambda = 550$ nm, temos $r_0 \approx 0{,}14$ m.

O disco padrão de CCZ-HS escolhido para servir de base para o projeto do espelho tem diâmetro de 150 mm e 25 mm de espessura (fig. 5.2). Com estes valores:

$$\frac{R^4}{h^2} = 506{,}25 \text{ e } \frac{2R}{h} = 6$$

Segundo Texereau (1961), até o limite da relação de Couder, um espelho pode ser apoiado por três pontos dispostos nos vértices de um triângulo eqüilátero exatamente inscrito no perímetro do espelho.

As dimensões escolhidas para o disco base são adequadas, permitindo desprezar qualquer preocupação quanto às deflexões por conta de seu próprio peso, assim como facilitam o projeto dos pontos de apoio, que poderão ser localizados até perto da borda do disco. Após o pupilamento que será descrito adiante, admitindo um seeing de 2", também se atinge o limite da condição de mínimo efeito da turbulência atmosférica,

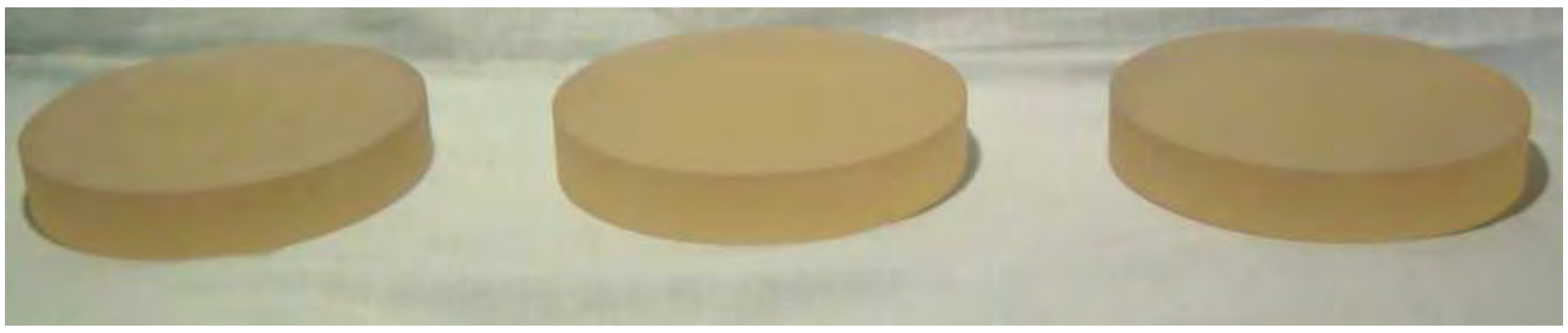

Fig. 5.2: Primeiros blocos de CCZ-HS adquiridos para o projeto.

### 5.1.2 - O projeto

Todo o projeto do espelho heliométrico final foi desenvolvido com software de desenho 3D, permitindo fácil estudo, análise e modificações de todos os detalhes e de seu conjunto, bem como das etapas de sua fabricação.

O conceito mestre é de ter os espelhos, em diedro, contidos lateralmente por um anel e ter este conjunto repousando sobre um plano óptico. Todo o conjunto é em CCZ-HS.

O anel de contenção é fabricado retirando-se o interior de um dos blocos de CCZ-HS. A espessura mínima para que seja segura a sua fabricação e, ao mesmo tempo, para que a peça ainda mantenha sua rigidez, é de 15 mm, figura 5.3, à esquerda.

Além disso, a superfície que faz contato com o plano óptico é rebaixada, diminuindo a sua altura em 1 mm, poupando apenas 3 regiões, separadas de 120º cada, para a fabricação dos calços de apoio inferiores, figura 5.3, à direita.

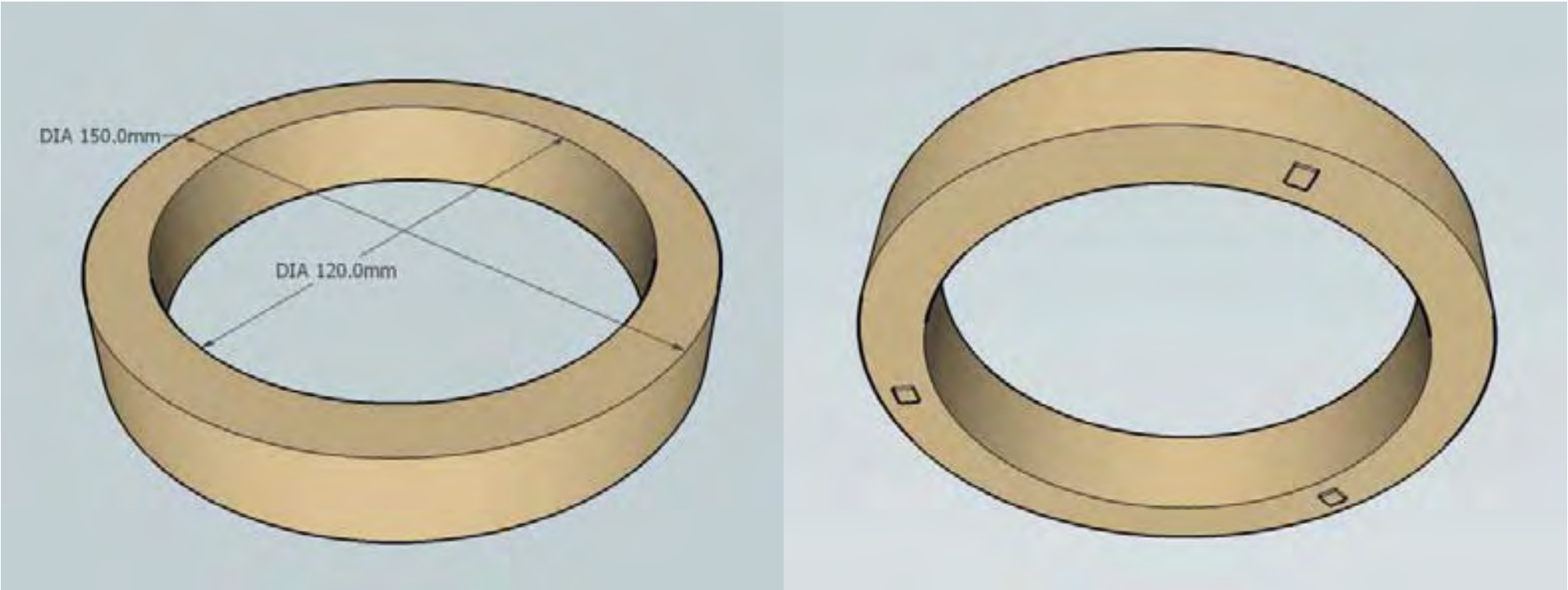

Fig. 5.3: Vistas, superior e inferior, do projeto do anel de contenção para os hemi-espelhos. Os calços, de 1mm de altura, desbastados na própria peça, podem ser vistos na superfície inferior do anel.

Conforme determinado no capítulo 4. o espelho depois de confeccionado é preso a uma mesa especial para ser cortado com serra diamantada.

Depois de cortado, nesta mesma mesa, as metades são deslocadas angularmente (pouco mais de 0,5º) e esta configuração é fixada mecanicamente, figuras 5.7-1 e 5.7-2.

Prendendo-se a mesa especial à uma mesa divisora, o fundo dos hemi-espelhos é desbastado para voltar a ser um plano horizontal (fig. 5.7-3 e 5.7-4). As laterais também são desbastadas para o conjunto recuperar a forma circular (fig. 5.7-5 e 5.7-6). Ainda preso à mesa especial, o fundo dos espelhos recebe um desbaste fino para se tornar um plano óptico (procedimento semelhante ao do espelho do protótipo, vide fig. 4.17).

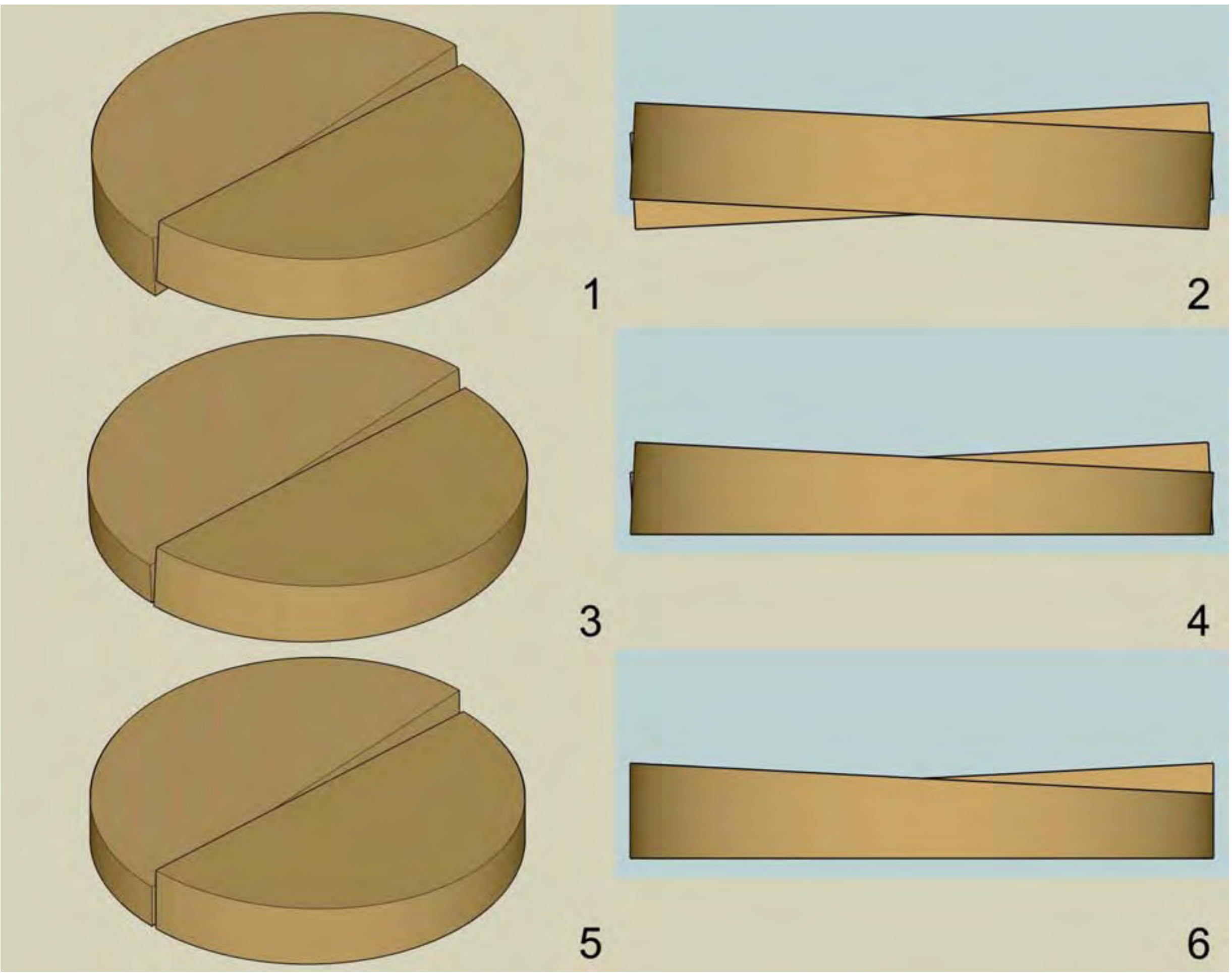

Fig. 5.4: Etapas do projeto de primeiro desbaste dos hemi-espelhos.

Mais tarde, os hemi-espelhos sofrem um segundo desbaste para diminuir o diâmetro do conjunto, permitindo serem encaixados dentro do anel de contenção (fig. 5.7), e um terceiro desbaste para escavar calços laterais e inferiores nas próprias peças.

Um furo central também é escavado no conjunto para acomodar a peça que segura os hemi-espelhos para estes últimos desbastes e que serve para a passagem de luz quando dos procedimentos de auto-colimação e comparação, figura 5.5.

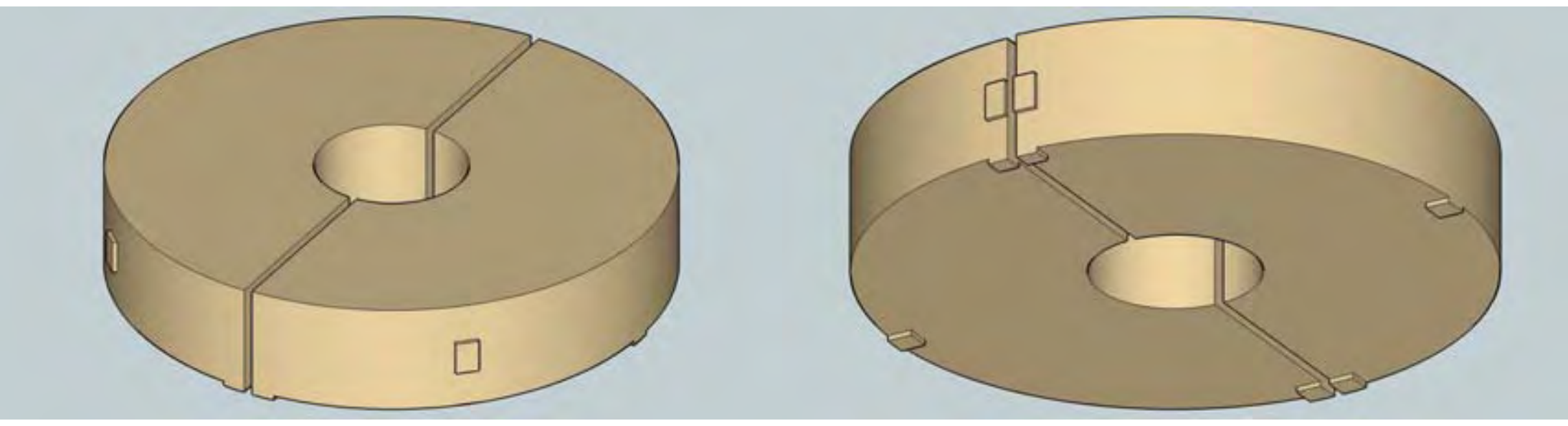

Fig. 5.5: Vistas do projeto final dos hemi-espelhos. Os calços laterais e inferiores, com 1 mm de espessura) serão desbastados na própria peça.

Os hemi-espelhos são forçados contra as paredes do interior do anel para conter seus movimentos laterais. Para servir de apoio espaçador entre os hemi-espelhos, é usada uma esfera cerâmica retificada, de baixo coeficiente de dilatação térmica, encaixada numa cavidade escavada em uma das peças (fig. 5.6 e 5.8). Finalmente, a fim de tensionar em direção às paredes, uma peça metálica, com efeito mola é adicionada, figura 5.6.

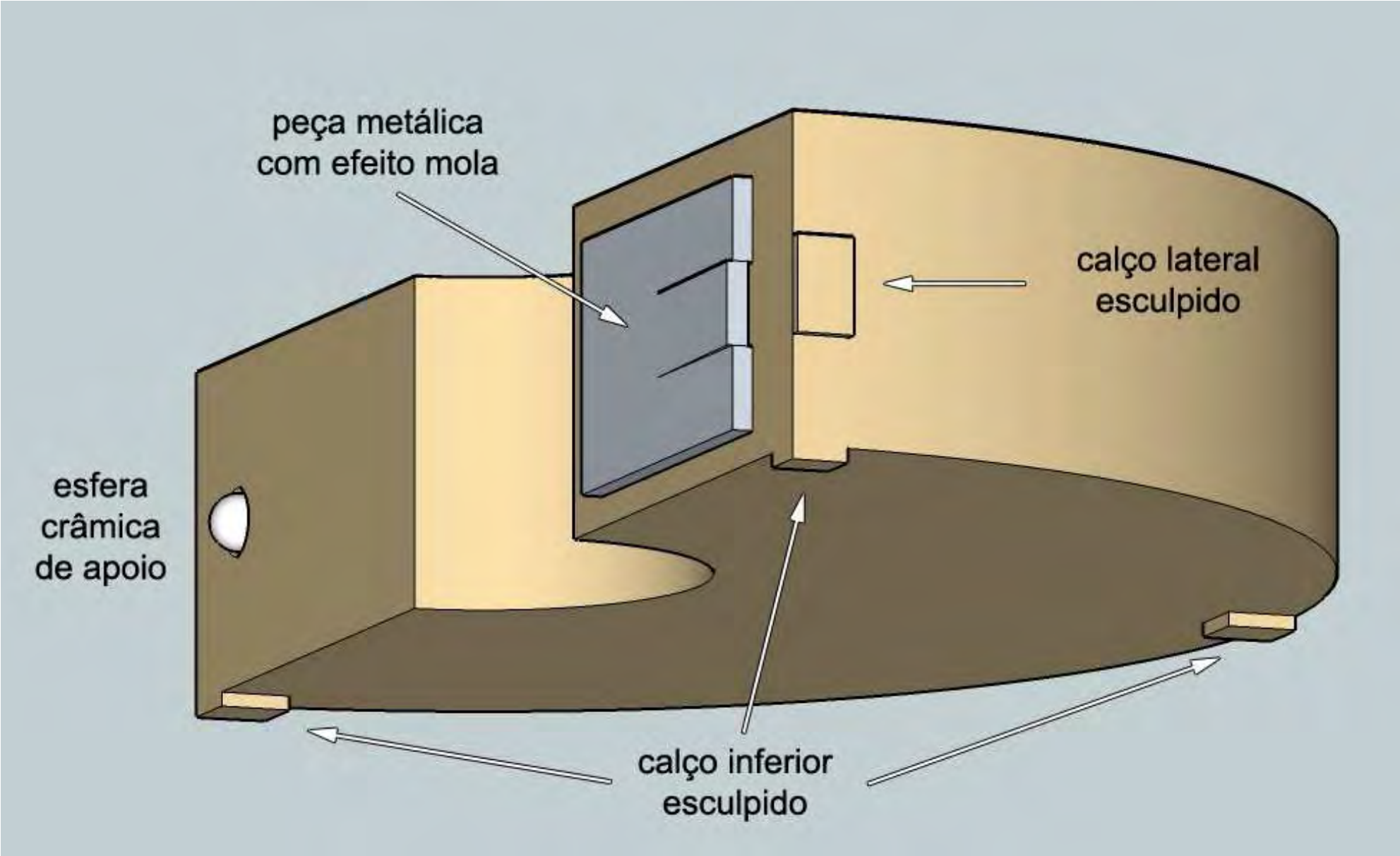

Fig. 5.6: Vista do projeto de apoio e contenção lateral dos hemi-espelhos.

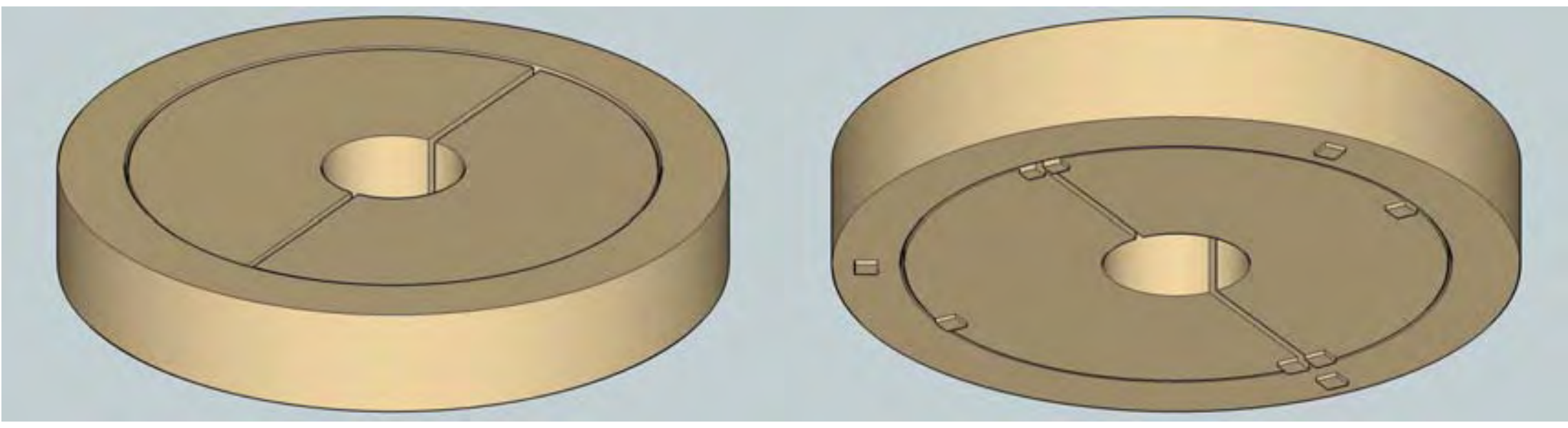

Fig. 5.7: Vistas do conjunto hemi-espelhos/anel de contenção.

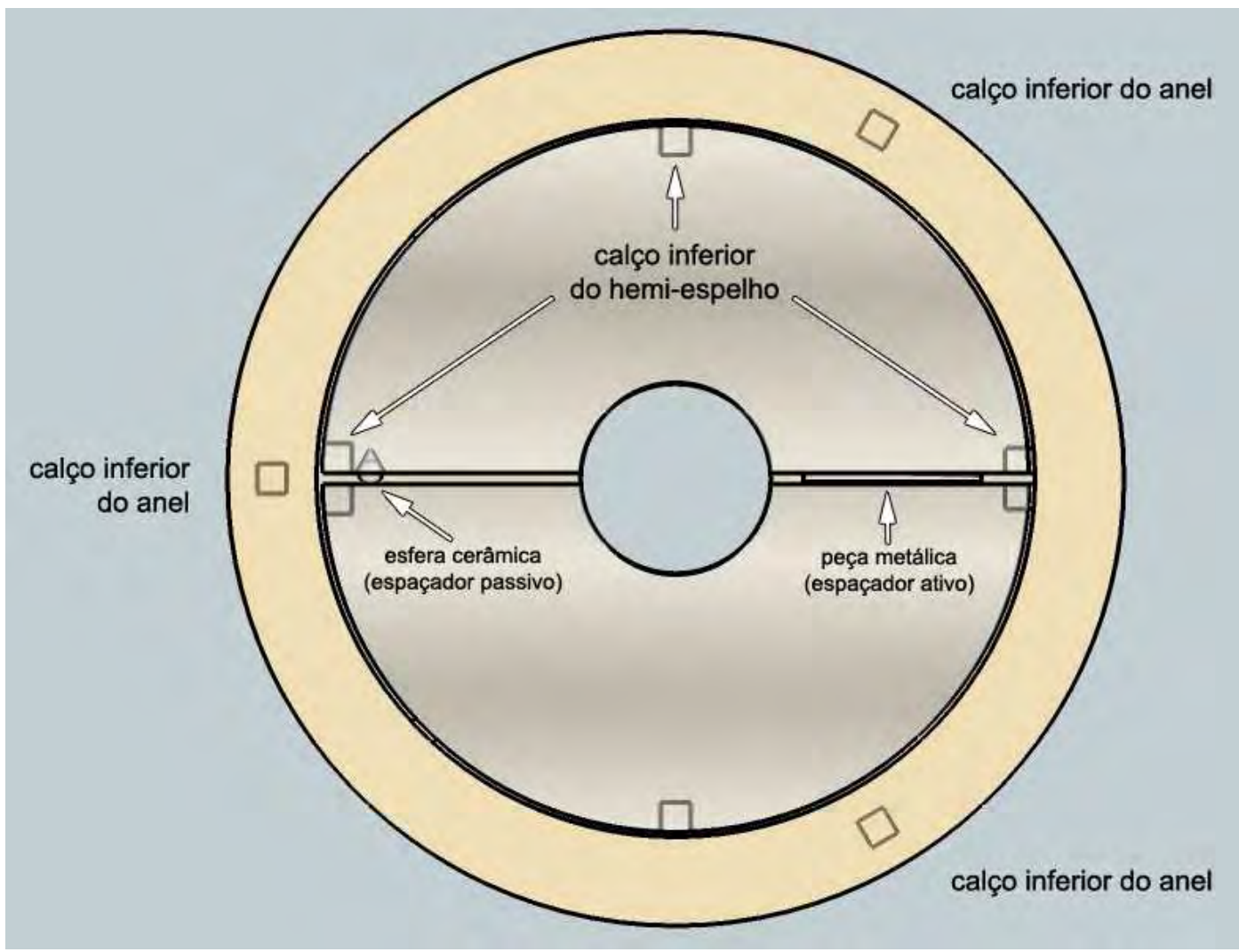

Fig. 5.8: Vista superior (transparente) do projeto de apoios do espelho heliométrico.

A base do conjunto hemi-espelhos/anel é um bloco inteiro de CCZ-HS, com as duas superfícies trabalhadas até se tornar um plano óptico de qualidade igual ou superior a $\lambda/4$, figura 5.9.

Detalhes do polimento e do teste óptico deste disco base são apresentados no Apêndice III.

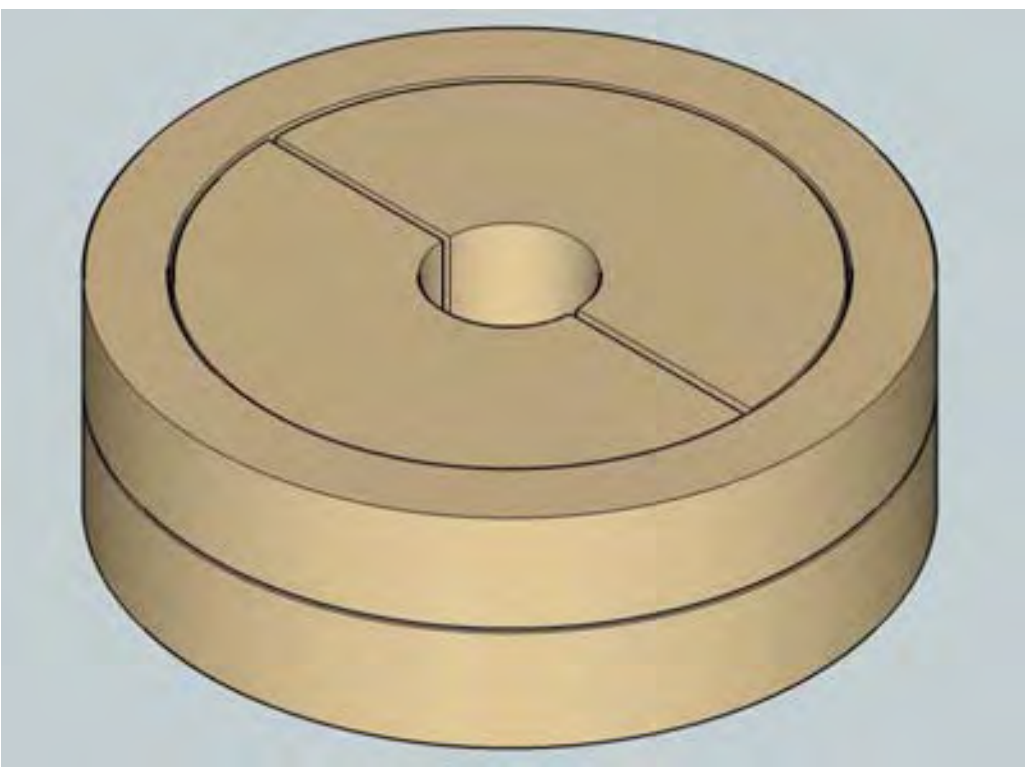

Fig. 5.9: Projeto do espelho heliométrico completo.

### 5.1.3 - A mesa especial de suporte para o corte e fabricação do diedro

A mesa especial de corte e fabricação do diedro é onde o espelho acabado é fixado com segurança, cortado ao meio (com perda mínima de material). Depois, ainda fixadas à mesa, as metades são deslocadas angularmente e travadas mecanicamente no ângulo escolhido.

Na figura 5.10 apresenta-se o projeto da mesa especial de corte e fabricação e sua execução final.

A base é dividida em dois conjuntos que basculam independentemente. Há um espaço entre eles suficiente para passar a serra de corte com segurança.

Na imagem da direita o nº 1 indica a base onde se fixa o espelho, 2, os parafusos e contra-parafusos para a movimentação angular e 3, os parafusos para a fixação mecânica do diedro.

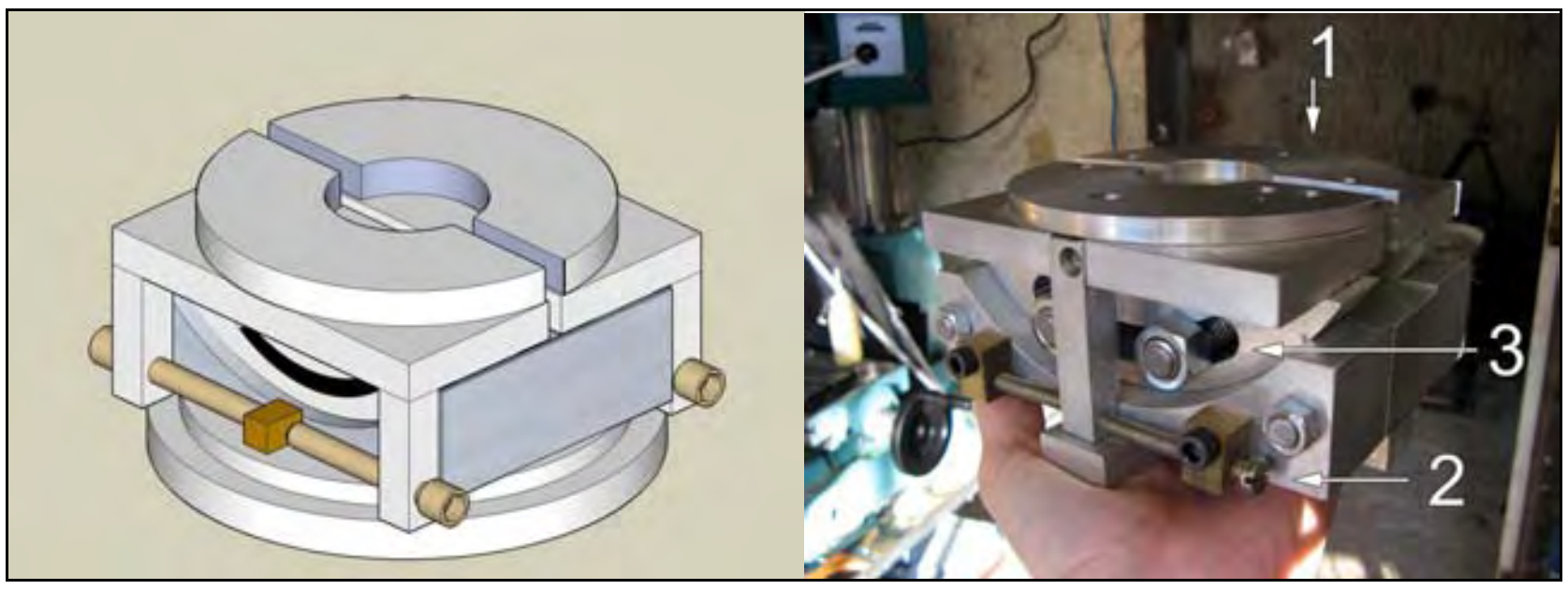

Fig. 5.10: Projeto 3D da base de corte do espelho e sua materialização em alumínio e bronze.

O espelho é fixado à mesa com sua superfície parabólica voltada para baixo. Os detalhes da construção desta mesa podem ser vistos no Apêndice IV.

### 5.1.4 - O espelho de CCZ-HS

As Características do espelho acabado são:

- Diâmetro = 150 mm
- Distância focal = 986 mm
- Relação focal f/D = 6,53. O limite mínimo de aceitabilidade para um espelho esférico de D = 150 mm uma é uma relação focal igual a 8. Abaixo deste limite uma superfície parabolóide é recomendada para garantir a qualidade óptica (Texereau, 1961).
- Superfície parabólica

- Qualidade óptica da superfície → λ/30 (erro pico-vale da frente de onda). Note-se que o limite de λ/8 já preenche o mínimo da condição óptica (Ceravolo *et al.*, 1992).
- Razão de Strehl = 0,99
- Defeito máximo da superfície = 20 nm

Os detalhes dos testes ópticos são apresentados no Apêndice V.

### 5.1.4.1 - Seqüência de corte e montagem do diedro

A mesa divisora é, por sua vez, fixada à fresadora e um relógio comparador é usado para a perfeita centragem da mesa especial, figuras 5.11 e 5.12.

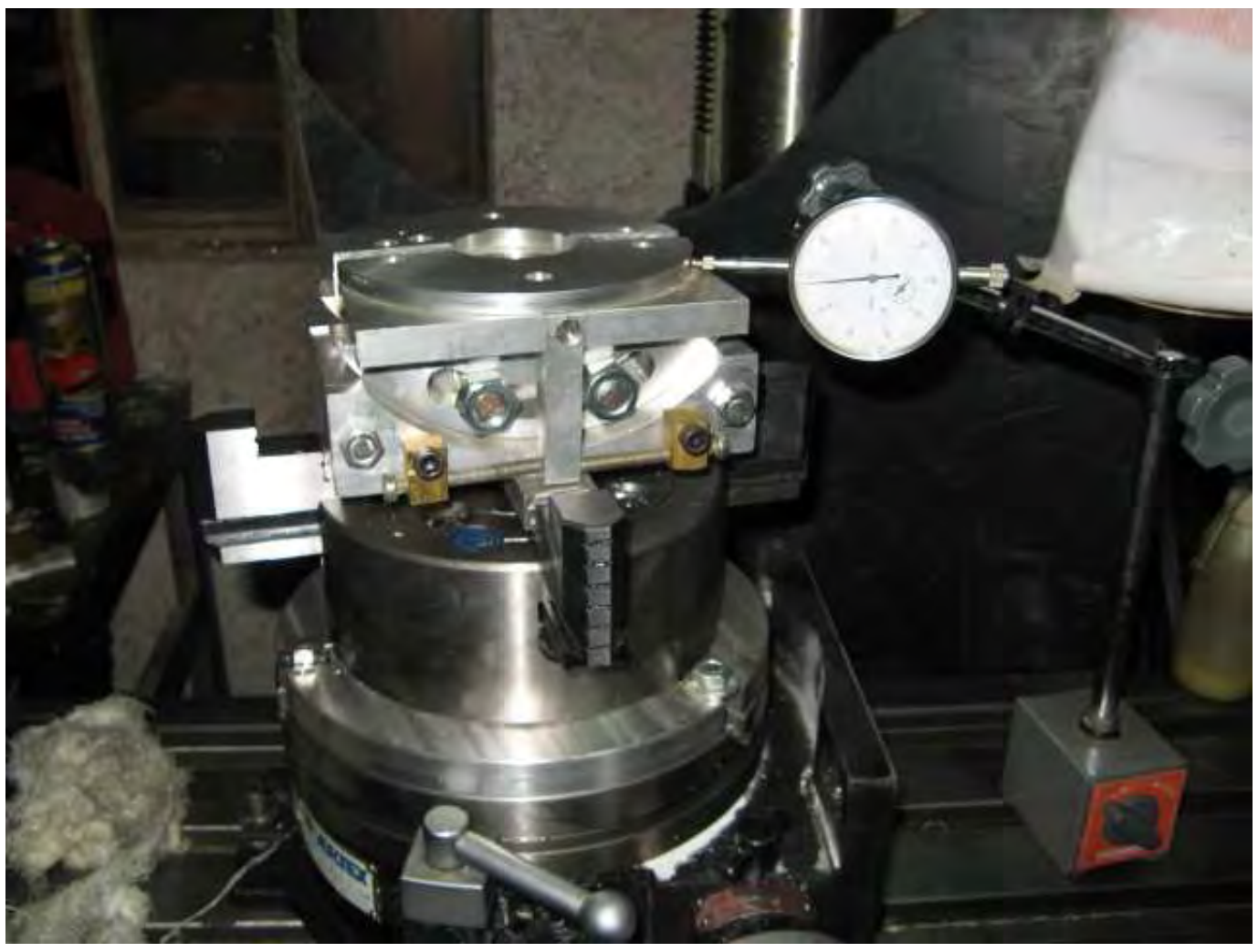

Fig. 5.11: Centragem da mesa especial antes da fixação do espelho.

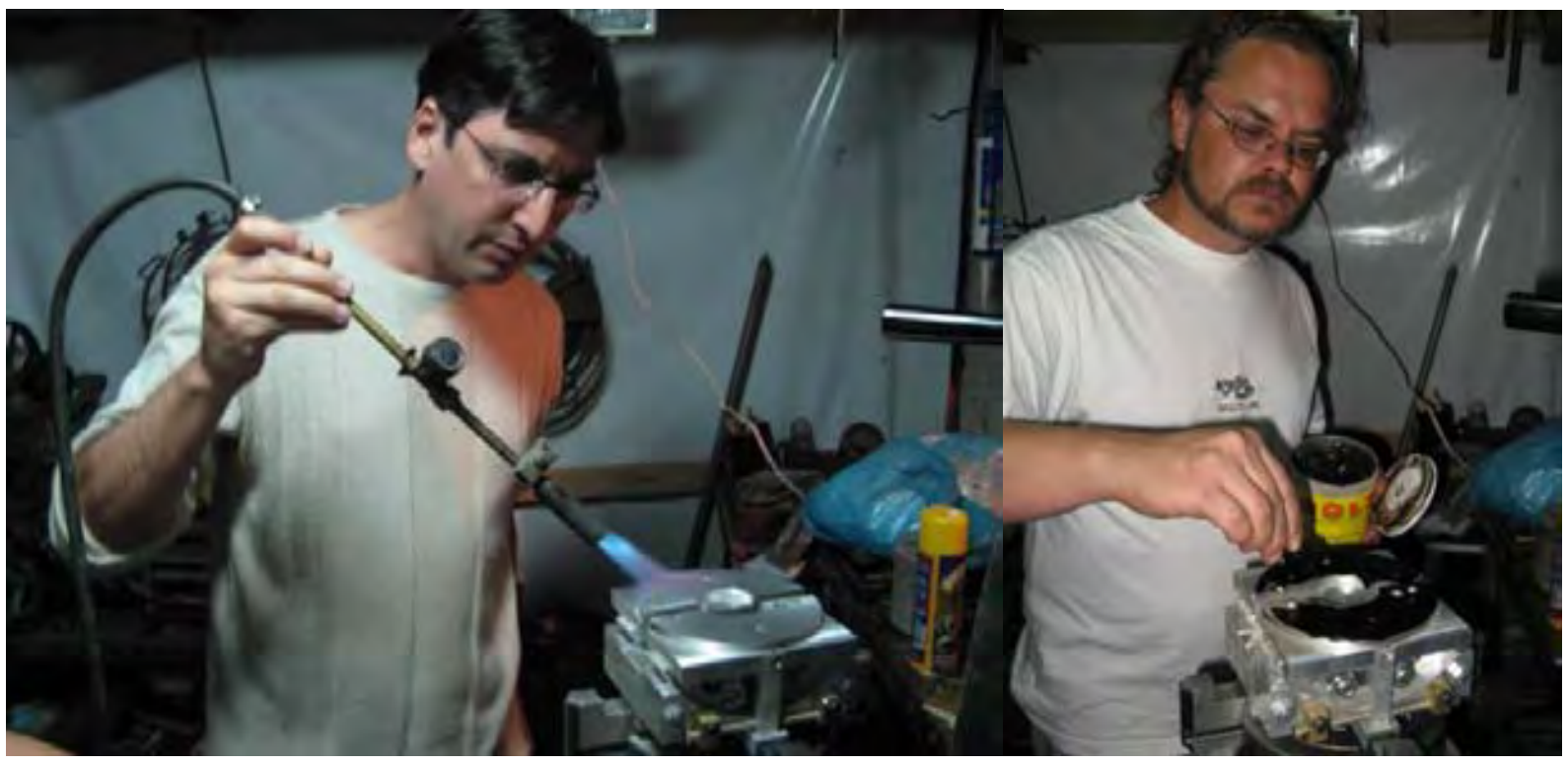

Fig. 5.12: Mesa sendo aquecida para a aplicação do piche.

O espelho é pressionado contra a mesa e depois centrado, utilizando-se, mais uma vez, o relógio comparador. A precisão desta centragem é de 30 μm, figuras 5.13, 5.14 e 5.15.

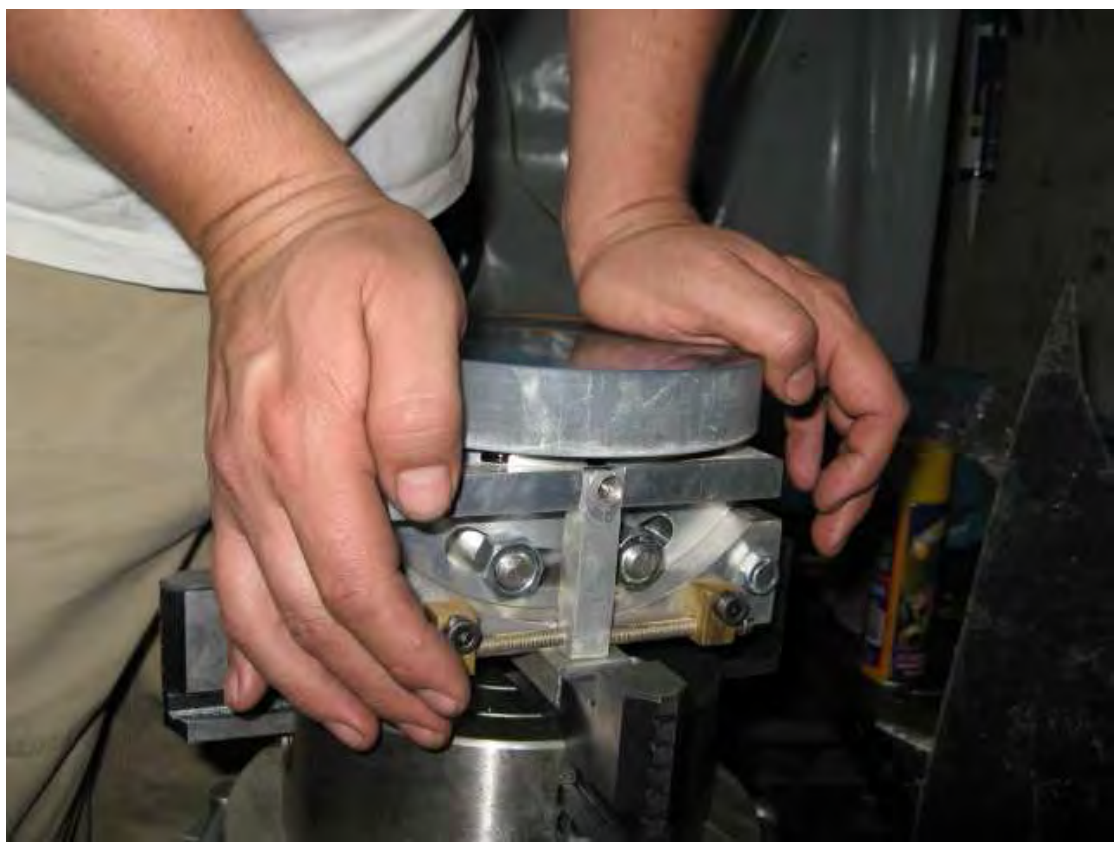

Fig. 5.13: Espelho sendo pressionado contra o piche viscoso.

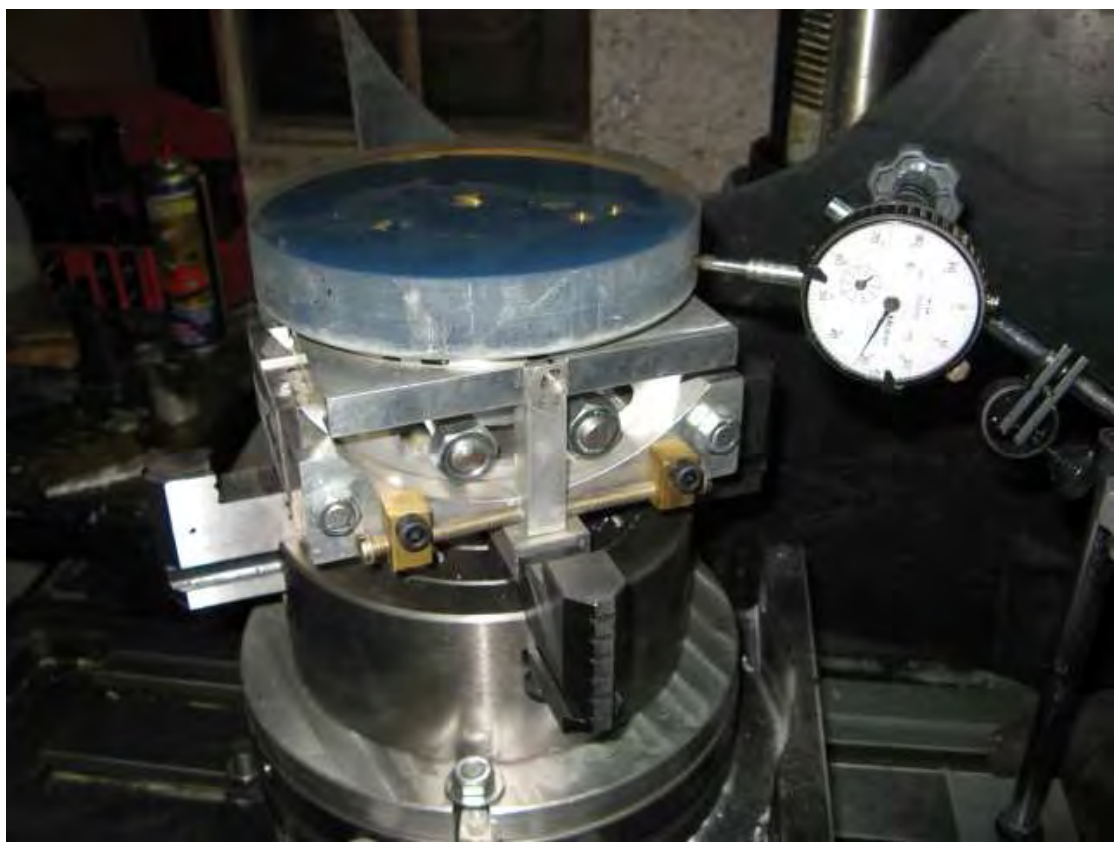

Fig. 5.14: Centragem do espelho no conjunto.

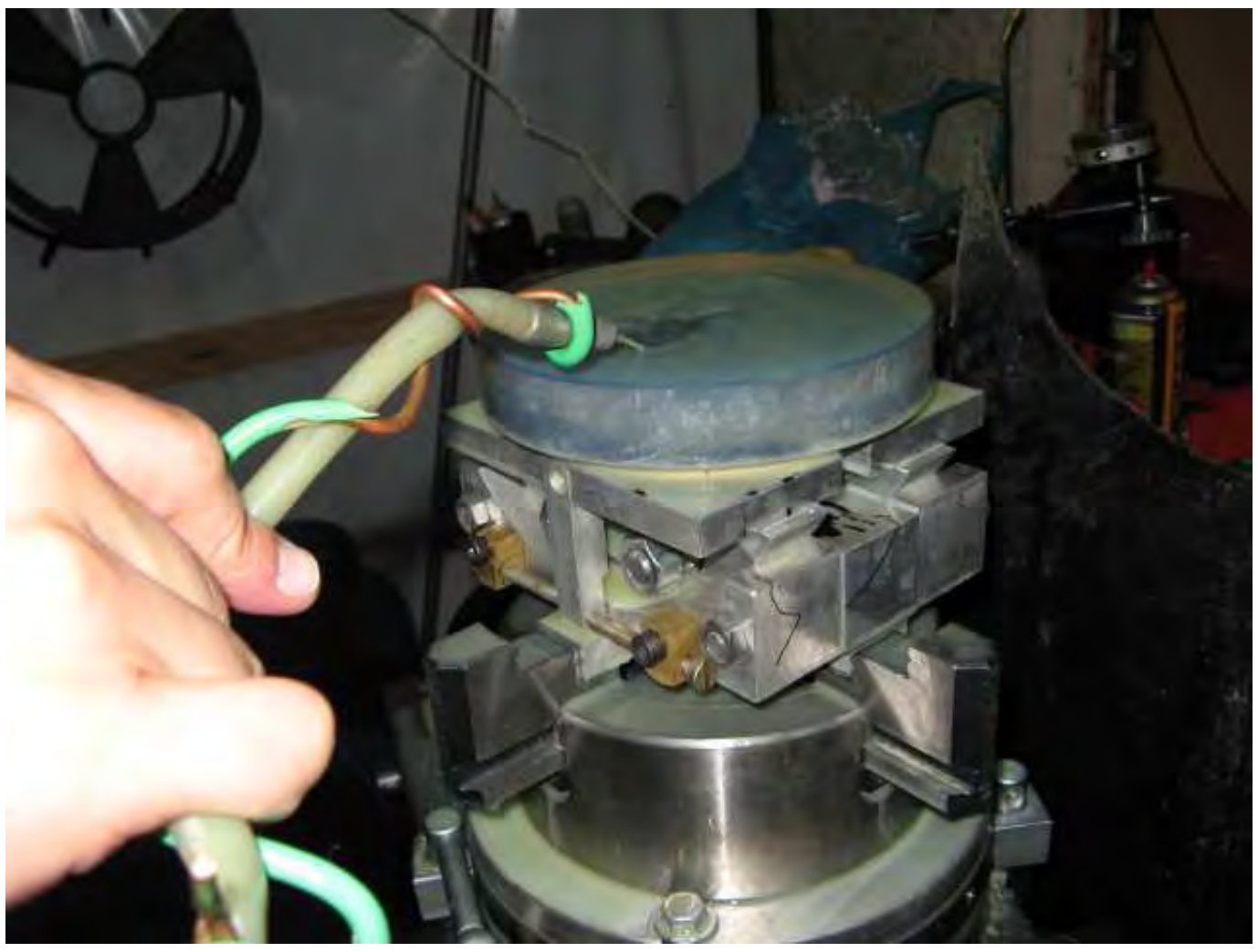

Fig. 5.15: Esfriando o conjunto com líquido refrigerante.

Todo o conjunto é, então, virado horizontalmente para o corte com a serra diamantada. Como a serra passa pelo espelho e entre as metades da mesa, é importante que este caminho esteja perfeitamente na horizontal. Mais uma vez o relógio comparador é usado para o ajuste, figura 5.16.

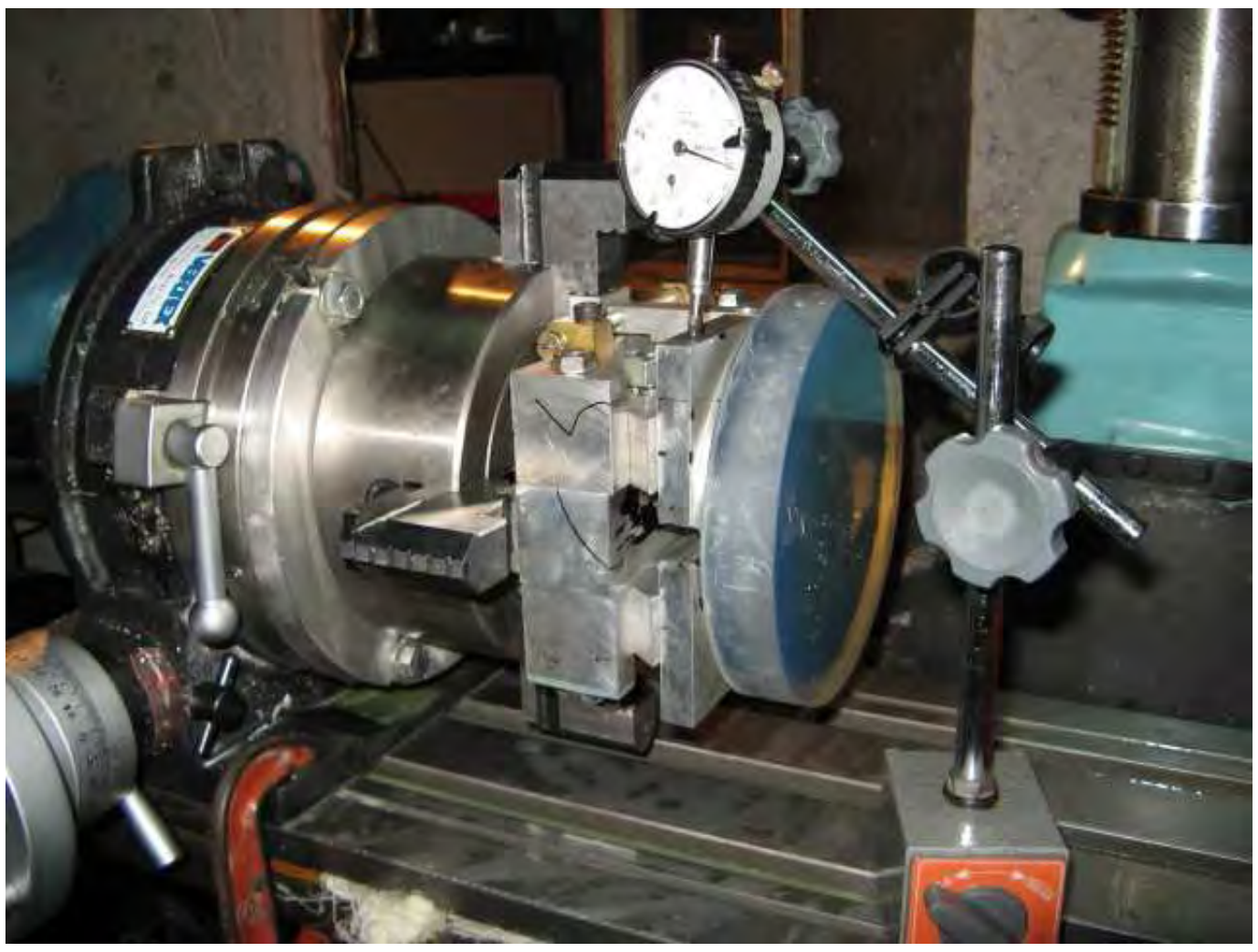

Fig. 5.16: Relógio comparador para ajuste antes do corte.

Depois, um paquímetro é usado para o ajuste da altura correta da serra no centro do disco. O corte é realizado com o avanço lento da serra de forma de forma a evitar fraturas ou trincas no corpo do espelho, como demonstrado no item 4.1.1.1, figuras 5.17 e 5.18.

O espaçamento entre as peças, devido à perda de material com o corte, é de pouco mais de 1,5 mm, figura 5.19.

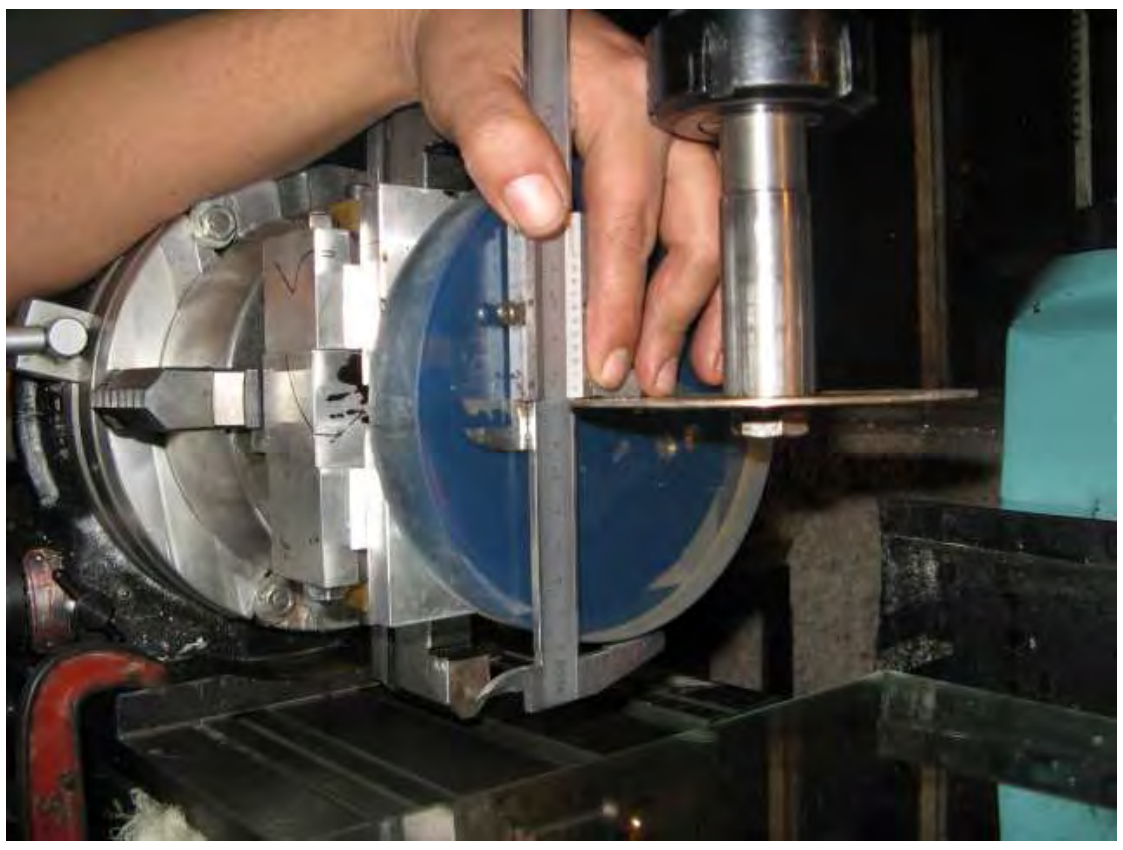

Fig. 5.17: Centragem da serra diamantada com um paquímetro.

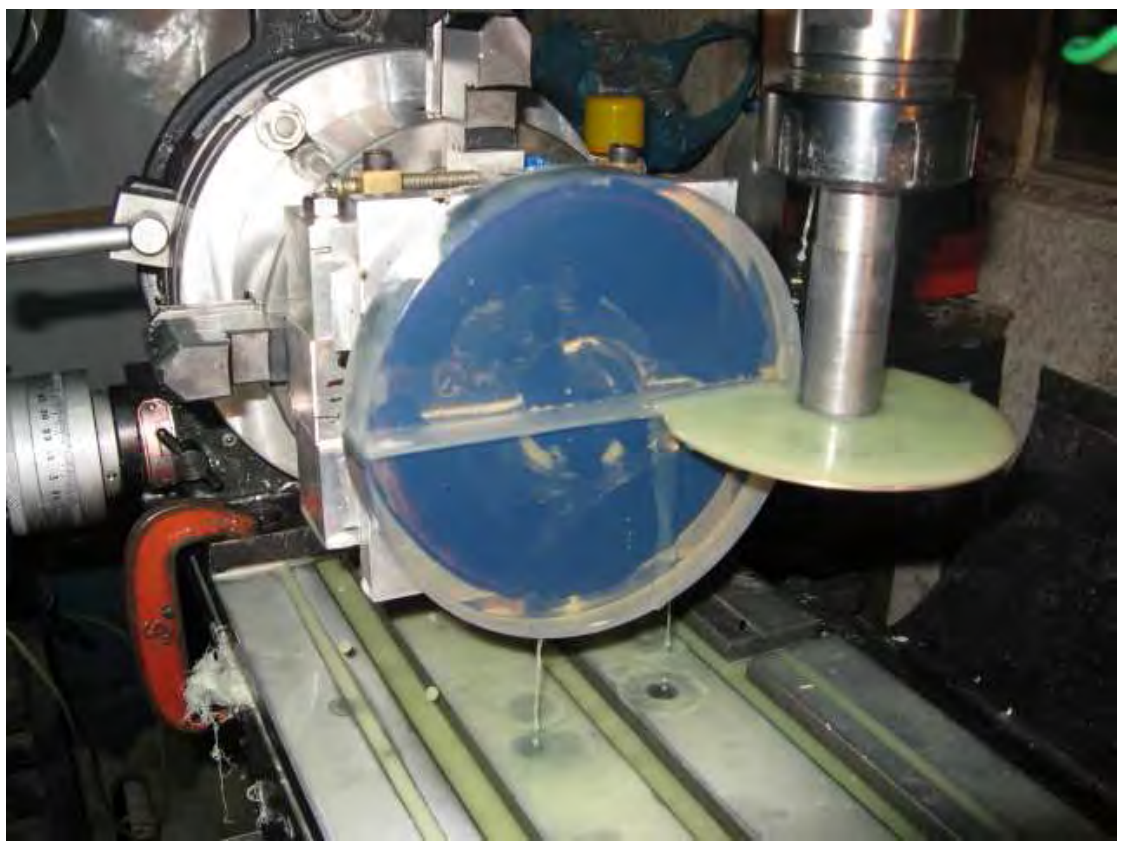

Fig. 5.18: Etapa final do corte do espelho e CCZ-HS.

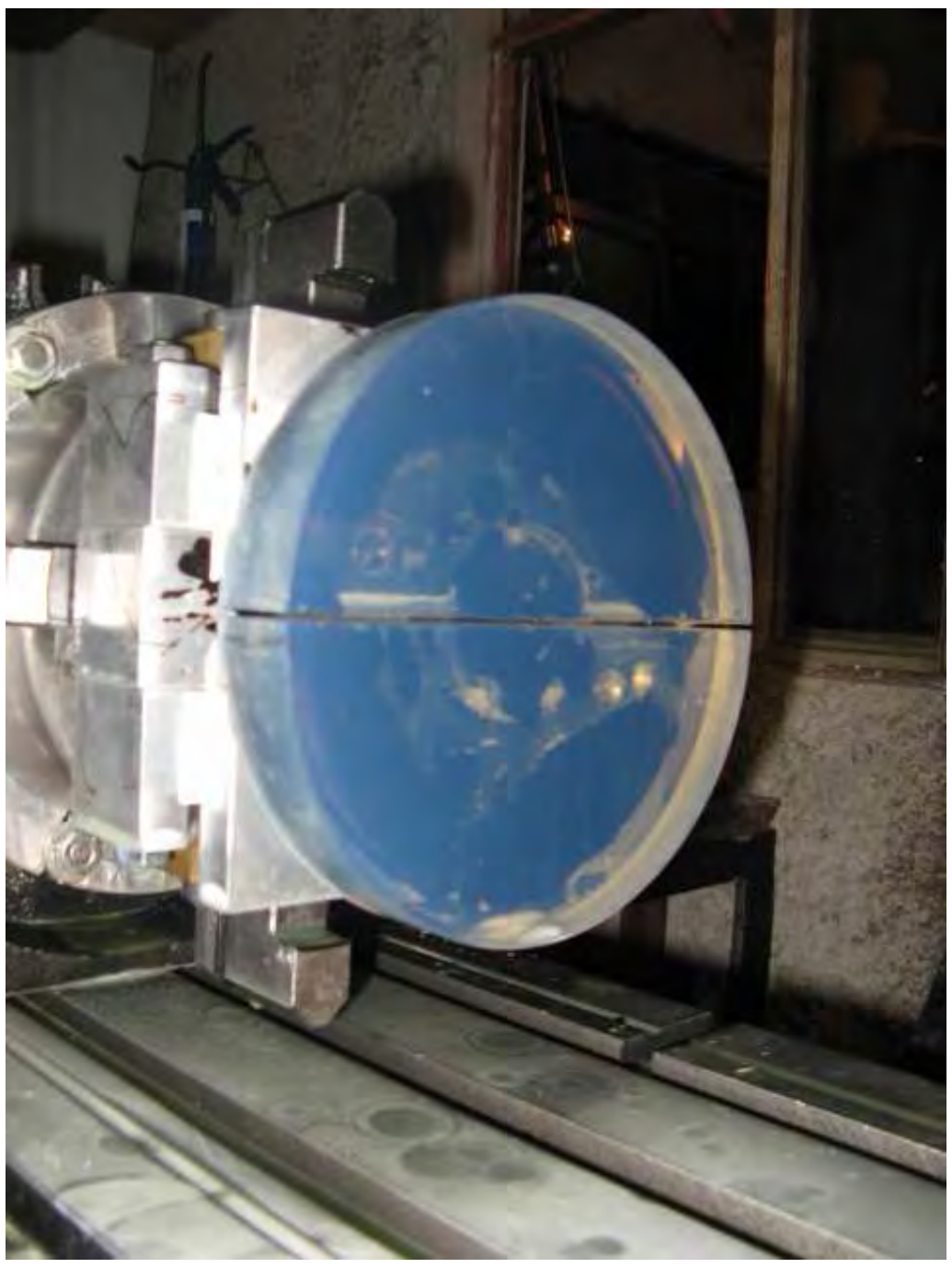

Fig. 5.19: Espelho totalmente bi-partido.

O diedro é realizado utilizando uma fonte laser, um divisor de feixe e um anteparo afastado. Somente um hemi-espelho é deslocado da sua posição inicial. A outra metade da mesa, portanto, não é mexida.

A fonte laser é montada numa base, tangenciando o eixo dos discos. O divisor de feixe é interposto no caminho do laser (fig. 5.20 e 5.21). A base da fonte laser pode ser girada de forma que os dois feixes paralelos possam atingir ora apenas um dos hemi-espelhos, ora os dois simultaneamente, figura 5.22.

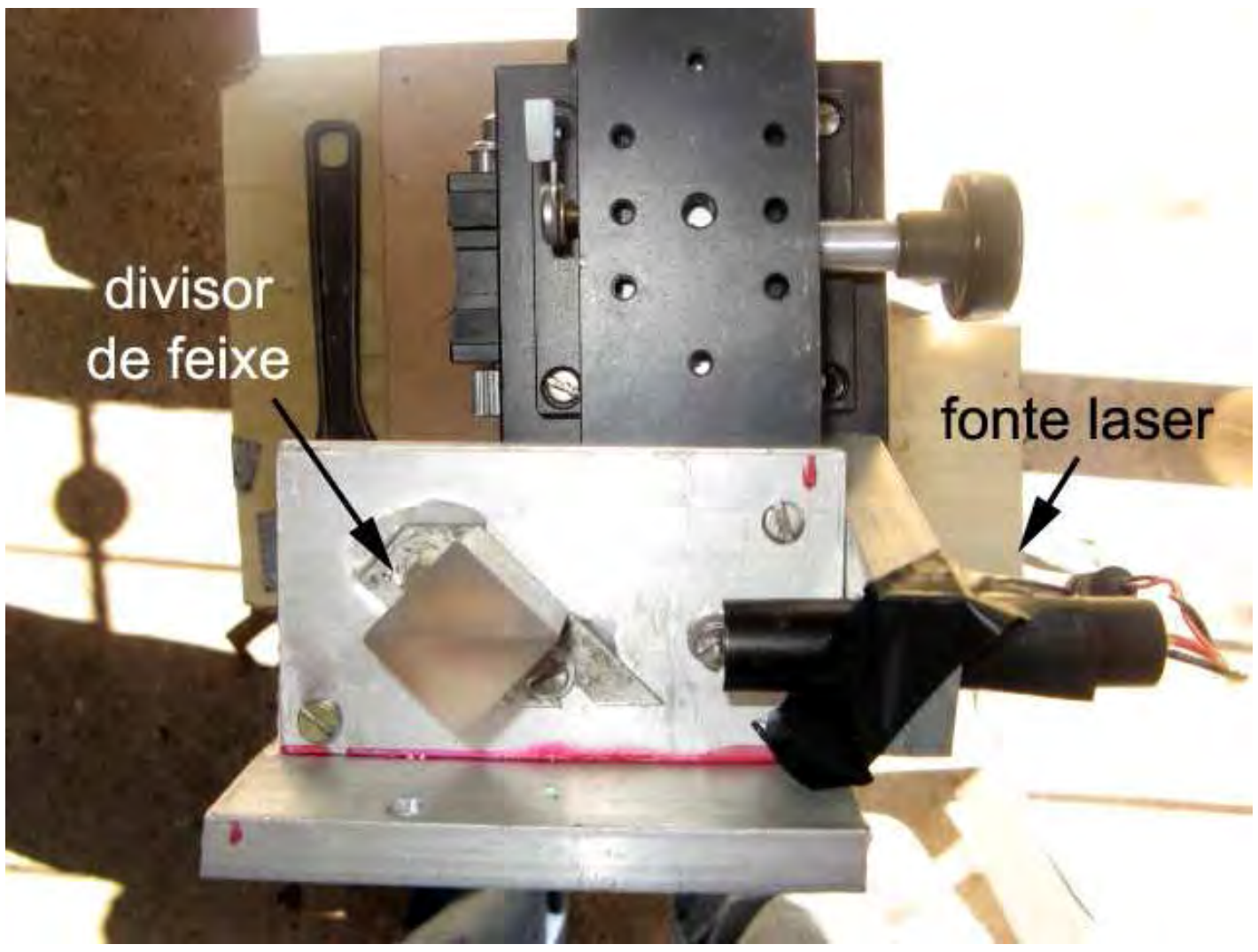

Fig. 5.20: Base para divisão do feixe de laser.

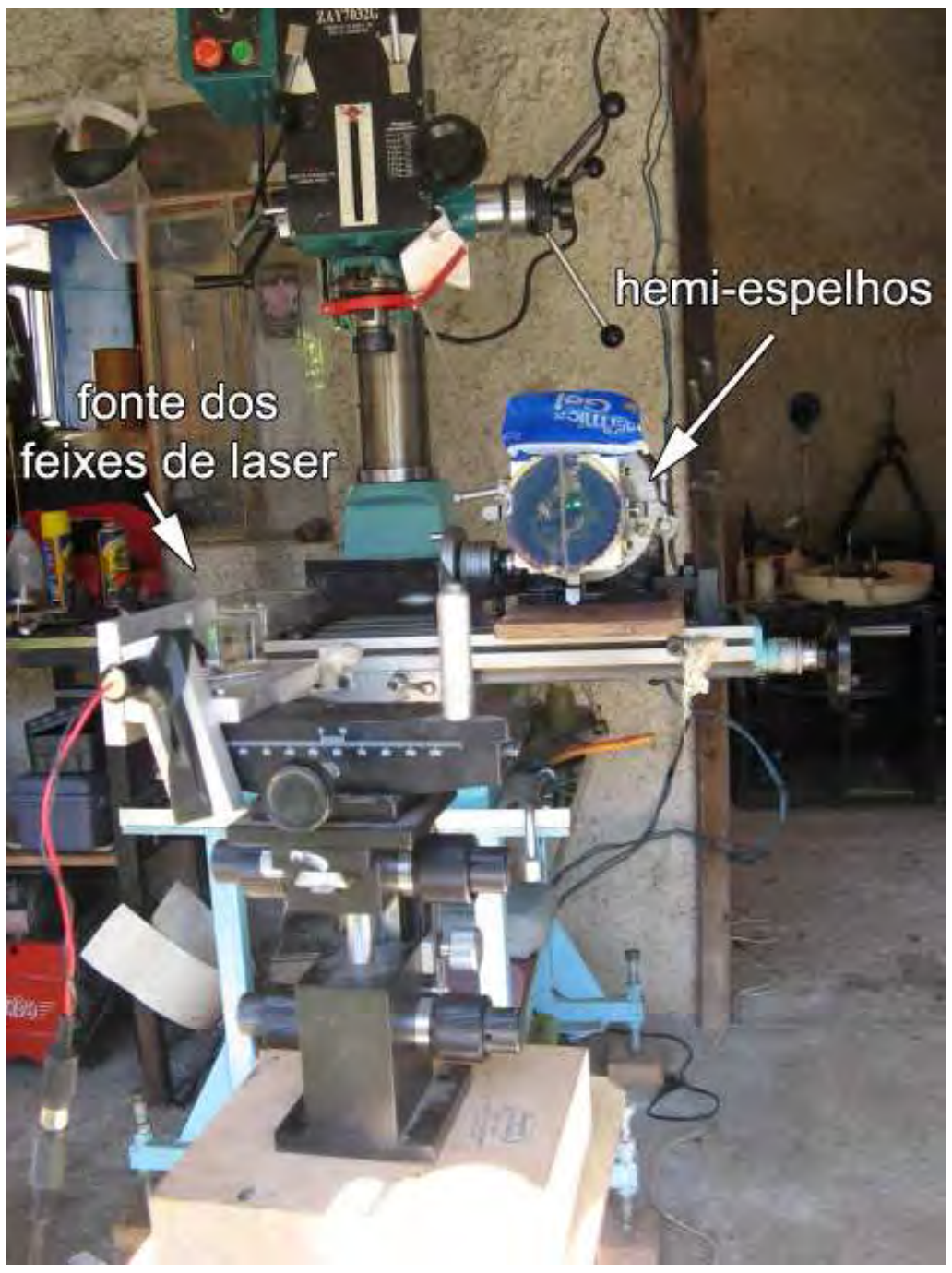

Fig. 5.21: Montagem instrumental para a criação do diedro. Na imagem se vê a fonte dos feixes posicionada em frente aos hemi-espelhos presos à mesa especial.

O dois feixes paralelos, com aproximadamente 8 mm de separação um do outro, emergem do divisor e atingem o conjunto mesa/espelho (fig. 5.22). São refletidos pelas superfícies planas da parte de trás dos hemi-espelhos para um anteparo localizado a pouco menos de 10 m de distância do conjunto, figura 5.23.

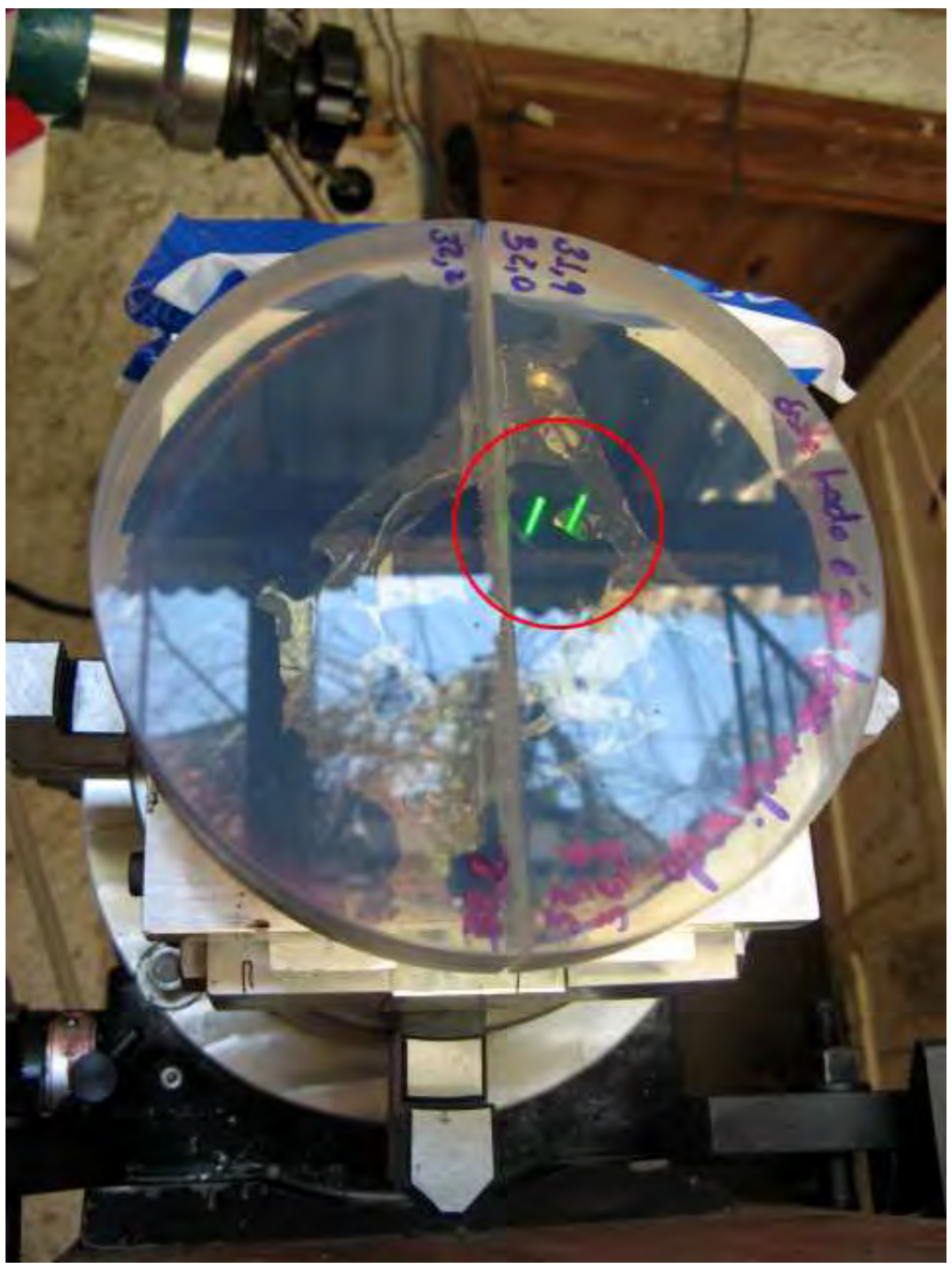

Fig. 5.22: O Círculo indica o lugar onde os dois feixes estão atingindo a parte de trás de um hemi-espelho. Quando os feixes são deslocados para a esquerda, cada feixe atinge um dos hemi-espelhos separadamente, permitido a medida do diedro no anteparo.

Na primeira configuração da experiência, os feixes são refletidos próximos entre si no anteparo. Na segunda configuração, um dos feixes se desloca, afastando-se do outro feixe. Na terceira, o outro feixe se desloca, aproximando-se novamente do primeiro. Medindo-se o deslocamento pode-se calcular o ângulo do diedro.

Uma folha de papel milimetrado é fixada num anteparo (fig. 5.23). A distância medida entre o anteparo e as superfícies refletoras vale 9365 mm. A esta distância, cada milímetro no papel corresponde a ~22". O valor mínimo do diedro a ser ajustado inicialmente é de 1980", paralelo à direção do corte, que neste caso é a direção vertical. No anteparo este ângulo corresponde a um deslocamento de 90 mm. Deslocamentos horizontais devem ser minimizados ao máximo, sendo tolerado um resíduo de até 1% do valor do diedro (~1 mm). Os feixes são alinhados para acertarem separadamente cada um dos hemi-espelhos e o diedro é criado ajustando-se o parafuso/trava (nº 2 da figura 5.13) de apenas um lado da mesa.

Com um paquímetro mede-se a posição dos centros dos círculos de luz refletida, figura 5.23.

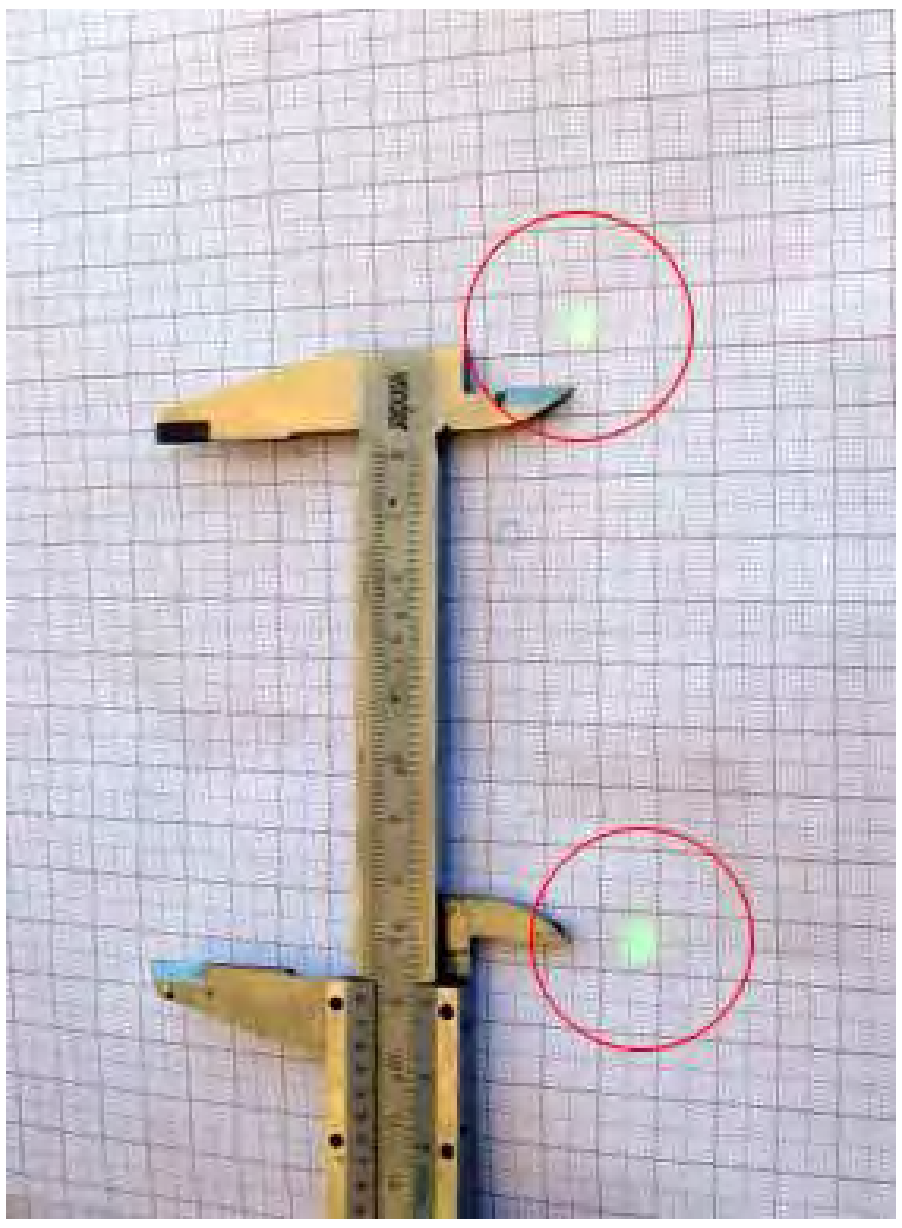

Fig. 5.23: Medindo os deslocamentos dos feixes com um paquímetro. Os círculos marcam os pontos onde os feixes atingem o papel milimetrado.

Para se fazer pequenas correções transversais no diedro (deslocamentos horizontais no anteparo), utiliza-se o parafuso lateral da mesa especial, figura 5.24.

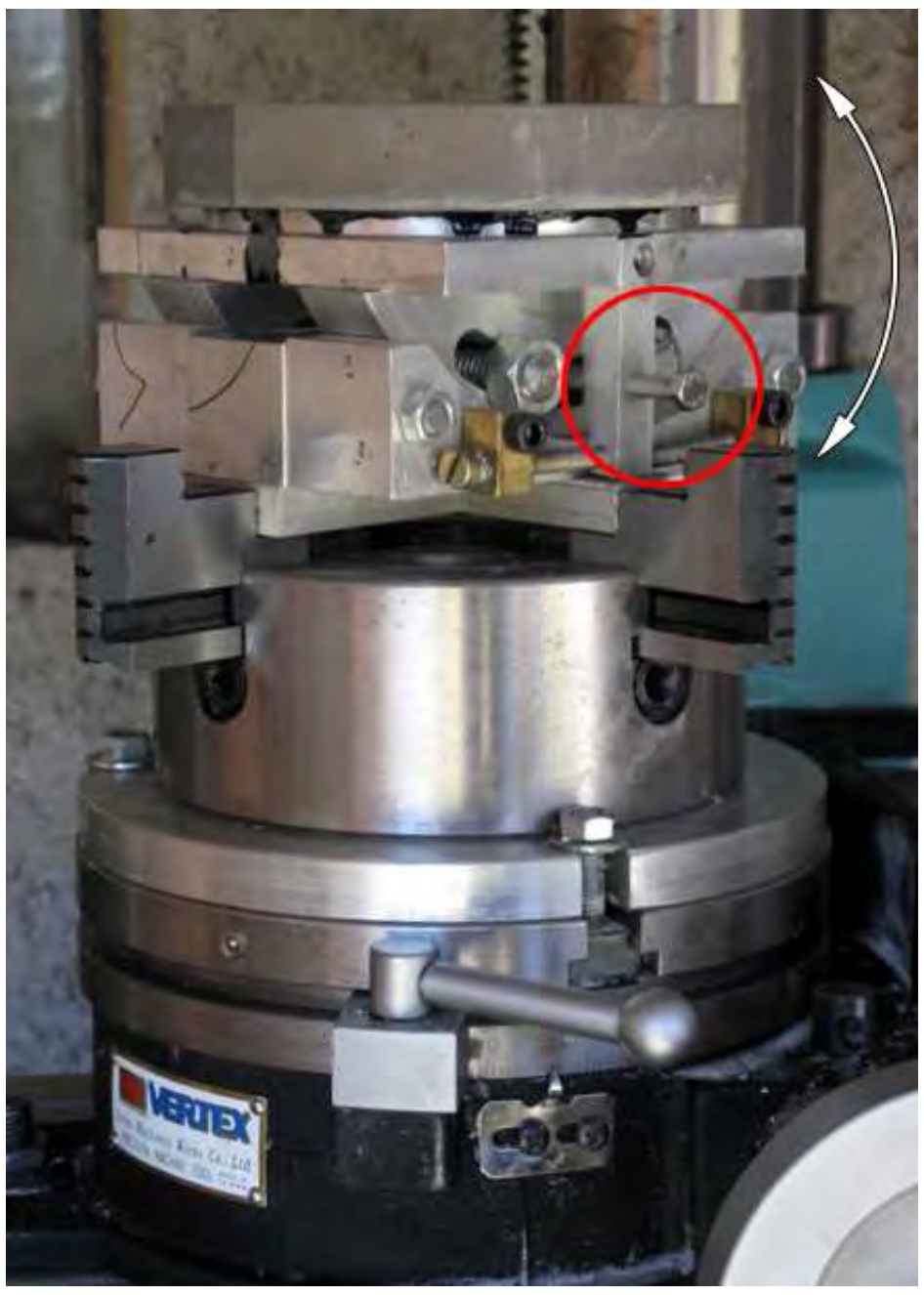

Fig. 5.24: O circulo indica o novo parafuso de ajuste acrescentado à mesa. A seta curva indica o movimento transversal que montagem passou a fazer.

A experiência de acertar com os feixes cada um dos hemi-espelhos e os dois separadamente é repetida várias vezes, a fim de se medir os deslocamentos diferenciais, de modo a minimizar qualquer erro sistemático da montagem experimental. Quando o diedro atinge o valor desejado, o parafuso nº 3 da figura 5.10 é travado.

Com o diedro travado mecanicamente, passa-se à fase do primeiro desbaste do conjunto (fig. 5.4-3 e 5.4-4). Em seguida, efetua-se o desbaste das laterais para a recuperação da forma circular do conjunto (fig. 5.4-5 e 5.4-6). Após estes desbastes, as bordas são chanfradas para evitar que lasquem, figuras 5.25, 5.26 e 5.27.

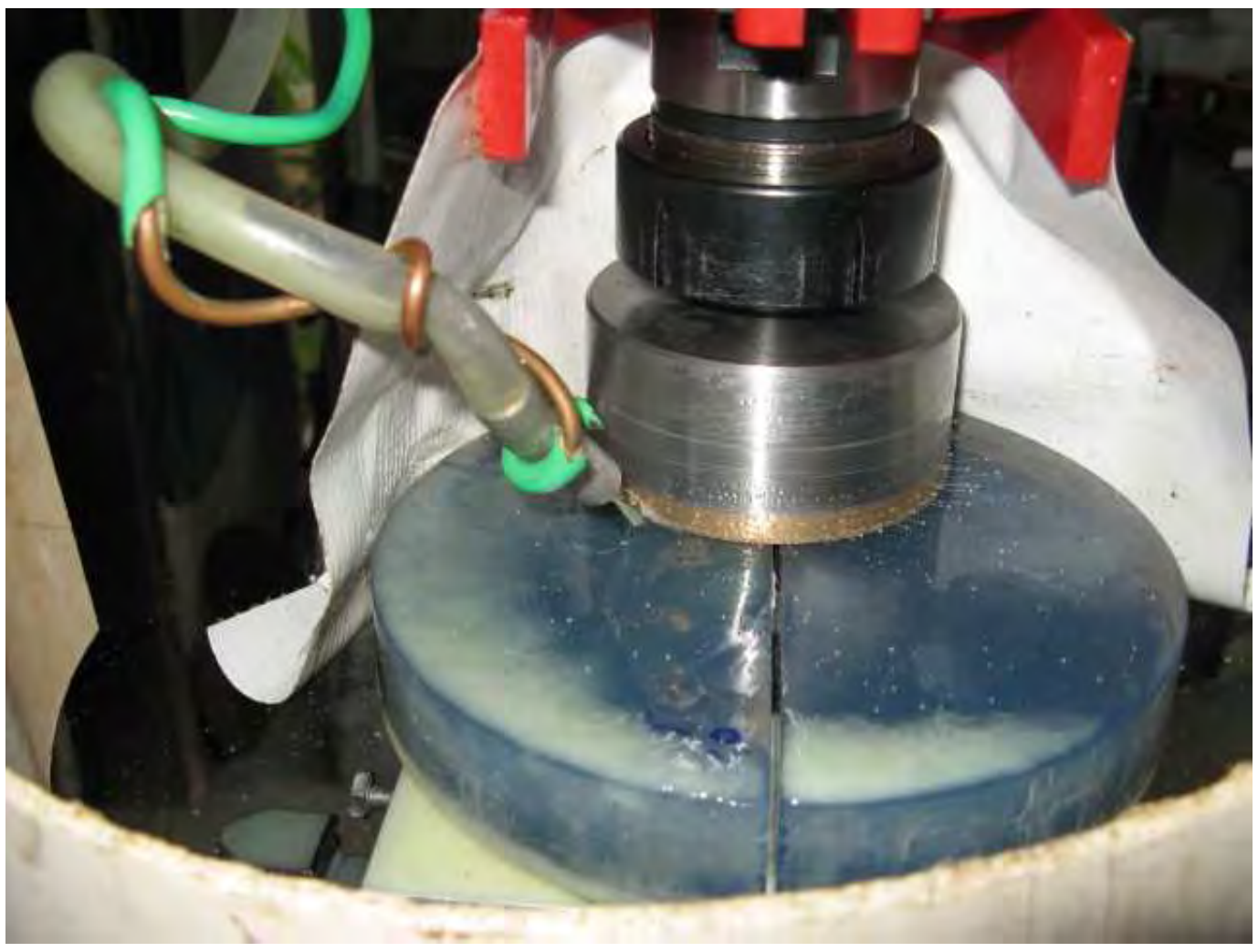

Fig. 5.25: Desbaste do fundo do espelho para planificá-los.

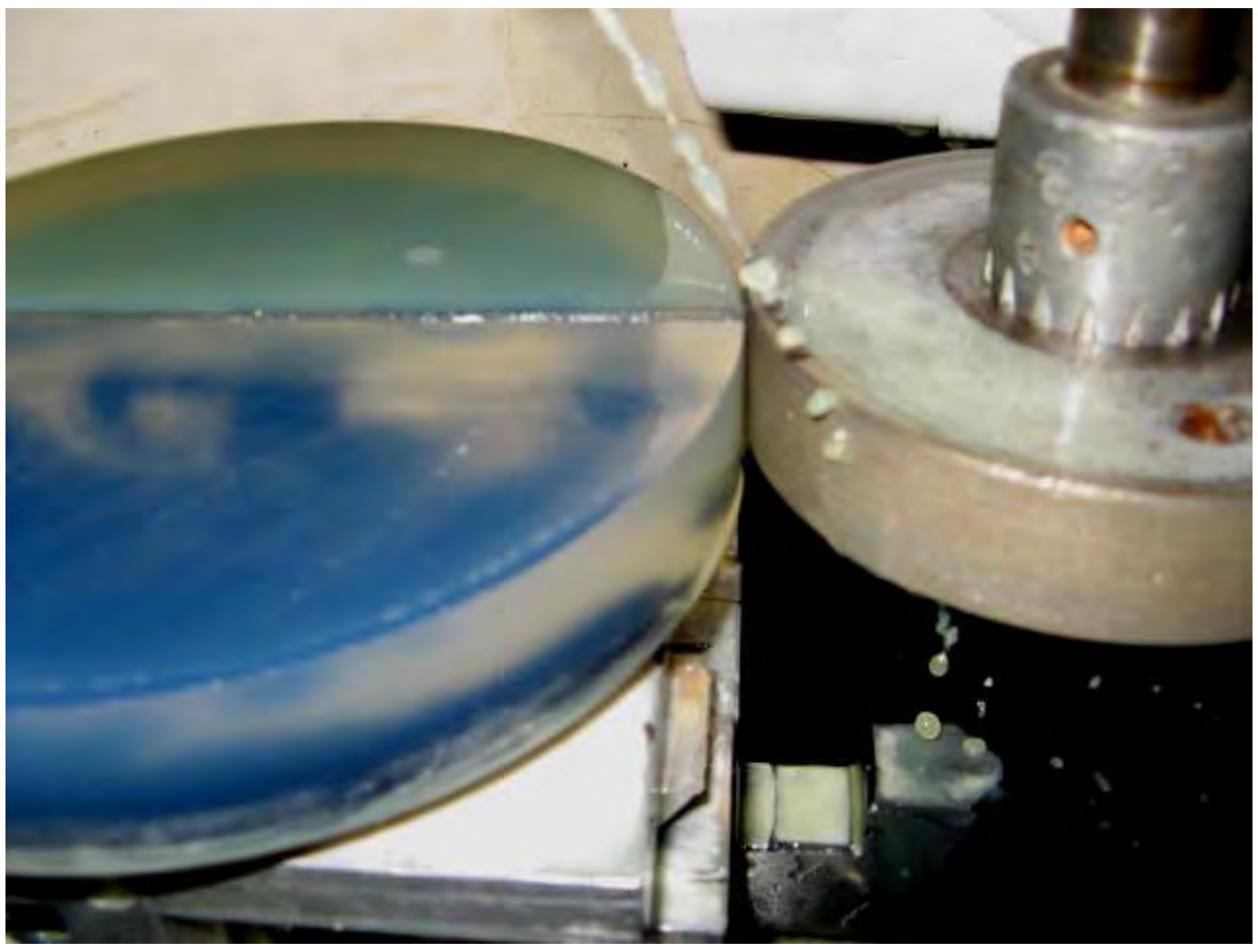

Fig. 5.26: Rebolo diamantado desbastando a lateral dos hemi-espelhos.

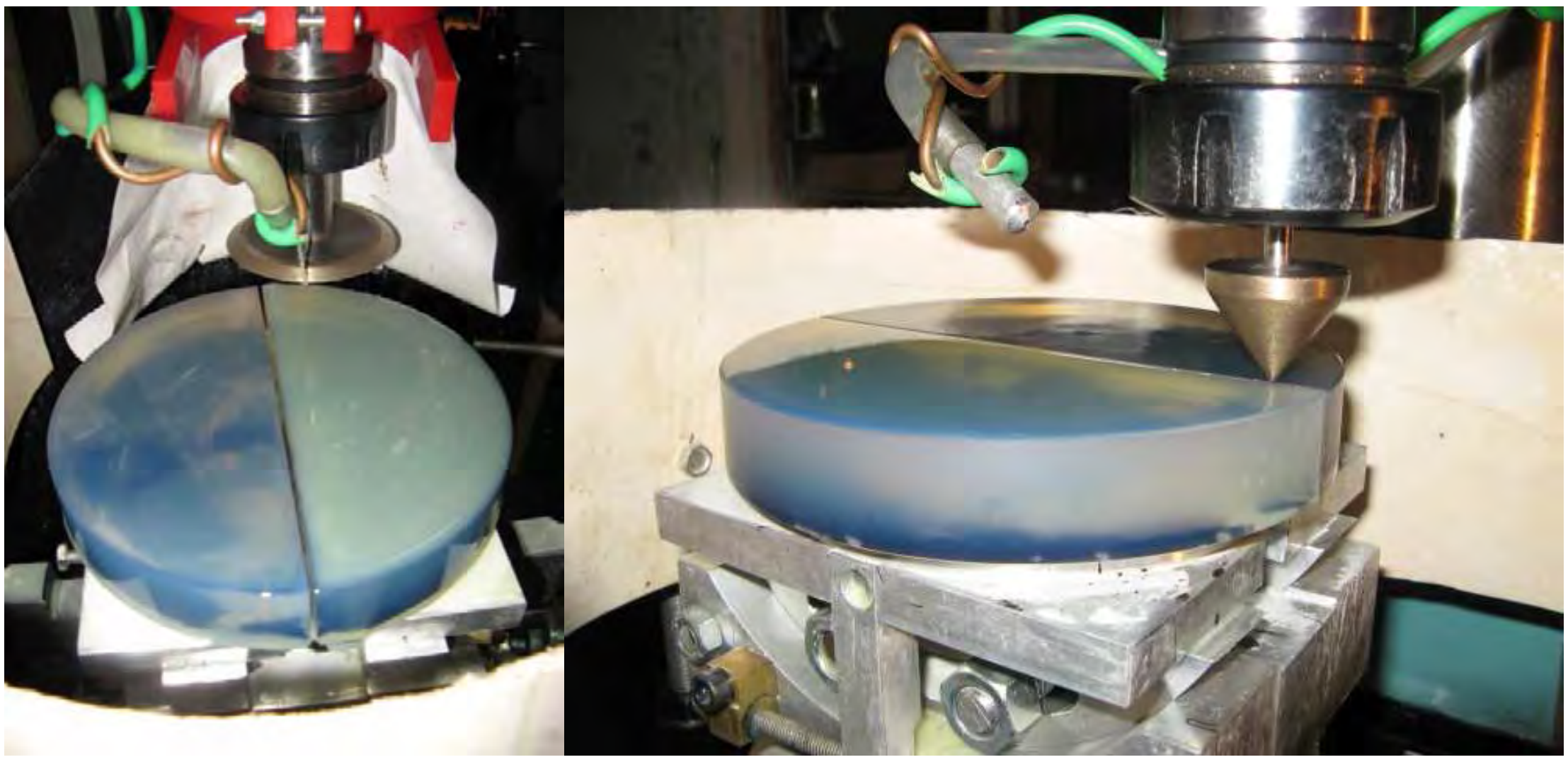

Fig. 5.27: Chanfrando as bordas.

O conjunto hemi-espelhos/mesa especial é solto da mesa divisora e a parte de trás dos hemi-espelhos é então desbastada para virar um plano óptico (fig. 5.28). Com um esferômetro, mede-se a planicidade das superfícies após cada sessão de desbaste fino (fig. 5.29). As sessões continuam até que o aparelho não mais acuse curvaturas. Após ser planificado, o conjunto é limpo e guardado em refrigeração para que, frios, os hemi-espelhos possam ser separados com segurança.

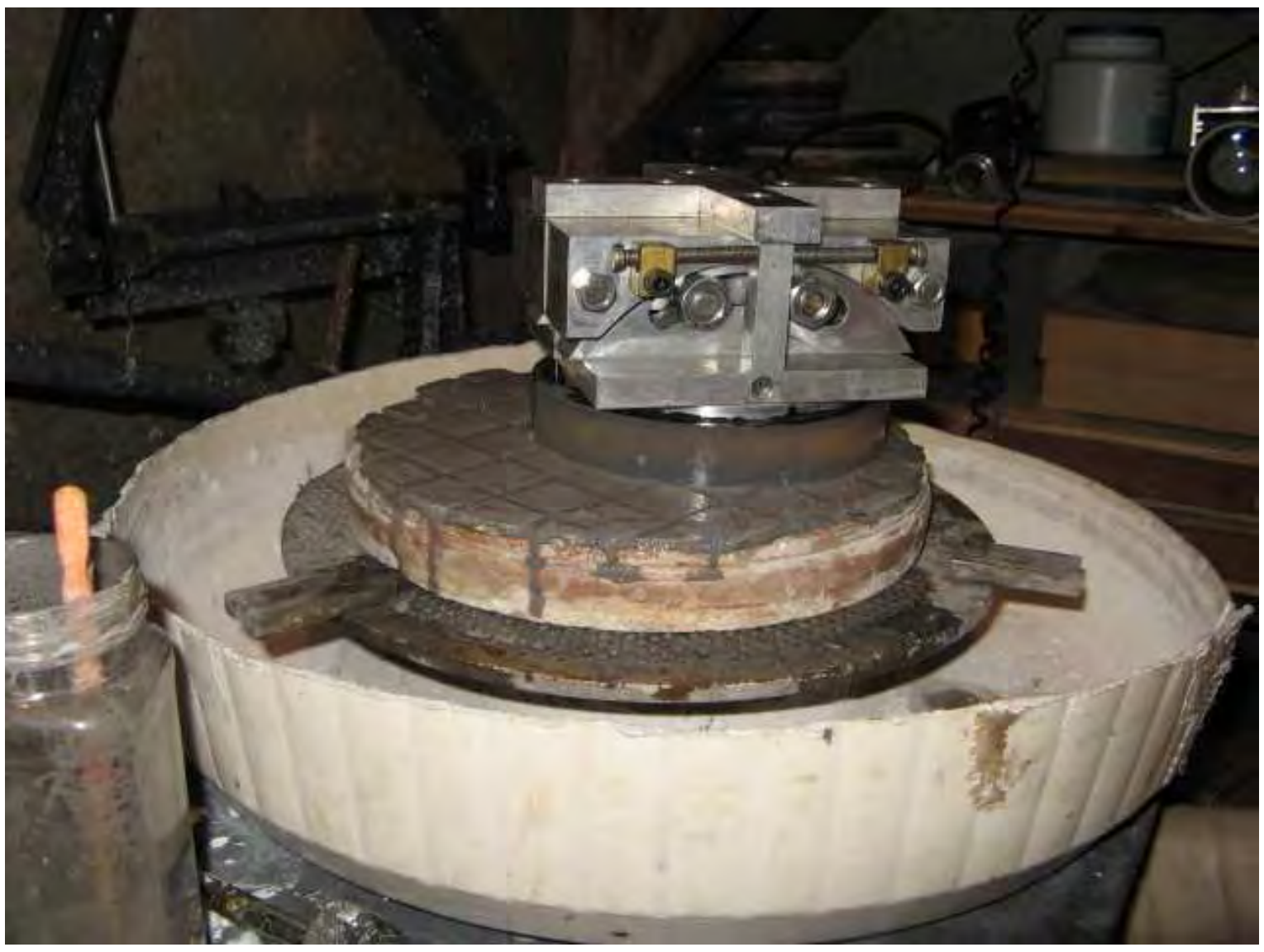

Fig. 5.28: Conjunto hemi-espelhos/mesa sobre a base de desbaste fino.

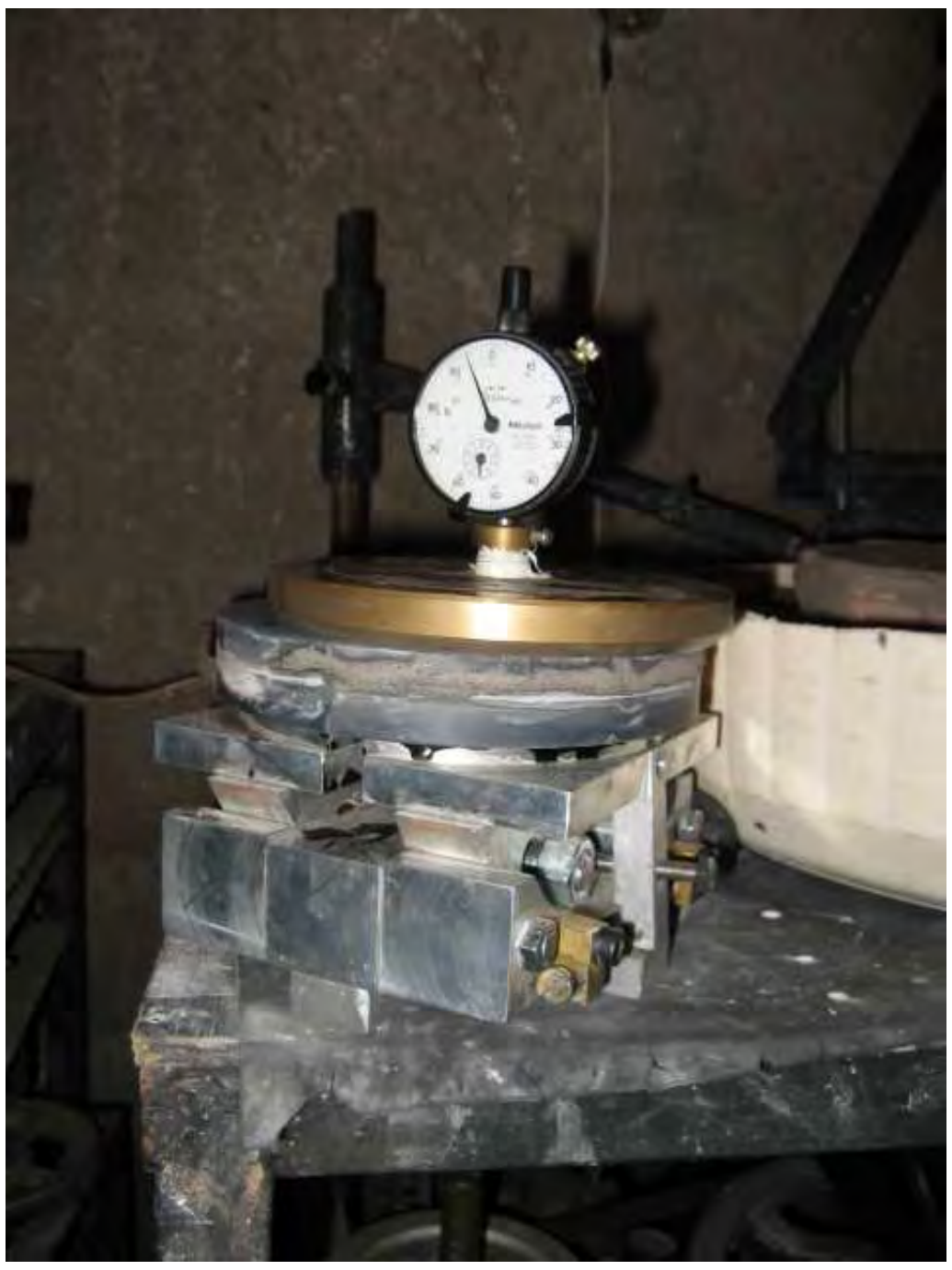

Fig. 5.29: Medição com um esferômetro após cada sessão de desbaste fino.

Na figura 5.30 vemos os hemi-espelhos posicionados sobre sua base óptica. Dois espaçadores de vidro são utilizados para preencher o espaço do material perdido no corte, restaurando o perímetro circular do espelho. Os hemi-espelhos, então fixados à base com uma fita são posicionados para uma inspeção visual do trabalho realizado, figura 5.31.

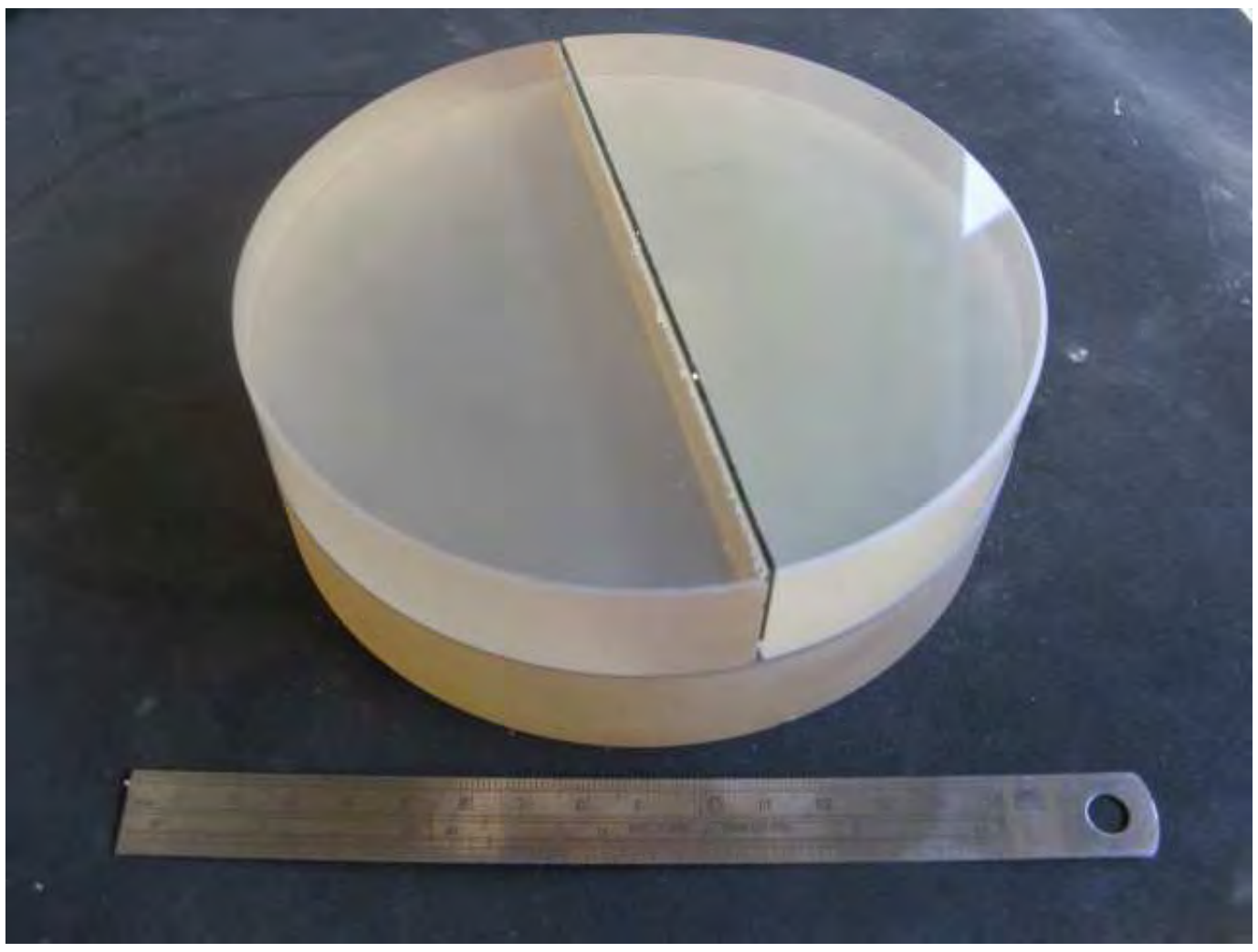

Fig. 5.30: Espelhos em diedro repousando sobre sua base óptica.

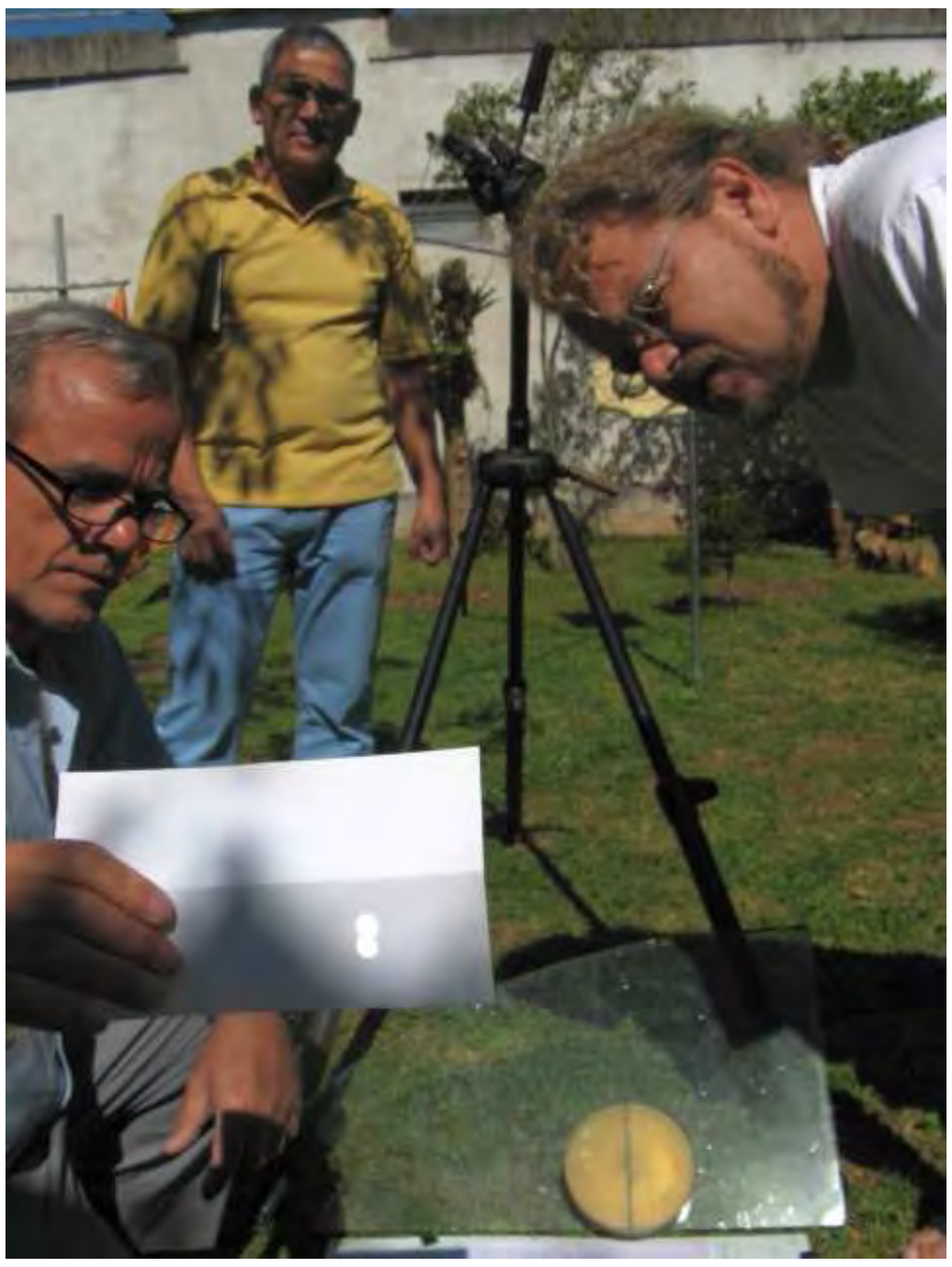

Fig. 5.31: A equipe técnica realizando os primeiros testes com o espelho heliométrico de CCZ.

O anel de contenção é confeccionado retirando-se a área central do disco com 120 mm de raio. Para isso é usada uma serra copo diamantada da mesma dimensão. Após o corte, a superfície interna do anel é desbastada para a retirada de pequenas imperfeições.

### 5.1.4.3 - Etapas finais da montagem do espelho heliométrico de CCZ.

Cada hemi-espelhos, na sua montagem final, toca a superfície interna do anel de contenção em apenas três pontos. Como o diâmetro externo do conjunto hemi-espelhos/espaçadores e o diâmetro interno do anel de contenção são iguais, a superfície externa dos hemi-espelhos foi rebaixada por desbaste, deixando apenas três regiões sem desbaste para servirem de calços laterais.

Primeiro, os hemi-espelhos são fixados entre duas peças metálicas. Uma superfície fina de borracha é usada para que este procedimento não danifique a superfície do espelho. Dois espaçadores são colocados entre os espelhos para que o perímetro do conjunto voltasse a ser circular.

A figura 5.32 traz a seqüência de fixação para o desbaste. Na 1ª imagem vemos os hemi-espelhos, separados, juntos às peças de alumínio de fixação. Na 2ª, vemos os hemi-espelhos, com os espaçadores, já posicionados sobre a peça de base. Na 3ª, as peças metálicas já aparafusadas, fixando os hemi-espelhos.

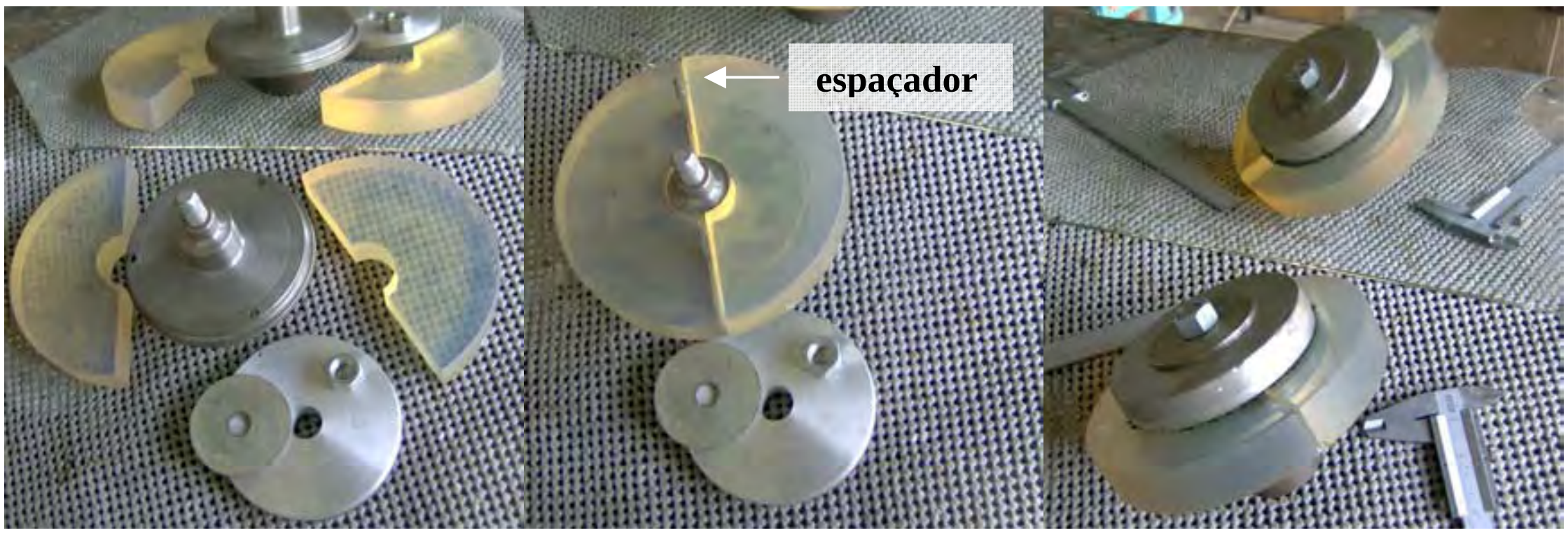

Fig. 5.32: Seqüência de fixação dos espelhos.

O conjunto é então preso a uma mesa divisora para o desbaste da lateral dos hemi-espelhos a fim diminuir seu diâmetro cerca de 1 mm (fig. 5.33). Na região lateral onde os calços devem ser criados, o rebolo é afastado da peça, preservando localmente o diâmetro original. Na figura 5.38, podem-se ver os hemi-espelhos, depois do desbaste, apoiados sobre sua base óptica. O anel de contenção é visto ao lado, figura 5.34.

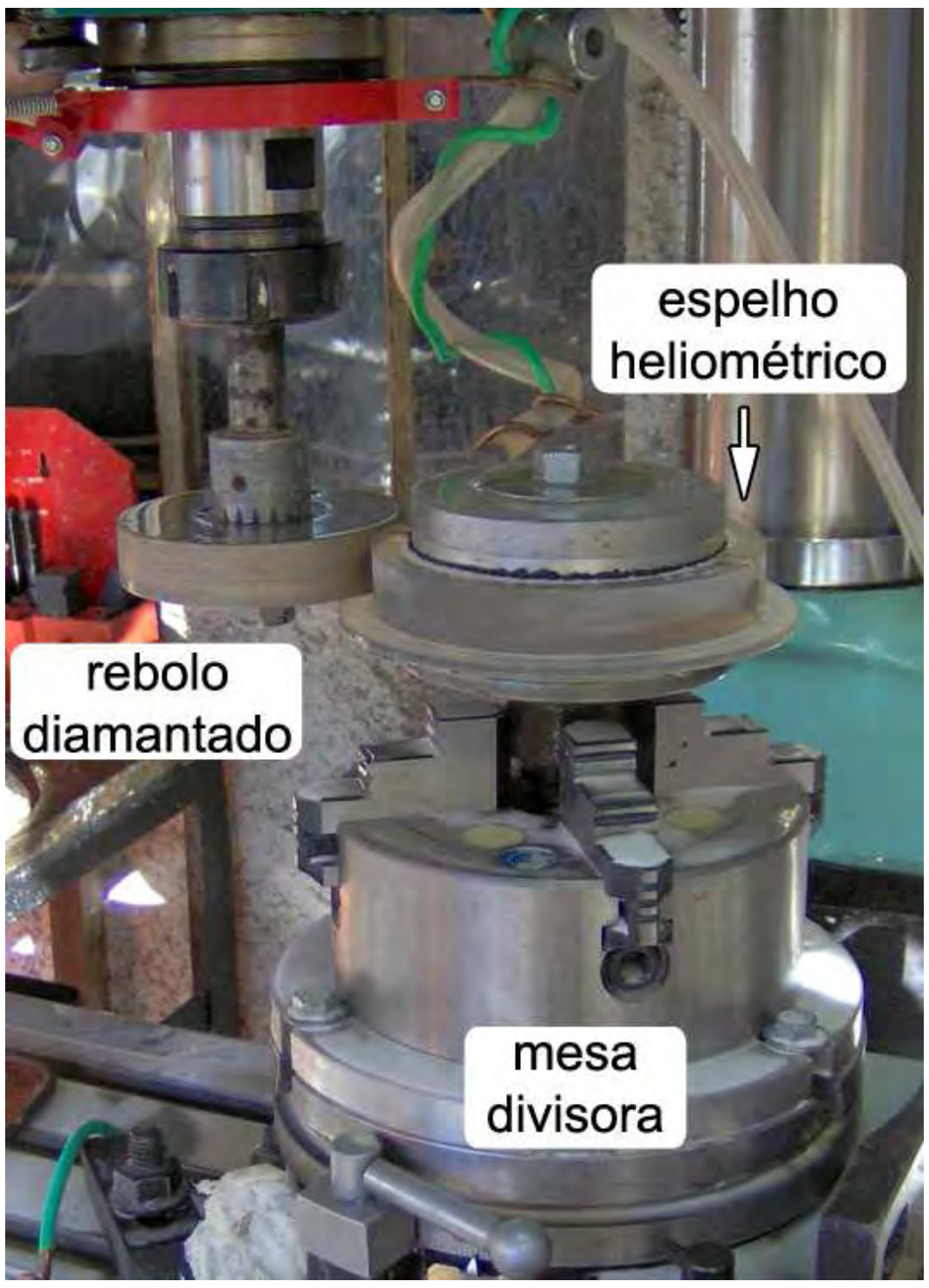

Fig. 5.33: Conjunto fixado à mesa divisora sendo desbastado.

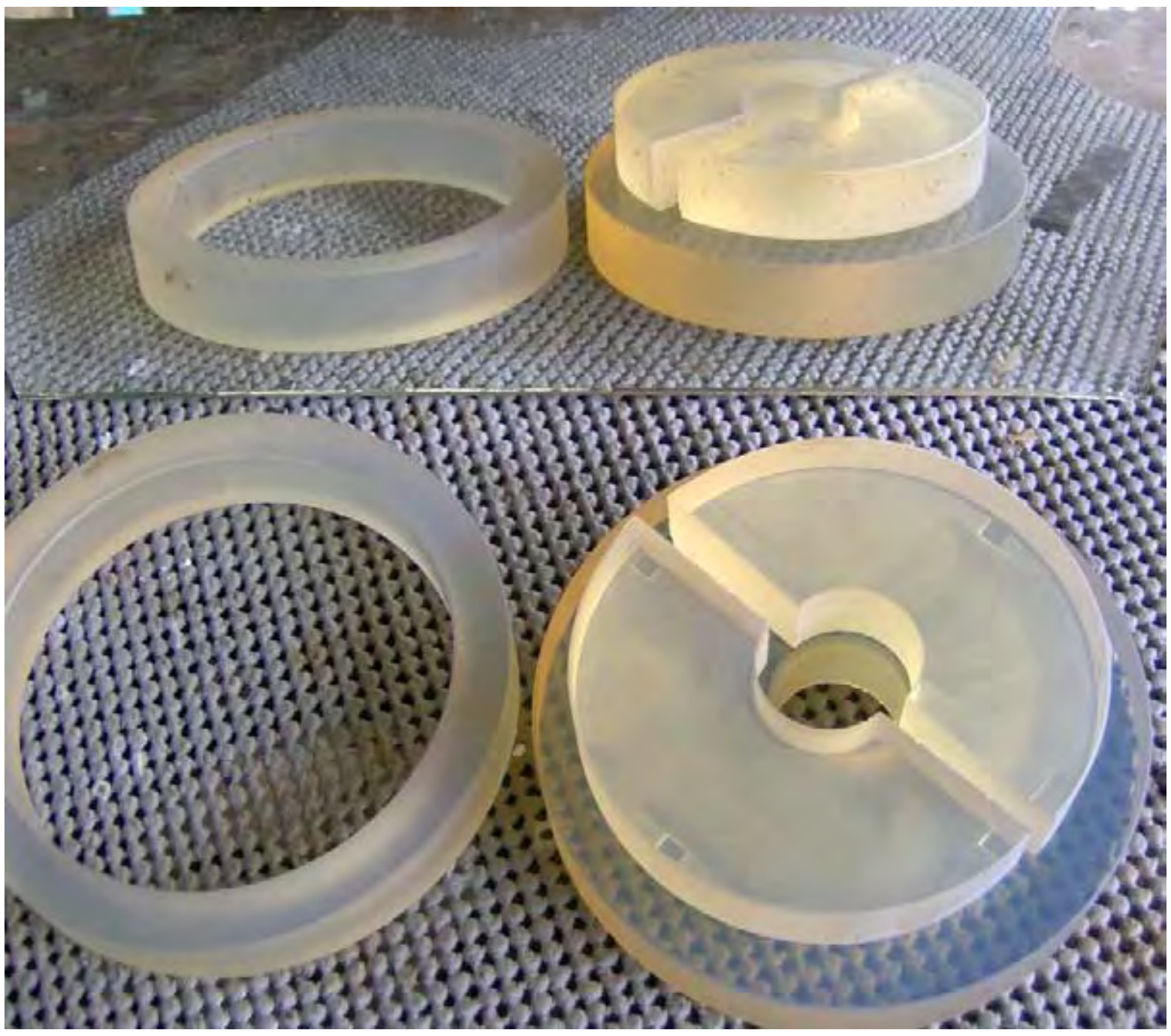

Fig. 5.34: Partes do espelho heliométrico praticamente pronto, dentro das especificações do projeto.

Um dos calços laterais pode ser visto na figura 5.35, comparado ao projeto inicial. Nas imagens seguintes é possível se ver os três calços inferiores e um calço lateral de um dos hemi-espelhos, figuras 5.36 e 5.37.

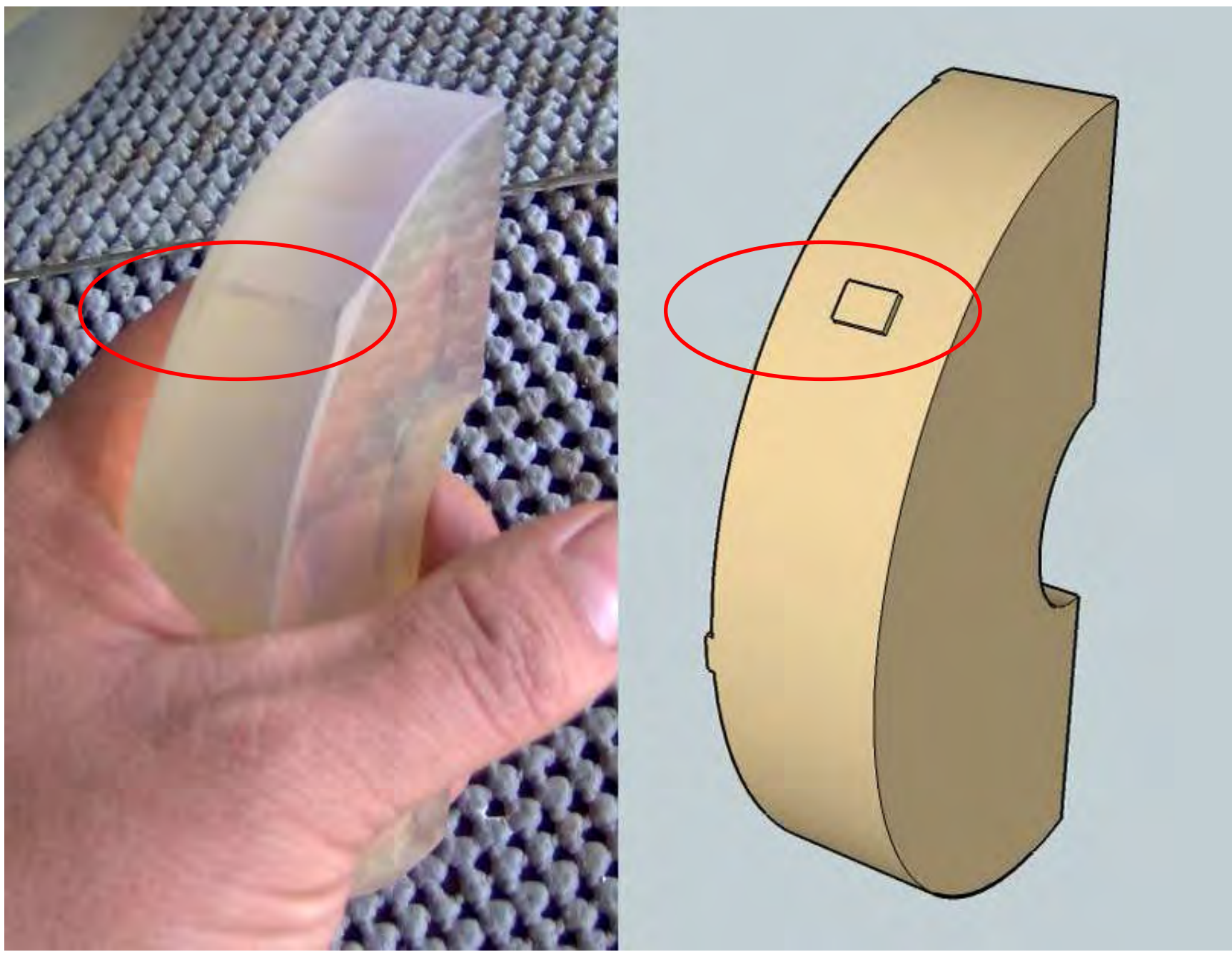

Fig. 5.35: Calço lateral pronto. À direita o desenho do projeto.

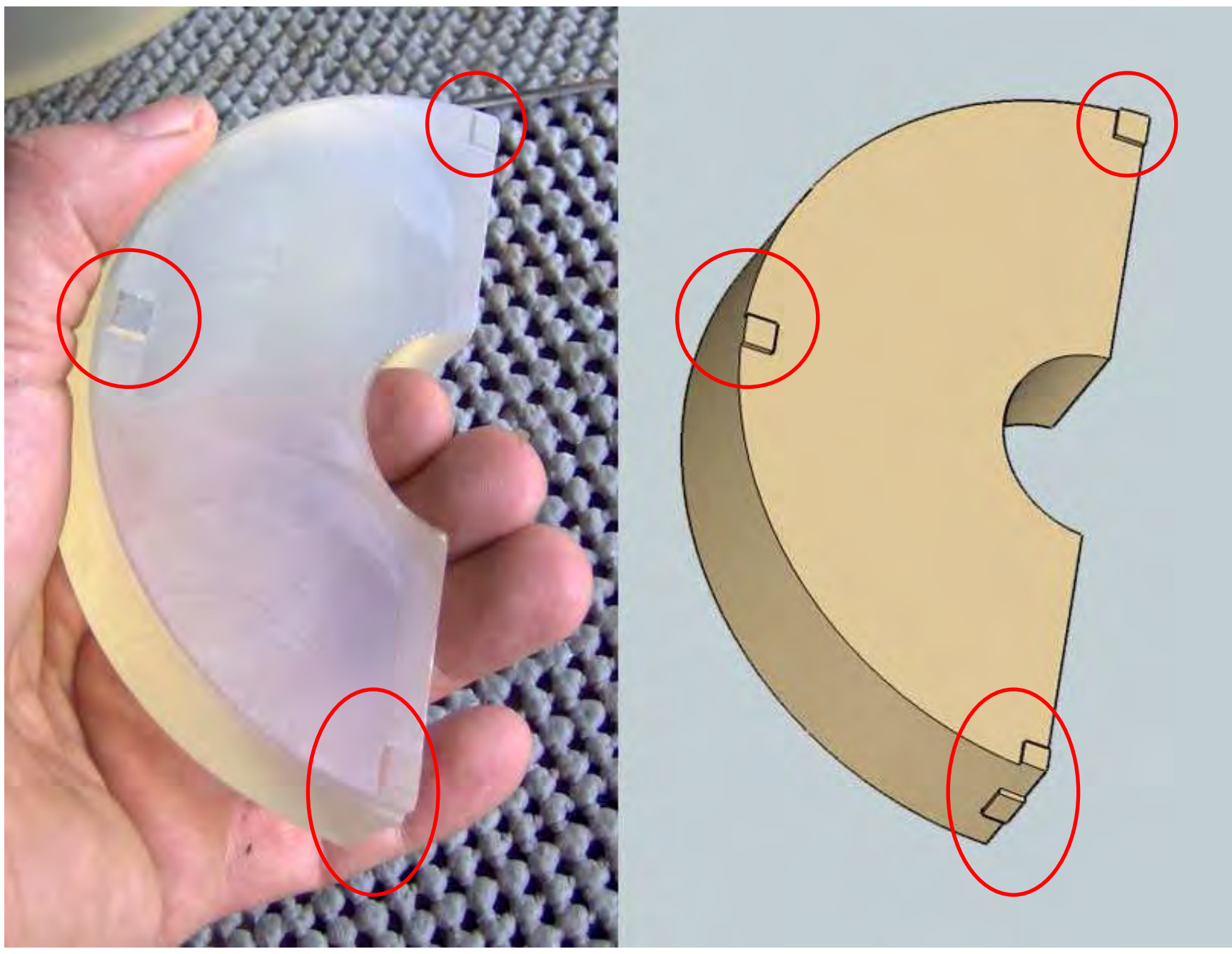

Fig. 5.36: Calços inferiores e lateral de um hemi-espelho. À direita o desenho do projeto.

O espelho heliométrico totalmente montado pode ser visto na figura 5.41, ao lado do esquema do projeto final. Os círculos vermelhos marcam as posições dos calços laterais.

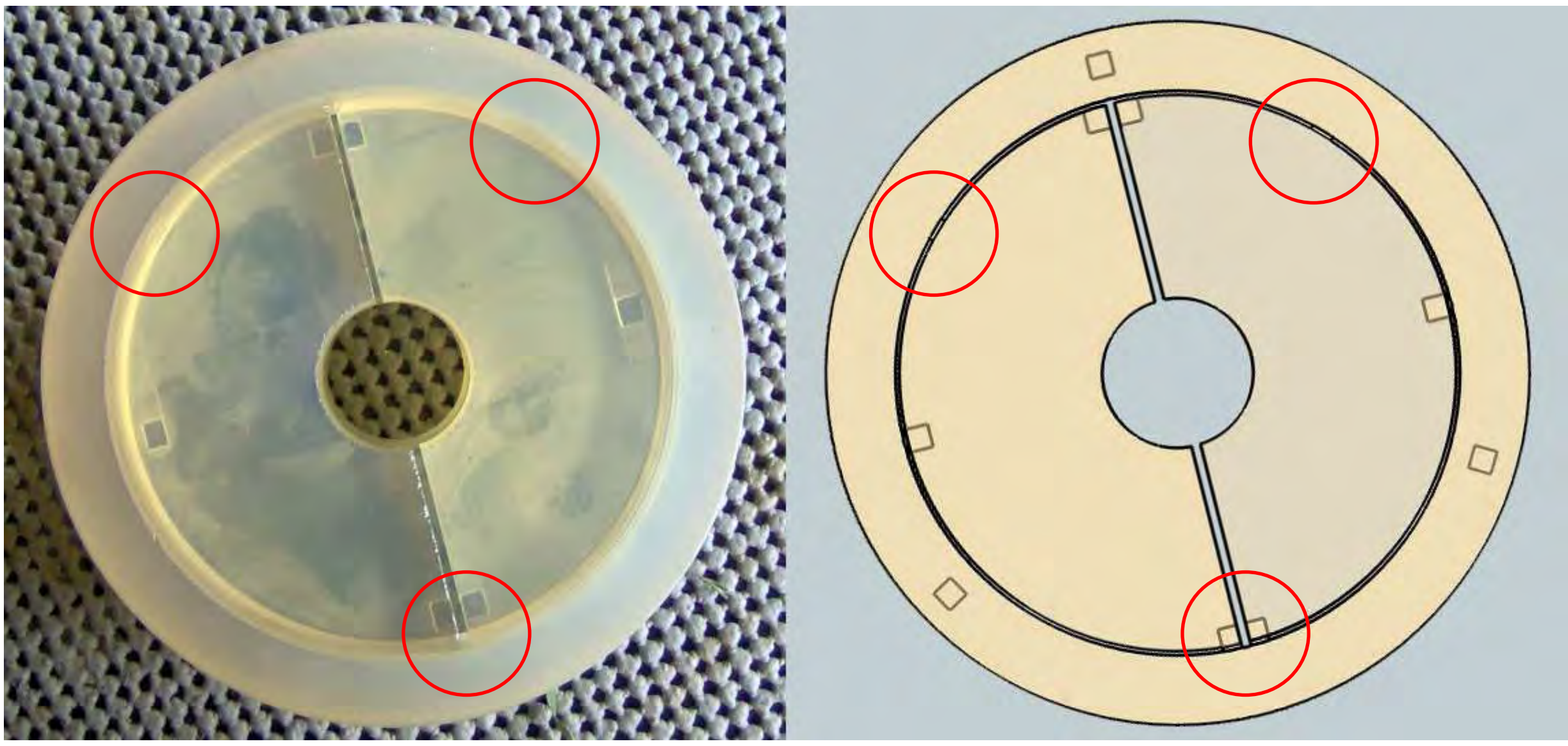

Fig. 5.37: Comparação entre o espelho heliométrico montado e seu projeto inicial.

### 5.1.4.4 - Ajuste final do diedro

Doravante chamamos de X o deslocamento das imagens que se dá paralelamente ao corte do espelho (desejado), e chamamos de Y qualquer outro deslocamento transversal (indesejado).

O objetivo é levar o diedro ao valor mais próximo possível escolhido para X e, ao mesmo tempo, anular o valor de Y. Note-se que uma pequena alteração ao valor nominal de X apenas varia de modo constante o afastamento entre as imagens. Para ajustar o diedro foi necessário desenvolver um teste com precisão da ordem do segundo de arco.

O ajuste do diedro envolve:

- montar todo o conjunto do espelho heliométrico;
- medir o diedro, de forma prática e suficientemente precisa;
- estudar o procedimento para a correção;
- retirar os espelhos de seu conjunto;
- corrigir desbastando os calços seletivamente;
- remontar o conjunto;
- medir novamente o diedro;
- repetir as operações até obter convergência, dentro da tolerância.

O conceito do procedimento é que um paquímetro posicionado no plano focal do espelho heliométrico terá duas imagens reais no infinito, criadas por cada hemi-espelho. Uma luneta, com objetiva suficientemente grande para interceptar os dois feixes de luz, recebe a imagem de dois paquímetros separados por uma distância análoga ao valor do diedro. O projeto da montagem experimental pode ser visto na figura 5.38.

A escala do instrumento é posicionada paralela ao corte do espelho, que é a direção de referência, pois assim, as partes internas dos bicos do paquímetro fiquem perpendiculares à direção do desdobramento das imagens (fig. 5.39). Com esta configuração, calculando-se a escala de placa do espelho, pode-se medir o diedro, abrindo e fechando o paquímetro até que os bicos opostos de uma e outra imagem se toquem, como se o paquímetro estivesse fechado. Se não houver deslocamentos transversais, as pontas dos bicos também estarão juntas.

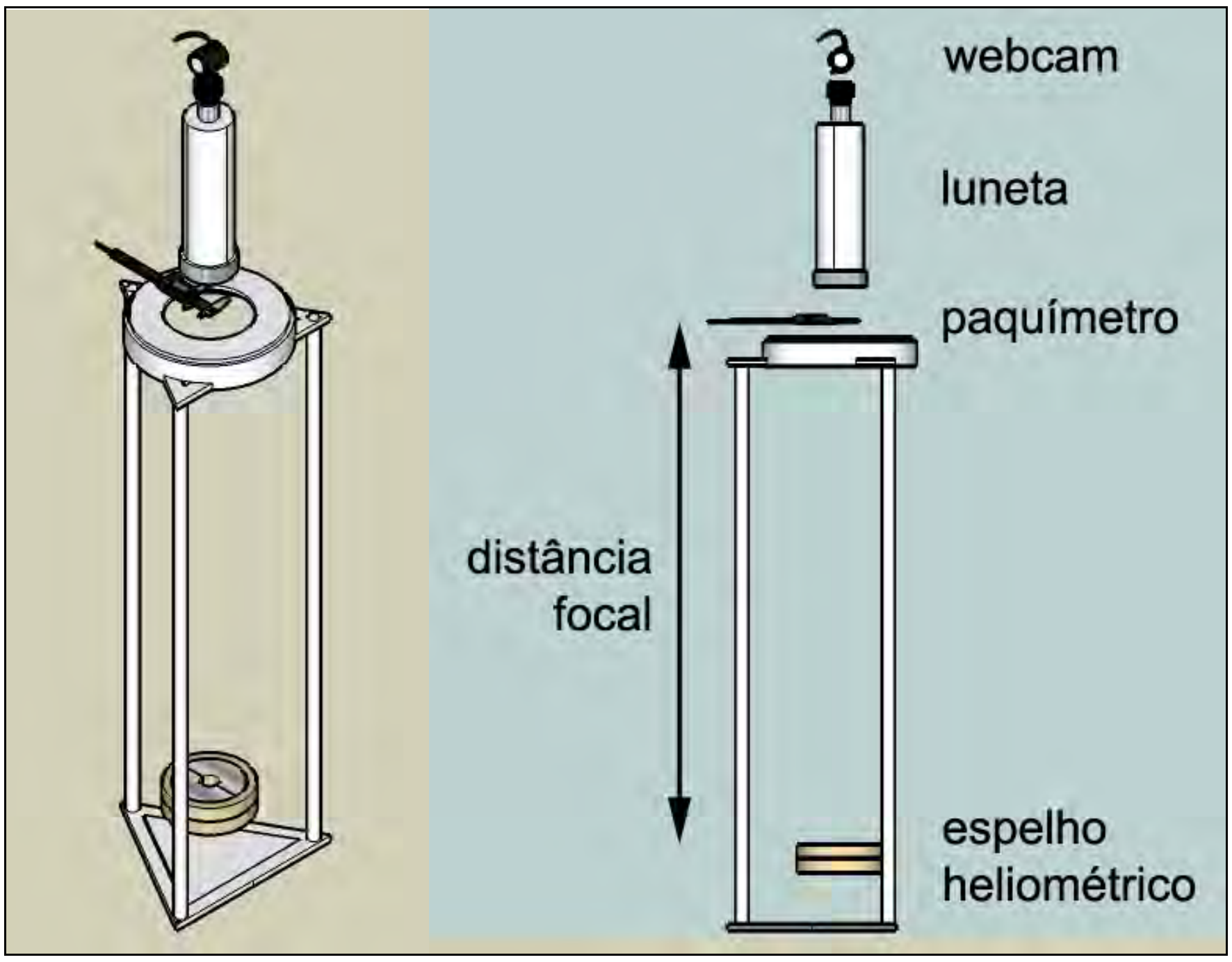

Fig. 5.38: Esquema experimental para o ajuste do diedro.

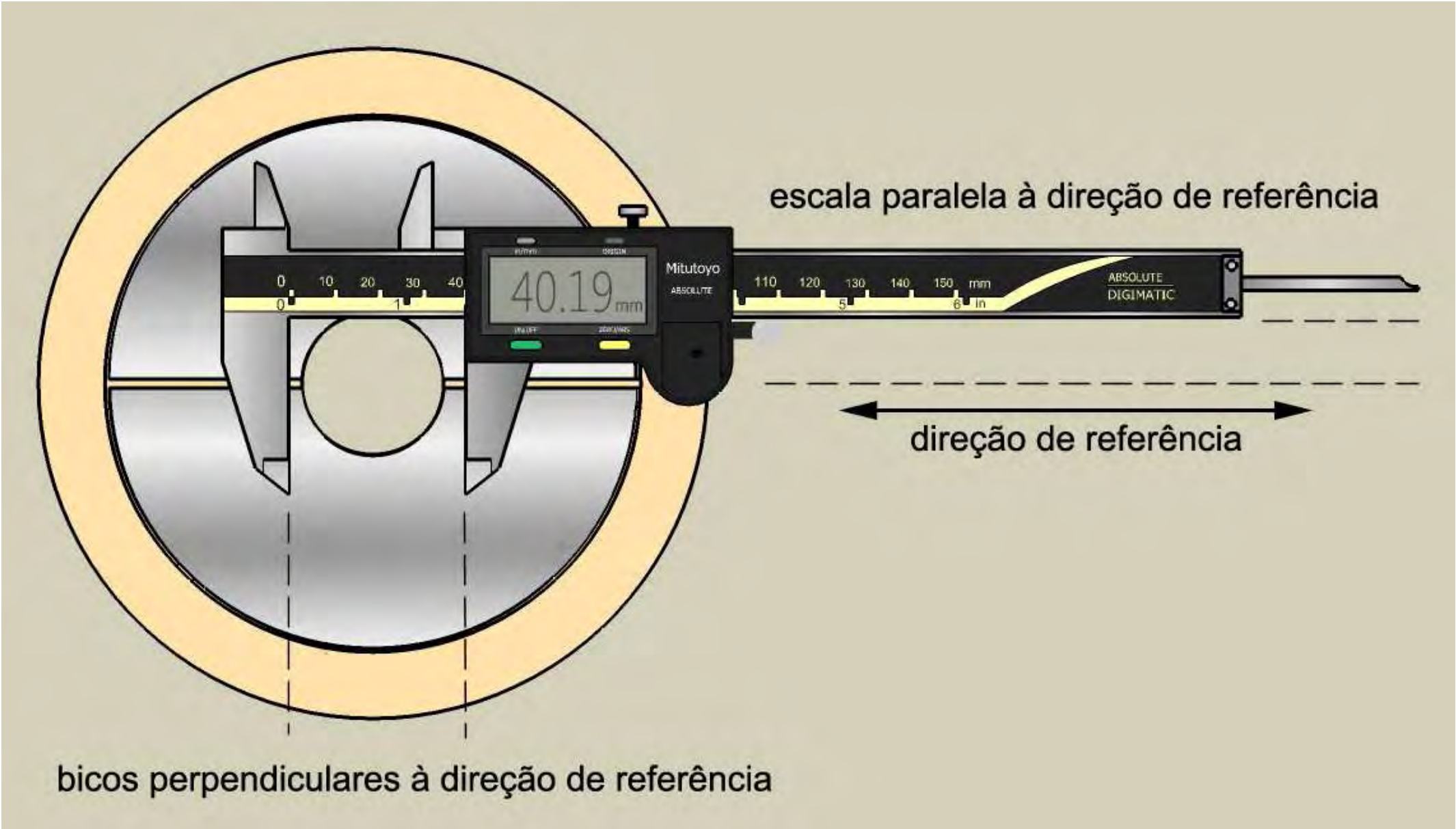

Fig. 5.39: Configuração para a medição do diedro.

A imagem da figura 5.40 exemplifica a medida de laboratório. Nela vemos a imagem heliométrica real do paquímetro, vista através da luneta e capturada por uma webcam. O deslocamento paralelo ao corte (X) pode ser medido diretamente pelo visor digital do paquímetro, o deslocamento transversal (Y), que deve ser minimizado, pode ser estimado por comparação.

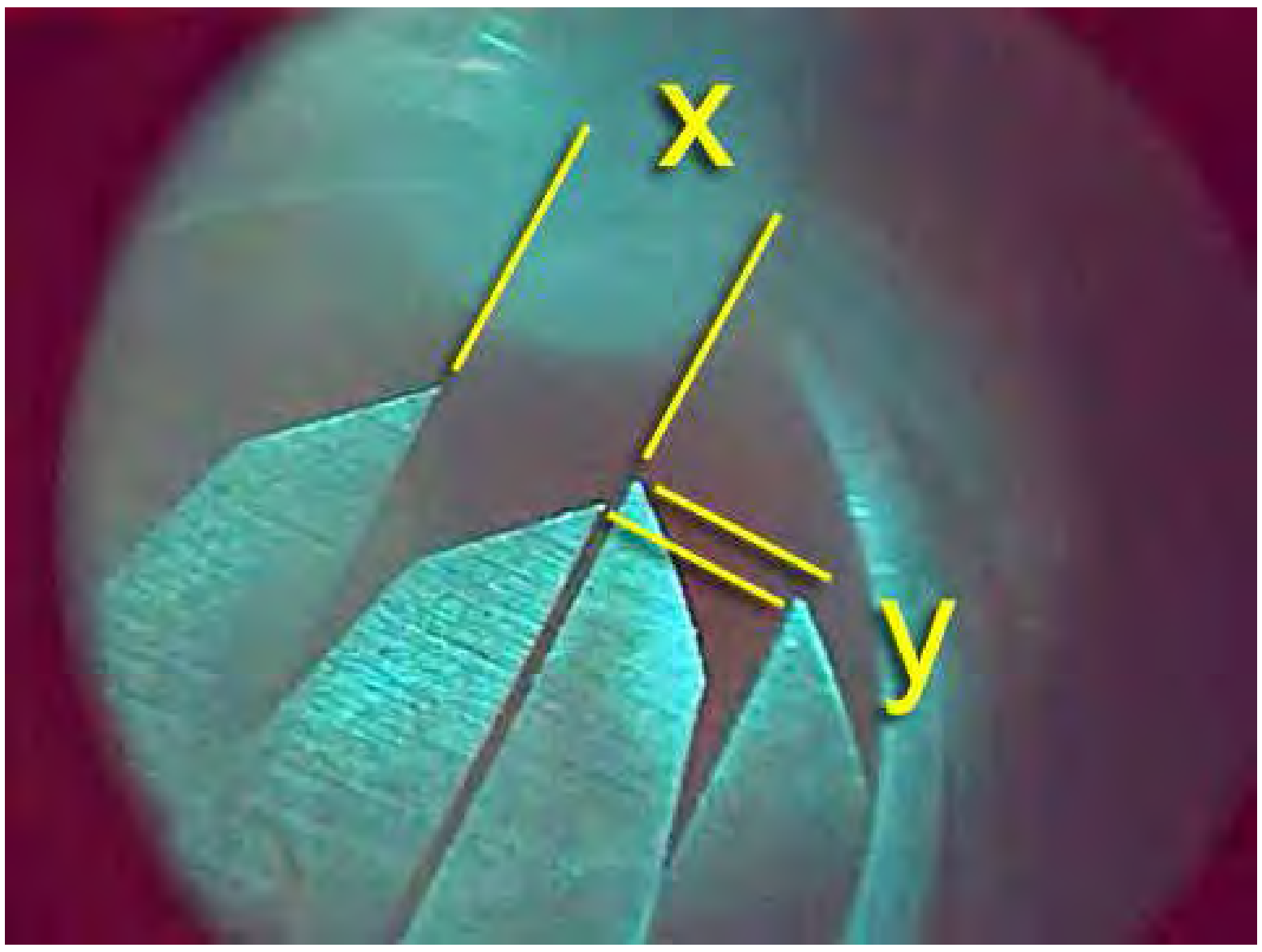

Fig. 5.40: Imagem heliométrica do paquímetro. X é a medida do diedro, na direção desejada e Y é a medida do diedro na direção oposta.

O cálculo do ângulo do diedro X é efetuado levando em conta que o espelho heliométrico possui uma distância focal de 986 mm. A esta distância, 1 mm no paquímetro é o equivalente a:

$$l = r\theta \Leftrightarrow \frac{\theta}{l} = \frac{1}{r} = \frac{1}{986mm}\left(\frac{180^o}{\pi}\right)\left(\frac{3600''}{1^o}\right) \cong 209{,}2''/mm$$

Um valor de X entre 9,6 mm e 9,7 mm é aceitável, pois significará que as imagens heliométricas estarão angularmente separadas entre ~2008,3" e ~2029,2", respectivamente. Esta faixa é suficiente para a separação completa dos discos solares, levando em consideração seu diâmetro aparente máximo no periélio com uma separação mínima, nesta situação, entre 28" e 50".

<u>Montagem instrumental</u>

Uma luneta, de aumento 8×, foi posicionada para observar as imagens heliométricas do paquímetro e uma câmera digital, tipo webcam, foi utilizada para capturar estas imagens, figuras 5.41 e 5.42.

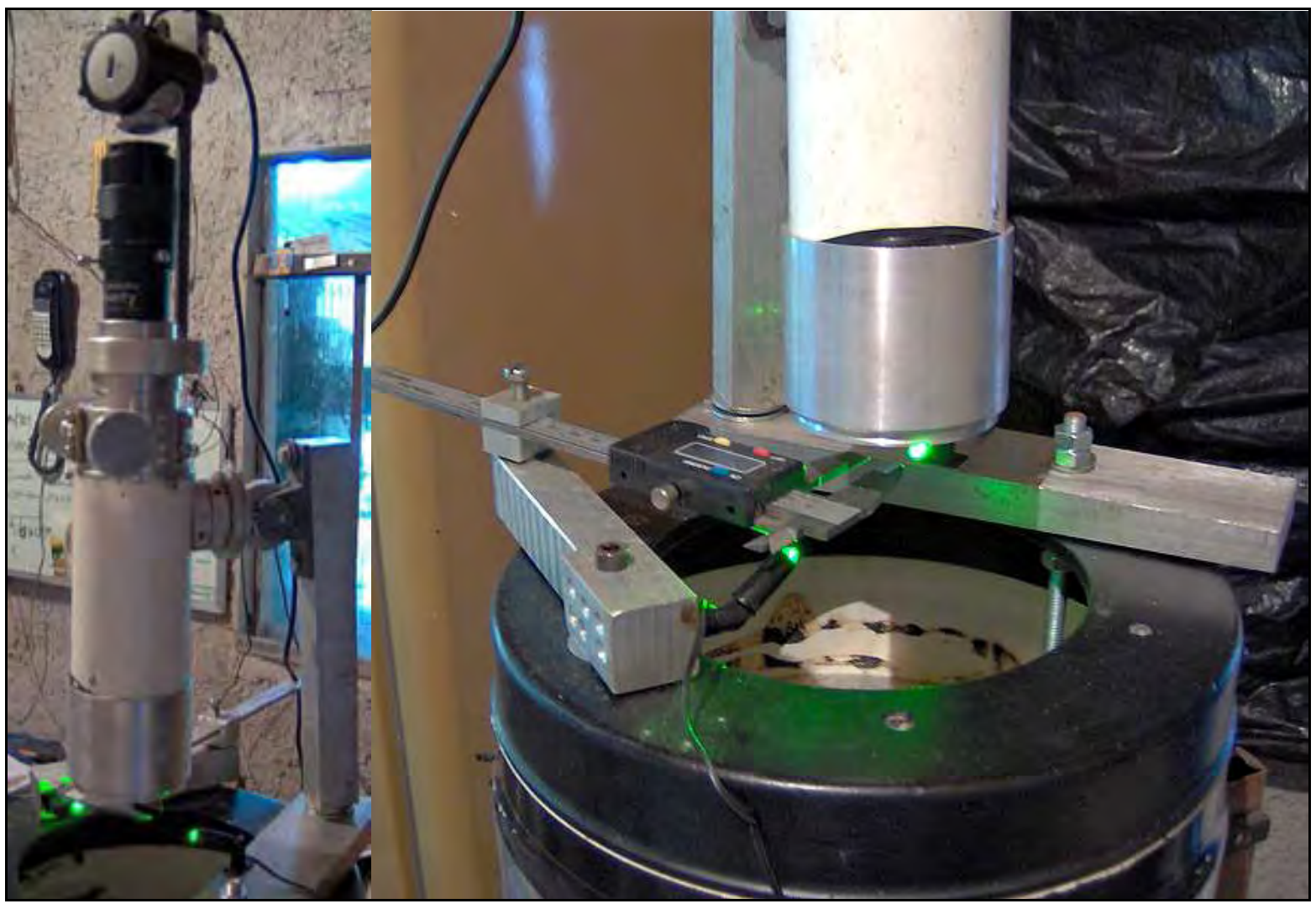

Fig. 5.41: Detalhes da montagem instrumental, mostrando o paquímetro, os leds de iluminação, a luneta e a webcam.

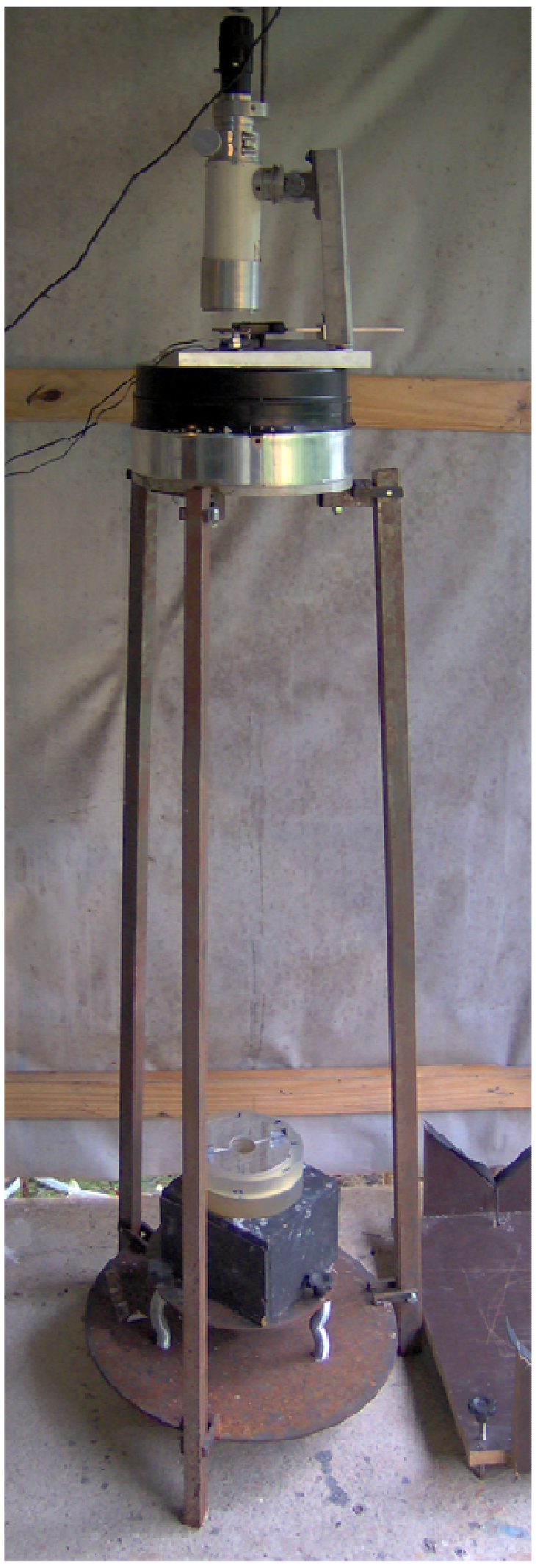

Fig. 5.42: Instalação instrumental completamente montada. Parafusos na base de apoio do espelho podiam ser ajustados para deixar o espelho na horizontal.

A imagem da figura 5.42 mostra a instalação experimental completamente montada para o ajuste de X. Sobre a base inferior do suporte de testes encontra-se o espelho do Heliômetro (hemi-espelhos, anel e base). No foco (986 mm acima do espelho) foi posicionado o paquímetro (iluminado artificialmente com LED´s de cor verde) e acima do paquímetro, em posição muito próxima da normal dos hemi-espelhos, a luneta e a câmera digital.

Como esperado, a primeira imagem formada pelos hemi-espelhos mostrou, de modo bastante claro, a necessidade de ajustes, tanto em X como em Y (fig. 5.40). Seguiu-se então uma primeira intervenção para corrigir o erro em X.

A identificação do tipo de desvio Y foi feita por ensaio. A colocação de um calço de espessura conhecida (papel alumínio com 0,02 mm) em um dos calços causava a alteração em Y de um dos hemi-espelhos e, conseqüentemente, em um dos bicos da imagem do paquímetro.

Na tarefa de correção dos diedros foram utilizadas uma placa de aço plana (mecanicamente plana, com erros inferiores a 0,001 mm sobre um diâmetro de 150 mm), uma mesa giratória (para variar as posições de contato dos calços inferiores do hemi-espelho a ser corrigido sobre seu plano óptico) e abrasivo (óxido de alumínio, grão 2000). Para tanto calculou-se a altura em que determinado pé do hemi-espelho deveria ser desgastado. A tarefa é particularmente delicada, pois a ordem de grandeza das correções era de centésimos de milímetro e qualquer ataque abrasivo muito localizado certamente acarreta desgaste desigual ou exagerado.

O desbaste dos calços inferiores de cada hemi-espelho foi feito por atrito simultâneo dos 3 pés sobre a base de aço encharcada com lama abrasiva, dando maior pressão sobre aquele calço que precisava ser mais desbastado. Cada sessão de ataque durava cerca de 20 segundos e após cada sessão, o espelho era completamente limpo e seco, e o conjunto todo remontado para nova análise visual. Foram necessárias 3 sessões, para cada direção, para se chegar ao ajuste final.

A imagem da figura 5.43 mostra o paquímetro como se estivesse fechado normalmente, indicando que o diedro está na configuração desejada: ~2010" (~9,6 mm, no paquímetro) e deslocamento transversal (Y) é desprezível.

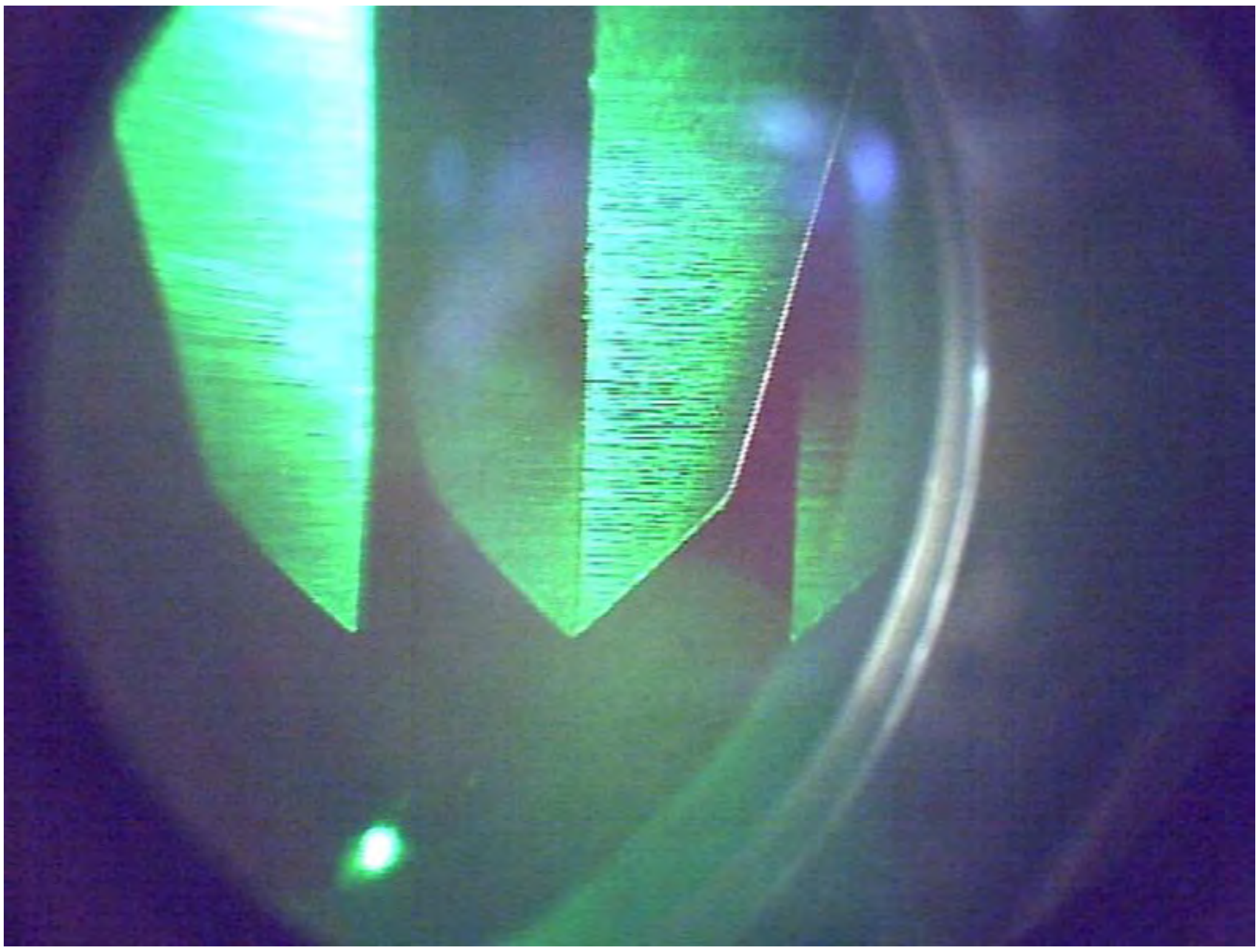

Fig. 5.43: Imagem heliométrica do paquímetro após o ajuste do diedro.

Concluídos os ajustes, letras e números de identificação foram gravados com ferramenta diamantada nas peças do espelho, de modo a marcar suas posições.

Para aferir se esta configuração final era mantida, o espelho foi repetidamente desmontado e remontado e examinado com o Teste do Paquímetro, confirmando a estabilidade do conjunto.

## 5.1.5 - Projeto da célula do espelho heliométrico de CCZ.

A célula suporte do espelho heliométrico de CCZ-HS precisa assegurar sua configuração final mecanicamente estável, figura 5.44.

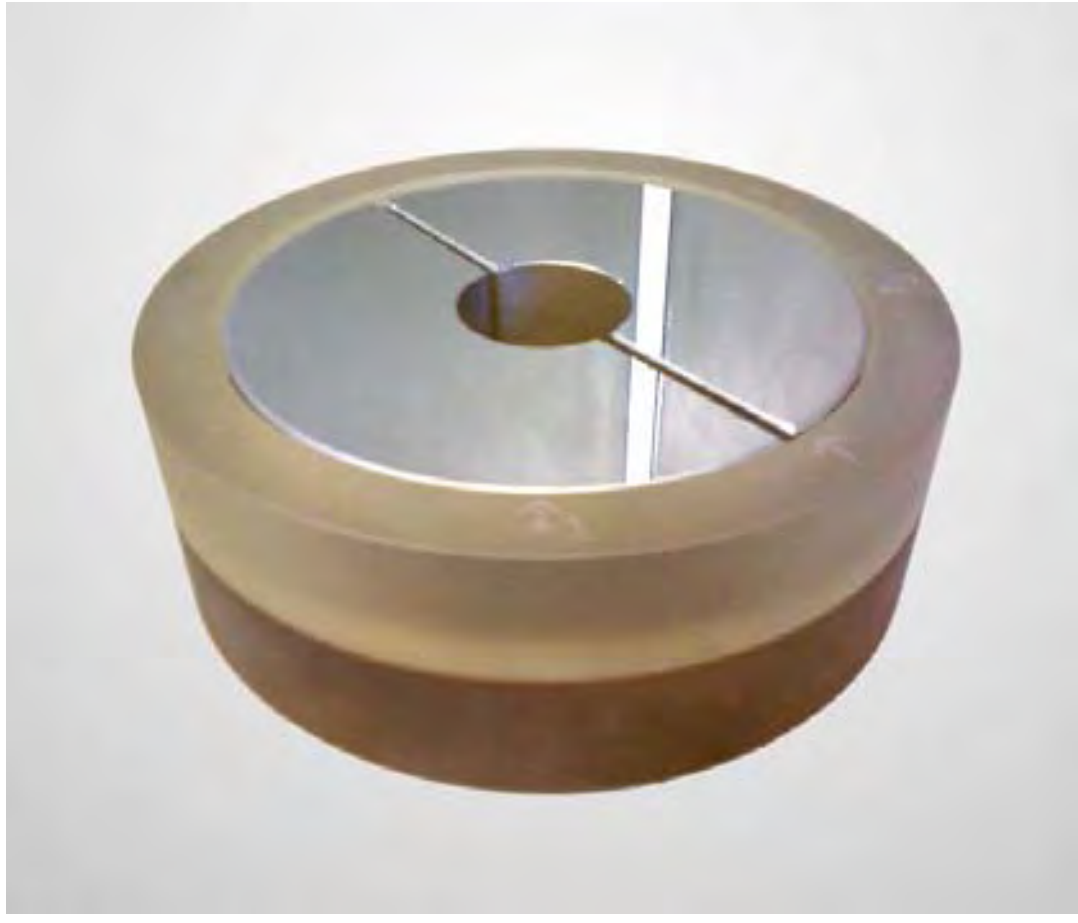

Fig. 5.44: Espelho heliométrico pronto.

Finalmente as seguintes condições são obtidas:

1. O plano óptico (base) precisa repousar sobre três calços, dispostos a 120º um do outro e ser contido lateralmente. Dois destes calços laterais são fixos e o terceiro, móvel, força a base contra os demais através de um efeito mola, produzido por uma placa de metal flexível, figuras 5.45 e 5.46.

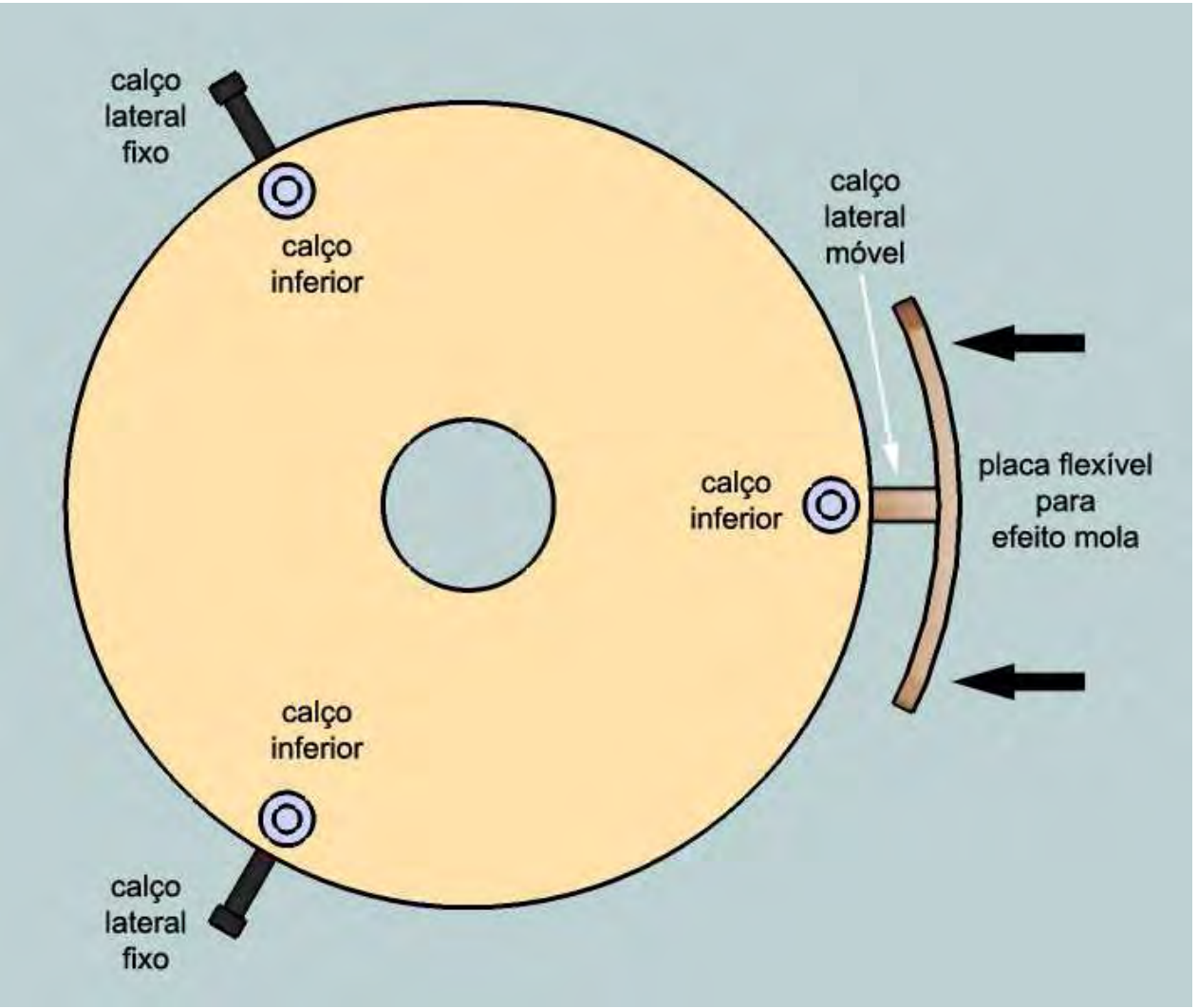

Fig. 5.45: Vista de baixo do projeto de apoio inferior da base e contenção lateral.

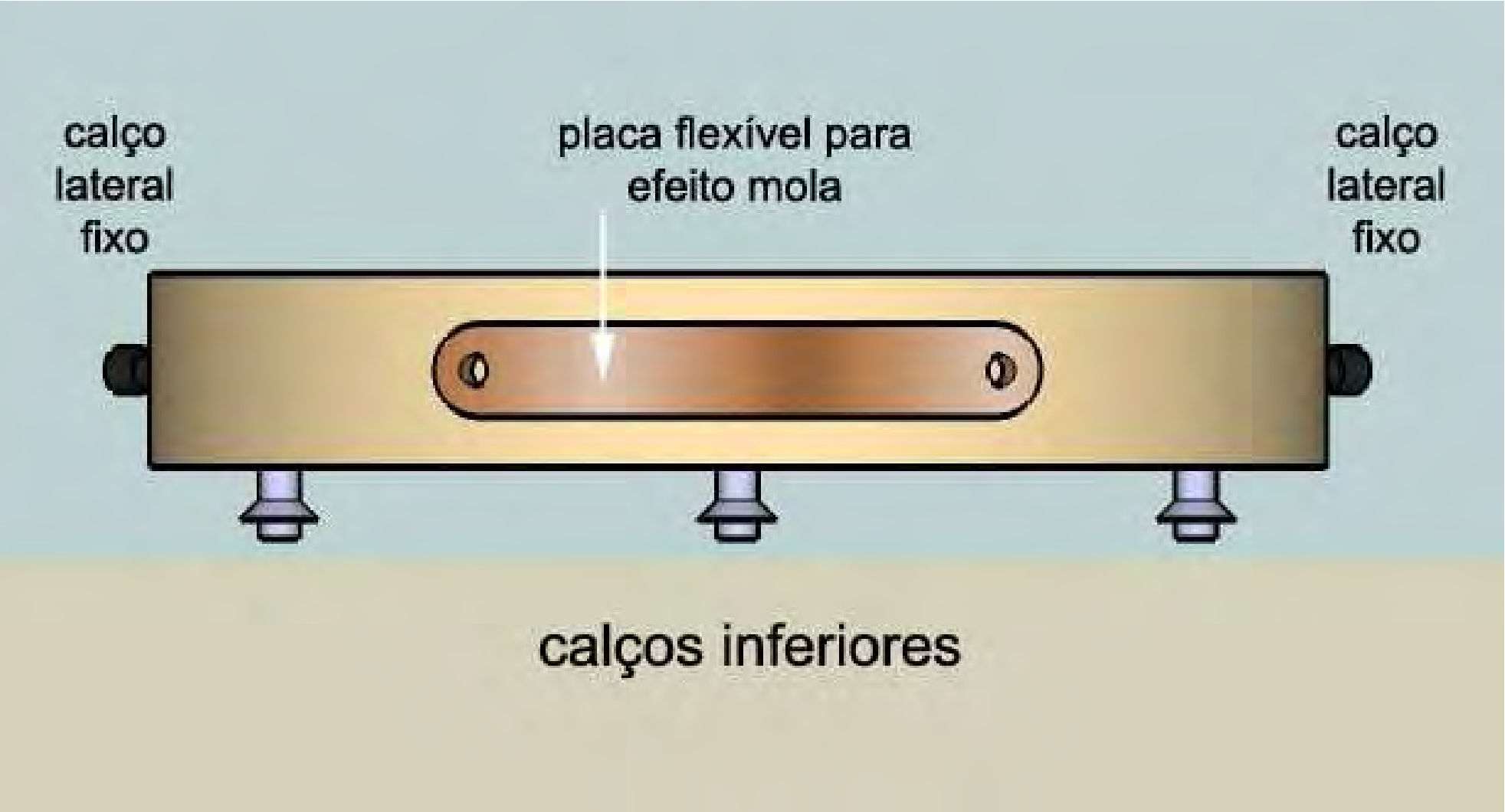

Fig. 5.46: Vista de frente do projeto.

2. O anel de contenção repousa sobre o plano óptico apoiado nos seus três calços esculpidos e é contido lateralmente pelo mesmo método utilizado na base, figura 5.47.

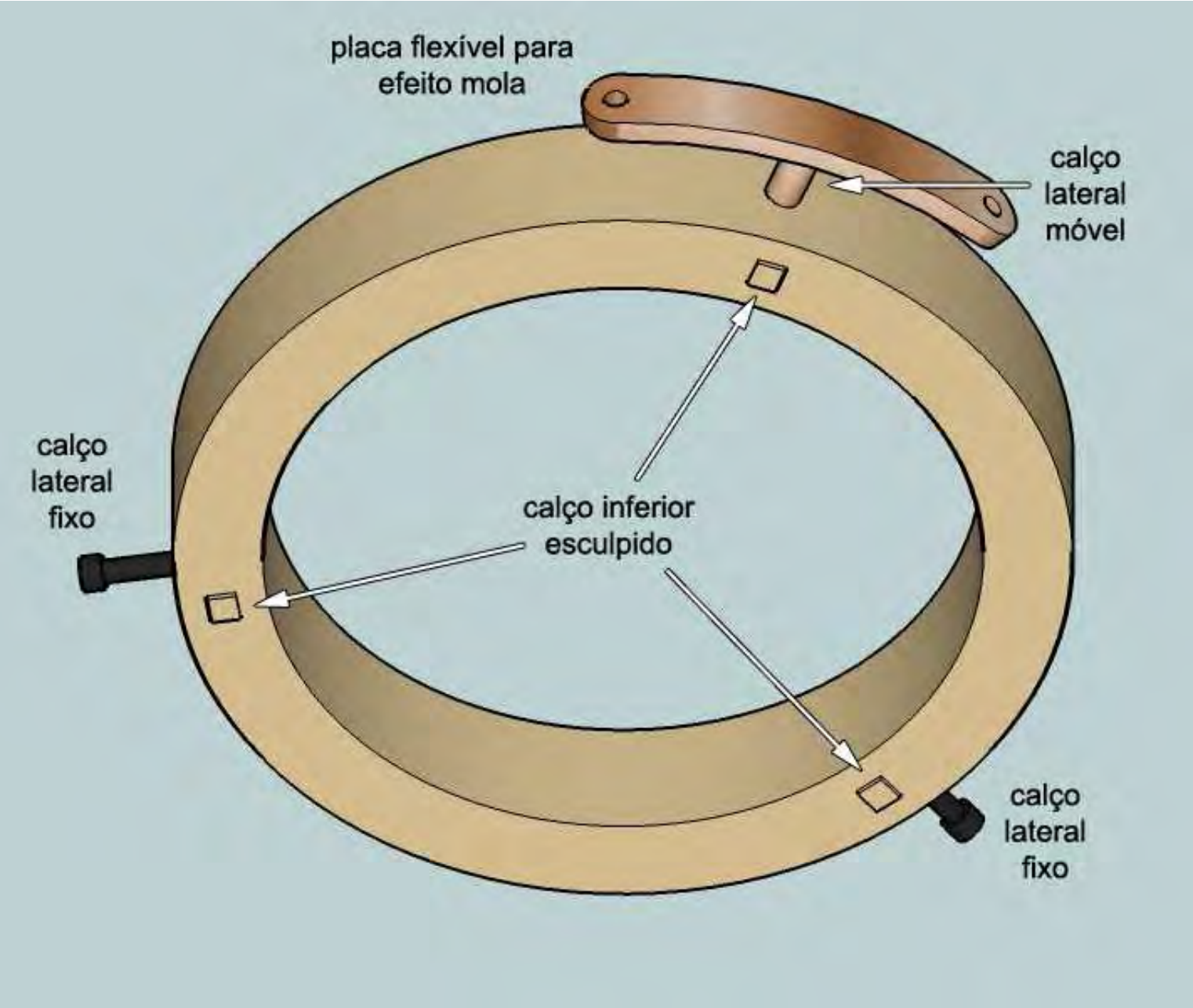

Fig. 5.47: Vista inferior do projeto para contenção do anel

3. Os calços esculpidos do anel ficam posicionados exatamente sobre os calços da base, figura 5.48.

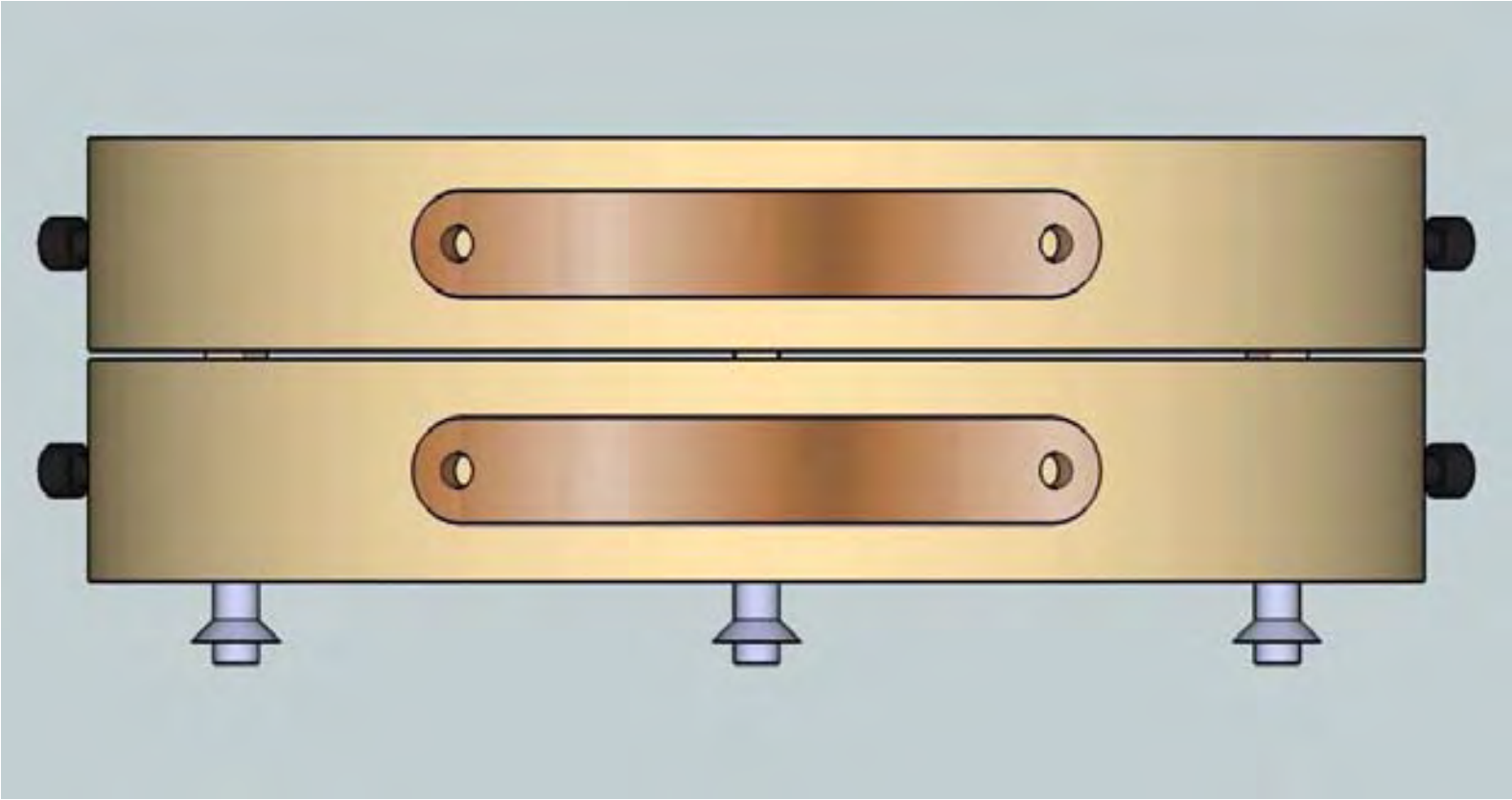

Fig. 5.48: Vista do projeto dos calços do anel e da base.

4. Os hemi-espelhos repousam sobre o plano óptico apoiados sobre seus três calços esculpidos. Cada hemi-espelho possui três calços laterais desbastados na própria peça. Estes calços são forçados contra a parede interna do anel por uma peça metálica com efeito mola localizada no espaço entre os hemi-espelhos. Simetricamente localizada e servindo como espaçador de apoio, fica a esfera cerâmica encaixada em um dos hemi-espelhos, numa cavidade também esculpida na peça (vide fig. 5.6 e 5.8).
5. Grampos metálicos servem de mola para forçar os hemi-espelhos contra o plano óptico, independentemente do anel de contenção. Seguindo a metodologia desenvolvida no item 4.2.1, estes pontos de força ficam na mesma vertical dos calços inferiores dos hemi-espelhos, figura 5.49.

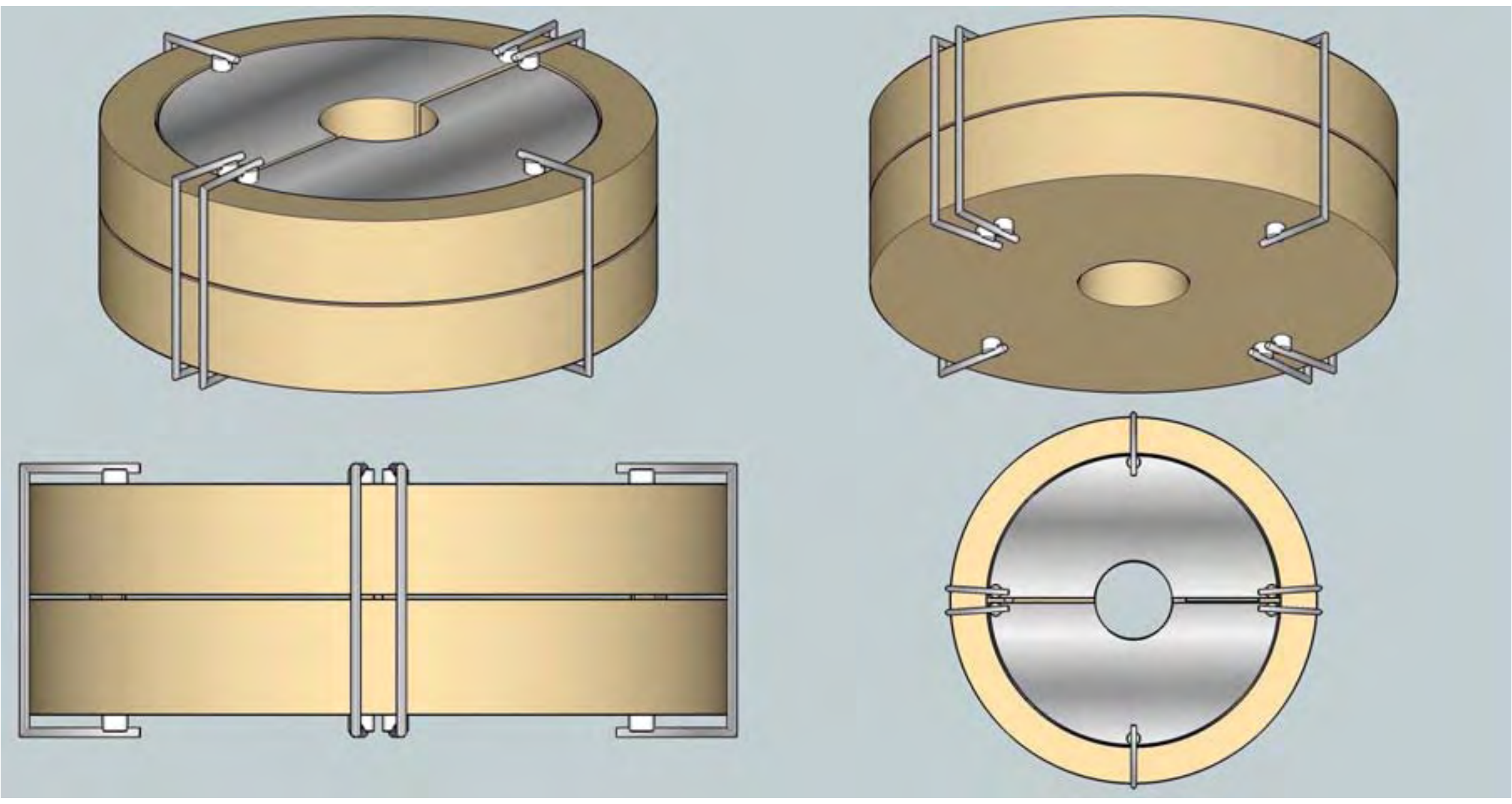

Fig. 5.49: Vistas do projeto para fixação dos hemi-espelhos ao plano óptico.

A célula que abriga o espelho heliométrico é fabricada em alumínio, com acabamento superficial em preto fosco, usinada em torno CNC (Comando Numérico Computadorizado) conforme as especificações técnicas encontradas no Apêndice VI, figura 5.50.

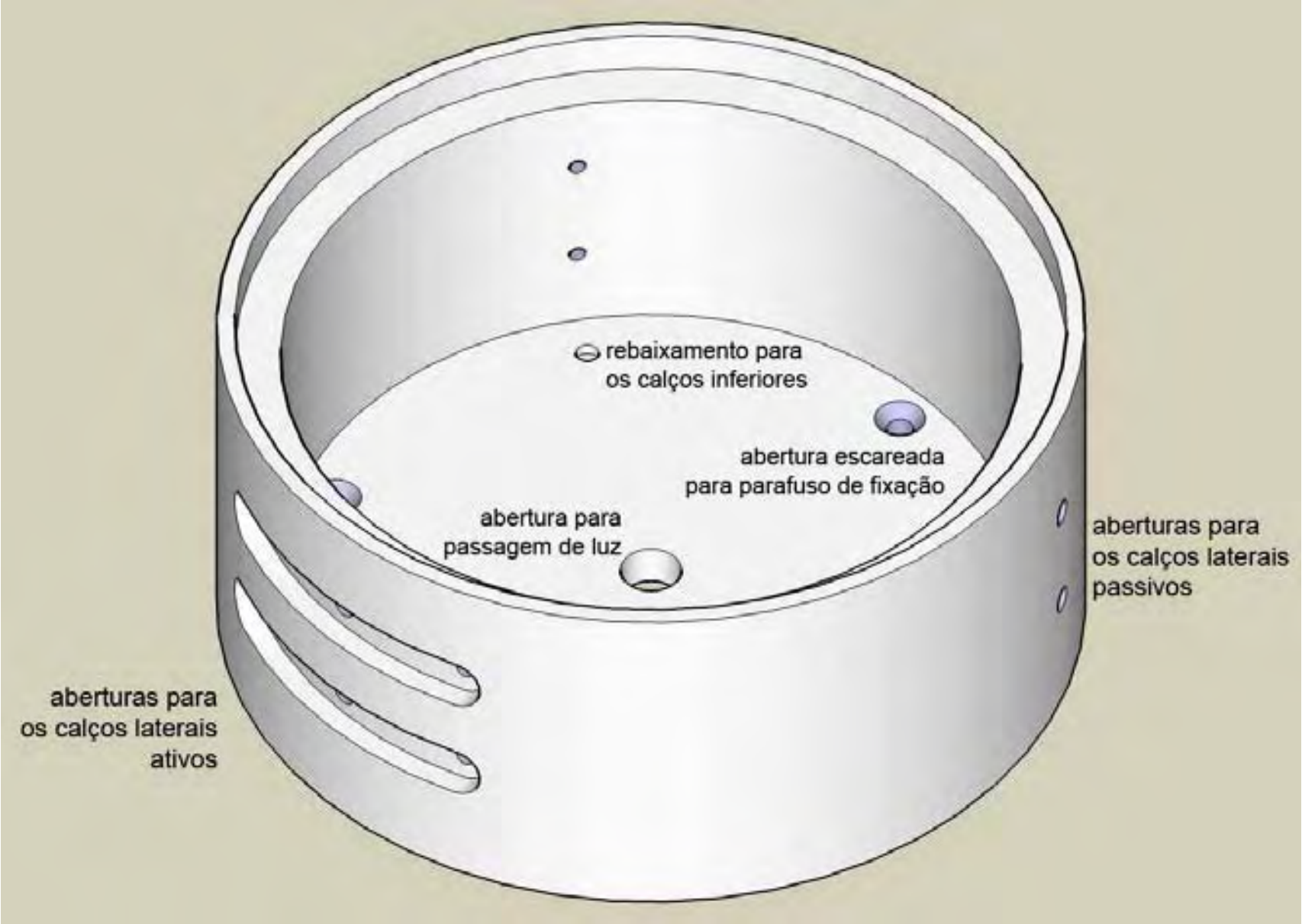

Fig. 5.50: Projeto 3D da célula do espelho heliométrico.

Um furo central serve para a passagem de luz para os testes de colimação e comparação, figura 5.51.

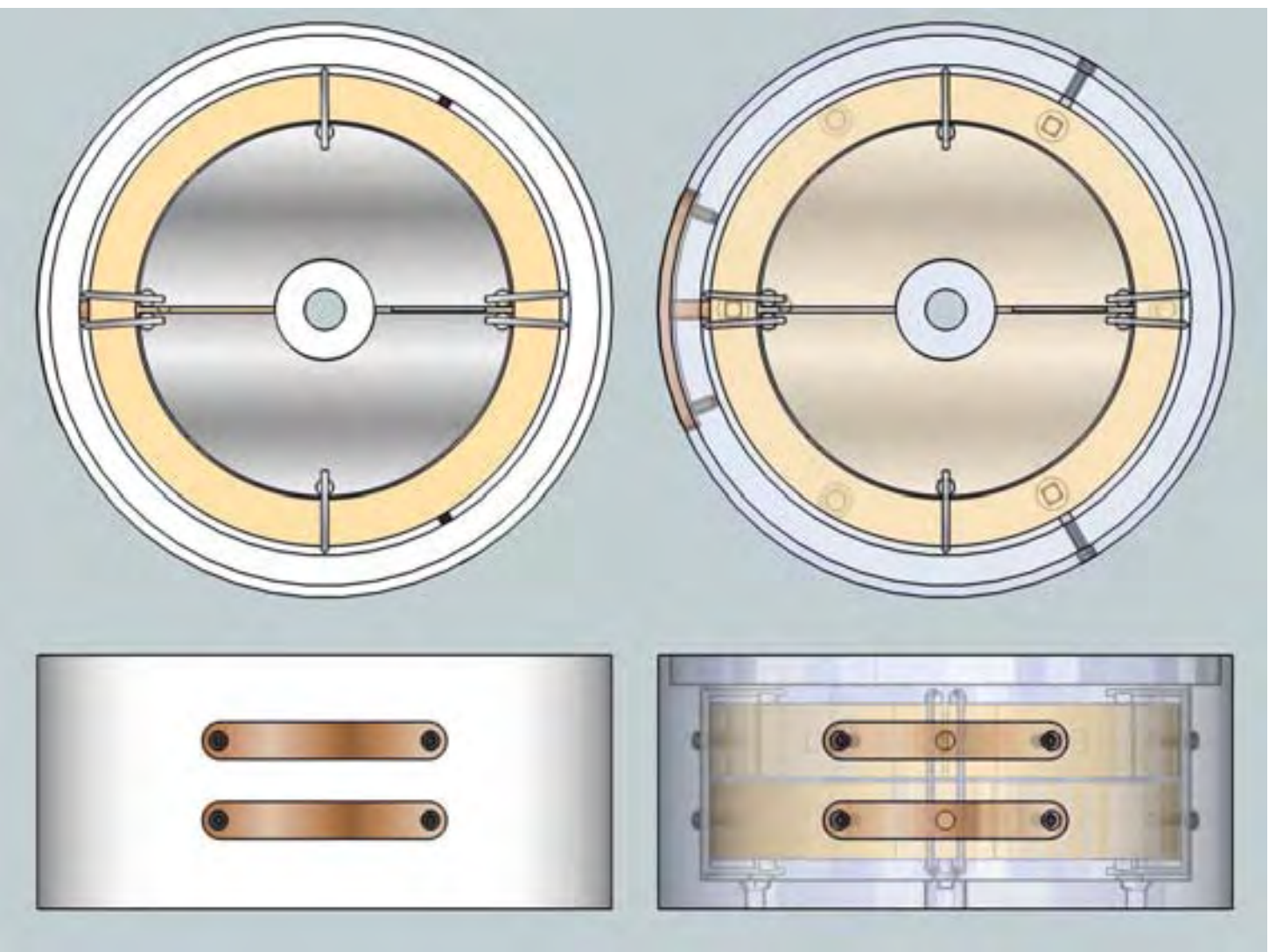

Fig. 5. 51: Vistas do projeto do espelho heliométrico dentro de sua célula.

## 5.1.6 – A Tampa forte e a pupila da célula suporte do espelho heliométrico

A tampa do suporte do espelho heliométrico se destina à fixação mecânica do conjunto (fig. 5.52) e ao mesmo tempo serve de suporte para a pupila dos hemi-espelhos. Adicionalmente esta tampa serve para a célula do espelho do teste de comparação. Esta peça precisa ter a espessura mínima de 6 mm para que parafusos possam ser atarraxados nela.

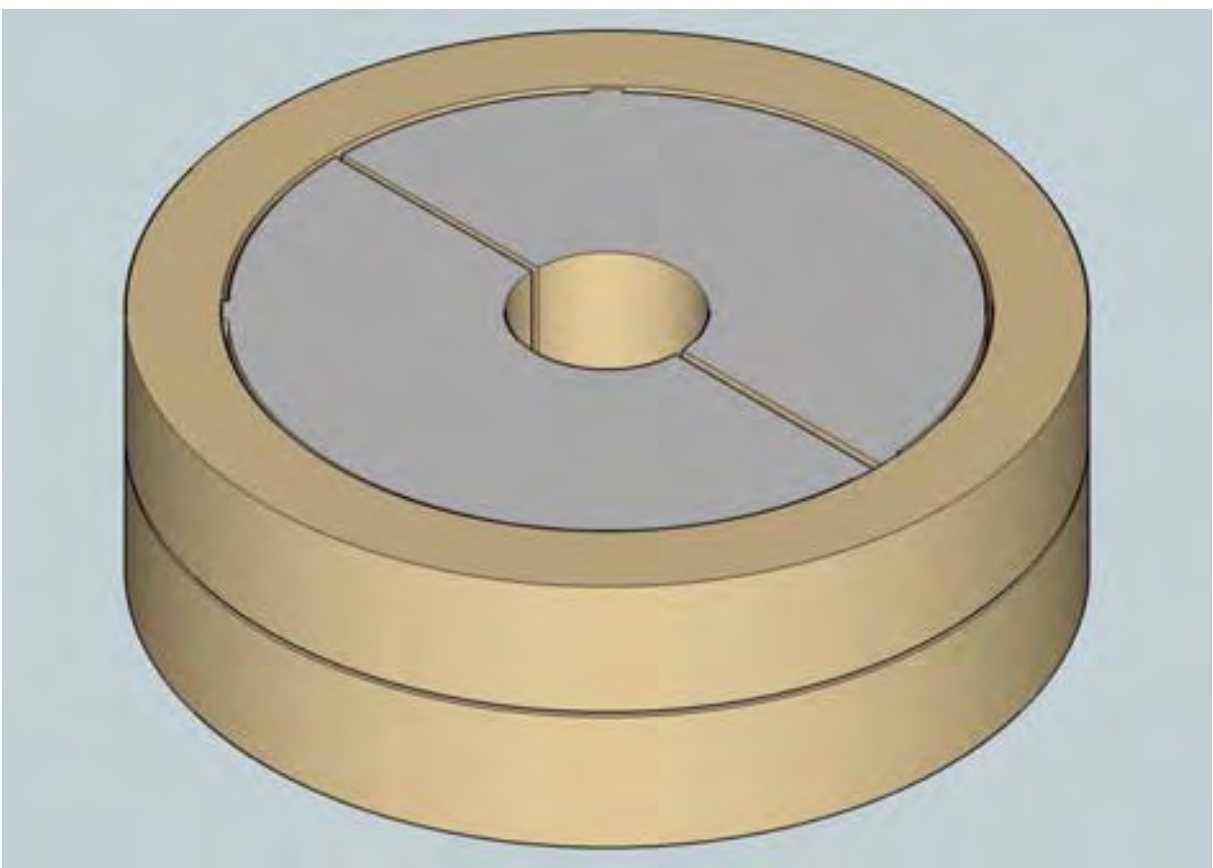

Fig. 5.52: Desenho do espelho completo do espelho heliométrico: dois hemi-espelhos, em diedro, dentro de um anel e este conjunto sobre um plano óptico, tudo feito de CCZ-HS.

O desenho da Tampa Forte seguiu o objetivo de tornar o feixe de luz refletido por cada hemi-espelho simétrico, tanto em relação ao eixo paralelo ao corte, quanto ao eixo ortogonal a este. O desenho mostrado na figura 5.53 mostra a abertura com a maior área possível para se obter esta simetria.

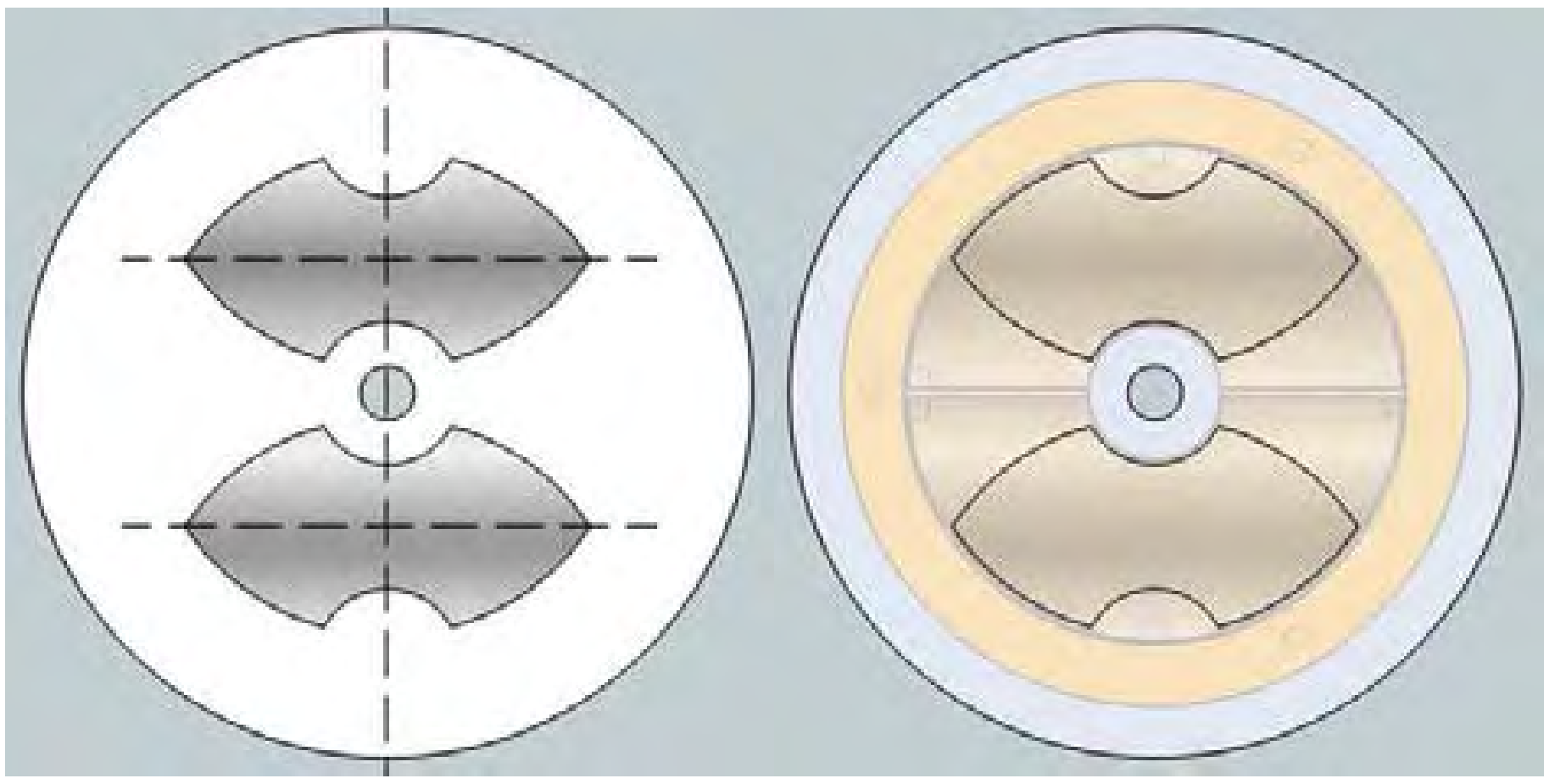

Fig. 5.53: À esquerda, os eixos de simetria e as aberturas da Tampa Forte dos espelhos heliométricos principal e de comparação. À direita, a representação do espelho heliométrico por baixo da tampa.

A Tampa Forte conta com uma abertura de 13 mm de diâmetro, feita em seu centro da para a fixação do suporte do cilindro de CCZ-HS que servirá de padrão (Sol fictício) para os testes de auto-colimação e comparação, figura 5.54.

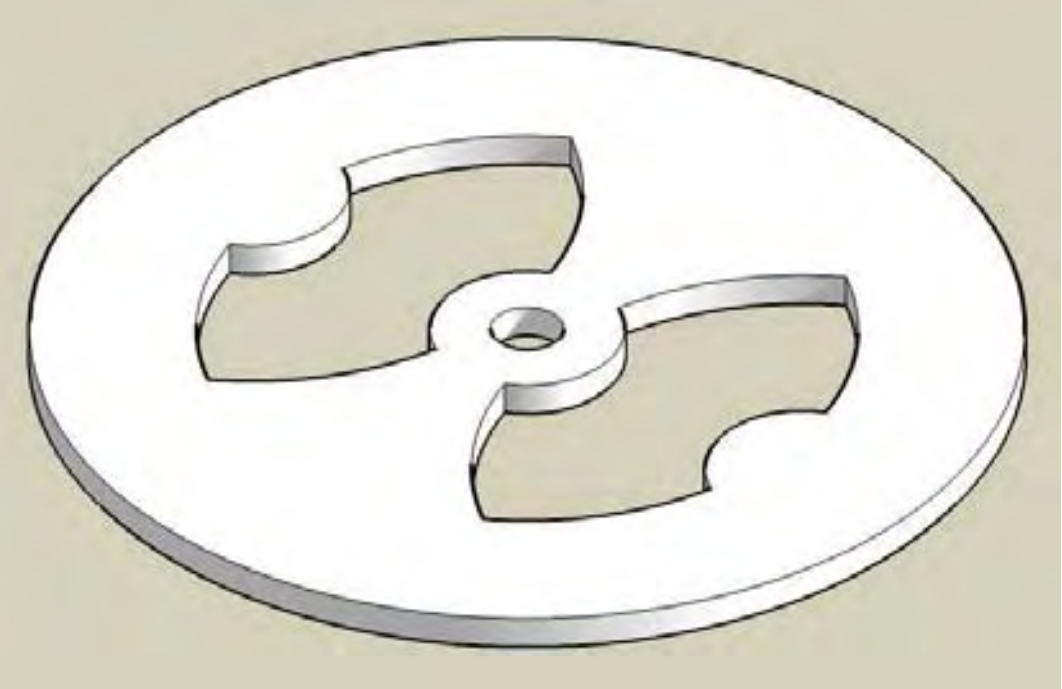

Fig. 5.54: Vista isométrica do desenho final da Tampa Forte.

A Pupila será uma peça fina, de 1 mm de espessura, que ficará sobre a Tampa Forte. Seu desenho manteve a simetria alcançada no desenho desta última. Como esta peça servirá somente à célula do espelho principal, levou-se em conta o deslocamento que os feixes de luz sofrem quando retornam ao espelho principal no teste de comparação. O desenho mostrado na figura 5.55 mostra a abertura com a maior área possível considerando este fator.

A abertura de 13 mm de diâmetro para o suporte do padrão é feita também nesta peça, figura 5.56.

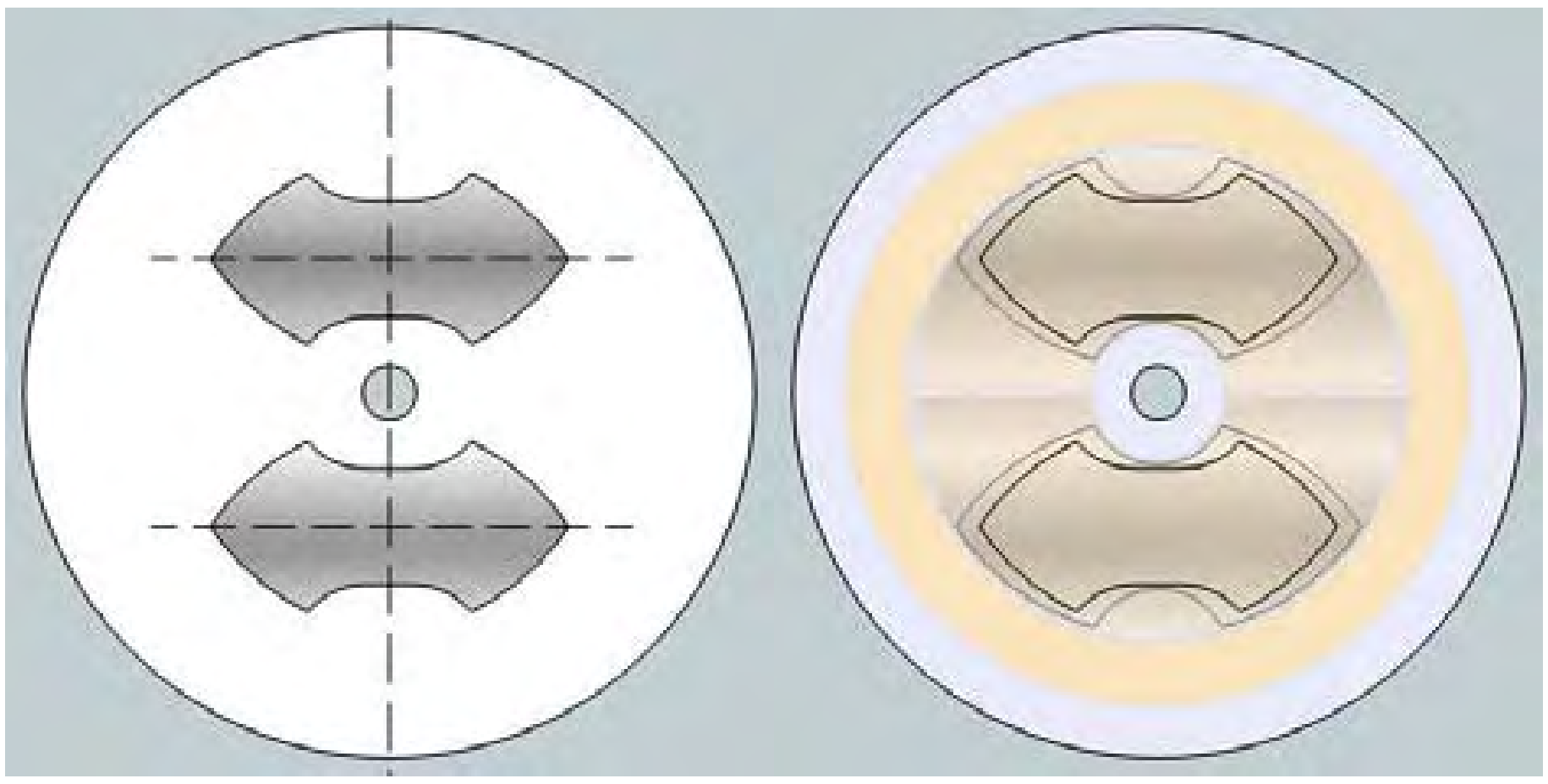

Fig. 5.55: À esquerda, os eixos de simetria e as aberturas da Pupila. À direita, a representação do espelho heliométrico por baixo da Pupila.

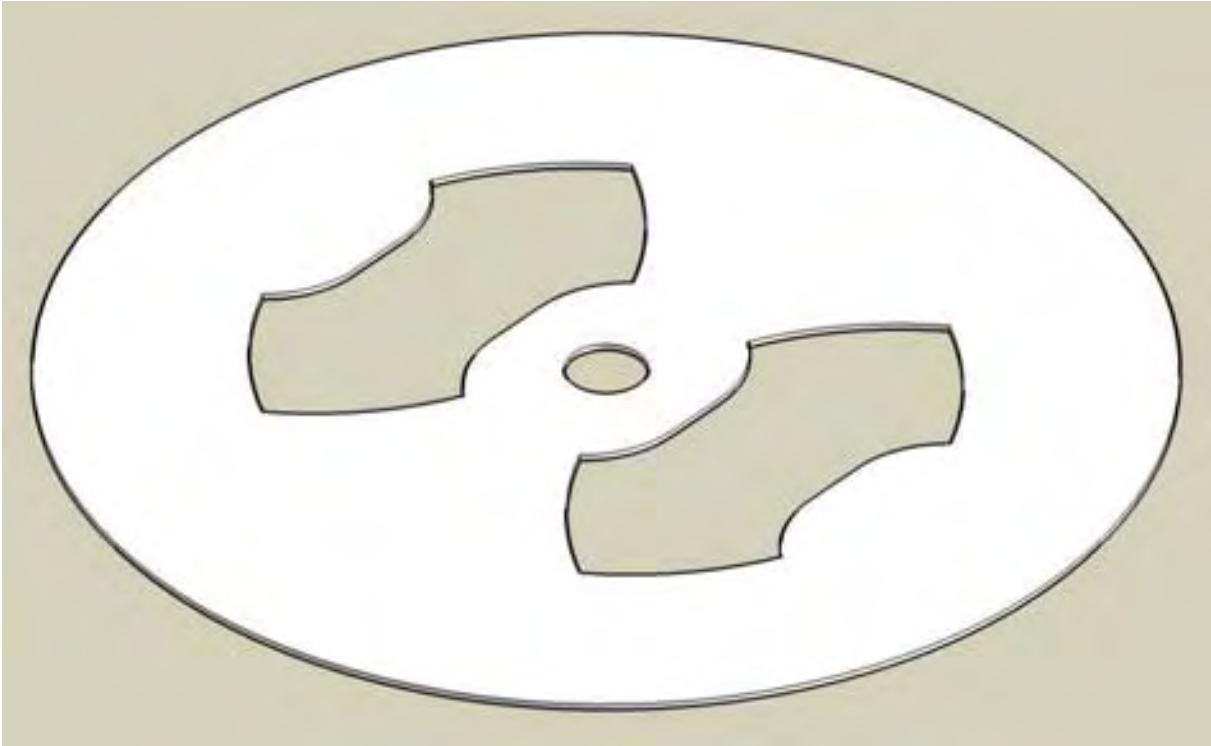

Fig. 5.56: Vista isométrica do desenho final da Pupila.

Presos à Tampa Forte, três parafusos-mola fixam a pupila e pressionam o anel de contenção contra o plano óptico (fig. 5.57). Estes parafusos ficam dispostos a 120° um do outro e atuam no anel verticalmente sobre seus os calços desbastados. As especificações técnicas da Pupila e da Tampa Forte podem ser encontradas no Apêndice VII.

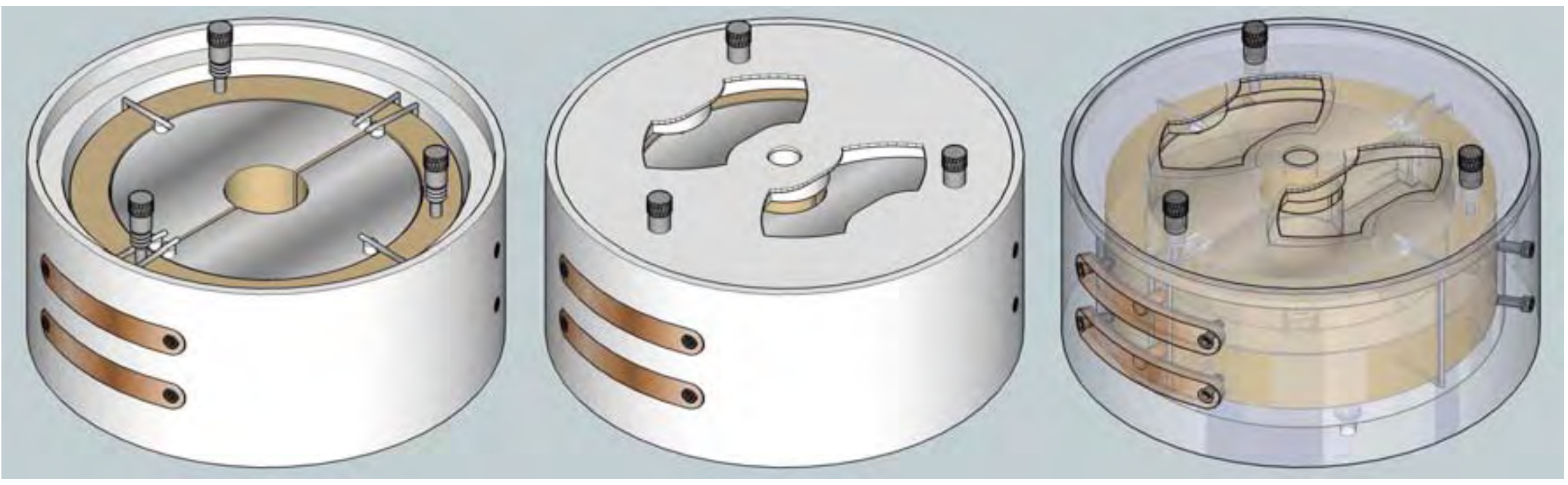

Fig. 5.57: Projeto completo da célula do espelho heliométrico.

A figura 5.58 mostra a Tampa Forte e a Pupila fabricadas em alumínio. As aberturas foram feitas a jato de água, por Comando Numérico Computadorizado, garantindo a precisão das medidas a menos de 0,3 mm.

A célula completa pode ser vista na figura 5.59, já montada em sua base de colimação. Na figura 5.60 podemos ver com mais detalhe o projeto do suporte do cilindro para os testes de colimação e comparação.

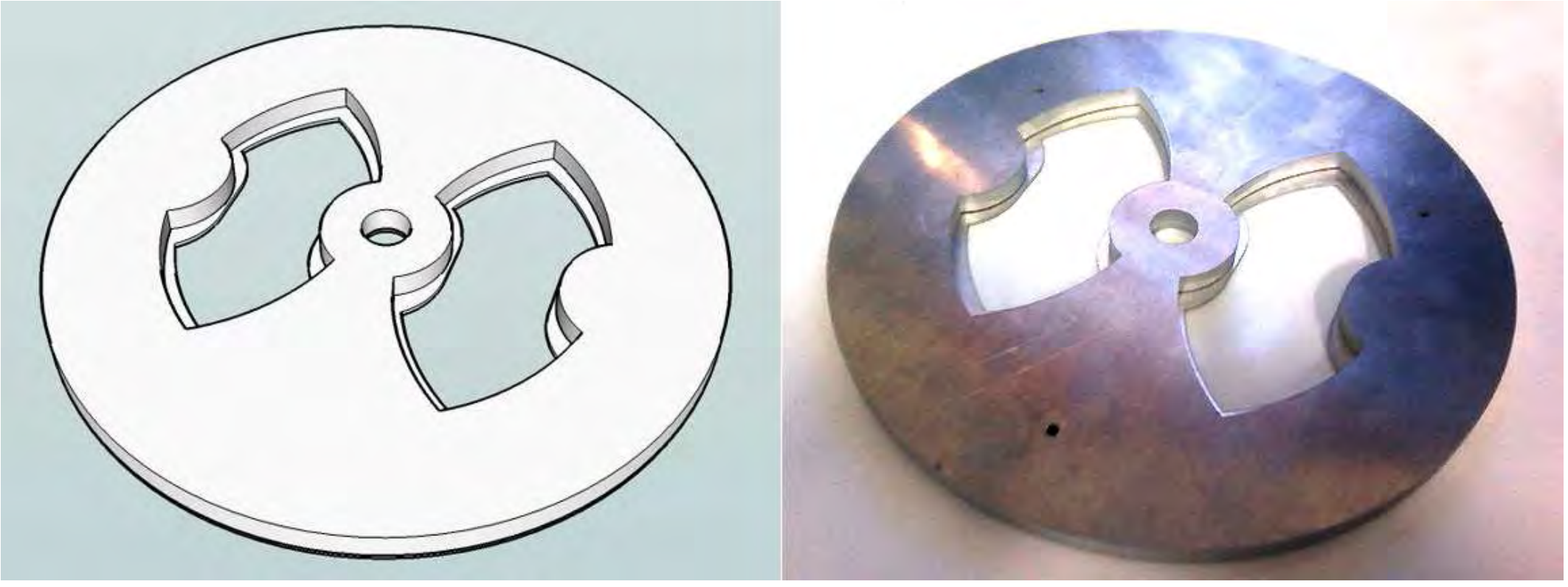

Fig. 5.58: Comparação entre projeto e realização da Pupila e da Tampa Forte.

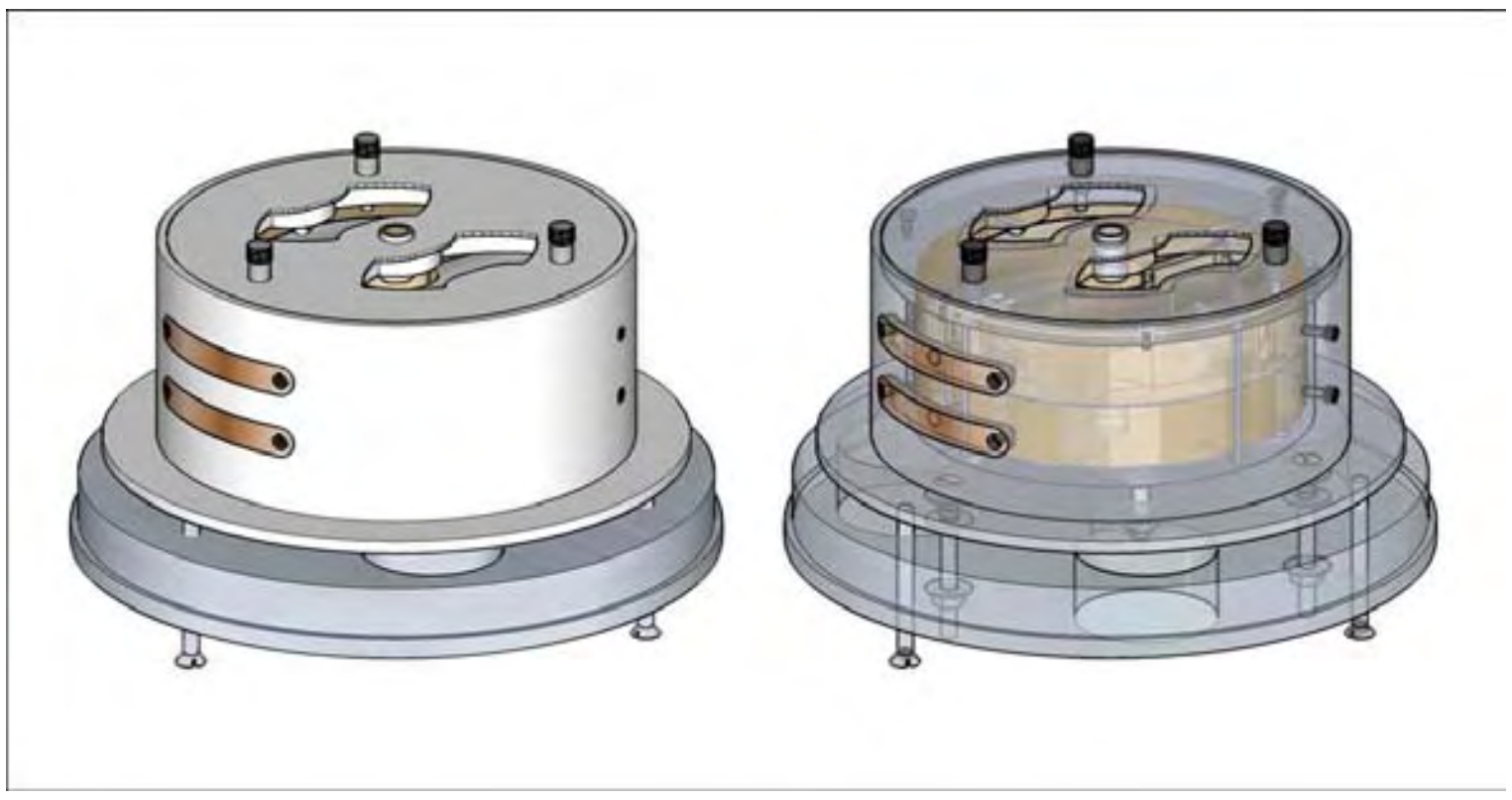

Fig. 5.59: Célula completa presa à base.

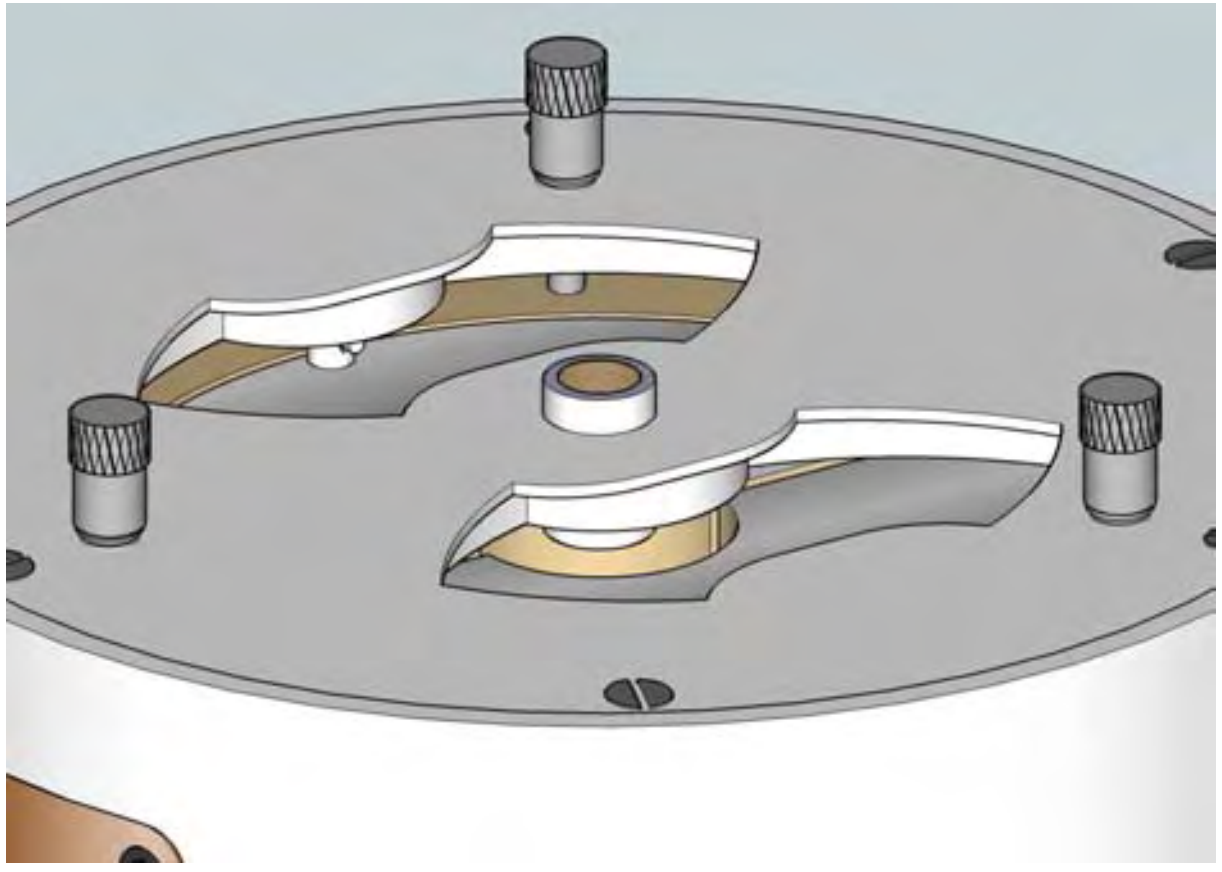

Fig. 5.60: Detalhe do projeto do suporte do cilindro que serve de padrão para os testes de auto-colimação e comparação.

## *5.2 - Partes eletro-mecânicas*

### 5.2.1 - Montagem do telescópio

A montagem em garfo do telescópio Celestron C14 é a mais robusta de todas as montagens do gênero e é capaz de suportar instrumentos de até 75 kg. Feita em alumínio extra-duro, é dotada de motorização em ascensão reta e declinação, com *encoders* em ambos os eixos para localização precisa de qualquer objeto celeste e *tracking*. Os motores são alimentados a 12 V DC, exigindo 10A para seu correto acionamento, figura 5.61.

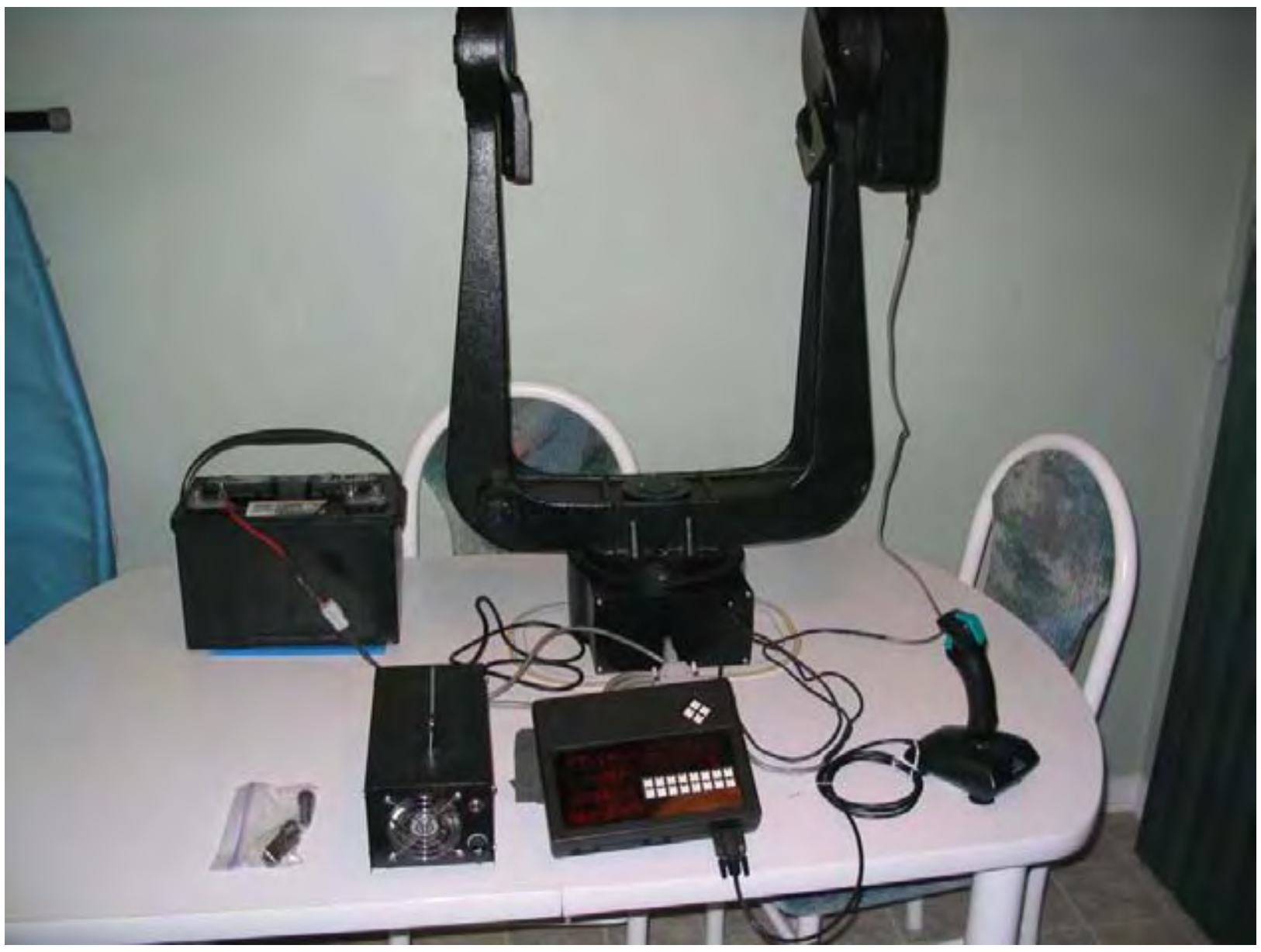

Fig. 5.61: Montagem e seus acessórios de controle.

As rodas dentadas são rolamentadas para maior rigidez e suavidade de movimento. Os parafusos sem-fim têm seu posicionamento ajustável em relação à roda dentada.

Uma adaptação do software SCOPE[3] (Dunna, 2009) foi utilizada para o controle completo desta montagem. Este programa é livre, em ambiente DOS, com perspetivas para Linux, figura 5.62.

Para a adaptação confeccionou-se um cabo especial para conexão, via porta serial, entre o computador e a caixa de controle dos motores (*Power Box*), assim como uma pequena caixa de comando, para acionamento manual da montagem.

[3] O SCOPE pode ser baixado diretamente da página de seu criador, M.Bartels.

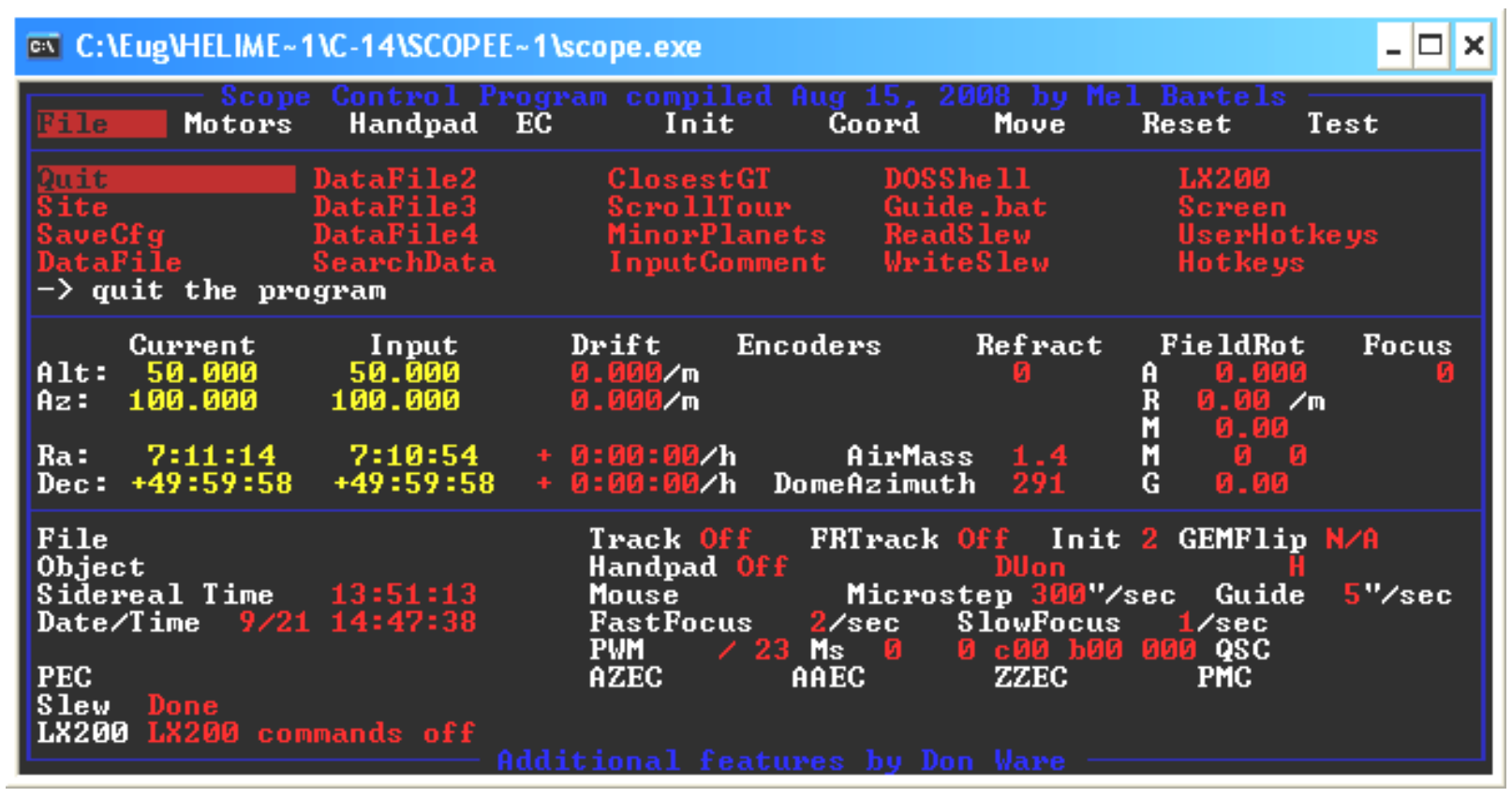

Fig. 5.62: Tela principal do programa de controle da montagem.

## 5.2.2 - Tubo de fibra carbono

O tubo do telescópio é de fabricação nacional, com padrão aeroespacial. Possui 250 mm de diâmetro, 1100 mm de comprimento e uma parede de 3 mm de espessura, figura 5.66.

As principais características para a escolha deste material para ser o corpo do instrumento foram:

- extraordinária resistência mecânica;
- elevada resistência a ataques químicos diversos;
- não afetado por corrosão, por se tratar de um produto inerte;
- rigidez extraordinária;
- estabilidade térmica e reológica;
- bom comportamento à fadiga e à atuação de cargas cíclicas;
- peso específico da ordem de 1,8 g/cm$^3$, o que lhe confere extrema leveza a ponto de não se considerar o seu peso próprio nos esforços. Seu peso específico é cerca de 5 vezes menor do que o do aço estrutural.

## 5.2.3 - O suporte do tubo

O suporte do tubo é feito em aço inox. Em seu desenho, tubos trançados (*truss*) garantem uma excelente estabilidade mecânica ao suporte (fig. 5.63). Ele é preso à montagem através de placas laterais, que podem deslizar pelo suporte a fim de que o ponto de equilíbrio do conjunto seja alcançado (fig. 5.64). Na parte inferior do suporte existe um rolamento autocentrante, que segura firmemente o telescópio pelo seu eixo, ao mesmo tempo em que permite seu livre giro, figura 5.65.

Para a centragem do tubo são usados parafusos com cabeças de nylon, presos por cantoneiras no anel superior do suporte.

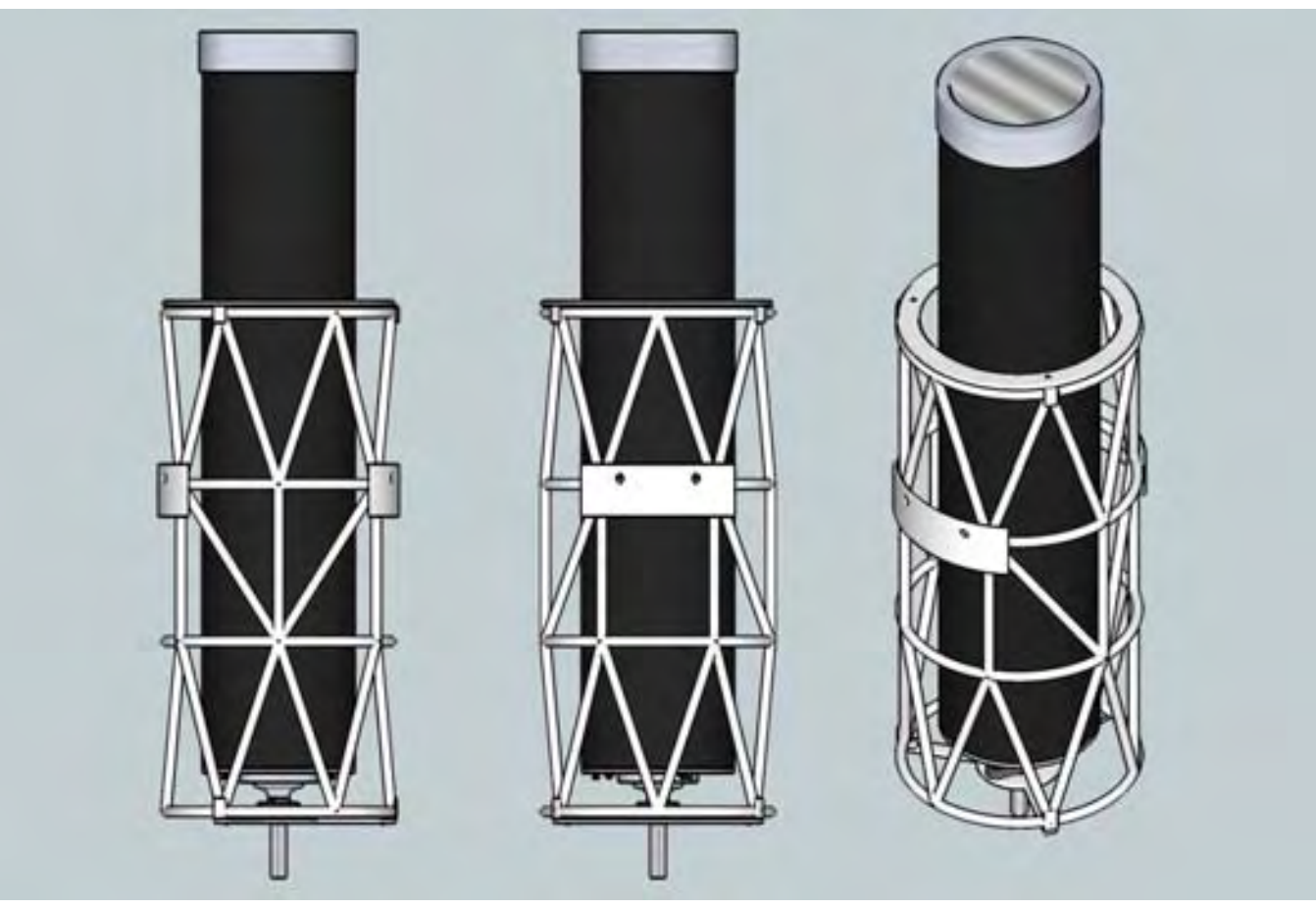

Fig. 5.63: Vistas do projeto do novo suporte do tubo.

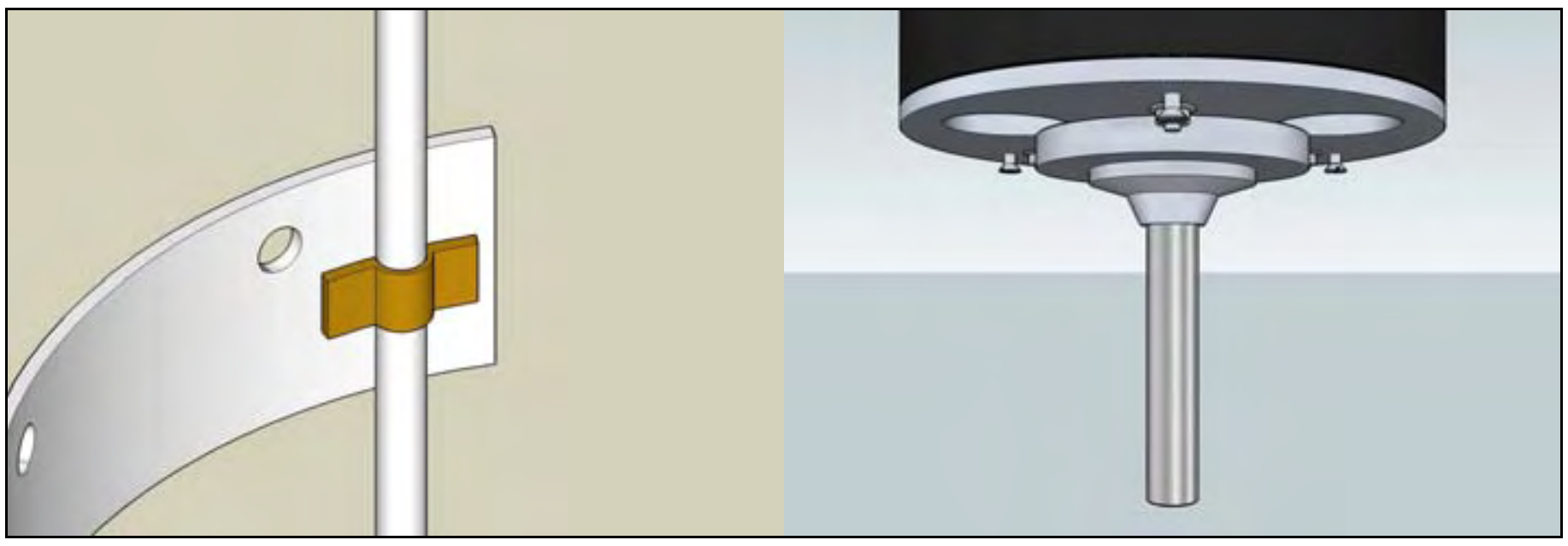

Fig. 5.64: À esquerda, o projeto da braçadeira que fixa a placa lateral do suporte à montagem. À direita, detalhe do projeto do eixo do telescópio preso à base de colimação.

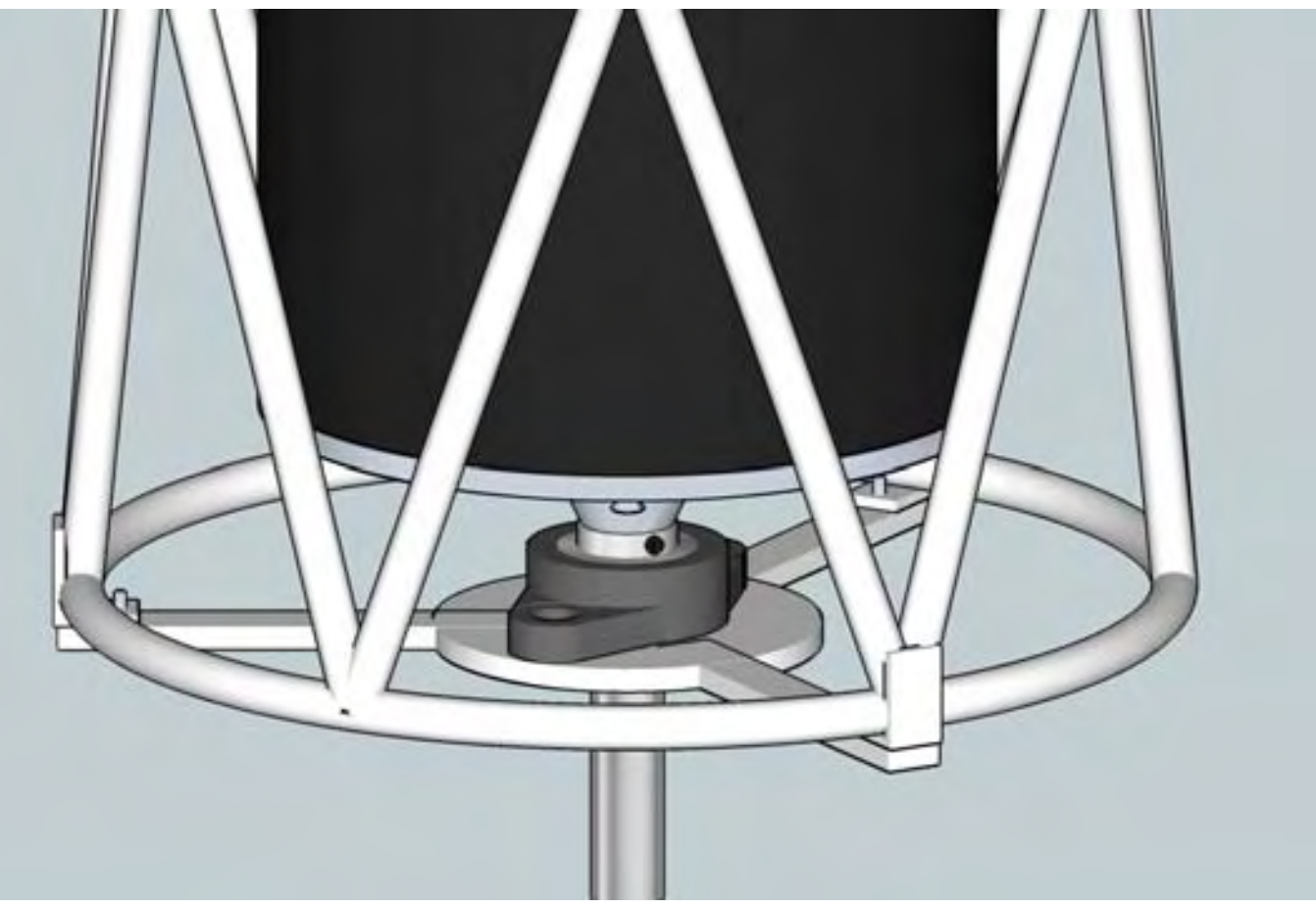

Fig. 5.65: Detalhe do projeto do rolamento autocentrante que prende o telescópio ao suporte do tubo.

A seguir vemos o tubo de fibra carbono adaptado para receber a base de colimação e o suporte *truss* do tubo fixado á montagem equatorial, ainda no laboratório, figura 5.66.

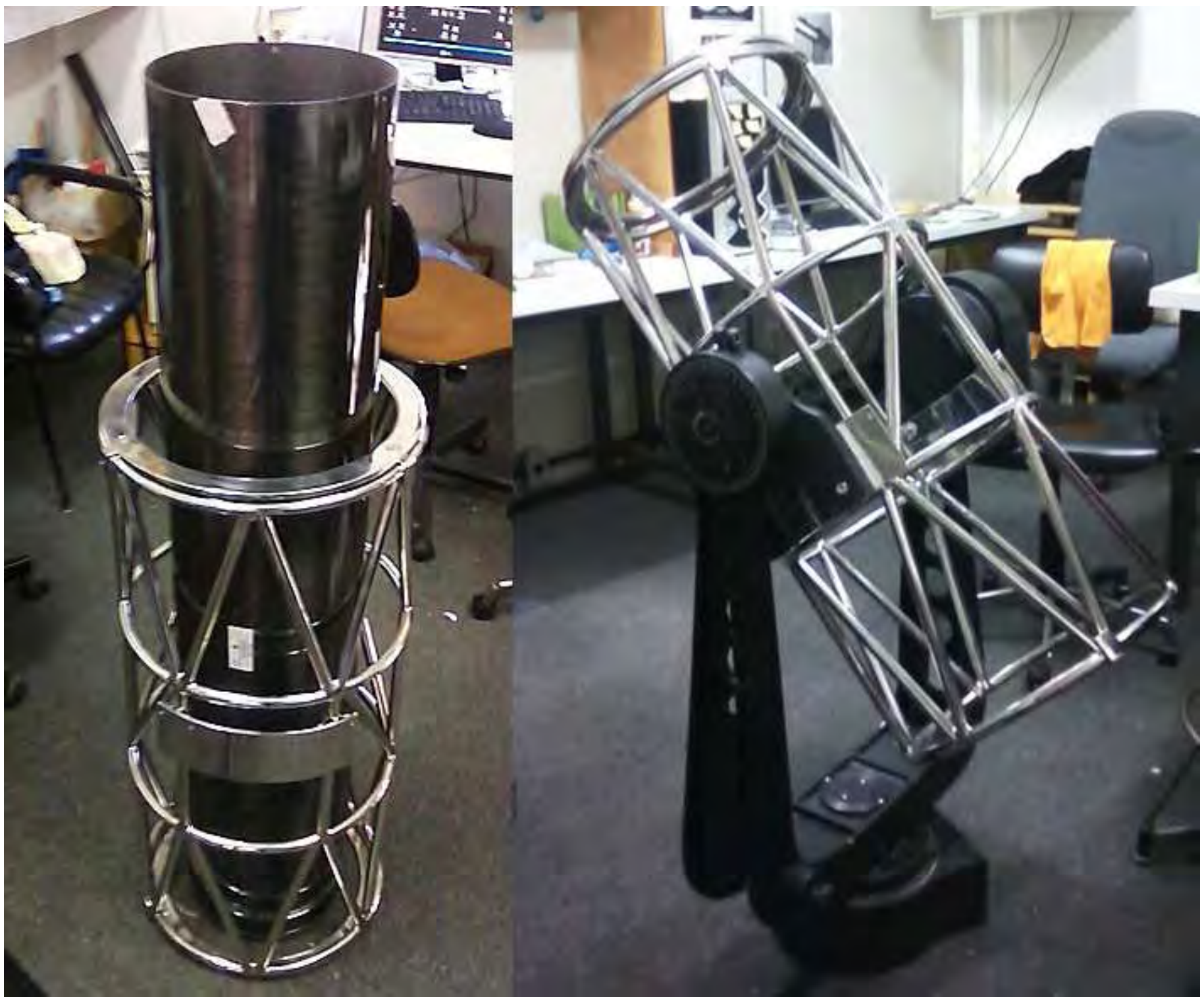

Fig. 5.66: Tubo de fibra carbono dentro do seu suporte e suporte preso à montagem.

## 5.2.4 - Suporte da câmara CCD

Para que a câmara CCD possa ser posicionada com mais estabilidade mecânica, não é usado um espelho secundário. A câmara ficará posicionada diretamente no plano focal do espelho, assim como foi no protótipo n°3.

Desta forma não há mais a necessidade de um grande e pesado focalizador, aliviando o tubo deste esforço. A não existência de um espelho secundário minimiza também, ao máximo, o número de superfícies ópticas envolvidas, representando um ganho na qualidade das imagens.

O suporte da câmara é denominado "placa-aranha", pois tem função semelhante à de uma aranha de telescópio, mas para o CCD. Esta placa é confeccionada em um disco fino de fibra carbono, com diâmetro de 243,5 mm com espessura de 3 mm. O uso da fibra garante a estabilidade mecânica e térmica. O desenho de suas aberturas segue o mesmo critério de simetria utilizado no espelho de CCZ, para o espelho heliométrico de testes, figura 5.67.

O CCD fica posicionado no centro da placa-aranha e possui liberdade de girar sobre seu eixo para o perfeito alinhamento entre sua matriz e o deslocamento das imagens heliométricas. Para o ajuste fino do posicionamento vertical do CCD e sua colimação existem três parafusos com molas, dispostos a 120° cada, que ficam presos ao tubo por cantoneiras, figura 5.68.

As imagens das figuras 5.68, 5.69 e 5.70 trazem a seqüência de montagem da placa-aranha ao tubo de fibra carbono.

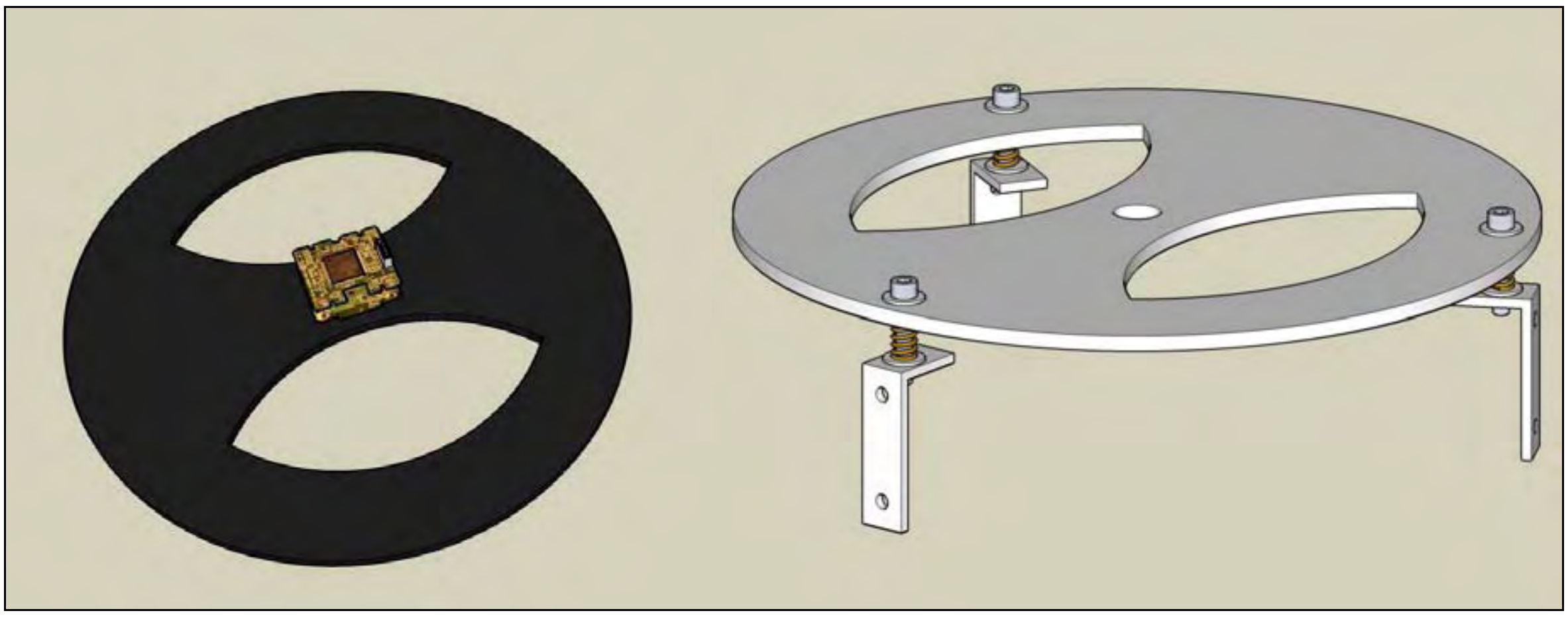

Fig. 5.67: Vistas do projeto da placa-aranha. À esquerda, se vê como o circuito do CCD ficará posicionado. À direita, se vê os parafusos de ajuste fino da posição da placa dentro do tubo.

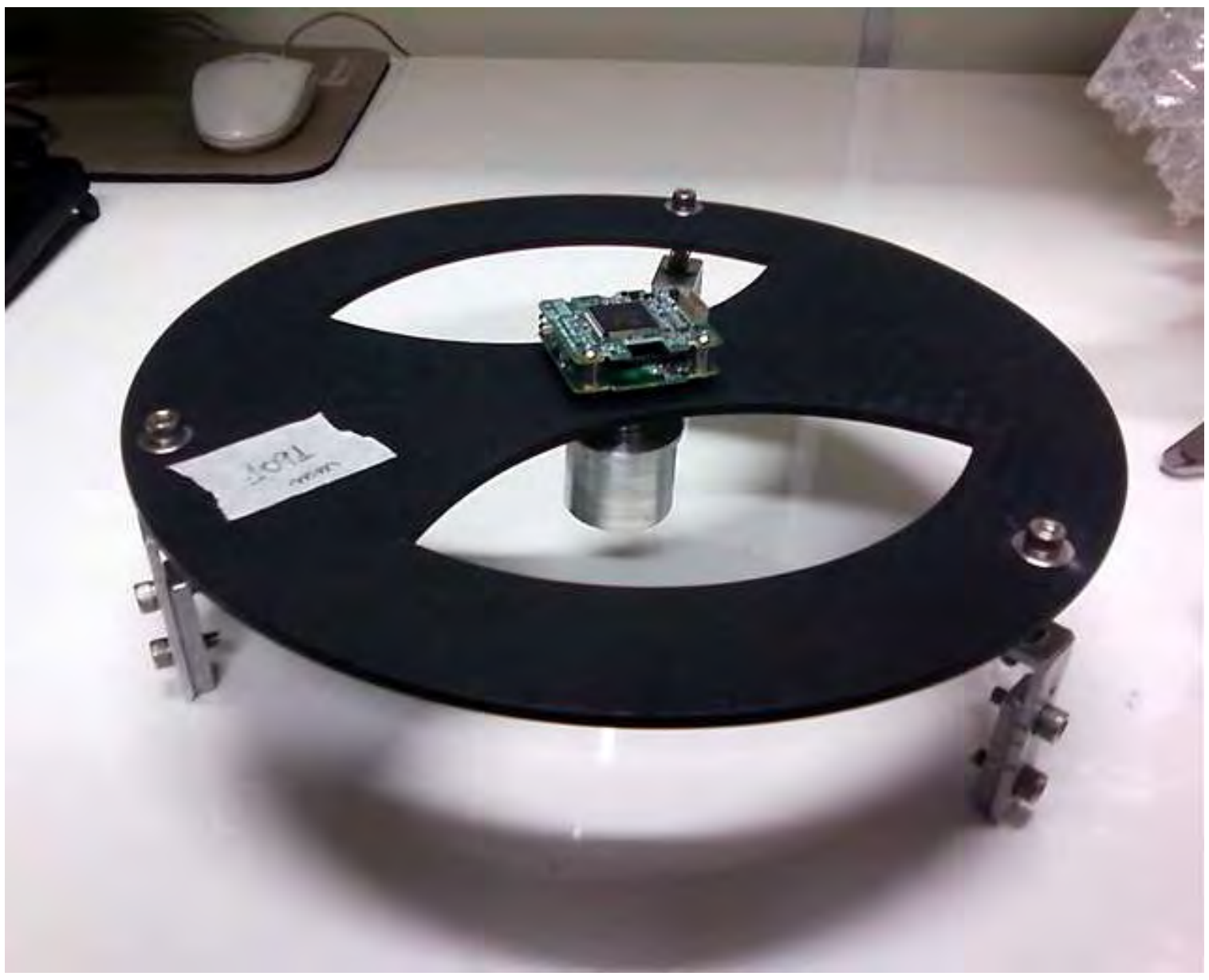

Fig. 5.68: Placa-aranha completamente confeccionada em fibra carbono, com o CCD em sua posição.

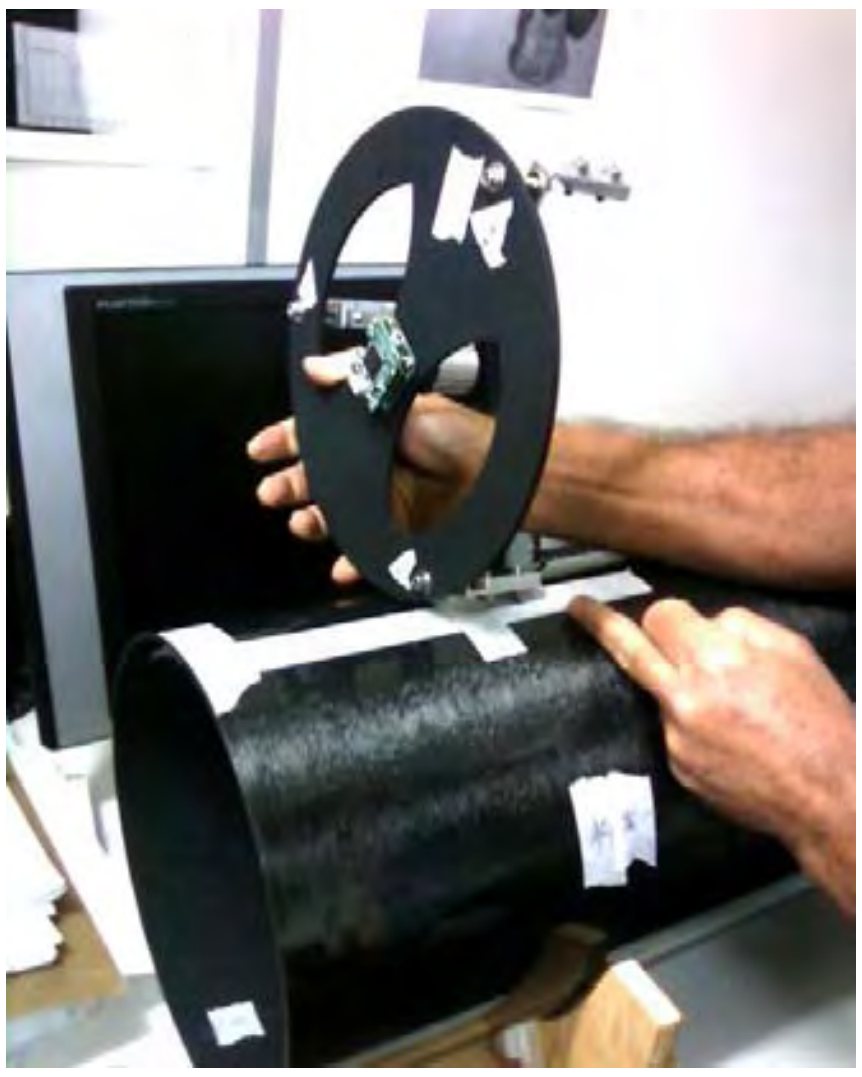

Fig. 5.69: Estudo do posicionamento da placa-aranha dentro do tubo.

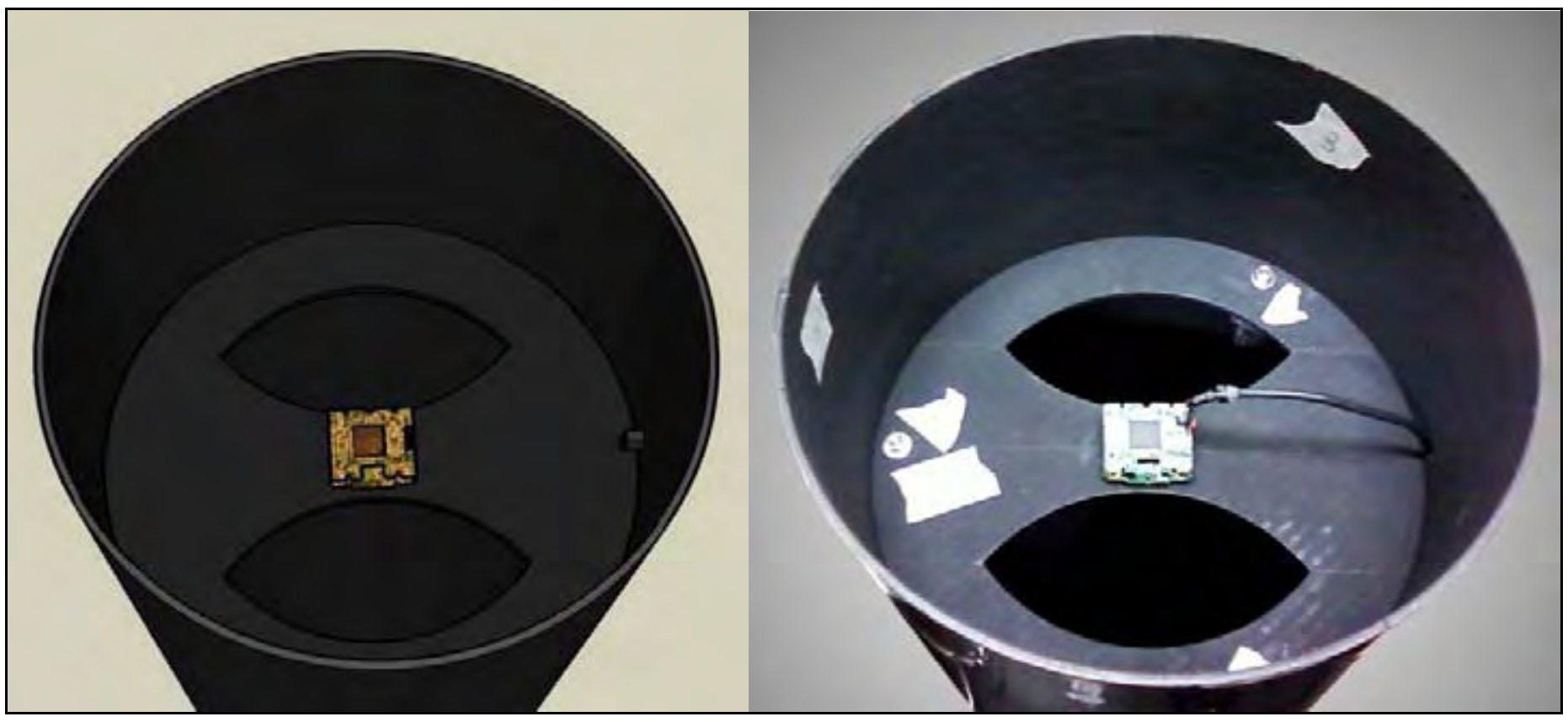

Fig. 5.70: Comparação entre o projeto do suporte no interior do tubo e sua efetiva montagem.

Na figura 5.71 podemos ver a comparação entre a placa-aranha da câmara teste e a placa-aranha da câmara definitiva, que será descrita do próximo item.

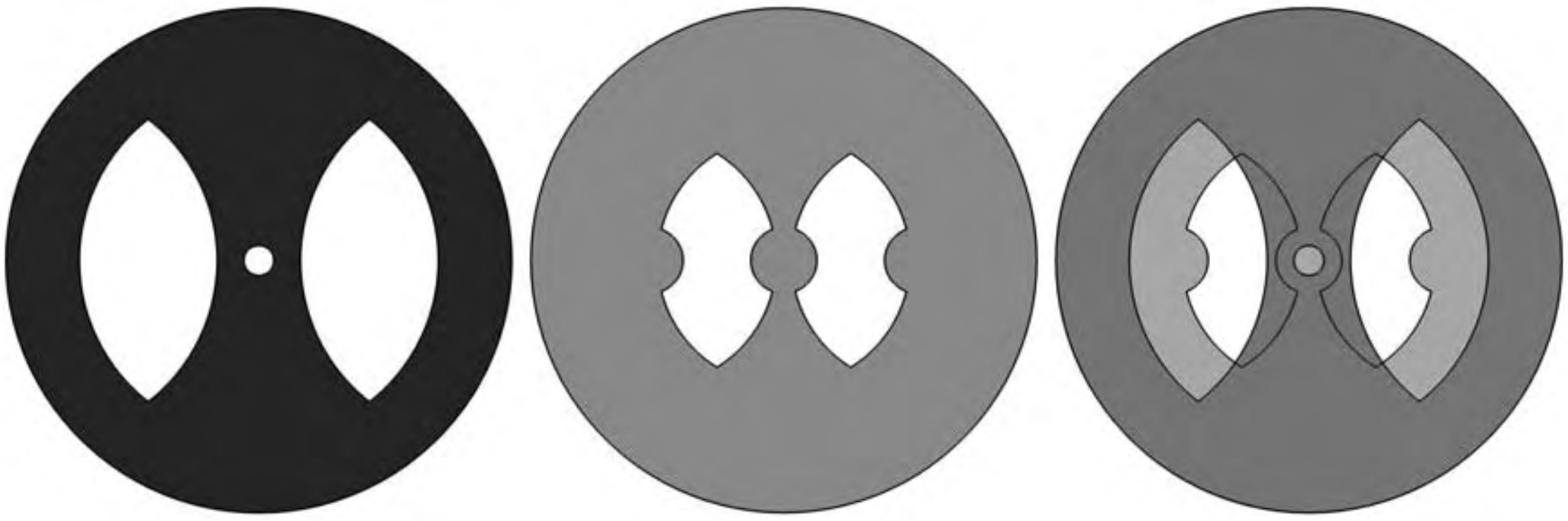

Fig. 5.71: À esquerda, a placa-aranha da câmara teste, no centro, a da câmara definitiva e .à direita a comparação entre as duas.

### 5.5.5 - Câmara CCD

A câmara CCD do Heliômetro, figura 5.72, é uma:

**DMK 31AU03.AS**

Câmara USB monocromática
Resolução 1024×768 *pixels* quadrados de 4,65 μm
CCD de 1/3", com até 30 imagens/s
Escala de placa 0,97"/*pixel*

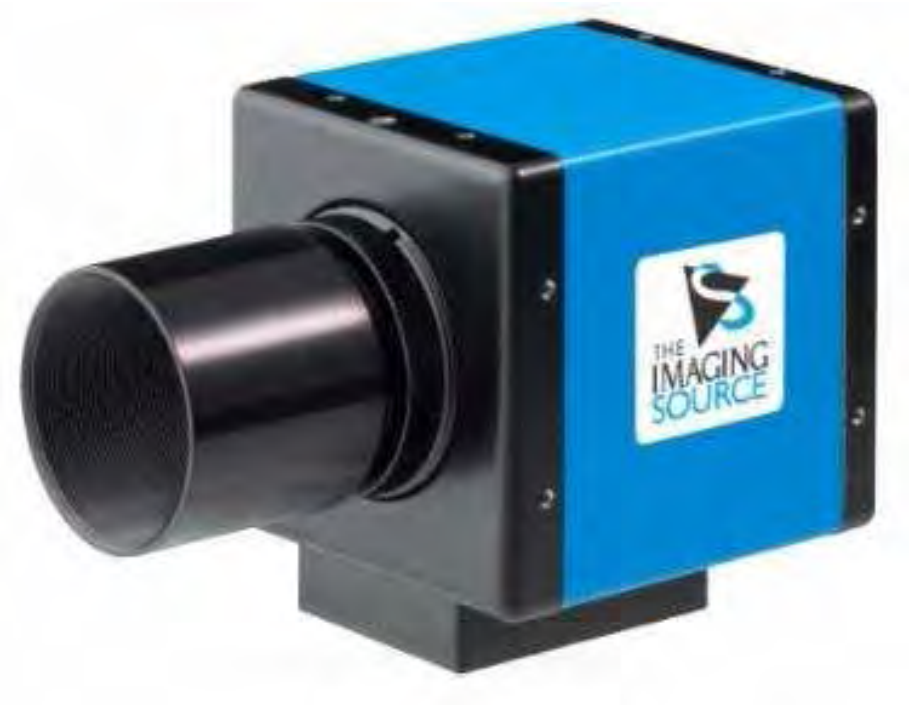

Fig. 5.72: Câmara CCD do Heliômetro.

Vemos na imagem da figura 5.73 a comparação entre os campos da DMK 31 e da câmara de teste. A imagem interior é real, obtida com a câmara de teste e a imagem exterior, uma simulação do campo da DMK 31.

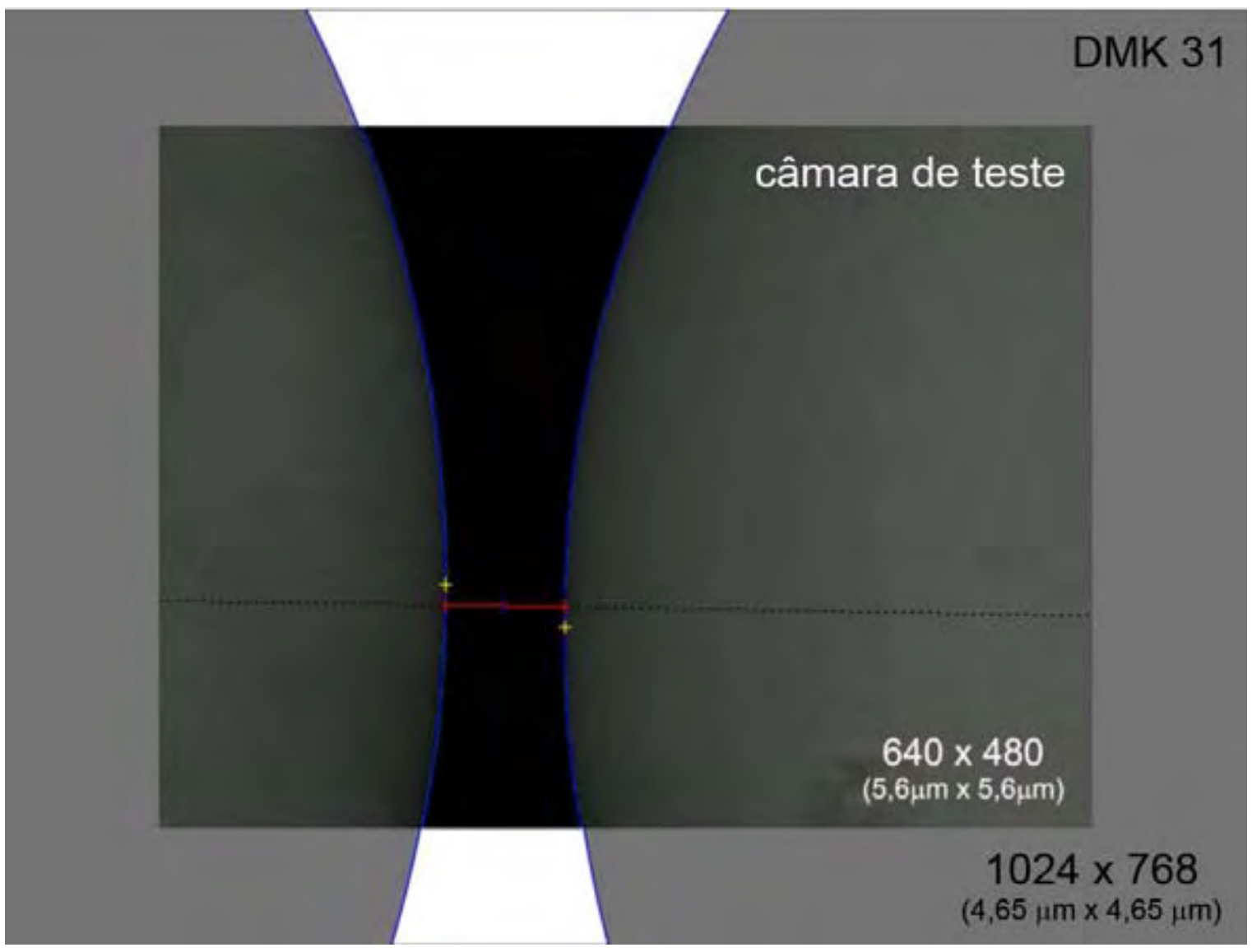

Fig. 5.73: Comparação entre os campos da câmara de teste e a câmara do Heliômetro.

Diferentemente da câmara de teste, a DMK possui um circuito de dimensões que impedem que ela seja posicionada diretamente no centro de sua placa-aranha. A solução encontrada é instalar apenas o chip da câmara no centro da placa, deixando seu circuito próximo à parede interna do tubo, sobre uma região sem aberturas.

Para desenvolver as adaptações eletrônicas e a tecnologia requerida para a retirada do chip, uma webcam Phillips SPC900NC, semelhante à câmara de teste, teve seu chip separado de seu circuito, e um circuito impresso foi especialmente fabricado para suporte do chip.

A imagem seguinte mostra a adaptação feita com sucesso na webcam. O sistema funciona perfeitamente e não houve perdas de sinal do CCD com o uso de um cabo de extensão, figura 5.74.

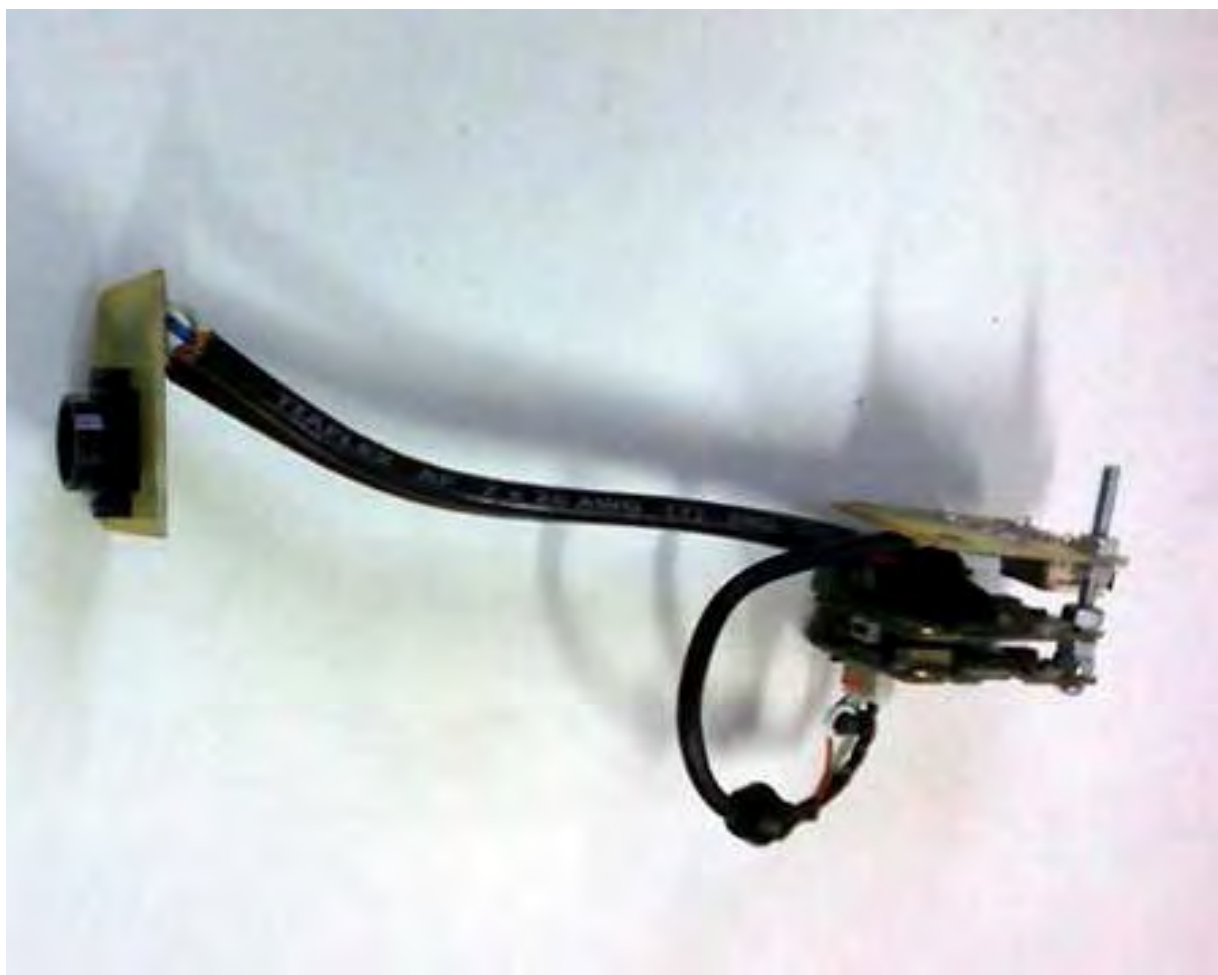

Fig. 5.74: CCD e circuitos separados.

## 5.5.6 - Construção do pilar

Com um desenho de curvas suaves, o pilar serve de base para a montagem equatorial do Heliômetro. O pilar é construído em concreto com vigas metálicas internas reforçando sua estrutura. Também possui dutos internos para a circulação de fiação elétrica, de telefonia, e de transporte de dados. O projeto e a seqüência de construção do pilar estão no Apêndice II.

## 5.5.7 - Teste do controle da montagem

Depois da montagem equatorial ser instalada no pilar, passou-se para a implementação do sistema de controle automático e manual dos seus motores de passo. Com o sistema de controle funcionando, o suporte do tubo é fixado à montagem, figuras 5.75, 5.76 e 5.77.

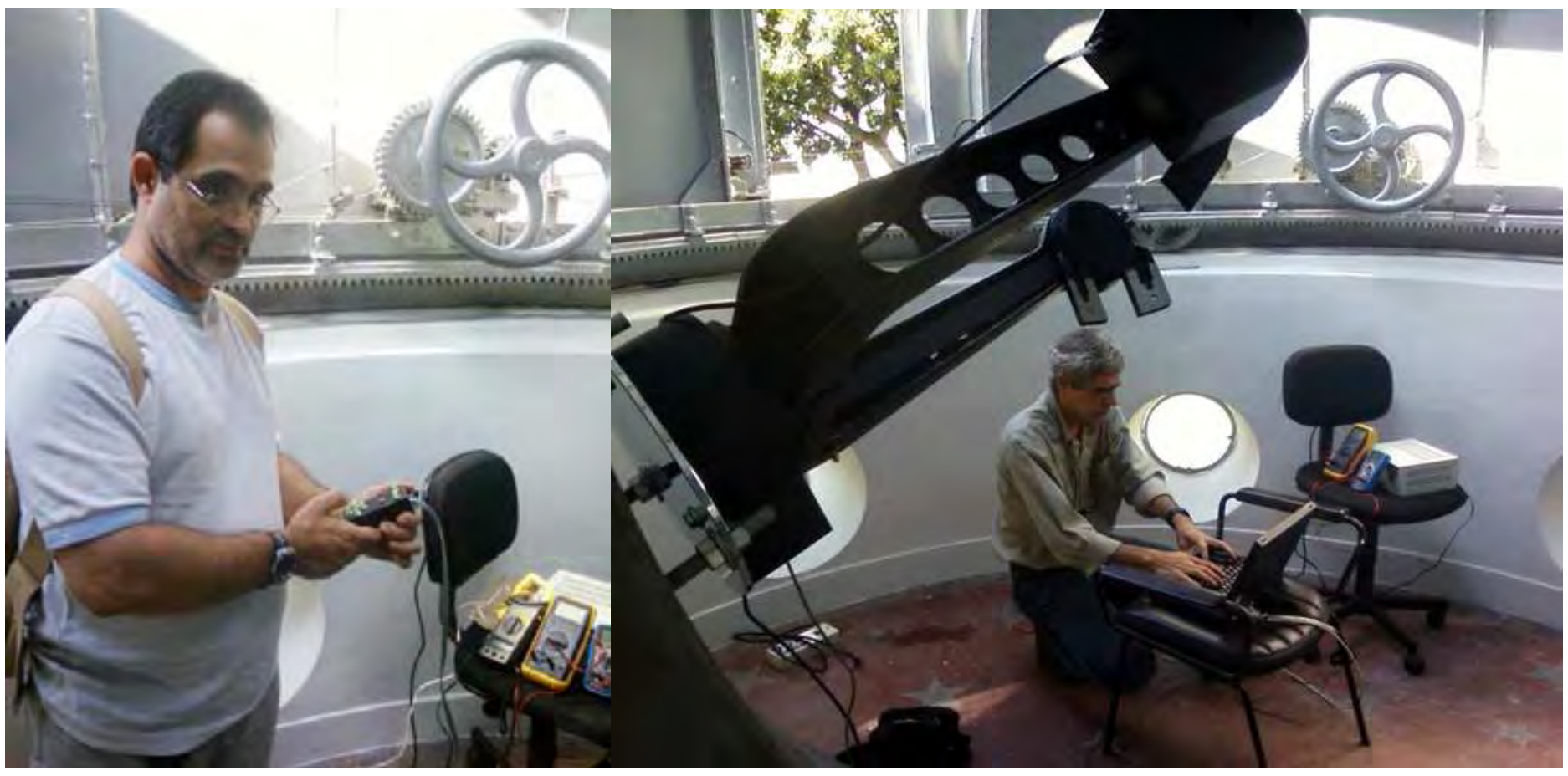

Fig. 5.75: Implementação do sistema de controle manual e automático da montagem.

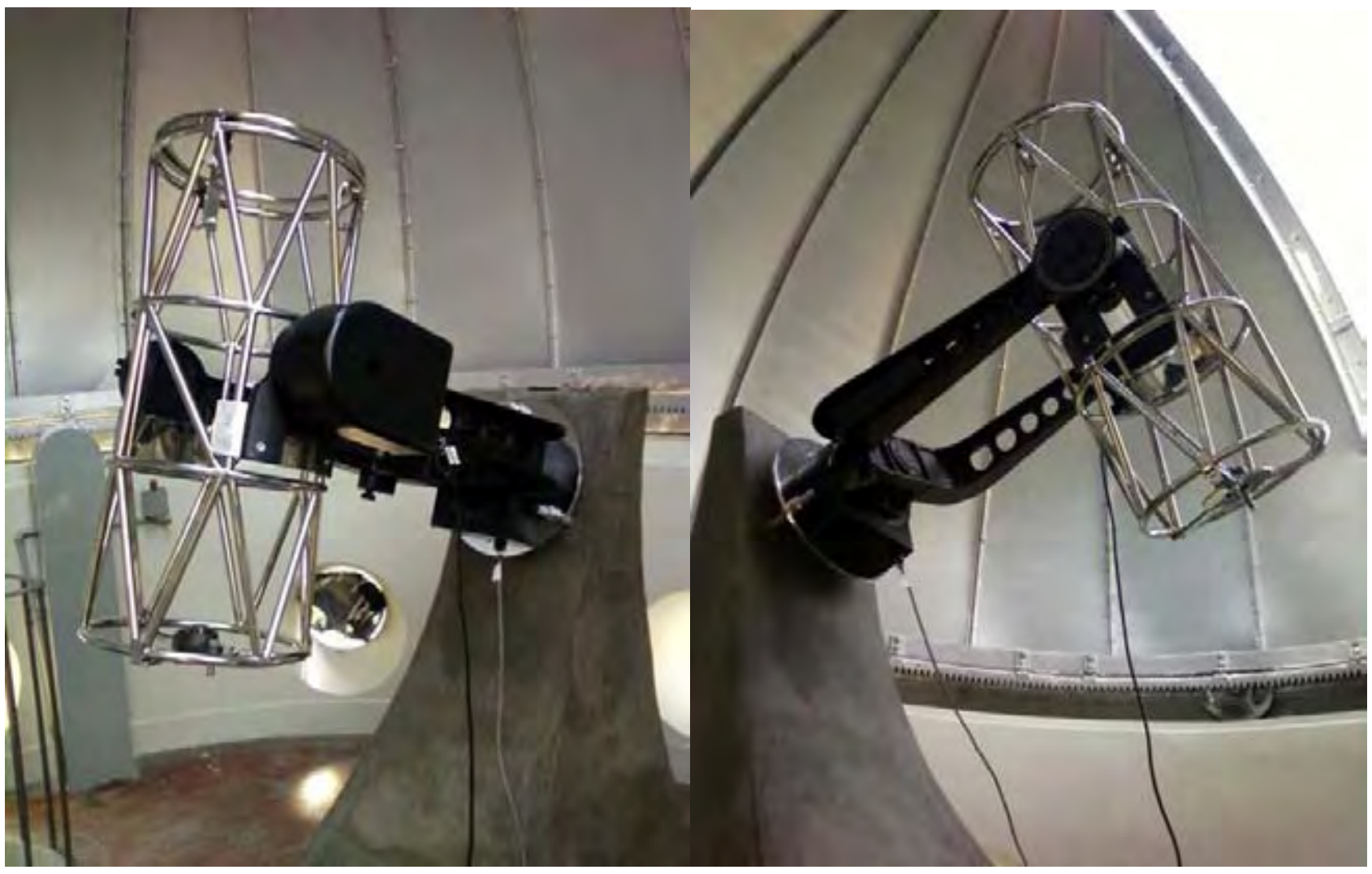

Fig. 5.76: Suporte do tubo fixado à montagem.

Por fim, o Heliômetro é instalado dentro em seu suporte e o conjunto balanceado.

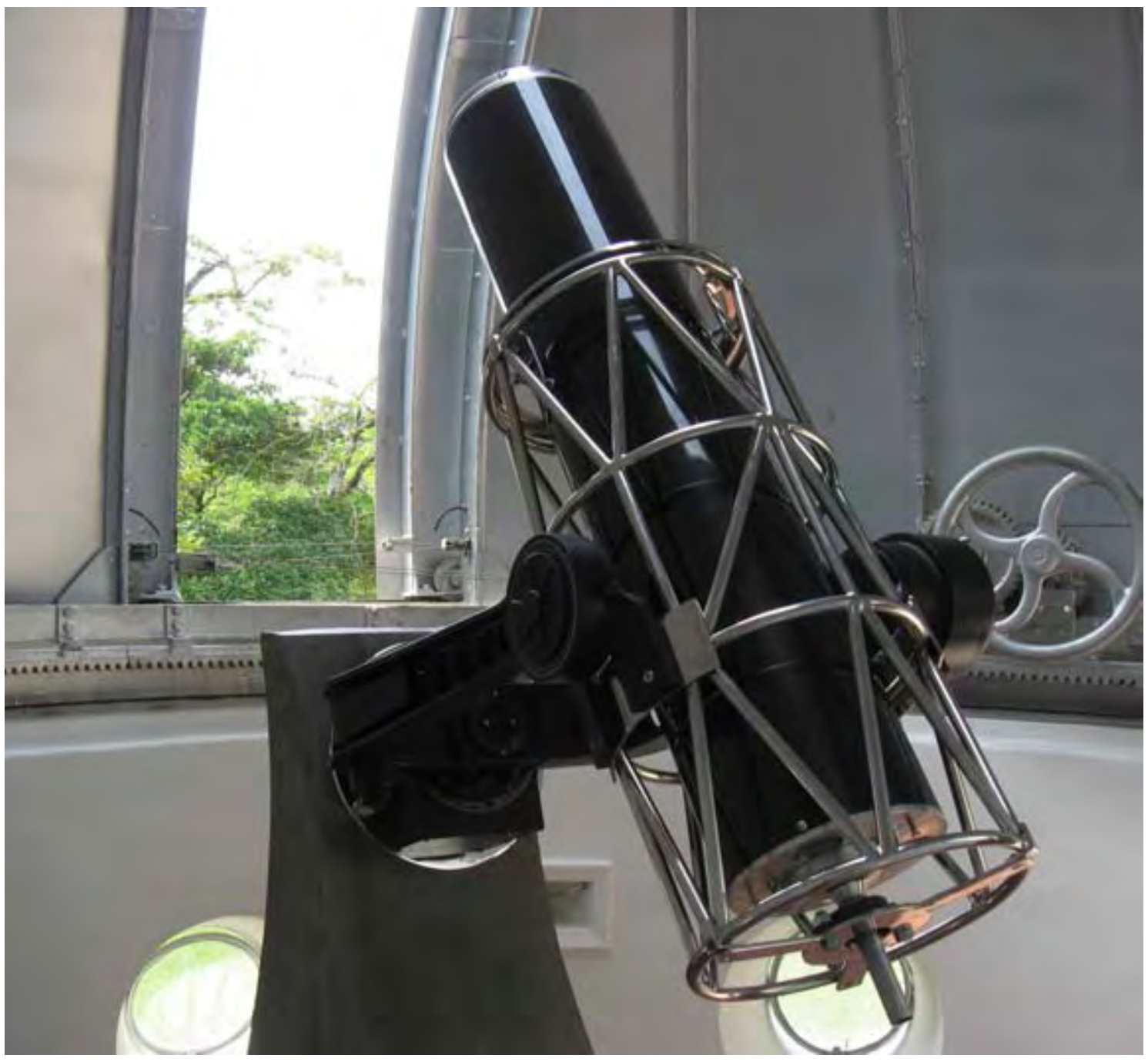

Fig. 5.77: Heliômetro instalado em sua cúpula.

A câmara CCD fica no interior do tubo, no plano focal do espelho heliométrico, fixada à placa-aranha feita também de fibra carbono. Esta placa possui aberturas com geometria semelhante às da pupila, com dimensões maiores apenas o suficiente para não causar vinhetagem nas imagens, de forma que sua rigidez mecânica é máxima, figura 5.78.

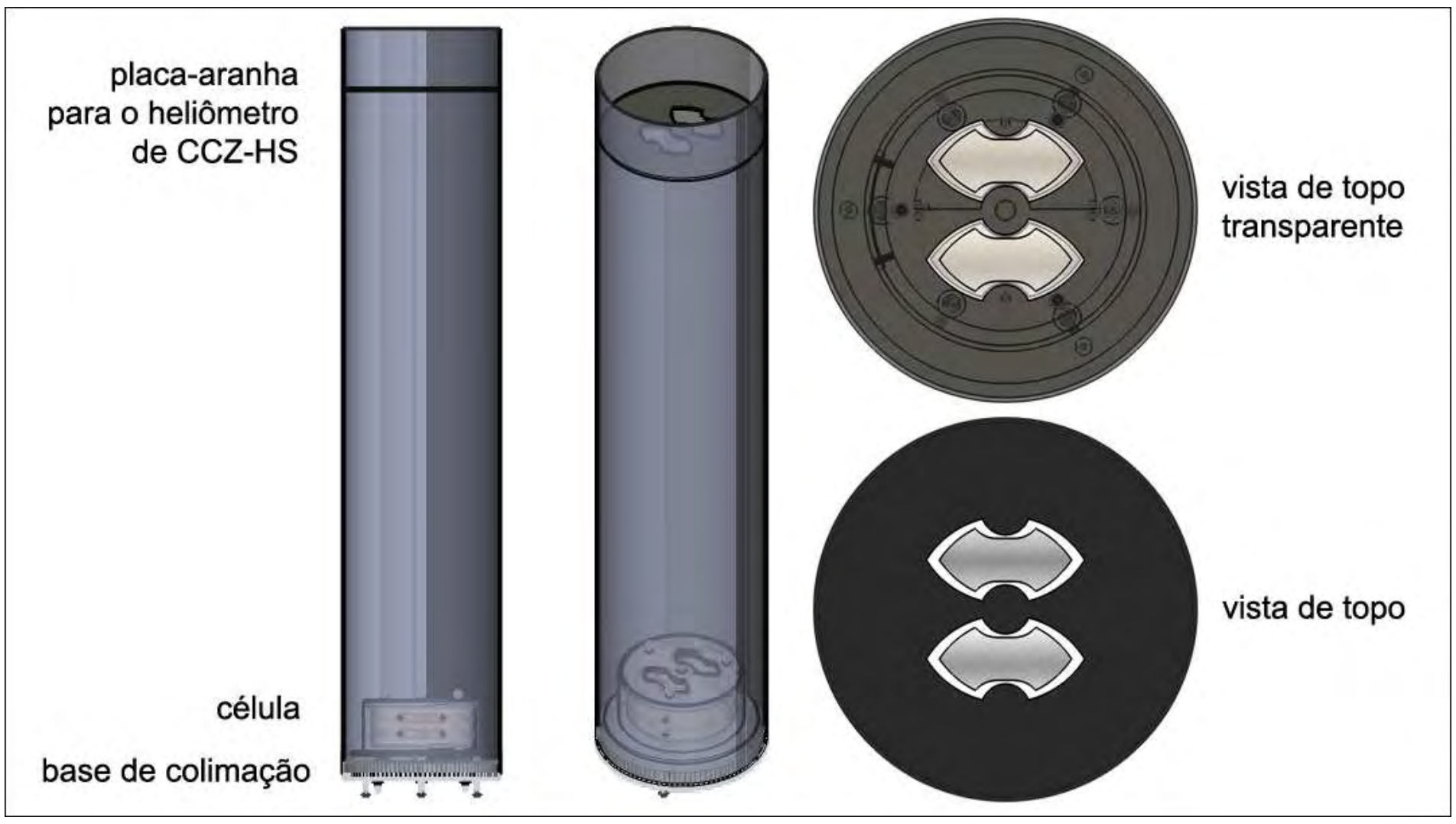

Fig. 5.78: Vistas do projeto da versão final do Heliômetro. Nem a câmara CCD, nem o filtro solar estão no desenho. A vista de topo mostra a área efetiva do espelho.

# Capítulo 6- Primeiras medidas com o Heliômetro

## 6.1 - O programa de aquisição das imagens

### 6.1.1 - Desenvolvimento da metodologia de captura das imagens

Como ponto de partida para aquisição das primeiras imagens heliométricas, ainda na fase de construção de protótipos, utilizou-se o programa QCfocus[4], *software* público para controle de câmaras para astrofotografia. Ele permite o ajuste completo dos parâmetros da câmera, tais como brilho, contraste, resolução, em *pixels*, da imagem, número de *frames* por segundo, entre outros, figura 6.1.

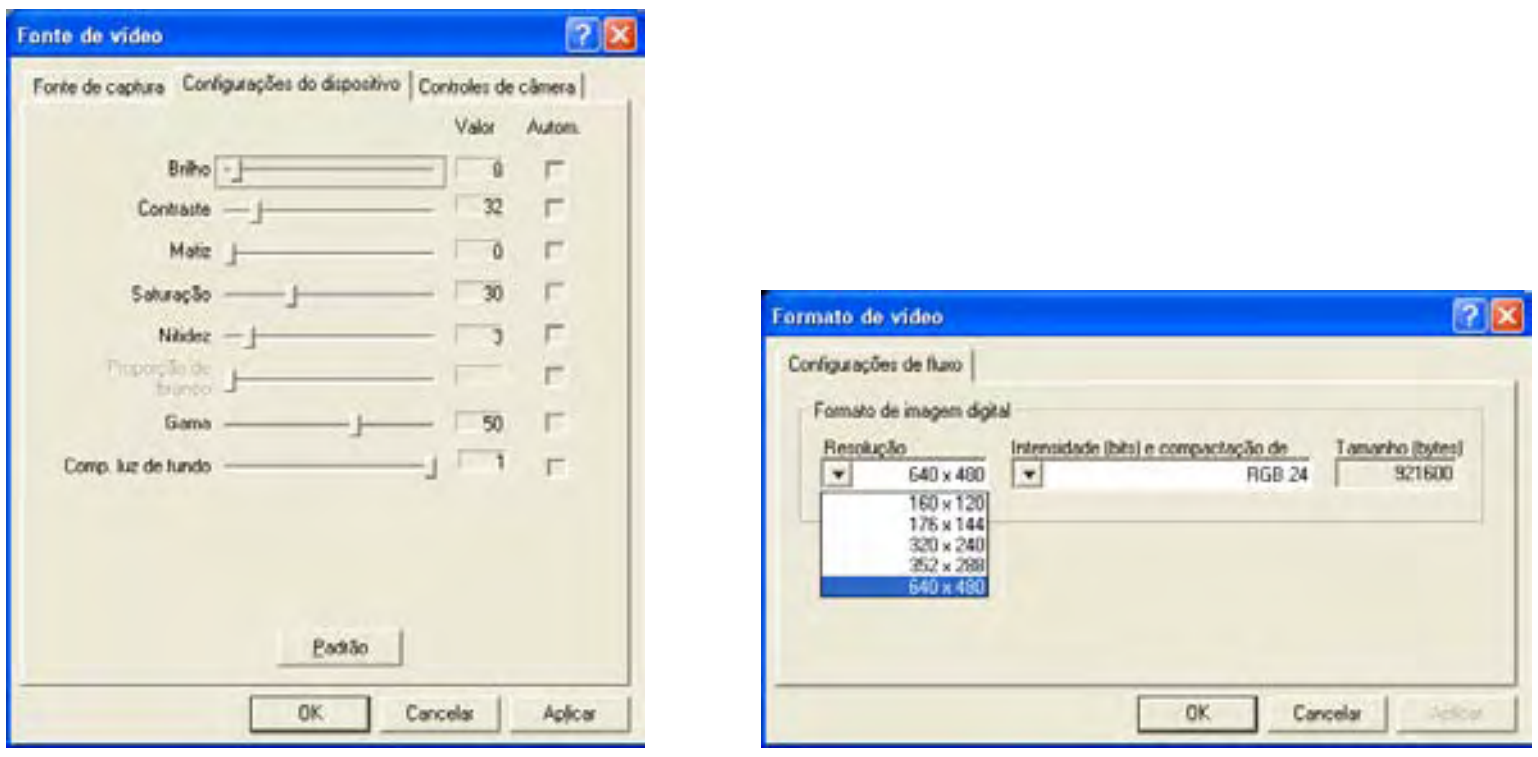

Fig. 6.1: Janelas para o ajuste das configurações da câmara.

O QCfocus também permite definir o número de imagens a serem obtidas (fig. 6.2), o diretório repositório e o nome-prefixo das imagens, figura 6.3.

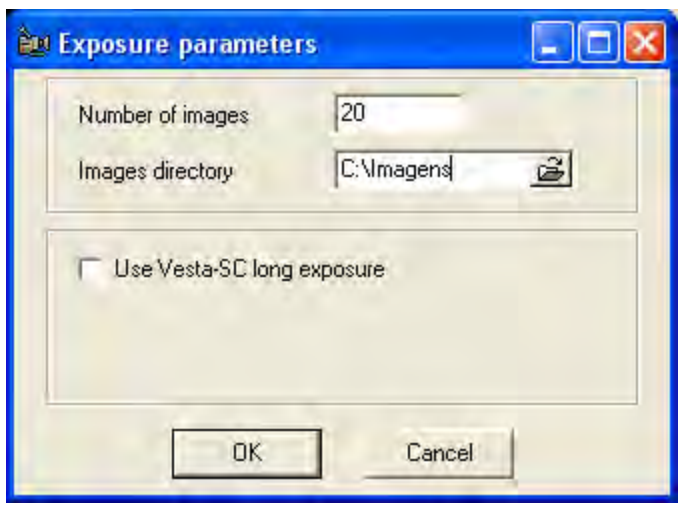

Fig. 6.2: janela para a escolha do número de imagens seqüenciais (20, neste exemplo) e o diretório de armazenamento.

[4] Detalhes do software podem ser encontrados neste endereço eletrônico: http://www.astrosurf.com/astropc/qcam/doc/uk_qcfocus1.html

A interface do programa pode ser vista na figura a seguir.

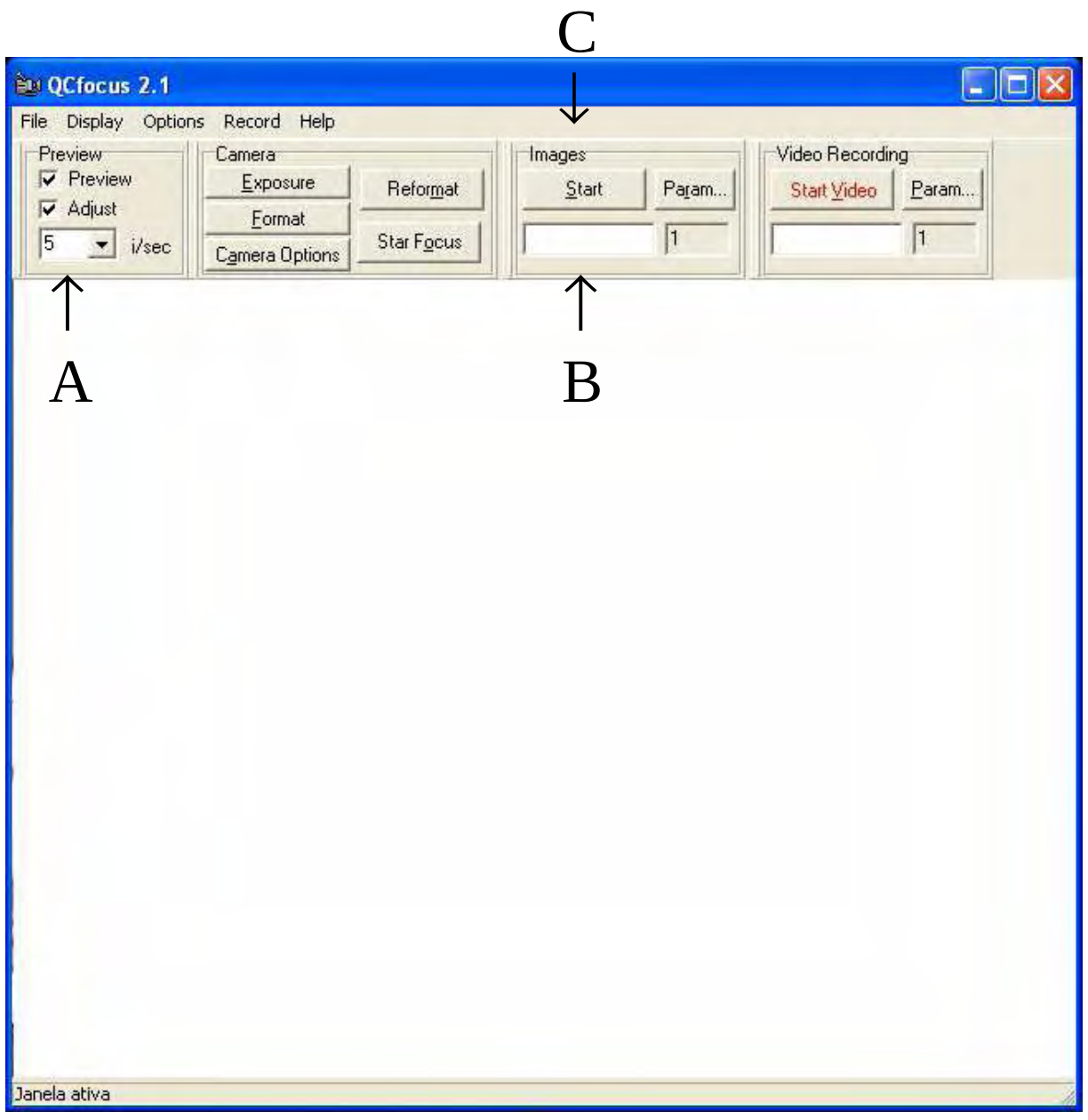

Fig. 6.3: Interface do QCfocus.. Quando o botão Start for acionado (letra A) define-se o número de imagens adquiridas. Se algum nome for escrito no campo abaixo do botão (letra B), então este nome servirá de prefixo para as imagens numeradas. O botão C inicia a aquisição.

Após cada sessão as imagens eram carregadas e examinadas, uma a uma, no primeiro programa de análise desenvolvido. O manuseio do QCfocus permitiu elencar uma série de itens desejados para o programa definitivo de aquisição, sobretudo quanto à datação da imagem adquirida. No entanto, as imagens adquiridas com o QCfocus foram de qualidade astrométrica (i.e., limitadas pelo ruído fotônico e da turbulência), permitindo o desenvolvimento do programa de aquisição (apresentado em 6.2).

### 6.1.2 - Estabelecimento da metodologia de aquisição das imagens

A fim de incluir a datação e automatizar os procedimentos de aquisição, adaptamos o código da rotina 'Acqwebcamera' (Chollet, 2007). A codificação original e a versão preparada para o Heliômetro estão em Delphi. A imagem da figura 6.4 apresenta a interface do programa de aquisição. Tanto a data, quanto o instante, são informações retiradas, em tempo real, do sistema operacional.

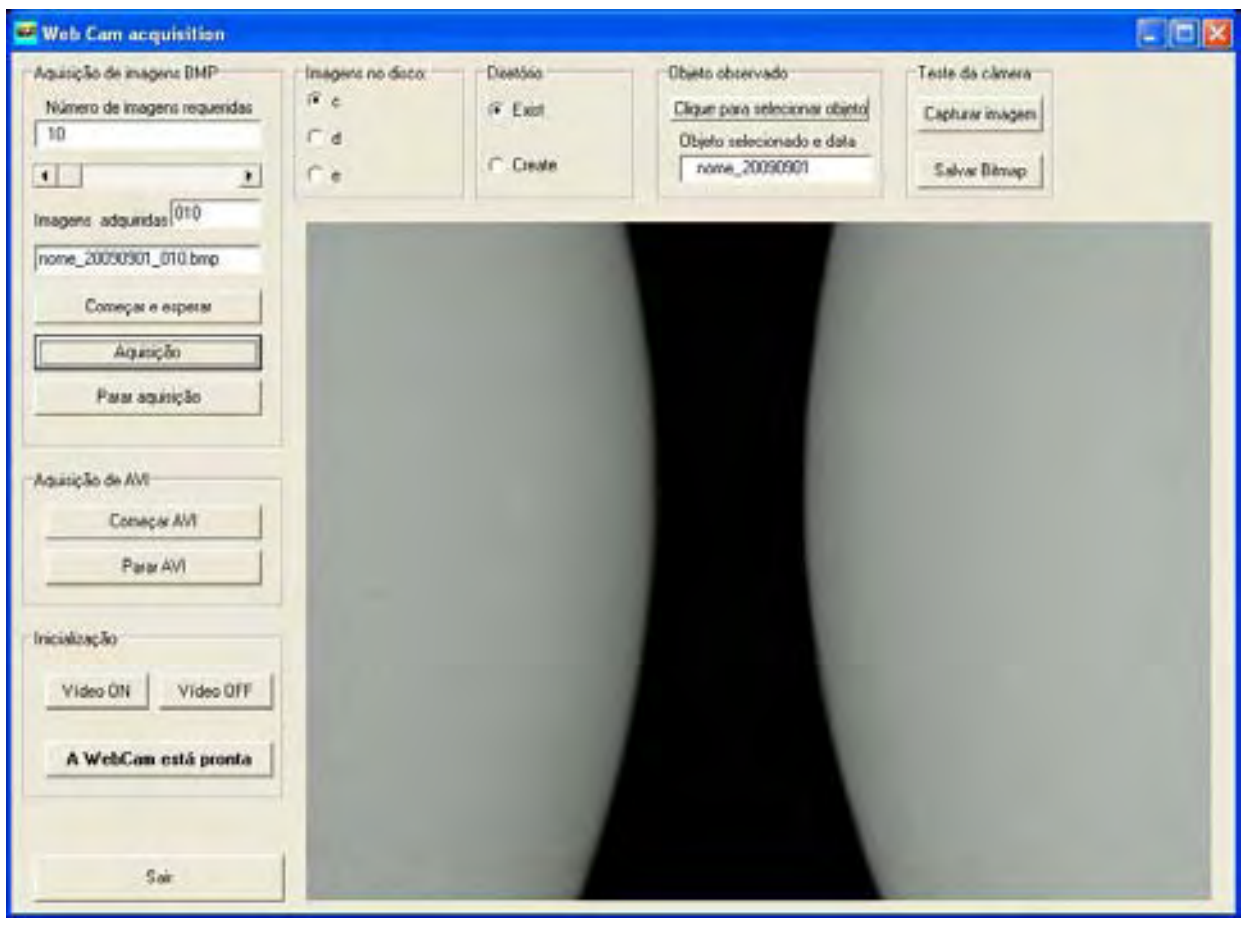

Fig. 6.4: Interface do programa de aquisição modificado.

A seqüência de comandos para a aquisição é mostrada na Tabela 6.1:

Tabela 6.1: seqüência de comandos.

| | |
|---|---|
| **1** | Escolha do disco para armazenar as imagens (c:\, d:\ ou e:\). No disco, diretório padrão chama-se 'Imagens', que, se não existe, é criado; |
| **2** | Escolha do nome do arquivo; |
| **3** | Ao 'selecionar objeto', a este nome escolhido é acrescentada a data corrente (ano, mês e dia); |
| **4** | Entrada do número de imagens requeridas; |
| **5** | Acionamento da câmara, no botão 'Começar e esperar'; |
| **6** | Aquisição. Ao nome e a data também é acrescido o instante da aquisição (hora, minuto, segundo e milésimo de segundo e o número do frame; |

No exemplo da figura 6.4, o nome completo do arquivo torna-se: nome_20090901_151737484_010.bmp, indicando que a 10ª imagem desta série foi armazenada em disco no dia 1º de setembro de 2009, às 15h 17min e 37,484s.

## 6.2 - O programa de análise das imagens

As coordenadas dos bordos solares são obtidas através das imagens adquiridas, definindo o limbo solar pelos pontos de inflexão da função luminosidade ao longo de cada linha da imagem. Esta é a definição utilizada no Astrolábio Solar e pela maior parte das equipes que efetuam medições do raio solar. A análise foi desenvolvida como mostrado a seguir e codificada em Borland Delphi.

Para cada linha é obtida a contagem de cada *pixel*. A intensidade correspondente a este *pixel* ($L_i$) é calculada através da média aritmética simples entre as contagens das cores primárias *RGB* (*Red*, *Green* e *Blue*):

$$L_i = (R_i + G_i + B_i)/3$$

Uma vez obtida a função luminosidade da linha, esta função é suavizada através de uma média corrida ponderada, cortando o ruído de altas freqüências:

$$L_i = (L_{i-4} + 2L_{i-3} + 3L_{i-2} + 4L_{i-1} + 5L_i + 4L_{i+1} + 3L_{i+2} + 2L_{i+3} + L_{i+4})/25$$

As abscissas dos pontos de inflexão da função de luminosidade da curva alisada serão aquelas onde a derivada primeira passar por extremos – um mínimo, para a imagem do disco solar da esquerda, e um máximo, para o da direita.

O valor da derivada em um ponto de abscissa *i*, da linha analisada, é encontrado segundo a fórmula da derivação melhorada de 3 pontos, com $n = 2$:

$$L'_i = \frac{L_{i+n} - L_{i-n}}{2n}$$

As figuras 6.5 e 6.6 mostram um exemplo da análise de uma das linhas da imagem heliométrica.

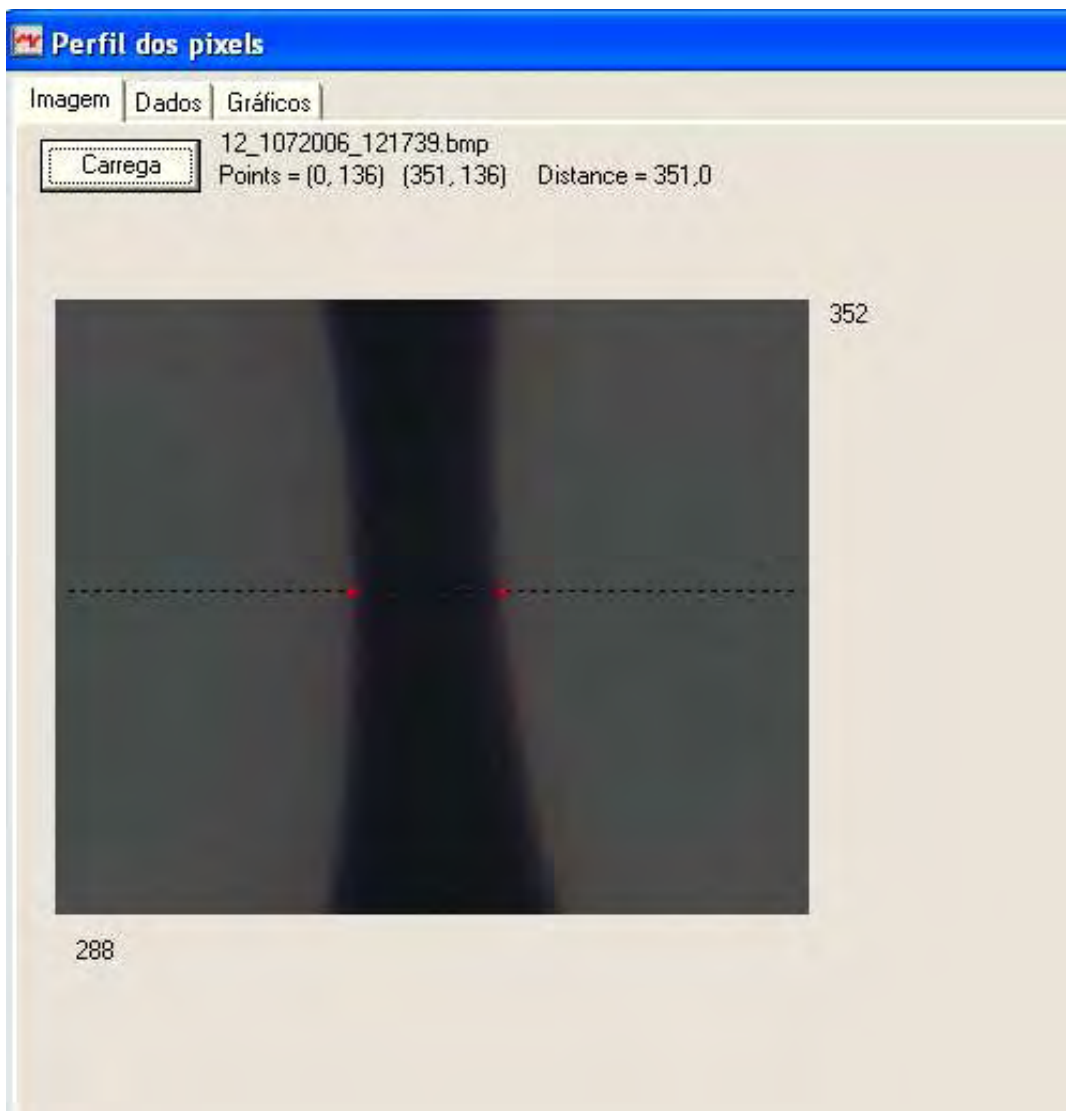

Fig. 6.5: Imagem dupla do disco solar, mostrando a linha que está sendo analisada (linha 136). Os pontos vermelhos mostram as coordenadas para os bordos nesta linha.

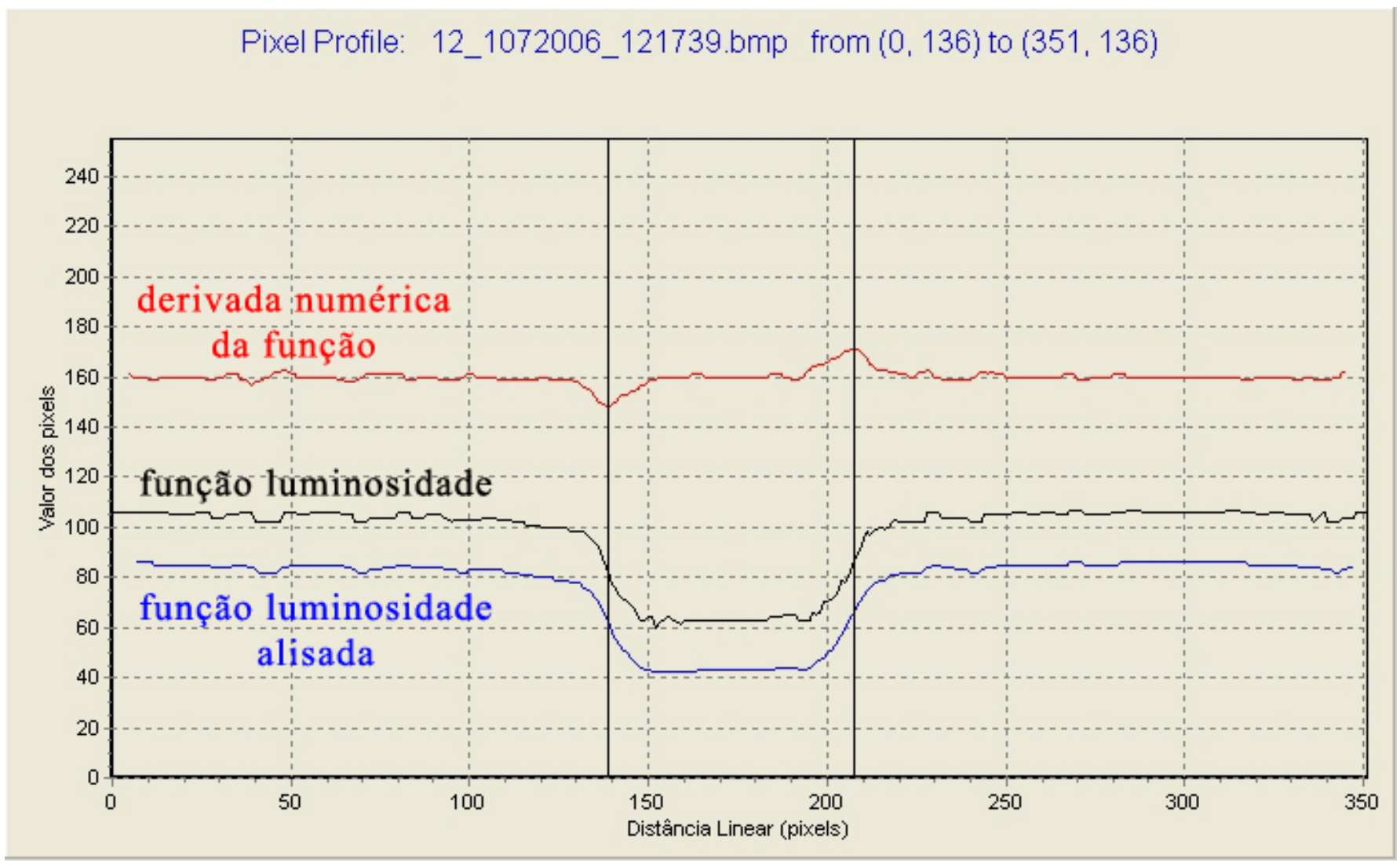

Fig. 6.6: Análise da linha 136 da imagem anterior. As linhas verticais indicam a localização dos pontos de inflexão da curva alisada. A escala de valores dos *pixels* não tem significado na figura, pois as curvas foram deslocadas verticalmente para melhor visualização.

Caso os extremos sejam formados por pontos seqüenciais de mesmo valor, o programa considera o ponto médio entre eles como a abscissa do ponto de inflexão.

Depois de identificar os dois conjuntos de pontos de inflexão em todas as linhas da imagem, o programa encontra as coordenadas dos centros e os raios dos discos solares, em primeira aproximação, pelo seguinte ajuste de funções parabólicas:

Equação da parábola, $x = f(y)$:

$x = a + by + cy^2$ Eq. 6.1

Equação da circunferência:

$$(y-y_0)^2 + (x-x_0)^2 = r_0^2 \Leftrightarrow (x-x_0)^2 = r_0^2 - (y-y_0)^2 \Leftrightarrow x - x_0 = \pm\sqrt{r_0^2 - (y-y_0)^2}$$

Escolhendo a raiz positiva:

$$x - x_0 = \sqrt{r_0^2 - (y-y_0)^2} \Leftrightarrow x = r_0\sqrt{1-\left(\frac{y-y_0}{r_0}\right)^2} + x_0$$

Na região onde $y \to y_0$, o segundo membro do radicando é << 1, então podemos fazer a seguinte aproximação:

$$x \cong r_0\left[1-\frac{1}{2}\left(\frac{y-y_0}{r_0}\right)^2\right] + x_0 = r_0 - \frac{y^2 - 2y_0 y + y_0^2}{2r_0} + x_0\,;$$

$$x \cong r_0 - \frac{y^2}{2r_0} + \frac{y_0 y}{r_0} - \frac{y_0^2}{2r_0} + x_0\,;$$

$$x \cong \left(r_0 + x_0 - \frac{1}{2}\frac{y_0^2}{r_0}\right) + \left(\frac{y_0}{r_0}\right)y + \left(-\frac{1}{2r_0}\right)y^2 \qquad \text{Eq. 6.2}$$

Comparando Eq. 6.1 e 6.2:

$$-\frac{1}{2r_0} = c \Leftrightarrow \boxed{r_0 = -\frac{1}{2c}}$$

$$\frac{y_0}{r_0} = b \Rightarrow \frac{y_0}{-1/2c} = b \Leftrightarrow \boxed{y_0 = -\frac{b}{2c}}$$

$$r_0 + x_0 - \frac{1}{2}\frac{y_0^2}{r_0} = a \Leftrightarrow x_0 = a + \frac{1}{2}\frac{y_0^2}{r_0} - r_0 \Rightarrow x_0 = a + \frac{1}{2}\frac{b^2/4c^2}{-1/2c} + \frac{1}{2c} \Rightarrow \boxed{x_0 = a - \frac{b^2}{4c} + \frac{1}{2c}}$$

→ $x_0$, $y_0$ são as coordenadas iniciais do centro da melhor circunferência que se ajusta à parábola e $r_0$, seu raio inicial.

Com estes pontos de partida, duas circunferências são ajustadas por mínimos quadrados, eliminando-se os pontos que estiverem além de ±3 σ do ajuste, corrigindo os valores iniciais e calculando a incerteza das medidas.

A reta que une os centros dos discos é ajustada e as coordenadas dos pontos onde esta reta faz interseção com as circunferências são calculadas. A separação entre estes pontos representa a menor distância entre os discos.

Na imagem da figura 6.7, as cruzes (em amarelo) indicam os pontos mais extremos dos discos. A linha tracejada representa a reta que une os centros dos discos e o segmento de reta (em vermelho) é a distância mínima encontrada. Em torno dos bordos (em azul), aparecem as circunferências ajustadas. A primeira linha de valores, embaixo da imagem, representa as coordenadas dos centros (abscissa e ordenada) para os discos, esquerdo e direito, respectivamente. Abaixo desta linha, entre os valores supracitados, está o raio de cada disco. Todos os valores estão em escala de *pixel*.

$A \longrightarrow$

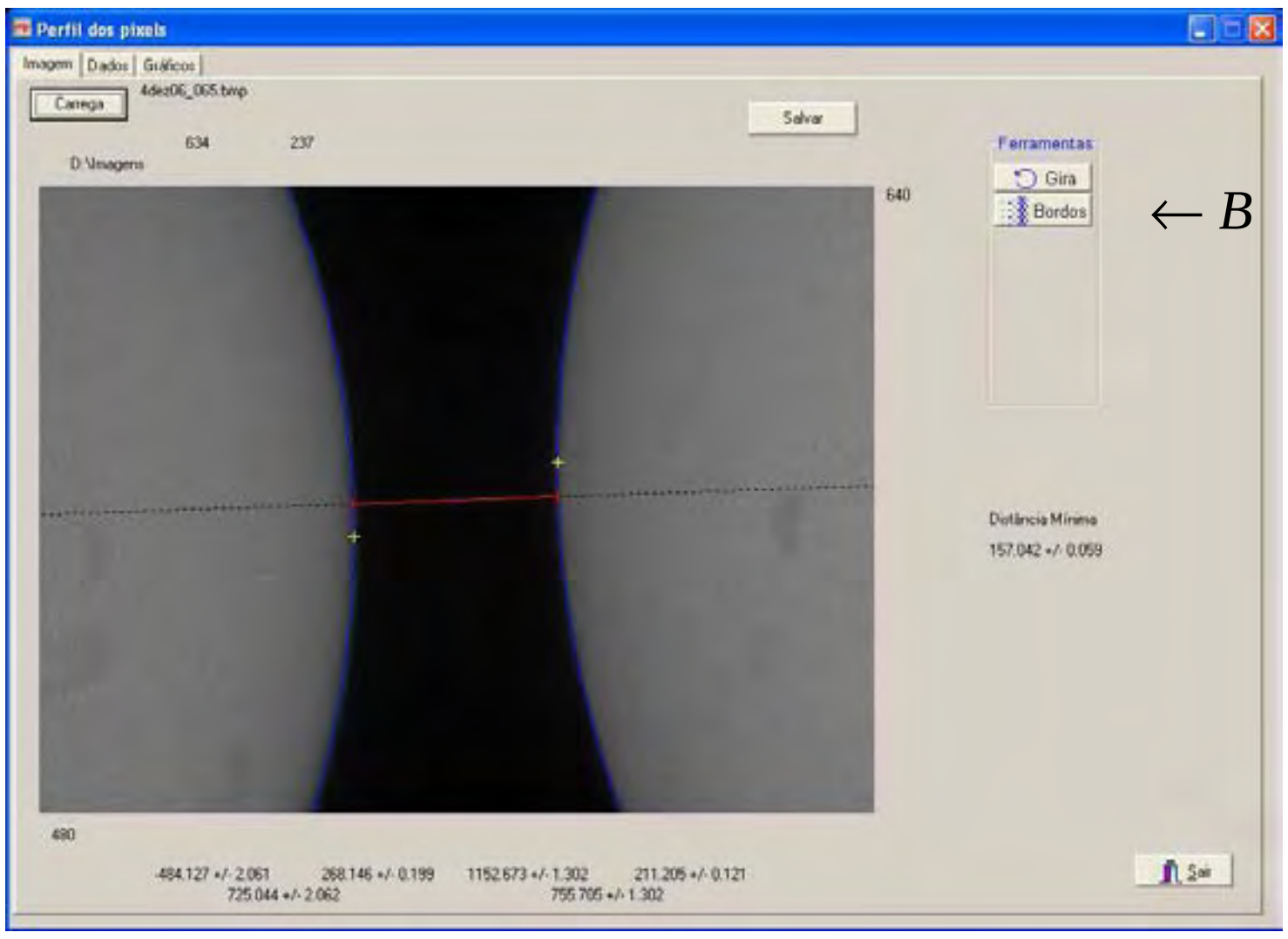

$\longleftarrow B$

Fig. 6.7: Interface do antigo programa de análise das imagens heliométricas. A letra *A* indica o botão para carregar as imagens, a letra *B*, o botão para o exame das imagens.

Ainda sobre a imagem anterior, a matriz do CCD não estava alinhada com o deslocamento dos centros dos discos solares, produzido pelos espelhos planos em diedro. Neste caso, os pontos extremos dos discos, indicados pelas cruzes, não são os pontos mais próximos entre os bordos. Isto só acontece quando há alinhamento com o deslocamento dos centros dos discos, seja qual for o método heliométrico de duplicação utilizado. A imagem da figura 6.8 ilustra este efeito:

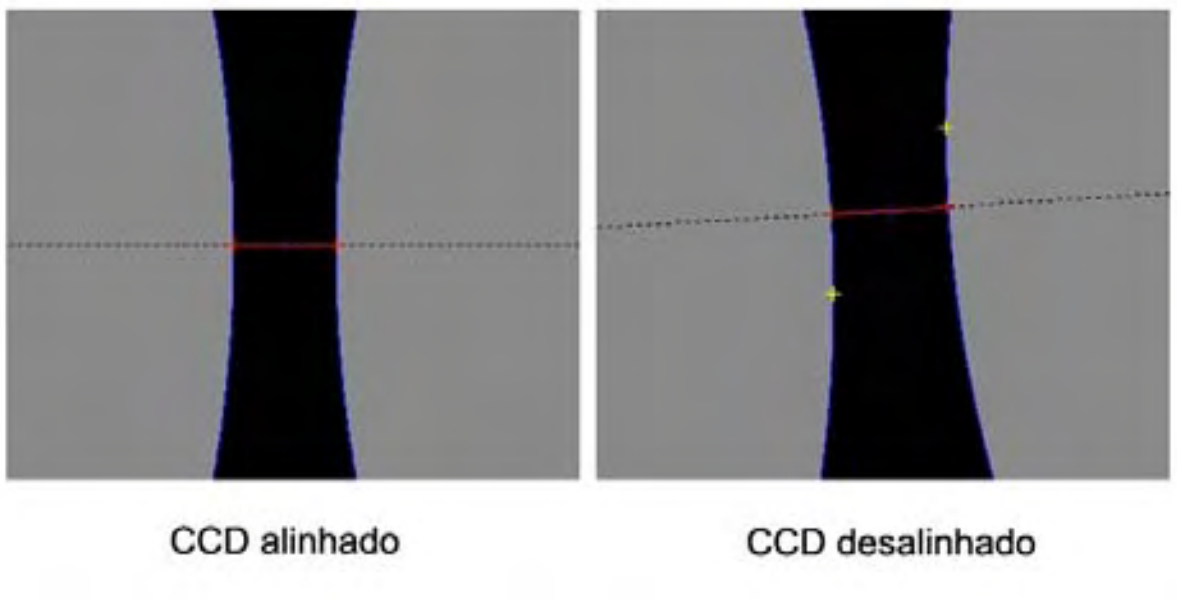

Fig. 6.8: À esquerda, os pontos extremos são também os mais próximos. À direita, não.

A incerteza no valor dos raios encontrados está correlacionada às incertezas dos valores das abscissas do centro do disco. Como este ajuste é não-linear, a retirada de pontos (além de $\pm 3\ \sigma$) no ajuste modifica enormemente seu valor.

A incerteza nos valores das ordenadas dos centros é cerca de 1 ordem de grandeza menor. A incerteza na distância mínima tem dependência com esta última, mas é cerca de 2 ordens de grandeza menor. Ou seja, para uma imagem, consegue-se determinar a distância mínima entre os discos solares com precisão de centésimos de *pixel*.

Não só a data, mas o instante da captura da imagem heliométrica, com precisão do milésimo de segundo, ou seja, 100 vezes a precisão requerida para o cálculo ds da escala de placa a partir das efemérides solares, são informações fundamentais que são armazenadas juntamente com a imagem. O CCD do Heliômetro tem dimensões angulares da ordem de 500 *pixels*. Como a distância focal é de 1 m, para um *pixel* 5,6 μm a escala de placa nominal é de 1,15"/*pixel*. Para utilizar o trânsito da imagem solar sobre o CCD, tipicamente a precisão temporal de 0,1 s/*pixel* deve estar disponível, logo a datação ao milésimo de segundo é suficiente e robusta.

### 6.2.1 - Processo de coleta automática das variáveis da imagem

A primeira etapa da análise automática das imagens consiste em armazenar as imagens, e nomeá-las incorporando a data e tempo da aquisição

O procedimento, ao mesmo tempo, permite a visualização e inspeção das imagens. Para mostrar as imagens de uma sessão observacional, dá-se um duplo-clique em um dos diretórios, no lado direito superior, e o aplicativo carregará todas as imagens do sub-diretório "imagem", dispostas seqüencialmente. O número total das imagens deste diretório aparece acima das mini-imagens, figura 6.9.

Ao clicar em qualquer mini-imagem, a imagem selecionada é mostrada em tamanho real, no lado esquerdo, acionando automaticamente o programa de análise, em vez de ficar clicando no botão "Bordos", como anteriormente, figura 6.9.

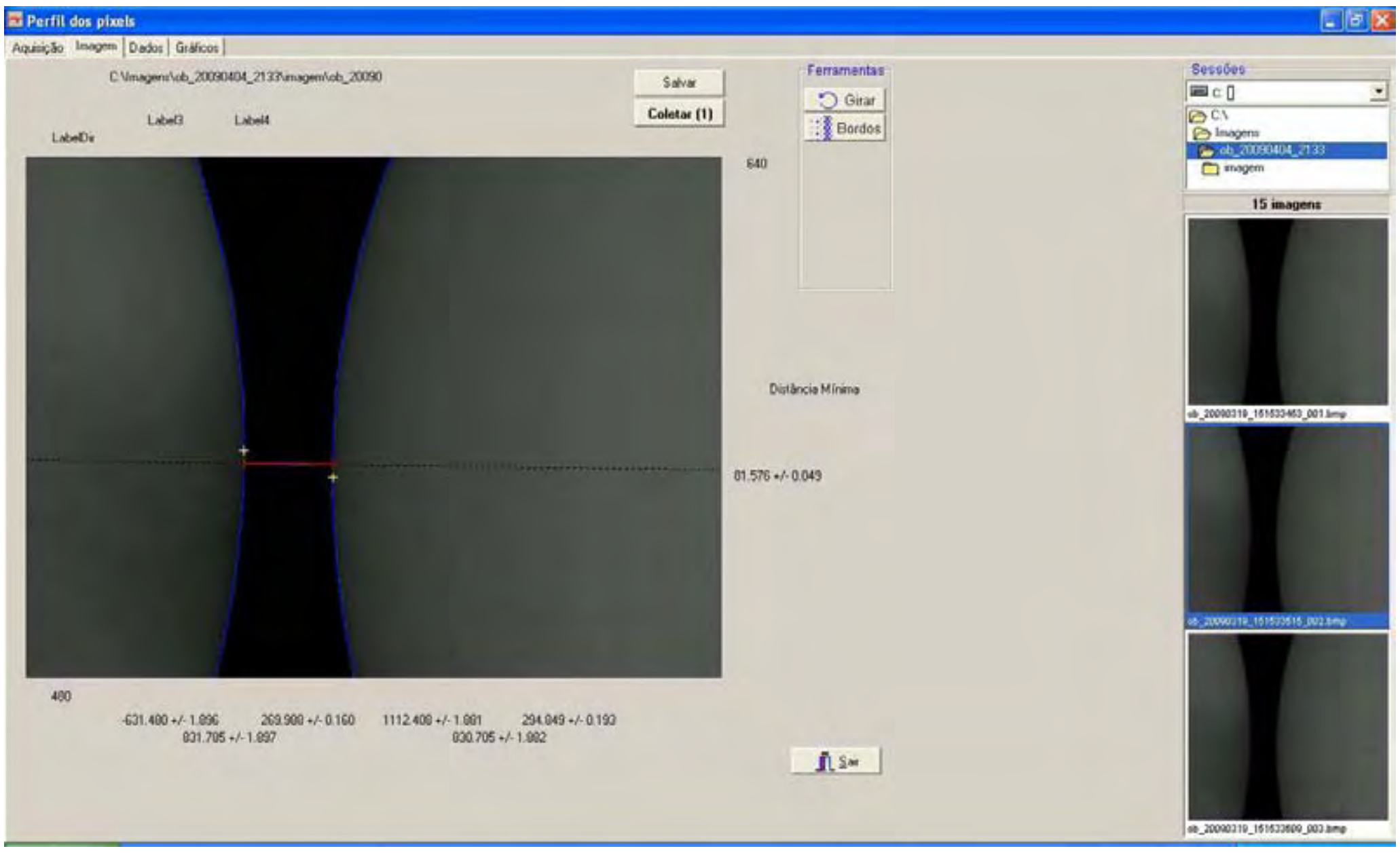

Fig. 6.9: Imagem heliométrica sendo analisada pelo novo programa de aquisição/análise.

Para realizar a coleta seqüencial das imagens da sessão escolhida, clica-se no botão "Coletar (1)" (fig. 6.9) e o programa passa a analisar ordenadamente todas as imagens presentes no diretório, gerando o arquivo texto de saída correspondente.

As principais características da coleta das imagens (fig. 6.10) são mostradas na Tab.6.2.

Tabela 6. 2: características da coleta das imagens

| | |
|---|---|
| **1** | O controle "inc", que define o valor de incremento da quantidade de imagens que serão selecionadas pelo controle "barra scroll", foi acrescentado. |
| **2** | Foi criado um limite para o número de imagens capturadas por sessão observacional (campo "max", da figura 6.10). |
| **3** | O campo "Número de imagens requeridas" pode ser atualizado manualmente independente do controle "barra scroll", sendo que os dois sempre terão o valor total de imagens requeridas para a captura de uma sessão observacional, um controle atualiza o outro. |
| **4** | O valor da quantidade total de imagens da última sessão observacional do campo "Número de imagens requeridas" é gravado e recuperado do arquivo PixelProfile.ini, quando reinicia-se o programa. |
| **5** | Na aba "imagem", foi adicionado um controle de "autoscroll" da imagem capturada selecionada, já prevendo que imagens com tamanho superior a 640×480 *pixels* possam ser visualizadas sem prejuízo de corte de partes da imagem. |
| **6** | Se por algum motivo a imagem carregada numa seqüência não puder ser analisada, o programa ignora esta imagem, passando para a seguinte, criando um arquivo tipo "log", contendo o número desta(s) imagem(ns). |

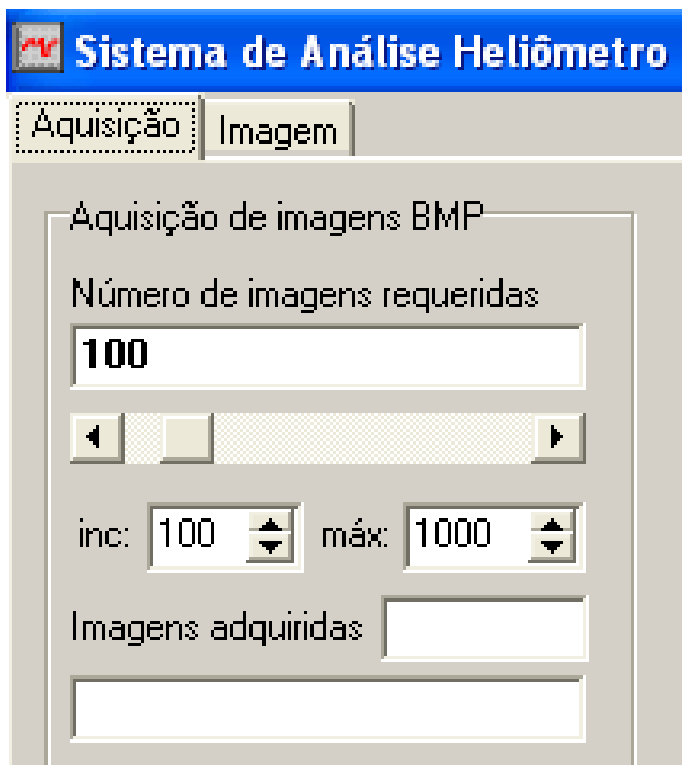

Fig. 6.10: Atualização do programa de captura das imagens heliométricas.

Depois da Campanha Observacional de maio de 2009, o programa foi atualizado para incluir no nome dos arquivos de imagem o Ângulo Polar da Câmara, a Temperatura e a Pressão Barométrica (fig. 6.11), informações relevantes para a correção da refração atmosférica, que será comentada mais adiante.

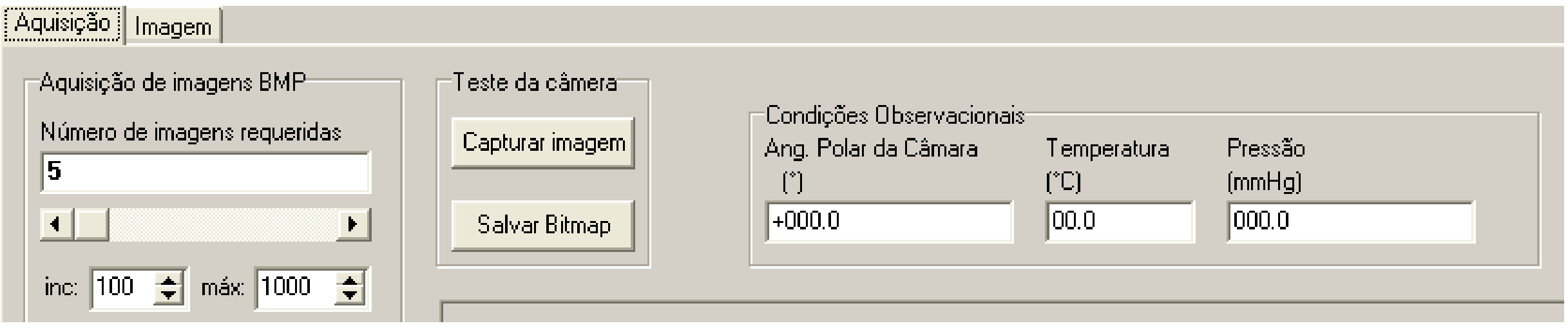

Fig. 6.11: Última atualização do programa de aquisição das imagens heliométricas.

As imagens, então, passaram a receber nomes como:

ob_20090604_120122546_-020.0_35.6_766.2_001.bmp.

No exemplo acima, esta imagem obtida no dia 4 de junho de 200 9, às 12 h, 01 min e 22,546 s, com ângulo polar da câmara de -20,0º, a temperatura de 35,6 ºC e sob pressão barométrica de 766,2 mmHg.

### 6.2.2 - Desenvolvimento do gerenciamento das imagens adquiridas

O programa de aquisição desenvolvido gera um arquivo de texto contendo informações completas para cada imagem analisada (Tab.6.3).

Na Tabela 6.3, da linha 4 até a linha 15 (primeiras linhas, grifadas em cinza), os valores se referem ao ajuste de uma circunferência para os pontos da borda do disco esquerdo da imagem, e da linha 16 até a 27 os valores se referem ao ajuste de uma circunferência para os pontos da borda do disco direito da imagem.

A 28ª linha é o principal valor procurado, a distância mínima entre os bordos.

À medida em que cresceu o número de imagens adquiridas e as rotinas de coleta e análise das imagens puderam ser exaustivamente testados, ambos as rotinas foram fundidas num único programa. O programa resultante é dito de "gerenciamento das imagens adquiridas", e seu próprio código foi depurado obtendo-se maior performance e eliminação de *bugs*.

O fluxograma abaixo detalha a estrutura pasta-sessão – "ob_20090404_2133"

```
|_ ob_20090404_2133 (ob = observação solar, 20090404 = aaaammdd, 2133 = hhmm)
  |
  |_ imagem
  |      |_ ob_20090319_151533453_001.bmp
  |      |_ ob_20090319_151533515_002.bmp
  |                        . . .
  |      |_ ob_20090319_151534406_nnn.bmp
  |
  |_ ob_20090404_2133_ED.txt
```

Tabela 6.3: Lista dos parâmetros de saída do programa de análise.

| VARIÁVEL | SIGNIFICADO |
|---|---|
| frame | nº da imagem |
| data | dia, mês e ano da observação |
| hora | hora, minuto, segundo e milissegundo da observação |
| NEF | nº de pontos por linha |
| NEV | nº de pontos utilizados (dentro de $\pm 3\sigma$) |
| DPE | desvio padrão |
| XEFinal | coordenada *X* do centro |
| SigXEFinal | incerteza da coordenada *X* do centro |
| YEFinal | coordenada *Y* do centro |
| SigYEFinal | incerteza da coordenada *Y* do centro |
| RaioEFinal | raio |
| SigRaioEFinal | incerteza do valor do raio |
| XextE | coordenada *X* do ponto extremo |
| YextE | coordenada *Y* do ponto extremo |
| SigXextE | incerteza da coordenada *X* do ponto extremo |
| NDF | nº de pontos por linha |
| NDV | nº de pontos utilizados (dentro de $\pm 3\sigma$) |
| DPD | desvio padrão |
| XDFinal | coordenada *X* do centro |
| SigXDFinal | incerteza da coord. *X* do centro |
| YDFinal | coordenada *Y* do centro |
| SigYDFinal | incerteza da coordenada *Y* do centro |
| RaioDFinal | raio |
| SigRaioDFinal | incerteza do valor do raio |
| XextD | coordenada *X* do ponto extremo |
| YextD | coordenada *Y* do ponto extremo |
| SigXextD | incerteza da coordenada *X* do ponto extremo |
| DistMin | distância mínima entre os bordos |
| SigDistMin | incerteza da distância mínima entre os bordos |
| Xmed | coordenada *X* do ponto médio entre os bordos |
| Ymed | coordenada *Y* do ponto médio entre os bordos |

Na imagem da figura 6.12 podemos ver como as interfaces dos programas de aquisição e análise se fundiram, com a eliminação das etapas como escolha do disco e do nome do arquivo.

Para inicializar a câmara clica-se em "Começar e esperar (1)". Quando as imagens heliométricas do Sol estão no campo, na posição escolhida, clica-se em "Aquisição (2)", e a seqüência programada de imagens é capturada e armazenada em disco, figura 6.12.

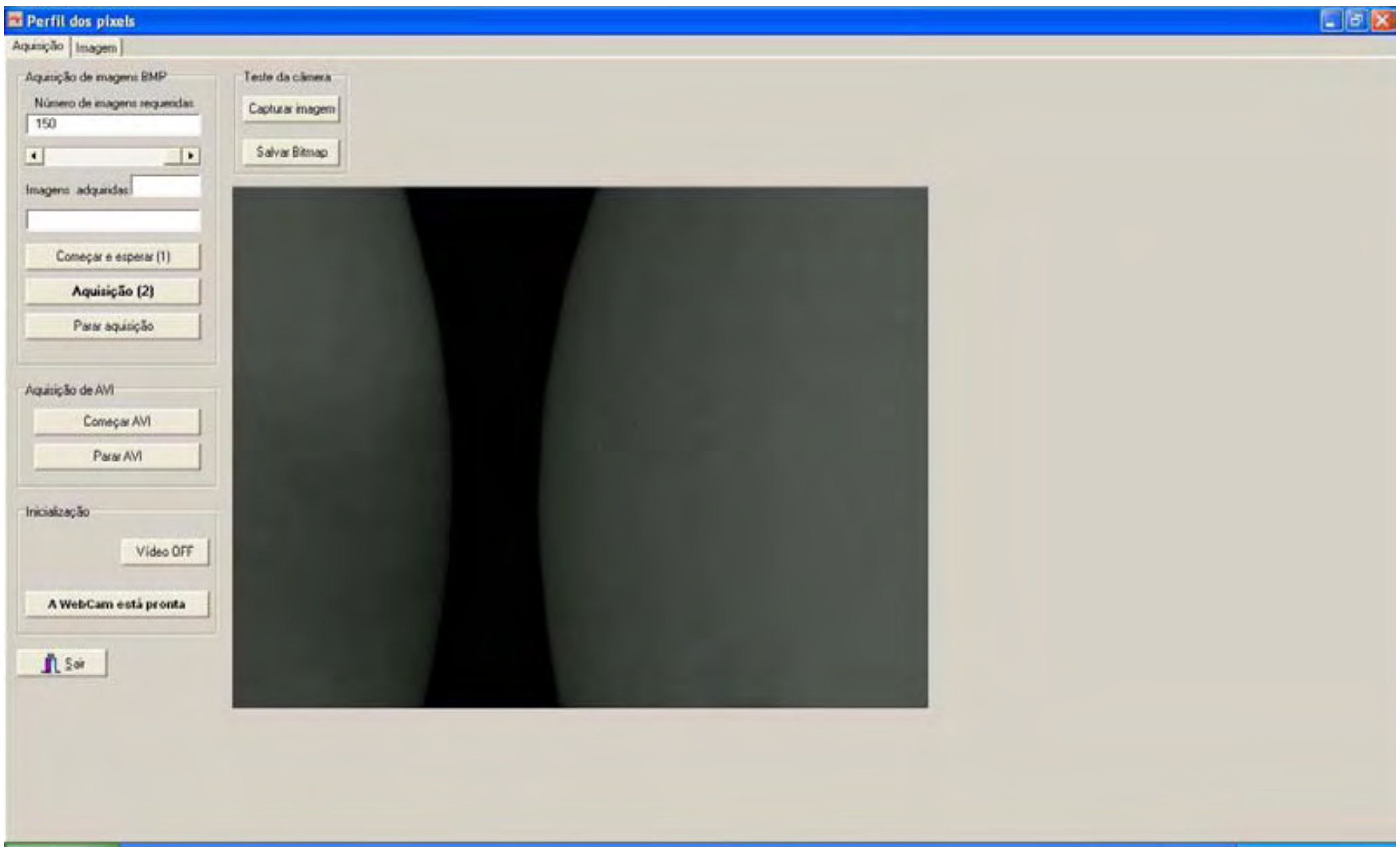

Fig. 6.12: Interface do programa de aquisição/análise das imagens heliométricas.

O gráfico da figura 6.13 traz a quantidade do número de imagens heliométricas obtidas ao longo da evolução dos protótipos e dos programas.

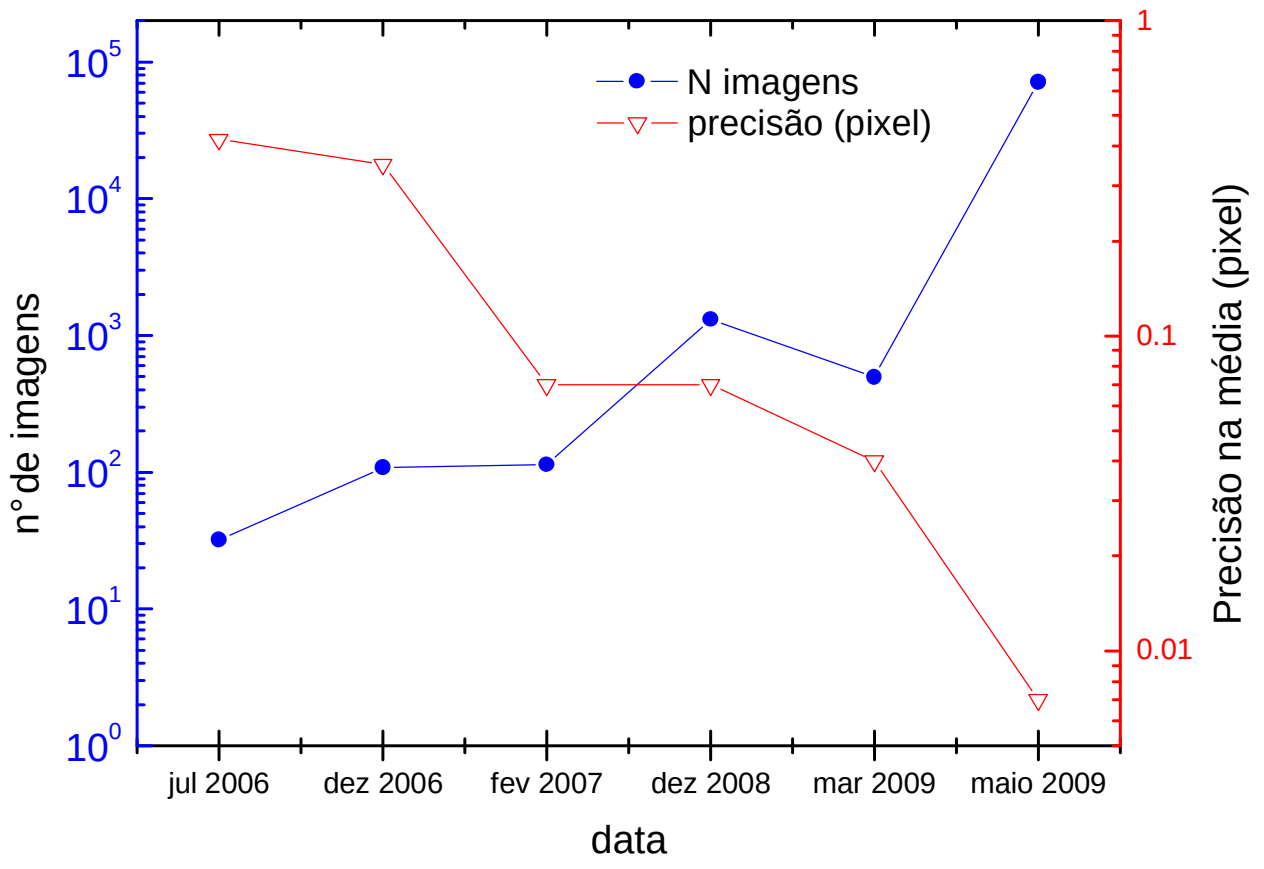

Fig. 6.13: Evolução do n° de imagens heliométricas.

## 6.3 - Correção da refração atmosférica diferencial.

### 6.3.1 - Refração atmosférica

Os termos diferenciais da astrometria da posição verdadeira de um corpo do sistema solar tem efeito completamente negligenciável para a determinação angular do diâmetro solar. Este não é o caso, contudo, para a refração atmosférica. Neste caso a correção às posições aparentes dos limbos devem ser compensadas. A consideração teórica é a usual. Quando a luz de um astro penetra na atmosfera terrestre, vai encontrando, progressivamente, camadas de ar cada vez mais densas e, conseqüentemente, com índices de refração crescente. O caminho percorrido pela luz não pode ser mais aproximado a uma linha reta, e vai, à medida que se aproxima do solo, curvando-se em direção à vertical.

Para o tratamento diferencial, é suficiente considerar uma atmosfera estratificada, negligenciando a curvatura da Terra[5], constituída de N camadas plano-paralelas, cada uma com seus respectivos índices de refração $n_0$, $n_1$, ... , $n_{N-1}$ e $n_N$, como mostra a figura 6.14. A última camada, já no espaço aberto, tem índice de refração unitário.

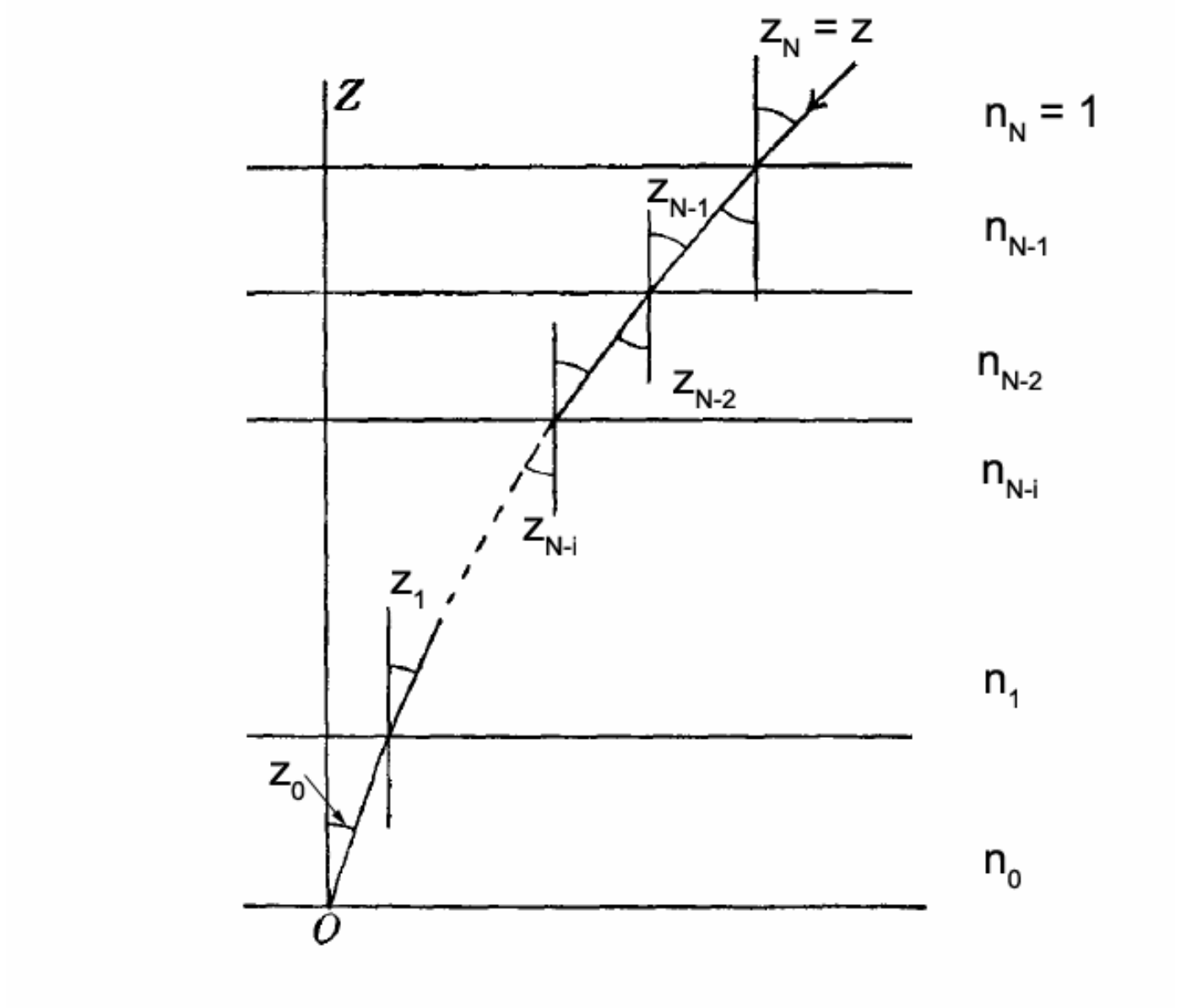

Fig. 6.14: Esquema do desvio do raio de luz devido à mudanças do índice de refração das camadas atmosféricas.

[5] Em valores absolutos, esta aproximação também é válida em pequenas distâncias zenitais. Levar-se aí em conta que o efeito da refração só é significativo na região mais densa da Atmosfera, a Troposfera, cuja escala de altura não ultrapassa 5km. O raio de curvatura da Troposfera é muitíssimo maior do que sua extensão vertical, portanto a aproximação do modelo de camadas da Atmosfera plano-paralelas é razoável próximo ao zênite.

Conforme mostra a figura 6.14, o ângulo de entrada que o raio de luz faz com a vertical é *z*. Ao penetrar na camada N-1, com índice de refração $n_{N-1}$, este ângulo modifica-se para $z_{N-1}$, e assim sucessivamente até atingir o solo num ângulo $z_0$, que será a distância zenital observada da fonte.

Aplicando-se a Lei de Snell, sucessivamente, em cada interface, temos:

$$\begin{cases} n_{N-1}\sin z_{N-1} = n_N \sin z_N = \sin z \\ n_{N-2}\sin z_{N-2} = n_{N-1}\sin z_{N-1} = \sin z \\ \vdots \\ n_1 \sin z_1 = n_2 \sin z_2 = \sin z \\ n_0 \sin z_0 = n_1 \sin z_1 = \sin z \end{cases}$$

Por fim, temos apenas: $n_0 \sin z_0 = \sin z$ Eq. 6.3

Este resultado independe da quantidade de camadas em que a Atmosfera foi dividida. Fica evidente que no limite de $N \rightarrow \infty$, a variação do índice de refração da atmosfera com a altura é contínua.

O ângulo *z* a distância zenital topocêntrica. Como $n_0 > 1$, a distância zenital observada ($z_0$) é menor do que a topocêntrica, com diferença crescente à medida que a distância zenital aumenta.

O ângulo $R = z - z_0$ é o ângulo de refração. Desta forma, substituindo na Eq. 6.3 e expandindo, temos:

$$z = z_0 + R$$

$$\Rightarrow n_0 \sin z_0 = \sin(z_0 + R) = \sin z_0 \cos R + \cos z_0 \sin R$$

Como *R* tem valor pequeno, podemos usar a aproximação para pequenos ângulos. Sendo assim:

$$\sin z_0 + \cos z_0 R = n_0 \sin z_0$$

$$\cos z_0 R = (n_0 - 1)\sin z_0$$

$$R = (n_0 - 1)\tan z_0$$

O resultado acima está expresso em radianos. Convertendo para segundos de arco, temos:

$R = K \tan z_0$, onde $K = 206265(n_0 - 1)$

Diferenciando para os bordos solares:

$\Delta R = K \sec^2 z_0 \, \Delta z$

Para distância zenital de 60°, a expressão corresponde a pouco mais de 2", de modo que os termos quadráticos abandonados na descrição da atmosfera paralela são da ordem de 15 *mas*. Sendo improvável que a experiência Heliométrica seja jamais conduzida a esta altura, ainda assim o erro cometido é inferior a componente diferencial da refração anômala. De toda forma, a expressão completa da refração, incluindo termos quadráticos será incluída na expressão final calculada. No restante deste capítulo ela é inútil.

O índice de refração $n_0$ tem dependência com as condições locais da atmosfera. Nas Condições Padrão de Temperatura e Pressão (CPTP), ou seja, 760 mmHg e 0°C, seu valor adotado é:

$n_0 = 1{,}0002927$

O que leva a um *K* = 60",4.

K que é a constante de refração, no comprimento de onda do visível..

O valor de *K* para condições não-padrão pode ser derivado da Lei de Dale-Glastone, que relaciona $(n_0 - 1)$ proporcionalmente à densidade do ar (Green, 1985).

$$(n_0 - 1) \propto \frac{P}{(273 + T)}$$

Usando os valores CPTP, da pressão e temperatura, a fórmula do ângulo de refração *R* fica:

$$R = 1 + 0{,}0002927 \frac{P/760}{(1 + T/273)} \tan z_0 \qquad \text{Eq. 6.4}$$

O método heliométrico baseia-se na medida da distância mínima entre bordos que estão a distâncias zenitais topocêntricas diferentes, de modo geral. Cada bordo sofrerá uma refração atmosférica diferente, sendo que seu efeito será maior para o bordo mais distante.

A figura 6.15 exemplifica o problema e resume a correção aplicada. Nela, **Z** representa a direção do zênite. Podemos ver que o deslocamento do bordo inferior, por estar mais distante, será maior do que o do inferior.

Ao deslocamento do bordo inferior chamaremos de $R(\underline{b})$ e ao do inferior, $R(\bar{b})$.

Se $d'$ é a distância observada e $d$, a distância desejada sem o efeito da refração, temos:

$$d = d' - R(\underline{b}) + R(\bar{b})$$

Onde,

$R(\bar{b}) =$ correção para o bordo superior.

$R(\underline{b}) =$ correção para o bordo inferior.

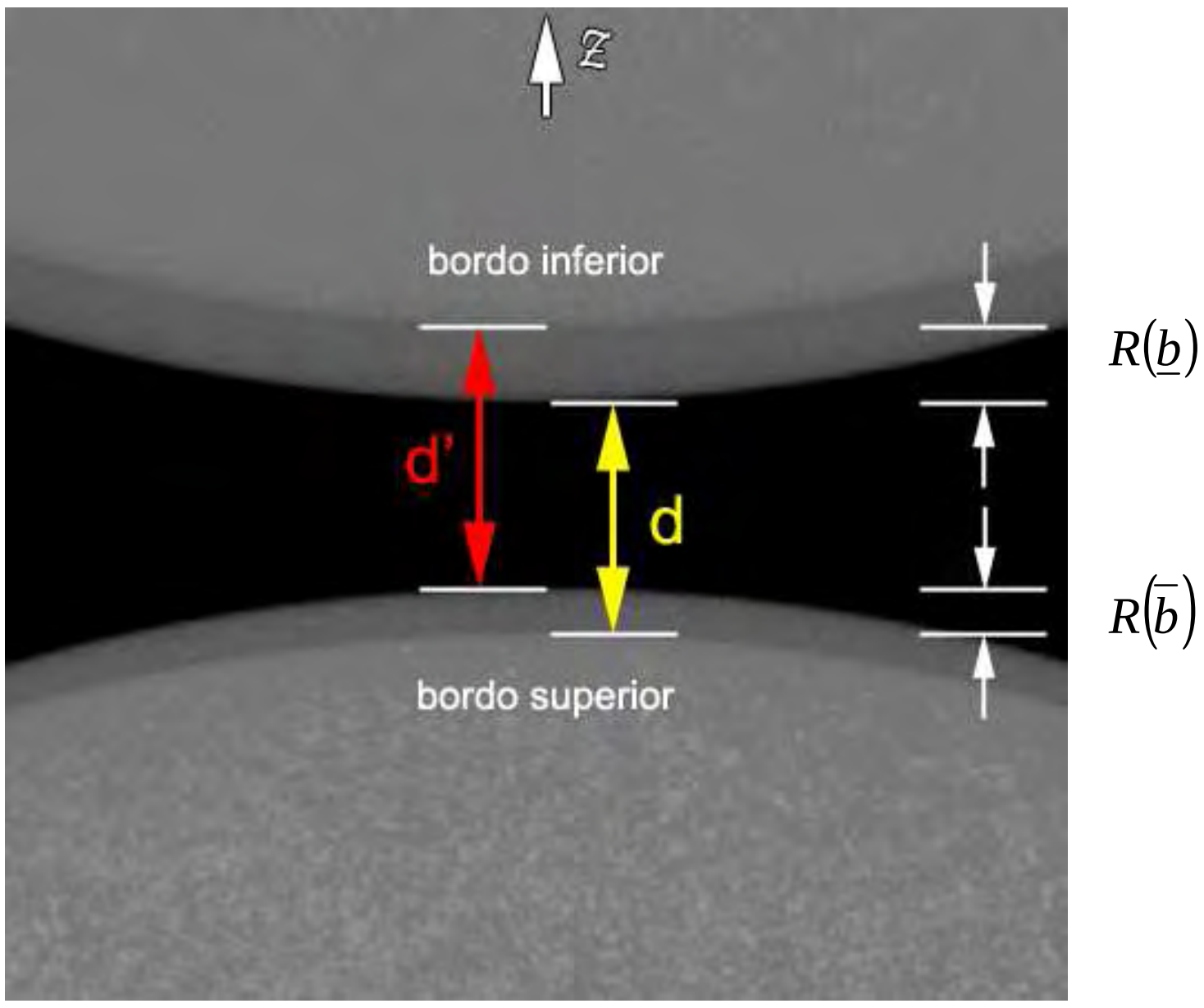

Fig. 6.15: Deslocamento (exagerado) dos bordos, superior e inferior, devido à refração atmosférica.

Se $\beta$ for o ângulo entre o deslocamento das imagens e o zênite, então a fórmula geral de correção é:

$$d = d' - \left[R(\underline{b}) + R(\bar{b})\right]\cos\beta \qquad \text{Eq. 6.5}$$

## 6.3.2 - Aplicação da correção de refração diferencial

A primeira convenção foi estabelecer uma referência angular para o desdobramento heliométrico das imagens. Foi chamado de θ, o ângulo polar da câmara CCD, contado a partir do norte como mostra a figura 6.16.

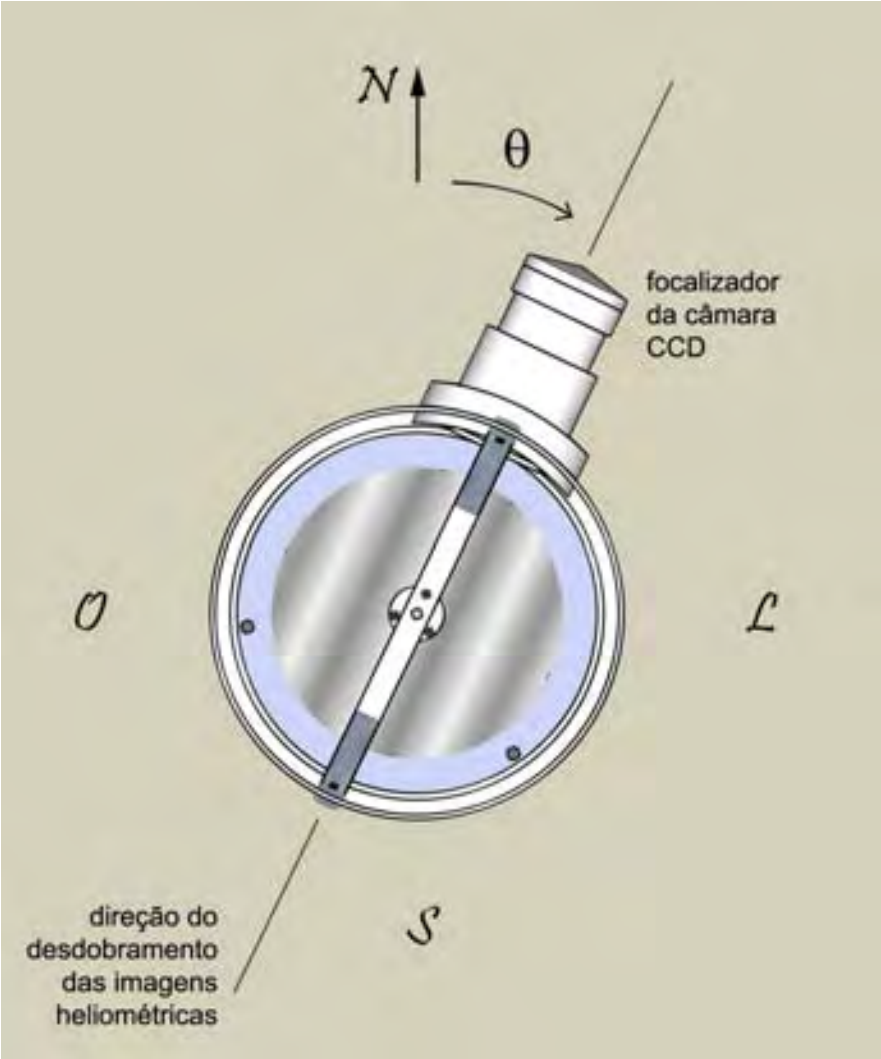

Fig. 6.16: Vista superior do Heliômetro e a convenção do ângulo polar da câmara CCD.

Os outros ângulos, figura 6.17, são:

- β, ângulo entre o zênite e câmara, ou seja a distância zenital do eixo do diedro do Heliômetro.
- S, ângulo entre o pólo e o zênite; ou seja a co-latitude
- APE, ângulo de posição do eixo do Sol, ou seja o complemento da inclinação do Sol.

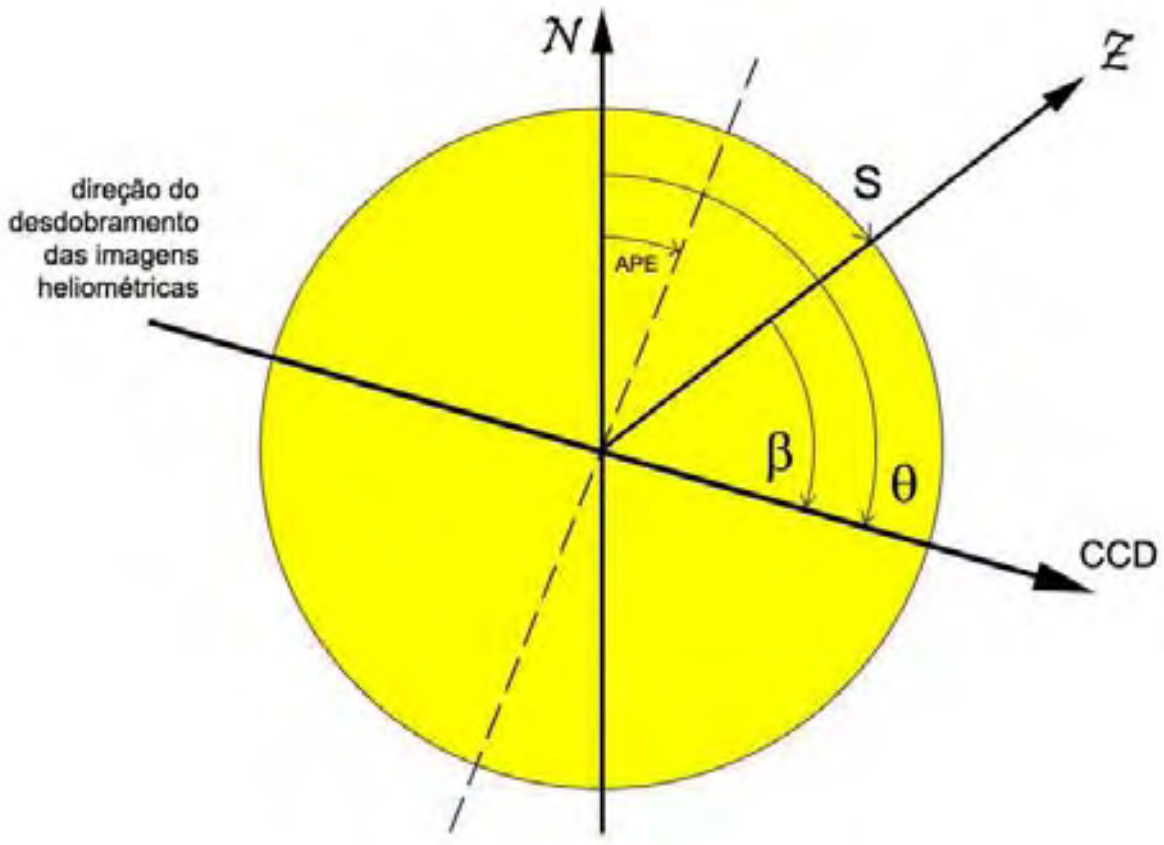

Fig. 6.17: Convenção dos ângulos envolvidos na correção.

O caso particular de $\theta$ = S ($\beta$ = 0) é foi mostrado na figura 6.15, onde o desdobramento da imagem se dá na direção do zênite.

O programa de correção utiliza as subrotinas do cálculo das efemérides, utilizadas no Anuário do Observatório Nacional.

Os parâmetros de entrada utilizados são:

- Ano, dia e mês;
- Hora, minuto, segundo e milisegundo;
- Ângulo polar do CCD, temperatura e pressão barométrica;
- Número do frame.

Estes parâmetros são escritos no próprio nome do arquivo, como no exemplo: ob_20090605_115719750_+166.4_28.7_766.4_010.bmp

Esta imagem heliométrica, por exemplo, foi feita no dia 5 de junho de 2009, às 11h, 57min e 19,750s, com ângulo polar do CCD de +166.4°, temperatura próxima ao instrumento de 28,7 °C e pressão barométrica de 766,4 mmHg.

Com estes dados, o programa calcula:

- Tempo Universal,
- Data Juliana;
- Ângulo horário local e distância zenital do centro do disco solar, sem efeito de refração;
- $\beta$ (ângulo do heliômetro a partir do zênite);
- Distância zenital dos bordos opostos, sem efeito de refração.

Chamando de $\bar{z}$ e $\underline{z}$, respectivamente, a distância zenital do bordo superior e a distância zenital do bordo inferior, temos pela Eq.2:

$$R(\bar{b}) = 1 + 0{,}0002927 \frac{P/760}{(1 + T/273)} \tan \bar{z}$$

$$R(\underline{b}) = 1 + 0{,}0002927 \frac{P/760}{(1+T/273)} \tan \underline{z}$$

Com estas correções, obtém-se a distância mínima correta aplicando-se a Eq. 6.5.

O valor do diâmetro do Sol é a diferença entre o valor do diedro e o valor da distância mínima corrigida, normalizado para 1 UA.

## *6.4 - Campanha observacional de maio de 2009*

Uma importante vantagem do método heliométrico vem do fato de podermos fazer medidas em qualquer heliolatitude, levando-se em conta a refração diferencial.

A figura 6.18 mostra como a imagem do disco do Sol é duplicada, para quatro posições do eixo de heliômetro. Neste exemplo, equador (posições B e D) e pólos solares (posições A e C).

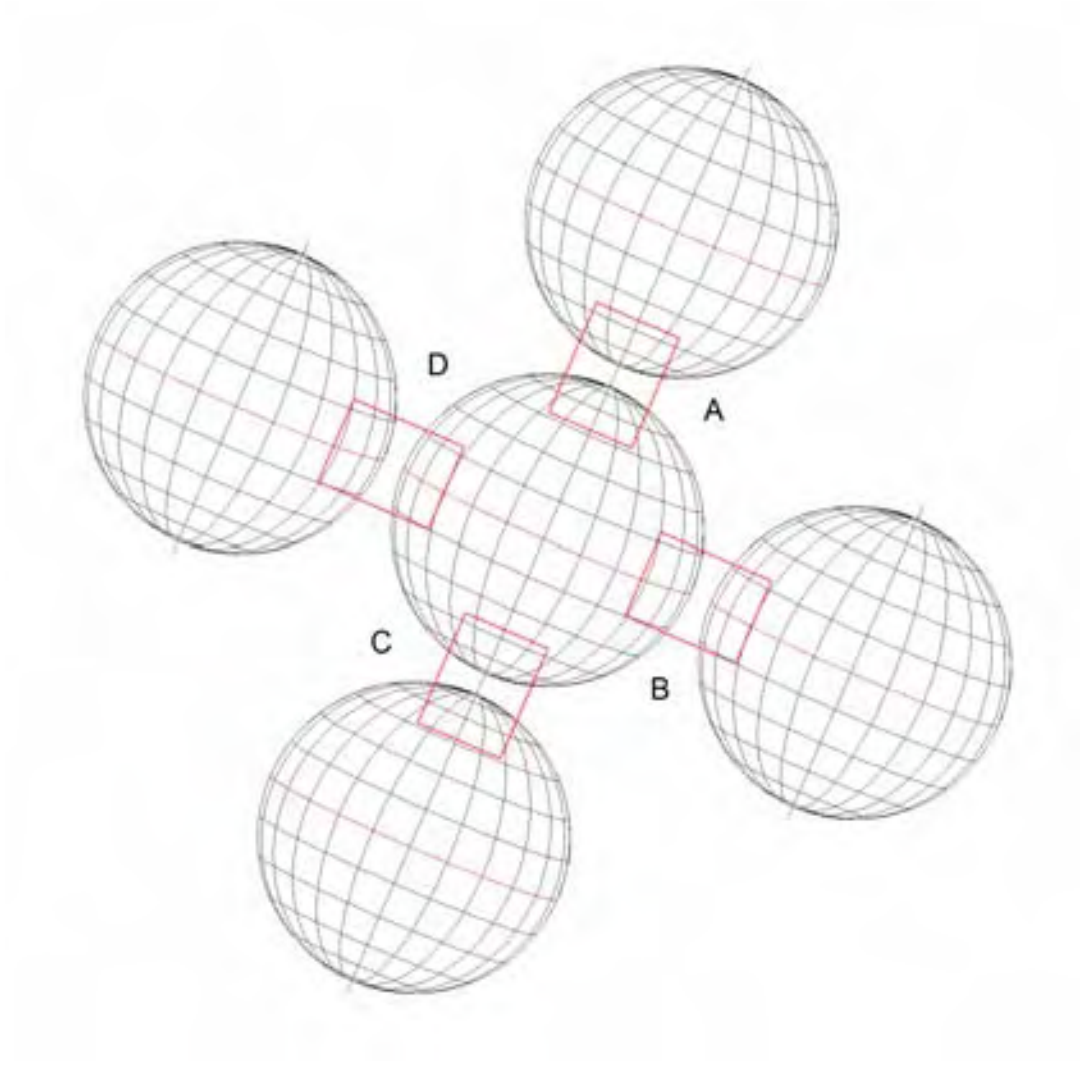

Fig. 6.18: Esquema do desdobramento das imagens solares para 4 posições. O retângulo central representa o campo do CCD.

Como todo o conjunto, espelho e CCD, é solidário, ao se girar o tubo do instrumento, pode-se ver pela figura 6.18 que a posição relativa entre as imagens no campo do CCD não muda e tudo se passa como se os discos solares girassem em sentidos opostos, figura 6.19.

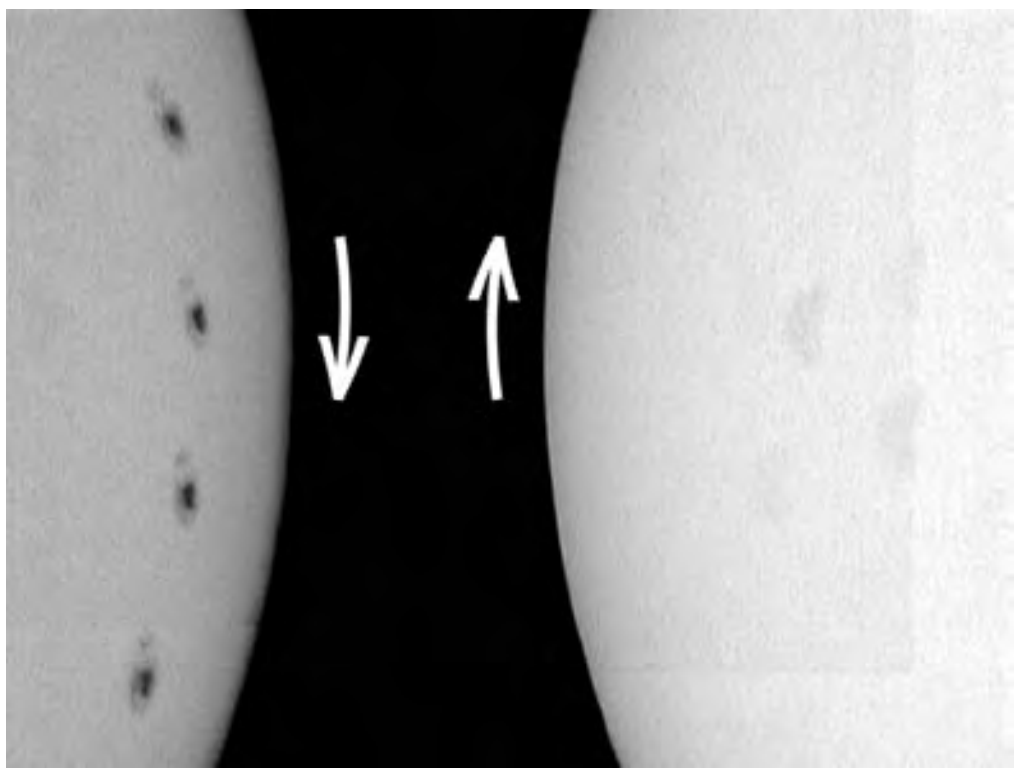

Fig. 6.19: Fotomontagem com 4 imagens dos discos solares, com 10º de diferença cada, simulando o giro do instrumento em seu eixo.

Nos dias 13, 14, 18, 19, 20, 21, 25, 26 e 27 de maio de 2009 foram realizadas observações heliométricas, no campus do Observatório Nacional, sempre em torno da passagem meridiana do Sol.

## 6.4.1 - Metodologia da observação

Cada dia de observação consistiu de 4 sessões consecutivas correspondentes a 4 heliolatitudes, separadas de ~90º cada[6]. Ajustava-se o ângulo polar da câmara CCD para o mais próximo possível do valor do APE, correspondendo ao semi-diâmetro polar e, na época, próximo ao máximo da refração diferencial. Esta era a 1ª heliolatitude observada.

Cada sessão, por sua vez, foi composta por 10 sub-sessões de 200 imagens cada, totalizando 2.000 imagens para cada heliolatitude. Sendo assim, foram obtidas 8.000 imagens para cada dia de observação e 72.000 imagens nestes nove dias do mês de maio de 2009.

A temperatura próxima ao telescópio e a pressão barométrica foram registradas para serem usadas no programa de correção da refração atmosférica das imagens solares.

[6] Aquela montagem não possuía nenhuma leitura da orientação do diedro heliométrico. A solução mais simples para contornar este problema foi estimar o 1º ângulo de posição calculando-se o quanto a linha de referência do tubo havia girado em relação ao seu próprio suporte. Esta posição no tubo era posteriormente girada de 90º em 90º.

## 6.4.2 - Resultados

A estimação inicial da qualidade das medidas foi através da verificação da distribuição dos valores das distâncias mínimas, sem correção alguma.

Como as imagens heliométricas foram obtidas em sessões com ângulos de desdobramento ora próximos à vertical, ora próximos à horizontal, era de se esperar que as medidas do primeiro grupo fossem mais afetadas pela refração atmosférica e apresentassem valores maiores do que os do segundo grupo.

Podemos ter uma idéia de qual deve ser esta diferença esperada, calculando a refração diferencial para uma distância zenital $z = 45^o$:

$R = K \tan z \Rightarrow \Delta R = K \sec^2 z \Delta z$ , onde $K = 60{,}4''$ e $\Delta z = 0{,}5^o \approx 8{,}72 \times 10^{-3}$ radianos

Substituindo-se os valores: $\Delta R \approx 1''$.

Esta diferença de valores realmente se verificou. Podemos ver na figura 6.20 que as medidas se reúnem em grupos separados.

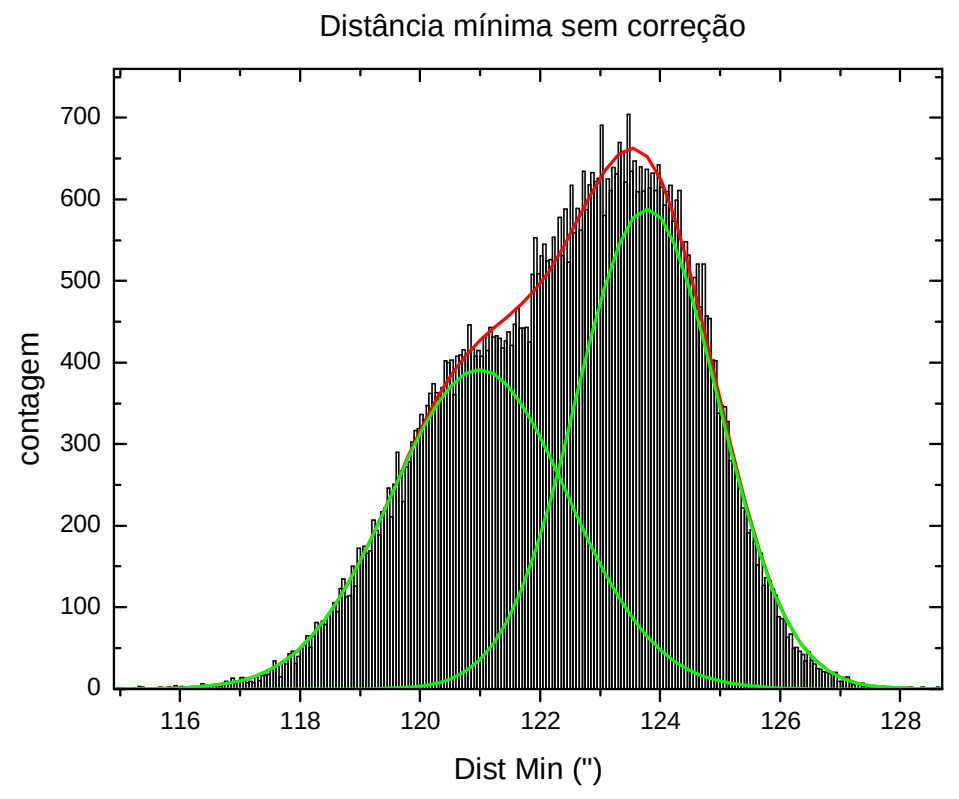

Fig. 6.20: Ajuste de gaussianas na distribuição dos valores da distância mínima, sem correção.

A separação entre os centros das gaussianas verifica qualitativamente o efeito esperado, sem a correção da refração vertical. Maiores diâmetros na direção horizontal do que na direção vertical. Na figura o efeito é amplificado pela não homegeneidade na distribuição das medidas, uma vez que a variação do diâmetro aparente no intervalo é de 5''.

A figura 6.21 traz a evolução do valor da distância mínima medida em comparação à distância mínima teórica (diedro de 2019" e valores do semi-diâmetro tirados do Anuário do Observatório Nacional, 2009). Os pontos pretos representam a média diária.

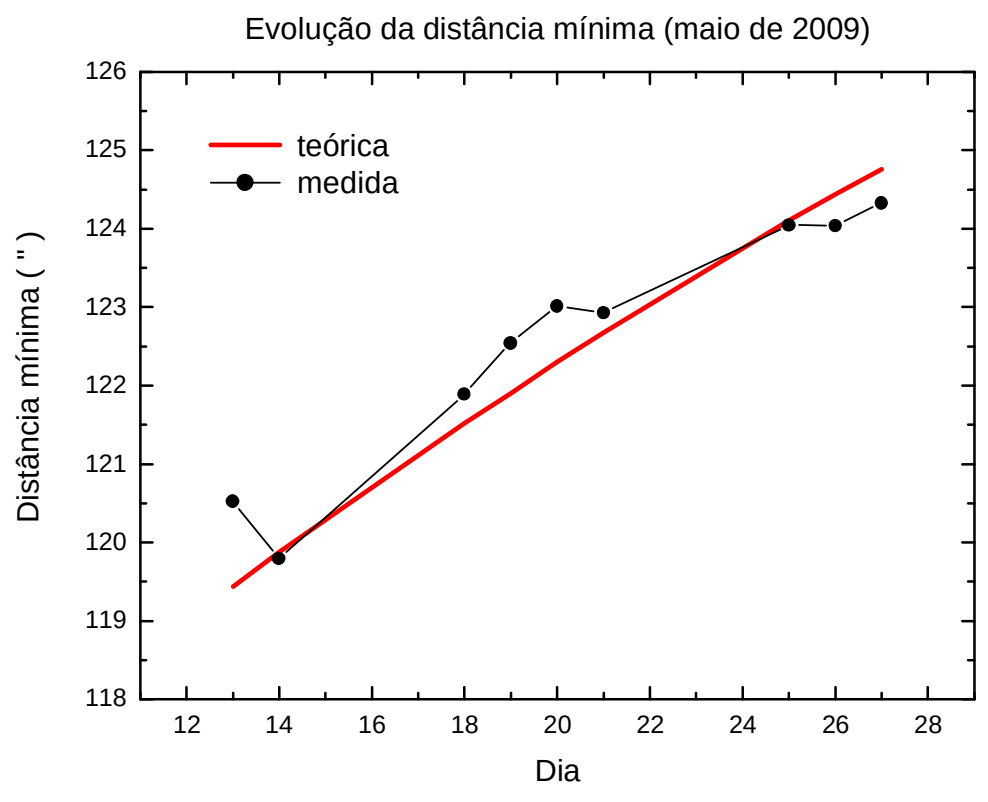

Fig. 6.21: Comparação entre a evolução do valor da distância mínima.

O gráfico mostra uma diferença entre o valor medido no dia 13 e o do dia 28 de ~5", o que explica a grande diferença entre as distribuições da figura 6.20.

Depois de corrigir as medidas da refração diferencial e normalizadas para 1 UA, o valor médio mensal das distâncias mínimas medidas foi de 118,247" ± 0,005", seguindo uma distribuição normal com $\sigma$ = 1,167".

A seguir o gráfico dos resíduos é apresentado da figura 6.22:

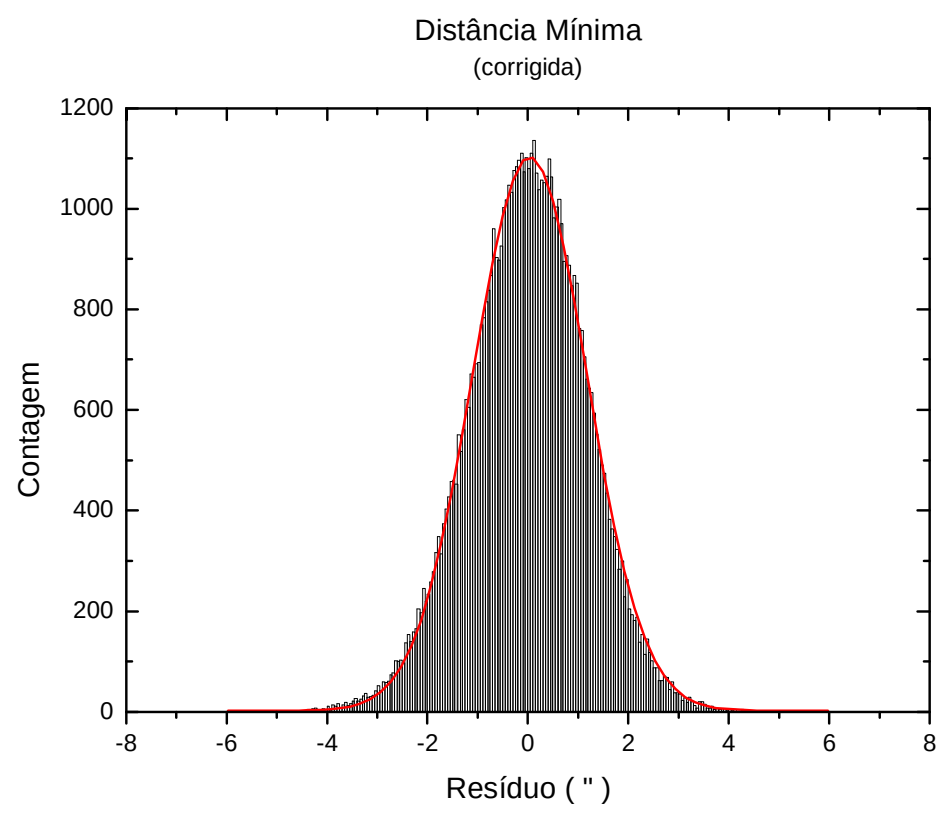

Fig. 6.22: Gráfico dos resíduos. Depois da correção as medidas seguem uma distribuição normal.

A tabela 6.4 traz os valores dos semi-diâmetros, normalizados para 1 UA, da campanha de 2009, para cada um dos dias e para o mês (fig. 6.23). Foram descartados todas as medidas além de ±3 σ da média do dia.

Tabela 6. 4: Média diária dos semi-diâmetros.

| **dia** | **nº de imagens** | **nº de imagens entre ± 3σ** | **<SD> (")** | **σ (")** |
|---|---|---|---|---|
| **13** | 7800 | 7726 | 959.084 ± 0.005 | 0.461 |
| **14** | 8200 | 8154 | 959.656 ± 0.006 | 0.559 |
| **18** | 8000 | 7924 | 959.440 ± 0.006 | 0.533 |
| **19** | 7800 | 7749 | 959.306 ± 0.005 | 0.459 |
| **20** | 8003 | 7941 | 959.367 ± 0.008 | 0.521 |
| **21** | 7997 | 7931 | 959.503 ± 0.007 | 0.639 |
| **25** | 7600 | 7553 | 959.646 ± 0.007 | 0.515 |
| **26** | 8000 | 7955 | 959.811 ± 0.006 | 0.501 |
| **27** | 8000 | 7952 | 959.82 ± 0.006 | 0.528 |
| Média da campanha | | | **959.519 ± 0,002** | **0.566** |

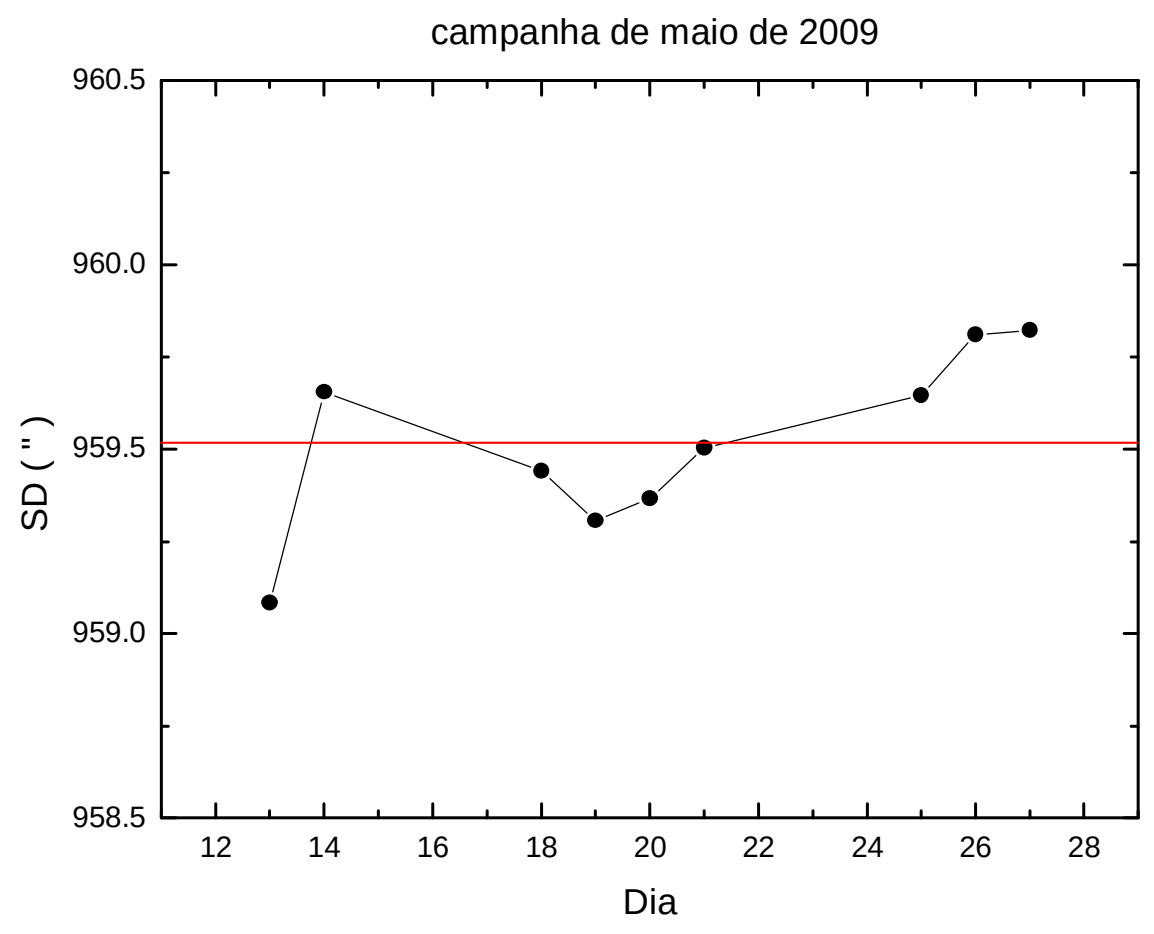

Fig. 6.23: Evolução da medida do semi-diâmetro. A linha horizontal representa o valor médio do mês.

A análise dos resíduos mostra que a hipótese nula (distribuição normal) é verificada, para α = 0,05, em todas as 9 sessões observacionais, figuras 6.24, 6.25 e 6.26.

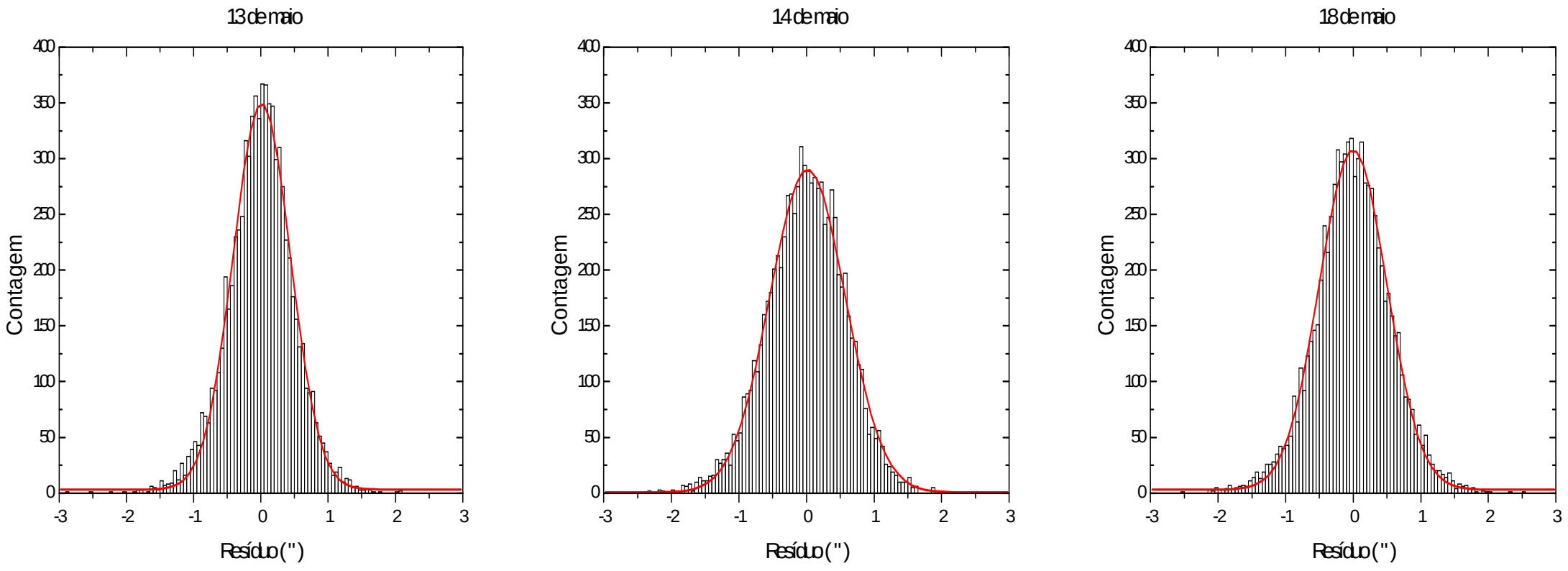

Fig. 6.24: Ajuste gaussiano para os resíduos das medidas do semi-diâmetro solar dos dias 13, 14 e 18 de maio.

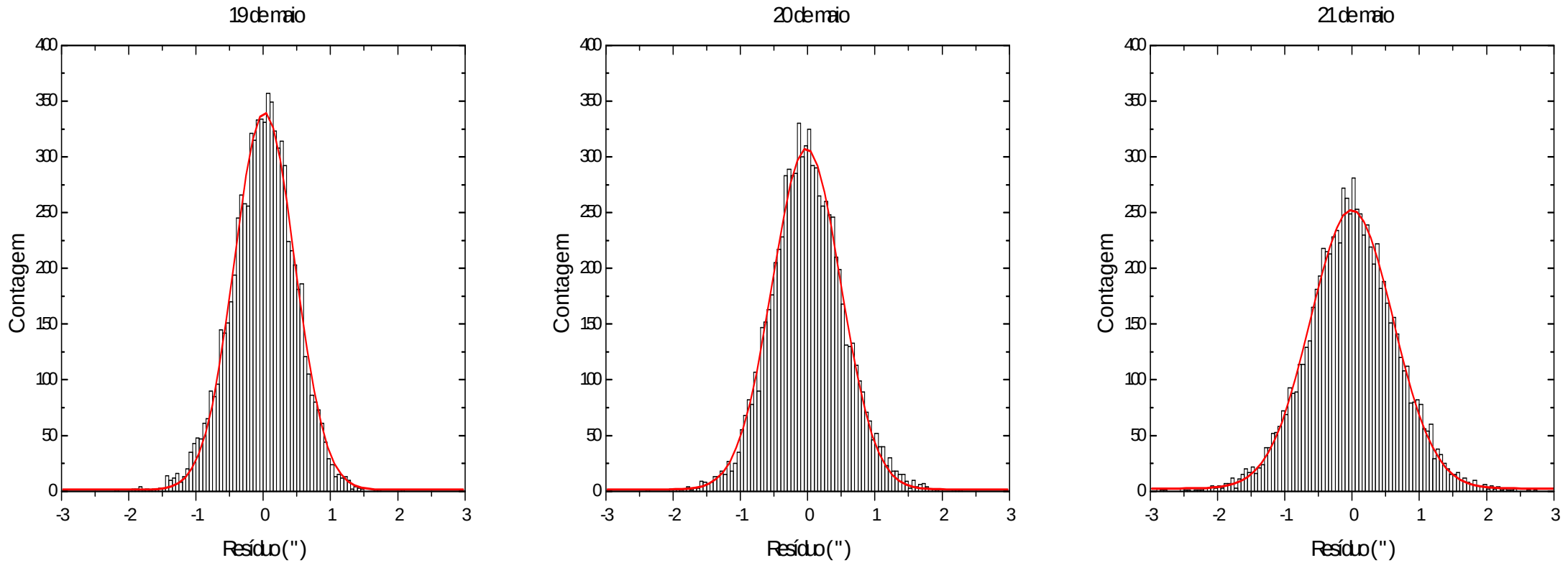

Fig. 6.25: Ajuste gaussiano para os resíduos das medidas do semi-diâmetro solar dos dias 19, 20 e 21 de maio.

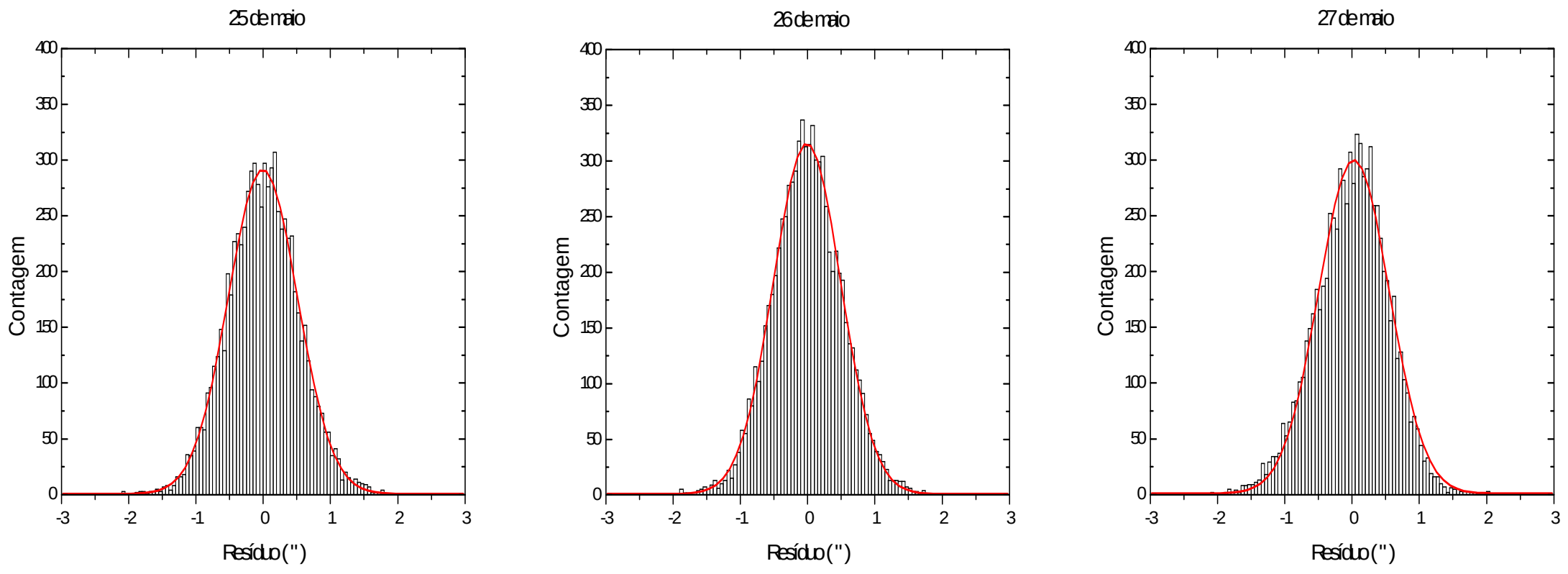

Fig. 6.26: Ajuste gaussiano para os resíduos das medidas do semi-diâmetro solar dos dias 25, 26 e 27 de maio.

Os resultados obtidos, ainda com este heliômetro de testes, mostraram que os valores dos semi-diâmetros seguem uma distribuição normal, com uma precisão de cerca de 0,5" para uma medida isolada (fig. 6.27). Isto indica a ausência de termos instrumentais significativos. Como o procedimento observacional é capaz de capturar e processar milhares de imagens heliométricas do Sol a cada sessão, as medidas podem, em princípio, atingir a acurácia de 0,005" em apenas um dia de observação.

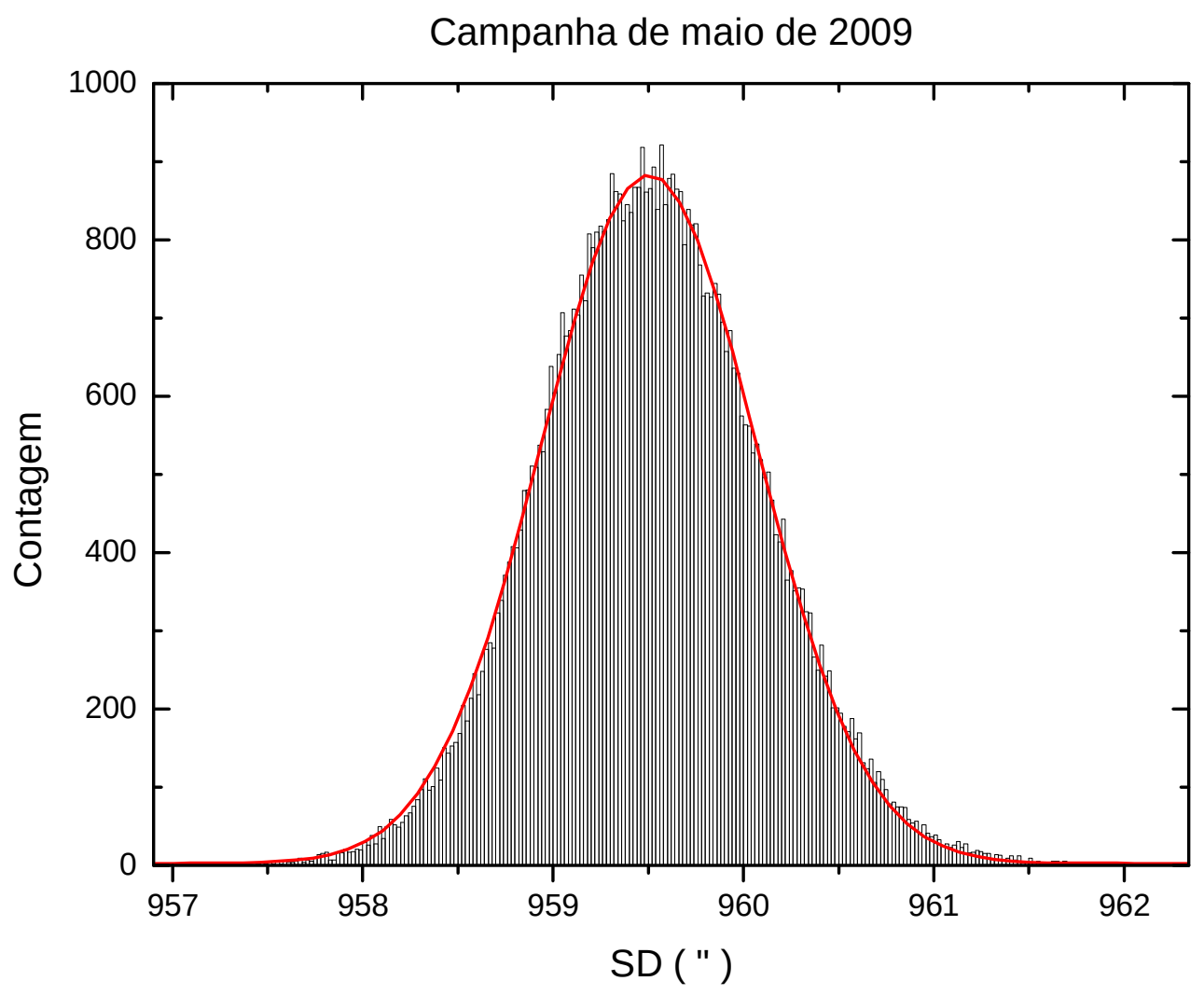

Fig. 6.27: Distribuição total dos semi-diâmetros medidos, com gaussiana superposta. As medidas seguem uma distribuição normal.

# Capítulo 7 – Correlação entre variações do semi-diâmetro solar e do campo geomagnético.

## *7.1 - Introdução ao problema*

A possibilidade de se fazer previsões de processos ou fenômenos naturais sempre foi um trabalho desafiador. Especialmente para aqueles fenômenos que ameaçam a vida humana.

Durante as últimas décadas, muito esforço em pesquisa tem sido dedicado por físicos estatísticos ao estudo deste problema, que inclui, entre muitos outros, terremotos (Sornette, 2004), tempestades magnéticas (Papa *et al.*, 2006 e Papa *et al.*, 2008) e reversões magnéticas (Merrill *et al.*, 1998; Dias *et al.*, 2008).

Alterações dos campos magnéticos na atmosfera do Sol resultam em variações nas emissões solares de partículas e radiação eletromagnética, assim como na reorganização da estrutura do vento solar e do campo magnético interplanetário. Estas mudanças, por sua vez, induzem variações aqui na Terra, como mudanças na composição da população de partículas na alta atmosfera e, dependendo da intensidade da onda de choque do vento solar, variações na magnetosfera terrestre, produzindo as tempestades magnéticas.

Estas tempestades têm períodos que duram de um a três dias, durante o qual o campo magnético local sofre variações rápidas. Elas são responsáveis por cerca de 1% da amplitude do campo magnético total que pode ser medido na superfície da Terra, no entanto, podem afetar gravemente atividades humanas, como provocar falha nas telecomunicações e induzir correntes em linhas de transmissão de energia. Este foi o provável motivo dos "apagões" em Quebec, Canadá, em 1989 e em Malmoe, no sul da Suécia, em 2003.

O estudo da correlação entre as séries temporais geomagnéticas e do diâmetro solar pode ajudar na busca de um método confiável de previsão deste tipo de evento, no médio e longo prazo.

Desde 1997, o grupo de estudos solares do Observatório Nacional/MCT realiza sistematicamente observações do Sol com o objetivo de registrar e estudar as variações de seu diâmetro (Jilinski *et al.*, 1998). Em 2002, uma rede internacional (*Réseau de Suivi au Sol du Rayon Solaire* – R2S3) foi formada para combinar os trabalhos análogos realizados em cinco países (Andrei *et al.*, 2003). Em 2006, o Observatório Nacional se associou ao Projeto

SCOSTEP/CAWSES para a investigação do Clima Espacial e questões ligadas ao sistema Sol-Terra (Andrei *et al.*, 2006).

Dentro deste projeto, o objetivo do trabalho realizado no Observatório Nacional é múltiplo: concretizar uma série de observações comensurável com o ciclo de atividade do Sol; digitalizar e estocar as séries coerentes (Astrolábio Solar e Heliômetro) de observações do diâmetro solar; contribuir para a compreensão da relação Terra-Sol, com ênfase em efeitos locais de curto prazo; usar estas séries e as de outros grupos da R2S3 para investigar a física da fotosfera e da zona de convecção do Sol; e, finalmente, atualizar e desenvolver a aquisição de dados, métodos de tratamento e instrumentos de observação.

Nesta parte do trabalho vamos investigar as correlações entre os picos de variação das contagens dos *flares*, os picos de variação do diâmetro solar e da intensidade do campo geomagnético medido pelas estações do Observatório Nacional, que neste caso devem ser associados a eventos de ejeção de massa coronal.

Embora estas observações sejam muito detalhadas, elas ainda não permitem tirar conclusões definitivas sobre o assunto, mas apontam para a possibilidade de uma forma de previsão destas complexas interações Sol-Terra.

## *7.2 - Série de dados*

Ainda não existem dados originados da última versão do Heliômetro e os dados das observações de maio de 2009 com o Heliômetro protótipo são escassos, por isso, a série dos semi-diâmetros tratada aqui foi a observada com o Astrolábio CCD do Observatório Nacional/MCT, de 2 de março de 1998 a 27 de novembro de 2003 (fig. 7.1). Esta serie é composta por mais de 18.000 observações, com erro interno médio de 0,20" e desvio-padrão de 0,569". Em média, são feitas 20 observações diariamente (considerando-se apenas os dias observados), bem distribuídas durante todo o ano. O pequeno intervalo sem observações, entre o dia 21 de setembro e 19 de dezembro de 2001, foi devido à parada do instrumento para manutenção.

As heliolatitudes observadas cobrem toda a figura solar em um ciclo semi-anual.

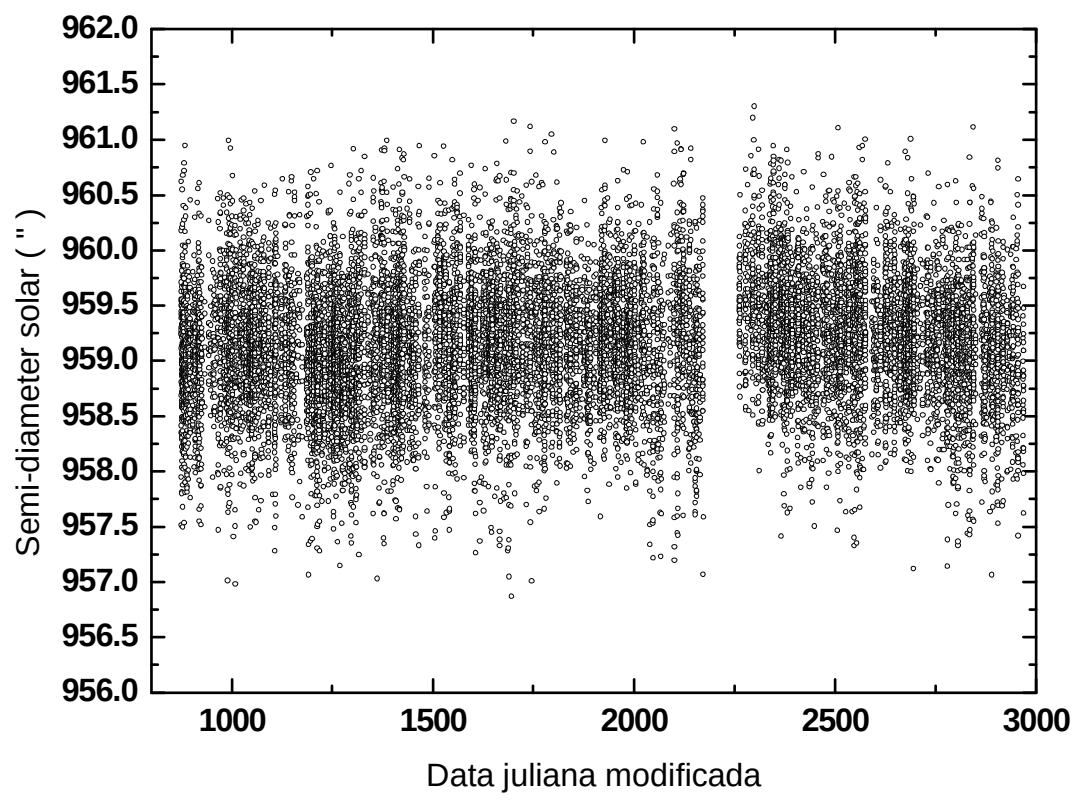

Fig. 7.1: Semi-diâmetro solar observado com o Astrolábio CCD – 1998/2003.

As observações geomagnéticas consistiram na medição da componente H e da intensidade total F do campo magnético, no Observatório Magnético de Vassouras. Os dados são gravados com a freqüência de 1 amostra/min, totalizando mais de 44.000 valores mensalmente. A precisão dos dados é de 1 nT e o erro relativo inferior a $10^{-4}$. As medidas geomagnéticas foram realizadas utilizando o Sistema INTERMAGNET (fluxgate 3 componentes, magnetômetro de prótons, plataforma de dados via satélite e periféricos), durante todo o dia.

Na figura 7.2 vemos a série da componente H do campo magnético, medida neste observatório, entre 1998 e 2003. Pode-se observar uma tendência decrescente com o passar do tempo. Esta tendência é causada por alterações na componente interna do campo magnético e foi retirada, por não ser o objeto do estudo. O mesmo foi feito com a componente direta, em torno de 19.400 nT, valor médio do campo produzido no interior da Terra (a amplitude das tempestades magnética mais severas é de cerca de 400 nT). Em um trabalho anterior Papa *et al.* (2006) mostraram que, após processos de filtragem apropriados tanto a distribuição da amplitude das perturbações geomagnéticas quanto dos eventos intermediários, seguem leis de potência.

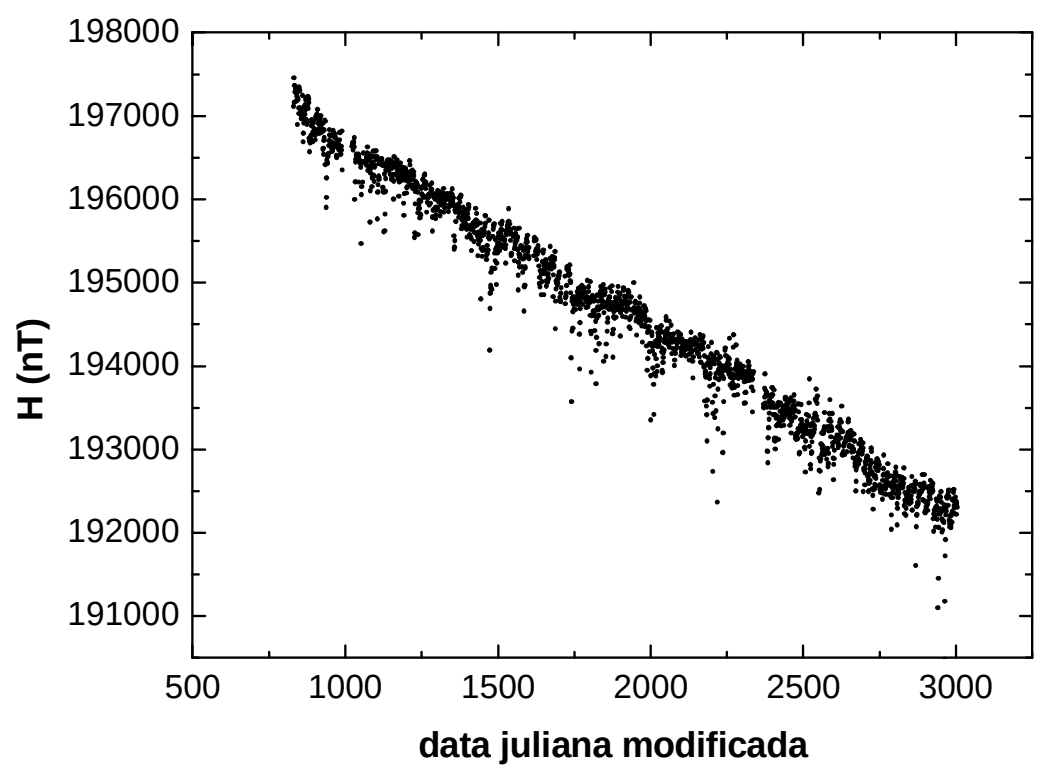

Fig. 7.2: Valores do campo magnético medido de março de 1998 a novembro de 2003, no Observatório Magnético de Vassouras.

Note-se que, em princípio, não há necessidade de procedimentos de filtragem mais sofisticados a fim de se comparar esta série com a do semi-diâmetro solar porque o período das medidas das principais perturbações geomagnéticas (24 horas da rotação da Terra), coincide perfeitamente com a freqüência de amostragem das medições do semi-diâmetro solar.

Para esta análise foram usadas as séries de contagem dos *flares*, assim como a série de contagem das manchas solares. Estes dados foram retirados do NGDC (*National Geophysical Data Center*). A contagem dos *flares* é dada pelo NGDC em valores de "$Q = i \times t$" (Kleczek, 1952), que quantifica a atividade diária dos *flares* durante 24 horas por dia. Esta relação dá, aproximadamente, o total da energia emitida pela explosão. Nessa relação, "i" representa a escala de intensidade de importância e "t" a duração (em minutos) do *flare*.

Neste caso, porém, decidiu-se usar o *Comprehensive Flare Index*, a fim de não introduzir qualquer viés nas correlações entre o diâmetro solar e a contagem das manchas solares, nem viés nos *flares* capazes de estar ligados ao desencadeamento das variações geomagnéticas medido na estação de campo Vassouras.

A distribuição destas duas séries, para o período tratado aqui, é apresentada nas figuras 7.3 e 7.4.

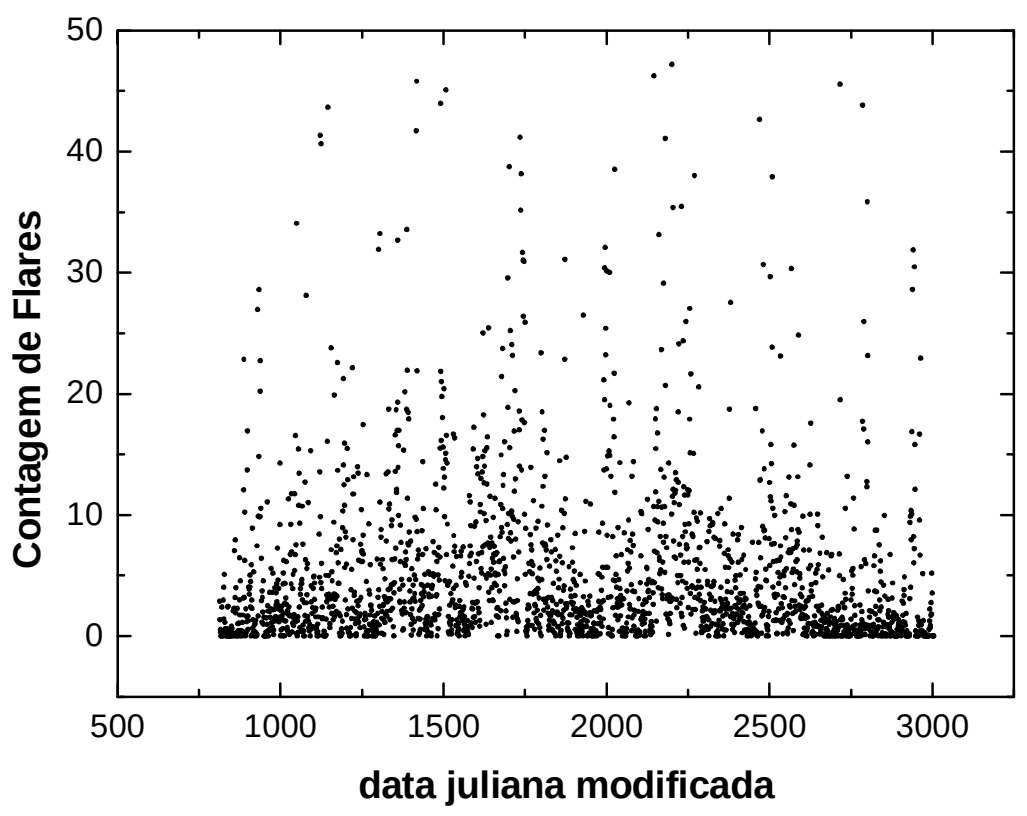

Fig. 7.3: Contagem de *Flares* do NGDC (*Comprehensive Flare Índex*).

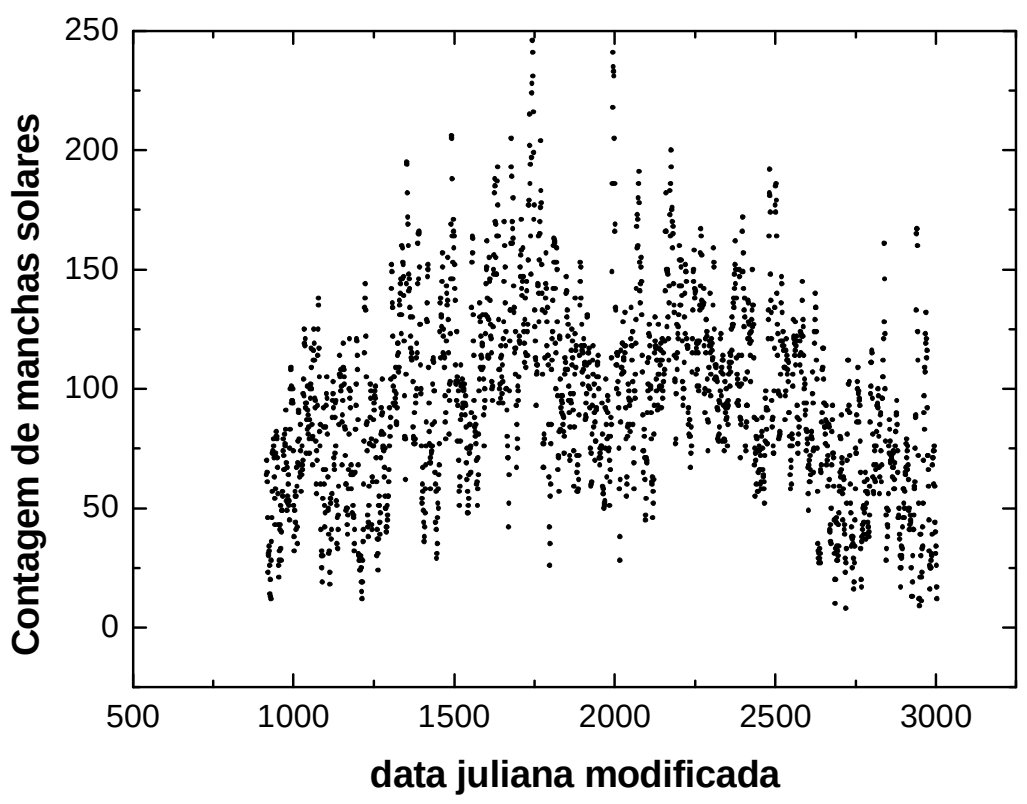

Fig. 7.4: Contagem das manchas solares do NGDC.

## 7.3 - Análise dos dados

As variações temporais do semi-diâmetro solar foram correlacionadas com estes indicadores da atividade solar utilizando-se os dados produzidos pelas observações no Observatório Nacional.

Embora não haja até o momento nenhum modelo globalmente aceito de causa e efeito para as variações observadas no diâmetro fotosférico, algumas hipóteses têm sido levantadas (Sofia *et al.*, 2005; Andrei *et al.*, 2006; Badache-Damiani *et al.*, 2007), todas partindo da evidência observacional para refinar a teoria solar.

Aqui, o elo foi investigado através do cálculo das correlações entre as séries do semi-diâmetro solar contra a da contagem de manchas solares e da contagem dos *flares*. Para cada par, diversas correlações foram calculadas, permitindo diferentes defasagens temporais entre as séries. Este método permite distinguir fenômenos que tenham uma relação causal, mas interligados com algum atraso de tempo entre eles. A fim de se obter um quadro mais amplo, correlações também foram calculadas entre a série do semi-diâmetro e outros estimadores da atividade solar, como o fluxo de 10,7 cm no rádio, a irradiância total, e intensidade do campo magnético total (Boscardin *et al.*, 2009 ). Os resultados são resumidos na Tabela 7.1.

Tabela 7.1: Máximo da correlação linear de Pearson entre a variação do semi-diâmetro solar e estimadores da atividade solar.

| **Par** | **SD-MS**[1] | **SD-CF**[2] | **SD-ER**[3] | **SD-IT**[4] | **SD-CM**[5] |
|---|---|---|---|---|---|
| **Correlação** | 0,80 | 0,66 | 0,88 | 0,78 | 0.62 |

Na tabela 7.1 são mostradas as correlações entre: semi-diâmetro medido contra variação da contagem de manchas solares (1), semi-diâmetro medido contra a contagem de *flares* (2), semi-diâmetro medido contra variação da emissão em 10,7 cm (3), semi-diâmetro medido contra variação da irradiância total (4) e semi-diâmetro medido contra variação da intensidade do campo magnético total (5).

Para as comparações discutidas aqui, as figuras 7.5 e 7.6 exibem a complexa interação entre a série das variações do semi-diâmetro com a série das variações das manchas solares e com a série de contagem de *flares*. Para ambas as comparações, é verificada a ocorrência de dois máximos, um próximo do tempo de atraso zero, e outro que exige um tempo da ordem do ano de atraso. Levando-se o estudo adiante, mas fora do escopo desta tese, verifica-se que quando os períodos em que ocorreram os picos de atividade solar são removidos da série, os valores máximos correspondentes ao tempo de atraso zero desaparecem. Isso sugere duas formas de resposta do semi-diâmetro em relação à atividade solar. Ao longo do ciclo de atividade solar o semi-diâmetro segue o ciclo com atraso. No entanto, quando ocorrem picos de atividade, também ocorre uma rápida variação do semi-diâmetro medido. Nestes casos, a variação do semi-diâmetro funciona como um indicador de atividade solar intensa.

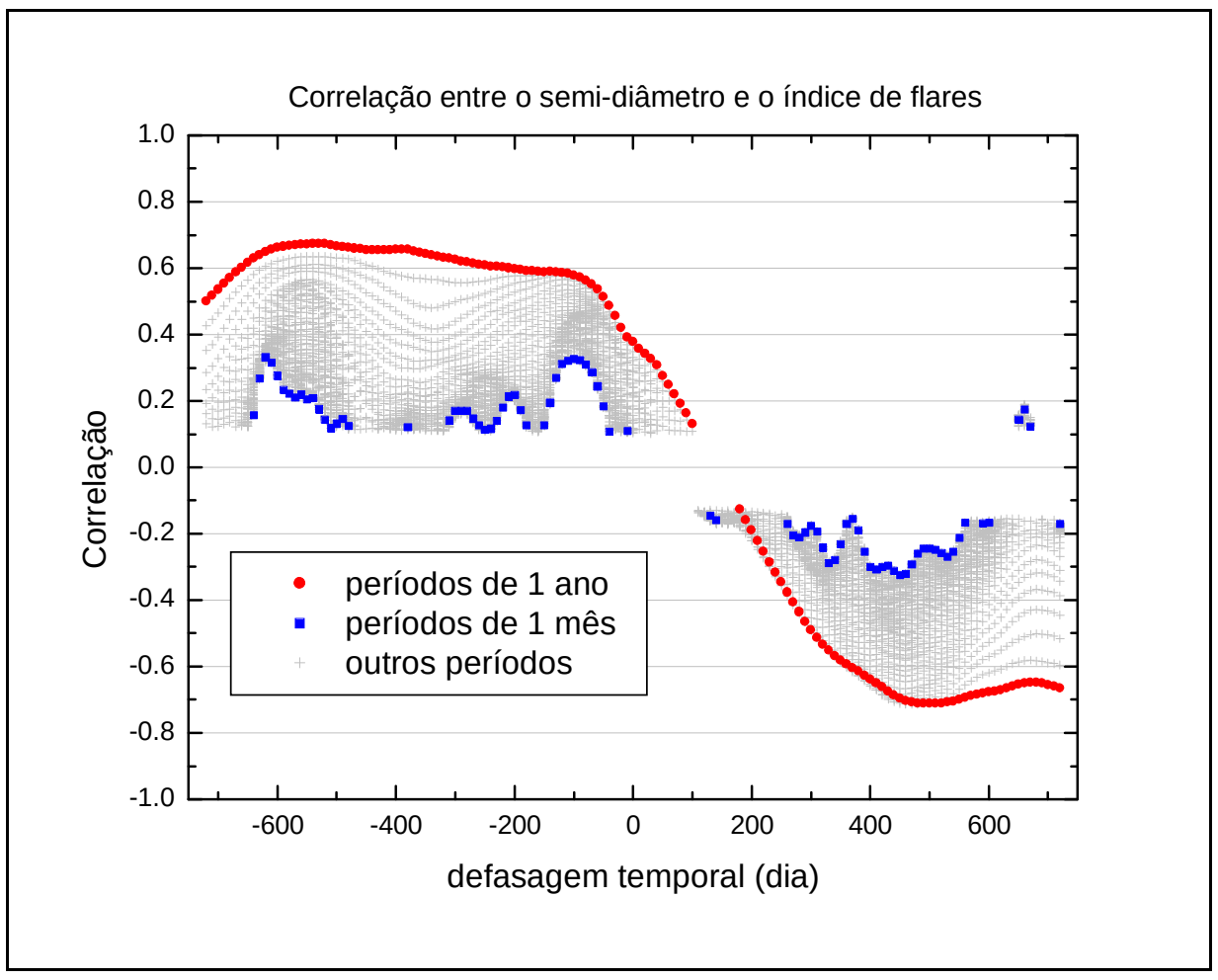

Fig. 7.5: Correlação entre a medida do semi-diâmetro e a série da contagem de *flares*. As linhas coloridas formam um envelope, entre a defasagem o anual (em vermelho) e a mensal (em azul).

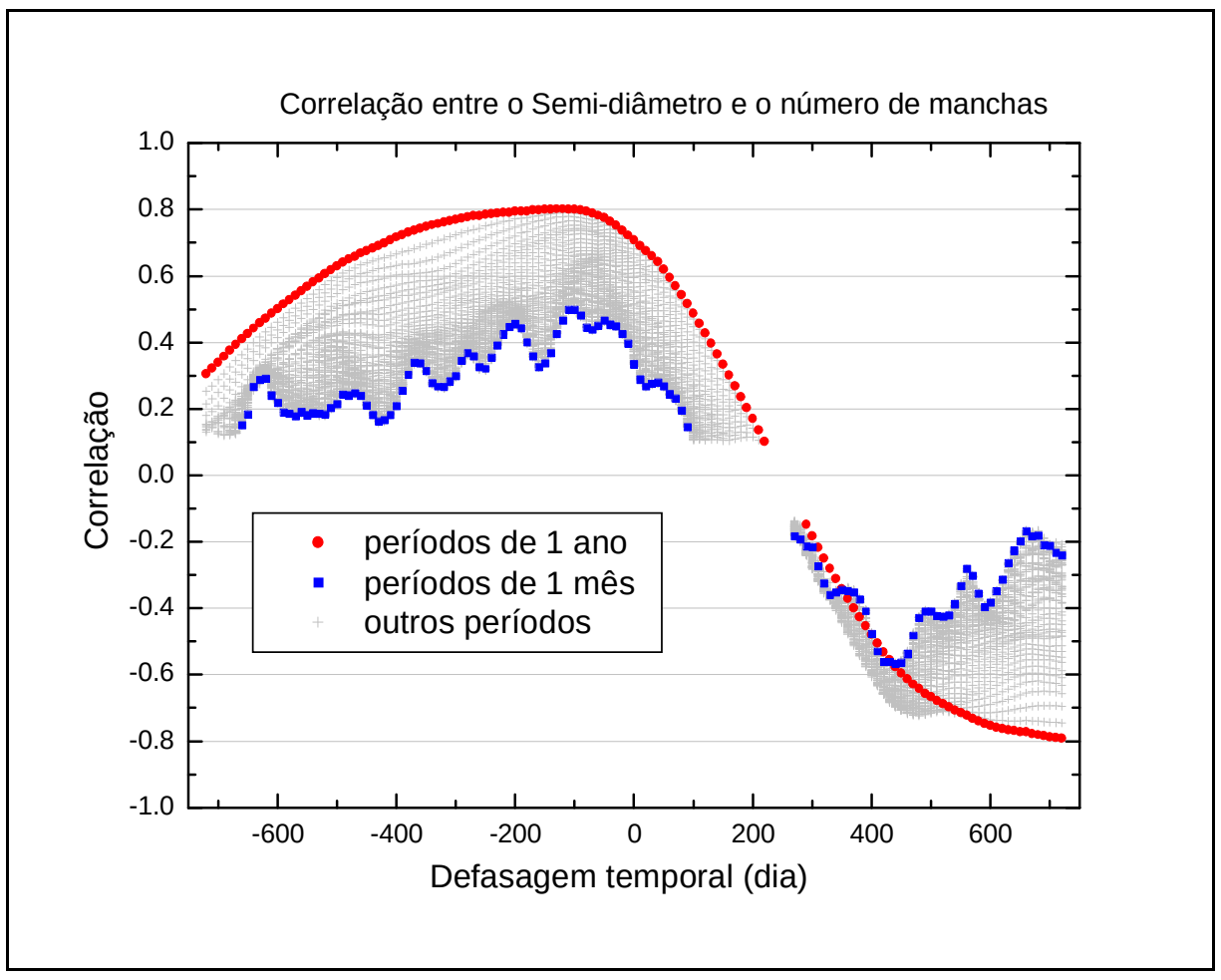

Fig. 7.6: Correlação entre a medida do semi-diâmetro e a série da contagem das manchas solares. As linhas mais coloridas formam um envelope, entre a defasagem o anual (em vermelho) e a mensal (em azul).

As variações das médias anuais do número de *flares* e do número de tempestades magnéticas revelam que fortes *flares* e fortes tempestades atingem seus máximos nos anos de máxima atividade solar. Suas séries temporais têm formas semelhantes e elevado coeficiente de correlação. A interpretação mais usual é que as variações do número de *flares* e do número de tempestades magnéticas podem ter uma causa comum. No entanto, a comparação entre os indicadores destes dois tipos de eventos não se correlacionam em uma escala menor do que um mês.

A explicação tem três partes: em primeiro lugar, apenas *flares* que apresentam emissões significativas em raios-X, ou de energia, provavelmente têm correlação com *CMEs* e as tempestades magnéticas; em segundo lugar, existe o aspecto direcional, que faz com que apenas uma fração das emissões solares observadas efetivamente atinja o campo geomagnético, e um número ainda mais reduzido seja realmente detectada e medida por alguma estação geomagnética; e em terceiro lugar, mas não menos importante para esta análise, existe o aspecto do atraso de vários dias entre o início de um *flare* e uma correspondente tempestade magnética.

Sendo assim, buscou-se rastrear os maiores episódios da variação do campo geomagnético, medido na estação de Vassouras, e os maiores episódios de variação encontrados na série do semi-diâmetro. As séries de contagem dos *flares* é usada como ponte entre o fator semi-diâmetro e a resposta geomagnética.

As séries de *flares* é dada em valores diários. Já a série de intensidade do campo geomagnético é muito mais detalhada e apresentada em valores medidos por minuto. No entanto, nem todas as medições de cada dia estão completas, por isso, apenas os dias em que pelo menos 25% das medidas foram efetivamente tomadas foram considerados. Um valor médio diário foi tomado a partir da média das medidas válidas. Para interpolar os valores que estavam faltando na série, foi usado um filtro FFT, com banda passante de 1 semana. O mesmo procedimento foi aplicado à série do semi-diâmetro, para que se pudessem adotar o mesmo número de pontos por dia e interpolar valores para dias sem observações. A banda passante de um semana foi escolhida porque, dentro da precisão de nossas medições, variações representativas no semi-diâmetro em intervalos menores apenas são distinguidas de forma comparativa.

Em seguida, as três séries foram normalizadas e uma busca automática foi feita para os picos máximos para as séries solares e mínimos para a série de intensidade do campo geomagnético. A fim de examinar as quatro séries dentro de janelas comuns, foi adotada uma largura de busca de três meses (o maior número de tempestades magnéticas para o período aqui analisado ocorreu no ano de 2000, com quatro tempestades). Isto equivale aproximadamente a uma janela móvel de largura igual a 5% do intervalo inteiro. O mesmo percentual foi estabelecido para a altura da janela de varredura, e como limite acima do qual máximos locais seriam aceitos.

Nas duas séries solares 14 máximos foram detectados. A figura 7.7 mostra as diferenças entre as datas dos picos. Como esperado da análise das séries completas, os picos de variação do semi-diâmetro e de contagem de *flares* estão bem correlacionados. A diferença média entre as datas é -9.6 ± 12,7 dias, portanto, não é significativa. Uma ligeira tendência negativa pode ser ajustada para a diferença, que aumenta gradativamente entre as duas séries durante períodos de Sol calmo. A Tabela 7.2 traz os valores máximos escolhidos para cada série. É interessante notar que tanto o pico de variações da contagem de *flares* e dos semi-diâmetros são maiores em torno do máximo do ciclo solar. O centro do máximo do ciclo solar 23 aconteceu em torno de janeiro de 2001, na DJM 52.000.

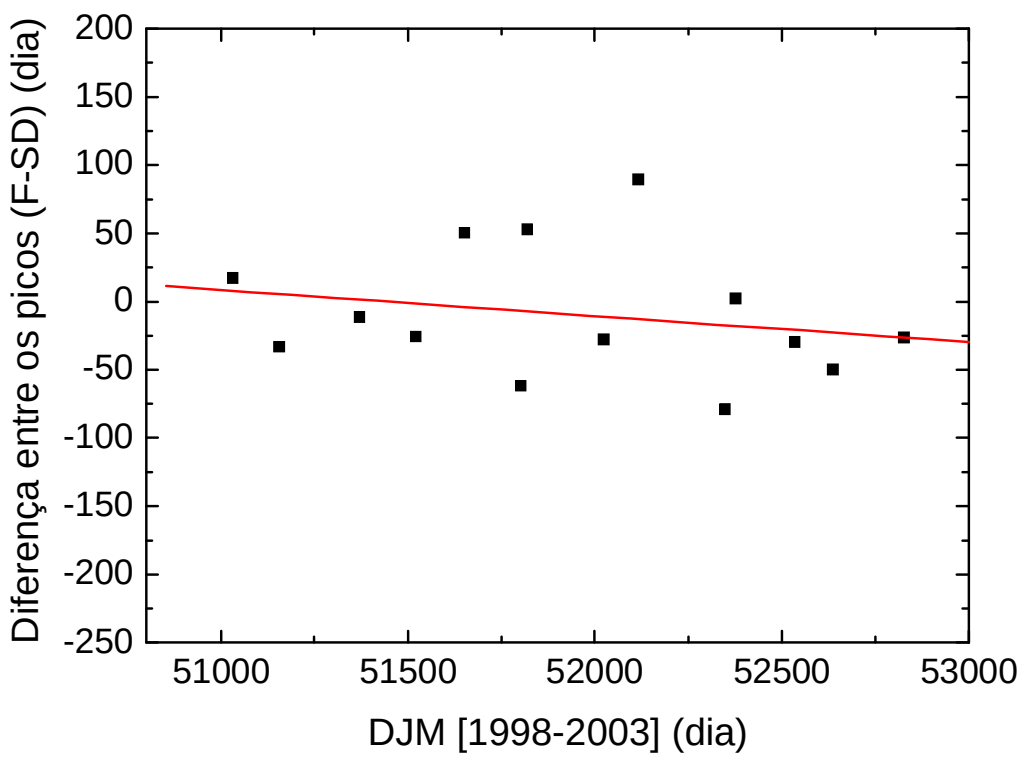

Fig. 7.7: Diferença de tempo entre os correspondentes picos dos eventos encontrados nas séries do semi-diâmetro e contagem de *flares*. A linha ajustada mostra uma diferença média de -9,6 ± 12,7 dias.

Tabela 7.2: Localização temporal dos principais eventos encontrados na série de intensidade do campo geomagnético medido na estação de Vassouras e nas séries do semi-diâmetro e dos *flares*, entre abril de 1998 e novembro de 2003. As datas são apresentadas em dia juliano modificado, como nos gráficos.

| Data | Pico GEOMAG[1] | Data | Pico SD[2] | Data | Pico F[3] |
|---|---|---|---|---|---|
| 51.053,5 | -2,44 | 51.032.4 | 1,21 | 51.049.5 | 0,47 |
| 51.128,5 | -2,05 | 51.156.7 | 1,78 | 51.123.5 | 1,29 |
| 51.240,5 | -1,34 | 51.371.0 | 1,48 | 51.359.5 | 1,93 |
| 51.445,5 | -1,43 | 51.522.2 | 1,96 | 51.496.5 | 4,38 |
| 51.587,5 | -1,66 | 51.652.3 | 1,86 | 51.702.5 | 3,30 |
| 51.742,5 | -3,65 | 51.803.4 | 1,78 | 51.741.5 | 6,90 |
| 51.877,5 | -1,33 | 51.820.6 | 1,94 | 51.873.5 | 0,81 |
| 52.011,5 | -2,90 | 52.025.4 | 1,76 | 51.997.5 | 2,47 |
| 52.184,5 | -2,50 | 52.117.2 | 1,96 | 52.206.5 | 2,23 |
| 52.384,5 | -2,76 | 52.348.7 | 2,24 | 52.269.5 | 1,47 |
| 52.554,5 | -2,59 | 52.377.4 | 3,03 | 52.379.5 | 0,45 |
| 52.600,5 | -1,07 | 52.536.2 | 1,13 | 52.506.5 | 1,81 |
| 52.730,5 | -1,79 | 52.637.6 | 0,95 | 52.587.5 | 0,54 |
| 52.871,5 | -1,73 | 52.827.0 | 1,69 | 52.800.5 | 0,84 |

Na tabela 7.2, para as séries temporais das medidas da intensidade do campo geomagnético da estação de Vassouras (1), do diâmetro solar da estação do Observatório Nacional (2) e da contagem de *flares* (3) os picos estão normalizados

Quando foram procurados os picos de variações geomagnéticas, adotando-se os mesmos critérios utilizados para a série dos parâmetros solares, 16 eventos foram encontrados. Esta quantidade é muito próxima aos dos 14 picos encontrados para as outras séries, o que é sugestivo uma vez que os critérios de pesquisa são puramente estatísticos. É interessante comentar que em procedimento auxiliar os picos positivos do campo geomagnético foram encontrados. Estes se referem aos *SSC* (início súbito) e apresentam concordância com os picos negativos. A fim de compatibilizar os eventos geomagnéticos intensos com os solares, recorreu-se a uma escolha estatística mantendo-se apenas os 14 menores eventos geomagnéticos. A Tabela 7.3 traz os eventos combinados. A figura 7.8 mostra a diferença de tempo entre os picos dos *flares* e menos os picos de intensidade geomagnética. Estes últimos, como esperado, ocorrendo após os primeiros.

Tabela 7.3: Localização temporal dos eventos principais encontrados nas séries do semi-diâmetro e dos *flares*, entre abril de 1998 e novembro de 2003. A data é apresentada em dia juliano modificado, como nos gráficos.

| Data | Pico SD | Data | Pico F |
|---|---|---|---|
| 51.032,4 | 1,21 | 51.049.5 | 0,47 |
| 51.156,7 | 1,78 | 51.123.5 | 1,29 |
| 51.371,0 | 1,48 | 51.359.5 | 1,93 |
| 51.522,2 | 1,96 | 51.496.5 | 4,38 |
| 51.652,3 | 1,86 | 51.702.5 | 3,30 |
| 51.803,4 | 1,78 | 51.741.5 | 6,90 |
| 51.820,6 | 1,94 | 51.873.5 | 0,81 |
| 52.025,4 | 1,76 | 51.997.5 | 2,47 |
| 52.117,2 | 1,96 | 52.206.5 | 2,23 |
| 52.348,7 | 2,24 | 52.269.5 | 1,47 |
| 52.377,4 | 3,03 | 52.379.5 | 0,45 |
| 52.536,2 | 1,13 | 52.506.5 | 1,81 |
| 52.637,6 | 0,95 | 52.587.5 | 0,54 |
| 52.827,0 | 1,69 | 52.800.5 | 0,84 |

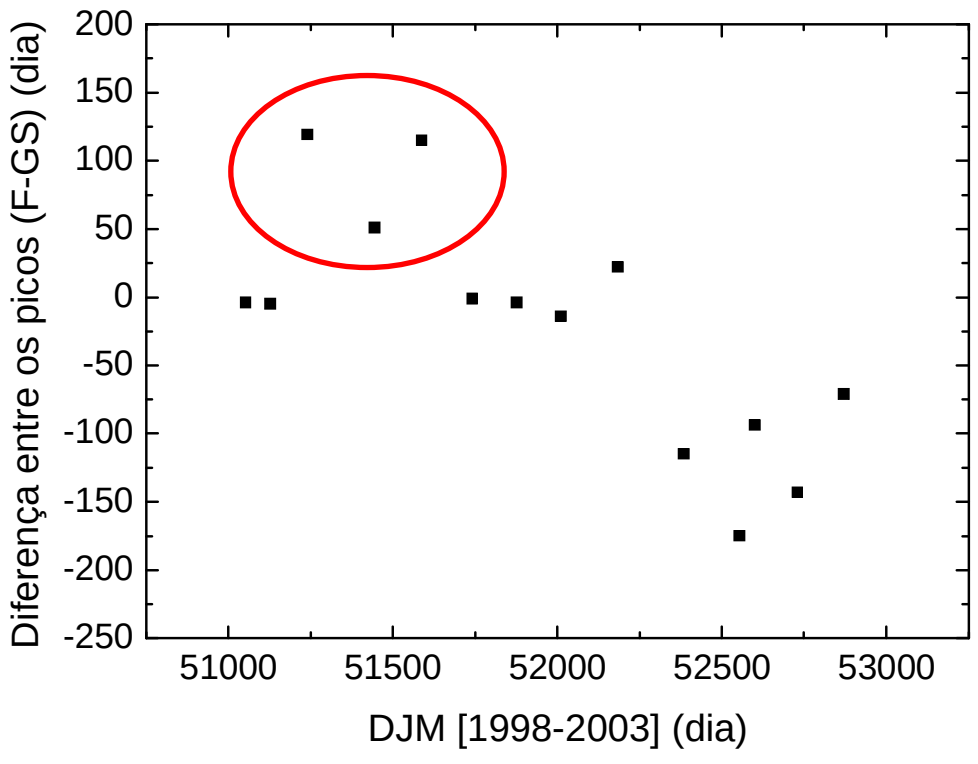

Fig. 7.8: Diferença de tempo dos máximos locais dos eventos de *flare* menos os dos eventos de intensidade do campo geomagnético. Em destaque os três eventos que aparecem ocorrendo depois da tempestade.

## 7.4 - Interpretação dos resultados

Nota-se na figura 7.8 que para três dos eventos combinados os *flares* parecem ter ocorrido depois da tempestade geomagnética (pontos em destaque no gráfico). Isto obviamente é um erro de combinação. Estes pontos foram deixados na análise para não introduzir ocorrências privilegiadas, uma vez que todas as combinações são de natureza estatística. Mesmo mantendo estas combinações errôneas, maiores atrasos médios significantes são obtidos depois do máximo solar (fig. 7.9). Antes do máximo o atraso médio é de 23,8 ± 18,6 dias (combinações errôneas tornando a estatística nula), enquanto que depois do máximo a média é de -61,0 ± 9,9 dias (pontos em destaque na figura 7.9). Isto pode se interpretado como se, à medida que o ciclo amortece, mais e mais *flares* começam a ser necessários para desencadear uma tempestade magnética.

Por outro lado, a dependência verificada entre os picos das variações do semi-diâmetro e da contagem de *flares*, conduz, de certa maneira, a uma conveniente influência mútua entre os picos das variações do semi-diâmetro solar e da intensidade do campo geomagnético, como mostrado na figura 7.9. Em períodos de Sol calmo, o variação do semi-diâmetro precede a correspondente tempestade magnética em 60 dias, enquanto existe pouco atraso quando o Sol está mais ativo (colocando de lado as já comentadas combinações erradas). Se novos dados confirmem tal comportamento, as medições do semi-diâmetro podem proporcionar uma técnica, de acesso fácil, de previsão de eventos potencialmente perigosos manifestados depois como picos de variação na intensidade do campo geomagnético.

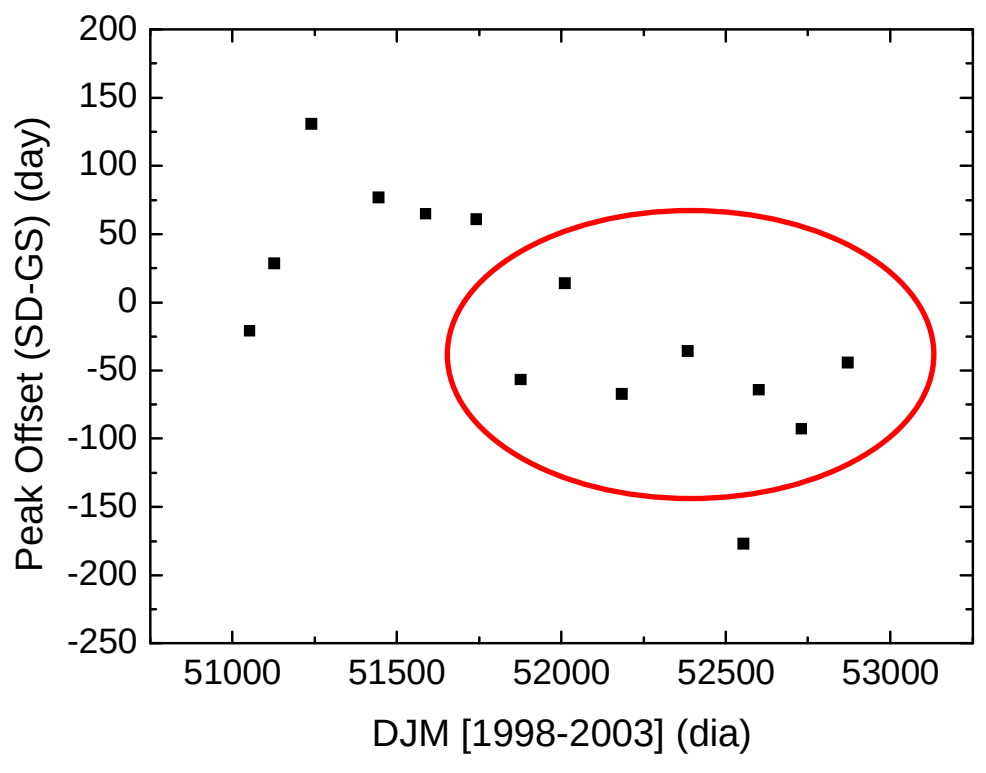

Fig. 7.9: Diferença de tempo dos máximos locais dos eventos do semi-diâmetro menos os dos eventos de intensidade do campo geomagnético. Destaque para os pontos em que estes eventos ocorrem, em média, com diferença de cerca de 60 dias.

# Capítulo 8 - Conclusões e Perspectivas

Os processos físicos prevalentes no Sol, na Terra, e no ambiente Sol-Terra variam continuamente. A alternância dos campos magnéticos na atmosfera solar, impulsionada pelo dínamo interno, é a causa principal da variabilidade solar. As alterações resultantes da produção de energia, sob a forma de plasma do vento solar, de radiação eletromagnética e de partículas, a partir da reorganização e reestrutura dos campos magnéticos solar e interplanetário, induzem variabilidade no meio ambiente terrestre, notadamente a parte mais próxima espaço, principalmente sob a forma de variações da temperatura e da composição das populações de partículas. Em suma, mudanças no meio ambiente terrestre podem resultar diretamente a partir da modulação do ciclo de 11 anos de produção de energia do Sol, bem como de efeitos cumulativos das variações de curto prazo associadas a eventos eruptivos que ocorrem em diferentes taxas e com amplitudes diferentes em diferentes fases do ciclo de atividade.

O diâmetro solar dá uma contribuição fundamental para elucidar a origem da atividade solar, relacionada com os campos magnéticos de superfície e da zona de convecção. A atividade solar define as principais variações do clima espacial, com seu impacto sobre o clima e a humanidade. Os impactos incluem: falhas eletrônicas em satélites; ou mesmo desvios orbitais por arrasto atmosférico, falhas de comunicação, problemas de navegação em aviões; risco de radiação para astronautas. O impacto geomagnético afeta sistemas elétricos desde o mau funcionamento de usinas geradoras, o rompimento de linhas de transmissão e distribuição, até a explosão de transformadores. As medidas do diâmetro solar se dão no contexto geofísico pelo estudo do Clima Espacial e pela análise da correlação com variações do campo geomagnético, portanto, sugerimos utilizar e experimentar esta técnica no contexto da Pesquisa do Clima Espacial.

A técnica heliométrica mostrou-se bastante consistente para fazer as medidas solares a que se propõe. Todos os protótipos construídos, com diferentes metodologias, são potencialmente capazes de produzir dados científicos. O modelo heliométrico escolhido e que foi desenvolvido nesta tese foi o do heliômetro refletor, com seu espelho principal bi-partido e com as metades fixamente montadas com uma diferença angular, ao longo do plano do corte do espelho, suficiente para separar os discos solares.

Uma vez definida a técnica, um heliômetro de testes foi montado, posto à prova e gradativamente aprimorado, tendo sempre como meta o desenvolvimento da metodologia necessária para a construção de um instrumento em seu estado da arte.

O programa desenvolvido especialmente para a aquisição e análise das imagens heliométricas do Sol foi aperfeiçoado ao longo do projeto atingindo a configuração automática atual. A medida da distância mínima entre os bordos dos discos é comprimento do segmento de reta que liga os pontos de interseção entre as circunferências ajustadas e a retas que une seus centros geométricos. Este procedimento torna esta medida muito mais estável, não obstante o fato de se usar ajustes não lineares, sensíveis à quantidade de pontos. Para uma imagem analisada, independentemente da incerteza das coordenadas dos centros dos discos (segundos de arco para as abscissas e décimos de segundos de arco para as ordenadas), consegue-se determinar a distância mínima entre os discos solares com precisão de centésimos de segundo de arco.

Os resultados obtidos, ainda com este heliômetro de testes, mostraram que os valores dos semi-diâmetros seguem uma distribuição normal, com uma precisão de cerca de 0,5 segundos de arco para uma medida isolada. Isto indica a ausência de termos instrumentais significativos. Como o procedimento observacional é capaz de capturar e processar milhares de imagens heliométricas do Sol a cada sessão, as medidas podem, em princípio, atingir a acurácia de 0,005 segundos de arco em apenas um dia de observação.

O sistema de giro do instrumento sobre seu eixo óptico funcionou perfeitamente para se observar o Sol em diferentes heliolatitudes. Não era esperado, no entanto, que este instrumento de testes chegasse a resultados significativos quanto a estas medidas. Primeiro, por se tratar de um protótipo, construído e montado com materiais comuns, como vidro, *webcam* e PVC. Segundo, com uma campanha de apenas um mês de observações não se alcança medidas suficientes para se obter, estatisticamente, diferenças de semi-diâmetro, em diferentes heliolatitudes, da ordem do milissegundo de arco.

O Heliômetro definitivo foi inteiramente projetado pensando-se sempre em estabilidade:

- Seu espelho em CCZ-HS é termicamente estável e tem as dimensões tais que não há preocupação quanto à flexão pelo seu próprio peso e com pontos de apoio, além de ficar perfeitamente acamado dentro de sua célula;
- A qualidade de sua superfície óptica é de $\lambda/30$, o que faz com que produza imagens com distorções em sua frente de onda não superior a $\lambda/15$.
- O seu corpo em fibra carbono garante a rigidez necessária para o suporte da célula do espelho e da câmara CCD, sem estar submetido a pressões ou trações. Ele também é termicamente estável;
- O desenho da pupila do espelho torna os feixes de luz, que constituirão a imagem, perfeitamente simétricos;
- A área útil dos espelhos sob a pupila garante imagens com boa resolução espacial, pois sua dimensão é comparável à dimensão da janela de coerência típica devido à turbulência atmosférica;
- O suporte do tubo, em barras de aço inox trançado, segura firmemente o tubo ao mesmo tempo em que o deixa livre para girar suave e precisamente em torno do seu eixo;
- Seu pilar de concreto reforçado e separado da estrutura da cúpula, não permite que vibrações passem para a montagem equatorial;
- O sistema de climatização diminuirá a turbulência atmosférica na entrada do tubo do instrumento.

Os pontos a ressaltar na performance do Heliômetro são: (1) a qualidade óptica de suas peças (superfícies corrigidas entre $\lambda/10$ e $\lambda/30$), a estabilidade térmica de seu espelho de CCZ ($\Delta L/L = 0{,}0 \pm 0{,}2 \times 10^{-7}$, entre 0ºC e 50ºC) e a rigidez mecânica de sua montagem (fibra carbono, suporte em *truss* e pilar) garantem que a qualidade das imagens geradas estarão limitas apenas pela atmosfera; (2) o desenho de sua pupila de entrada deixa as figuras de difração perfeitamente simétricas e sua abertura garante uma ótima relação entre a área útil do espelho e a janela de coerência atmosférica, estabilizando ainda mais as imagens; (3) seu sistema é capaz de processar milhares de imagens diariamente, o que, em tese, melhora a precisão de seus resultados; (4) sua montagem giratória permite o detalhado recobrimento em heliolatitudes; (5) observação em diferentes comprimentos de onda, com fácil adaptação de uma roda de filtros; (6) sua motorização permite diferentes métodos observacionais, como por exemplo, com acompanhamento ou estático; (7) sua versatilidade permite o mapeamento e modelagem da dependência em distância zenital.

Quanto ao primeiro estudo da inter-relação entre o diâmetro solar, medido pelo Astrolábio CCD, e as medidas geomagnéticas, podemos dizer que é encontrada correspondência entre os picos das séries ligadas à atividade solar, qual sejam, a do semi-diâmetro solar, dos *flares*, e da contagem de manchas e os picos negativos da série da intensidade do campo geomagnético. No entanto esta correspondência é complexa, requerendo modos diferenciados de resposta e fase dependendo da etapa do ciclo de atividade solar.

A interpretação direta da evidência observacional, indica que o semi-diâmetro solar parece apresentar variações significativas antes das variações correspondentes do campo geomagnético. Devido à combinação da seqüência de respostas, desde a deflagração de manchas solares até a detecção de um pico na variação do campo geomagnético, esta anterioridade se apresenta sobretudo na fase descendente do ciclo solar, ao redor de até 60 dias.

Se novos dados confirmem tal comportamento, as medições do semi-diâmetro podem proporcionar uma técnica, de acesso fácil, de previsão de eventos potencialmente perigosos manifestados depois como mínimos na intensidade do campo geomagnético.

Para isto, faz-se necessária a continuidade do acompanhamento do diâmetro solar através, de pelo menos, mais um ciclo de atividade, mantendo a mesma metodologia. Ou metodologia equivalente, mas de maior precisão, diretividade e acuidade temporal.

O Heliômetro configura tal técnica. Ele não só será mais preciso que o Astrolábio nesta tarefa, como também, com milhares de medidas diárias, abreviará o processo de análise, permitindo o estudo com mais detalhes às respostas rápidas, inclusive de forma mais local, uma vez que o instrumento tem a liberdade de fazer um detalhado recobrimento em heliolatitude. O Astrolábio, no entanto, continuará com suas observações solares em paralelo às observações do Heliômetro. Estas duas séries independentes de observações do semi-diâmetro do Sol serão superpostas e comparadas num futuro próximo.

# Capítulo 9 - Bibliografia

# Apêndice I - Desenvolvimento do teste de auto-colimação e de comparação.

A auto-colimação e a comparação são fundamentais para o monitoramento constante da estabilidade das medidas heliométricas.

Um padrão foi estabelecido e uma rotina de medições deste padrão deverá ser seguida em conjunto com as medidas do instrumento, de forma que se houver variações no valor deste padrão, estas servirão para se estabelecer um modelo de correção das séries.

O padrão utilizado para o teste foi um pequeno cilindro, com diâmetro de 9,60 mm e altura de 6,05 mm, feito de CCZ-HS (o que garante sua estabilidade dimensional para toda a faixa de temperaturas operacionais). Este cilindro tem suas faces despolidas e fica posicionado próximo ao espelho principal, figuras A-I.1 e A-I.2.

## *Teste de auto-colimação*

### Montagem instrumental da experiência.

Para o teste de auto-colimação, um espelho parabólico, de distância focal semelhante à do espelho heliométrico é posicionado na entrada do tubo do telescópio, com seu eixo óptico alinhado ao eixo do tubo. O cilindro estará exatamente na distância focal do espelho de colimação. Quando iluminado, a luz refletida pelo espelho parabólico será paralela ao seu eixo, de forma que a imagem real de sua face despolida estará infinitamente distante do espelho heliométrico. Esta configuração reproduz a geometria da observação do Sol: dois discos separados, com diâmetros semelhantes ao do disco solar, serão captados pela câmara CCD, figura A-I.1.

Deste modo, o programa desenvolvido para analisar as imagens heliométricas do Sol pode ser usado da mesma maneira para medir as imagens dos discos de CCZ-HS, permitindo que as medidas reais possam ser calibradas pela medida de auto-colimação, em intervalos pré-programados.

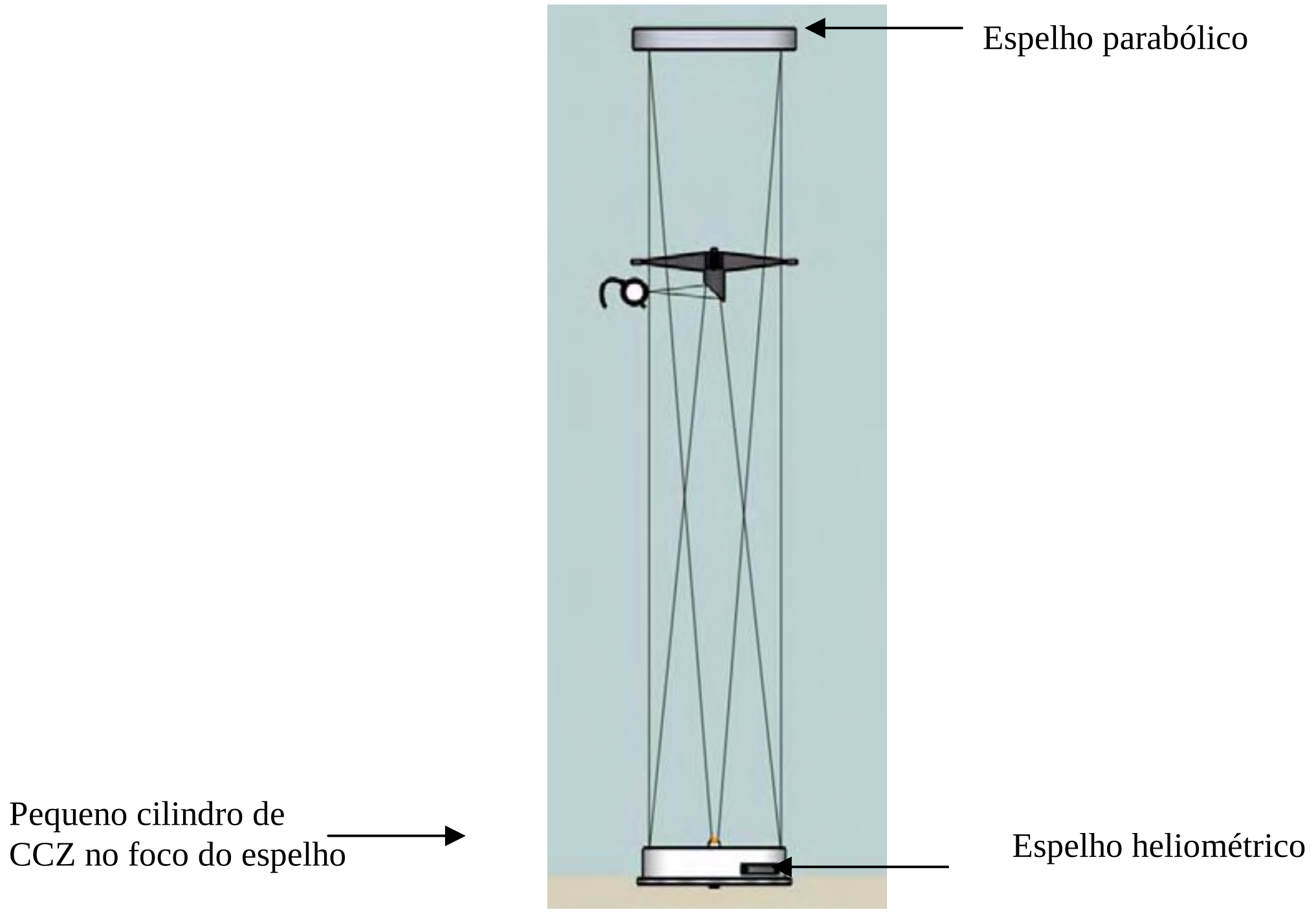

Fig. A-I.1: Caminho óptico da montagem instrumental.

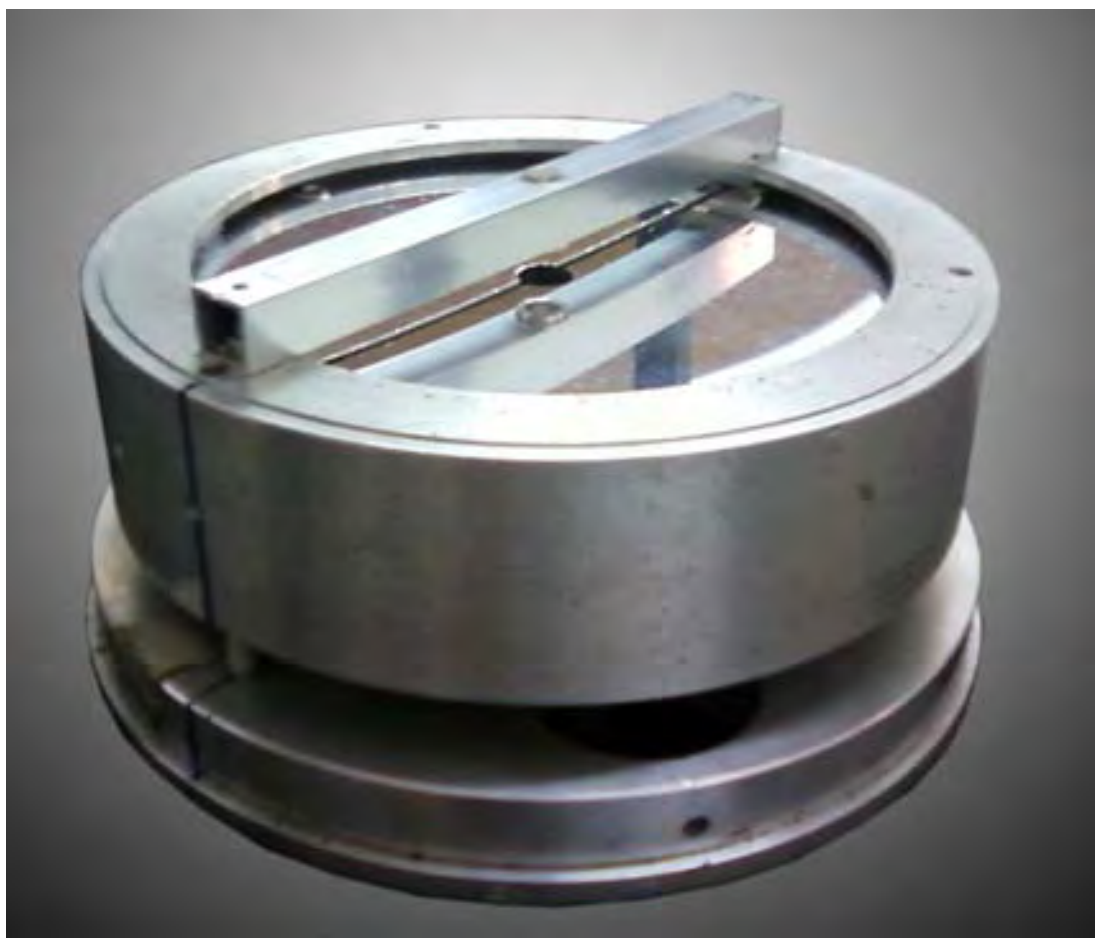

Fig. A-I.2: Suporte do cilindro sobre o protótipo do espelho heliométrico. A abertura central entre os hemi-espelhos permite que o cilindro seja iluminado.

Uma cruz feita com fios de nylon, de 15 μm de espessura, é posicionada sobre a superfície do disco, de forma que uma das direções fique alinhada com a direção de corte do espelho (referência para o deslocamento das imagens), figura A-I.3.

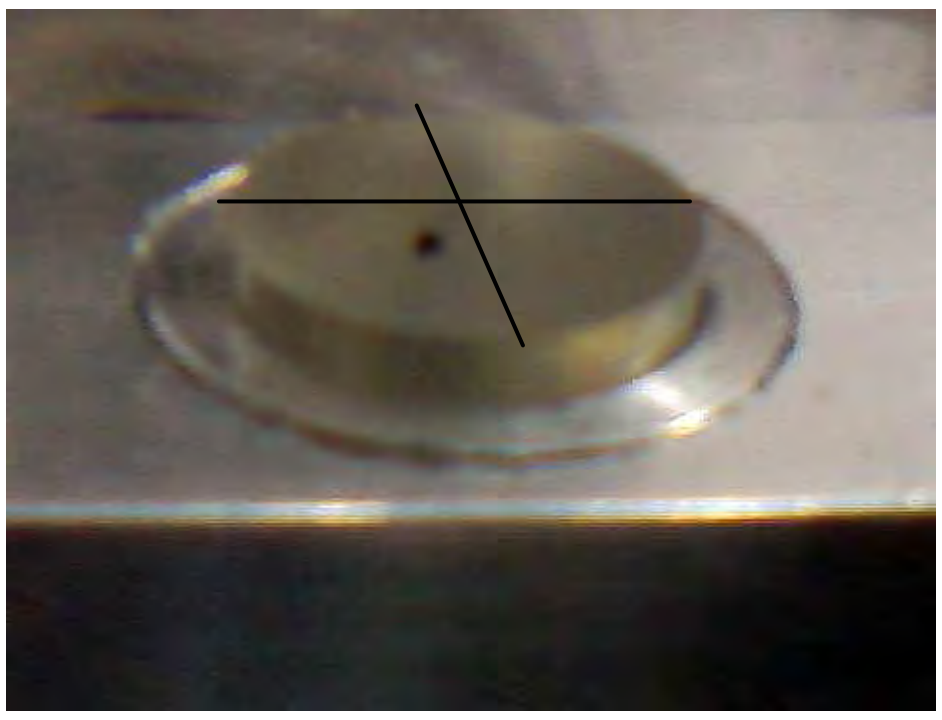

Fig. A-I.3: Detalhe do disco despolido preso ao suporte. As linhas sobre o cilindro representam a cruz de nylon não apareceu nitidamente na foto.

O espelho parabólico é colocado num suporte com parafusos de colimação para seu ajuste. O suporte é confeccionado de forma a se adaptar à abertura do tubo, figura A-I.4.

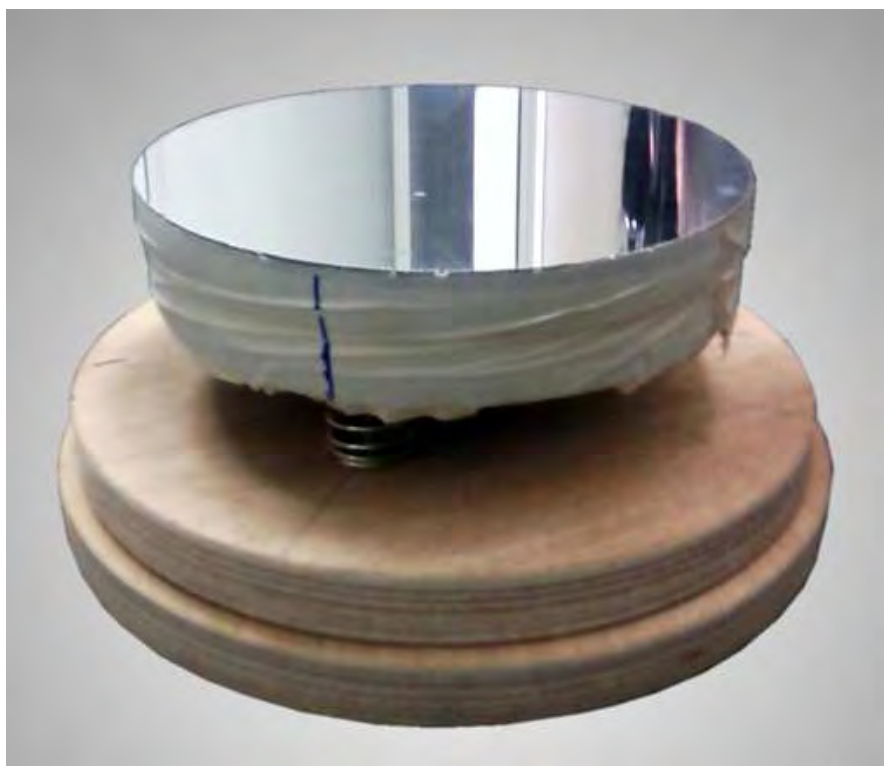

Fig. A-I.4: Espelho parabólico protótipo de auto-colimação fixo à sua base.

Com esta montagem mede-se o deslocamento das imagens provocado pelo diedro do hemi-espelhos em duas direções ortogonais, figura A-I.5:

- O deslocamento angular ao longo da direção paralela ao corte do espelho é determinado pela distância entre as bordas dos discos (fig. A-I.6, à direita). O monitoramento desta distância é usado na calibração.

- O deslocamento angular ao longo da direção ortogonal ao corte é avaliado pelo deslocamento vertical da imagem do fio (fig. A-I.6, à esquerda).

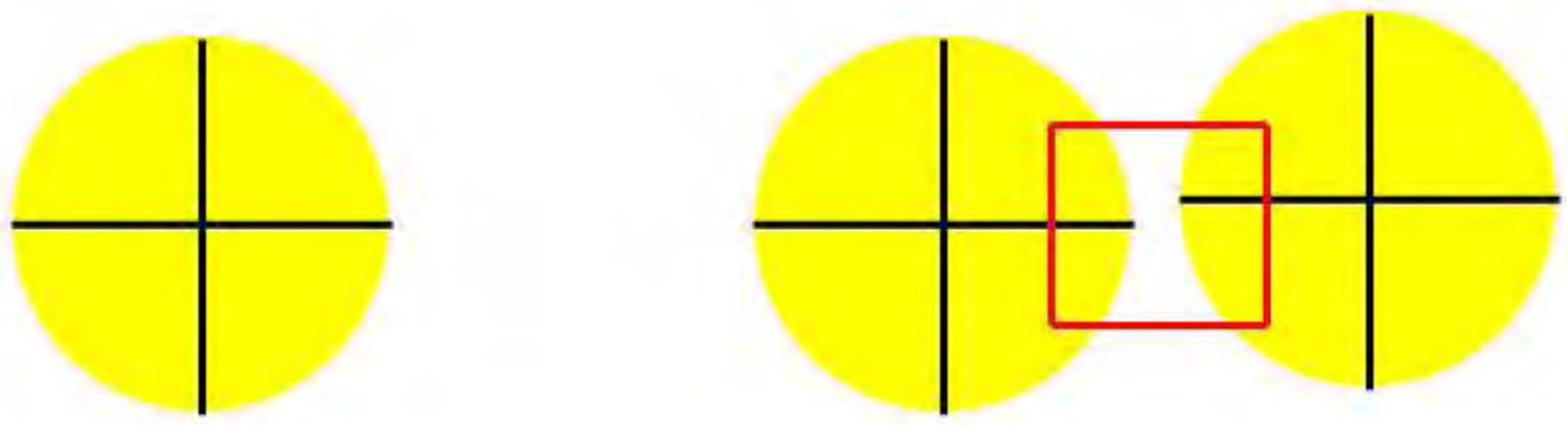

Fig. A-I.5: Representação do cilindro com a cruz de nylon e sua imagem heliométrica final. O retângulo, na última imagem, representa a área que é vista pelo CCD.

Na figura A-I.6 pode-se ver a imagem obtida pela câmara CCD do cilindro de CCZ-HS, com a linha ao longo de sua superfície (o cruzamento das linhas se dá próximo ao centro do disco, e não aparece neste procedimento de teste). Na imagem da esquerda, as setas indicam o deslocamento ortogonal à direção da medida. Na imagem à direita pode-se ver o resultado do programa de análise utilizado nas imagens heliométricas do disco padrão. Para que o programa funcione convenientemente, toda a superfície da imagem dos discos é alisada (cada ponto é a média ponderada dentro de uma área de 10 *pixels* de raio centrada no próprio ponto), restando apenas uma estreita faixa de 10 *pixels* ao longo das bordas. A pequena linha entre os discos indica a distância mínima entre eles.

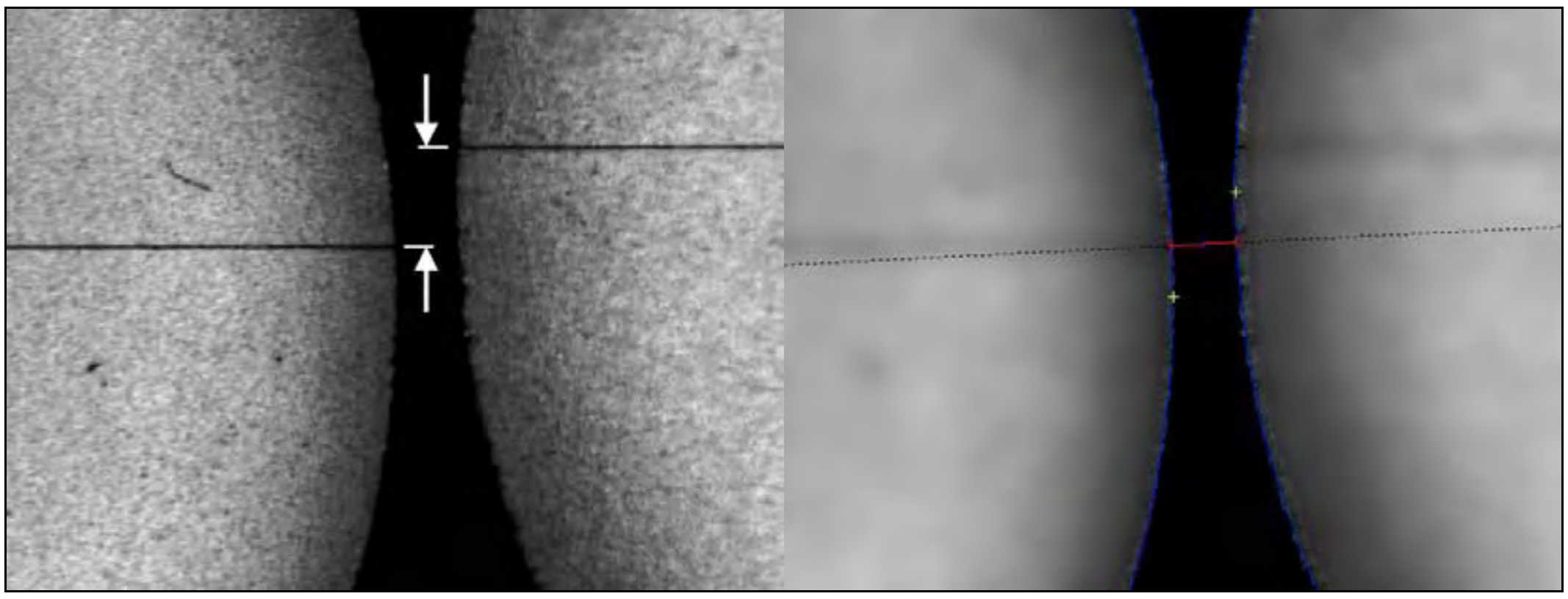

Fig. A-I.6: Imagem do CCD do cilindro de CCZ e do programa de análise das imagens.

Na auto-colimação, o tamanho das imagens depende da estabilidade das distâncias ao longo do eixo do instrumento. Para garantir esta estabilidade, o espelho de auto-colimação também é de CCZ-HS. Uma célula para contê-lo foi projetada e confeccionada em alumínio com pontos de contato de forma a se encaixar sempre da mesma maneira na entrada do tubo do instrumento.

O tubo do instrumento sendo de fibra carbono assegura a estabilidade mecânica necessária para a colimação.

## *Teste de comparação*

### **Montagem instrumental da experiência.**

No procedimento de comparação é usado um segundo espelho heliométrico para quase anular a ação do espelho principal, posicionado da mesma maneira que o espelho parabólico do teste de auto-colimação, figura A-I.7.

Como o cilindro está no foco do espelho heliométrico de comparação, desta vez, duas imagens reais de sua face despolida, afastadas uma da outra do valor do diedro daquele, estarão infinitamente distantes do espelho principal.

Se os espelhos fossem iguais em todas as suas características ópticas, o espelho principal recomporia as duas imagens em uma só no seu plano focal. Como isso é improvável, as imagens são recombinadas com pequenos deslocamentos residuais.

A estabilidade dos deslocamentos angulares residuais nos dá informações sobre a estabilidade de ambos os heliômetros.

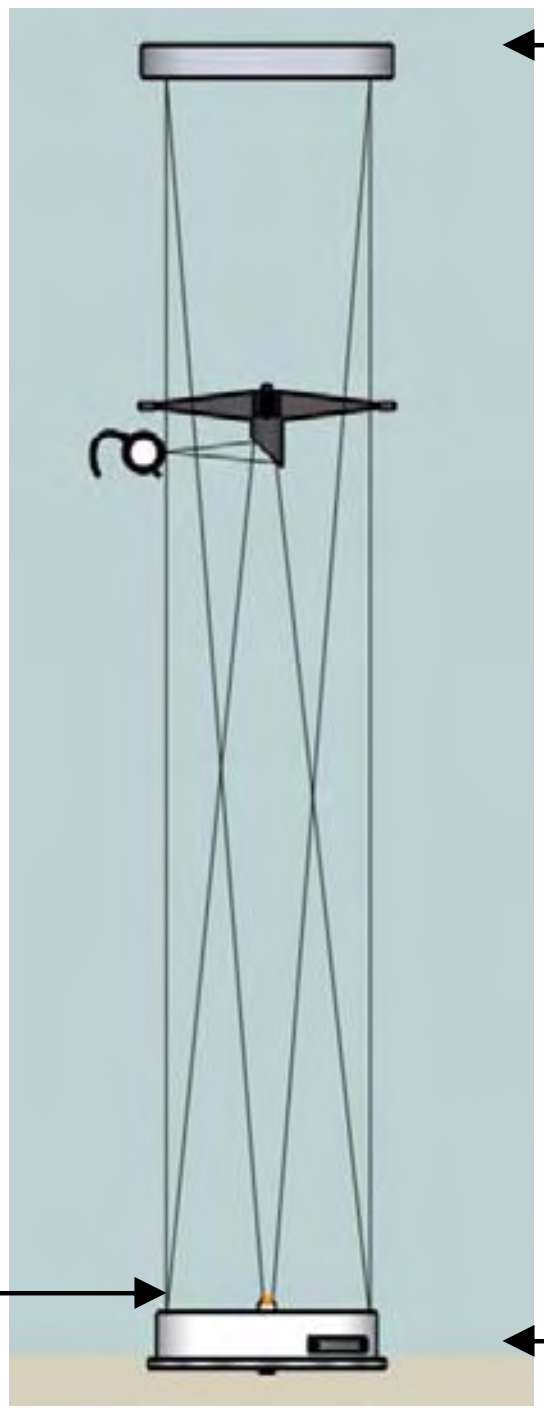

Fig. A-I.7: Caminho óptico da segunda montagem instrumental.

O diedro do espelho heliométrico de comparação foi confeccionado com pequenos calços (seu valor foi conferido pelo teste descrito na sessão 5.1.4.4) para ficar o mais próximo possível do valor do diedro do espelho heliométrico principal. A fixação do diedro segue o mesmo princípio da técnica anteriormente usada: molas comprimindo verticalmente os hemi-espelhos sobre os seus calços inferiores e calços laterais para conter os deslocamentos horizontais. O conjunto é fixado a um suporte com parafusos de colimação para seu ajuste. Este suporte é confeccionado de forma a se encaixar na abertura do tubo, figura A-I.8.

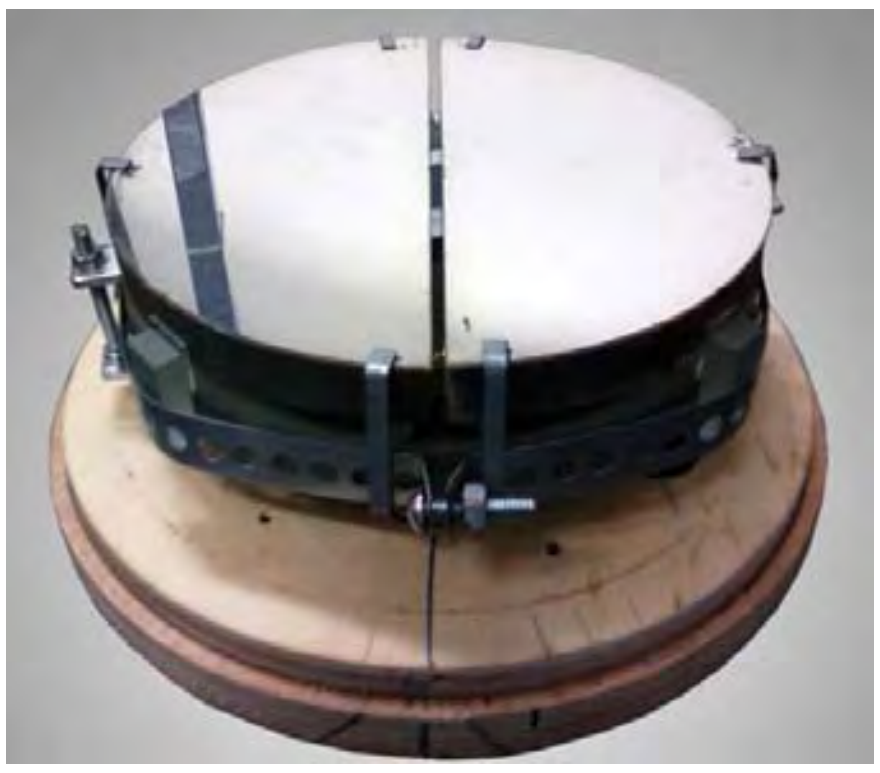

Fig. A-I.8: Espelho heliométrico de comparação.

A figura A-I.9 traz uma imagem de simulação do teste de comparação. Na imagem da figura A-I.10 se vê a porção central da face despolida do cilindro duplicada. A cruz de fios de nylon também aparece duplicada e deslocada em direção ao canto superior direito da imagem.

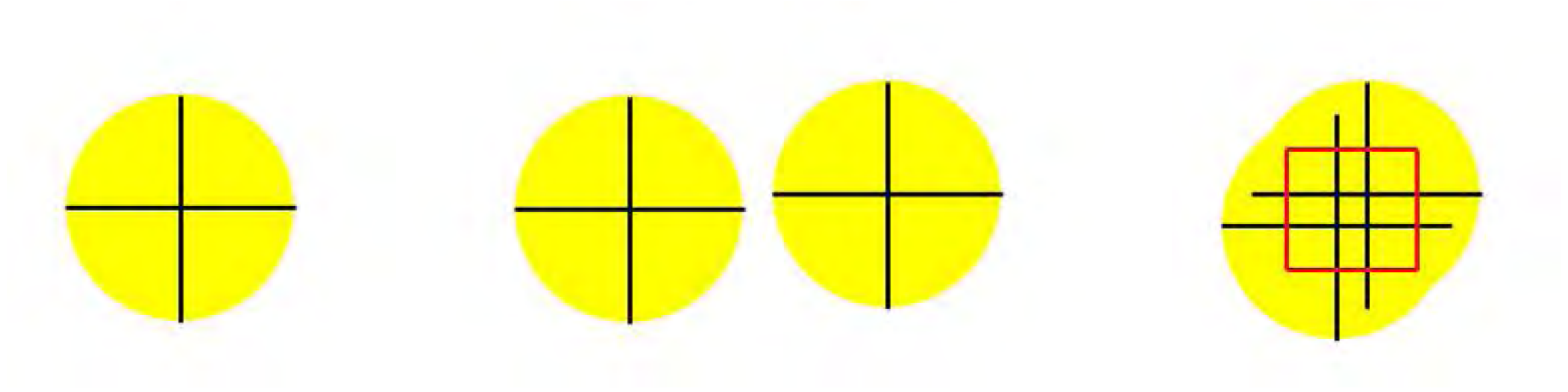

Fig. A-I.9: Representação do cilindro com a cruz de nylon e a imagens heliométricas, intermediária e final. O retângulo, na última imagem, representa o que é visto pelo CCD.

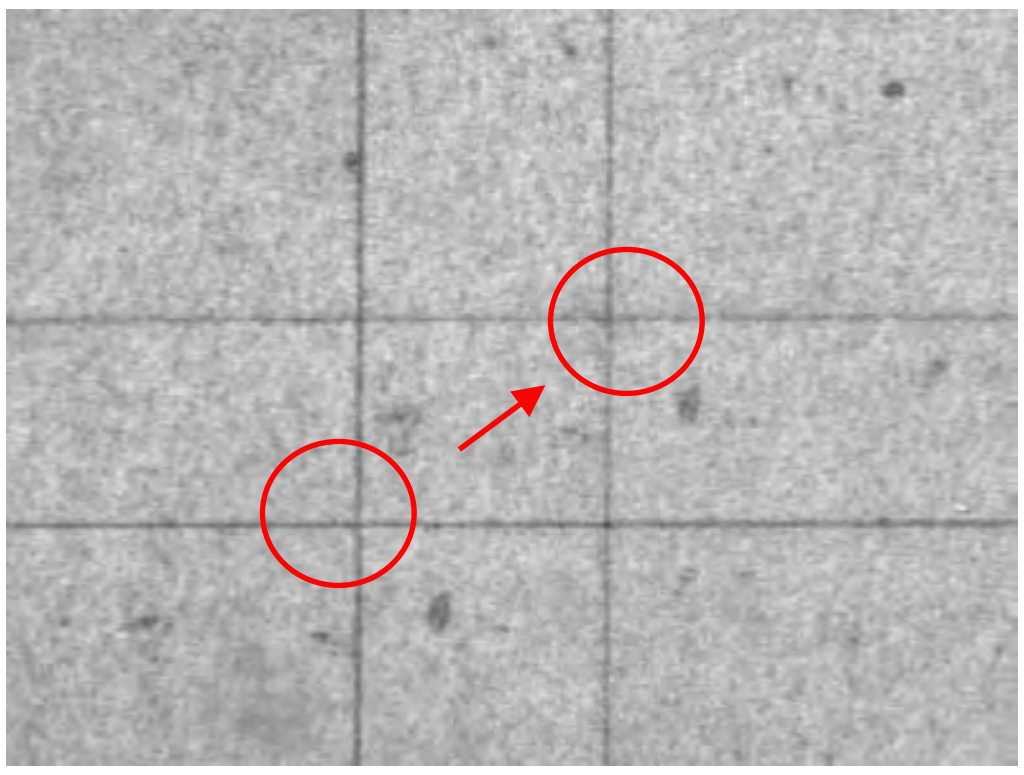

Fig. A-I.10: Imagem do CCD do cilindro de CCZ. As indicações mostram o deslocamento residual da cruz central.

Com esta montagem também é possível avaliar este deslocamento residual ao longo das duas direções ortogonais.

Para garantir a estabilidade dimensional, o espelho heliométrico de comparação também é feito em CCZ-HS e acamado em sua célula com a mesma metodologia usada no espelho principal.

Na imagem da figura A-I.11, se vê o registro de uma característica interessante e muito vantajosa do método heliométrico: a notável independência de suas medidas em relação às variações da posição focal (como explicado no Capítulo 2).

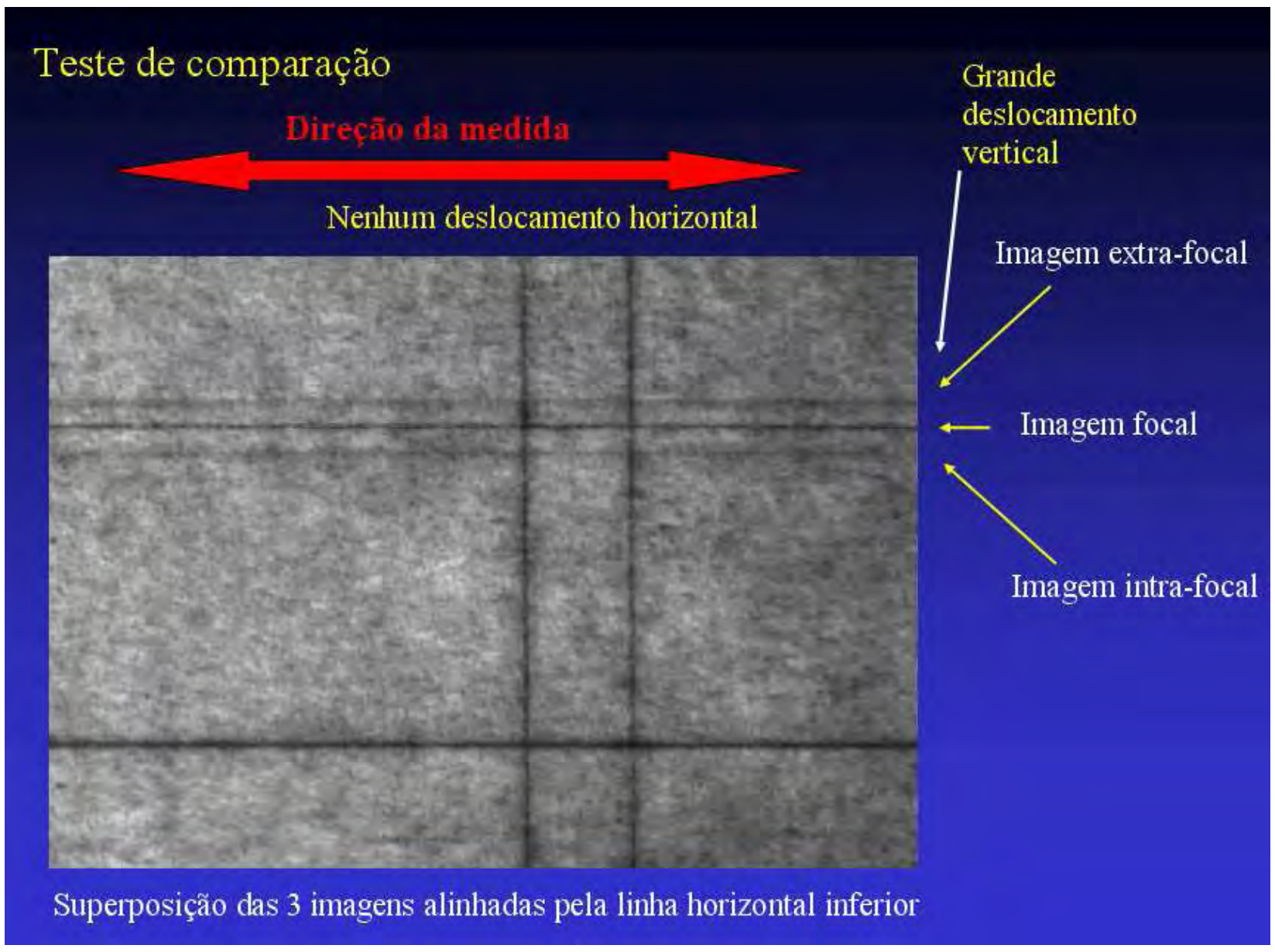

Fig. A-I.11: Superposição de imagens mostrando a estabilidade das imagens na direção em que é feita a medida heliométrica.

Aqui temos três imagens sobrepostas correspondentes à colocação da câmera CCD, em três condições, em relação ao espelho principal:

1. No plano focal;
2. Em uma posição extra-focal;
3. Em uma posição intra-focal.

As imagens foram alinhadas pela linha horizontal inferior para facilitar a visualização deste efeito. Percebe-se que, apesar do deslocamento aplicado à câmara, a distância entre os fios verticais não se alterou. Ao longo da vertical, por outro lado, há uma forte dependência com o processo de focalização, como se pode ver pelo grande deslocamento dos fios horizontais.

## *Algumas suposições*

### Suposição 1

Agora vamos supor que a inevitável turbulência atmosférica esteja fazendo o plano focal do espelho variar aleatoriamente nas direções intra e extra-focal.

Que efeito isto teria sobre a medida da distância mínima entre as imagens heliométricas do Sol?

Pelo descrito acima, podemos esperar que os discos não se afastem horizontalmente, mas que se movam em direções opostas verticalmente, aumentando o valor da distância mínima.

No cálculo a seguir usamos os seguintes valores, baseados na experiência com o Heliômetro de testes, para avaliar este erro na medida:

Escala de placa: 1,15"/*pixel*

Diedro: 2017"

Coordenada x fixa do centro do disco esquerdo: -561 *pixels*

Coordenada x fixa do centro do disco direito: +1201 *pixels*

Coordenada y variável do centro do disco esquerdo: [+234,5;+244,5] *pixels*

Coordenada y variável do centro do disco esquerdo: [+244,5;+234,5] *pixels*

Obs.: A origem das coordenadas é o canto superior esquerdo da imagem do CCD.

O resultado é apresentado na tabela A-I.1:

Tabela A-I.1: Diferença no valor da distância mínima com o deslocamento vertical dos centros dos discos.

| Deslocamento vertical entre os centros (") | Diferença no valor medido da distância mínima (") |
|---|---|
| -11,5 | 0,033 |
| -9,2 | 0,021 |
| -6,9 | 0,012 |
| -4,6 | 0,005 |
| -2,3 | 0,001 |
| 0 | 0 |
| 2,3 | 0,001 |
| 4,6 | 0,005 |
| 6,9 | 0,012 |
| 9,2 | 0,021 |
| 11,5 | 0,033 |

O erro devido à turbulência é, em sua definição, não direcional. O experimento mostra que, além disso, para pequenas variações de orientação, o efeito é da ordem $10^{-3}$ da própria variação em uma medida isolada.

## Suposição 2

O espelho heliométrico de comparação, assim como o espelho principal, possui uma direção de referência que é a direção do corte, a direção do desdobramento da imagem. Temos que supor que ao ser encaixado na frente do tubo, para o teste, exista uma flutuação do ângulo entre as direções dos cortes dos dois espelhos.

Que efeito isto tem sobre a medida das imagens de comparação?

O tubo do telescópio tem 125 mm de raio. Se presumirmos que a precisão do encaixe mecânico é de 0,1 mm, isto leva a um desalinhamento da ordem de ~3 minutos de arco.

A figura A-I.12 mostra a geometria deste problema: um ponto **p** da imagem é deslocado pelo primeiro espelho de cerca de 2000" e retorna ao ponto **p'** pelo segundo espelho, com uma diferença de ângulo θ.

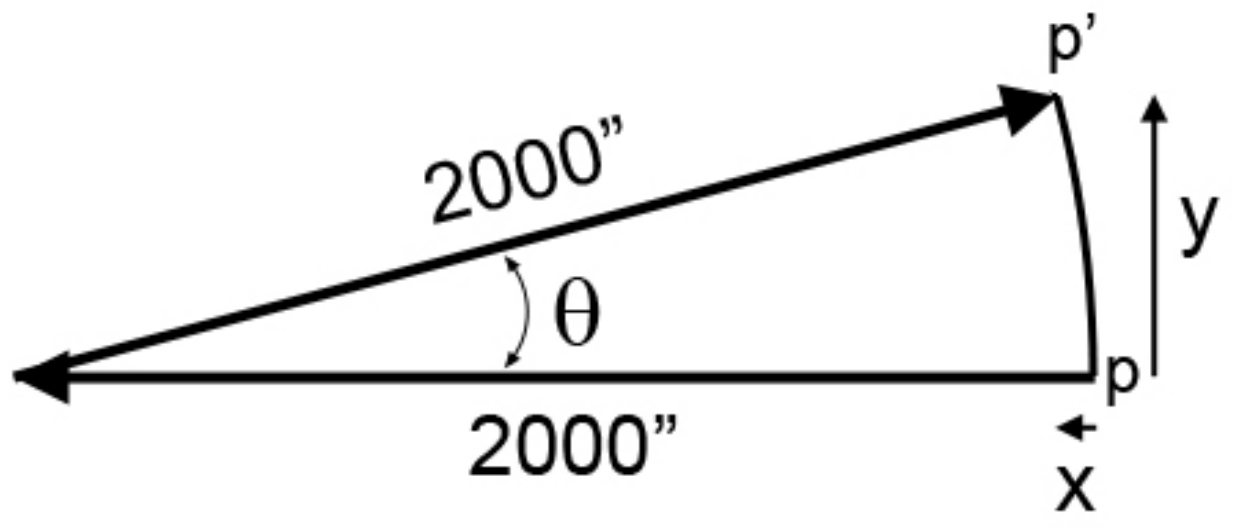

Fig. A-I.12: Geometria do erro de alinhamento entre os espelhos heliométricos.

O deslocamento x, na direção da medida heliométrica para um θ = 3' será:

$$x = 2000''\left[1 - \cos(3')\right] \approx 0{,}001''$$

Ao contrário do efeito ligado à turbulência, aqui, um erro de desalinhamento deve ser suposto como de forte tendência sistemática. No entanto, como a mecânica fina pode fornecer um encaixe significativamente melhor que o aqui exemplificado, o erro cometido sobre a medida do ângulo diedro é desprezível.

# Apêndice II - Projeto do Pilar do Heliômetro

O pilar foi projetado em conjunto com a equipe técnica do Observatório Nacional, responsável pela sua construção.

## *O Projeto*

Com um desenho de curvas suaves, ele serve de base para a montagem equatorial do Heliômetro, figura A-II.1.

Uma placa suporte, presa ao corpo da montagem, se encaixa em três barras rosqueadas de 7\8", separadas de 120° uma da outra. Estas barras, por construção, estão alinhadas com o eixo polar da Terra, mas uma pequena folga no encaixe da placa permite o ajuste fino e regulagem do alinhamento da montagem com a direção do pólo.

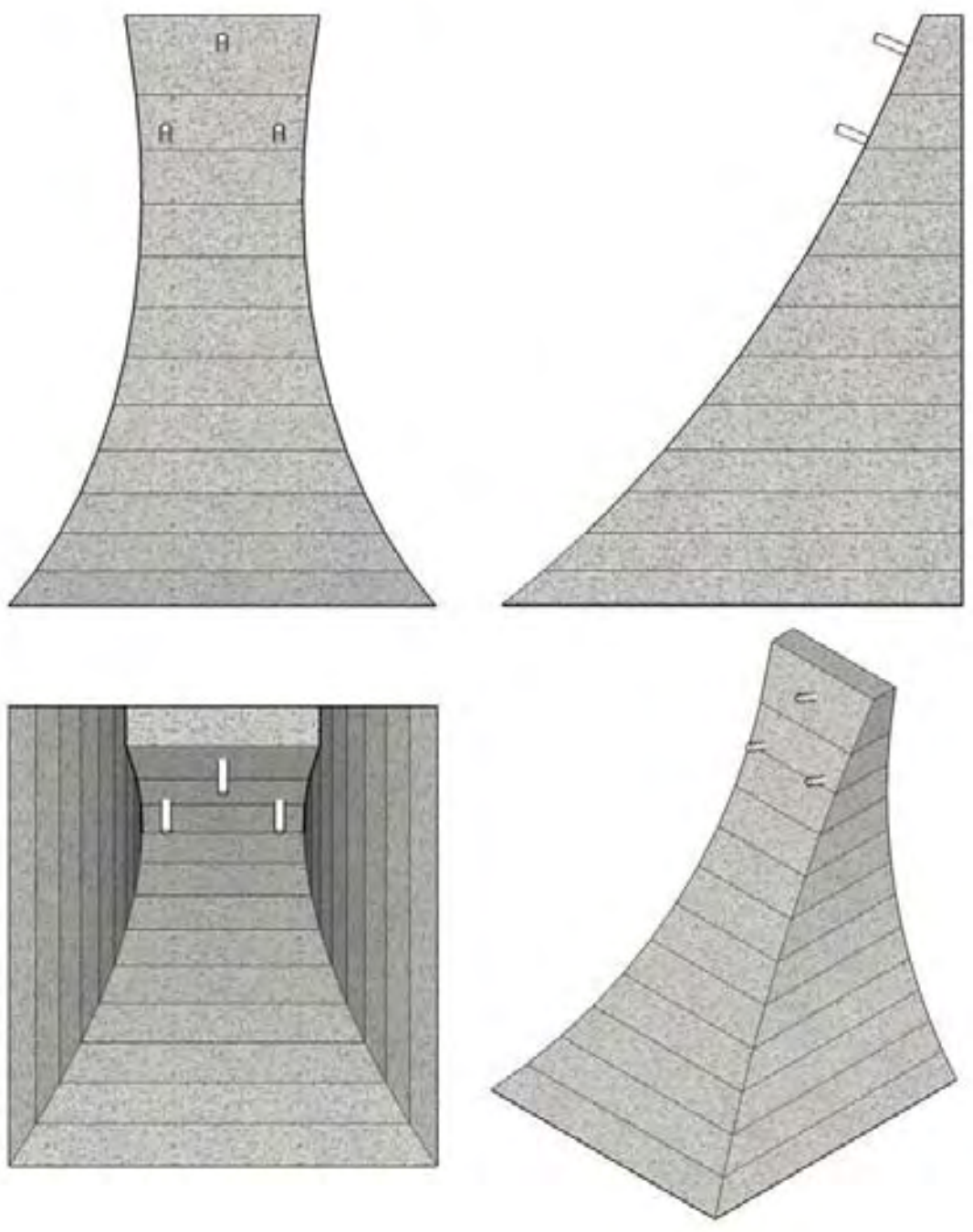

Fig. A-II.1: Vistas do projeto do pilar do Heliômetro.

O pilar é construído em concreto. Internamente vigas metálicas reforçam sua estrutura (fig. A-II.2). As cotas de toda a estrutura, assim como a escolha dos materiais, ficaram a cargo da equipe técnica do Observatório Nacional. O requerimento instrumental ordenado foi: durabilidade na escala de década, carga superior a duas vezes o peso total do instrumento e estabilidade comensurável com a base do pilar, pré-existente e incrustada na rocha, além de compatibilidade com a recomendação dos Institutos do Patrimônio Histórico (IPHAN e INEPAC) de adequação ao estilo arquitetônico original. Dutos internos prevêem a circulação de fiação elétrica, de telefonia, e de transporte de dados.

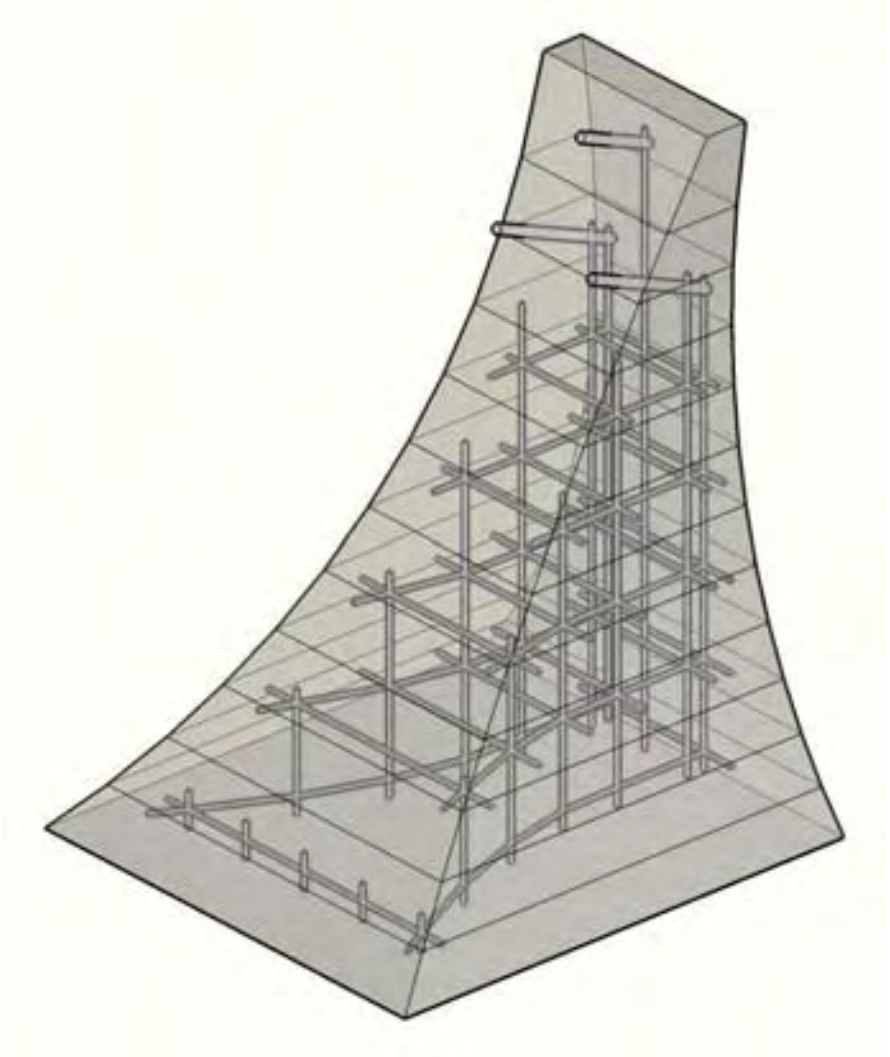

Fig. A-II.2: Vista transparente do projeto do pilar mostrando sua estrutura interna de reforço.

## *Seqüência de construção*

### 5 de junho de 2009

A montagem equatorial do heliômetro protótipo foi totalmente desmontada, figura A-II.3.

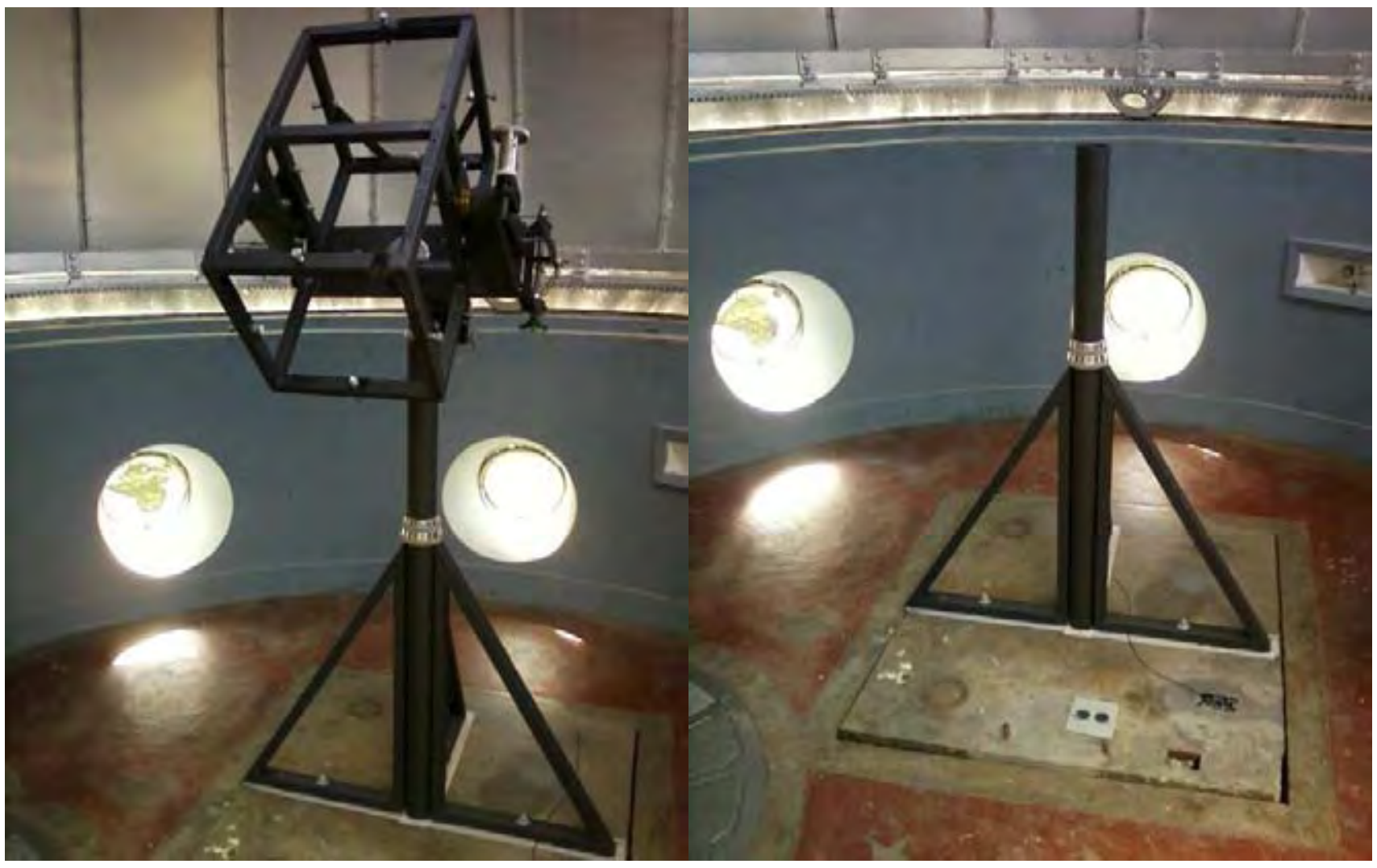

Fig. A-II.3: Retirada da montagem do heliômetro de testes.

## 16 de junho de 2009

A estrutura de vigas internas para o reforço estrutural foi montada com os dutos para a passagem dos cabos elétricos, de telefonia e de rede, figura A-II.4.

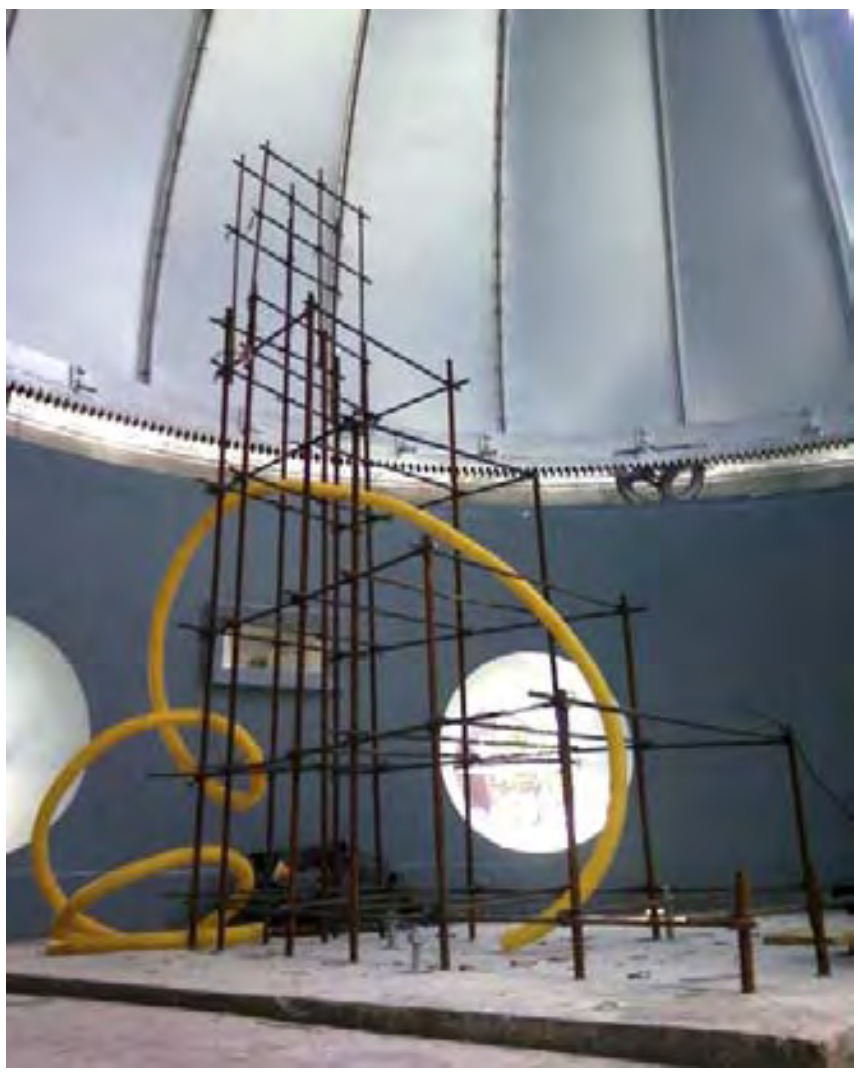

Fig. A-II.4: Foto da estrutura de reforço do pilar.

## 17 de junho de 2009

Um molde em madeira foi confeccionado na oficina do Observatório Nacional e depois levado para a cúpula, onde "vestiu" a estrutura metálica, figuras A-II.5 e A-II.6.

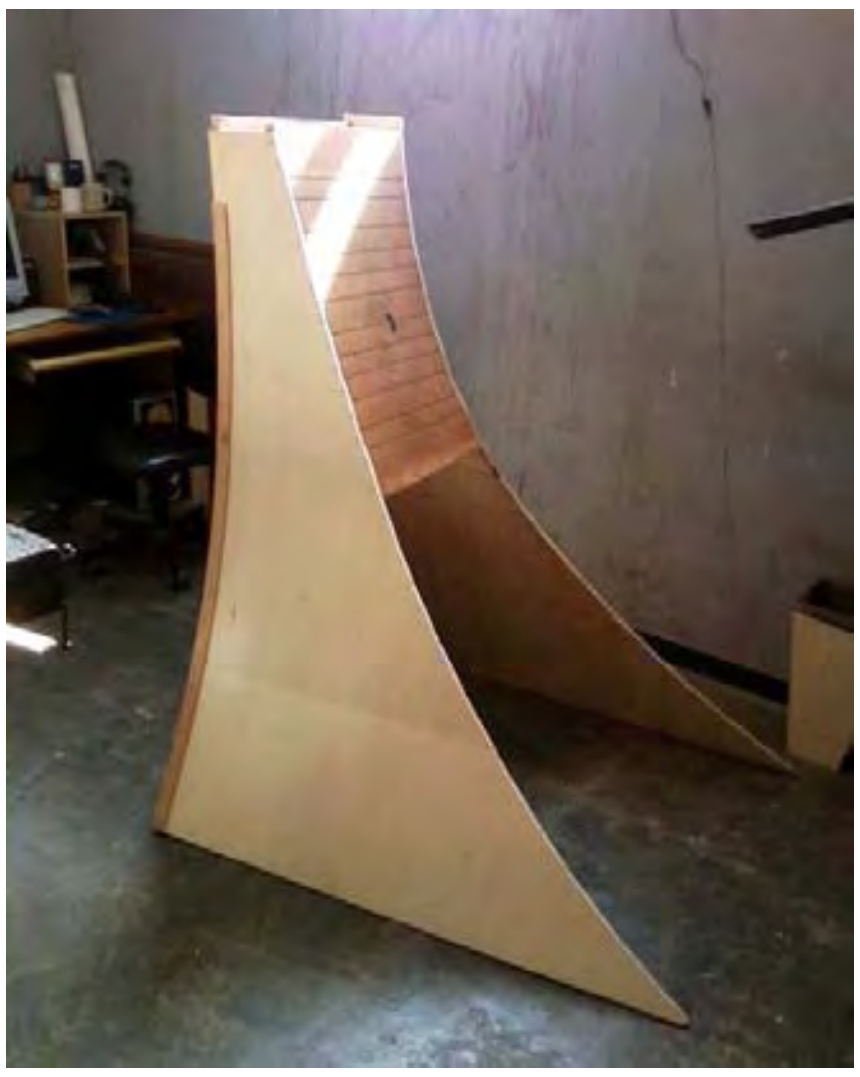

Fig. A-II.5: Molde em madeira do pilar.

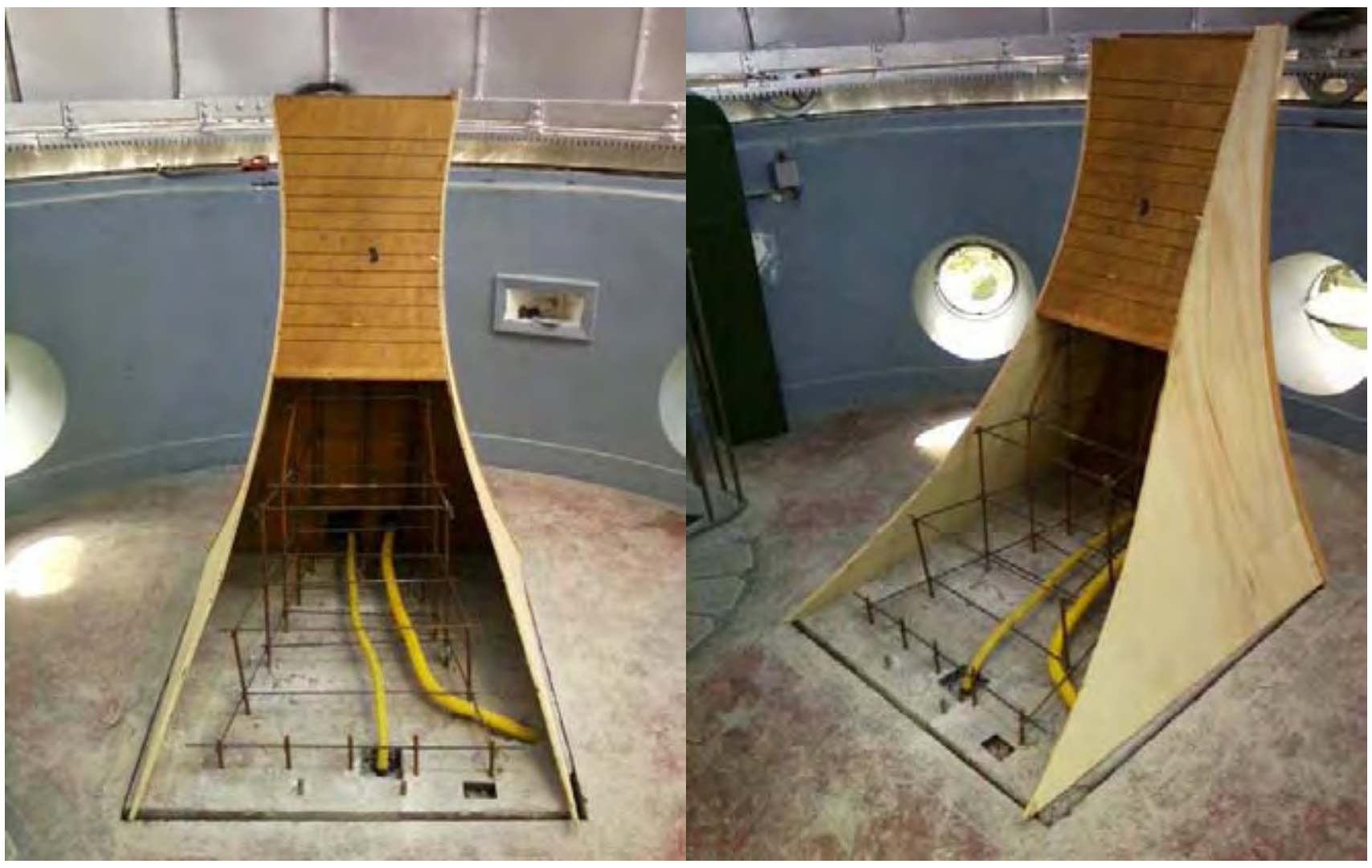

Fig. A-II.6: Molde já posicionado sobre a estrutura de reforço.

## 19 de junho de 2009

Um gabarito de madeira, simulando a placa suporte, foi feito para posicionar corretamente as barras que sustentam a montagem e mantê-las na posição correta durante a concretagem e sua cura (fig. A-II.7). Estas barras foram firmemente amarradas às ferragens internas.

A parte de baixo anterior do molde foi fechada para receber a primeira camada de concreto, figura A-II.8.

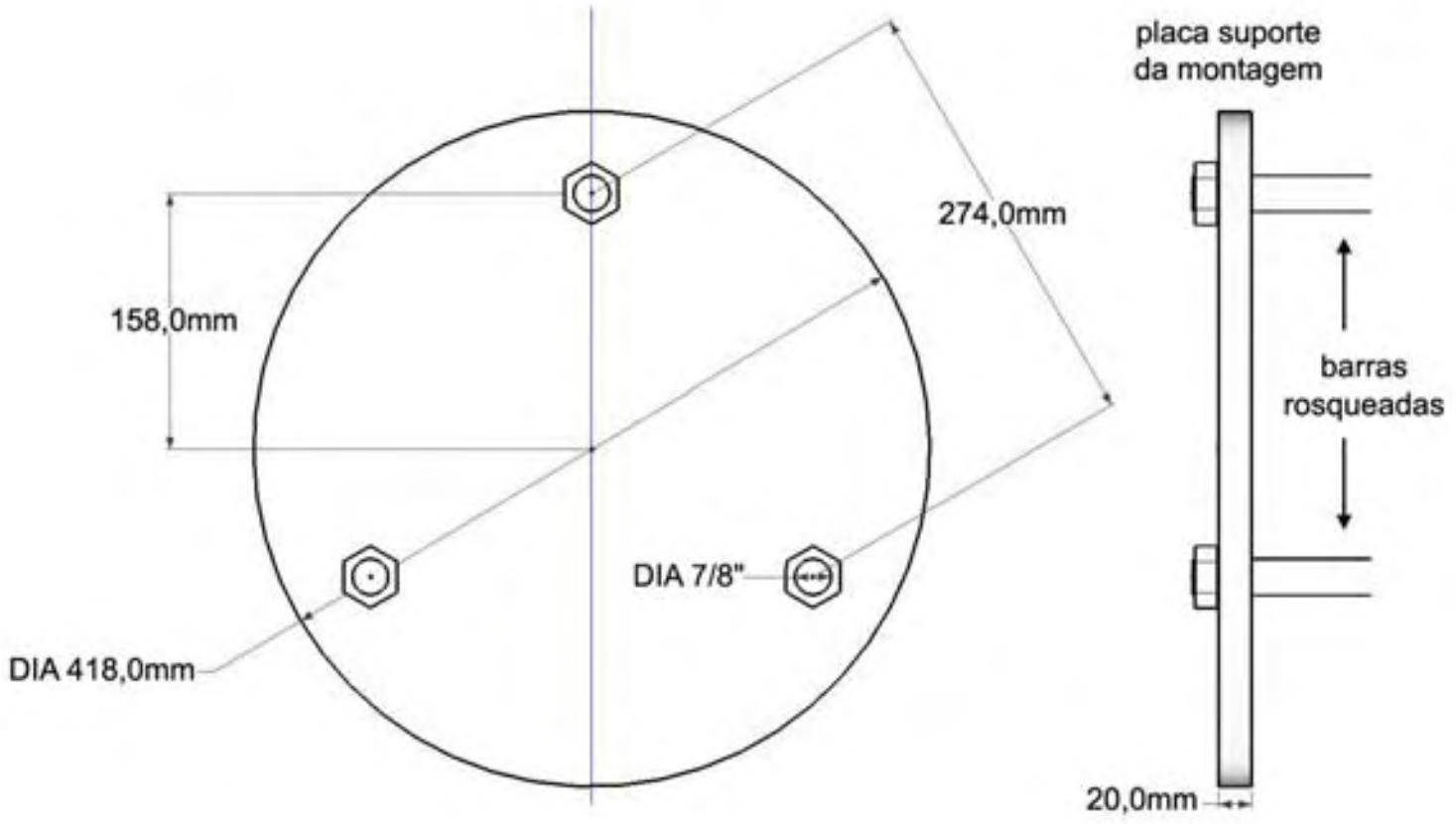

Fig. A-II.7: Projeto da placa suporte da montagem. Feito de ferro fundido, é cromada para o acabamento superficial.

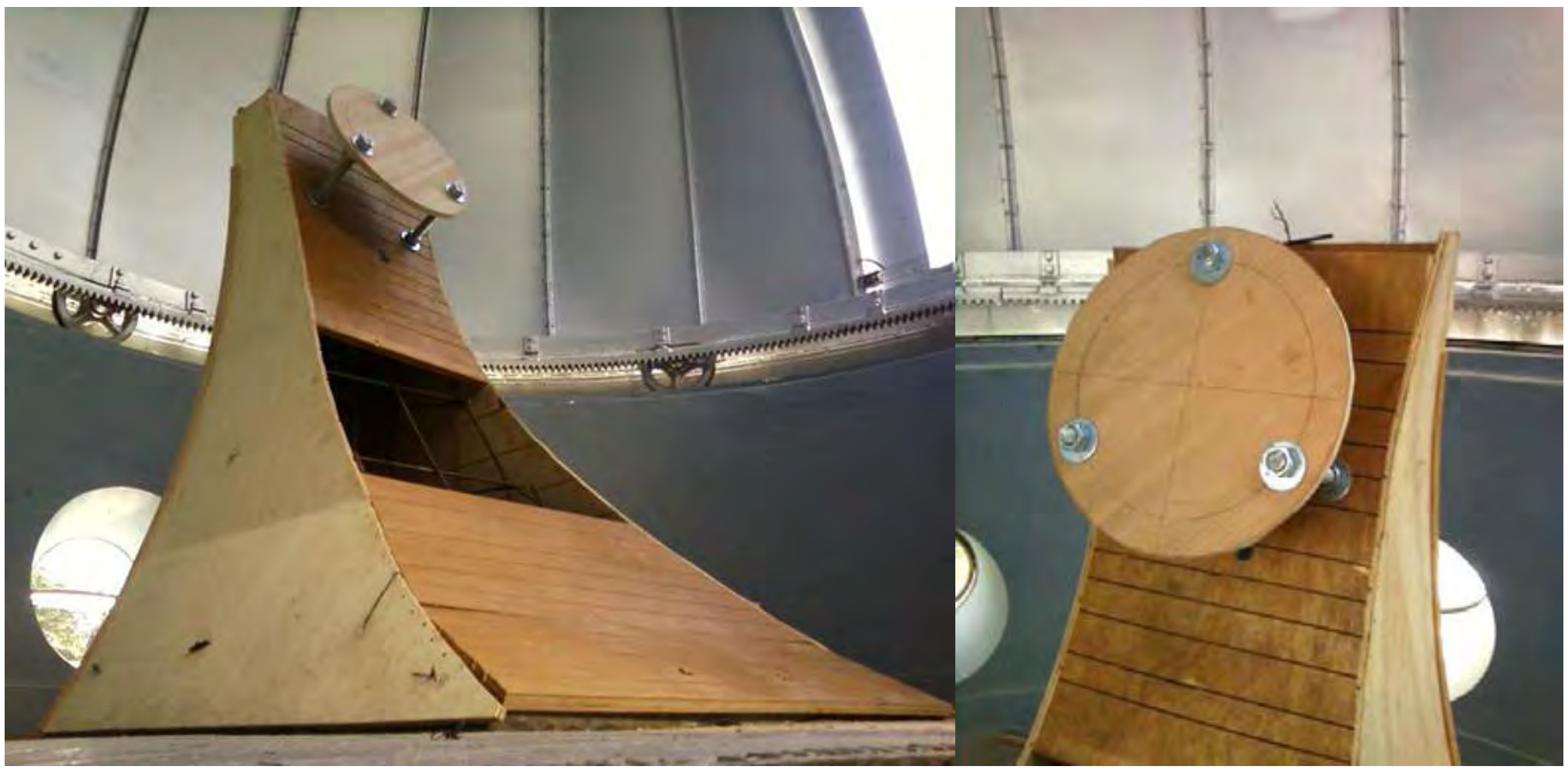

Fig. A-II.8: Gabarito de posicionamento das barras rosqueadas e a parte de baixo do molde fechada.

## 22 de junho de 2009

A primeira camada de concreto foi aplicada na parte da manhã e, à tarde, a abertura do molde foi fechada para ser preenchido completamente, figura A-II.9 e A-II.10.

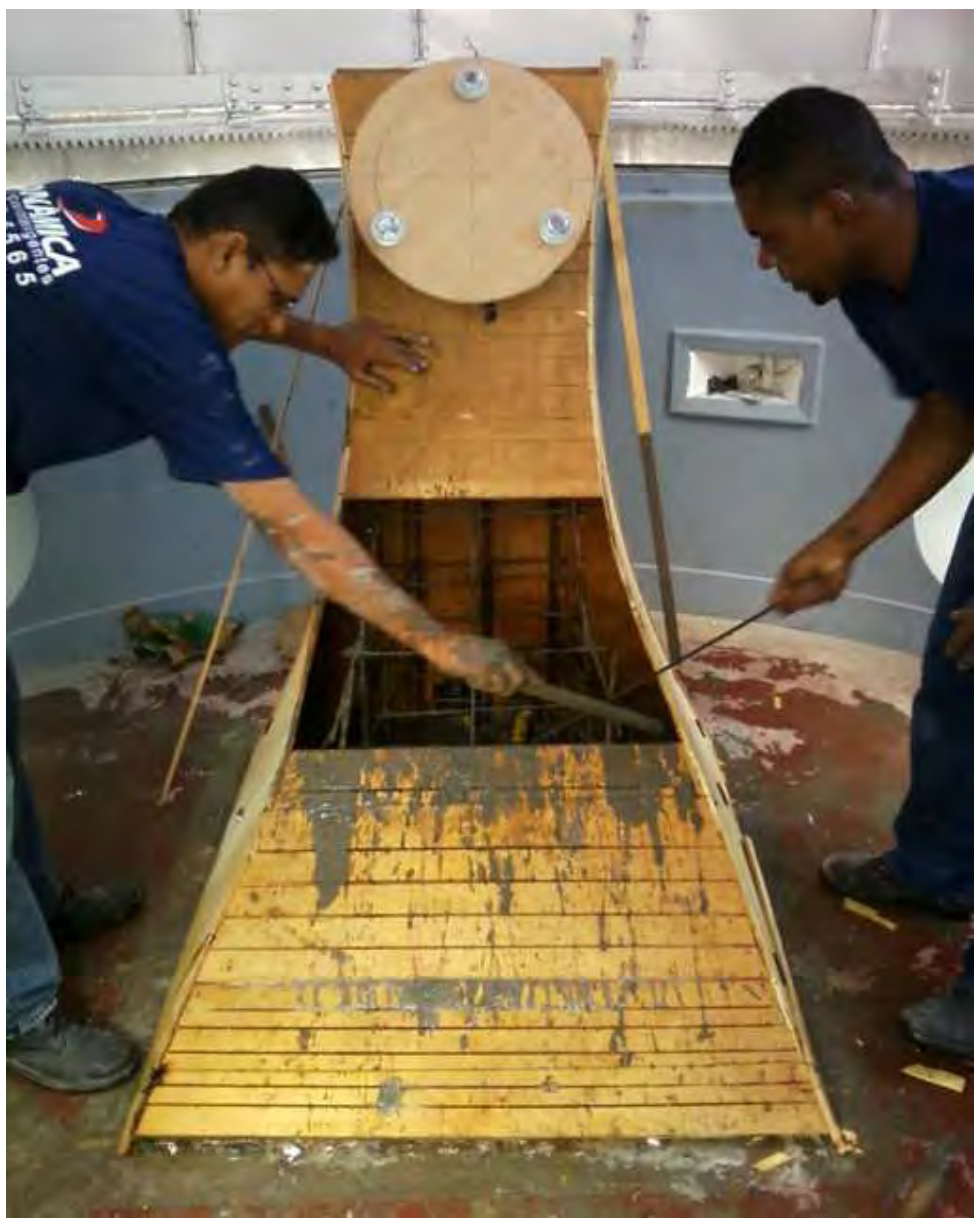

Fig. A-II.9: Aplicação da primeira camada de concreto.

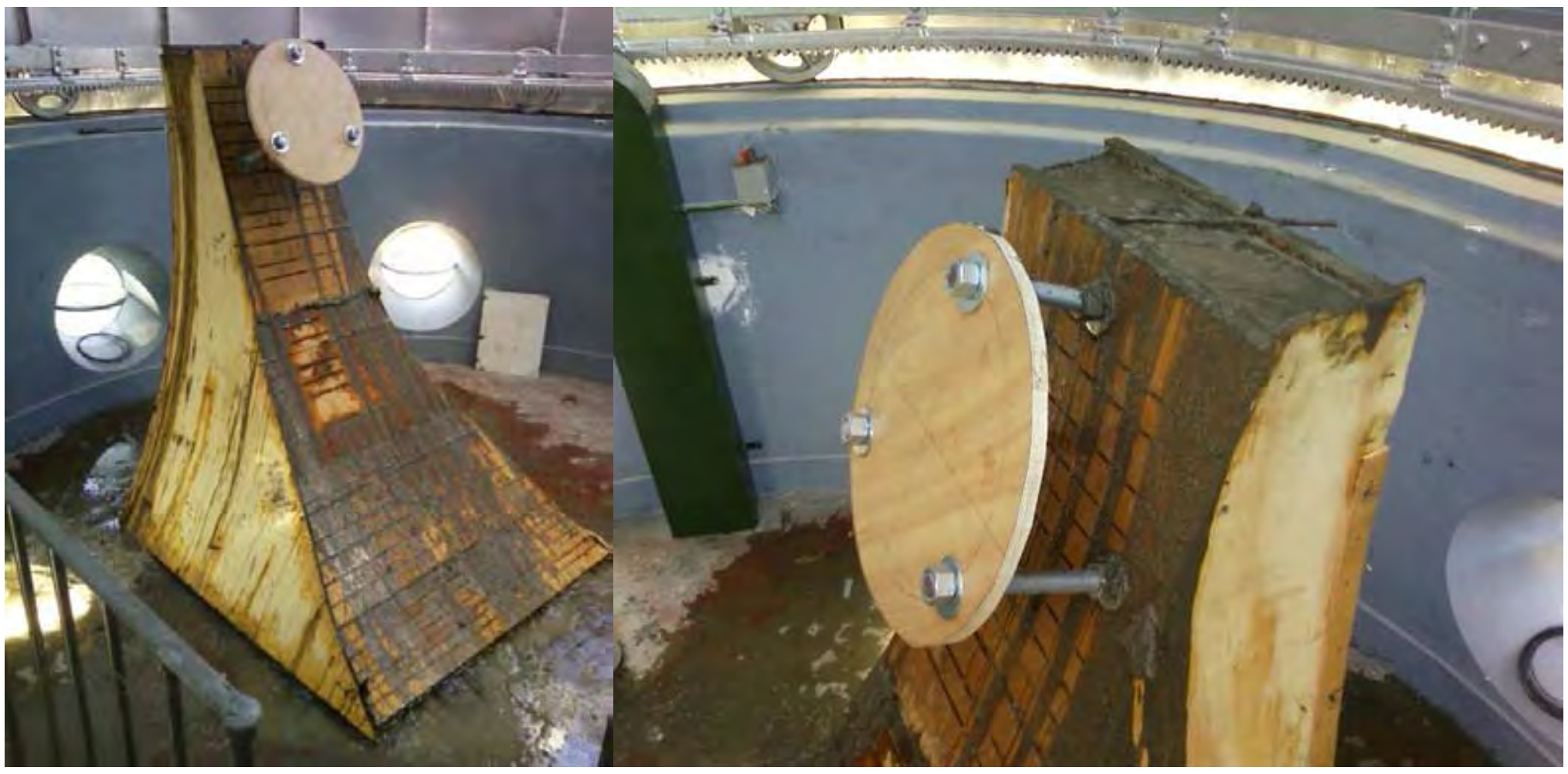

Fig. A-II.10: Molde completamente preenchido por concreto.

## 30 de junho de 2009

Depois da cura completa do concreto, o molde foi retirado, expondo o pilar que, então, recebeu acabamentos superficiais, figura A-II.11.

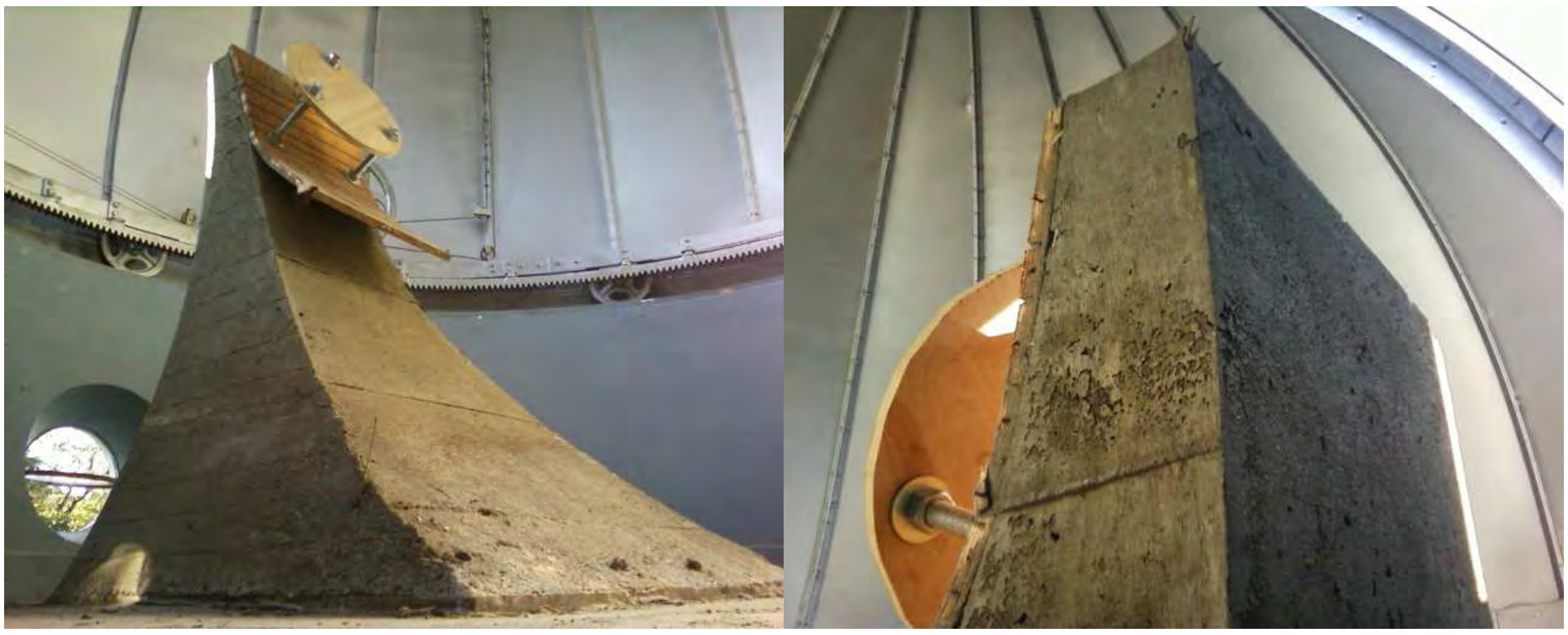

Fig. A-II.11: Superfície do concreto após a cura.

## 24 de julho de 2009

A base da montagem foi aparafusada à placa suporte e o conjunto foi encaixado nas barras rosqueadas, figura A-II.12.

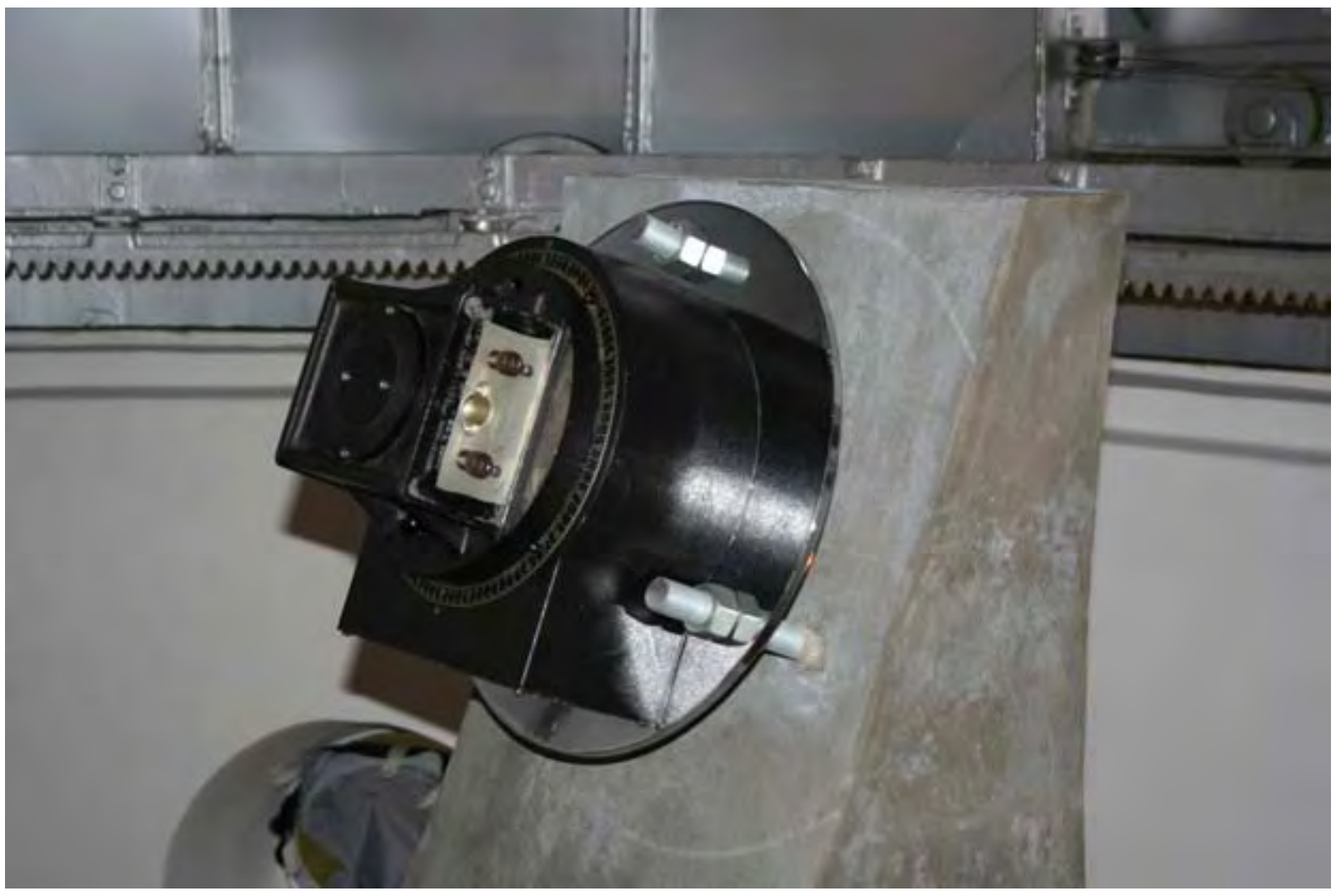

Fig. A-II.12: Base da montagem e placa suporte fixadas ao pilar.

Uma vez fixada a base da montagem, os braços do garfo foram aparafusados e a montagem ficou completa, figura A-II.13.

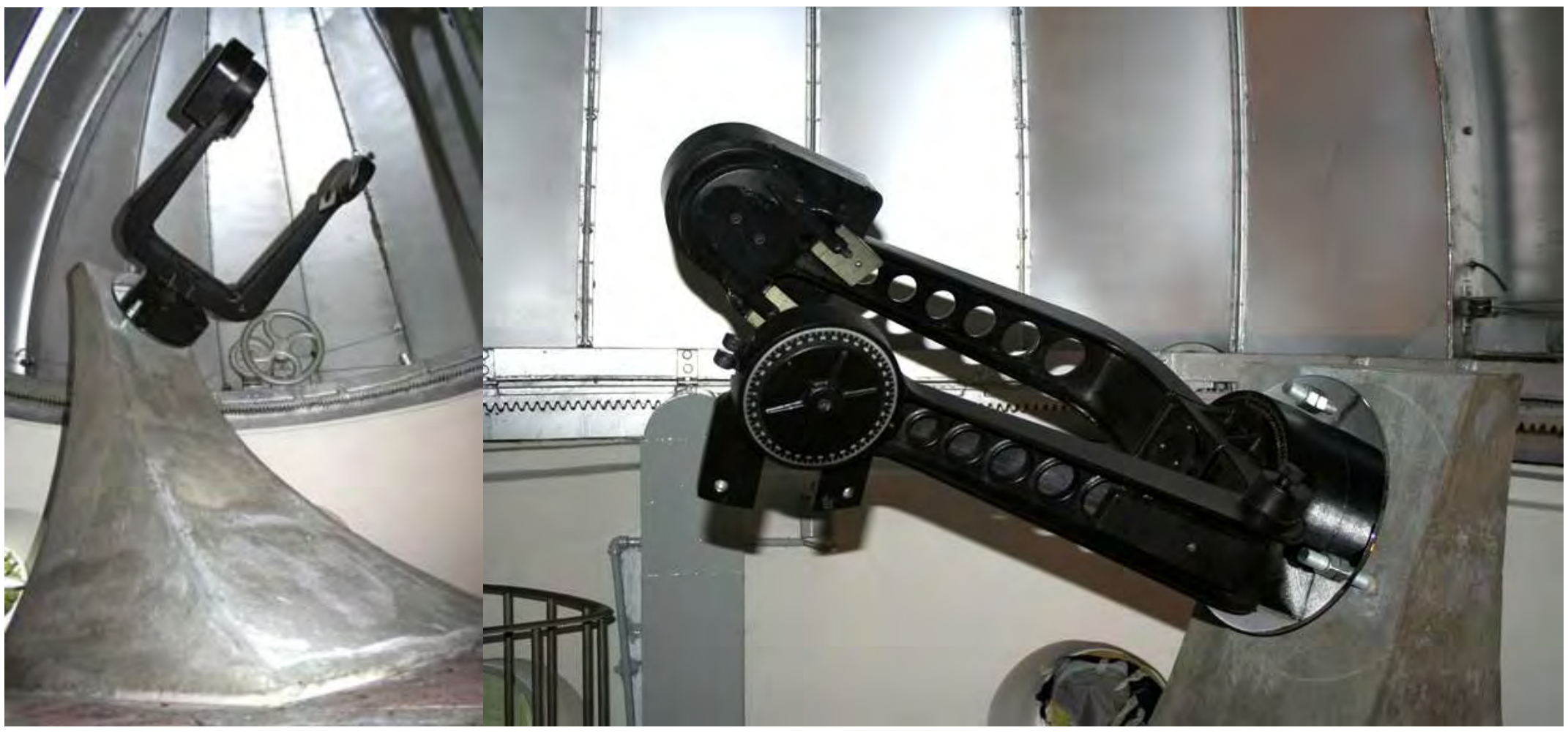

Fig. A-II.13: Montagem equatorial de garfo em seu pilar definitivo.

# Apêndice III - Desbaste e polimento das faces planas dos discos de CCZ-HS

Dois discos de CCZ-HS têm apenas uma das faces aplainadas (desbastadas e polidas). O terceiro disco, que serve de plano óptico, tem suas duas faces aplainadas, figura A-III.1.

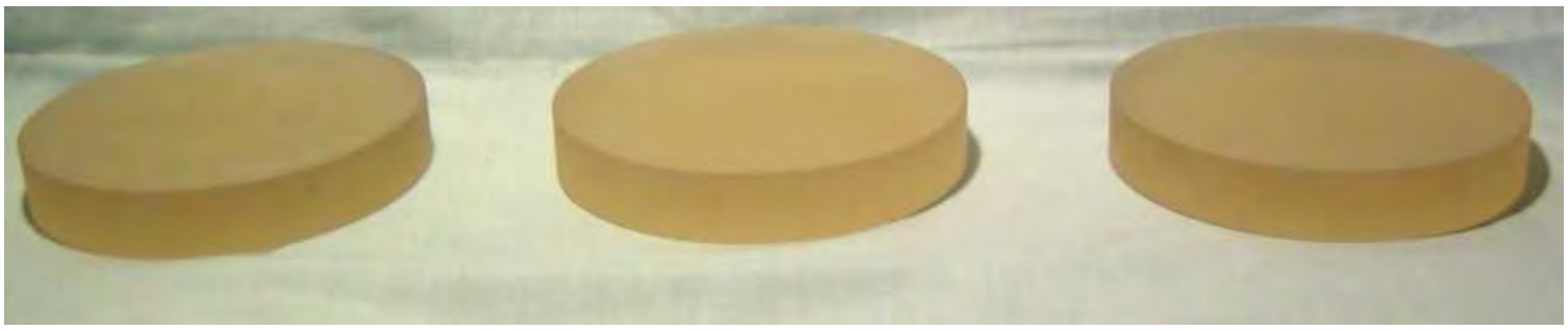

Fig. A-III.1: Discos de CCZ-HS originais do fabricante.

## *Desbaste*

Para o desbaste das faces foi usado grão de esmeril 2000, cujo diâmetro do grão é de ~12 μm. Empregando equipamento servo-mecânico, os discos foram desbastados contra uma base de aço mecanicamente plana, ou seja, de imperfeições inferiores a 1 μm, figura A-III.2.

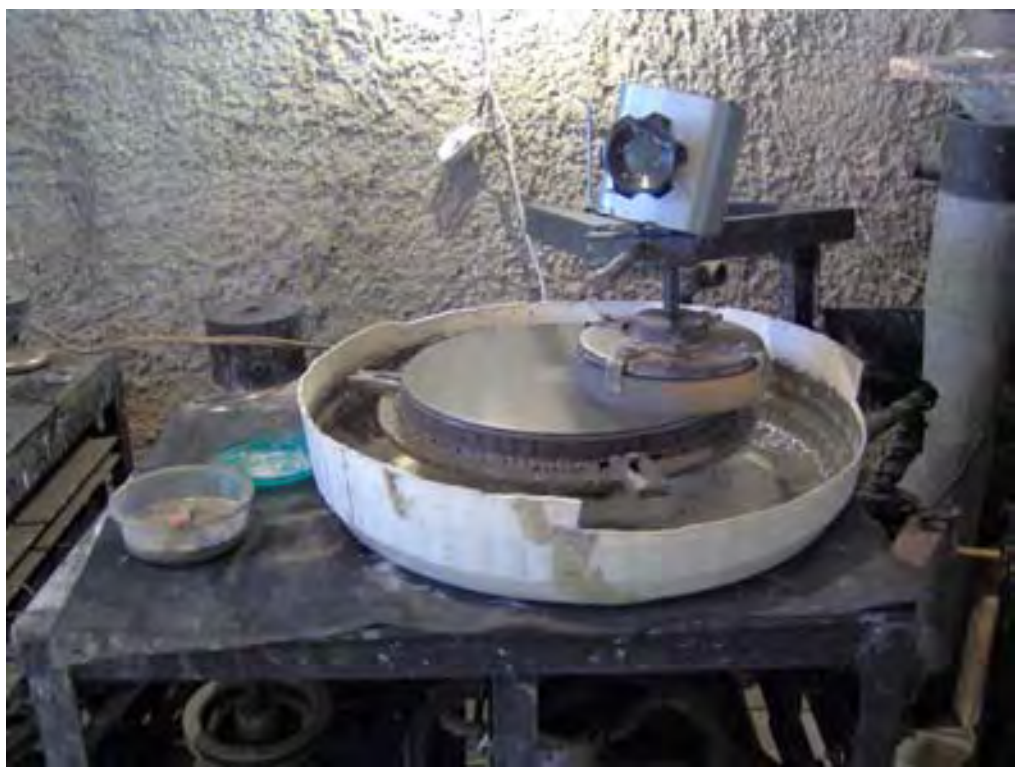

Fig. A-III.2: Disco de CCZ-HS preso à maquina de desbaste/polimento.

O processo é continuado, intercalado por sessões de verificação, até que o esferômetro não apontasse mais curvatura nas faces, com uma precisão de 5 μm, figura A-III.3.

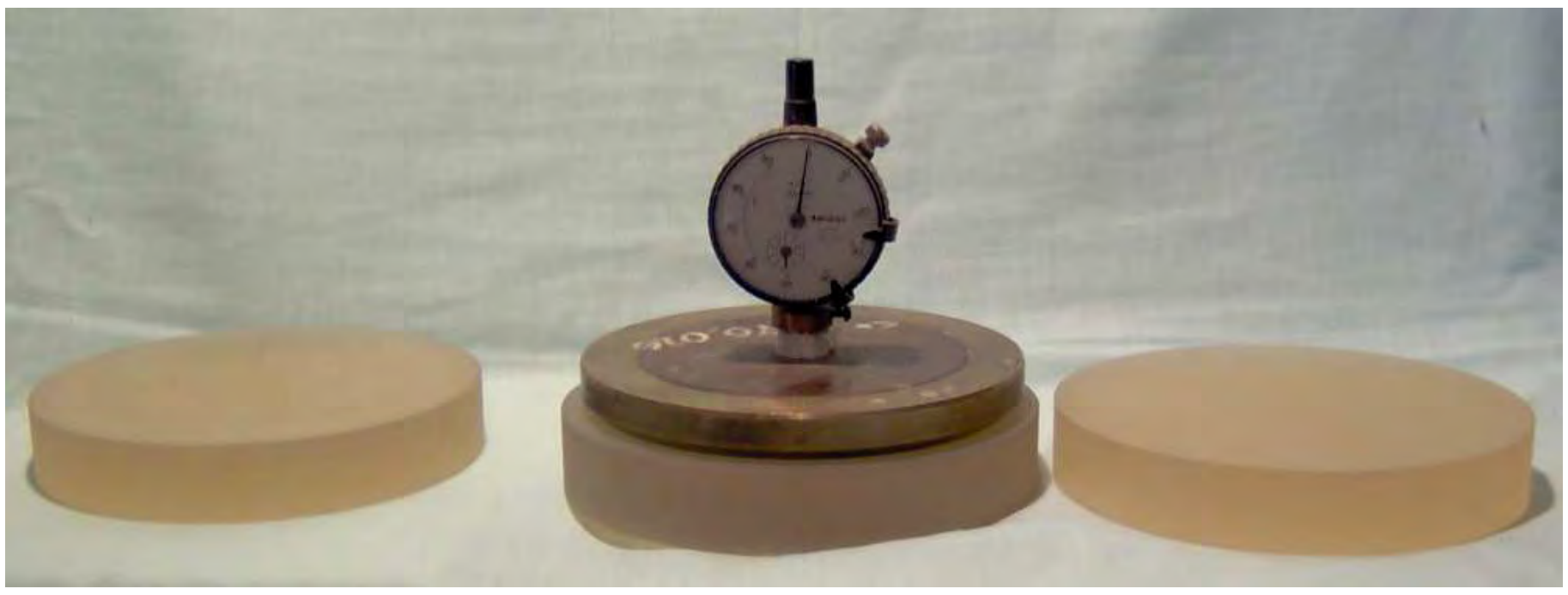

Fig. A-III.3: Esferômetro medindo a marca-zero de curvatura em um dos discos.

## *Teste óptico do plano base dos hemi-espelhos*

Para este teste de laboratório foi usado um Interferômetro de Newton, constituído de um plano óptico de referência (com 6" de diâmetro, feito em quartzo de 25,4 mm de espessura e planicidade de $\lambda/20$) e um fonte de luz verde (~530 nm), figura A-III.4.

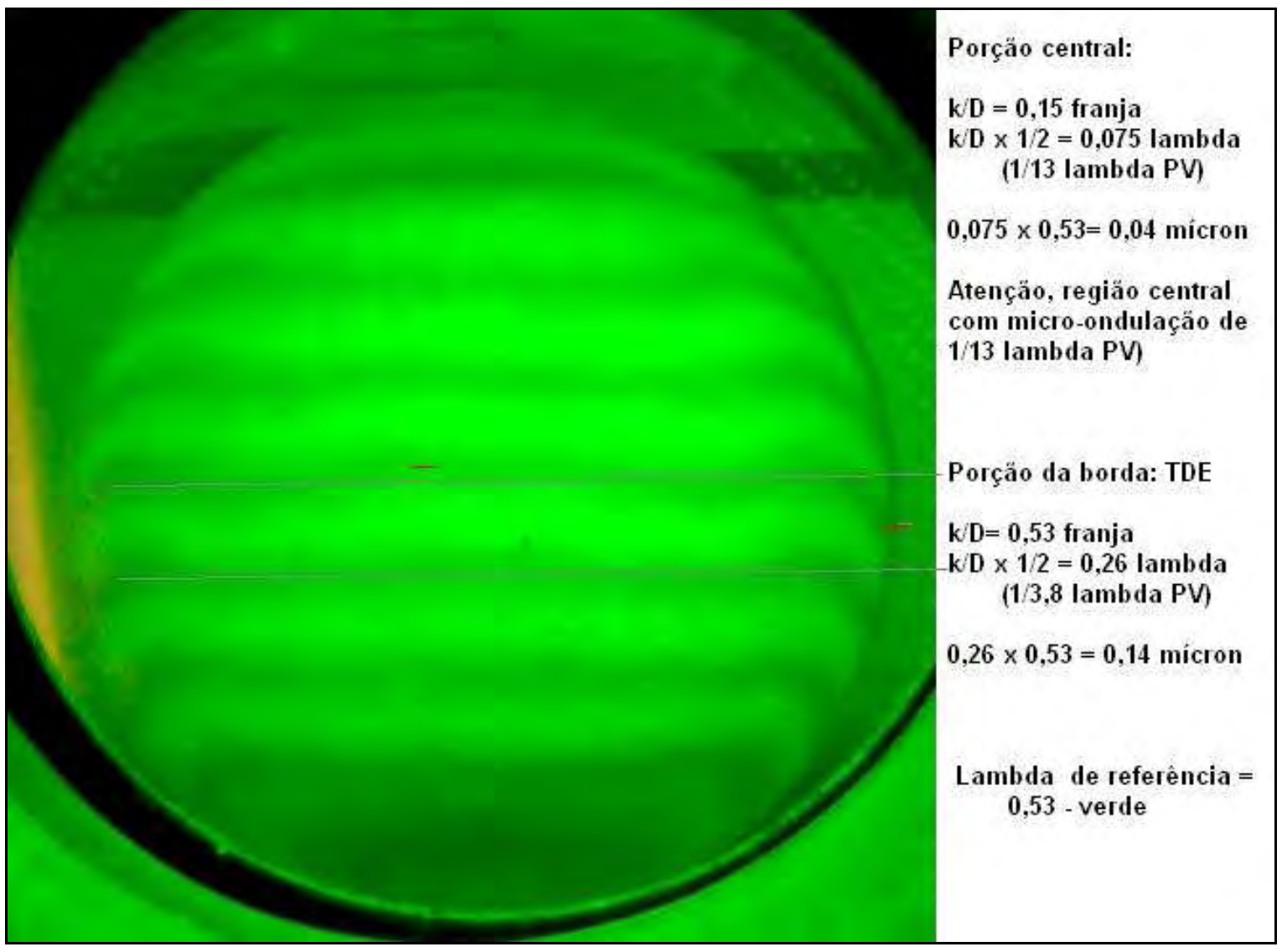

Fig. A-III.4: Teste interferométrico no disco base dos hemi-espelhos.

O resultado final aponta um erro pico-vale na porção central da superfície em torno de $\lambda/13$ (~0,04$\mu$m) e um erro pico-vale na região da borda em torno de ~$\lambda/4$ (~0,13$\mu$m).

Em conclusão, este plano óptico, que é a base onde os hemi-espelhos do heliômetro são assentados, atende de sobremaneira às especificações e exigências do projeto.

# Apêndice IV - Fabricação da mesa especial para o corte do espelho do Heliômetro e para a fabricação do diedro

Tendo em vista a especificidade do objetivo, quanto a dimensões, ausência de tensionamento e fissuras, foi projetada uma mesa especial para que o corte do espelho objetivo.

Depois do corte, cada metade é inclinada, no mesmo plano do corte, com suavidade e precisão, sem tencionar o raio de curvatura da superfície do espelho. Uma vez atingida a inclinação desejada, todo o conjunto é travado mecanicamente.

A figura A-IV.1 traz a concepção do projeto:

- a mesa é presa a uma peça convexa, que desliza sobre um suporte côncavo;
- os parafusos horizontais inclinam as mesas, forçando contra uma peça fixa;
- os parafusos de travamento não são vistos no desenho.

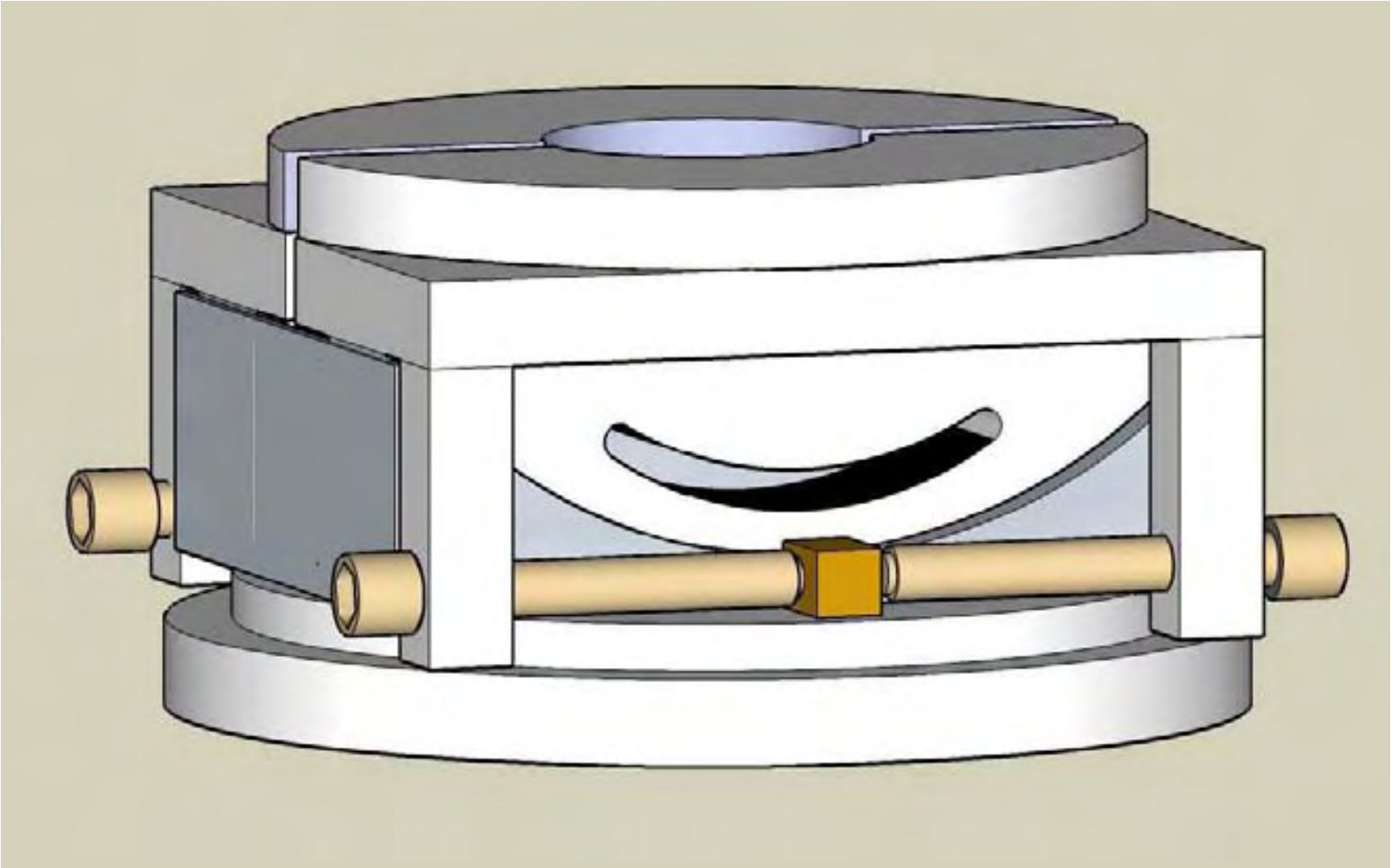

Fig. A-IV.1: Projeto final da mesa de corte e inclinação.

## *Seqüência de construção das peças e montagem da mesa*

Inicialmente, o disco de alumínio que recebe o espelho do Heliômetro foi usinado, figuras A-IV.2, A-IV.3 e A-IV.4.

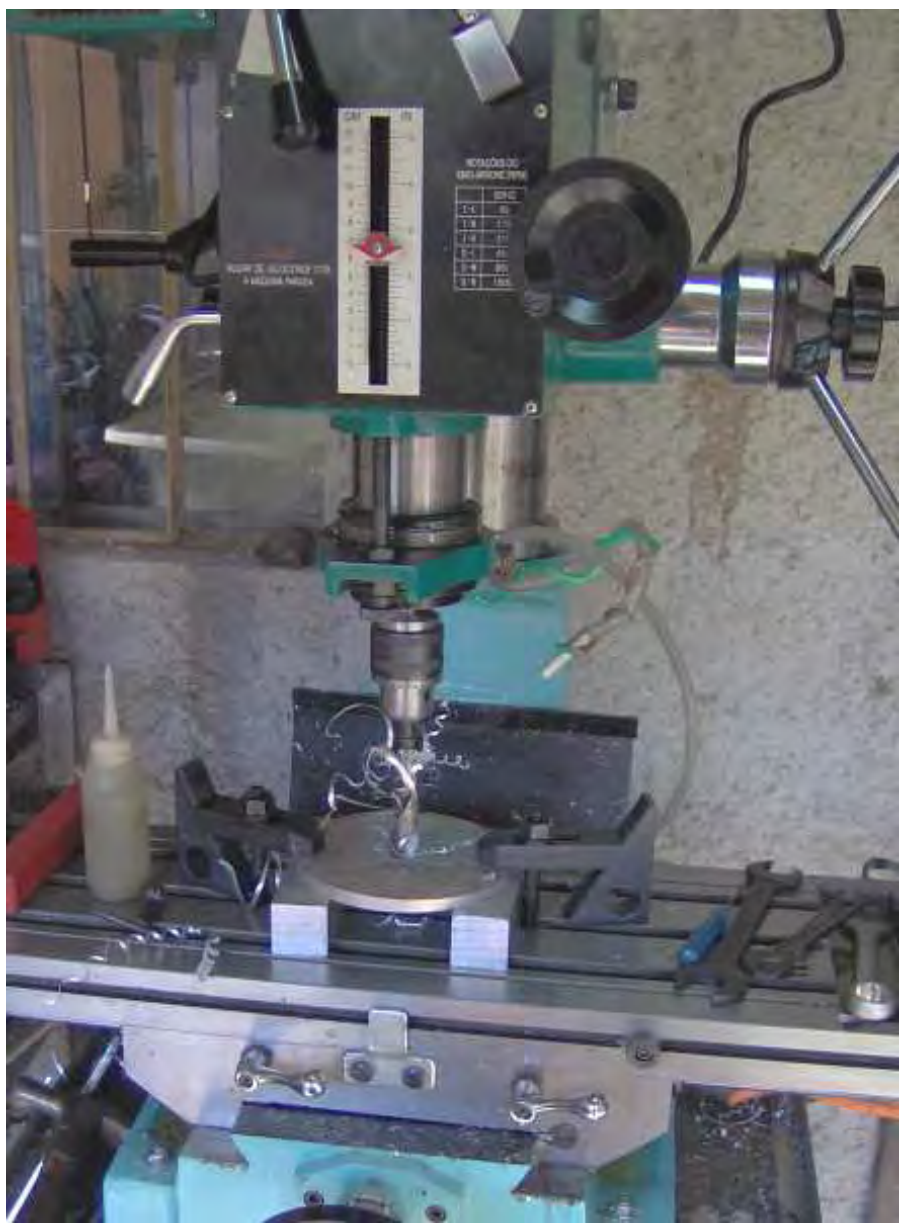

Fig. A-IV.2: Furo inicial de 20mm.

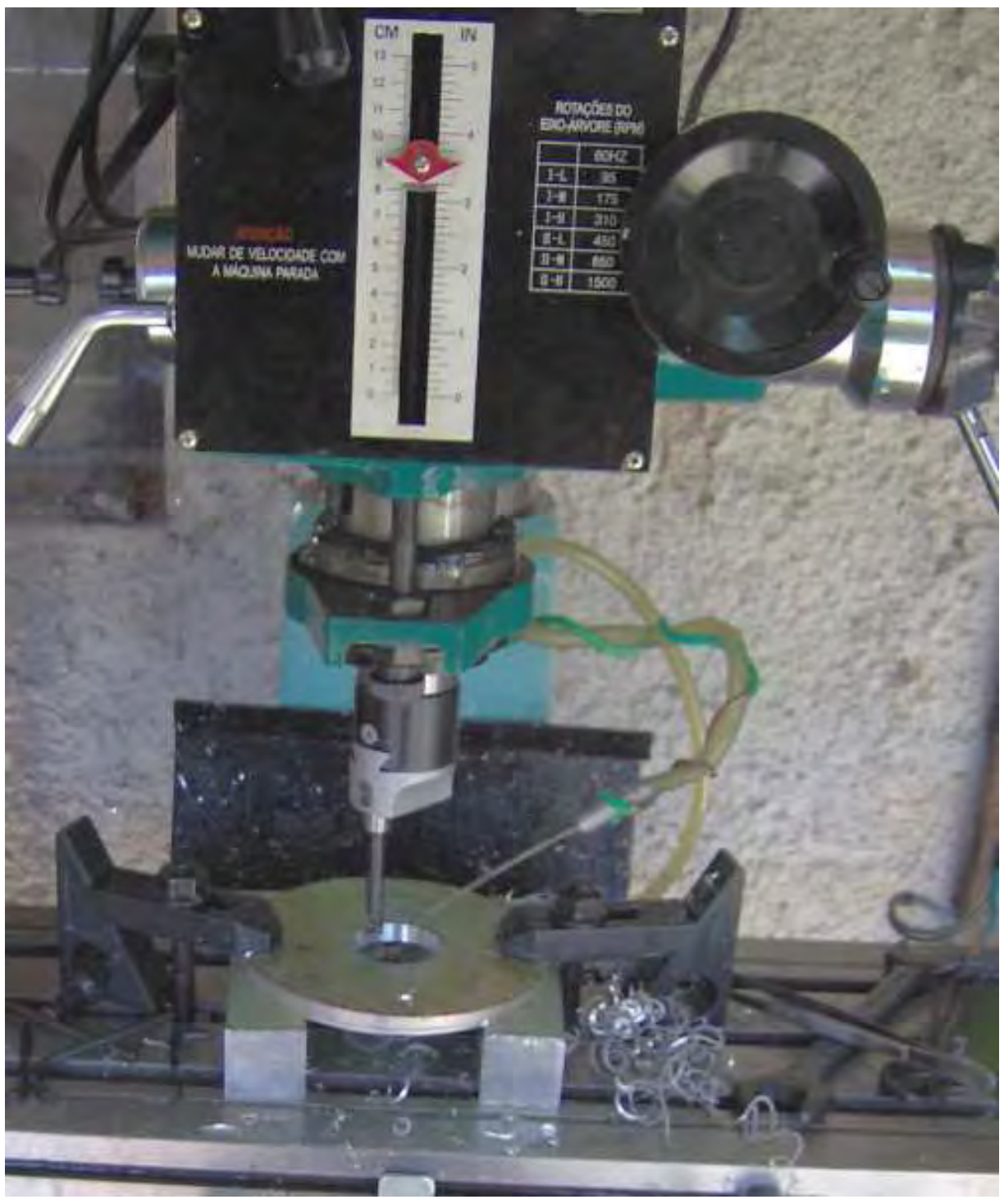

Fig. A-IV.3: Alargamento do furo central, usando a ferramenta de brocagem variável, até o valor projetado de 50mm.

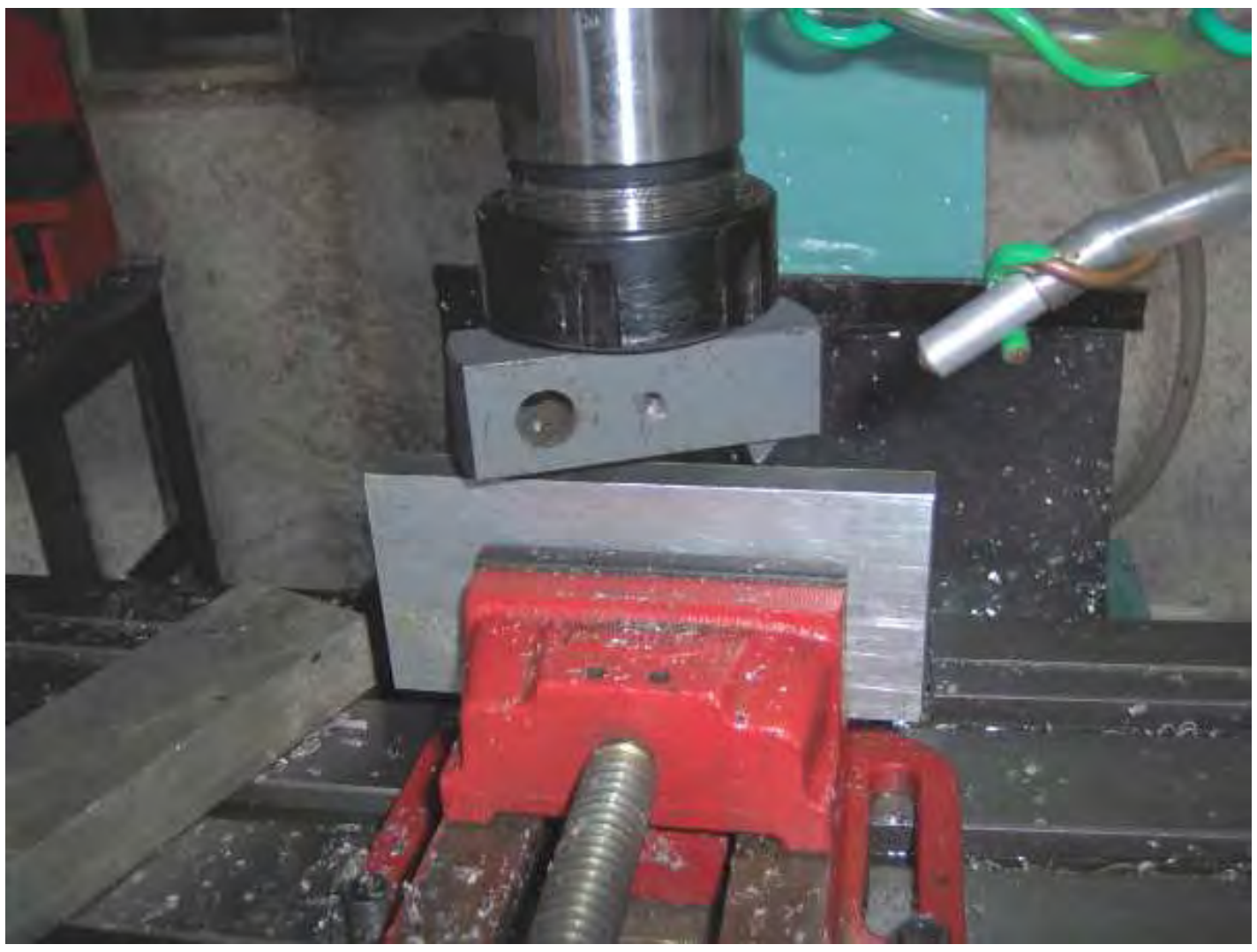

Fig. A-IV.4: Acabamento das peças pré-cortadas usando uma fresa de um corte só.

## Fabricação da peça com curvas convexas e rasgos curvos.

Um disco de 165 mm de diâmetro de 27 mm de espessura foi torneado para ficar com suas faces planas e colocado na mesa divisora.

Para a centragem da peça foi usado um relógio comparador centesimal, com erro máximo de 0,02 mm, figura A-IV.5.

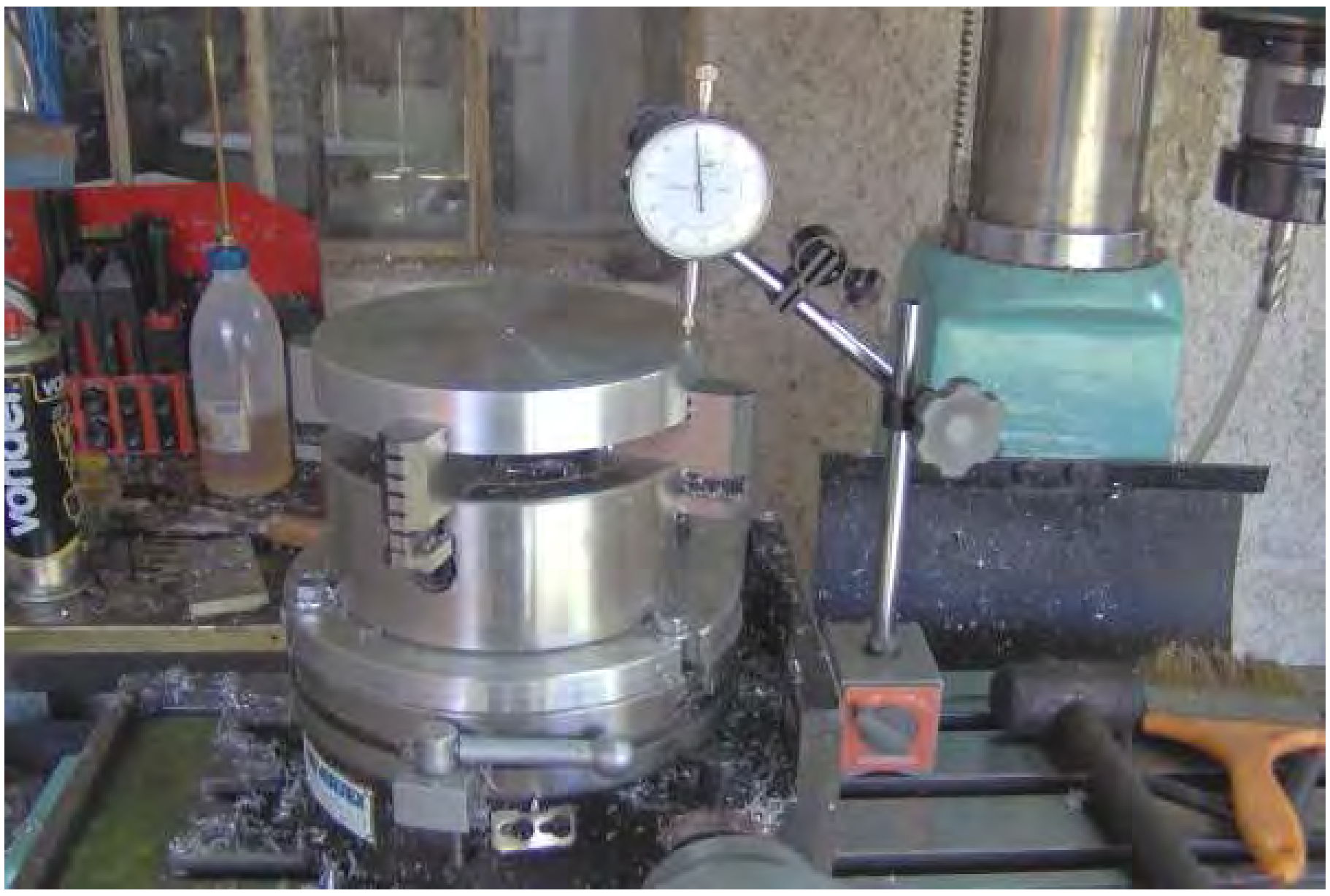

Fig. A-IV.5: Disco de alumínio preso à mesa divisora.

As localizações dos cortes e das posições dos rasgos curvos foram marcadas na peça para a fresagem e corte, figuras: A-IV.6 a A-IV.11.

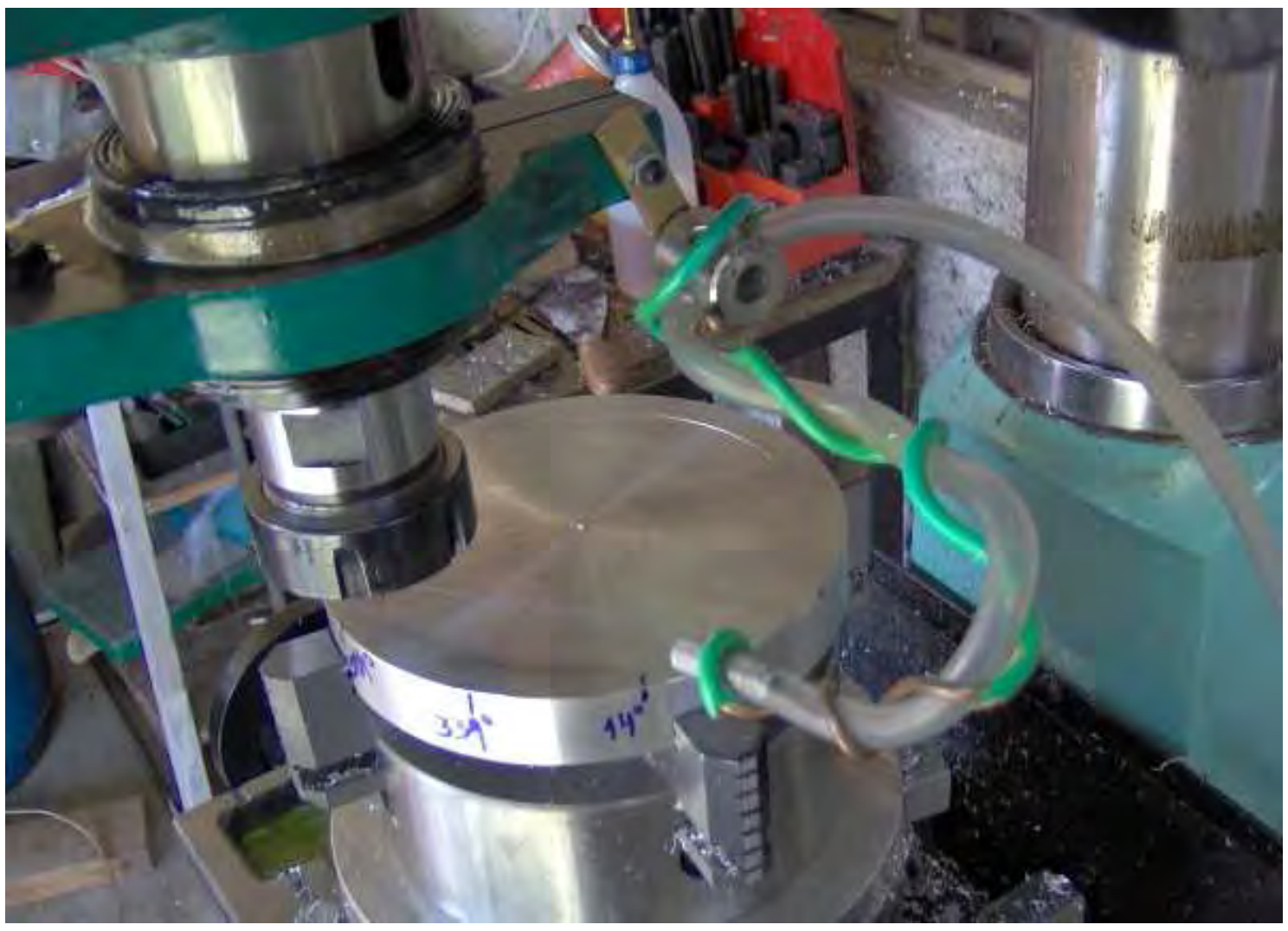

Fig. A-IV.6: Peça pronta para ser fresada.

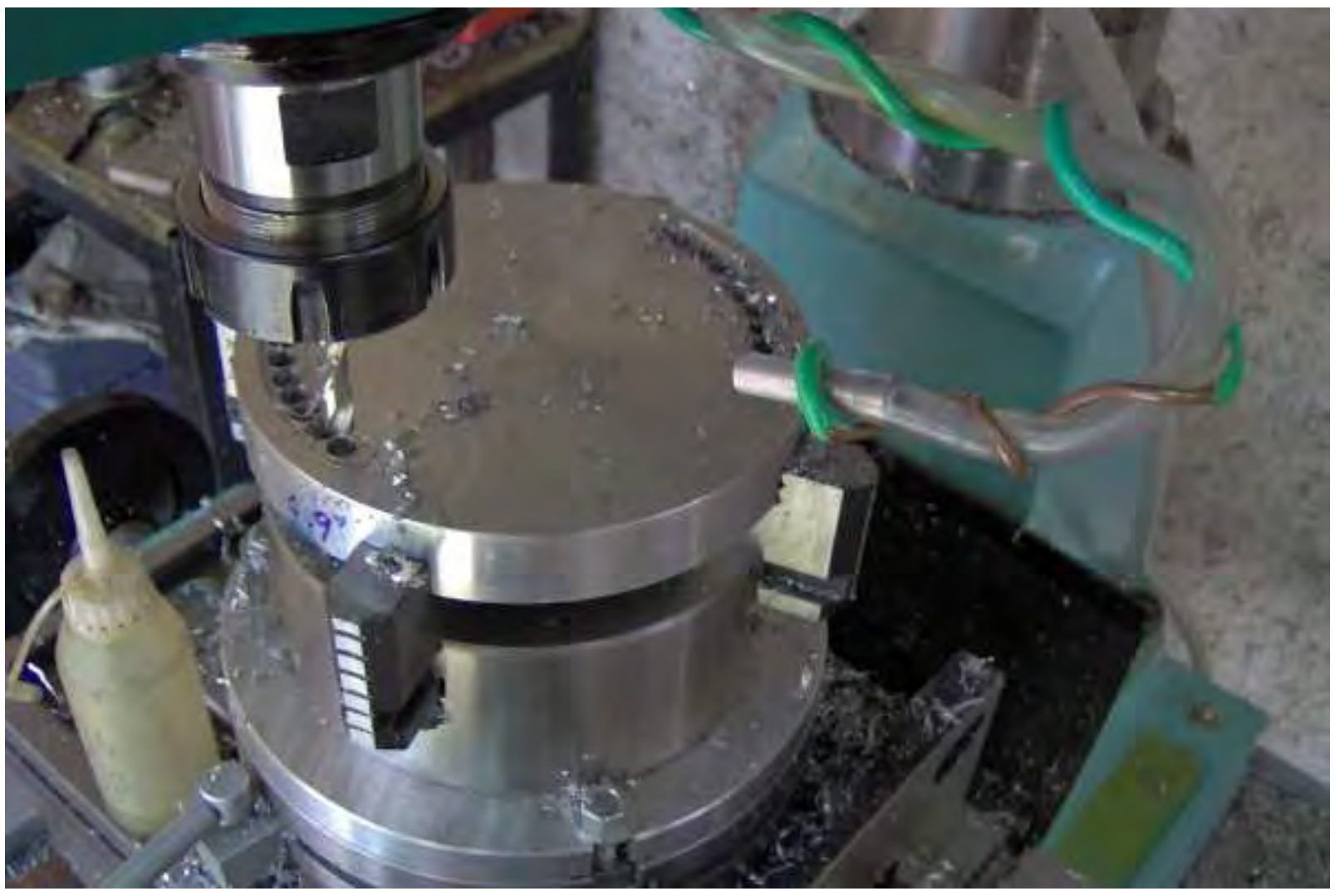

Fig. A-IV.7: Pré-furos antes de se abrir os rasgos.

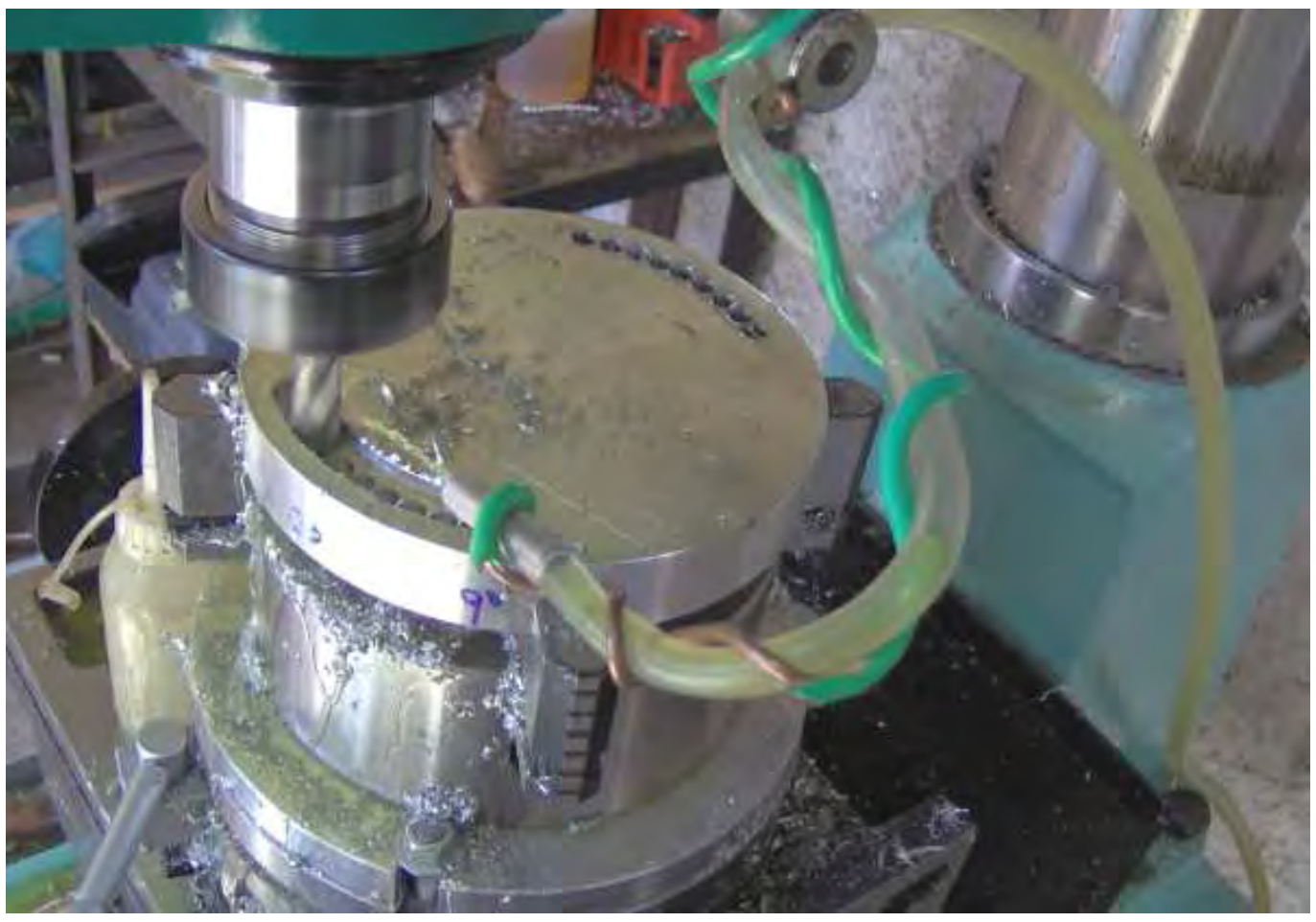

Fig. A-IV.8: Fresagem dos rasgos curvos.

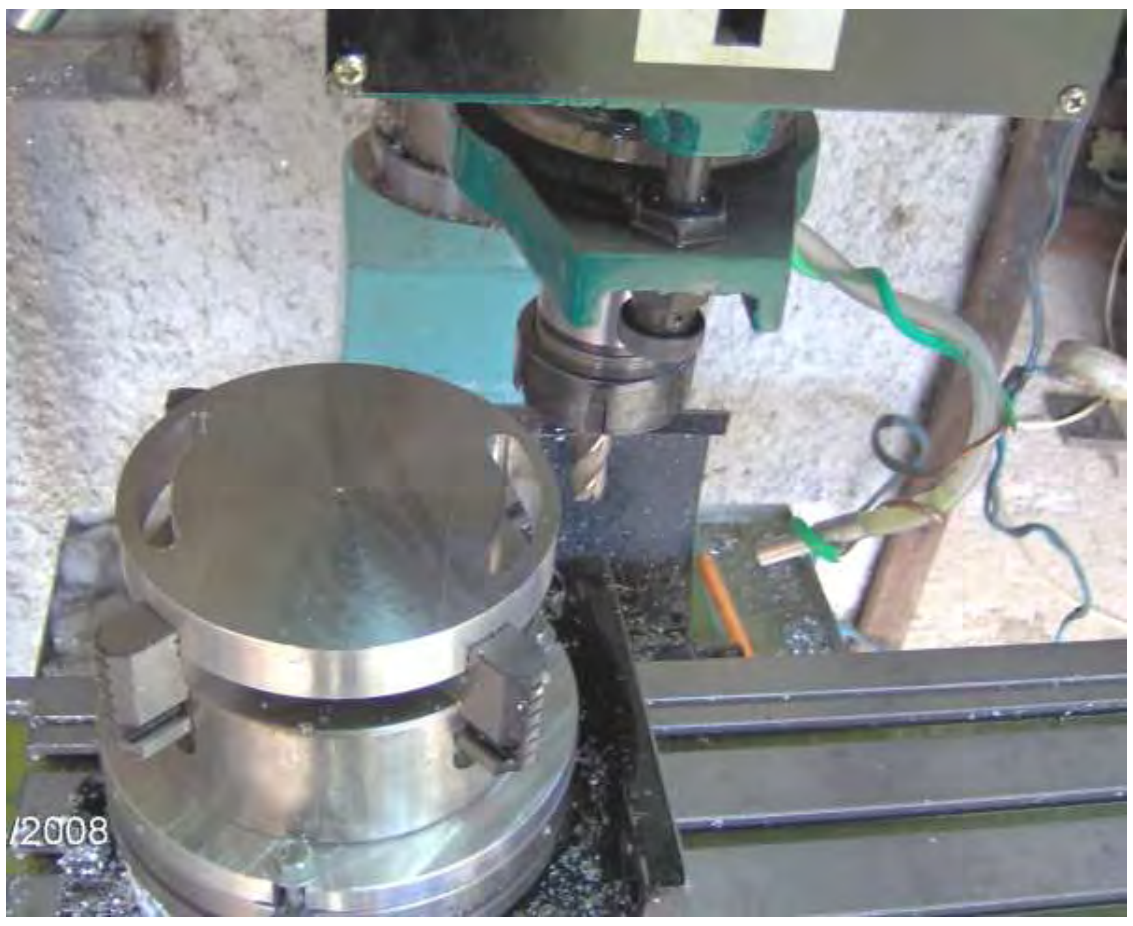

Fig. A-IV.9: Rasgos curvos abertos na peça.

A peça, então, foi preparada para o corte.

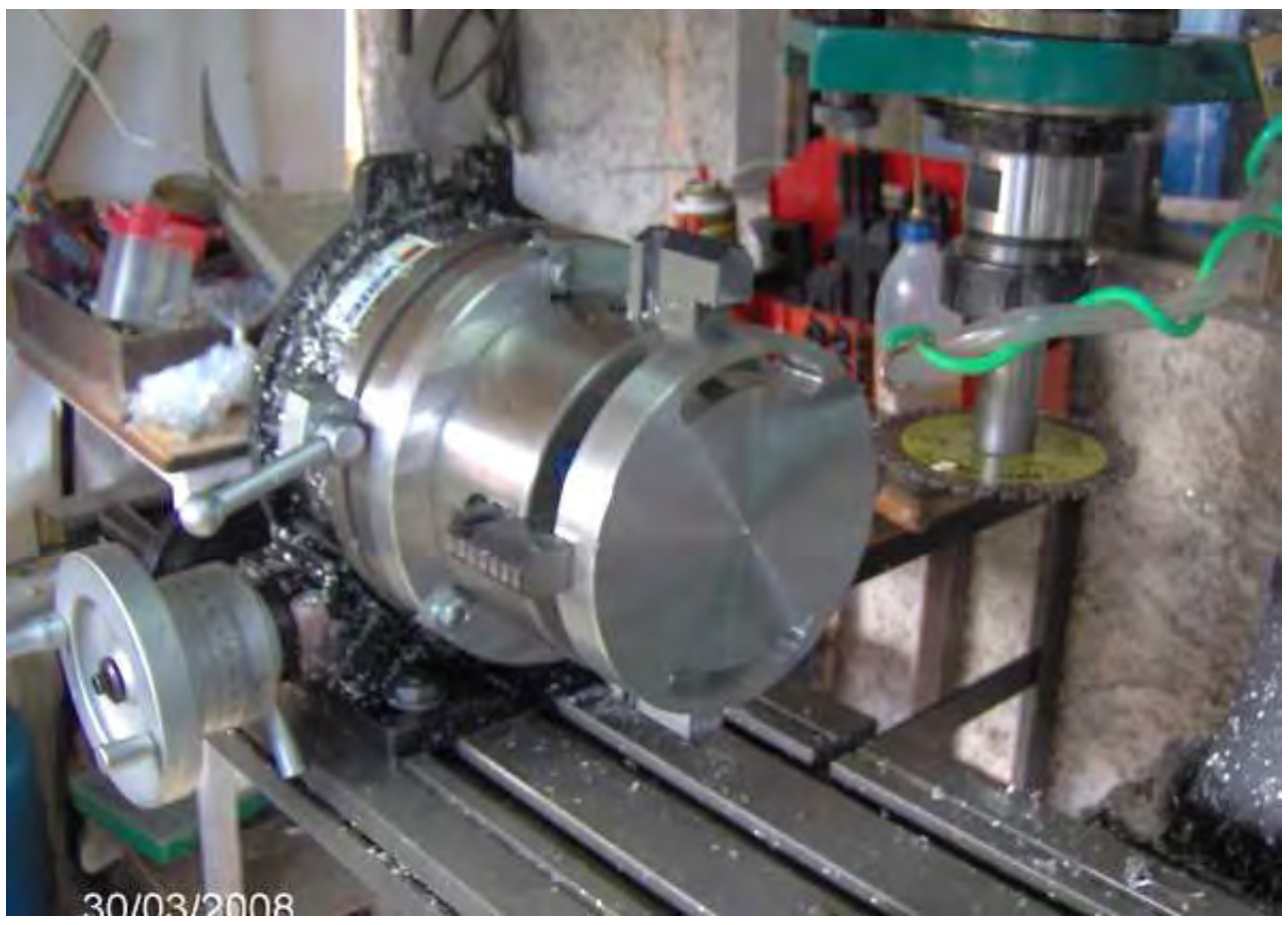

Fig. A-IV.10: Peça pronta para ser cortada.

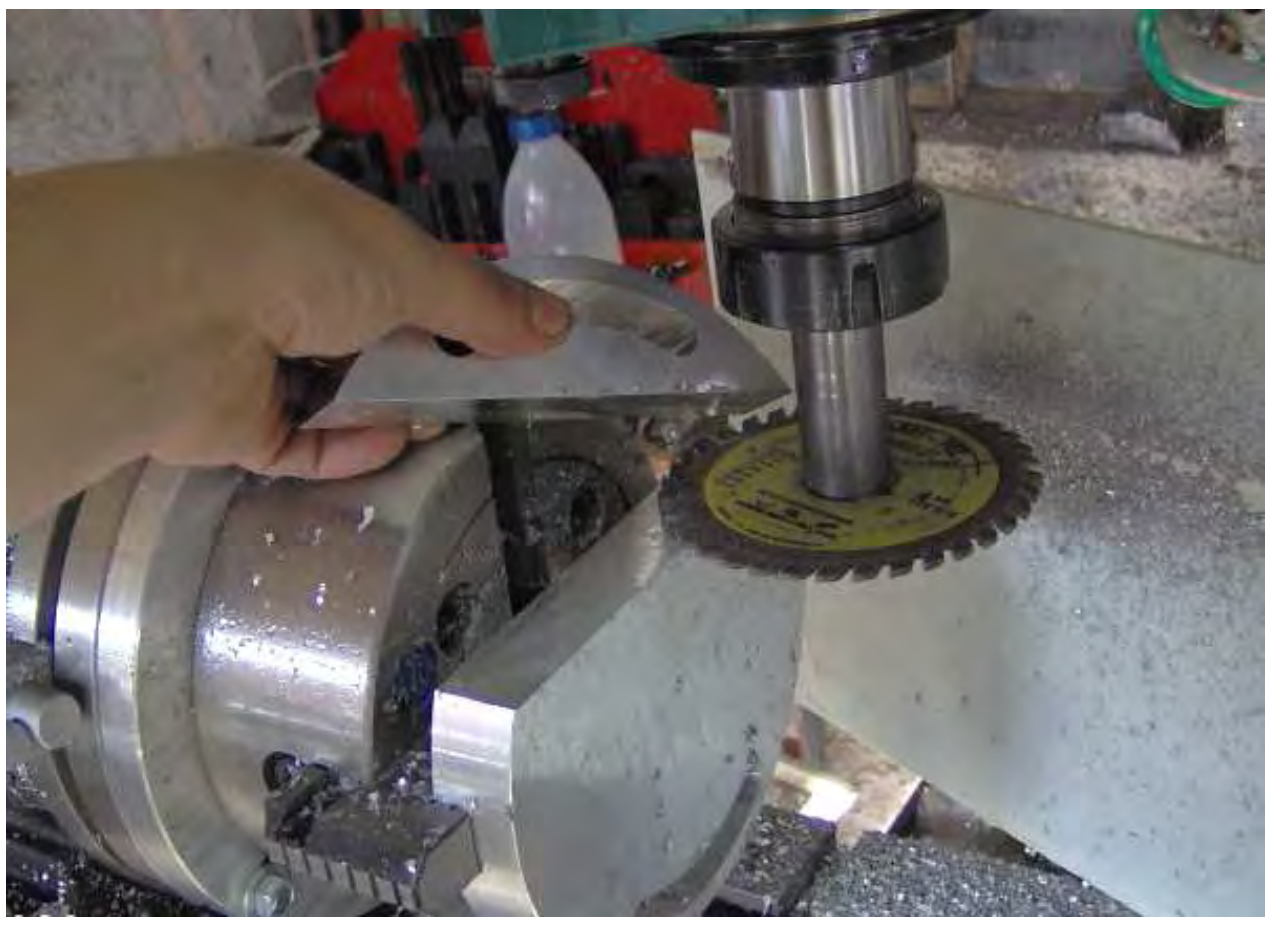

Fig. A-IV.11: Primeira peça convexa cortada.

Para obter as duas peças côncavas, 2 blocos quadrados de alumínio, de 1,5" de espessura e cerca de 130 mm de comprimento, foram unidos por parafusos para serem trabalhados sob as mesmas condições e simultaneamente. O raio de curvatura do corte foi cerca de 0,1 mm menor do que o raio da peça convexa (82,5 mm), figura A-IV.12.

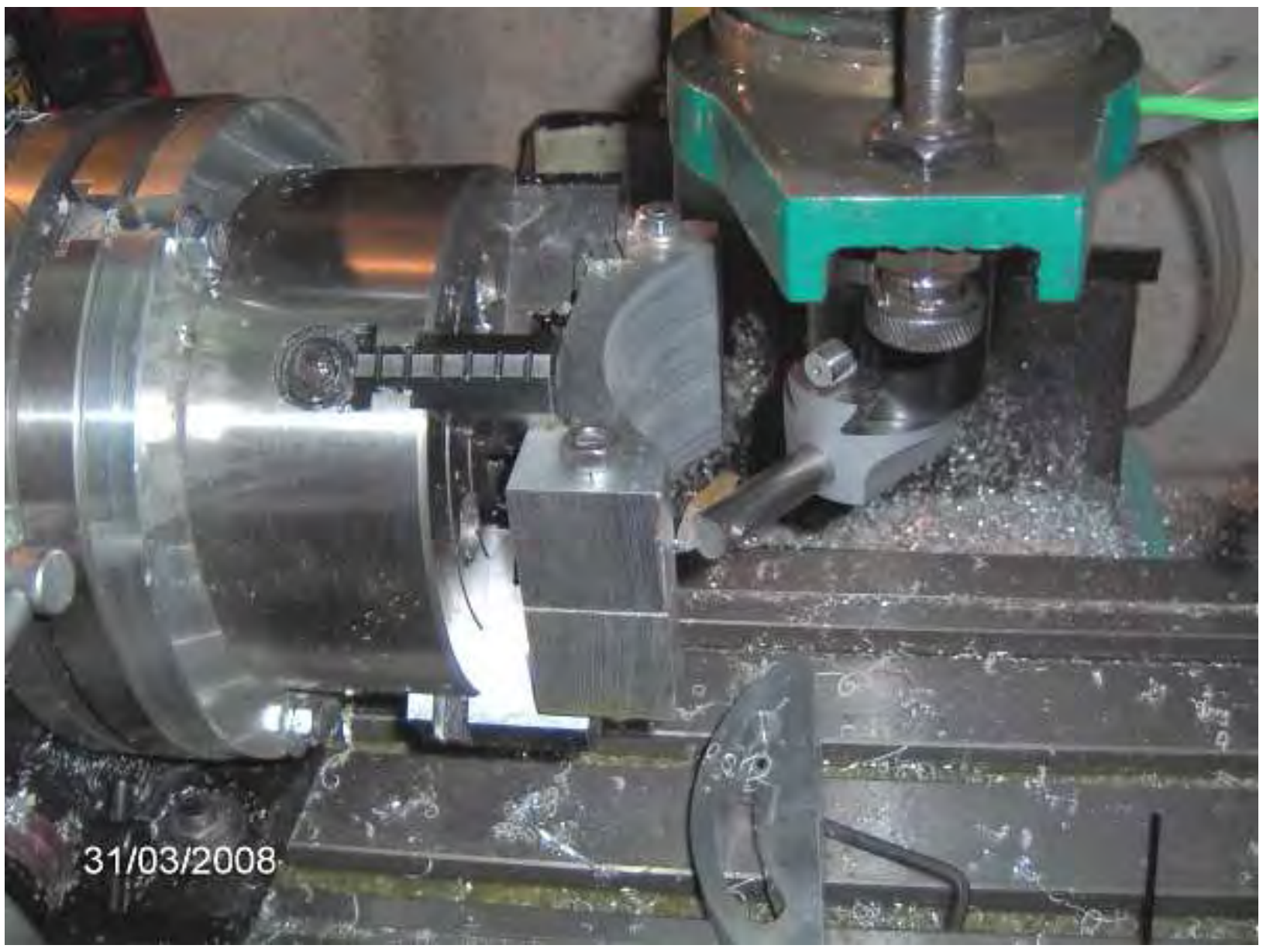

Fig. A-IV.12: Escavação simultânea das peças côncavas.

As imagens a seguir mostram as peças acabadas, figuras A-IV.13 a A-IV.15.

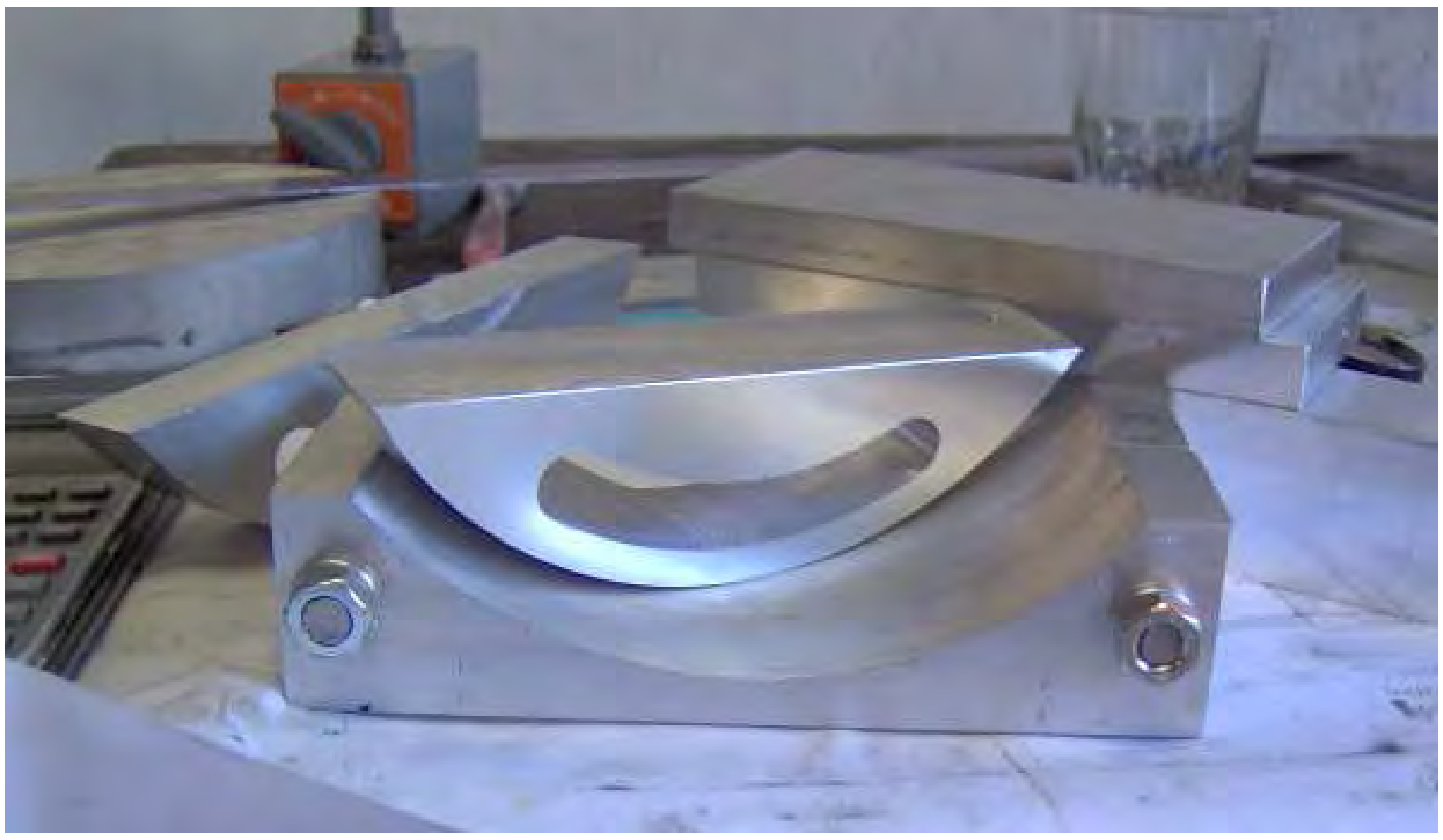

Fig. A-IV.13: Peças curvas do suporte prontas.

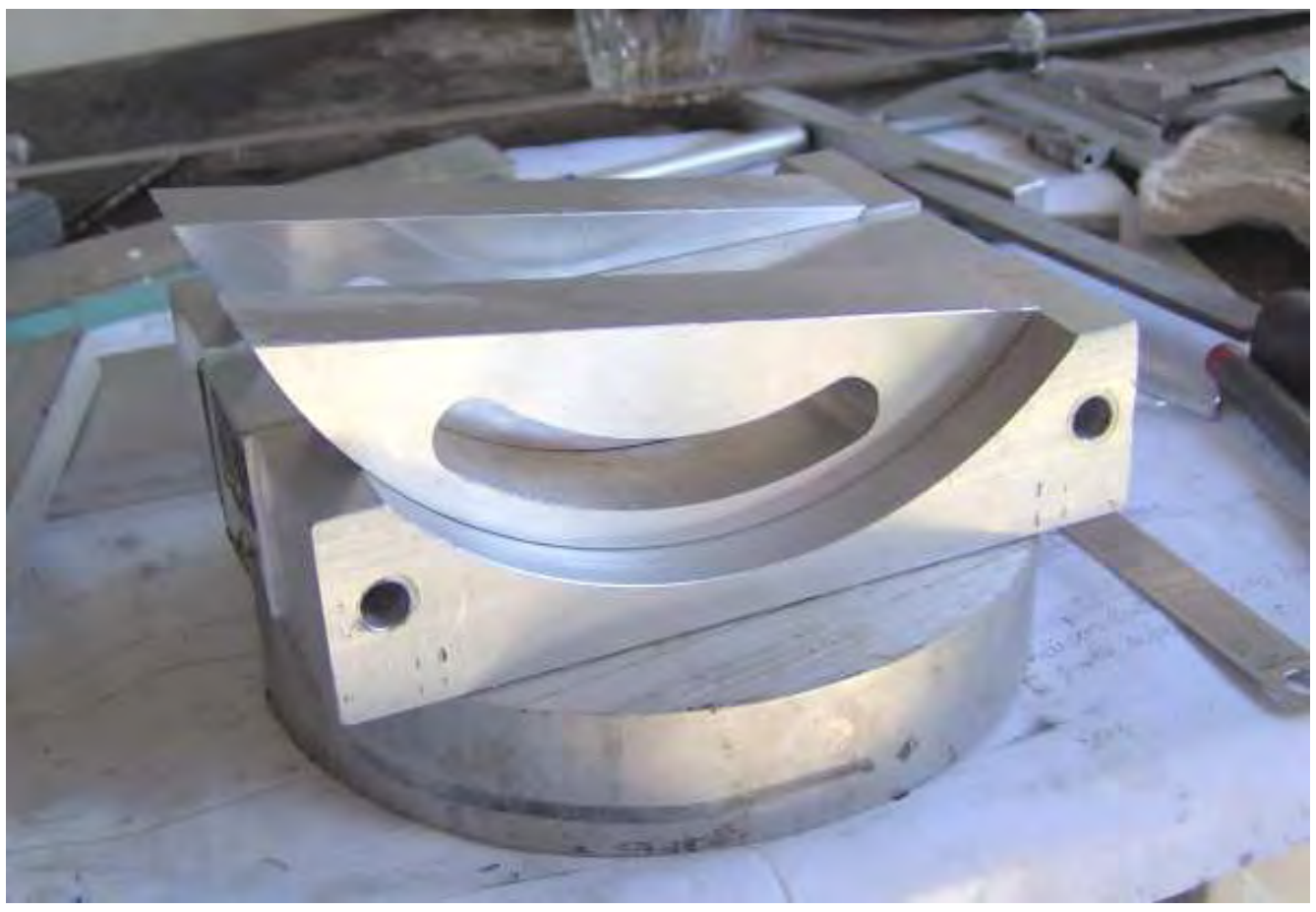

Fig. A-IV.14: As peças deslizam com suavidade.

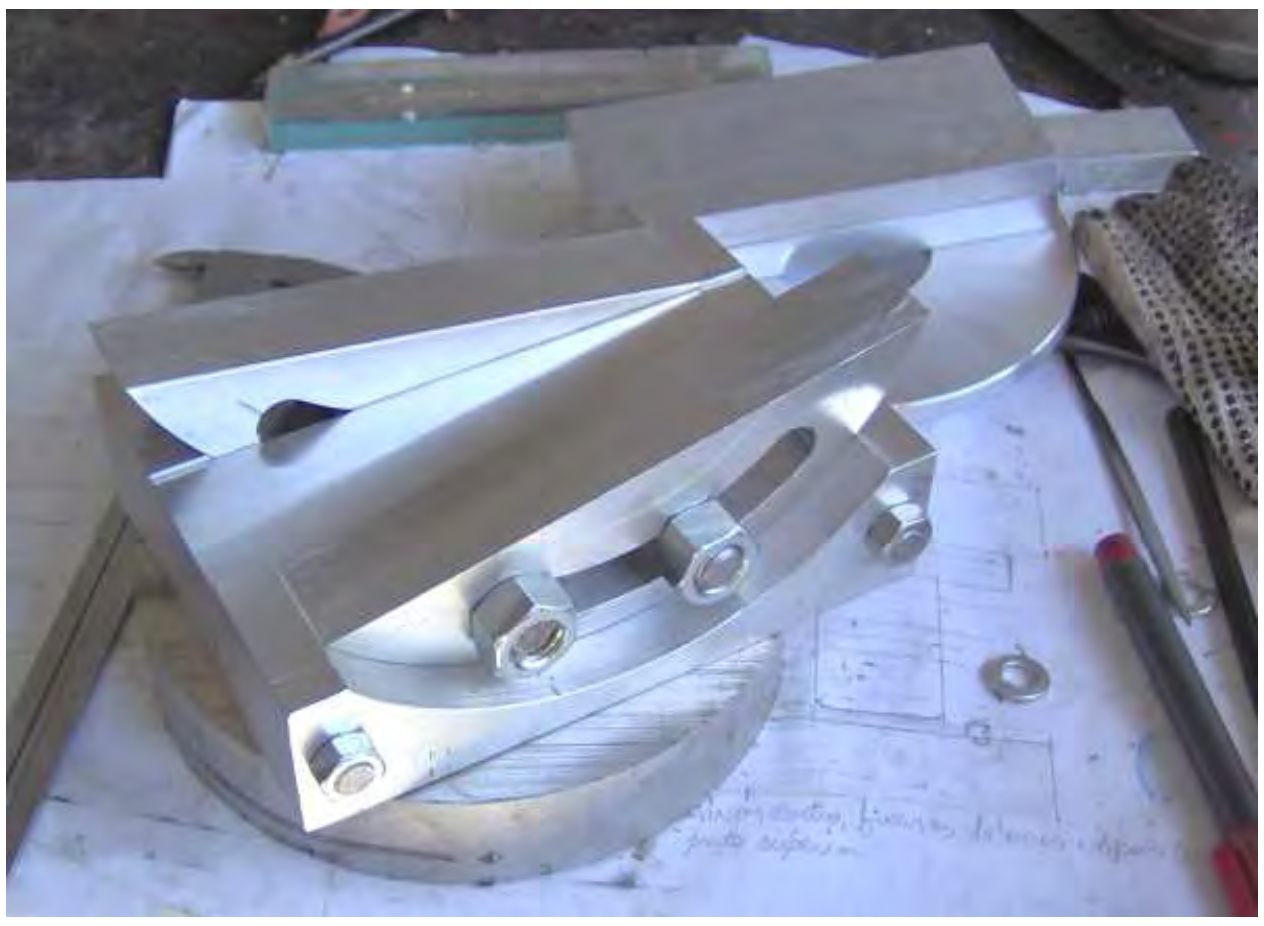

Fig. A-IV.15: Peças aparafusadas. Os parafusos de travamento já estão no seu lugar.

O conjunto foi montado de novo na mesa divisora para a geração, no disco superior do suporte, de uma calota esférica para a acomodação da superfície do espelho, figura A-IV.16.

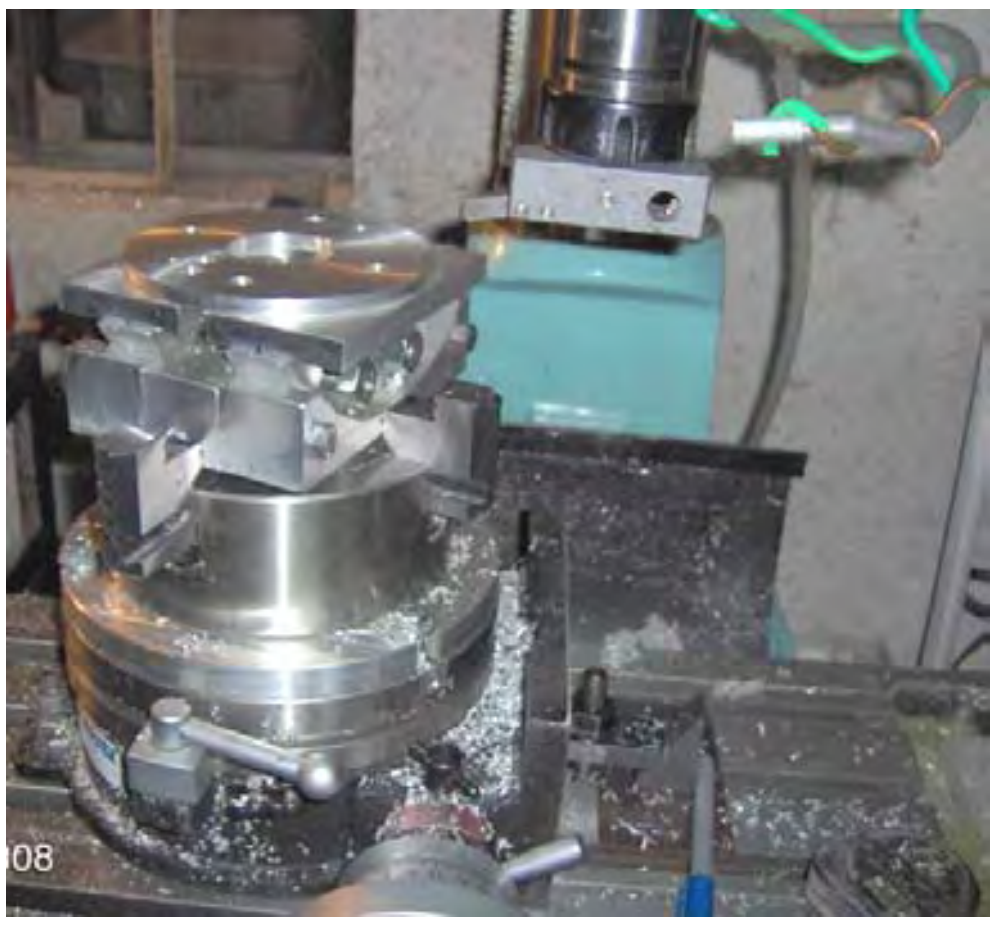

Fig. A-IV.16: Peça circular para a fixação do espelho pronta.

O disco superior foi retirado e um rasgo-encaixe na placa inferior foi fresado. Neste rasgo foi fixado um bastão de alumínio de ½" quadrado para servir de guia, figura A-IV.17.

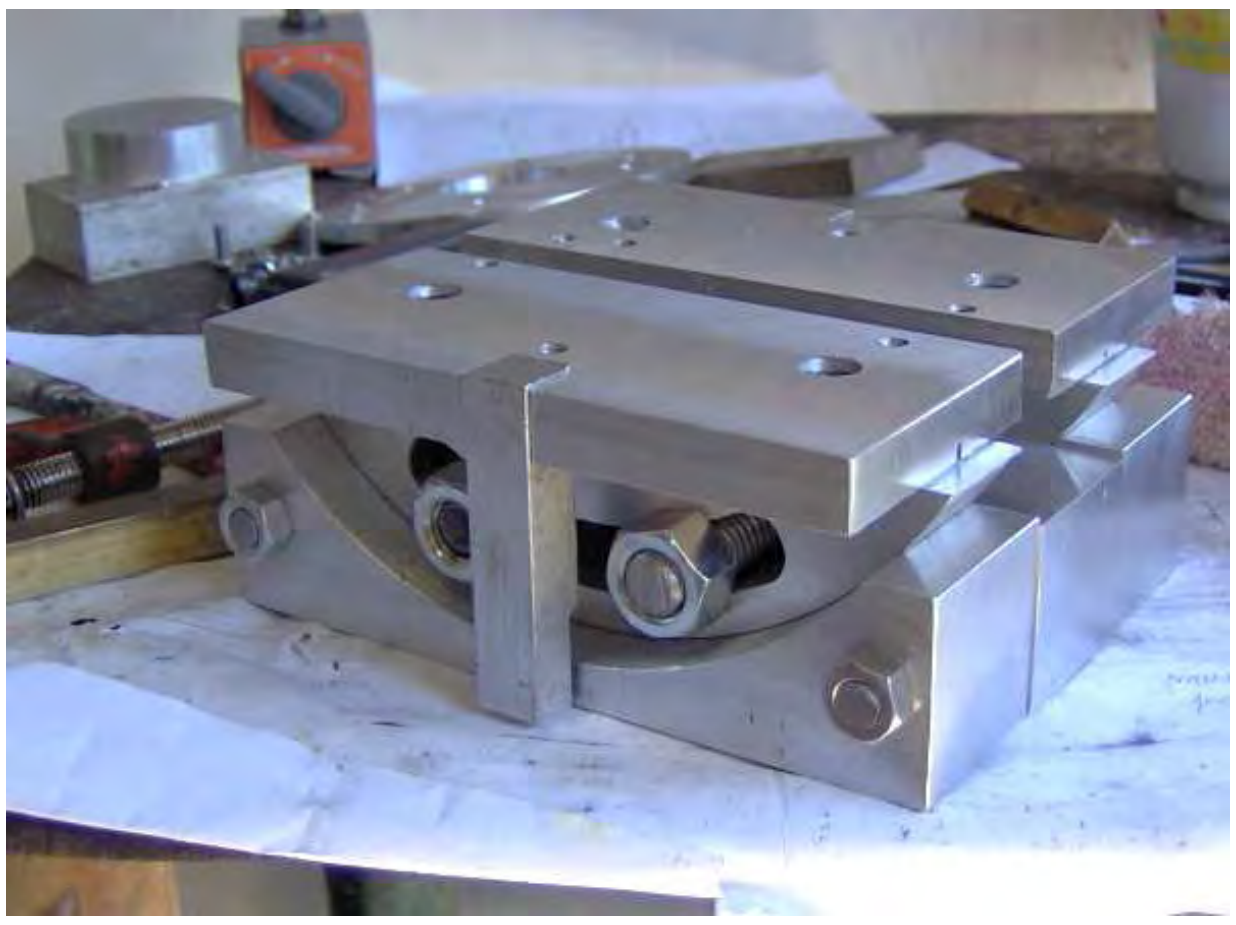

Fig. A-IV.17: Bastão a ser acionado por parafusos (puxa-empurra) fixados no suporte.

O bastão de alumínio é acionado por 2 parafusos de cada lado, um empurra e outro alivia a tensão causada pelo primeiro. A resposta é a inclinação precisa de cada lado do suporte, ao nível do segundo.

Nos dois cubos de latão na base inferior estão rosqueados os dois parafusos que permitem a inclinação das peças, figuras A-IV.18 e A-IV.19.

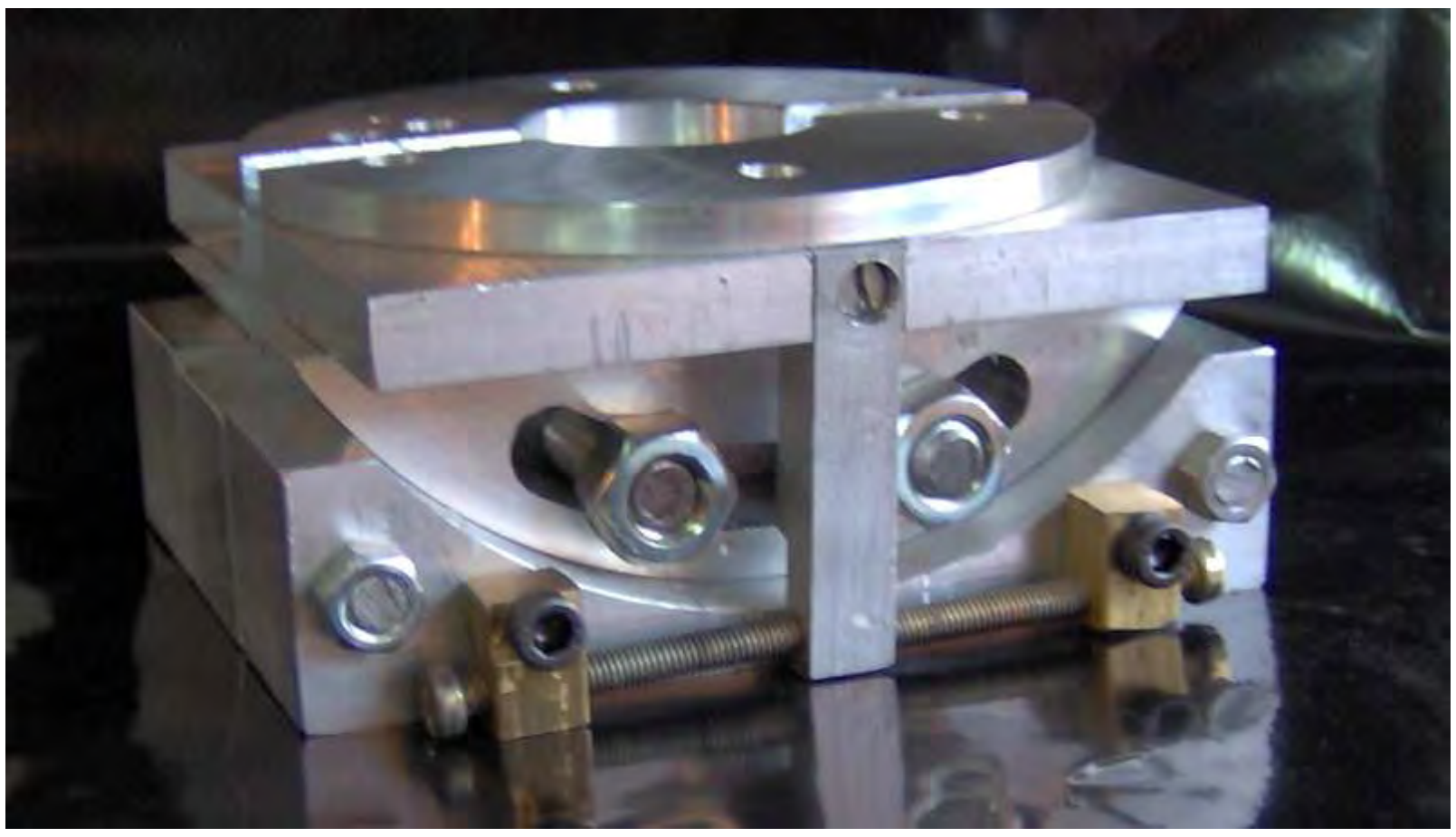

Fig. A-IV.18: Vista dos parafusos de ajuste do ângulo de inclinação da mesa.

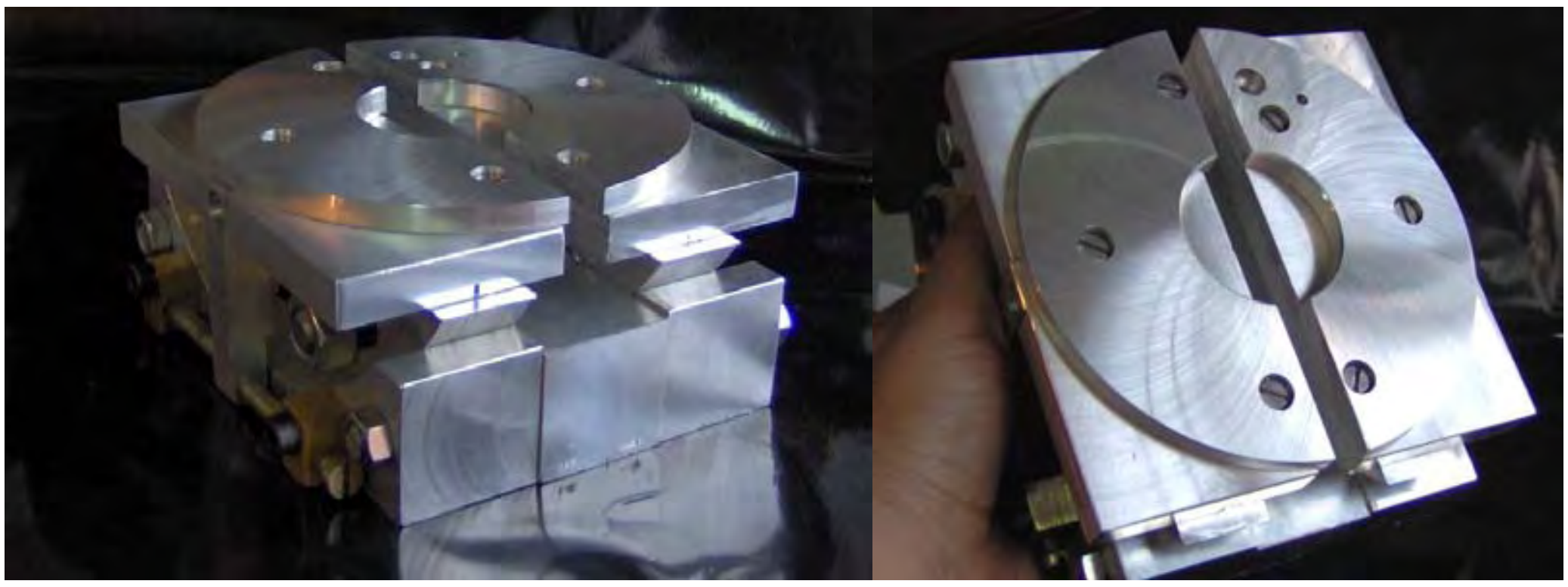

Fig. A-IV.19: Vista superior da mesa.

# Apêndice V - Testes ópticos do espelho do Heliômetro, antes e depois do corte

Características físicas do espelho acabado:

- Diâmetro = 150 mm
- Distância focal = 986,0 mm
- Relação focal f/D = 6,53
- Superfície parabólica

## *Teste de Ronchi (qualitativo)*

Posição intra-focal, grade de 5 linhas por milímetro, fonte monocromática filiforme com 18 μm de espessura, colocada no centro de curvatura do espelho, figura A-V.1.

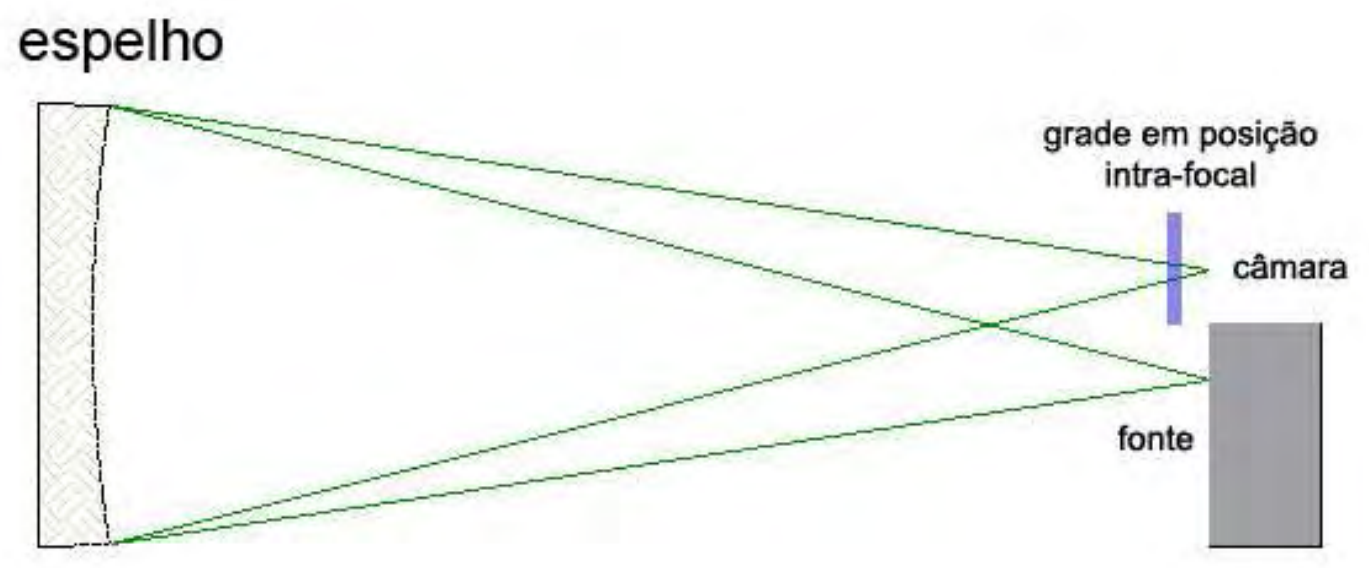

Fig. A-V.1: Desenho esquemático do teste de Ronchi intra-focal.

Neste teste, um espelho esférico produz sombras retas, um parabólico, sombras curvadas, que é o caso deste. Como as curvas das sombras são suaves e não apresentam grandes distorções, pode-se dizer que, qualitativamente, o espelho do Heliômetro não possui grandes defeitos zonais, figura A-V.2.

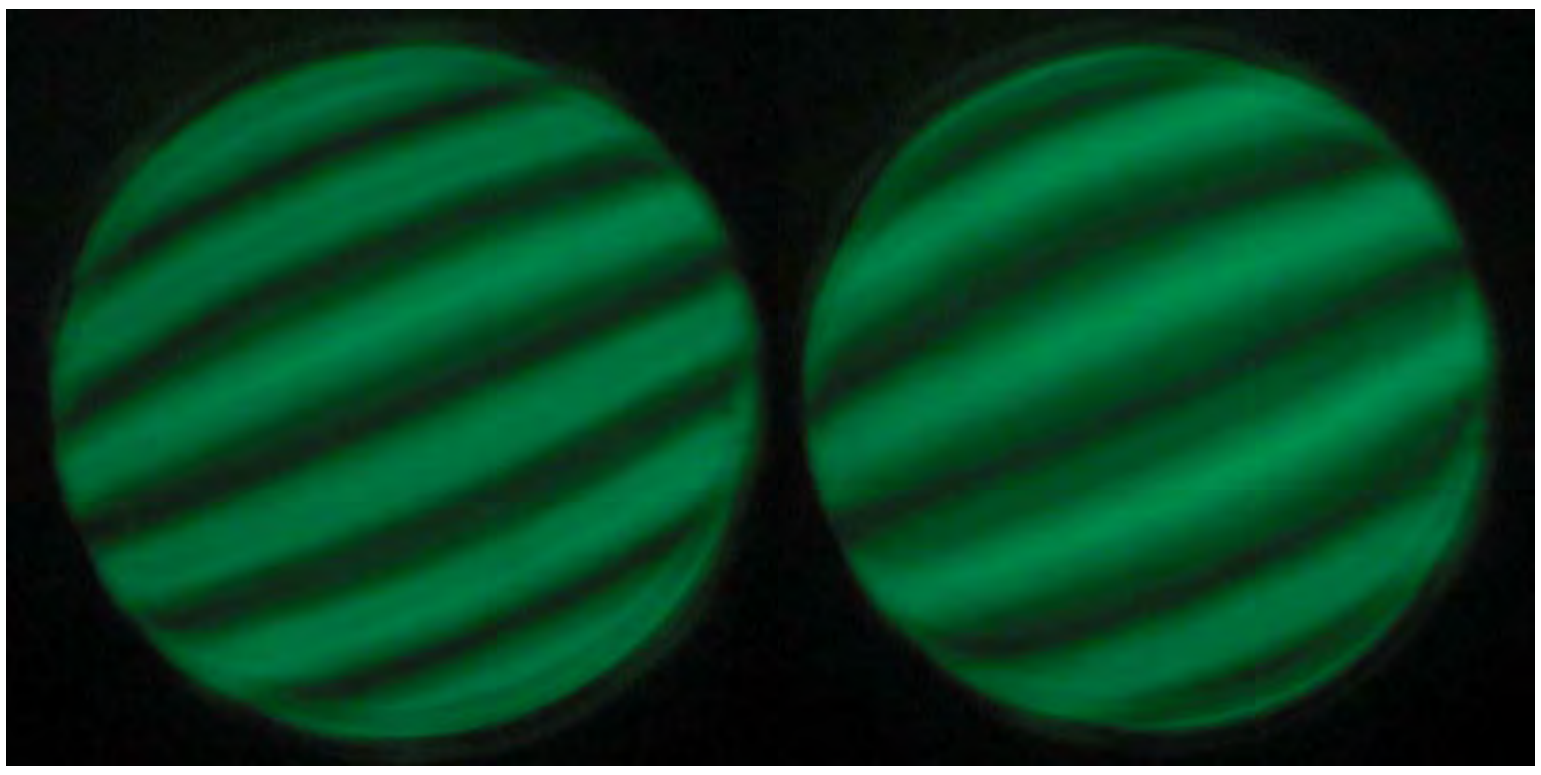

Fig. A-V.2: Imagens do teste de Ronchi intra-focal.

## *Teste de Foucault (quantitativo)*

O espelho, depois de parabolizado, passou por testes de Foucault com auxilio de uma máscara de Couder de 4 zonas, figura A-V.3.

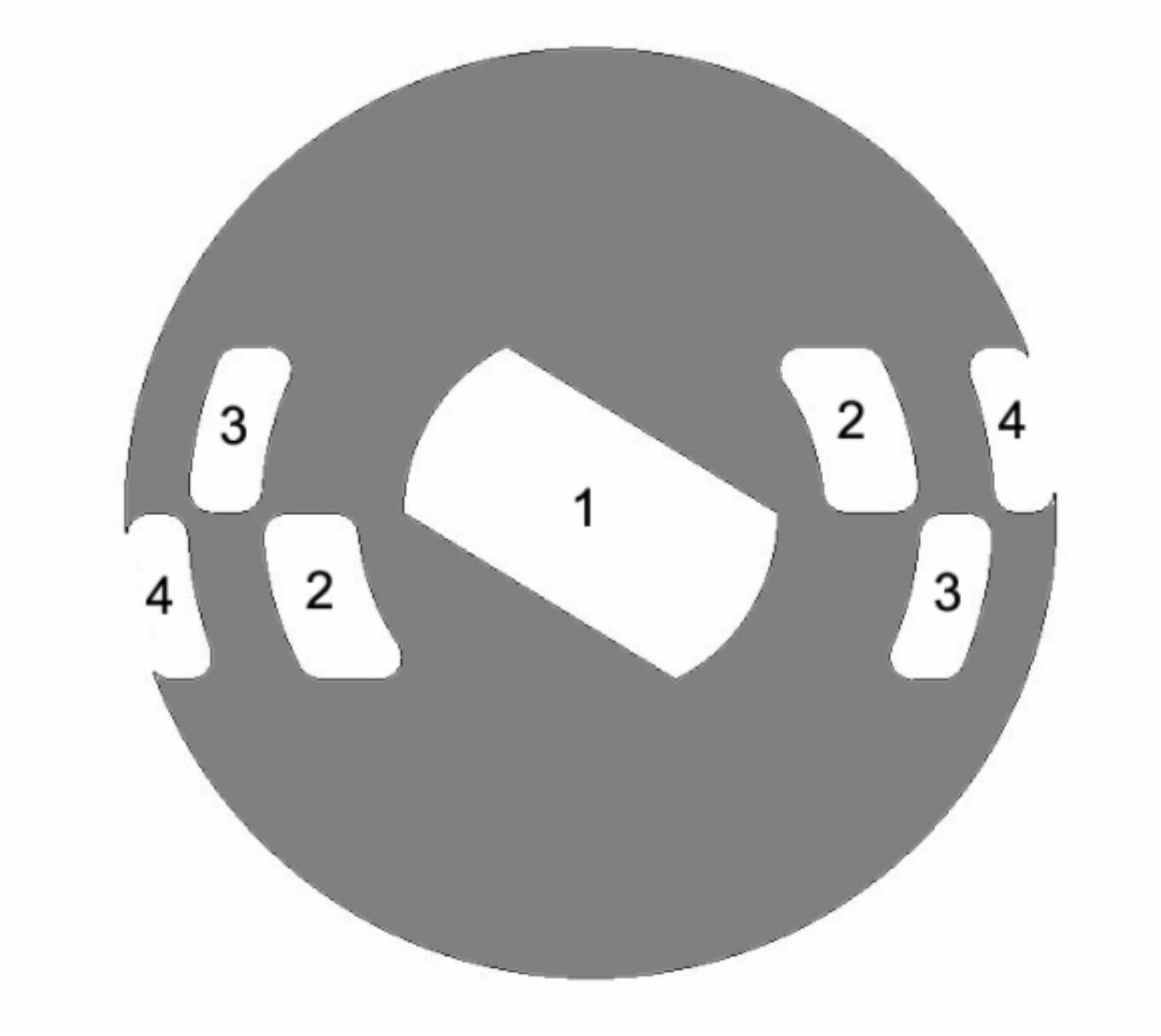

Fig. A-V.3: Exemplo de uma máscara de Couder de 4 zonas.

Para a redução dos dados foi usada uma planilha eletrônica para a parabolização de superfícies, baseada no método de Zambuto (Dunna, 2008).

Os dados da máscara de Couder usados para dividir a calota parabólica em 4 zonas a serem medidas através do Foucault e os resultados finais da frente de onda em valores Pico-Vale, RMS e Razão de Strehl são exemplificados na figura A-V.4.

| | A | B | C | D | E | F | G | H | I |
|---|---|---|---|---|---|---|---|---|---|
| 2 | | Units = | mm | Focal Ratio = 6,530 | | | | Zones = | 4 |
| 3 | | Mirror D = | 151,000 | Mirror Focal Length = 986,000 | | | | ROC Adj.= | -16 |
| 4 | | Optical D = | 150,0 | Maximum RTA = 0,800 | | | | p(Ro)= | 4,3998E-03 |
| 5 | | ROC= | 1972,00 | PeakValley WF error = 1/30,2 λ | | | Deformation coef. = | | -1,0000 |
| 6 | | Lambda = | 5,4864E-04 | WaveFront RMS = 1/65,955 λ | | | Locks ML and RTA at Z= | | 0 |
| 7 | Foc. Source = | | Moving | Strehl Ratio = 0,9909 | | | Corrections | Center.70%= | 94,33% |
| 8 | Zone Center= | | 50,0% | | Current : Heliometro | | | Edge.70%= | 89,16% |
| 9 | Couder Mask | | | | | | | Total= | 92,45% |
| 10 | Z | Inside | Outside | Effective Radius | Knifedge Deviation | Astig. Eval. | % Correction | Theoretical KE shift | Corrected Meas. |
| 11 | 1 | 0,01 | 28,00 | 14,005 | 0,09624 | | – | | 6,7800 |
| 12 | 2 | 28 | 49,00 | 38,5 | 0,02405 | | 88,9% | 0,65219 | 7,3600 |
| 13 | 3 | 49 | 60,50 | 54,75 | 0,01564 | | 98,9% | 0,76841 | 8,1200 |
| 14 | 4 | 60,5 | 75,00 | 67,75 | -0,07192 | | 89,2% | 0,80756 | 8,8400 |

Fig. A-V.4: Fragmento da planilha redutora de dados. Os dados da máscara de Couder usados para dividir a calota parabólica em 4 zonas a serem medidas através do Foucault estão nas células das linhas 11 a 14, colunas B e C. As leituras finais de cada zona da máscara de Couder estão nas células das linhas 11 a 14, coluna I. Os resultados finais da frente de onda em valores Pico-Vale, RMS e Razão de Strehl são lidos nas células das linhas 5 a 7, coluna E.

As características ópticas mais importantes extraídas da planilha são:

1. Erro Pico-Vale da frente de onda = $\lambda/30{,}2$
2. RMS da frente de onda = $\lambda/65{,}9$
3. Razão de Strehl = 0,99

A razão de Strehl é a razão da intensidade do pico observado no plano de detecção de um telescópio, ou outro sistema de imagens, a partir de uma fonte pontual em relação ao pico de intensidade máxima teórica de um sistema de imagem perfeita, trabalhando no limite de difração (Strehl, 1895, 1902 *apud* Hardy, 1998, p. 114). Este valor significa que 99% da luz de uma estrela coletada por este espelho cai sobre a área do disco mínimo teórico de difração de 4,4 micra de diâmetro (célula I4, fig. A-V.4) no ponto de melhor foco.

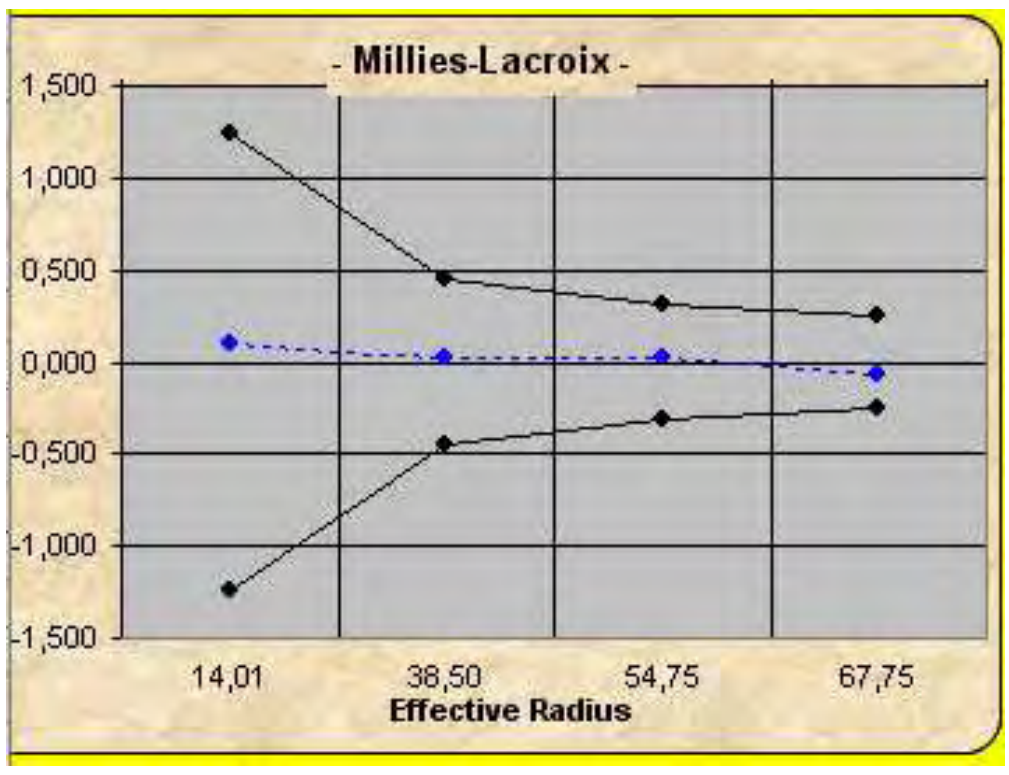

Fig. A-V.5: Limites de Millies-Lacroix

No gráfico da figura A-V.5, a escala vertical é a diferença em mm entre a posição medida na faca, no aparelho de medição (Foucault), e a sua posição ideal. A escala horizontal, em mm, corresponde aos raios do centro das quatro zonas de medição do espelho. As curvas externas marcam os limites aceitáveis para esta diferença, de forma que o espelho mantenha a qualidade da imagem.

Por este mesmo gráfico da figura A-V.5, por exemplo, este espelho poderia ter uma diferença entre a curva teórica e a real, a 38,5 mm de seu centro, de até ±0,5 mm, sem que isso comprometesse sua qualidade óptica, dentro de determinados limites de tolerância, neste caso limites de Millies-Lacroix.

A linha central, entre os limites, mostra que as medidas deste espelho seguem a curva parabólica ideal com desvios não superiores à ± 0,1 mm.

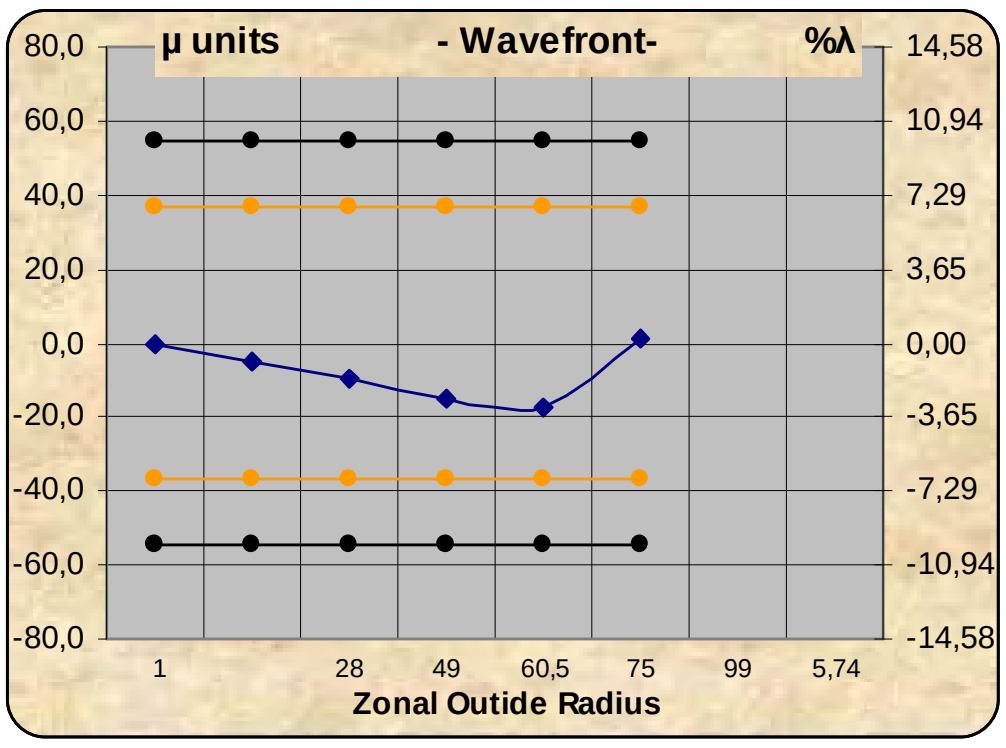

Fig. A-V.6: Desvio da frente de onda em função da superfície do espelho.

O gráfico da figura A-V.6 informa o valor do desvio médio máximo percentual da frente de onda em função do desvio da superfície parabólica do espelho de seu valor ideal.

A escala vertical da esquerda corresponde a diferenças de altura entre as superfícies, real e teórica, expressas em nanômetros. A escala horizontal informa as zonas do semidiâmetro do espelho sob teste, a partir do centro da peça, em mm. E a escala vertical da direita, o percentual da frente de onda que está sendo desviado do seu valor ideal, em porcentagens do comprimentos de onda.

Segundo o gráfico da figura A-V.6, este espelho apresenta um defeito médio máximo de -20 nm, a 60,5 mm do centro. O desvio máximo da frente de onda que este defeito causa não é superior a 0,04λ.

## *Teste de Ronchi por auto-colimação (qualitativo)*

As condições de contorno são: posição intra-focal, grade de 5 linhas por milímetro e fonte extensa. O plano óptico utilizado tem 210 mm de diâmetro e um furo central de 40 mm. A fonte, situada no foco do espelho parabólico tem sua imagem real projetada no infinito. O espelho plano intercepta os raios paralelos e os reflete para o espelho parabólico novamente. Este por sua vez os converge para o foco onde é analisado.

A dupla incidência no espelho (fig. A-V.7) faz com que qualquer defeito em sua superfície se apresente 2 vezes maior do é realmente.

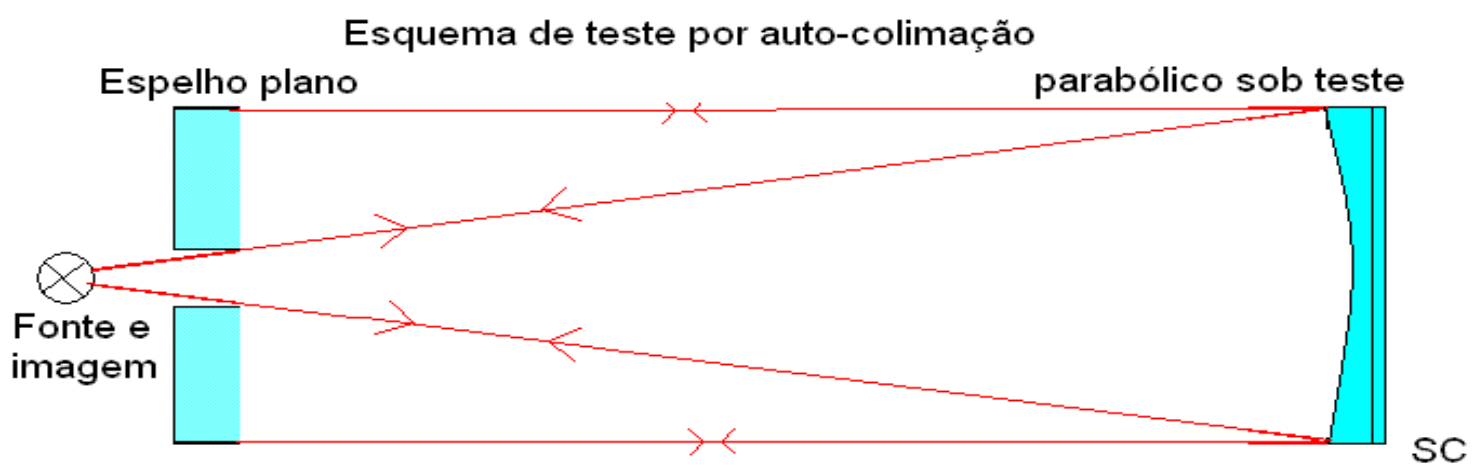

Fig. A-V.7: Caminhos ópticos do teste, mostrando a dupla incidência no espelho testado.

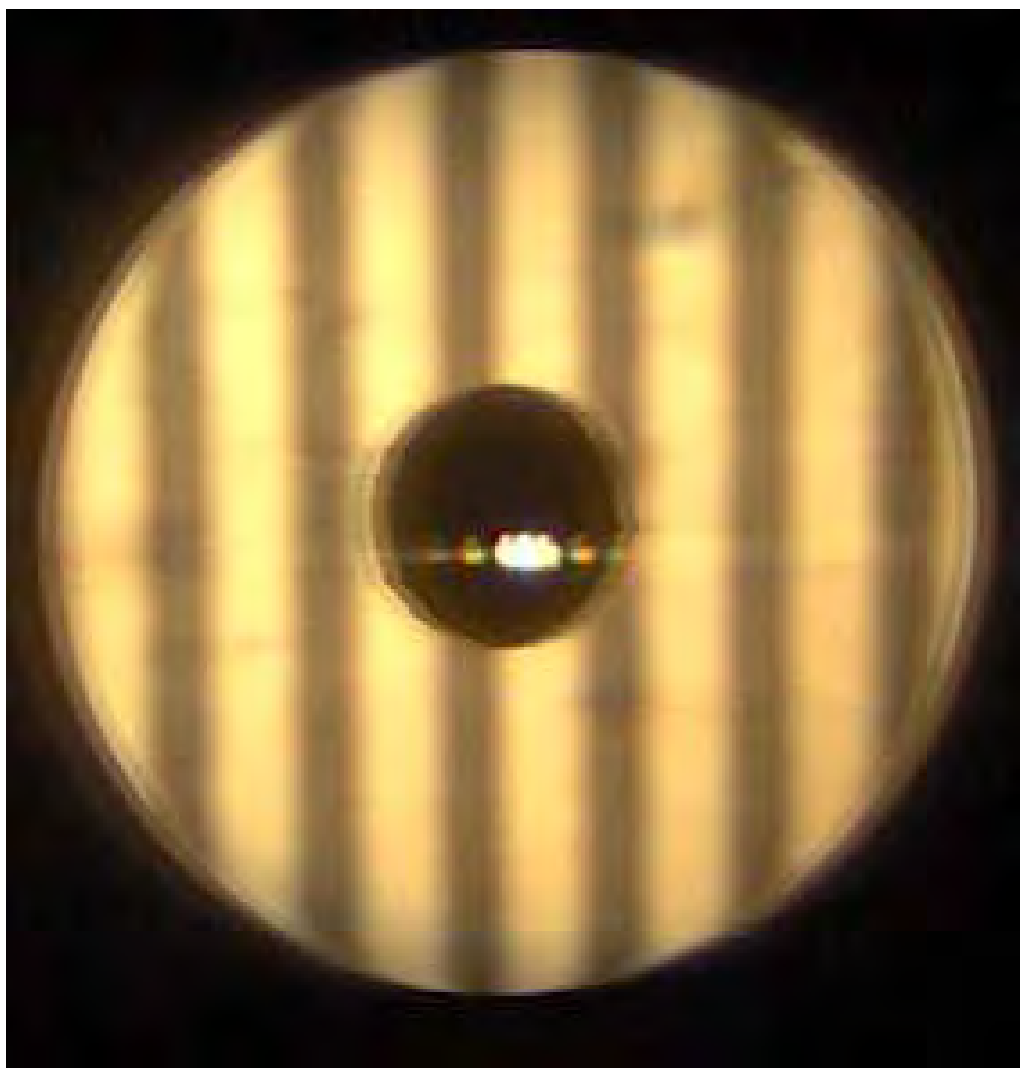

Fig. A-V.8: Imagem de teste de Ronchi por auto-colimação (intra-focal).

Neste teste, as sombras de uma superfície parabólica devem se apresentar absolutamente retas, figura A-V.8.

O leve encurvamento das sombras visualizado nas bordas do espelho (fig. A-V.8) indicam que seu bordo está abatido (TDE). Como o espelho é posteriormente bi-seccionado e lapidado nas bordas, o pequeno defeito óptico desaparece.

Apenas a título de comparação, a figura A-V.9 a seguir traz a imagem deste mesmo teste feito em um espelho de um telescópio comercial, de grau de qualidade semi-profissional. Note-se que neste caso as sombras se apresentam ligeiramente curvadas.

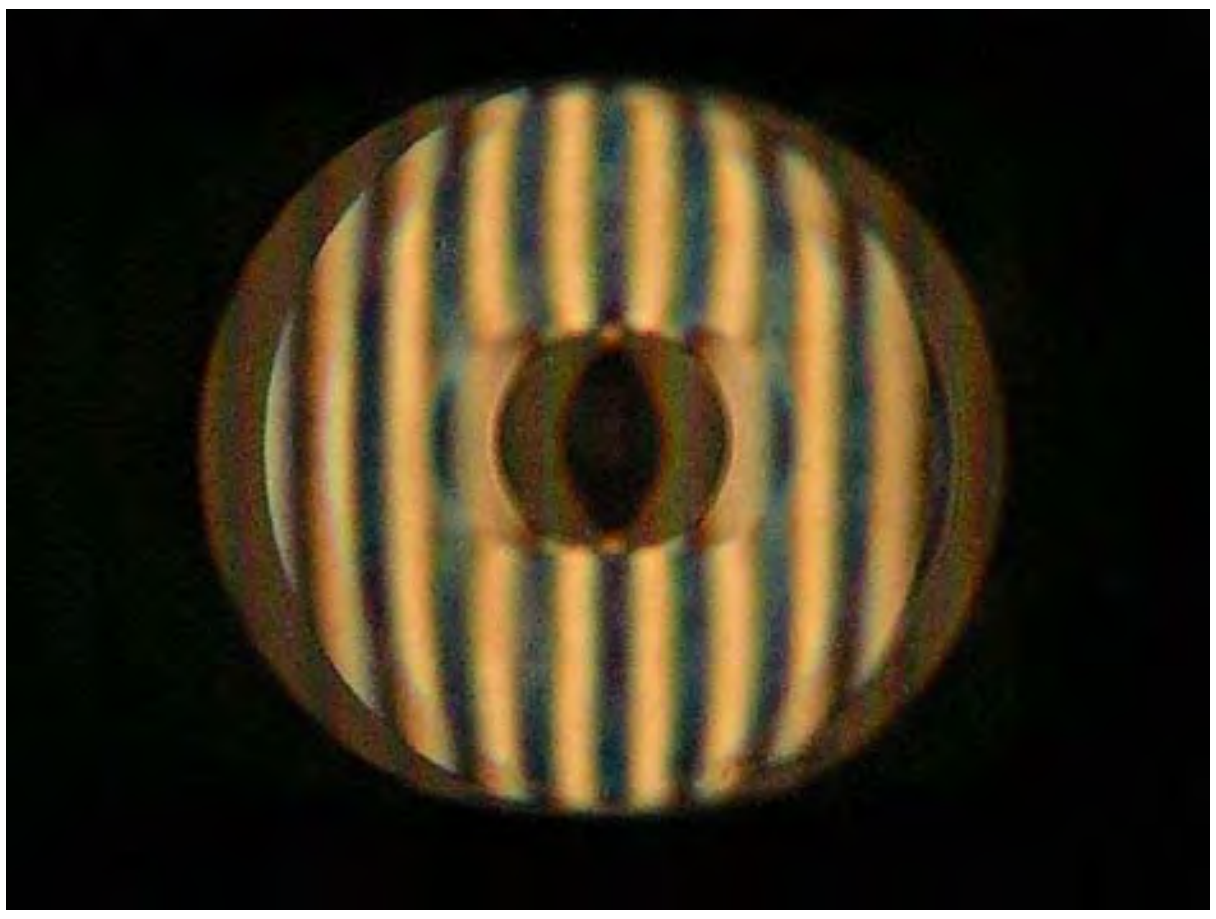

Fig. A-V.9: Imagem para referência qualitativa. Teste de Ronchi por auto-colimação (intra-focal) em um Takahashi.

Os testes a seguir foram feitos com o espelho cortado e aluminizado.

## *Teste de Ronchi por auto-colimação (qualitativo)*

As condições de contorno são: posição intra-focal e grade de 5 linhas por milímetros.

Os testes foram repetidos para 3 posições distintas em relação à linha de corte, em cada hemi-espelho. Em todas as posições, as sombras se apresentem absolutamente retas, verificando a condição de uma superfície de excelente qualidade óptica, figuras: A-V.10, A-V.11 e A-V.12.

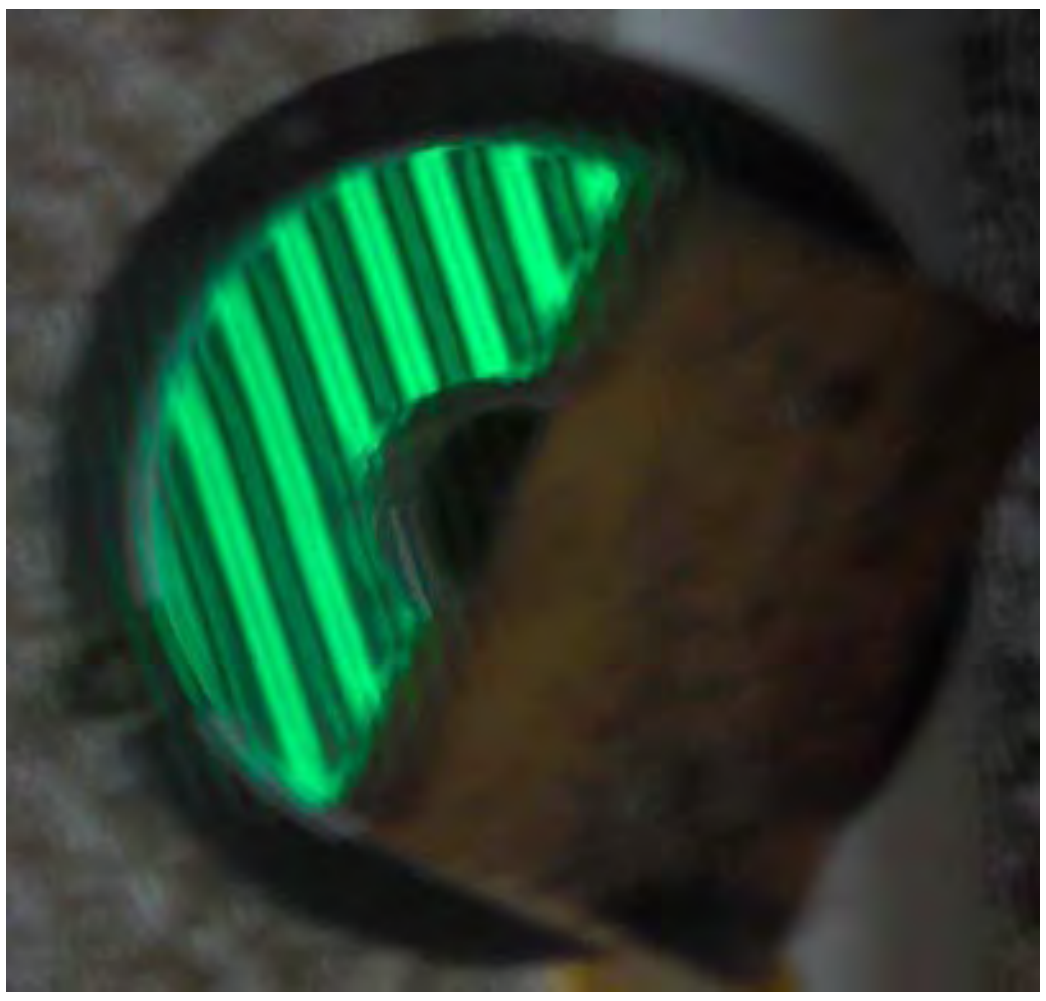

Fig. A-V.10: Sombras de Ronchi em posição intra-focal a 45° em relação ao corte.

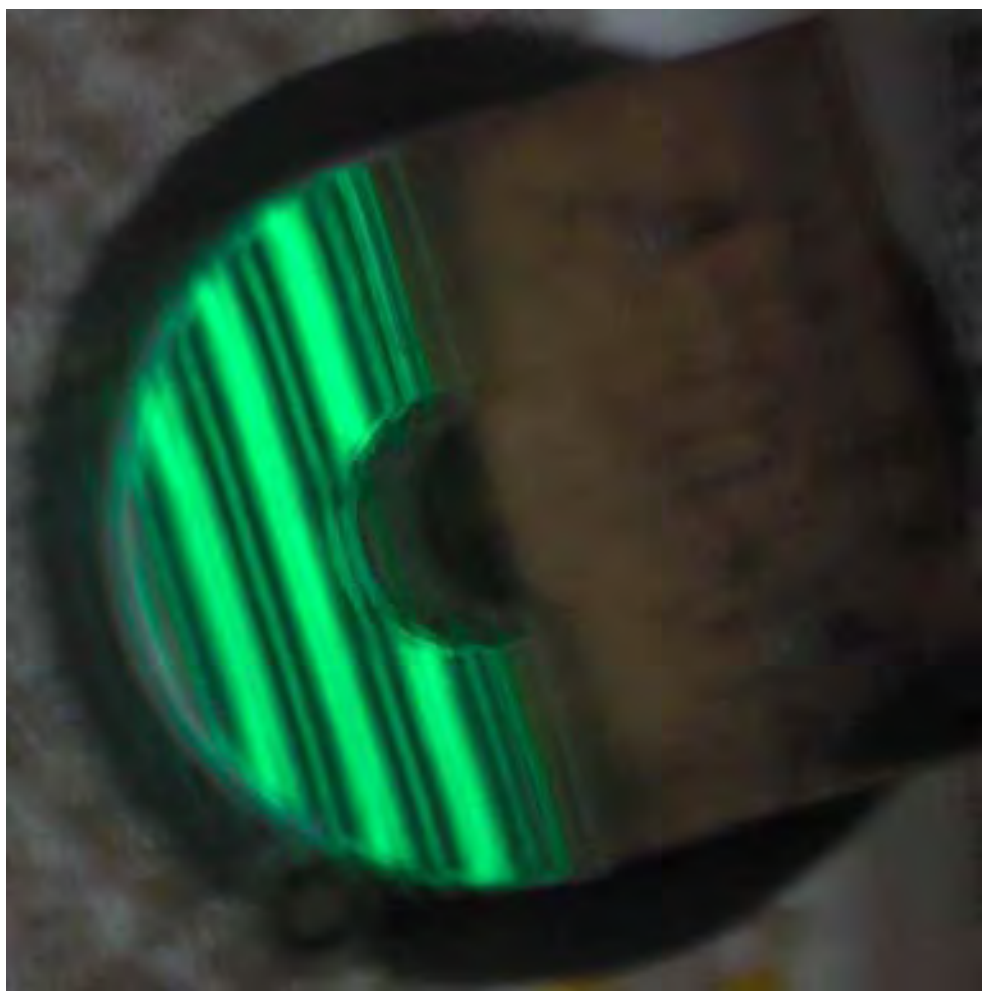

Fig. A-V.11: Sombras de Ronchi em posição intra-focal paralelas em relação ao corte.

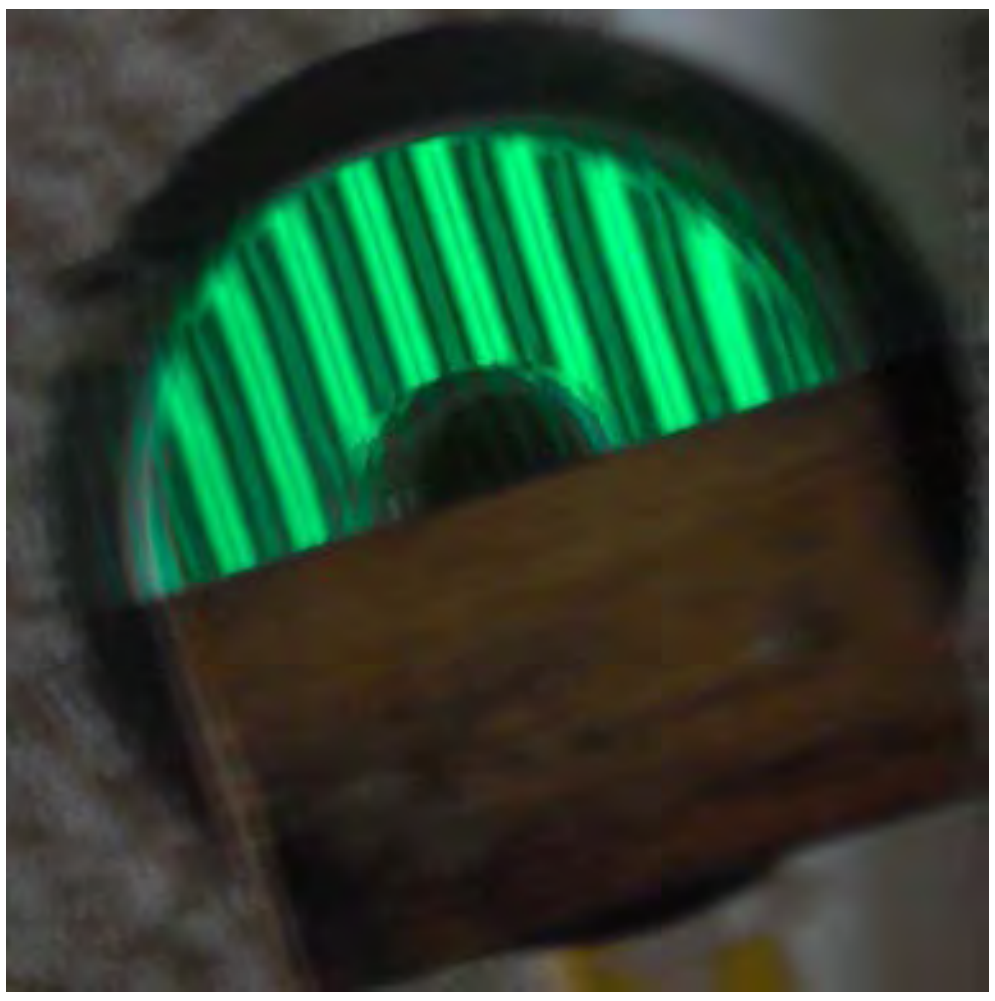

Fig. A-V.12: Sombras de Ronchi em posição intra-focal perpendiculares em relação ao corte.

Imagens idênticas foram obtidas com o outro hemi-espelho e por isso não são mostradas.

## *Teste de Wire por auto-colimação (quantitativo)*

Neste teste é utilizado um fio delgado (fio de seda de 18~25 μm de espessura) e uma fonte fendiforme monocromática. (18 μm). A figura de difração do fio pode ser vista como uma sombra central ladeada por pequenas outras sombras. A segunda sombra está aproximadamente a 3λ/4 de distância da sombra central (Texereau, 1961).

Este teste é quantitativo sobre a região em que a sombra central se encontra figura A-V.13.

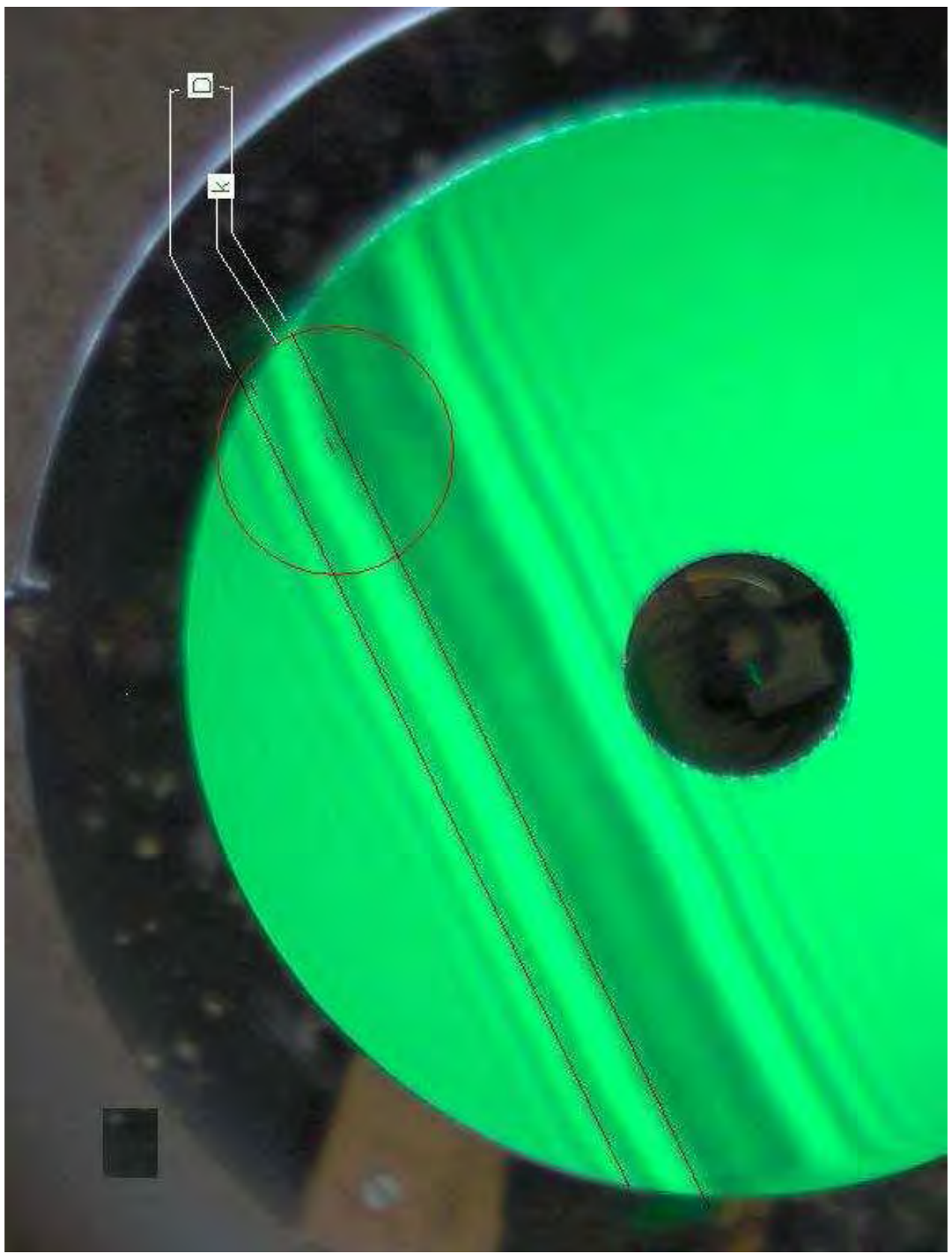

Fig. A-V.13: Exemplo de imagem do teste de wire em um espelho com defeito em sua superfície.

Na figura A-V.13, chamando de "D" a distância entre a sombra central e a 1ª franja escura, e de "k" a distância máxima do desvio que afaste da linha reta, a relação k/D nos dá diretamente o valor de desvio em fração de franja. Multiplicando-se por 0,75, obtemos o valor do desvio em fração de comprimento de onda.

Como no teste de auto-colimação há dupla incidência no espelho testado (fig. A-V.7), temos que dividir esta relação por 2.

As imagens trazem o teste de wire feito num dos hemi-espelhos, figuras A-V.14 e A-V.15.

A distância *D* entre a sombra central e a 1ª sombra é de 17 *pixel*. Como visivelmente não há desvios, admitiremos $k = 1$ *pixel*. Sendo conservadores, usaremos 0,5 como fator multiplicativo de $\lambda$, ao invés de 0,375 (0,75/2).

Temos como defeito máximo na frente de onda de: $\frac{k}{D}(0{,}5\lambda) = \frac{1}{17}\frac{\lambda}{2} = \frac{\lambda}{34} \cong 0{,}03\lambda$

Que concorda com o resultado mostrado no gráfico da figura A-V.4.

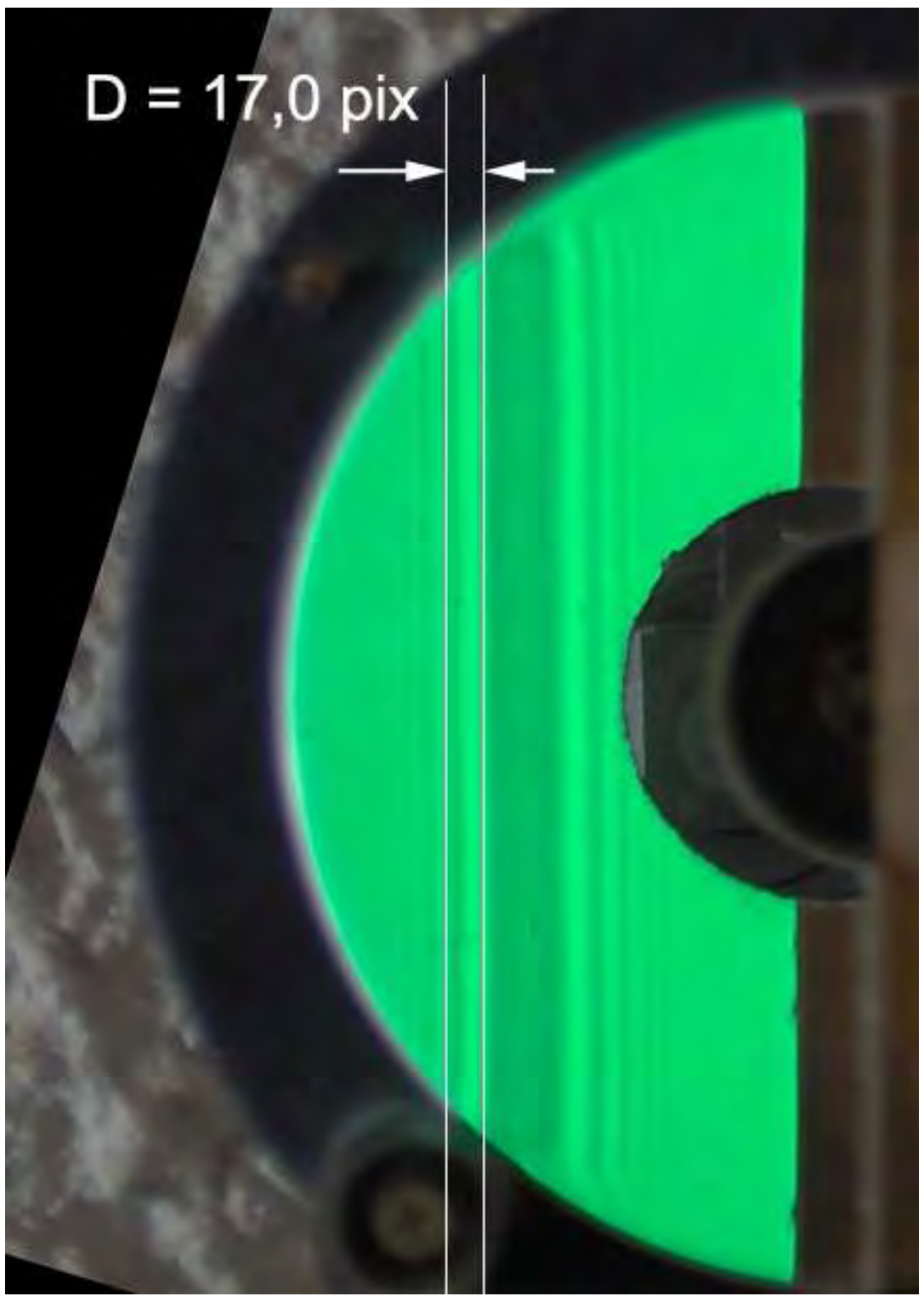

Fig. A-V.14: Teste de wire quantitativo em um dos hemi-espelhos.

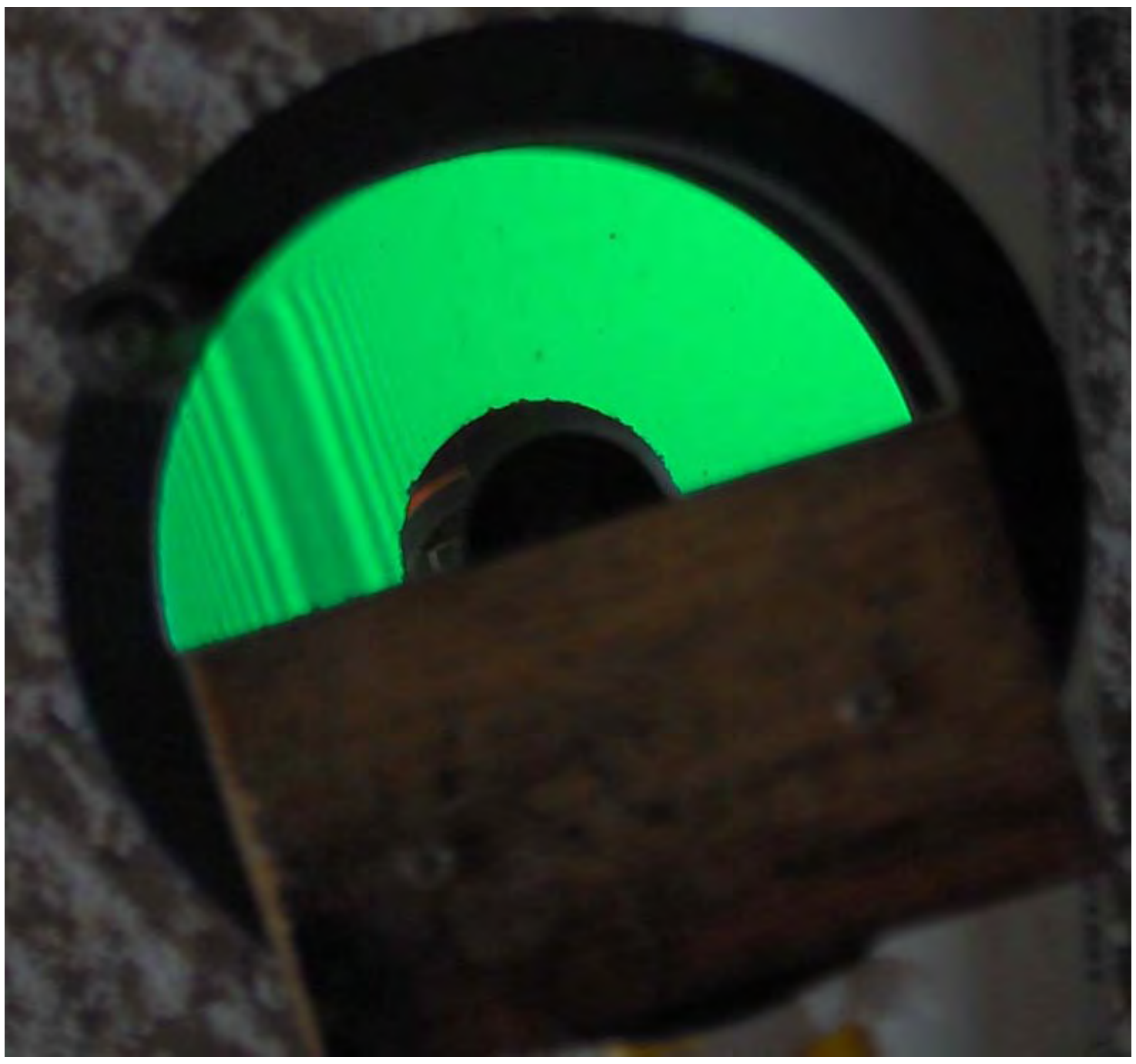

Fig. A-V.15: Teste de wire ortogonal ao corte, em um dos hemi-espelhos.

Imagens idênticas e, portanto, resultados idênticos foram obtidos com o outro hemi-espelho e por isso as imagens não são mostradas aqui.

# Apêndice VI - Especificações técnicas da célula do espelho heliométrico de CCZ-HS

Esta célula foi especialmente projetada para servir de suporte mecânico do espelho do Heliômetro.

O espelho fica perfeitamente acamado em seu interior. Apoios estáticos e dinâmicos, atuam sobre ele lateralmente, na base e no anel de contenção, de forma que o conjunto fica completamente estabilizado, figuras A-VI.1 a A-VI.5.

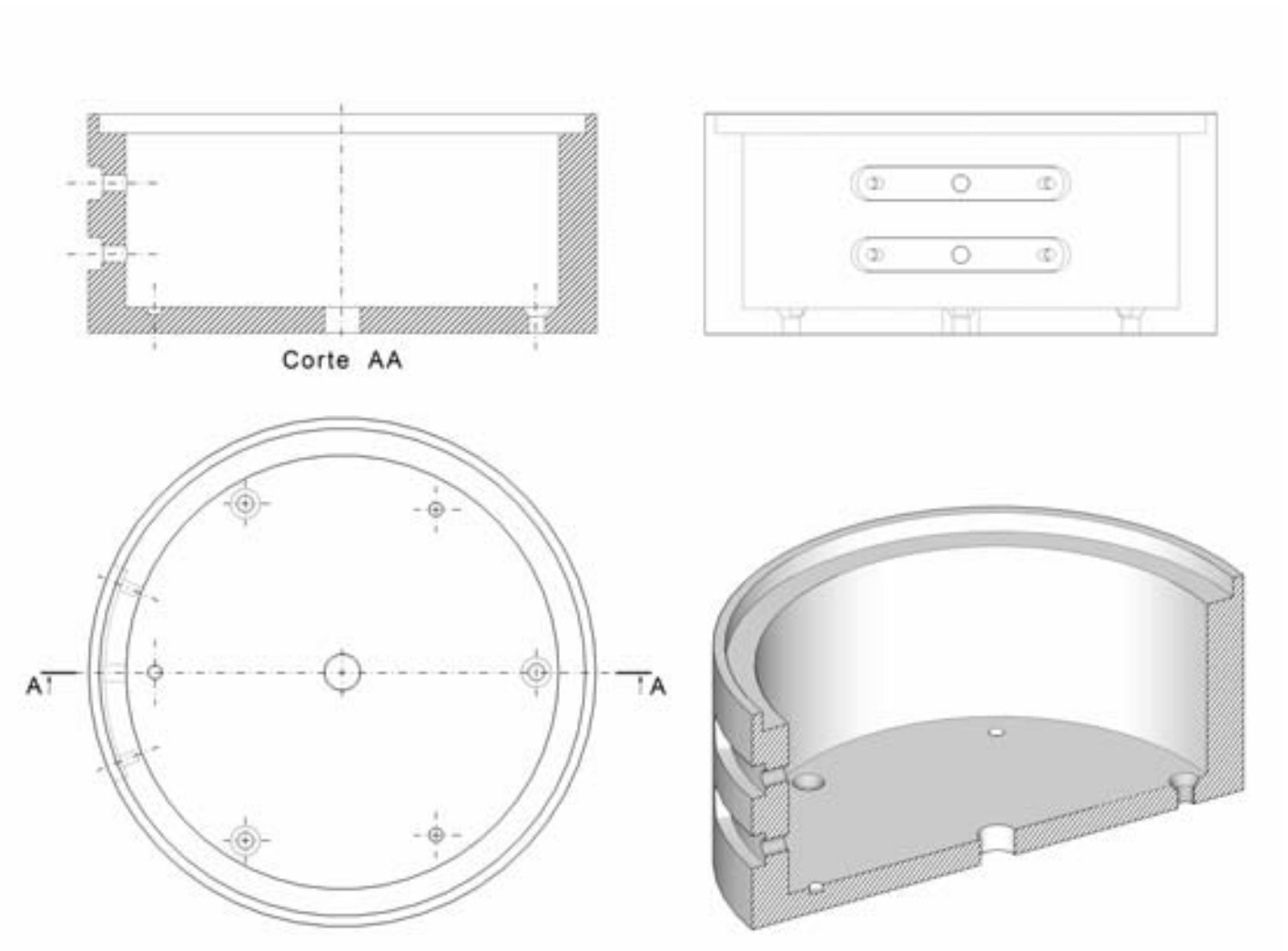

Fig. A-VI.1: Projeto da célula, com corte AA, que passa pelo plano de referência do Heliômetro.

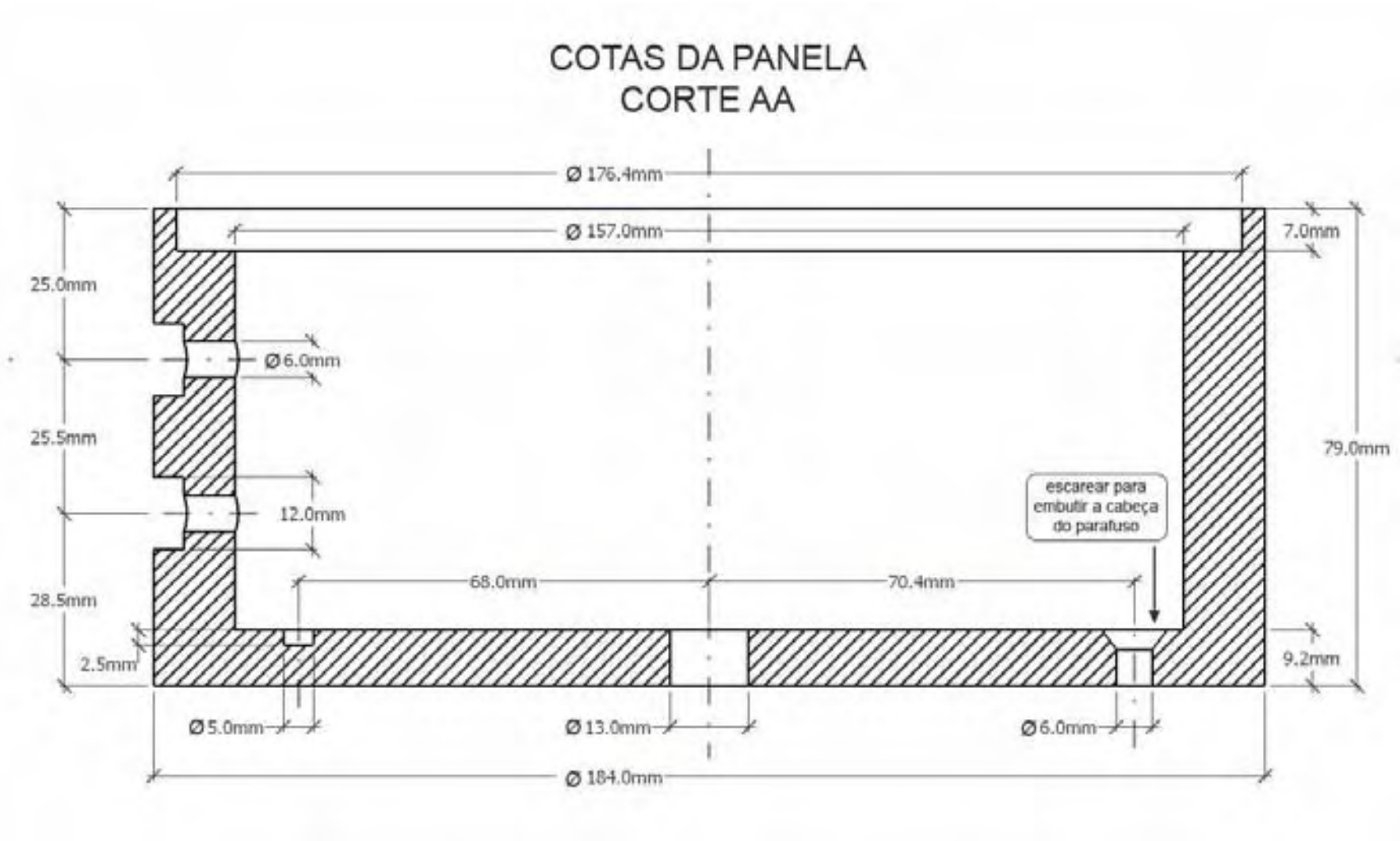

Fig. A-VI.2: Cotas referentes ao corte AA.

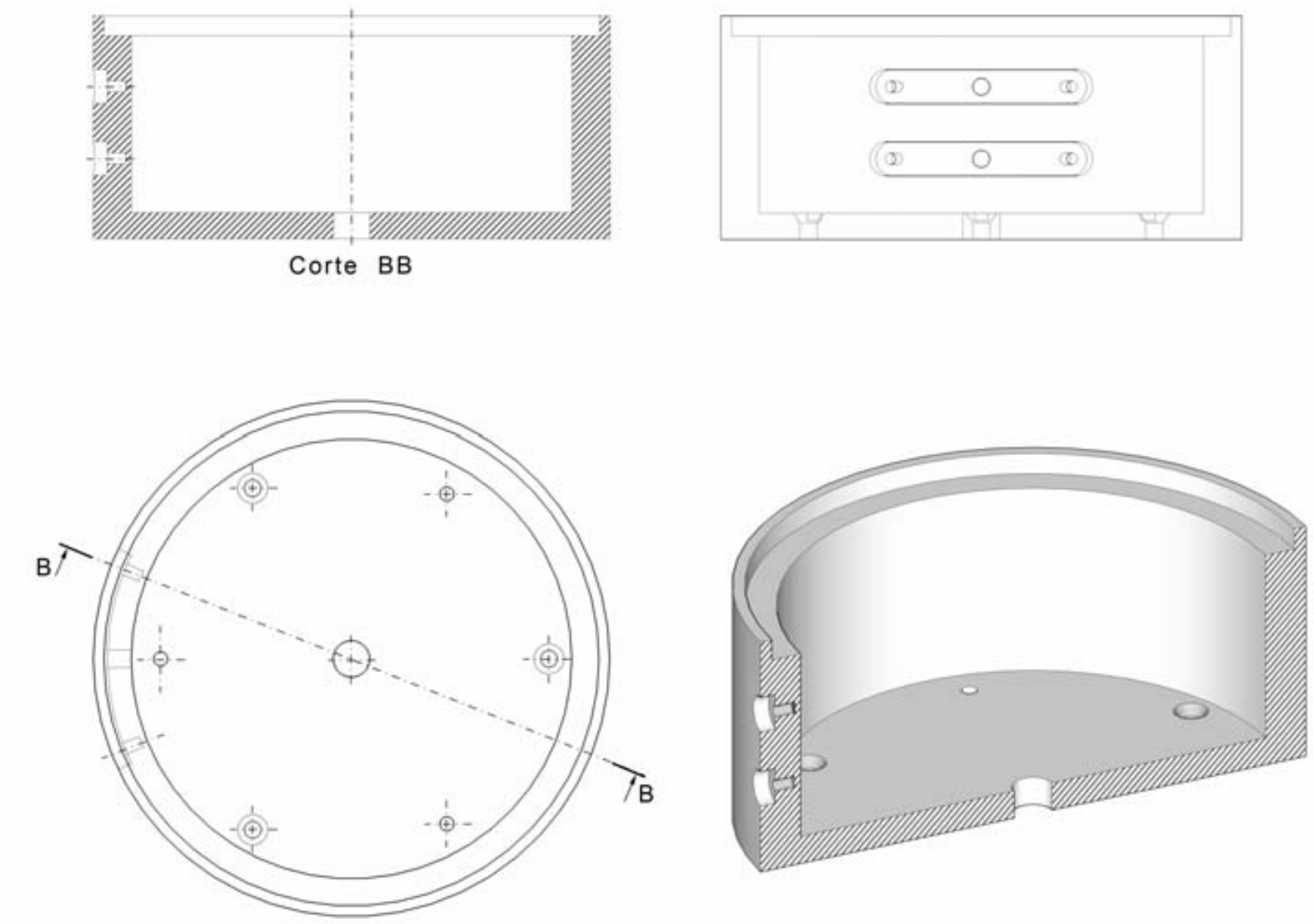

Fig. A-VI.3: Projeto da célula, com corte BB, que passa pelo plano dos parafusos da placa metálica.

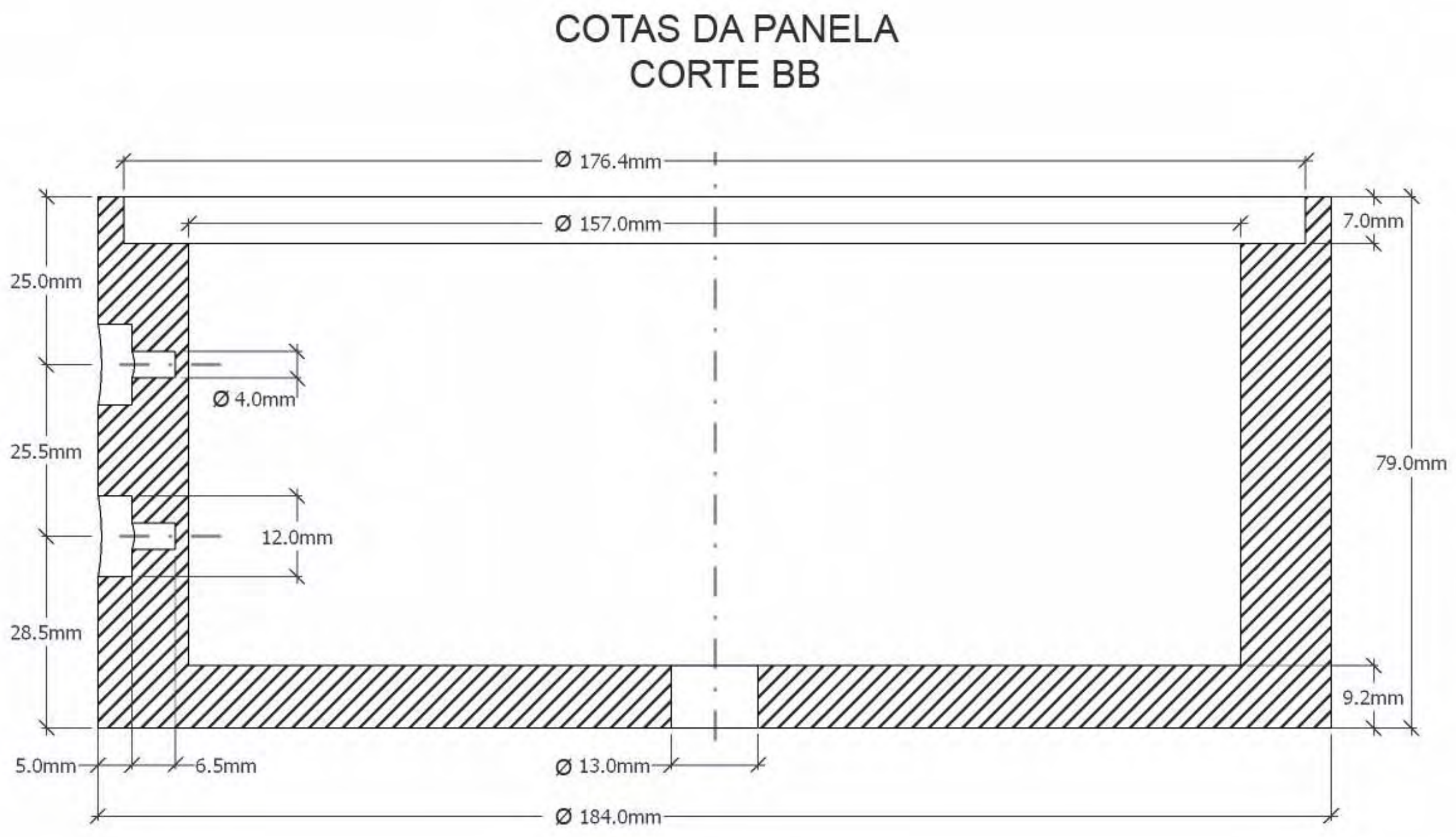

Fig. A-VI.4: Cotas referentes ao corte acima.

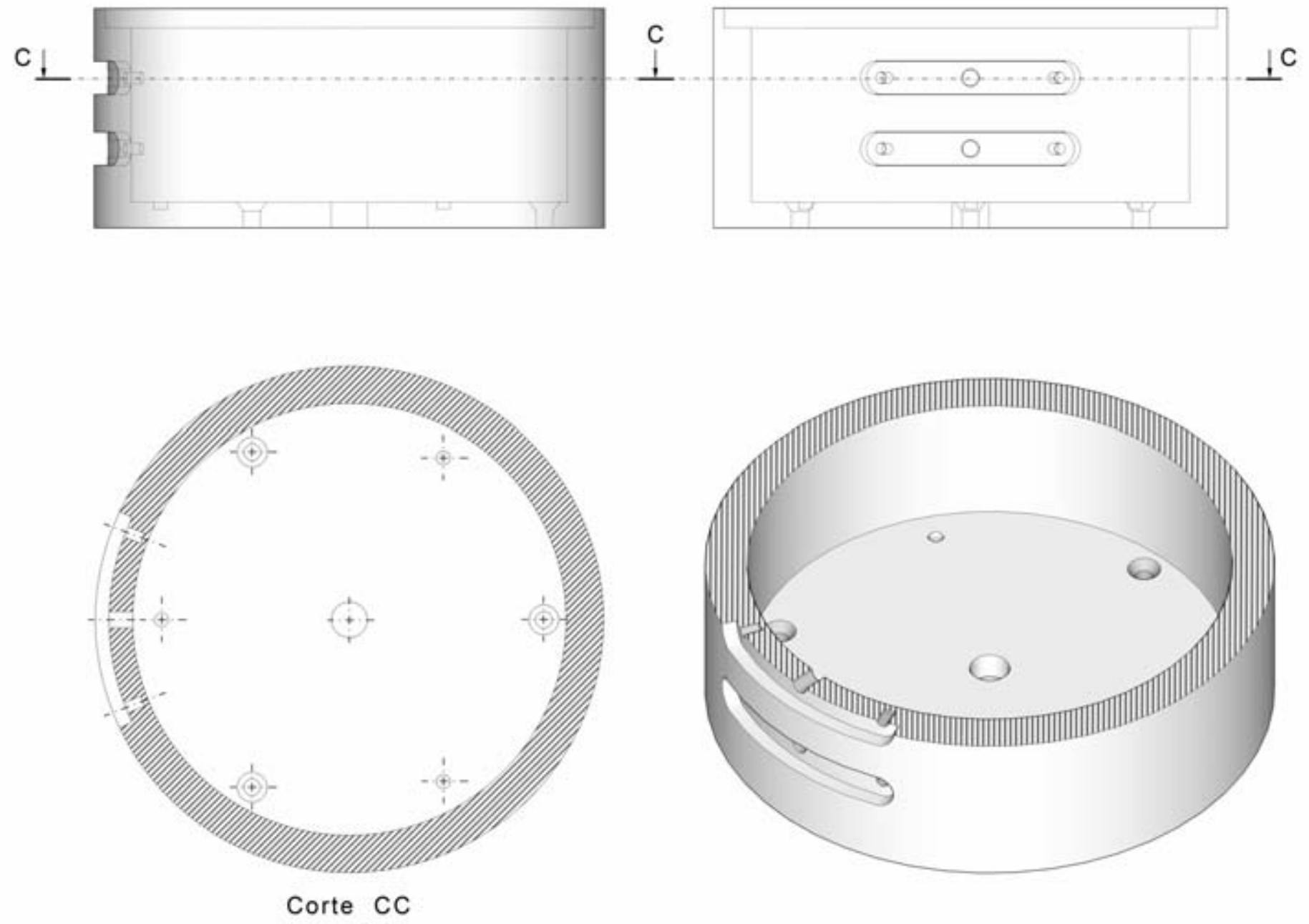

Fig. A-VI.5: Projeto da célula, com corte CC, que passa pelo plano paralelo aos parafusos da placa metálica.

O desenho técnico a seguir foi executado, conforme as especificações do Projeto, pela CENIC, empresa nacional que atua em projetos aeroespaciais, responsável pela fabricação do tubo de fibra carbono do Heliômetro, figura A-VI.6.

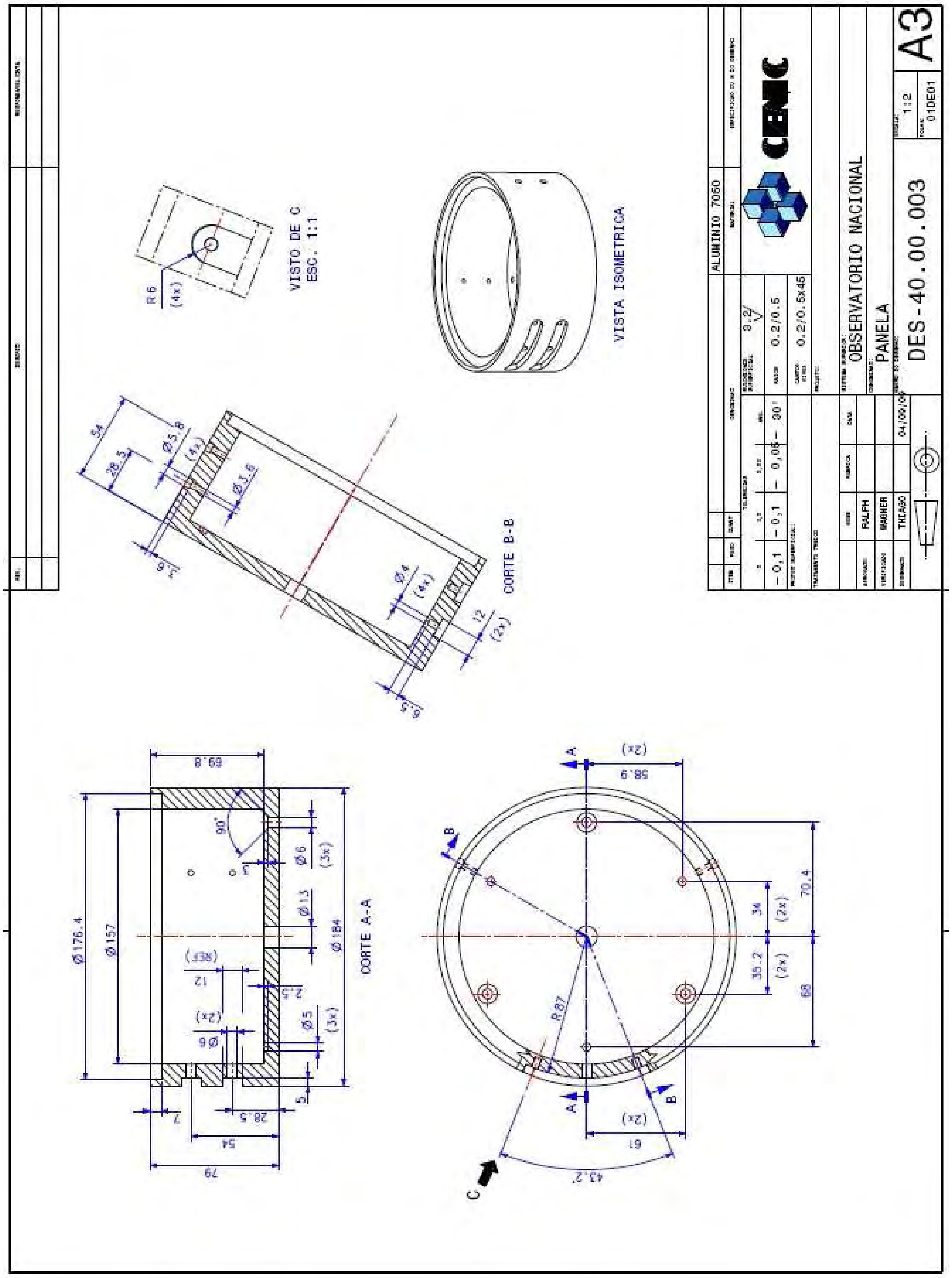

Fig. A-VI.6: Desenho técnico da célula preparado pela CENIC.

# Apêndice VII - Projeto da Tampa Forte e da Pupila da célula suporte do espelho heliométrico

## *Tampa forte*

O projeto começa com o desenho de um disco de 176 mm de diâmetro. Em seguida as circunferências que representam a borda externa de um dos hemi-espelhos e da abertura central são desenhadas. Seus raios foram reduzidos de 1 mm, para formar uma margem.

No passo seguinte as hemi-circunferências são rebatida verticalmente, de forma que a área útil do espelho forme uma figura com simetria, figura A-VII.1, à esquerda.

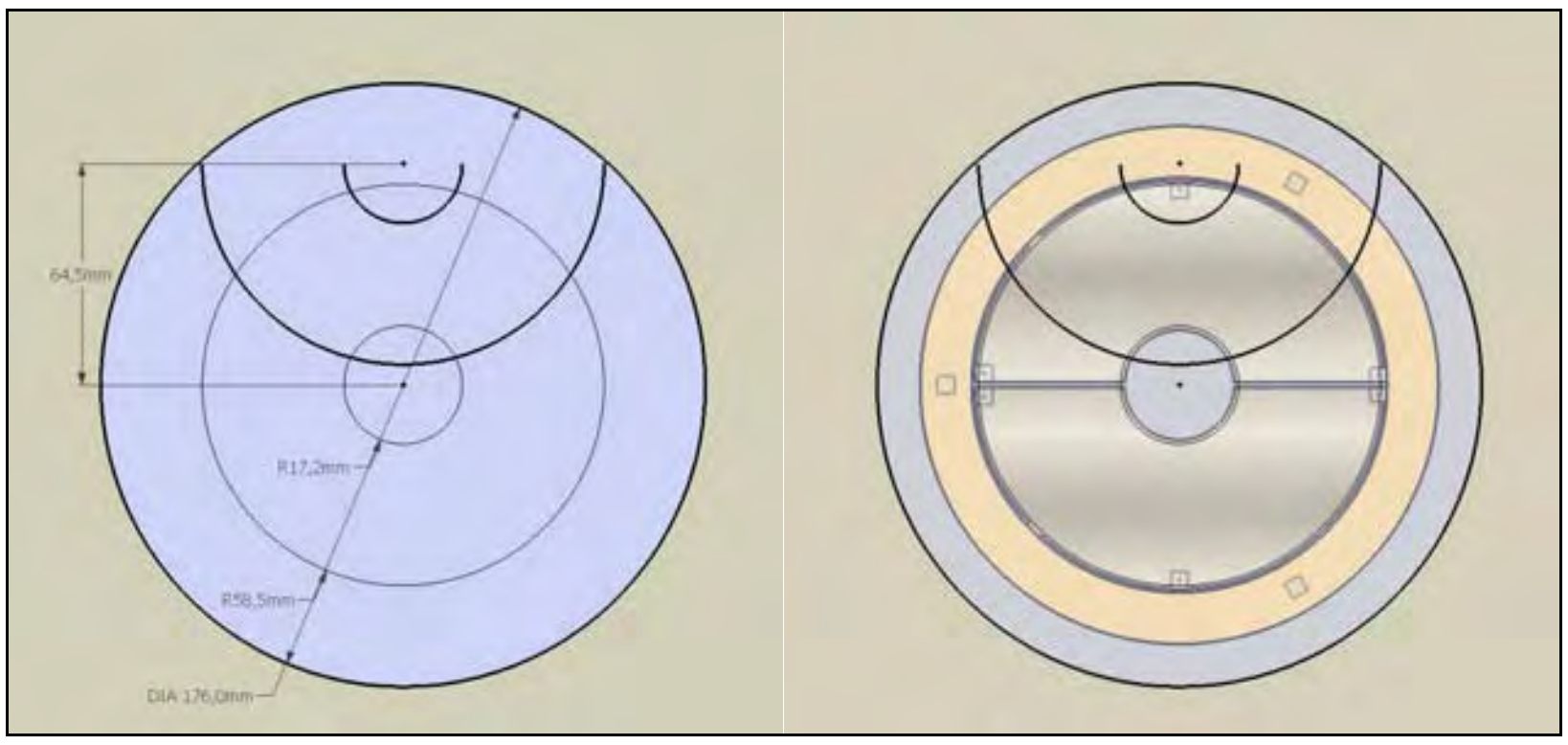

Fig. A-VII.1: Vista superior da tampa do berço do espelho do Heliômetro. A imagem da esquerda traz as cotas e a da direita, a representação do espelho heliométrico por baixo da tampa.

Somente a área referente à intercessão das linhas foi selecionada (fig. A-VII.2). Esta área representa a abertura feita na tampa, figura A-VII.3.

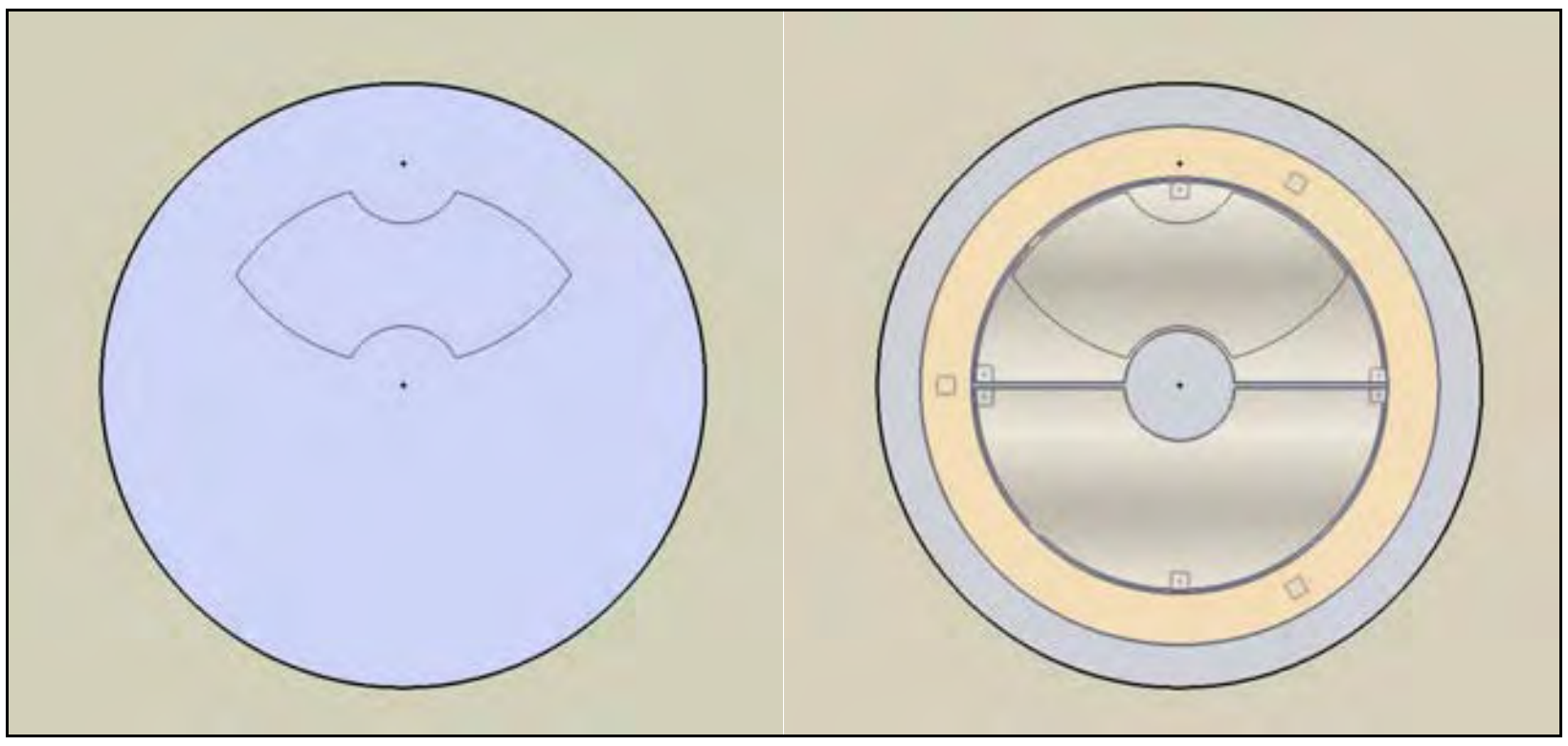

Fig. A-VII.2: À esquerda, a área de intercessão entre as circunferências e à direita, a representação do espelho heliométrico por baixo da tampa.

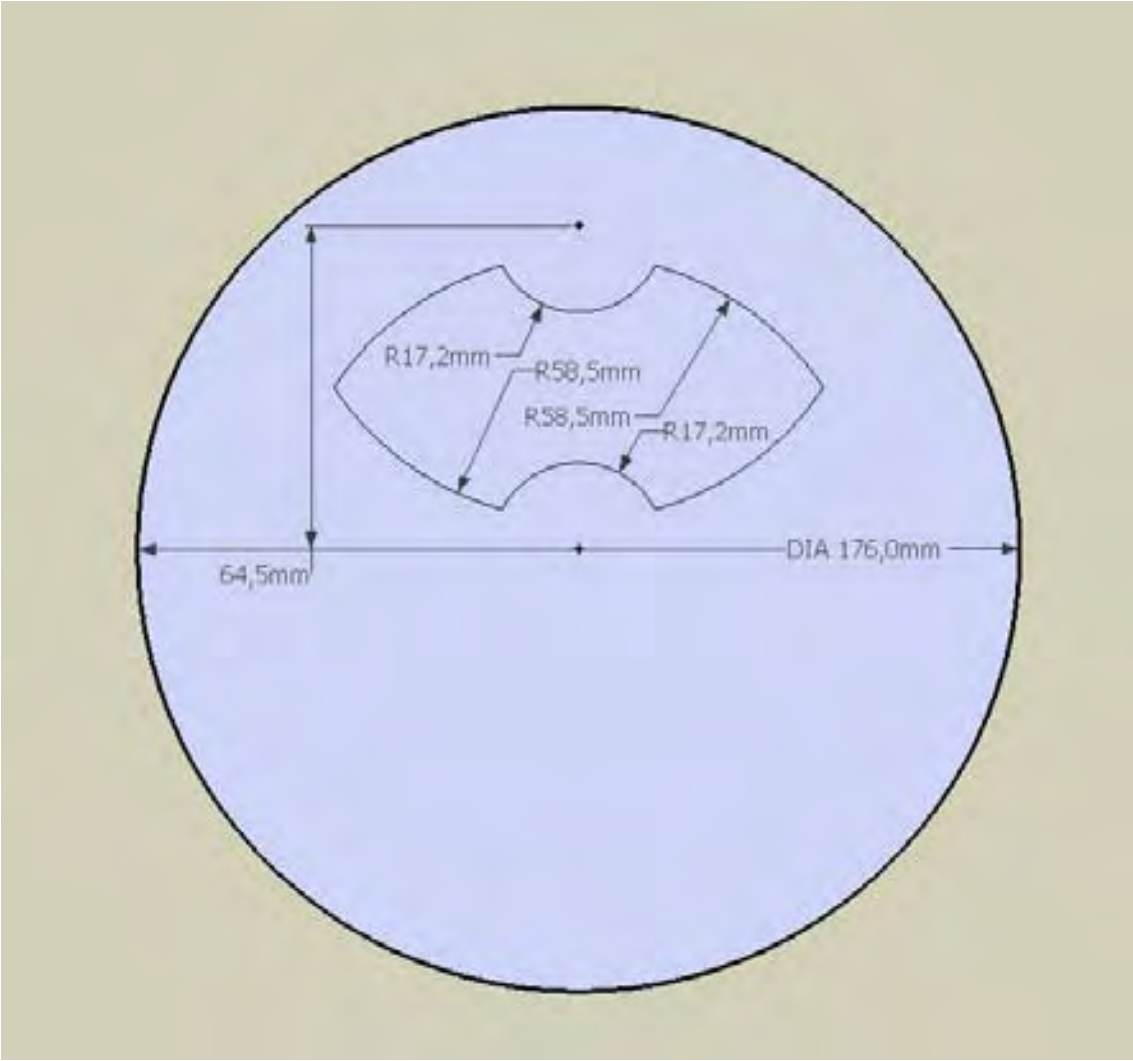

Fig. A-VII.3: Cotas para a abertura da tampa.

Em seguida o desenho desta área foi rebatido para o outro hemisfério, formando aberturas iguais com simetria em relação a seus eixos de referência: o corte do espelho e a direção ortogonal a ele, figura A-VII.4.

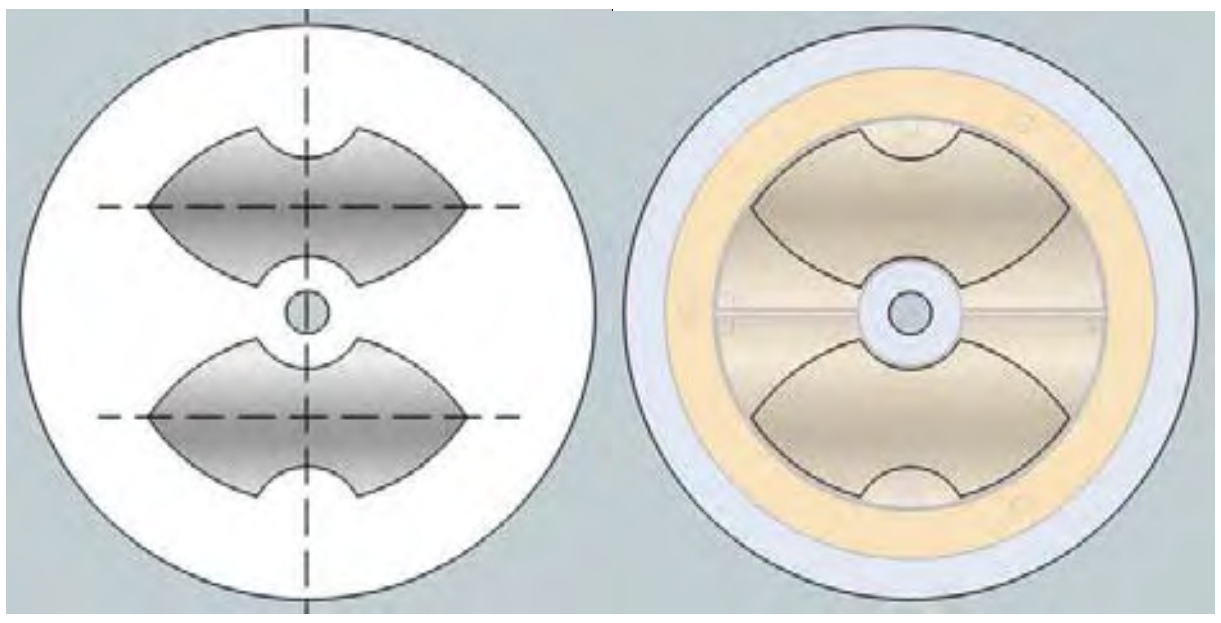

Fig. A-VII.4: À esquerda, as aberturas com a maior área com simetria da Tampa Forte dos espelhos heliométricos principal e de comparação. À direita, a representação do espelho heliométrico por baixo da tampa.

Finalmente, uma abertura de 13 mm de diâmetro é feita no centro da peça para o encaixe do suporte do cilindro de CCZ-HS, que configura o "Sol Padrão" para os testes, figura A-VII.5.

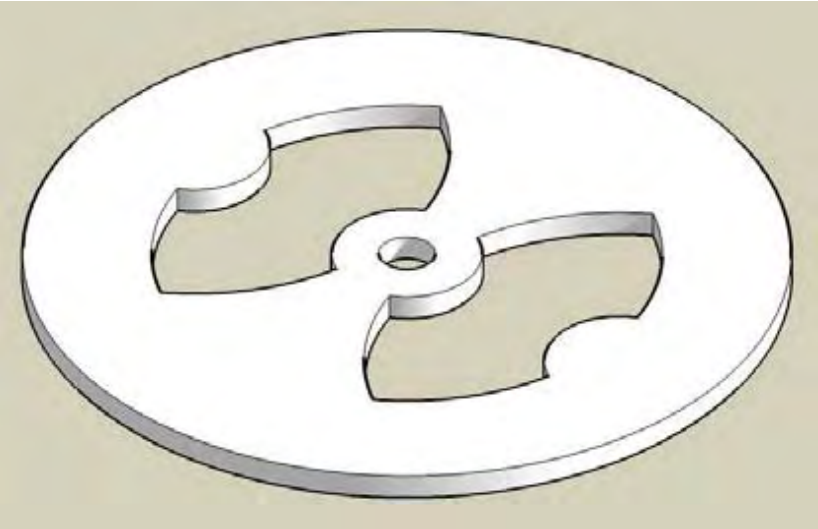

Fig. A-VII.5: Desenho final do projeto da Tampa Forte, com 6 mm de espessura e abertura para o suporte do Sol Padrão.

## *Pupila*

No teste de comparação, o feixe paralelo de luz, de cada hemi-espelho, retorna ao espelho principal com um deslocamento, na mesma direção do corte, de pouco mais de 2 mm. Como a pupila servirá somente à célula do espelho principal, levou-se em conta este deslocamento para o desenho final da peça. O perímetro da pupila é interno ao da Tampa Forte, com uma distância mínima de 1 mm. Levando todos estes fatores em consideração, o deslocamento, em ambos os sentidos ao longo do corte do espelho é de 5,7 mm.

O desenho final da abertura da Tampa Forte foi a base do desenho inicial da abertura da Pupila, com um aumento de 1 mm no valor dos raios dos círculos menores e posterior deslocamento da imagem no valor supracitado, figuras A-VII.6 e A-VII.7.

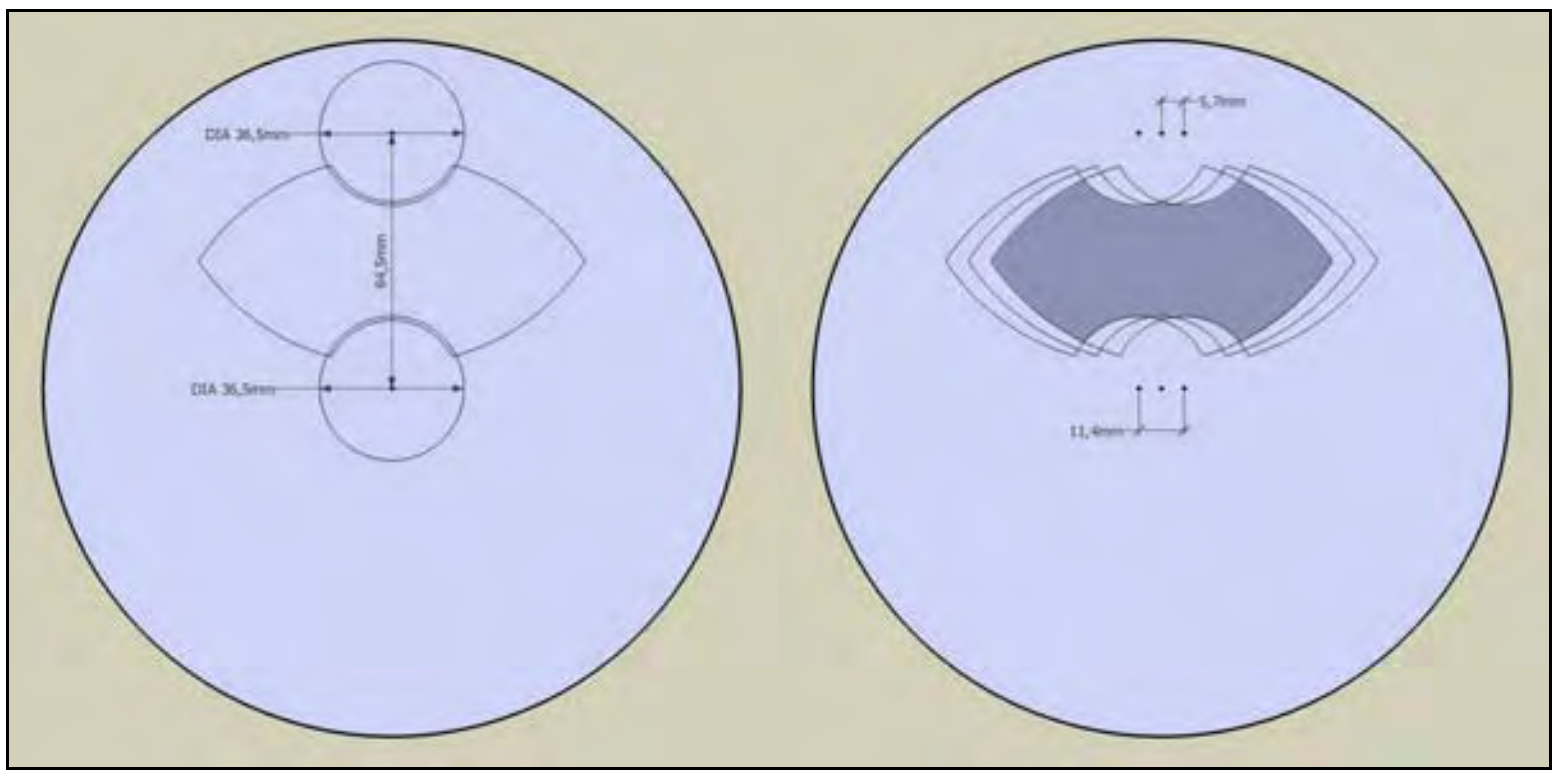

Fig. A-VII.6: À esquerda, os primeiros passos para o projeto da pupila. À direita, o deslocamento da imagem e a utilização apenas da região de intercessão (área mais escura).

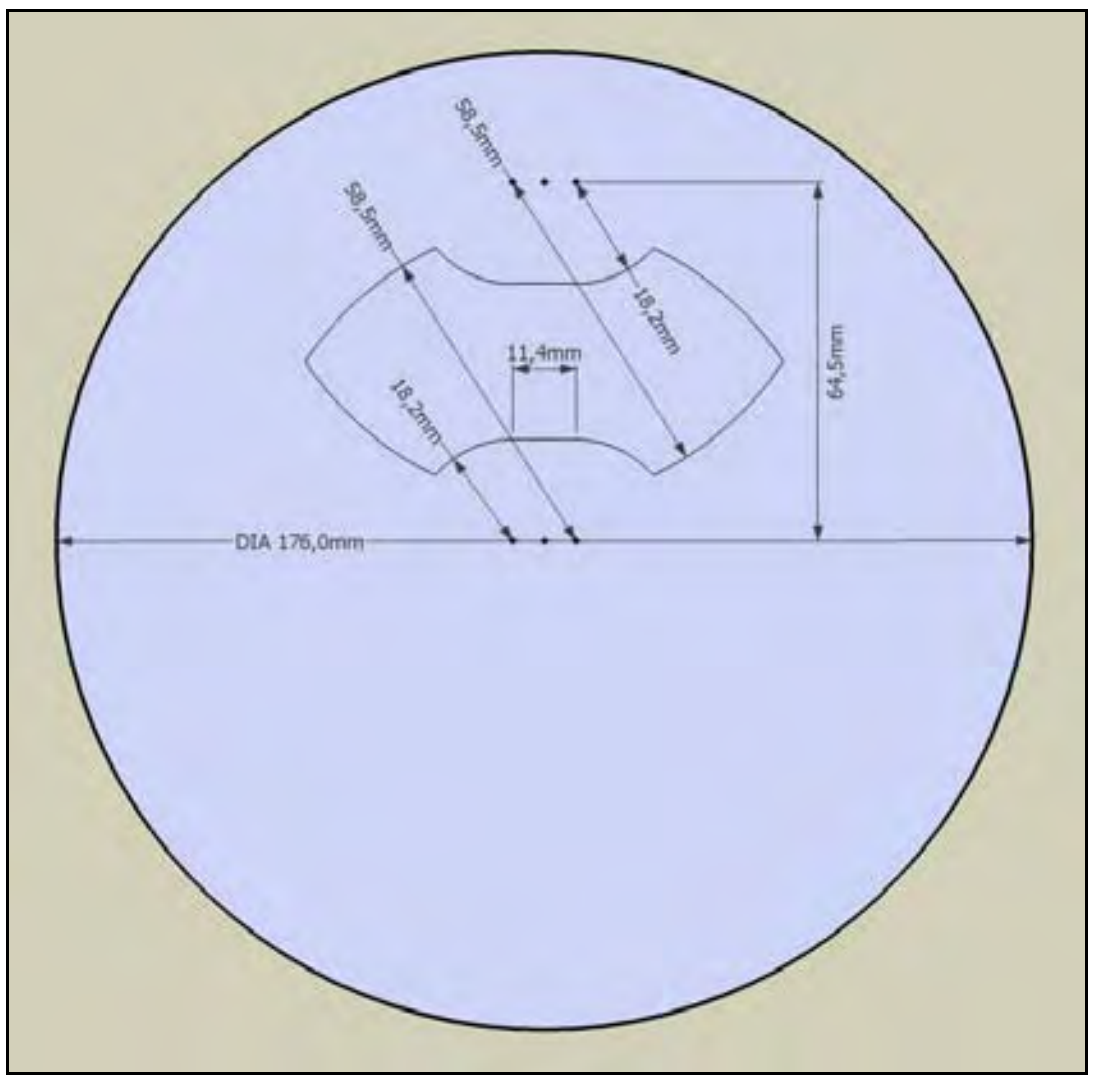

Fig. A-VII.7: Cotas para a abertura da pupila.

Como no projeto anterior, o desenho desta área é rebatido para o outro hemisfério, formando aberturas iguais e com simetria em relação a seus eixos de referência: o corte do espelho e a direção ortogonal a ele, figura A-VII.8.

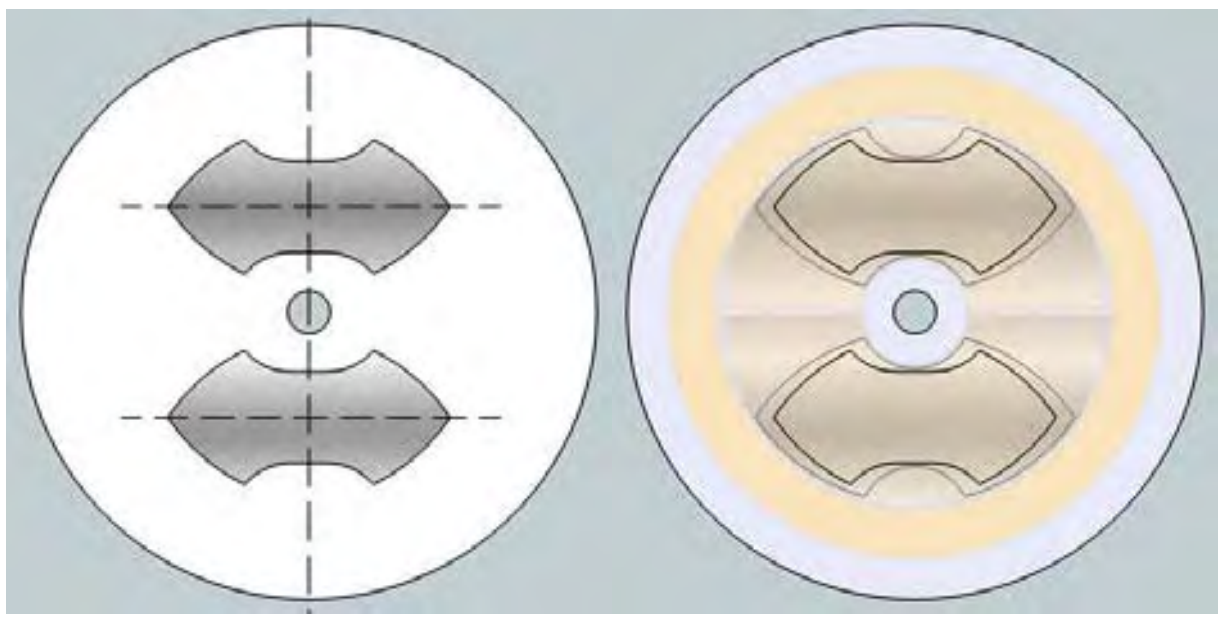

Fig. A-VII.8: À esquerda, os eixos de simetria e as aberturas da Pupila. À direita, a representação do espelho heliométrico por baixo da Pupila.

No final, a abertura de 13 mm de diâmetro para passar o suporte do cilindro de CCZ-HS que serve de "Sol Padrão" para os testes, figura A-VII.9.

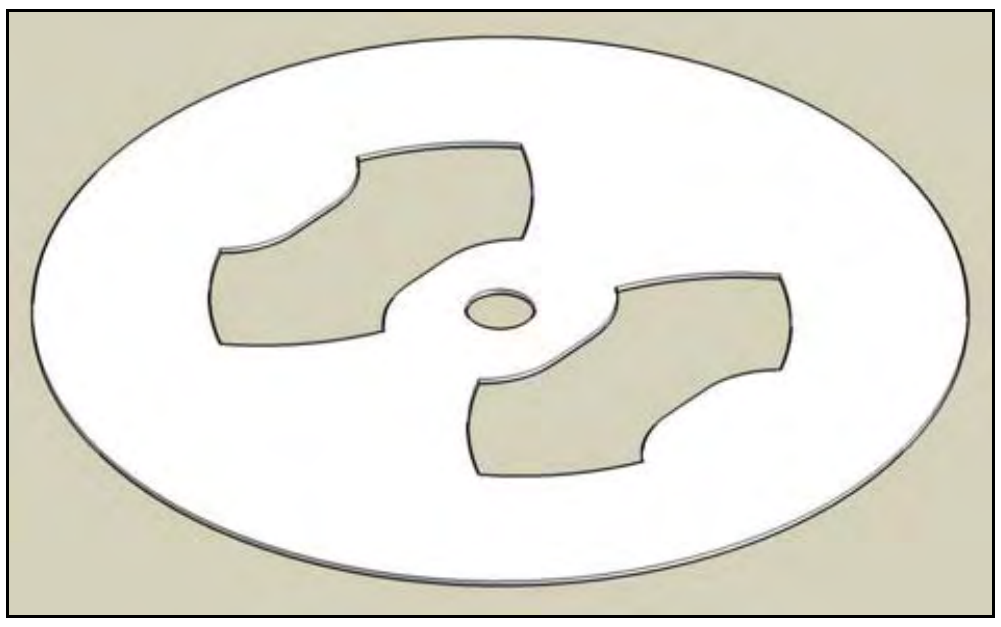

**Fig. A-VII.9: Aparência final da pupila, com 1 mm de espessura e abertura para o suporte do Sol Padrão.**

## *Desenhos técnicos*

Os desenhos técnicos das figuras A-VII.10 e A-VII.11 foram executados, conforme as especificações do Projeto, pela CENIC, empresa nacional que atua em projetos aeroespaciais, responsável pela fabricação do tubo de fibra carbono do Heliômetro.

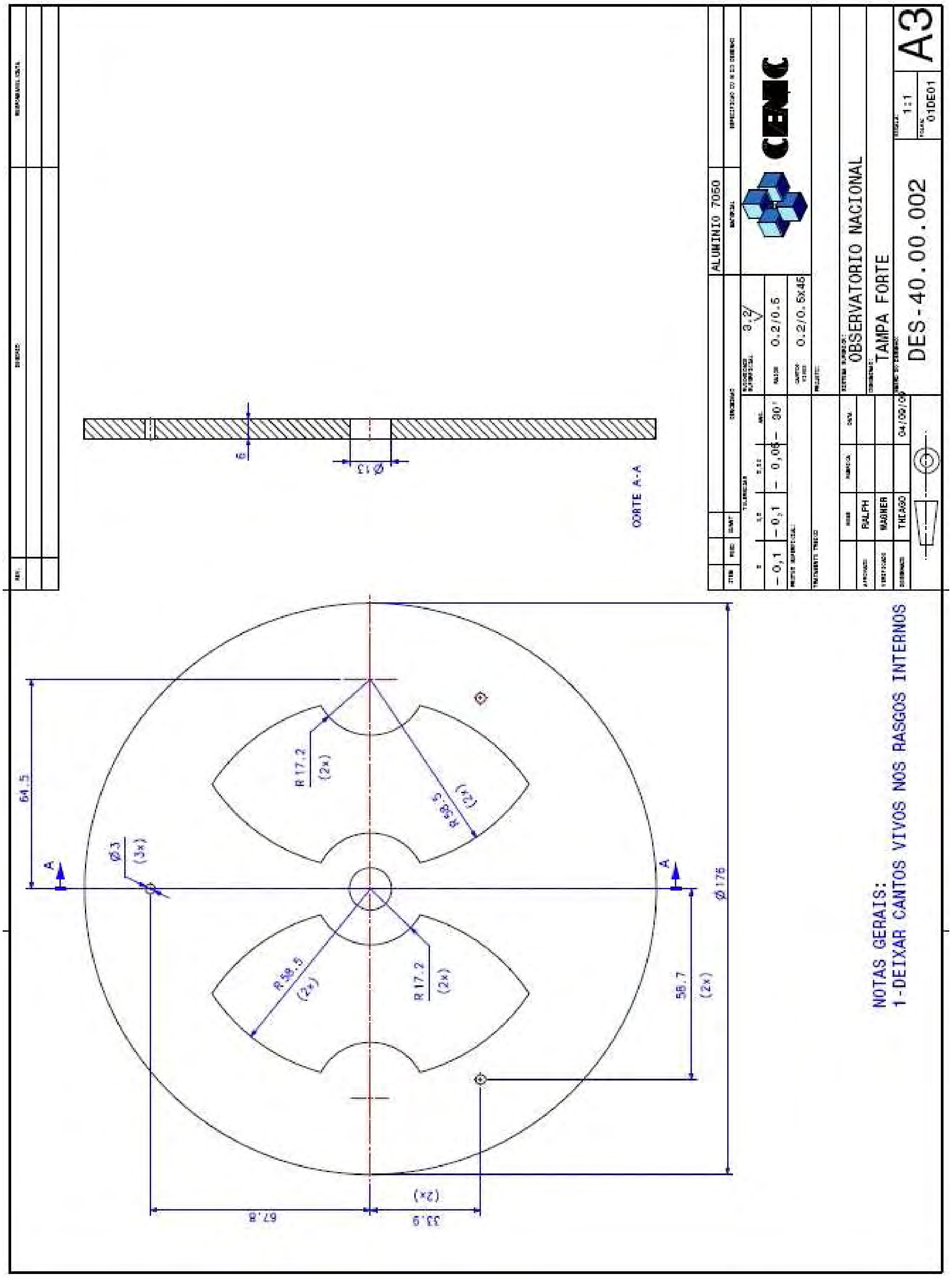

Fig. A-VII.10: Desenho técnico da tampa-forte da célula do espelho heliométrico.

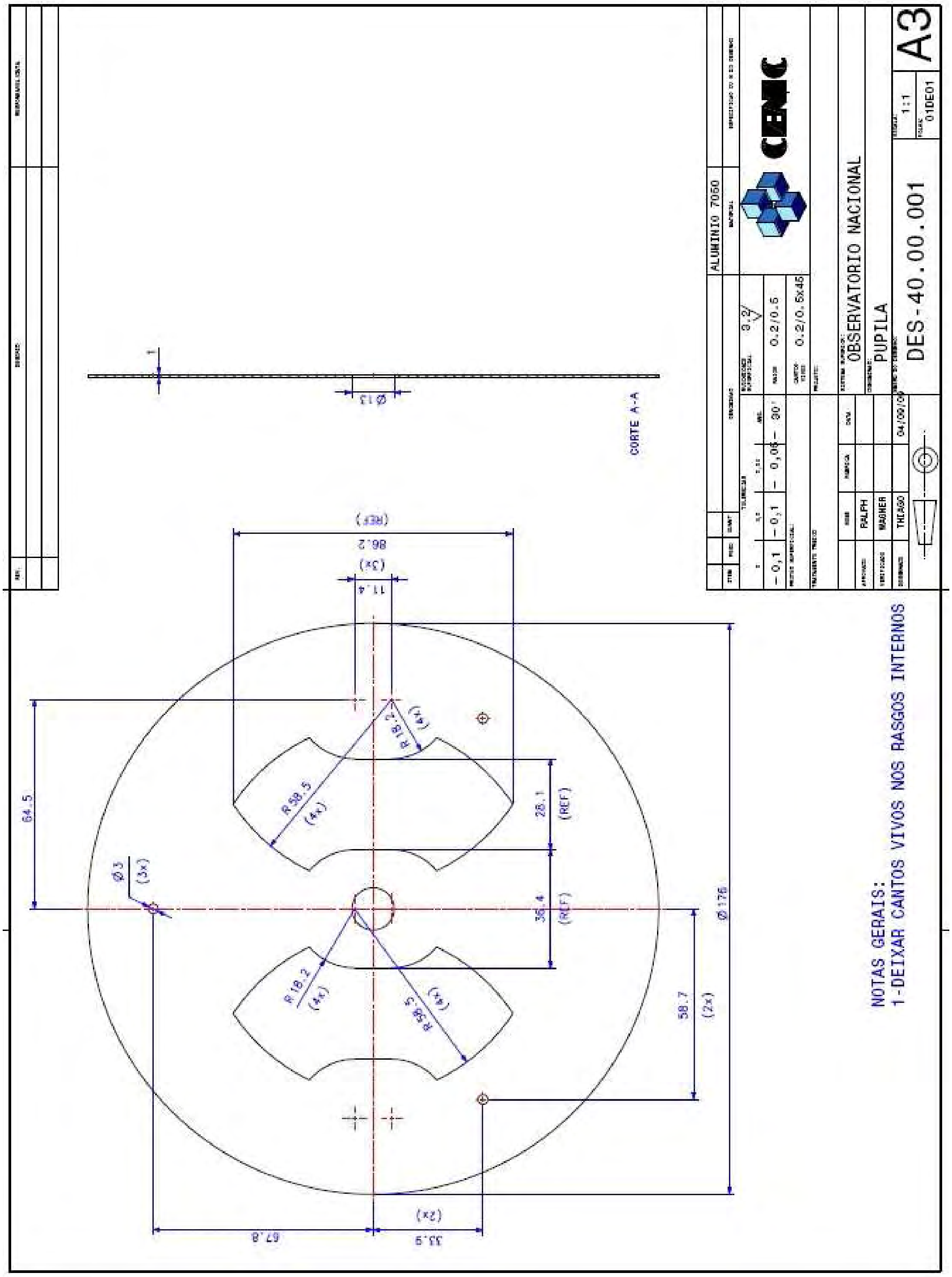

Fig. A-VII.11: Desenho técnico da pupila da célula do espelho heliométrico.

# Índice das figuras

# Índice das tabelas